\documentclass[11pt,a4paper,english]{scrbook}
\usepackage[T1]{fontenc}
\usepackage[latin9]{inputenc}
\usepackage{textcomp}
\usepackage{amsthm}
\usepackage{amsmath}
\usepackage{graphicx}
\usepackage{amssymb}
\usepackage{esint}
\usepackage{eso-pic}
\usepackage{fancyhdr}

\makeatletter

\newenvironment{abstract}%
{\cleardoublepage\null \vfill\begin{center}%
\bfseries \abstractname \end{center}}%
    {\vfill\null}

\numberwithin{equation}{section}

\def\maketitle{%
  \null
  \thispagestyle{empty}%
  \vfill
  \begin{center}\leavevmode
    \normalfont
    \fontseries{b}
    {\Huge \@title\par}%
    \vskip 1cm
    \fontseries{\seriesdefault}
    {\LARGE by}%
    \vskip 0.5cm
    \fontseries{b}
    {\huge \@author\par}%
    \vskip 1cm
    \fontseries{\seriesdefault}
    {\LARGE School of Physics and Astronomy,} %
    \vskip 0.25cm
    {\LARGE University of St Andrews} %
    \vskip 9cm
    {\LARGE Submitted for the degree of} %
    \vskip 0.25cm
    {\LARGE Doctor of Philosophy} %
    \vskip 1cm
    {\LARGE \@date\par}%
  \end{center}%
  \vfill
  \null
  \cleardoublepage
  }

\@ifundefined{showcaptionsetup}{}{%
 \PassOptionsToPackage{caption=false}{subfig}}
\usepackage{subfig}
\makeatother

\usepackage{babel}

\newcommand\AlCentroPagina [1]{%
\AddToShipoutPicture*{\AtPageCenter{%
\makebox(-50,0){\includegraphics%
[width=5cm]{#1}}}}}

\begin{document}
\captionsetup{justification=raggedright}

\title{Hawking Radiation in Dispersive Media}
\author{Scott James Robertson}
\date{May, 2011}
\AlCentroPagina{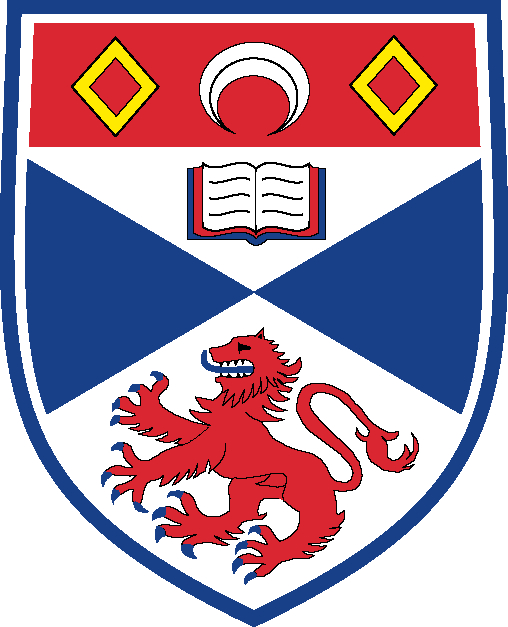}

\maketitle


\begin{center}
\textbf{I. Candidate's declarations}
\end{center}
\vskip 0.2cm
I, Scott James Robertson, hereby certify that this thesis, which is approximately 40,000 words in length,
has been written by me, that it is the record of work carried out by me and that it has not been
submitted in any previous application for a higher degree.
\vskip 0.3cm
I was admitted as a research student in September 2006 and as a candidate for the degree of PhD
in September 2006; the higher study for which this is a record was carried out in the
University of St Andrews between 2006 and 2010.
\vskip 0.5cm
\textit{Date}
\hskip 4cm
\textit{Signature of candidate}
\vskip 1cm
\begin{center}
\textbf{II. Supervisor's declaration}
\end{center}
\vskip 0.2cm
I hereby certify that the candidate has fulfilled the conditions of the Resolution and Regulations
appropriate for the degree of PhD in the University of St Andrews and that the candidate is qualified
to submit this thesis in application for that degree.
\vskip 0.5cm
\textit{Date}
\hskip 4cm
\textit{Signature of supervisor}
\vskip 1cm
\begin{center}
\textbf{III. Permission for electronic publication}
\vskip 0.2cm
\end{center}
In submitting this thesis to the University of St Andrews I understand that I am giving permission for it to be made
available for use in accordance with the regulations of the University Library for the time being in force, subject to any
copyright vested in the work not being affected thereby.  I also understand that the title and the abstract will be
published, and that a copy of the work may be made and supplied to any bona fide library or research worker, that
my thesis will be electronically accessible for personal or research use unless exempt by award of an embargo as
requested below, and that the library has the right to migrate my thesis into new electronic forms as required to ensure
continued access to the thesis.  I have obtained any third-party copyright permissions that may be required in order to allow
such access and migration, or have requested the appropriate embargo below.
\vskip 0.3cm
The following is an agreed request by candidate and supervisor regarding the electronic publication of this thesis:
\vskip 0.3cm
\textit{Access to printed copy and electronic publication of thesis through the University of St Andrews.}
\vskip 0.5cm
\textit{Date}
\hskip 4cm
\textit{Signature of candidate}
\vskip 0.5cm
\hskip 4.9cm
\textit{Signature of supervisor}

\begin{abstract}
Hawking radiation, despite its presence in theoretical physics for
over thirty years, remains elusive and undetected. It also suffers,
in its original context of gravitational black holes, from conceptual
difficulties. Of particular note is the trans-Planckian problem, which
is concerned with the apparent origin of the radiation in absurdly
high frequencies. In order to gain better theoretical understanding
and, it is hoped, experimental verification of Hawking radiation,
much study is being devoted to systems which model the spacetime geometry
of black holes, and which, by analogy, are also thought to emit Hawking
radiation. These analogue systems typically exhibit dispersion, which
regularizes the wave behaviour at the horizon but does not lend itself
well to analytic treatment, thus rendering Hawking's prediction less
secure. A general analytic method for dealing with Hawking radiation
in dispersive systems has proved difficult to find.

This thesis presents new numerical and analytic results for Hawking
emission spectra in dispersive systems. It examines two black-hole
analogue systems: it begins by introducing the well-known acoustic
model, presenting some original results in that context; then, through
analogy with the acoustic model, goes on to develop the lesser-known
fibre-optical model. The following original results are presented
in the context of both of these models:
\begin{itemize}
\item an analytic expression for the low-frequency temperature is found
for a hyperbolic tangent background profile, valid in the entire parameter
space; it is well-known that the spectrum is approximately thermal
at low frequencies, but a universally valid expression for the corresponding
temperature is an original development;
\item an analytic expression for the spectrum, valid over almost the entire
frequency range, when the velocity profile parameters lie in the regime
where the low-frequency temperature is given by the Hawking prediction;
previous work has focused on the low-frequency thermal spectrum and
the characterization of the deviations from thermality, rather than
a single analytic expression; and
\item a new unexplored regime where no group-velocity horizon exists is
examined; the Hawking spectra are found to be non-zero here, but also
highly non-thermal, and are found, in the limit of small deviations, to vary with the square of the maximum deviation;
the analytic expression for the case with a horizon is found to carry over to
this new regime, with appropriate modifications.
\end{itemize}
Furthermore, the thesis examines the results of a classical frequency-shifting
experiment in the context of fibre-optical horizons. The theory of
this process is presented for both a constant-velocity and a constantly-decelerating
pulse, the latter case taking account of the Raman effect. The resulting
spectra are at least qualatitively explained, but there is a discrepancy
between theory and experiment that has not yet been accounted for.

\end{abstract}

\tableofcontents{}

\listoffigures

\part{Introduction\label{par:Introduction}}

\chapter{Hawking Radiation\label{sec:Theoretical-Origins-of-Hawking-Radiation}}

Black holes - so-called because of the apparent impossibility of escape
from them - are not entirely black. That was the surprising claim
made by Hawking \cite{Hawking-1974,Hawking-1975} over thirty years
ago. Examining the behaviour of quantum fields in the vicinity of
a black hole, he showed that, far from being emission-free, it should
emit a steady flux of thermal radiation, with a temperature proportional
to $\kappa$, the gravitational field strength at the event horizon:\begin{equation}
k_{B}T=\frac{\hbar\kappa}{2\pi c}=\frac{\hbar c^{3}}{8\pi GM}\,.\label{eq:black_hole_temperature}\end{equation}
With this remarkable result, Hawking completed a thermodynamic treatment
of black holes that had been developed previously by Bekenstein \cite{Bekenstein-1973,Bekenstein-1974},
who had concluded that the temperature of a Schwarzschild black hole should be proportional to $\kappa$ \cite{Bekenstein-1973}
(though his analysis is based mainly on entropy, and he did not consider the possibility of thermal emission).
The work of Bekenstein and Hawking - and of others \cite{Hartle-Hawking-1976,Unruh-1976,Damour-Ruffini-1976,
Sanchez-1978,Fredenhagen-Haag-1990,Parikh-Wilczek-2000},
rederiving Hawking's result in various ways - brings together the normally disparate
areas of gravity, quantum theory and thermodynamics; a glance at the
various fundamental constants appearing in Eq. (\ref{eq:black_hole_temperature})
makes this fusion clear. Thirst for understanding of the underlying
connections between these mighty realms of physics provides ample
motivation for the study of what has come to be known as Hawking radiation.

A pre-requisite for any such study must be the acknowledgement that
Hawking radiation is not without its own problems, both practical
and conceptual. On the practical side, the predicted temperature -
at least in the gravitational context in which it was first derived
- is virtually untestable. A solar mass black hole would, according
to Eq. (\ref{eq:black_hole_temperature}), have a temperature of about
$10^{-6}\,\mathrm{K}$ - six orders of magnitude smaller than the
temperature of the cosmic microwave background (CMB). Any radiation from
the black hole would be drowned out by the CMB. Therefore, experimental
verification of black hole radiation would seem to require an unusually
light black hole, orders of magnitude lighter than the Sun - a very
unlikely object.

Conceptually, there is the trans-Planckian problem \cite{tHooft-1985,Jacobson-1991,Brout-et-al-Primer},
which has to do with the validity of the derivation of Hawking radiation.
Let us briefly explain the problem; it is discussed further in §\ref{sub:Event-horizon}.
In most derivations, the spacetime is assumed to collapse, as in a
star collapsing to form a black hole. Modes of the quantum vacuum
are incident from infinity, propagating through the collapsing spacetime
and out to infinity again, experiencing a gravitational redshift as
they climb out of the ever-deepening gravitational well. The steady
thermal flux seen at late times can be traced back to those vacuum
modes which just manage to escape the event horizon, slowed and redshifted
to greater and greater degrees. Thus, any low-frequency mode seen
in the late-time thermal spectrum can be traced back to an incident
vacuum mode of ever-increasing frequency; indeed, to an \textit{exponentially}
increasing frequency! The frequencies of these incident modes very
quickly exceed the Planck scale \cite{Brout-et-al-Primer}, widely believed to be a fundamental
quantum limit. We cannot justify the use of quantum field theory at
such scales, yet it appears that Hawking radiation is dependent on
the existence of these initial trans-Planckian frequencies. Whatever
is the correct physics, can we be sure that it will preserve Hawking
radiation?

These difficulties can be tackled by appealing to artificial event
horizons, or physical systems possessing horizons analogous to those
of gravitational black holes \cite{ArtificialBlackHoles,Schutzhold-Unruh}. This idea
was first proposed by Unruh \cite{Unruh-1981}, who found that perturbations
of a stationary background fluid flow behave just as a scalar field
in Lorentzian spacetime \cite{Unruh-1981,Visser-1993}. In particular,
if the flow velocity crosses the speed of sound, the surface where
it does so is entirely analogous to a black-hole event horizon, and
on quantizing the perturbation field, one predicts analog Hawking
radiation in such a system. Of course, this model is subject to the
same trans-Planckian problem as the gravitational case. In reality,
however, the trans-Planckian problem tends not to arise, because sound
waves in fluids - or, indeed, propagating waves in any of the experimentally-realisable
analog systems - exhibit \textit{dispersion}. That is, the behaviour of waves
changes at different scales by mechanisms which are better-understood
than quantum gravity; for example, the sizes of atoms or molecules
places a fundamental limit on the wavelengths of sound waves. Unruh
showed numerically that even after taking high-frequency dispersion
into account (and assuming that the dispersion is not too strong),
Hawking radiation is still predicted, with the same temperature as in the
dispersionless model \cite{Unruh-1995}. The conclusion is that the trans-Planckian problem is a mathematical artifact,
while the Hawking radiation exists regardless of the physics at the high-energy scale. This discovery has prompted a
great deal of interest in a range of black-hole analogue systems: in Bose-Einstein
condensates \cite{Garay-et-al-2000,Garay-et-al-2001,Barcelo-Liberati-Visser-2001-arXiv,Barcelo-Liberati-Visser-2001,Giovanazzi-et-al-2004},
in ultracold fermions \cite{Giovanazzi-2005}, in superfluid Helium-3 \cite{Jacobson-Volovik-1998,Volovik-1999,Fischer-Volovik-2001,Volovik2001195,HeliumDroplet},
in water \cite{Schutzhold-Unruh-2002,Rousseaux-et-al-2008,Rousseaux-et-al-2010},
and in optics \cite{Reznik-1997,Leonhardt-Piwnicki-1999,Leonhardt-Piwnicki-2000,Leonhardt-2002}.
These black-hole analogies might not teach us about quantum gravity
directly, but they can give insight into the particle creation process,
hopefully pointing towards the real origin of Hawking radiation \cite{Jacobson-1996}.

As we shall see, Hawking radiation derives from the conversion of
incident \textit{positive}-norm vacuum fluctuations into outgoing
modes with \textit{negative} norm. Determination of the resultant
spectrum is achieved through the calculation of the relevant coefficients
(known as \textit{Bogoliubov coefficients}). In the absence of dispersion,
Hawking's original result is obtained. Unsurprisingly, the introduction
of dispersion complicates the derivation, and its general effects
on the Hawking radiation are not yet clear. It has been shown that,
in the limit of a low velocity gradient or weak dispersion (the \textit{adiabatic} limit),
Hawking's prediction is preserved in the
low-frequency regime. This was first demonstrated numerically by Unruh
\cite{Unruh-1995}, using a finite-difference time-domain (FDTD) algorithm
for wavepacket propagation and introducing a high-frequency truncation
to the dispersion relation. The result was explained analytically
by Brout \textit{et al.} \cite{Brout-et-al-1995} by solving the wave
equation in momentum space, where, in the vicinity of the horizon,
the equations become more tractable. Corley and Jacobson considered
stationary modes of a single frequency, reducing the wave equation
to an ordinary differential equation. Thus, they were able to calculate
entire spectra numerically \cite{Corley-Jacobson-1996,Corley-1997},
and to characterise their deviations from thermality. Corley used
equations in momentum space to solve for the stationary modes analytically
\cite{Corley-1998}, using a different dispersion relation from that
used by Unruh but coming to the same conclusion that the Hawking prediction
remains valid for low frequencies in the adiabatic limit. He also
found that the spectrum remains thermal in the non-adiabatic limit,
i.e., in the limit of very large velocity gradient, but that this can
still be solved for analytically \cite{Corley-1997}; a similar result
has been demonstrated more recently in the context of Bose-Einstein condensates \cite{Recati-2009}.
Further numerical \cite{Jacobson-Mattingly-1999,Macher-Parentani-2008,Carusotto-et-al-2008}
and analytic \cite{Himemoto-Tanaka-2000,Saida-Sakagami-2000,Unruh-Schutzhold-2005,Schutzhold-Unruh-2008} works
have extended the context in various directions, but all are agreed that, in the adiabatic limit and low-frequency regime,
Hawking radiation behaves as it does in the dispersionless case.
While deviations and lowest-order approximations for the high-frequency regime have been studied
\cite{Saida-Sakagami-2000,Macher-Parentani-2008},
a general expression for the spectrum valid over the entire frequency range has never been found; nor has
such an expression for the low-frequency temperature, connecting the adiabatic limit with the
step-discontinuous limit. Furthermore, the regime which contains no horizon in the dispersionless limit
has not yet been explored.

In this thesis, we shall examine two analogue systems. In Part
II we discuss Unruh's original model of acoustic waves in a moving fluid; in Part III, we focus
on the lesser-known analogy of light in optical fibres. We shall see that Hawking radiation is
robust, remaining when high-frequency modes are suppressed, and even
when there is no event horizon (in the usual sense); that said, thermality
is not so robust, for only in the dispersionless case is the Hawking
spectrum strictly thermal. We shall see that, in the cases examined, we can find general analytic expressions
for the Hawking spectrum and for the low-frequency temperature. For the rest of this introduction, we shall use the acoustic model to
examine how black hole analogies arise in theory and how, in the strictly dispersionless case,
Hawking radiation derives from trans-Planckian vacuum modes.

\section{The Schwarzschild metric as a moving medium\label{sub:Schwarzschild_metric_moving_medium}}

General Relativity identifies gravity with curvature of spacetime,
described by the spacetime metric $ds^{2}$. It has a unique spherically
symmetric vacuum metric: the Schwarzschild metric \cite{Misner-Thorne-Wheeler}, which in coordinates
$\left(t_{S},r,\theta,\phi\right)$ takes the form\begin{equation}
ds^{2}=\left(1-\frac{r_{S}}{r}\right)c^{2}dt_{S}^{2}-\left(1-\frac{r_{S}}{r}\right)^{-1}dr^{2}-r^{2}d\Omega^{2}\,,\label{eq:Schwarzschild_metric}\end{equation}
where $d\Omega^{2}=d\theta^{2}+\sin^{2}\theta\, d\phi^{2}$ is the
angular line element, $t_{S}$ is the Schwarzschild time coordinate
and $r_{S}=2GM/c^{2}$ is the Schwarzschild radius. This describes,
for example, the spacetime around a star or planet with relatively
low rotation rate. As $r_{S}/r\rightarrow0$, the Schwarzschild metric
approaches the flat Minkowski metric, so the coordinates $\left(t_{S},r,\theta,\phi\right)$
correspond to the usual spherical coordinates of flat spacetime for
an observer at infinity. However, Eq. (\ref{eq:Schwarzschild_metric})
contains two singularities, at $r=0$ and $r=r_{S}$. The singularity
at $r=0$ is a genuine singularity of Schwarzschild spacetime \cite{Misner-Thorne-Wheeler}; we
shall not be concerned with it here. It is the surface $r=r_{S}$
- the event horizon - that is of interest to us.

\begin{figure}
\includegraphics[width=0.8\columnwidth]{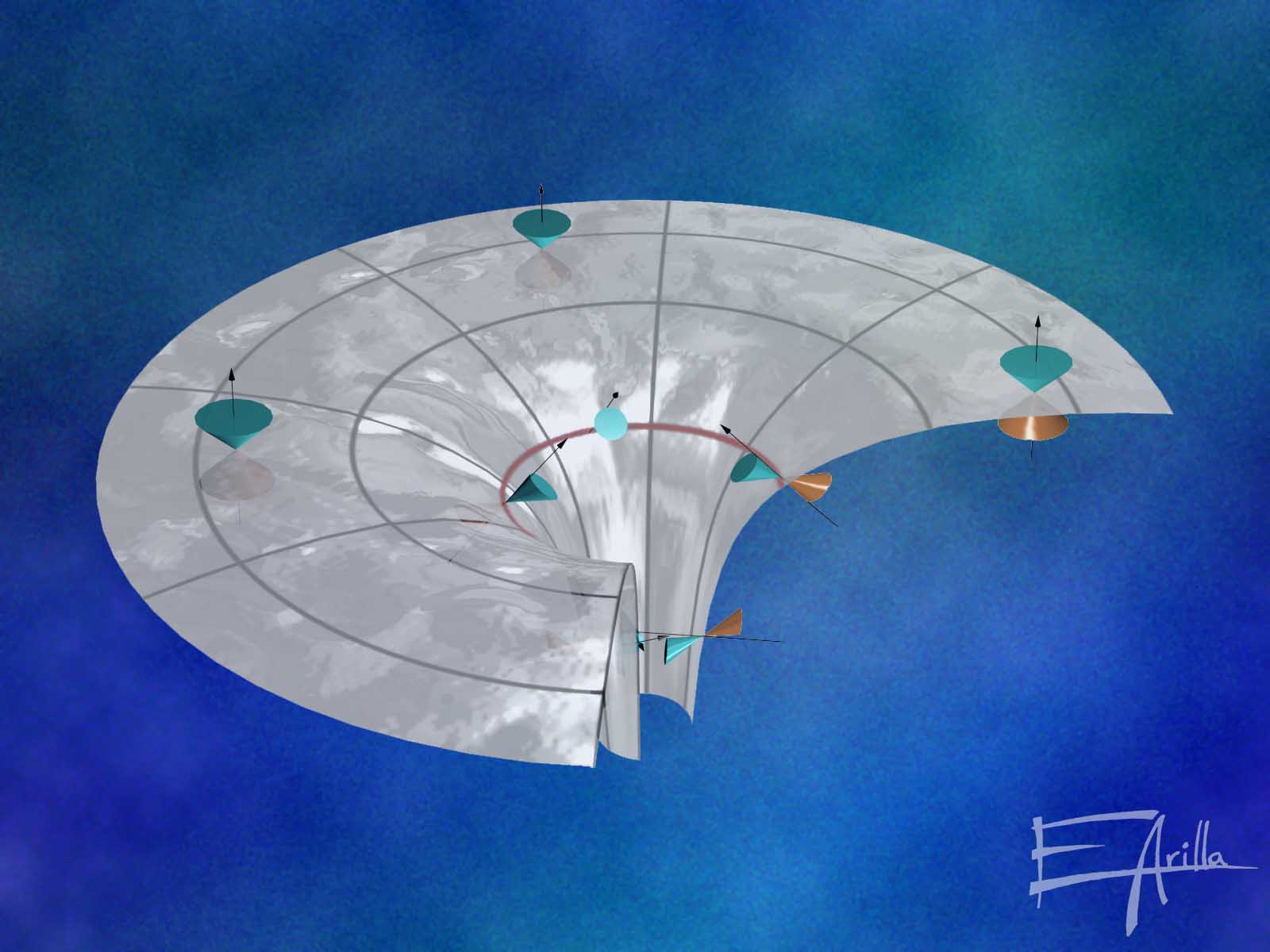}

\caption[\textsc{Black hole spacetime}]{\textsc{Black hole spacetime}: The curvature of spacetime near a
black hole is such that all future-directed paths are ingoing within
the event horizon. [Credit: Enrique Arilla]}

\end{figure}

Let us briefly review the effects of the event horizon by examining
light trajectories, or null curves, with $ds^{2}=0$. Since the spacetime
is spherically symmetric, we shall not concern ourselves with angular
trajectories, so we also set $d\Omega^{2}=0$. This leaves us with
a differential equation for the null curves:\begin{equation}
\frac{dt_{S}}{dr}=\pm\frac{1}{c}\left(1-\frac{r_{S}}{r}\right)^{-1}\,.\label{eq:Schwarzschild_radial_null_curves}\end{equation}
Far from the Schwarzschild radius, where $r\gg r_{S}$, $\left|dt_{S}/dr\right|\rightarrow1/c$,
so that light behaves just as it does in flat spacetime. However,
as we approach the Schwarzschild radius, $\left|dt_{S}/dr\right|$
diverges: light takes longer and longer to travel any distance, and,
if travelling towards the event horizon, can never reach it in a finite
time $t_{S}$. The time $t_{S}$ comes to a standstill at the event
horizon.

Despite this peculiar behaviour, the event horizon is not a genuine
singularity of Schwarzschild spacetime; it simply appears as such
in the coordinates $\left(t_{S},r\right)$, which are unsuited to
the region $r\leq r_{S}$. The Schwarzschild time $t_{S}$ is the
main culprit, for we may follow in the footsteps of Painlev\'e \cite{Painleve-1921}, Gullstrand \cite{Gullstrand-1922} and Lema\^itre \cite{Lemaitre-1933} by defining
a new time,\begin{equation}
t=t_{S}+2\frac{\sqrt{r_{S}r}}{c}+\frac{r_{S}}{c}\ln\left(\frac{\sqrt{r/r_{S}}-1}{\sqrt{r/r_{S}}+1}\right)\,;\label{eq:Lemaitre_time}\end{equation}
this has the differential\begin{equation}
dt=dt_{S}+\frac{1}{c}\sqrt{\frac{r_{S}}{r}}\left(1-\frac{r_{S}}{r}\right)^{-1}dr\,,\label{eq:Lemaitre_time_differential}\end{equation}
which, when substituted in Eq. (\ref{eq:Schwarzschild_metric}), yields
the metric\begin{equation}
ds^{2}=c^{2}dt^{2}-\left(dr+\sqrt{\frac{r_{S}}{r}}c\, dt\right)^{2}+r^{2}d\Omega^{2}\,.\label{eq:Lemaitre_metric}\end{equation}
In the coordinates $\left(t,r,\theta,\phi\right)$, the metric clearly
has no singularity at $r=r_{S}$. (It is true that the coefficient
of $dt^{2}$ vanishes there, but this is easily remedied by considering
a new position coordinate $q$ such that $dq=c\, dt+\sqrt{r/r_{S}}dr$
, so that $ds^{2}=c^{2}dt^{2}-\left(r/r_{S}\right)dq^{2}+r^{2}d\Omega^{2}$.
This is not too important, however, and we can gain a more intuitive
understanding of the metric by keeping $r$ as our position coordinate.)
Notice that, while the definition of $t$, Eq. (\ref{eq:Lemaitre_time}),
is applicable only when $r>r_{S}$, the metric of Eq. (\ref{eq:Lemaitre_metric})
is easily extendable to all values of $r$ greater than zero. The
new coordinate $t$ has opened up a previously inaccessible region
of the spacetime. Keeping $t$ fixed while decreasing $r$, we see
from Eq. (\ref{eq:Lemaitre_time}) that, as we approach the Schwarzschild
radius, we must have $t_{S}\rightarrow\infty$ to compensate for the
divergence of the logarithm. Thus, with respect to our original coordinates,
the transformation to the coordinates $\left(t,r\right)$ is accompanied
by an extension of the spacetime into the infinite future. Our discussion
of light trajectories anticipated this: since an ingoing light ray
approaches the horizon at an infinitely slower rate (with respect
to $t_{S}$), crossing the horizon requires $t_{S}\rightarrow\infty$.

\begin{figure}
\subfloat{\includegraphics[width=0.45\columnwidth]{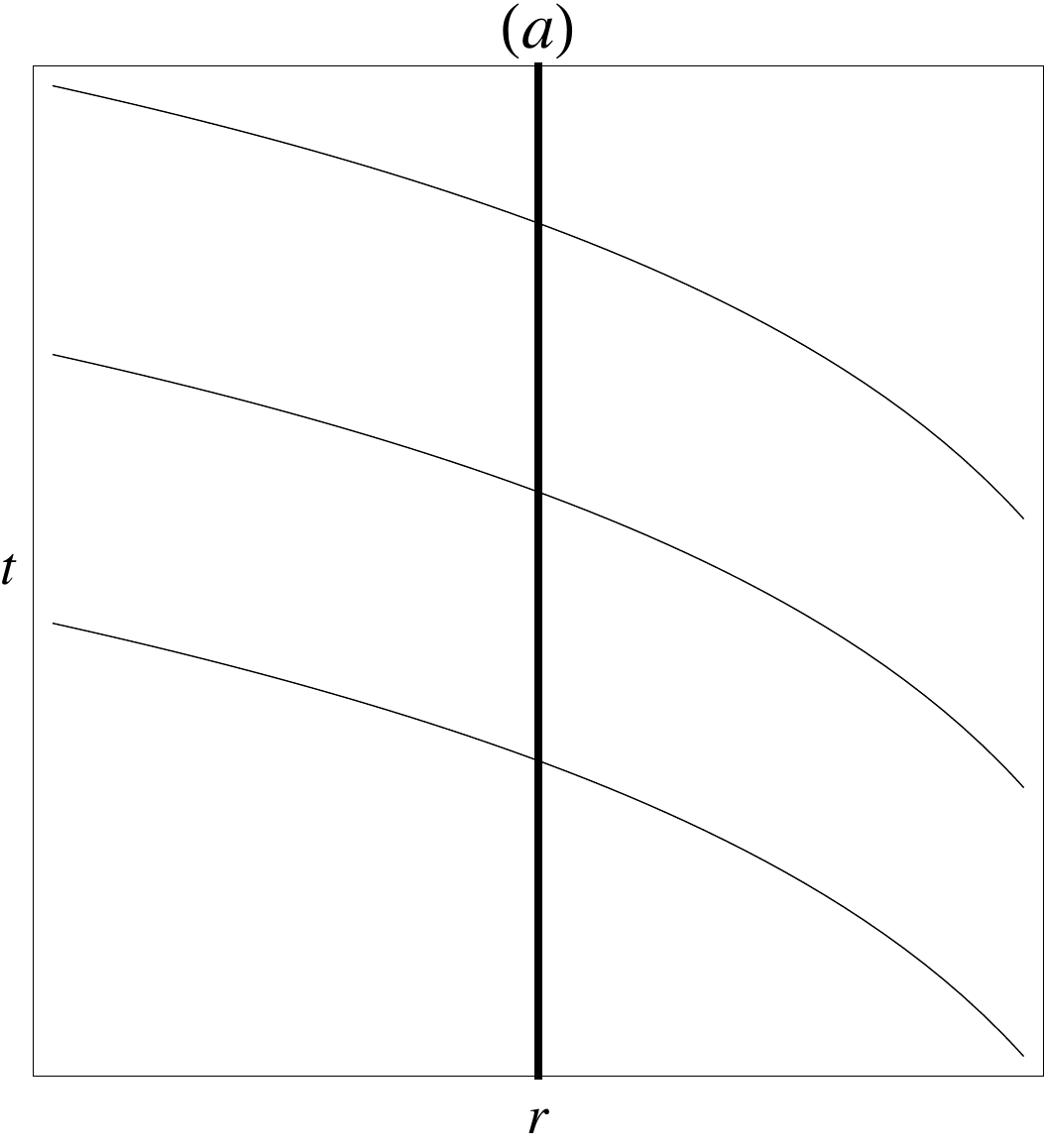}} \subfloat{\includegraphics[width=0.45\columnwidth]{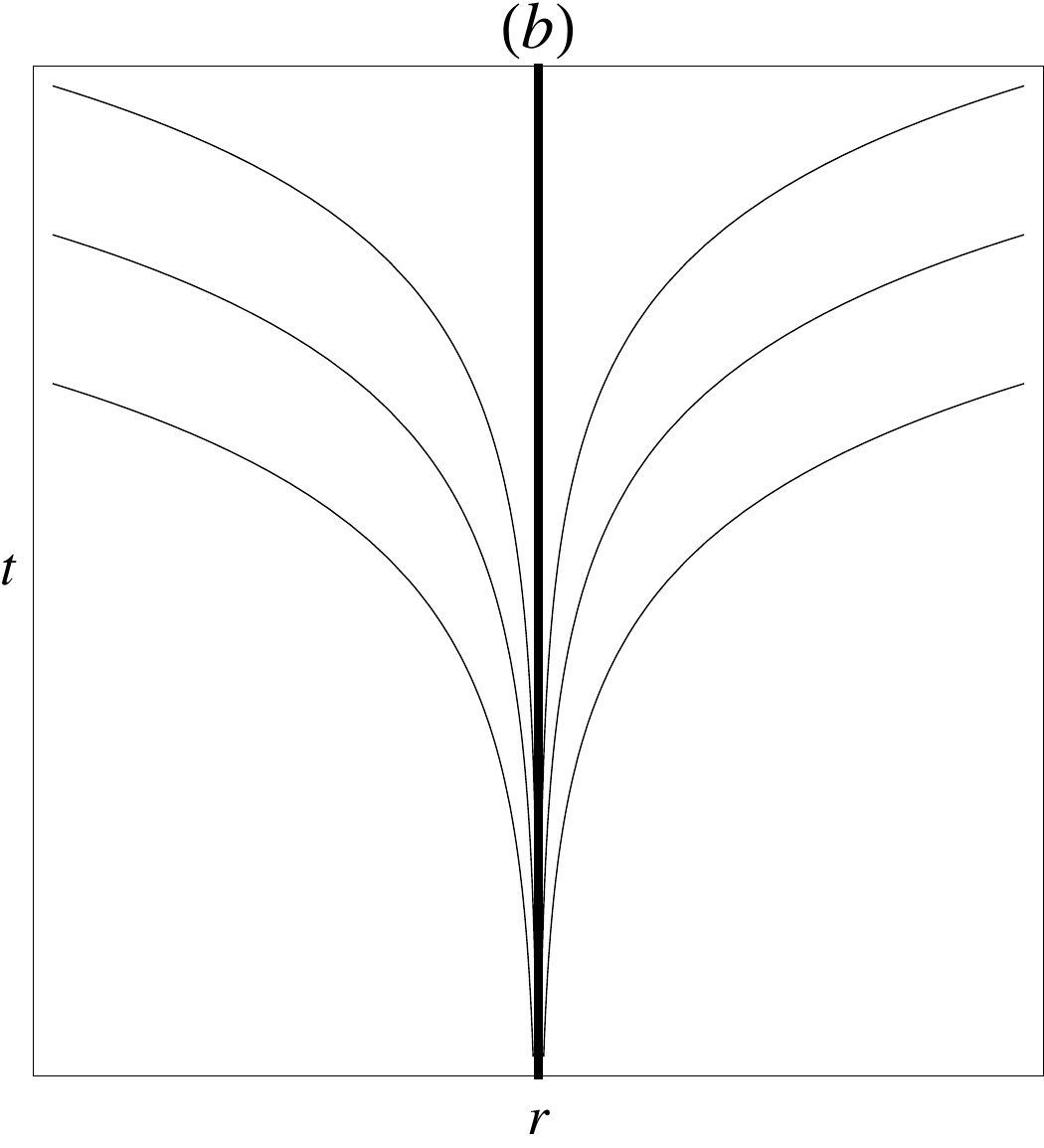}}

\caption[\textsc{Radial null curves near a black hole horizon}]{\textsc{Radial null curves near a black hole horizon}: Spacetime
diagrams of light trajectories in the metric of Eq. (\ref{eq:Lemaitre_metric}),
where the thick centre line represents the horizon $r=r_{S}$. Figure
$\left(a\right)$ shows the trajectories which are left-moving with
respect to the imagined fluid, obeying the first equation of (\ref{eq:Lemaitre_radial_null_curves});
these experience nothing unusual at the horizon. Figure (b) shows
the trajectories which are right-moving with respect to the imagined
fluid, obeying the second equation of (\ref{eq:Lemaitre_radial_null_curves});
as these are propagated further back in time, they come to a standstill
at the horizon.\label{fig:Lemaitre-Null-Curves}}

\end{figure}

Let us take a moment to interpret the metric (\ref{eq:Lemaitre_metric}).
Again, we shall consider only radial trajectories, setting $d\Omega^{2}=0$.
The key point to note is that, if $dr/dt=-c\sqrt{r_{S}/r}$, the metric
reduces to $ds^{2}=c^{2}dt^{2}$. $t$, then, is the proper time measured
by an observer on the trajectory $dr/dt=-c\sqrt{r_{S}/r}$. Since
this condition clearly maximises $ds^{2}$, these trajectories are
geodesics, and $t$ measures proper time along them. It is as though
space consists of a fluid \cite{Jacobson-PTP-1999,Hamilton-Lisle-2008}, flowing inwards with velocity $-c\sqrt{r_{S}/r}$
to converge on the point $r=0$. The geodesics just defined are those
which are stationary with respect to this fluid. The coordinate $q$,
defined above, is constant on such a geodesic, and can be thought
of as the proper spatial coordinate; its differential, combining $dr$
and $dt$, is derived from simple addition of velocities (which can
be done in the Galilean fashion since all velocities are measured
with respect to the proper time $t$). The fluid should be thought
of as the fabric of space itself, for the speed of light is measured
with respect to it. At the Schwarzschild radius, the fluid flows inward
with speed $c$; anything that falls beneath this radius will inexorably
be dragged towards the centre, $r=0$. This view is reinforced by
looking at the radial null curves in the coordinates $\left(t,r\right)$.
Setting $ds^{2}=0$ and $d\Omega^{2}=0$, we find two possible trajectories
for light:\begin{alignat}{1}
\frac{dt}{dr}=-\frac{1}{c}\left(1+\sqrt{\frac{r_{S}}{r}}\right)^{-1}\,,\qquad & \mathrm{or}\qquad\frac{dt}{dr}=\frac{1}{c}\left(1-\sqrt{\frac{r_{S}}{r}}\right)^{-1}\,.\label{eq:Lemaitre_radial_null_curves}\end{alignat}
The forms of these trajectories near the point $r=r_{S}$ are shown
in Figure \ref{fig:Lemaitre-Null-Curves}. The first possibility represents
rays which are ingoing with respect to the fluid; the total velocity
is $-c-c\sqrt{r_{S}/r}$, the sum of the light's velocity and the
fluid's velocity. It is perfectly regular at the horizon, and tilts
over as it propagates, travelling faster and faster as it moves to
ever smaller radii. The second possibility represent rays which are
outgoing with respect to the fluid, having a total velocity of $c-c\sqrt{r_{S}/r}$.
This is not regular at the horizon, nor should it be: there, the competing
velocities of the fluid and the light exactly cancel, giving a total
velocity of zero. Rays at higher radii will have a positive total
velocity, and will eventually escape to infinity; rays at smaller
radii will have a negative total velocity, unable to overcome the
fluid flow, and will propagate inwards to $r=0$.

\begin{figure}
\includegraphics[width=0.8\columnwidth]{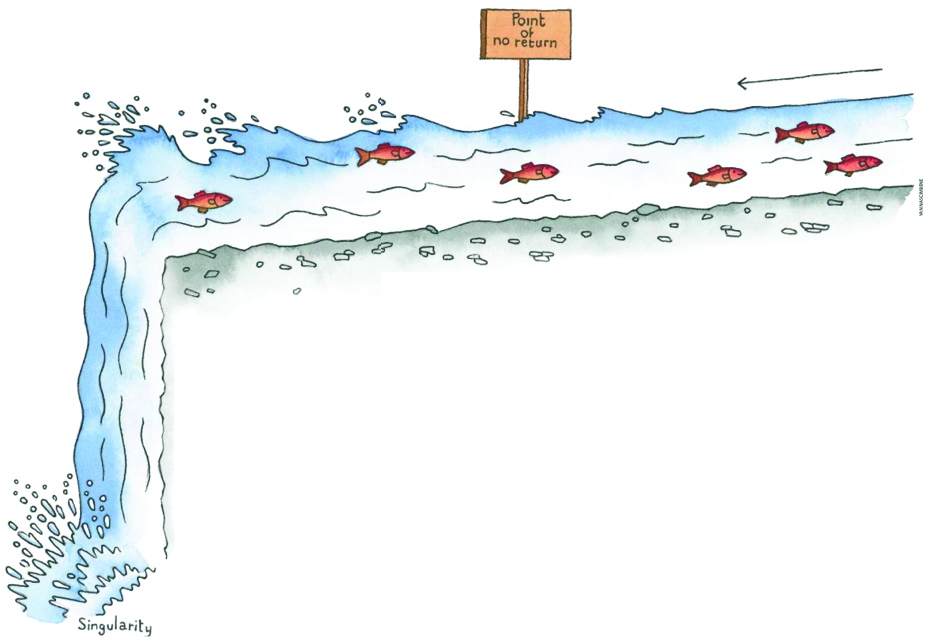}

\caption[\textsc{Black hole horizon in a river}]{\textsc{Black hole horizon in a river}: At the {}``Point of no return'',
the flow speed of the water is exactly equal to the maximum speed
of the fish. Any fish that travel further downstream than this point
will be dragged inexorably towards the waterfall. [Credit: Yan Nascimbene]\label{fig:Fishy-Black-Hole}}

\end{figure}

This analogy with a moving medium forms the basis of artificial black
holes and event horizon analogies. We may simply replace $-c\sqrt{r_{S}/r}$
in Eq. (\ref{eq:Lemaitre_metric}) with the more general velocity
profile $V\left(x\right)$ to obtain, in $1+1$-dimensional spacetime,
the metric\begin{equation}
ds^{2}=c^{2}dt^{2}-\left(dx-V\left(x\right)dt\right)^{2}\,,\label{eq:generalized_Lemaitre_metric}\end{equation}
where now $c$ is to be interpreted as the velocity with respect to
the medium in question (not necessarily the speed of light). For example,
this generalized metric may be applied to a system so far removed
from astrophysics as a river flowing towards a waterfall, so that
the flow speed increases in the direction of flow, as illustrated
in Figure \ref{fig:Fishy-Black-Hole}. Imagine this river is populated
by fish who can swim only up to a maximum speed $c$; this speed is,
of course, with respect to the water, not the coordinate $x$. Then
the above metric suffices to describe the trajectories of fish in
this river. Fish who are far from the waterfall, where the current
is low, are free to swim around as they please, experiencing no significant
resistance in either direction. However, as the current increases,
there may be a point at which $\left|V\right|=c$. As the fish approach
this point, they will find it increasingly difficult to swim back
upstream; passing this point, motion upstream is impossible, for the
current is so strong that the fish, no matter how hard or in which
direction they swim with respect to the water, are doomed to be swept
over the waterfall. The point where $\left|V\right|=c$ is the event
horizon, and the trajectories of fish swimming at exactly the speed
$c$ are analogous to the trajectories of light near a black hole
horizon (see Eqs. (\ref{eq:Lemaitre_radial_null_curves}) and Fig.
\ref{fig:Lemaitre-Null-Curves}).

\begin{figure}
\includegraphics[width=0.8\columnwidth]{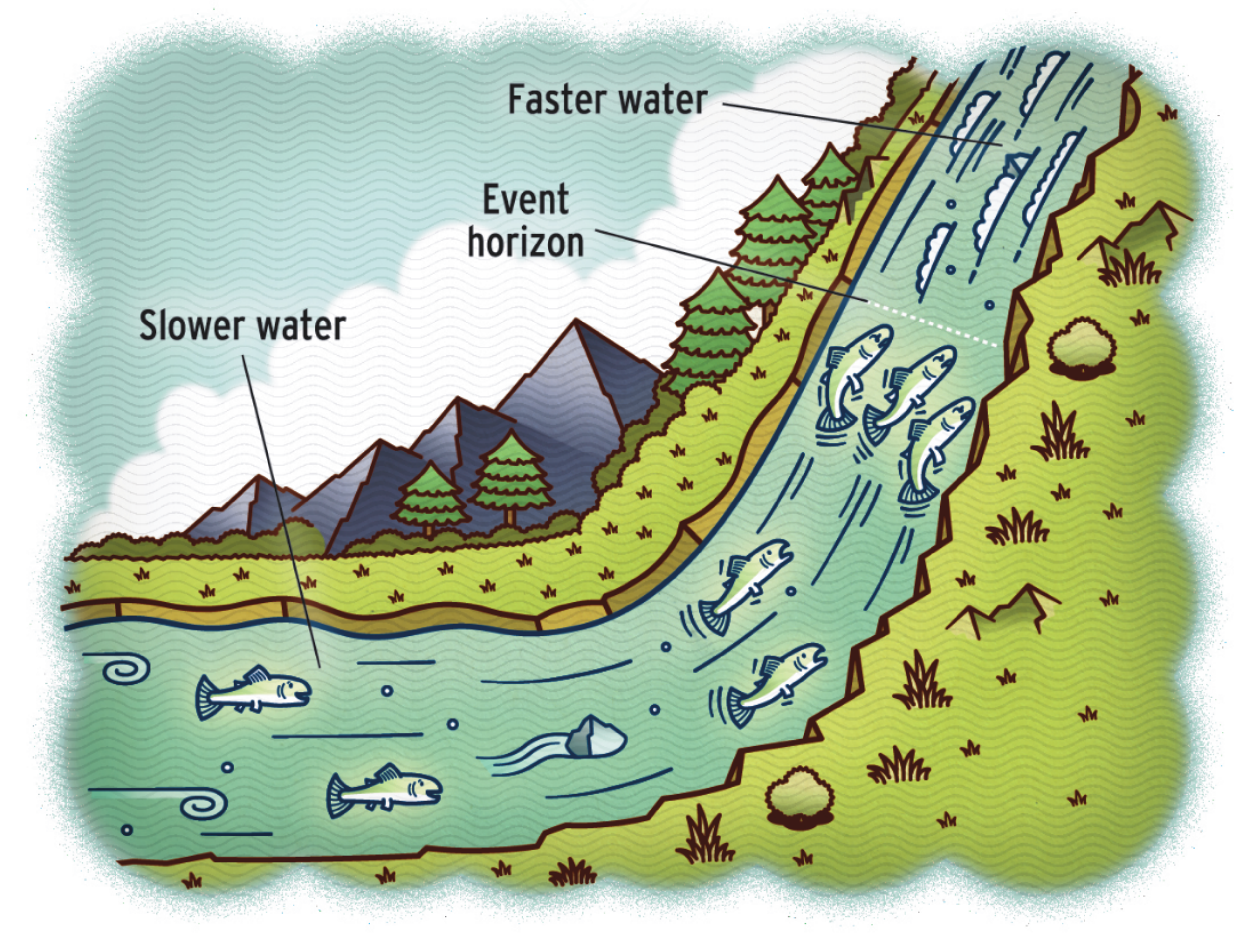}

\caption[\textsc{White hole horizon in a river}]{\textsc{White hole horizon in a river}: As in Figure \ref{fig:Fishy-Black-Hole},
the event horizon is the point at which the flow speed is exactly
equal to the maximum speed of the fish. Here, however, the flow speed
\textit{decreases} in the direction of flow, so that the fish may
swim arbitrarily close to the event horizon, but may never cross it. [Credit: A. Cho, Science \textbf{319}, 1321 (2008)]\label{fig:Fishy-White-Hole}}

\end{figure}

It is important to note that the existence of a point at which $\left|V\right|=c$
does not necessarily imply that the metric (\ref{eq:generalized_Lemaitre_metric})
is equivalent to a black hole spacetime. Nothing can escape from a
black hole once the horizon is crossed; more precisely, the region
where the flow is superluminal ($\left|V\right|>c$) flows \textit{away}
from the horizon. But the opposite is also possible: starting with
a black hole-like spacetime and performing the transformation $V\rightarrow-V$
- or, equivalently, $t\rightarrow-t$ - we find that the superluminal
region flows \textit{towards} the horizon. This is a white hole spacetime,
the time-reverse of a black hole spacetime, and white hole trajectories
are simply time-reversed black hole trajectories. Therefore, all light
rays initially in the superluminal region are dragged towards the
horizon. Those moving counter to the flow will come to a standstill
there: whether on the superluminal or subluminal side, they will approach
the horizon asymptotically, never actually reaching it. In the river
analogy, this is equivalent to water flowing \textit{from} a waterfall,
so that its flow is initially very fast but slows as it travels; this
is shown in Figure \ref{fig:Fishy-White-Hole}. The horizon is again
where $\left|V\right|=c$. Fish far from the waterfall, where the
current is low, may come and go as they please; but as they travel
upstream, they will find it increasingly difficult to continue, and
must come to a standstill at the horizon.

The white hole is of limited use in astrophysics. It \textit{is} encoded
in Schwarzschild spacetime, from which it is derived by changing the
sign of the $r$-dependent terms in Eqs. (\ref{eq:Lemaitre_time})-(\ref{eq:Lemaitre_metric}),
but it extends the spacetime into the inifinite past. Therefore, its
validity requires that the Schwarzschild metric is valid for $t_{S}\rightarrow-\infty$,
which is seldom the case, for black holes are believed to form from
gravitational collapse. However, turning away from gravity to the
more general case of moving media, white holes are just as realisable
as black holes, and we shall make use of them in Part III when we
discuss light in optical fibres.

\section{The acoustic field and its wave equation\label{sub:Acoustic_field_wave_equation}}

Having now shown the spacetime metric (\ref{eq:generalized_Lemaitre_metric})
of a moving medium to be analogous to that of a Schwarzschild black
hole, let us study the behaviour of fields in such a spacetime. For
simplicity, we shall assume a massless scalar field; in the context
of a fluid, such a field may arise from small perturbations in the
background flow \cite{Unruh-1981,ArtificialBlackHoles}. If the fluid
is irrotational (as it must be in one spatial dimension), the flow
velocity may be expressed as the derivative of a potential: $V\left(x\right)=-\partial\Phi/\partial x$.
Variations in pressure and density (i.e., sound waves) give rise to
small deviations $\phi=\delta\Phi$. We shall treat the background
flow as constant in time, and the small variations as the independent
field $\phi$, which will be referred to as the acoustic field. Being
massless, the acoustic field has speed $c$ with respect to the medium;
since this holds for all frequencies, this implies that the fluid
is dispersionless. (Dispersion is introduced in Part II.)

Let us begin with the Principle of Least action: the acoustic field
$\phi\left(t,x\right)$ varies from one configuration to another in
such a way that the action integral is an extremum (usually a minimum).
The action is the integral of the Lagrangian:\begin{equation}
S=\int\int dx\, dt\, L\left(\phi,\partial_{t}\phi,\partial_{x}\phi\right),\label{eq:action}\end{equation}
so all the physics of the model is contained in the Lagrangian, $L$.
An extremum of the action is found by infinitesimally varying the
field $\phi$ and its derivatives, $\partial_{t}\phi$ and $\partial_{x}\phi$,
then setting the resulting variation in $S$ to zero. This yields
the Euler-Lagrange equation:\begin{equation}
\frac{\partial L}{\partial\phi^{\star}}-\frac{\partial}{\partial t}\left(\frac{\partial L}{\partial\left(\partial_{t}\phi^{\star}\right)}\right)-\frac{\partial}{\partial x}\left(\frac{\partial L}{\partial\left(\partial_{x}\phi^{\star}\right)}\right)=0\,.\label{eq:Euler-Lagrange}\end{equation}
Being scalar and massless, the Lagrangian for the acoustic field is
\cite{Birrell-Davies}\begin{equation}
L=\frac{1}{2}\sqrt{-g}g^{\mu\nu}\partial_{\mu}\phi^{\star}\partial_{\nu}\phi\,.\label{eq:scalar_massless_Lagrangian}\end{equation}
$g^{\mu\nu}$ is the inverse metric tensor, and $g$ is the determinant
of the metric tensor $g_{\mu\nu}$. Using the metric (\ref{eq:generalized_Lemaitre_metric}),
the Lagrangian becomes\begin{equation}
L=\frac{1}{2c}\left(\left|\left(\partial_{t}+V\left(x\right)\partial_{x}\right)\phi\right|^{2}-c^{2}\left|\partial_{x}\phi\right|^{2}\right)\,,\label{eq:scalar_massless_Lagrangian_moving_fluid}\end{equation}
and plugging this into the Euler-Lagrange equation yields the wave
equation\begin{equation}
\left(\partial_{t}+\partial_{x}V\right)\left(\partial_{t}+V\partial_{x}\right)\phi-c^{2}\partial_{x}^{2}\phi=0\,.\label{eq:dispersionless_acoustic_wave_equation}\end{equation}

Solutions of the wave equation are more easily derived once its conservation
laws have been identified. These are directly related to symmetries
of the Lagrangian. There are two nontrivial symmetries of the Lagrangian
(\ref{eq:scalar_massless_Lagrangian_moving_fluid}). One is time translation:
$L$ is invariant under the transformation $t\rightarrow t+\Delta t$.
This leads to conservation of frequency; more precisely, it implies
the existence of stationary modes of the form $\phi_{\omega}\left(x\right)e^{-i\omega t}$,
which are examined in §\ref{sub:Stationary_modes} below. The other
symmetry is phase rotation: $L$ is invariant under the transformation
$\phi\rightarrow\phi\, e^{i\alpha}$, where $\alpha$ is a real constant.
This implies conservation of the scalar product, defined as\begin{eqnarray}
\left(\phi_{1},\phi_{2}\right) & = & i\int_{-\infty}^{+\infty}dx\left\{ \phi_{1}^{\star}\left(\partial_{t}+V\partial_{x}\right)\phi_{2}-\phi_{2}\left(\partial_{t}+V\partial_{x}\right)\phi_{1}^{\star}\right\} \nonumber \\
 & = & i\int_{-\infty}^{+\infty}dx\left\{ \phi_{1}^{\star}\pi_{2}-\phi_{2}\pi_{1}^{\star}\right\} \,,\label{eq:scalar_product}\end{eqnarray}
where $\phi_{1}$ and $\phi_{2}$ are both solutions of the wave equation
(\ref{eq:dispersionless_acoustic_wave_equation}), and the canonical
momentum\begin{equation}
\pi=\frac{\partial L}{\partial\left(\partial_{t}\phi^{\star}\right)}=\left(\partial_{t}+V\left(x\right)\partial_{x}\right)\phi\,.\label{eq:canonical_momentum}\end{equation}
The scalar product of a solution with itself is of particular importance,
and is called the \textit{norm}.

A general solution to the acoustic wave equation (\ref{eq:dispersionless_acoustic_wave_equation})
is easily found. We define new variables $u$ and $v$ as follows:\begin{alignat}{1}
u=t-\int^{x}\frac{dx^{\prime}}{c+V\left(x^{\prime}\right)}\,,\qquad & v=t+\int^{x}\frac{dx^{\prime}}{c-V\left(x^{\prime}\right)}\,.\label{eq:defining_U_and_V}\end{alignat}
Then, after some algebra, we find that the wave equation in the coordinates
$\left(u,v\right)$ is simply\begin{equation}
\partial_{u}\partial_{v}\phi=0\,.\label{eq:U_V_wave_equation}\end{equation}
So, in the absence of horizons where $\left|V\right|=c$, $\phi$
is simply a sum of two arbitrary functions, one a function of $u$
only, the other a function of $v$ only:\begin{eqnarray}
\phi & = & \phi_{u}\left(u\right)+\phi_{v}\left(v\right)\nonumber \\
 & = & \phi_{u}\left(t-\int^{x}\frac{dx^{\prime}}{c+V\left(x^{\prime}\right)}\right)+\phi_{v}\left(t+\int^{x}\frac{dx^{\prime}}{c-V\left(x^{\prime}\right)}\right)\,.\label{eq:general_solution}\end{eqnarray}
With respect to the fluid, $\phi_{u}$ is right-moving while $\phi_{v}$
is left-moving. The fact that the two functional forms maintain their
shapes is a consequence of the absence of dispersion: all wave components
have the same velocity, $c$, with respect to the fluid, and so does
the waveform as a whole. The only ambiguity is in the direction of
travel; thus the solution splits into a right-moving and a left-moving
part.

If horizons are present, the coordinates $u$ and $v$ defined in Eqs.
(\ref{eq:defining_U_and_V}) do not cover the entire spacetime, so
that extra coordinates - and hence extra functions in the general
solution (\ref{eq:general_solution}) - must be added. This is examined
in greater detail in §\ref{sub:Event-horizon}.

\section{Stationary modes\label{sub:Stationary_modes}}

The general solution (\ref{eq:general_solution}) ceases to apply
when dispersion is introduced (although the labels $u$ and $v$,
representing waves which are right- and left-moving with respect to
the fluid, continue to be useful). However, even with dispersion,
the symmetries of the Lagrangian described above remain valid, as
do their corresponding conservation laws. We shall present here a
more comprehensive form for the general solution: a linear superposition
of stationary modes.

As previously mentioned, the invariance of the Lagrangian under time
translation implies the existence of solutions of the form\begin{equation}
\phi\left(t,x\right)=e^{-i\omega t}\phi_{\omega}\left(x\right)\,,\label{eq:stationary_solution}\end{equation}
called \textit{stationary modes}. In the present case of no dispersion,
Eq. (\ref{eq:general_solution}) implies that the spatial part of
the solution takes the form\begin{eqnarray}
\phi_{\omega}\left(x\right) & = & C_{u}\,\exp\left(i\omega\int^{x}\frac{dx^{\prime}}{c+V\left(x^{\prime}\right)}\right)+C_{v}\,\exp\left(-i\omega\int^{x}\frac{dx^{\prime}}{c-V\left(x^{\prime}\right)}\right)\nonumber \\
 & = & C_{u}\,\phi_{\omega}^{u}\left(x\right)+C_{v}\,\phi_{\omega}^{v}\left(x\right)\,.\label{eq:stationary_mode_U_V}\end{eqnarray}
It thus splits into two independent parts, one right-moving and one
left-moving with respect to the fluid. Defining the local wavevector
$k\left(x\right)$ such that $\phi_{\omega}^{u/v}\left(x\right)=\exp\left(i\int^{x}dx^{\prime}\, k_{\omega}^{u/v}\left(x^{\prime}\right)\right)$,
we see that\begin{alignat}{1}
k_{\omega}^{u}\left(x\right)=\frac{\omega}{c+V\left(x\right)}\,,\qquad & k_{\omega}^{v}\left(x\right)=-\frac{\omega}{c-V\left(x\right)}\,,\label{eq:U_V_wavevectors}\end{alignat}
which are summarized by the relation\begin{equation}
\left(\omega-Vk\right)^{2}=c^{2}k^{2}\,.\label{eq:Doppler_formula}\end{equation}
This formula is very intuitive once we recognise that $\omega$ is
related to the free-fall frequency $\omega_{\mathrm{ff}}$ (i.e.,
the frequency measured in the frame in which the medium is at rest)
via the Doppler formula: $\omega=\omega_{\mathrm{ff}}+Vk$. Equation
(\ref{eq:Doppler_formula}), then, is simply $\omega_{\mathrm{ff}}^{2}=c^{2}k^{2}$
- the dispersion relation in the free-fall frame. Comparing with Eqs.
(\ref{eq:U_V_wavevectors}), we see that\begin{equation}
\omega_{\mathrm{ff}}=\omega-Vk=\begin{cases}
ck & \mathrm{for}\: u\mathrm{-modes}\\
-ck & \mathrm{for}\: v\mathrm{\mathrm{-modes}}\end{cases}\,.\label{eq:free-fall-freq_u-and-v}\end{equation}

Consider the stationary modes\begin{equation}
\phi_{\omega}^{u/v}\left(t,x\right)=\exp\left(-i\omega t\pm i\omega\int^{x}\frac{dx^{\prime}}{c\pm V\left(x^{\prime}\right)}\right)\,,\end{equation}
with canonical momenta\begin{equation}
\pi_{\omega}^{u/v}\left(t,x\right)=-i\frac{\omega c}{c\pm V}\phi_{\omega}^{u/v}\left(t,x\right)\,.\end{equation}
The scalar product between $u$-modes is\begin{eqnarray}
\left(\phi_{\omega_{1}}^{u},\phi_{\omega_{2}}^{u}\right) & = & i\int dx\left\{ \phi_{\omega_{1}}^{u\star}\pi_{\omega_{2}}^{u}-\phi_{\omega_{2}}^{u}\pi_{\omega_{1}}^{u\star}\right\} \nonumber \\
 & = & \left(\omega_{1}+\omega_{2}\right)e^{i\left(\omega_{1}-\omega_{2}\right)t}\int dx\frac{c}{c+V\left(x\right)}\exp\left(-i\left(\omega_{1}-\omega_{2}\right)\int^{x}\frac{dx^{\prime}}{c+V\left(x^{\prime}\right)}\right)\nonumber \\
 & = & 4\pi\, c\,\omega_{1}\,\mathrm{sgn}\left(c+V\right)\,\delta\left(\omega_{1}-\omega_{2}\right)\,,\label{eq:norm_u-modes}\end{eqnarray}
where the last line is found by making the substitution $z=\mathrm{sgn}\left(c+V\right)\,\int^{x}dx^{\prime}/\left(c+V\left(x^{\prime}\right)\right)$,
and the integration is taken over the whole real line of $z$. If
$c+V$ is nowhere zero, there is a one-to-one correspondence between
the $z$- and $x$-axes, and the integration thus extends over all
$x$. (When $c+V=0$ for some $x$, the mode must be split between
separate regions; this is dealt with in §\ref{sub:Event-horizon}.)
Similarly, the scalar product between $v$-modes is\begin{eqnarray}
\left(\phi_{\omega_{1}}^{v},\phi_{\omega_{2}}^{v}\right) & = & i\int dx\left\{ \phi_{\omega_{1}}^{v\star}\pi_{\omega_{2}}^{v}-\phi_{\omega_{2}}^{v}\pi_{\omega_{1}}^{v\star}\right\} \nonumber \\
 & = & \left(\omega_{1}+\omega_{2}\right)e^{i\left(\omega_{1}-\omega_{2}\right)t}\int dx\frac{c}{c-V\left(x\right)}\exp\left(i\left(\omega_{1}-\omega_{2}\right)\int^{x}\frac{dx^{\prime}}{c-V\left(x^{\prime}\right)}\right)\nonumber \\
 & = & 4\pi\, c\,\omega_{1}\,\mathrm{sgn}\left(c-V\right)\,\delta\left(\omega_{1}-\omega_{2}\right)\,,\label{eq:norm_v-modes}\end{eqnarray}
where the last line is found via the substitution $z=\mathrm{sgn}\left(c-V\right)\int^{x}dx^{\prime}/\left(c-V\left(x^{\prime}\right)\right)$,
and the integration is taken over the entire $z$-axis. (Again, if
$c-V=0$ for some $x$, the mode must be split into regions; see §\ref{sub:Event-horizon}.)
Finally, we remark that the $u$- and $v$-modes are orthogonal to
each other, as are modes and their conjugates:\begin{equation}
\left(\phi_{\omega_{1}}^{u},\phi_{\omega_{2}}^{v}\right)=\left(\phi_{\omega_{1}}^{u},\phi_{\omega_{2}}^{u\star}\right)=\left(\phi_{\omega_{1}}^{v},\phi_{\omega_{2}}^{v\star}\right)=0\,.\end{equation}
Notice, from Eqs. (\ref{eq:norm_u-modes}) and (\ref{eq:norm_v-modes}),
that the sign of the norm is equal to the sign of the wavevector $k$
for $u$-modes, but to minus the sign of $k$ for $v$-modes. More
succinctly, with the use of Eq. (\ref{eq:free-fall-freq_u-and-v}),
the sign of the norm is equal to the sign of the free-fall frequency:\begin{equation}
\mathrm{sgn}\left[\left(\phi_{\omega}^{u/v},\phi_{\omega}^{u/v}\right)\right]=\mathrm{sgn}\left[\omega_{\mathrm{ff}}\right]\,.\label{eq:norm_and_free-fall-freq}\end{equation}
This correspondence continues to hold in dispersive systems.

We may redefine the stationary modes so that they are normalized:
for $\omega>0$,\begin{alignat}{3}
\phi_{\omega}^{u}\left(t,x\right) & = & \frac{e^{-\mathrm{sgn}\left(c+V\right)\, i\omega u}}{\sqrt{4\pi\, c\,\omega}} & = & \frac{1}{\sqrt{4\pi\, c\,\omega}}\exp\left(-\mathrm{sgn}\left(c+V\right)\, i\omega\left[t-\int^{x}\frac{dx^{\prime}}{c+V\left(x^{\prime}\right)}\right]\right)\,,\label{eq:normalized_U_modes}\\
\phi_{\omega}^{v}\left(t,x\right) & = & \frac{e^{-\mathrm{sgn}\left(c-V\right)\, i\omega v}}{\sqrt{4\pi\, c\,\omega}} & = & \frac{1}{\sqrt{4\pi\, c\,\omega}}\exp\left(-\mathrm{sgn}\left(c-V\right)\, i\omega\left[t+\int^{x}\frac{dx^{\prime}}{c-V\left(x^{\prime}\right)}\right]\right)\,,\label{eq:normalized_V_modes}\end{alignat}
so that\begin{equation}
\left(\phi_{\omega_{1}}^{u},\phi_{\omega_{2}}^{u}\right)=\left(\phi_{\omega_{1}}^{v},\phi_{\omega_{2}}^{v}\right)=\delta\left(\omega_{1}-\omega_{2}\right)\,.\label{eq:positive_norm}\end{equation}
A complete set of modes must include the complex conjugates of these.
According to Eq. (\ref{eq:scalar_product}), complex conjugation changes
the sign of the scalar product:\begin{equation}
\left(\phi_{\omega_{1}}^{u\star},\phi_{\omega_{2}}^{u\star}\right)=\left(\phi_{\omega_{1}}^{v\star},\phi_{\omega_{2}}^{v\star}\right)=-\delta\left(\omega_{1}-\omega_{2}\right)\,.\label{eq:negative_norm}\end{equation}
Finally, then, any solution of the wave equation (\ref{eq:dispersionless_acoustic_wave_equation})
can be written as a linear superposition of modes:\begin{equation}
\phi\left(t,x\right)=\int_{0}^{\infty}d\omega\left\{ a_{\omega}^{u}\phi_{\omega}^{u}\left(t,x\right)+b_{\omega}^{u}\phi_{\omega}^{u\star}\left(t,x\right)+a_{\omega}^{v}\phi_{\omega}^{v}\left(t,x\right)+b_{\omega}^{v}\phi_{\omega}^{v\star}\left(t,x\right)\right\} \,,\label{eq:decomposed_acoustic_field}\end{equation}
where\begin{alignat}{1}
a_{\omega}^{u}=\left(\phi_{\omega}^{u},\phi\right)\,,\qquad & b_{\omega}^{u}=-\left(\phi_{\omega}^{u\star},\phi\right)\,,\nonumber \\
a_{\omega}^{v}=\left(\phi_{\omega}^{v},\phi\right)\,,\qquad & b_{\omega}^{v}=-\left(\phi_{\omega}^{v\star},\phi\right)\,.\label{eq:field_expansion_coefficients}\end{alignat}

\section{Quantization of the acoustic field\label{sub:Quantization}}

Having expressed the acoustic field in terms of orthonormal modes,
it is readily quantized via the usual methods of Quantum Field Theory
\cite{Birrell-Davies}. Firstly, however, we note that the physical
acoustic field is a real-valued quantity, so that we may write Eq.
(\ref{eq:decomposed_acoustic_field}) in the form\begin{eqnarray}
\phi\left(t,x\right) & = & \int_{0}^{\infty}d\omega\left\{ a_{\omega}^{u}\phi_{\omega}^{u}\left(t,x\right)+a_{\omega}^{u\star}\phi_{\omega}^{u\star}\left(t,x\right)+a_{\omega}^{v}\phi_{\omega}^{v}\left(t,x\right)+a_{\omega}^{v\star}\phi_{\omega}^{v\star}\left(t,x\right)\right\} \nonumber \\
 & = & \int_{0}^{\infty}d\omega\left\{ a_{\omega}^{u}\phi_{\omega}^{u}\left(t,x\right)+a_{\omega}^{v}\phi_{\omega}^{v}\left(t,x\right)+\mathrm{c.c.}\right\} \,,\label{eq:real_valued_decomposed_field}\end{eqnarray}
where $\mathrm{c.c.}$ stands for complex conjugate. We may now promote
the real-valued field variable $\phi$ to the Hermitian operator $\hat{\phi}$,
the coefficient variables $a_{\omega}^{u/v}$ to annihilation operators
$\hat{a}_{\omega}^{u/v}$ and the complex conjugate coefficient variables
$\left(a_{\omega}^{u/v}\right)^{\star}$ to creation operators $\left(\hat{a}_{\omega}^{u/v}\right)^{\dagger}$:\begin{eqnarray}
\hat{\phi}\left(t,x\right) & = & \int_{0}^{\infty}d\omega\left\{ \hat{a}_{\omega}^{u}\phi_{\omega}^{u}\left(t,x\right)+\hat{a}_{\omega}^{u\dagger}\phi_{\omega}^{u\star}\left(t,x\right)+\hat{a}_{\omega}^{v}\phi_{\omega}^{v}\left(t,x\right)+\hat{a}_{\omega}^{v\dagger}\phi_{\omega}^{v\star}\left(t,x\right)\right\} \nonumber \\
 &  & \int_{0}^{\infty}d\omega\left\{ \hat{a}_{\omega}^{u}\phi_{\omega}^{u}\left(t,x\right)+\hat{a}_{\omega}^{v}\phi_{\omega}^{v}\left(t,x\right)+\mathrm{h.c.}\right\} \,,\label{eq:quantized_acoustic_field}\end{eqnarray}
where $\mathrm{h.c.}$ stands for Hermitian conjugate. We also have
a Hermitian operator for the canonical momentum:\begin{equation}
\hat{\pi}\left(t,x\right)=\int_{0}^{\infty}d\omega\left\{ \hat{a}_{\omega}^{u}\pi_{\omega}^{u}\left(t,x\right)+\hat{a}_{\omega}^{v}\pi_{\omega}^{v}\left(t,x\right)+\mathrm{h.c.}\right\} \,.\label{eq:quantized_canonical_momentum}\end{equation}
These may be inverted, as in (\ref{eq:field_expansion_coefficients}),
to find expressions for the annihilation and creation operators:\begin{equation}
\hat{a}_{\omega}^{u}=i\int_{-\infty}^{+\infty}dx\left\{ \phi_{\omega}^{u\star}\left(t,x\right)\hat{\pi}\left(t,x\right)-\pi_{\omega}^{u\star}\left(t,x\right)\hat{\phi}\left(t,x\right)\right\} \,,\label{eq:annihilation_operator}\end{equation}
where the Hermitian conjugate gives the operator $\hat{a}_{\omega}^{u\dagger}$
and a similar expression holds for the $v$-mode operators.

To complete the quantization, the commutators of the various operators
must be specified. This is done for $\hat{\phi}$ and $\hat{\pi}$,
the other operators following suit. We define\begin{alignat}{1}
\left[\hat{\phi}\left(\tau,x\right),\hat{\pi}\left(\tau,x^{\prime}\right)\right]=i\,\delta\left(x-x^{\prime}\right)\,,\qquad & \left[\hat{\phi}\left(\tau,x\right),\hat{\phi}\left(\tau,x^{\prime}\right)\right]=\Big[\hat{\pi}\left(\tau,x\right),\hat{\pi}\left(\tau,x^{\prime}\right)\Big]=0\,.\label{eq:phi_pi_commutators}\end{alignat}
Then the annihilation and creation operators obey the Bose commutation
relations\begin{equation}
\left[\hat{a}_{\omega_{1}}^{u},\hat{a}_{\omega_{2}}^{u\dagger}\right]=\left[\hat{a}_{\omega_{1}}^{v},\hat{a}_{\omega_{2}}^{v\dagger}\right]=\delta\left(\omega_{1}-\omega_{2}\right)\,,\label{eq:Bose_commutation_relations}\end{equation}
all other commutators being zero.

\section{Event horizon\label{sub:Event-horizon}}

Suppose there exists a point at which $V=-c$. Since the flow is to
the left at precisely the wave speed with respect to the flow, the
total speed of right-moving waves here is zero. This is an event horizon
for right-moving waves.

Let us suppose that the event horizon is situated at the origin, $x=0$;
and, moreover, that the derivative of $V$ is non-zero there. Then
we may approximate the flow velocity to first-order in $x$ as follows:\begin{equation}
V\left(x\right)\approx-c+\alpha x\,.\label{eq:linearized_flow_velocity}\end{equation}
If $\alpha>0$, the magnitude of the velocity increases in the direction
of flow; this is equivalent to a black hole horizon. However, if $\alpha<0$,
the speed decreases in the direction of flow, and we have a white
hole horizon. We shall assume in what follows that $\alpha>0$, so
that we have a black hole. It should be borne in mind that the white
hole is entirely analogous, and behaves as the time reverse of a black
hole.

Equation (\ref{eq:general_solution}) states that waves which are
right-moving with respect to the fluid are described by the arbitrary
function $\phi_{u}\left(u\right)$, where $u$ is given by\begin{equation}
u=t-\int^{x}\frac{dx^{\prime}}{c+V\left(x^{\prime}\right)}=t-\frac{1}{\alpha}\log\left(\frac{\alpha}{c}\left|x\right|\right)\,;\label{eq:U_horizon}\end{equation}
the second equality holds in the vicinity of the horizon. It is evident
from Eq. (\ref{eq:U_horizon}) that a horizon is located at $x=0$,
for the space is divided into two separate regions, with $x=0$ marking
the boundary between them. To see this, imagine first that we have
a wavepacket centred at a certain value of $u$, and that this is
located, at a certain time, at a positive value of $x$. If $t$ increases,
then $\log\left(\alpha\left|x\right|/c\right)/\alpha$ must increase
by exactly the same amount, and so $\left|x\right|$ increases. Since
$x$ is positive, $x$ itself increases, and the wavepacket moves
to the right. Similarly, if we trace the wavepacket back in time by
decreasing $t$, $\log\left(\alpha\left|x\right|/c\right)/\alpha$
must also decrease by the same amount, and $x$ decreases; the wavepacket
has come from the left. But the logarithm diverges to $-\infty$ at
the origin; this means that, no matter how far back in time we look,
$\log\left(\alpha\left|x\right|/c\right)/\alpha$ can be decreased
by a corresponding amount \textit{without $x$ ever becoming negative}.
The wavepacket must have originated arbitrarily close to the event
horizon, moving very slowly forwards at first, picking up speed the
further it travels. (The trajectories are precisely those plotted
on the right-hand side of Fig. \ref{fig:Lemaitre-Null-Curves}$\left(b\right)$.)
If the horizon has existed since the infinite past, then the wavepacket
could never have been in the left-hand region (where $x$ is negative).
Since this applies to the entire wavepacket, it also implies that
the wavepacket must initially have been arbitrarily thin; or, equivalently,
that it must have originated from arbitrarily short wavelengths. This
is precisely the trans-Planckian problem.

Exactly the same analysis holds in the left-hand region, but in the
opposite direction. Any wavepacket that exists there moves to the
left, and must have originated arbitrarily close to the left-hand
side of the horizon. If the horizon has existed since the infinite
past, then at no point in the past could the wavepacket have been
located at a positive value of $x$. Moreover, the further back in
time the wavepacket is traced, the thinner it becomes. (This corresponds
to the left-hand side of Fig. \ref{fig:Lemaitre-Null-Curves}$\left(b\right)$.)

Since a complete set of modes must be able to account for the existence
of waves throughout all space, there must be two separate $u$-modes
for each frequency, one for each of the regions separated by the horizon:\begin{eqnarray}
\phi_{\omega,R}^{u}\left(t,x\right) & = & \theta\left(x\right)\frac{1}{\sqrt{4\pi c\omega}}\exp\left(-i\omega u\right)\,,\label{eq:stationary_modes_horizon_R}\\
\phi_{\omega,L}^{u}\left(t,x\right) & = & \theta\left(-x\right)\frac{1}{\sqrt{4\pi c\omega}}\exp\left(i\omega u\right)\,.\label{eq:stationary_modes_horizon_L}\end{eqnarray}
The modes of Eqs. (\ref{eq:stationary_modes_horizon_R}) and (\ref{eq:stationary_modes_horizon_L})
are normalized according to Eq. (\ref{eq:positive_norm}). The change
in sign of the exponent between the right-hand and left-hand regions
comes about because of the change in sign of $c+V$ (see Eq. (\ref{eq:normalized_U_modes}));
it ensures that both $\phi_{\omega,R}^{u}$ and $\phi_{\omega,L}^{u}$
have positive norm.

\section{Out-modes and in-modes with an event horizon\label{sub:Out-modes-and-in-modes}}

The modes $\phi_{\omega,R}^{u}$ and $\phi_{\omega,L}^{u}$ - defined
above for the case of an event horizon for $u$-modes - are not unique.
Each has the same frequency (in the observer's frame) as the complex
conjugate of the other. Therefore, any linear combination of one with
the complex conjuate of the other is itself a stationary mode. In
order to normalize such a linear combination, we simply employ the
linearity of the scalar product:\begin{multline*}
\left(\alpha\phi_{\omega_{1},R}^{u}+\beta\phi_{\omega_{1},L}^{u\star},\alpha\phi_{\omega_{2},R}^{u}+\beta\phi_{\omega_{2},L}^{u\star}\right)\\
=\alpha^{\star}\alpha\left(\phi_{\omega_{1},R}^{u},\phi_{\omega_{2},R}^{u}\right)+\alpha^{\star}\beta\left(\phi_{\omega_{1},R}^{u},\phi_{\omega_{2},L}^{u\star}\right)+\beta^{\star}\alpha\left(\phi_{\omega_{1},L}^{u\star},\phi_{\omega_{2},R}^{u}\right)+\beta^{\star}\beta\left(\phi_{\omega_{1},L}^{u\star},\phi_{\omega_{2},L}^{u\star}\right)\\
=\left(\left|\alpha\right|^{2}-\left|\beta\right|^{2}\right)\delta\left(\omega_{1}-\omega_{2}\right)\,.\end{multline*}
So, if $\left|\alpha\right|^{2}-\left|\beta\right|^{2}=1$, the linear
combination $\alpha\phi_{\omega,R}^{u}+\beta\phi_{\omega,L}^{u\star}$
is normalized with positive norm; it is also automatically orthogonal
to any such combination involving a different frequency or the modes
$\phi_{\omega,L}^{u}$ and $\phi_{\omega,R}^{u\star}$, since the
individual modes are orthogonal to these. Therefore, any set of modes\begin{alignat}{1}
\phi_{\omega,1}^{u}=\alpha_{\omega,1}\phi_{\omega,R}^{u}+\beta_{\omega,1}\phi_{\omega,L}^{u\star}\,,\qquad & \phi_{\omega,2}^{u}=\alpha_{\omega,2}\phi_{\omega,L}^{u}+\beta_{\omega,2}\phi_{\omega,R}^{u\star}\,,\label{eq:generalized_U_modes}\end{alignat}
where\begin{equation}
\left|\alpha_{\omega,j}\right|^{2}-\left|\beta_{\omega,j}\right|^{2}=1\,,\label{eq:normalization_of_U_modes}\end{equation}
form a complete set of orthonormal, positive-norm $u$-modes. A transformation
between different sets of modes is called a \textit{Bogoliubov transformation}.

Despite the limitless possibilities, very few of these sets of modes
are useful. We should deal only with those sets of modes which correspond
to possible measurements. Consider the simplest example: $\alpha_{\omega,j}=1$,
$\beta_{\omega,j}=0$. This, of course, is simply the set of localized
modes given in Eqs. (\ref{eq:stationary_modes_horizon_R}) and (\ref{eq:stationary_modes_horizon_L}).
As we have seen, these modes correspond to single outgoing wavepackets.
An outgoing phonon travelling to the right with frequency $\omega$
is precisely an excitation of the mode $\phi_{\omega,R}^{u}$; similarly,
a phonon travelling to the left with frequency $\omega$ is an excitation
of $\phi_{\omega,L}^{u}$. They differ in this respect from the other
linear combinations, which correspond to two outgoing wavepackets
rather than just one. In light of this property, the set of modes
$\phi_{\omega,R}^{u}$ and $\phi_{\omega,L}^{u}$ are termed \textit{out-modes}:
they correspond to a single outgoing wave.

We expect that there should also exist a set of modes that correspond
to a single ingoing wave. However, the behaviour of the modes already
found would seem to belie this expectation. As noted in §\ref{sub:Event-horizon}
above, all $u$-modes drift away from the horizon; how can we possibly
form an ingoing wavepacket from these? We cannot, of course; not while
there is a horizon. There is a trick, however, due to Unruh and others,
which we borrow from the gravitational case \cite{Unruh-1976,Damour-Ruffini-1976,Hawking-1976}.
There, the analysis normally begins by assuming that a horizon has
not always existed, but forms at some finite time from a collapsing
spacetime. The in-modes, then, are well-defined, for they are constructed
in a spacetime where a horizon has not yet formed. It is found that
the results of this analysis, in which in-modes are defined as positive-norm
with respect to the initial time coordinate, are identical to the
results obtained by ignoring the formation process and defining the
in-modes as positive-norm with respect to the Kruskal coordinate.
Now, near $x=0$, the unit vector defined by the Kruskal coordinate
is proportional, but opposite in direction, to the unit vector defined
by the $x$-coordinate. Therefore, purely positive-norm modes with
respect to the Kruskal coordinate must be analytic in the upper-half
$x$-plane near $x=0$. This completely defines the in-modes: $x$,
on traversing the upper-half complex plane to $-x$, simply picks
up a phase factor of $e^{i\pi}$, so that the relative amplitude of
the left-hand mode compared to the right-hand mode is $e^{-\pi\omega/\alpha}$.
After normalizing, we have\begin{eqnarray}
\phi_{\omega,R}^{u,\mathrm{in}} & = & \frac{1}{\sqrt{2\,\sinh\left(\frac{\pi\omega}{\alpha}\right)}}\left(e^{\frac{\pi\omega}{2\alpha}}\phi_{\omega,R}^{u,\mathrm{out}}+e^{-\frac{\pi\omega}{2\alpha}}\phi_{\omega,L}^{u,\mathrm{out}\star}\right)\,,\nonumber \\
\phi_{\omega,L}^{u,\mathrm{in}} & = & \frac{1}{\sqrt{2\,\sinh\left(\frac{\pi\omega}{\alpha}\right)}}\left(e^{\frac{\pi\omega}{2\alpha}}\phi_{\omega,L}^{u,\mathrm{out}}+e^{-\frac{\pi\omega}{2\alpha}}\phi_{\omega,R}^{u,\mathrm{out}\star}\right)\,.\label{eq:in-modes_out-modes}\end{eqnarray}

The in-modes and out-modes - together with their complex conjugates
- are distinct sets of orthonormal $u$-modes. Taken together with
the $v$-modes, they form a complete set of orthonormal modes which
are solutions of the wave equation (\ref{eq:dispersionless_acoustic_wave_equation}).
Therefore, we have two natural ways of decomposing the $u$ part of
the field operator in Eq. (\ref{eq:quantized_acoustic_field}) when
a horizon is present:\begin{eqnarray}
\negthickspace\negthickspace\negthickspace\hat{\phi}\left(t,x\right) & = & \int_{0}^{\infty}d\omega\left\{ \hat{a}_{\omega,R}^{u,\mathrm{in}}\phi_{\omega,R}^{u,\mathrm{in}}\left(t,x\right)+\hat{a}_{\omega,L}^{u,\mathrm{in}}\phi_{\omega,L}^{u,\mathrm{in}}\left(t,x\right)+\hat{a}_{\omega}^{v}\phi_{\omega}^{v}\left(t,x\right)+\mathrm{h.c.}\right\} \label{eq:field_in-mode_decomposition}\\
 & = & \int_{0}^{\infty}d\omega\left\{ \hat{a}_{\omega,R}^{u,\mathrm{out}}\phi_{\omega,R}^{u,\mathrm{out}}\left(t,x\right)+\hat{a}_{\omega,L}^{u,\mathrm{out}}\phi_{\omega,L}^{u,\mathrm{out}}\left(t,x\right)+\hat{a}_{\omega}^{v}\phi_{\omega}^{v}\left(t,x\right)+\mathrm{h.c.}\right\} \,.\label{eq:field_out-mode_decomposition}\end{eqnarray}
Given that these two representations are equivalent, and that we know
the Bogoliubov transformation that relates the in- and out-modes,
we also find that an analogous transformation exists between the quantum
amplitude operators. Plugging the transformation into Eq. (\ref{eq:field_in-mode_decomposition})
and comparing with Eq. (\ref{eq:field_out-mode_decomposition}), we
find:\begin{eqnarray}
\hat{a}_{\omega,R}^{u,\mathrm{out}} & = & \frac{1}{\sqrt{2\,\sinh\left(\frac{\pi\omega}{\alpha}\right)}}\left(e^{\frac{\pi\omega}{2\alpha}}\,\hat{a}_{\omega,R}^{u,\mathrm{in}}+e^{-\frac{\pi\omega}{2\alpha}}\,\hat{a}_{\omega,L}^{u,\mathrm{in}\dagger}\right)\,,\nonumber \\
\hat{a}_{\omega,L}^{u,\mathrm{out}} & = & \frac{1}{\sqrt{2\,\sinh\left(\frac{\pi\omega}{\alpha}\right)}}\left(e^{\frac{\pi\omega}{2\alpha}}\,\hat{a}_{\omega,L}^{u,\mathrm{in}}+e^{-\frac{\pi\omega}{2\alpha}}\,\hat{a}_{\omega,R}^{u,\mathrm{in}\dagger}\right)\,.\label{eq:operator_transformation}\end{eqnarray}

\section{Spontaneous creation of phonons\label{sub:Spontaneous-creation}}

The quantum amplitudes $\hat{a}$ and $\hat{a}^{\dagger}$ are to
be interpreted as annihilation and creation operators, respectively.
That is, the operator $\hat{a}^{\dagger}$ has the effect of exciting
its corresponding mode by a single quantum, while the operator $\hat{a}$
has the opposite effect. The \textit{vacuum state} is that in which
no modes are excited: it is the absence of {}``particles''. Since
one cannot annihilate an excitation from the vacuum state, it must
vanish when acted on by any annihilation operator. Mathematically
speaking, then, the vacuum state is uniquely defined as the eigenstate
of all annihilation operators with eigenvalue zero:\begin{equation}
\hat{a}\left|0\right\rangle =0\qquad\forall\:\hat{a}\,,\label{eq:definition_of_vacuum}\end{equation}
where $\left|0\right\rangle $ denotes the vacuum state.

The vacuum state would seem to depend on the specific set of annihilation
operators - and, hence, on the specific set of orthonormal modes -
used to define it. However, if the positive-norm modes of one set
are formed from linear combinations of positive-norm modes of the
other set - and, equivalently, one set's negative-norm modes are linear
combinations of the other set's negative-norm modes - then the annihilation
operators of each set are also linear combinations of each other.
In this case, the vacuum states defined by the two sets of modes are
identical, for all annihilation operators, of either set, give zero
when acting on $\left|0\right\rangle $. Thus, the vacuum state is
uniquely defined by the space spanned by the positive-norm modes,
and is not so sensitive to the individual modes themselves.

The corollary of the previous statement is perhaps more interesting:
two sets of modes, whose positive-norm elements span different spaces,
define different vacuum states. Although the entire space, combining
both positive- and negative-norm modes (that is, the field operator
$\hat{\phi}$), is exactly the same, a state which is empty with respect
to one set of modes can look non-empty with respect to another, provided
the elements of one set combine positive- \textit{and} negative-norm
modes of the other.

This is precisely the situation with the in- and out-modes of Eq.
(\ref{eq:in-modes_out-modes}). We conclude that the vacua defined
by these two sets of modes - called, respectively, the \textit{in-vacuum}
and the \textit{out-vacuum} - are different. The in-vacuum is that
state which contains no phonons in the infinite past; since all wavepackets
are incoming (travelling towards the horizon) in the infinite past,
this is the same as saying that there are no incoming phonons. Conversely,
the out-vacuum contains no phonons in the infinite future, and since
all wavepackets are outgoing there, this is equivalent to there being
no outgoing phonons. The inequality of the two vacua means that, if
there are no incoming phonons - the in-vacuum $\left|0_{\mathrm{in}}\right\rangle $
being the usual, and most natural, assumption for the state - then
there \textit{will} be outgoing phonons. The event horizon leads to
spontaneous particle creation.

We may show this explicitly using Eq. (\ref{eq:operator_transformation}).
The expectation value of right-moving outgoing phonons in the in-vacuum
is\begin{eqnarray}
\left\langle n_{\omega,R}^{u,\mathrm{out}}\right\rangle  & = & \left\langle 0_{\mathrm{in}}\right|\hat{a}_{\omega,R}^{u,\mathrm{out}\dagger}\hat{a}_{\omega^{\prime},R}^{u,\mathrm{out}}\left|0_{\mathrm{in}}\right\rangle \nonumber \\
 & = & \left\langle 0_{\mathrm{in}}\right|\frac{1}{\sqrt{2\,\sinh\left(\frac{\pi\omega}{\alpha}\right)}}\left(e^{\frac{\pi\omega}{2\alpha}}\hat{a}_{\omega,R}^{u,\mathrm{in}\dagger}+e^{-\frac{\pi\omega}{2\alpha}}\hat{a}_{\omega,L}^{u,\mathrm{in}}\right)\frac{1}{\sqrt{2\,\sinh\left(\frac{\pi\omega^{\prime}}{\alpha}\right)}}\left(e^{\frac{\pi\omega^{\prime}}{2\alpha}}\hat{a}_{\omega^{\prime},R}^{u,\mathrm{in}}+e^{-\frac{\pi\omega^{\prime}}{2\alpha}}\hat{a}_{\omega^{\prime},L}^{u,\mathrm{in}\dagger}\right)\left|0_{\mathrm{in}}\right\rangle \nonumber \\
 & = & \frac{1}{\sqrt{4\,\sinh\left(\frac{\pi\omega}{\alpha}\right)\,\sinh\left(\frac{\pi\omega^{\prime}}{\alpha}\right)}}e^{-\frac{\pi\left(\omega+\omega^{\prime}\right)}{2\alpha}}\left\langle 0_{\mathrm{in}}\right|\hat{a}_{\omega,L}^{u,\mathrm{in}}\hat{a}_{\omega^{\prime},L}^{u,\mathrm{in}\dagger}\left|0_{\mathrm{in}}\right\rangle \nonumber \\
 & = & \frac{1}{\sqrt{4\,\sinh\left(\frac{\pi\omega}{\alpha}\right)\,\sinh\left(\frac{\pi\omega^{\prime}}{\alpha}\right)}}e^{-\frac{\pi\left(\omega+\omega^{\prime}\right)}{2\alpha}}\left\langle 0_{\mathrm{in}}\right|\hat{a}_{\omega^{\prime},L}^{u,\mathrm{in}\dagger}\hat{a}_{\omega,L}^{u,\mathrm{in}}+\delta\left(\omega-\omega^{\prime}\right)\left|0_{\mathrm{in}}\right\rangle \nonumber \\
 & = & \frac{1}{2\,\sinh\left(\frac{\pi\omega}{\alpha}\right)}e^{-\frac{\pi\omega}{\alpha}}\,\delta\left(\omega-\omega^{\prime}\right)\nonumber \\
 & = & \frac{1}{e^{\frac{2\pi\omega}{\alpha}}-1}\,\delta\left(\omega-\omega^{\prime}\right)\,.\label{eq:thermal_spectrum}\end{eqnarray}
Remarkably, the spectrum of phonons emitted is precisely a bosonic
thermal distribution with temperature $\alpha/\left(2\pi\right)$;
replacing fundamental constants,\begin{equation}
k_{B}T=\frac{\hbar\alpha}{2\pi}\,.\label{eq:Hawking_temperature}\end{equation}
Although the result in Eq. (\ref{eq:thermal_spectrum}) has been derived
as a number expectation value, the appearance of the delta function
shows that it is really a density. In §\ref{sub:Spontaneous-creation-II}
(and Ref. \cite{Corley-Jacobson-1996}), it is shown that this is
proportional to the spectral flux density,\begin{equation}
\frac{d^{2}N}{d\omega\, dt}=\frac{1}{2\pi}\frac{1}{e^{\frac{2\pi\omega}{\alpha}}-1}\,,\end{equation}
the number of phonons emitted per unit frequency interval per unit
time.

Due to the symmetry of the transformation (\ref{eq:operator_transformation}),
the expectation value of left-moving outgoing phonons is exactly equal
to the thermal spectrum of (\ref{eq:thermal_spectrum}). There is
a deeper significance to this than just symmetry, however. The transformation
of the in- and out-operators allows us to write the in-vacuum explicitly
in terms of the out-vacuum. Using the Fock basis, in which the annihilation
and creation operators behave in the standard way:\begin{alignat}{1}
\hat{a}_{\omega}\left|n\right\rangle _{\omega}=\sqrt{n}\left|n-1\right\rangle _{\omega}\,,\qquad & \hat{a}_{\omega}^{\dagger}\left|n\right\rangle _{\omega}=\sqrt{n+1}\left|n+1\right\rangle _{\omega}\,,\label{eq:annihilation_creation_Fock_basis}\end{alignat}
the in-vacuum state is given by\begin{eqnarray}
\left|0_{\mathrm{in}}\right\rangle  & = & Z^{-\frac{1}{2}}\prod_{\omega}\sum_{n=0}^{\infty}\frac{1}{n!}\left(e^{-\frac{\pi\omega}{a}}\,\hat{a}_{\omega,R}^{u,\mathrm{out}\dagger}\,\hat{a}_{\omega,L}^{u,\mathrm{out}\dagger}\right)^{n}\left|0_{\mathrm{out}}\right\rangle \nonumber \\
 & = & Z^{-\frac{1}{2}}\prod_{\omega}\sum_{n=0}^{\infty}e^{-\frac{n\pi\omega}{a}}\left|n\right\rangle _{\omega,R}^{u,\mathrm{out}}\left|n\right\rangle _{\omega,L}^{u,\mathrm{out}}\,,\label{eq:in-vacuum_out-Fock-basis}\end{eqnarray}
where $Z$ is a normalizing prefactor defined such that $\left\langle 0_{\mathrm{in}}\right.\left|0_{\mathrm{in}}\right\rangle =1$.
The fact that the out-creation operators appear only in $R$-$L$
pairs shows that the radiation, though it looks thermal in the right
and left sides separately, is strongly correlated \textit{between}
the two sides. Phonons are emitted in pairs, one in the subsonic,
the other in the supersonic region; and a measurement of the number
of phonons in any state on one side of the horizon infers that there
are equally many phonons in the corresponding state on the other side.
The left and right systems, separated by the horizon, are \textit{entangled}.

\section{Conclusion and discussion\label{sub:Discussion}}

The above derivation is quite general, in that it applies to any spacetime
that can be described by the metric (\ref{eq:generalized_Lemaitre_metric}).
Of course, this includes gravitational black holes (see Eq. (\ref{eq:Lemaitre_metric})):
using $V\left(r\right)=-c\sqrt{r_{S}/r}$, we find that, at the Schwarzschild
radius, $V^{\prime}\left(r_{S}\right)=c/\left(2r_{S}\right)=c^{3}/\left(4GM\right)$,
and plugging into Eq. (\ref{eq:Hawking_temperature}), we have $k_{B}T=\hbar c^{3}/\left(8\pi GM\right)$
- exactly Hawking's original formula, Eq. (\ref{eq:black_hole_temperature}).

In the absence of dispersion, the event horizon is a well-defined
point; for there is only one wave speed, $c$, and it is precisely
where $\left|V\right|=c$ that a horizon is established. Hawking radiation
is derived by linearizing $V$ around this point; and since, as discussed
in §\ref{sub:Event-horizon}, any wavepacket can be traced
back in time to arbitrary thinness and arbitrary closeness to the
horizon, it does not matter how small is the region over which this
linearized velocity profile is valid. Thus, the Hawking temperature
can depend only on the first derivative, $\alpha$, of $V$ at the
horizon.

This is no longer true once dispersion is taken into account. In that case, the wave velocity
with respect to the medium is not constant but varies with frequency. The horizon,
then, is no longer so well-defined, and it is impossible to pick out
a single parameter which determines the Hawking spectrum. Moreover, as has been demonstrated
by Unruh \cite{Unruh-1995} and as will be shown in Chapter \ref{sec:Methods-of-Acoustic-Model},
an outgoing wavepacket does not originate from an arbitrarily thin wavepacket squeezed up against
its own horizon.  Instead, dispersive effects cause a change in the speed of the wavepacket as its wavelength
is reduced, and the wavepacket turns around at a shifted frequency, originating from infinity. 
Thus, with the introduction of dispersion, we are inclined to ask: does the Hawking spectrum survive?
And if so, what is its shape, and on what parameters does it depend?

In the next chapter, we formulate the theoretical framework for a quantized acoustic field, much as in the present
chapter, but with dispersion taken into account.  Chapter \ref{sec:Methods-of-Acoustic-Model} examines some ways
of finding the spectrum in specific circumstances, and Chapter \ref{sec:Results-for-Acoustic-Model} presents some numerical
results.  Later chapters extend the analogy to the case of optical fibres.

\pagebreak{}

\part{Hawking Radiation in Dispersive Fluids\label{par:Horizons-in-Dispersive-Fluids}}

\chapter{The Acoustic Model\label{sec:The-Acoustic-Model}}

In Chapter \ref{sec:Theoretical-Origins-of-Hawking-Radiation}, we
examined the acoustic model in the absence of dispersion. We shall
now deal with the generalization of this model to a dispersive fluid.
Much of the following treatment is directly analogous to the analysis
of Chapter \ref{sec:Theoretical-Origins-of-Hawking-Radiation}. However,
the solutions are more complicated: dispersion leads to the new phenomenon
of frequency shifting, whereby a wavepacket is reflected from a group-velocity
horizon at a different frequency. The trans-Planckian problem is in
this manner avoided, but the solutions are not so tractable as Eq.
(\ref{eq:general_solution}), nor is the Bogoliubov transformation
of Eqs. (\ref{eq:in-modes_out-modes}) and (\ref{eq:operator_transformation})
so simple. The present chapter describes the theoretical model; the
methods used to determine its solutions, and some numerical results
of these methods, are left to Chapters \ref{sec:Methods-of-Acoustic-Model}
and \ref{sec:Results-for-Acoustic-Model}.

\section{The wave equation and its solutions\label{sub:The-wave-equation}}

Dispersion - which describes a frequency-dependent wave velocity -
manifests itself as a breakdown of Lorentz invariance. The wave behaviour
can no longer be described by a spacetime metric, and we must instead
take the Lagrangian as the starting point. Dispersion is modelled
by the appearance of higher-order derivatives in the Lagrangian, so
we must now write the action integral as\begin{equation}
S=\int\int dx\, dt\, L\left(\phi,\partial_{t}\phi,\partial_{x}\phi,\ldots\right).\label{eq:action_with_dispersion}\end{equation}
The Lagrangian, given in the absence of dispersion by Eq. (\ref{eq:scalar_massless_Lagrangian_moving_fluid}),
is now generalized to \cite{Unruh-1995,Corley-Jacobson-1996}\begin{equation}
L=\frac{1}{2}\left|\left(\partial_{t}+V\partial_{x}\right)\phi\right|^{2}-\frac{1}{2}\left|F\left(-i\partial_{x}\right)\phi\right|^{2}\,.\label{eq:Lagrangian_with_dispersion}\end{equation}
$F\left(k\right)$ describes the dispersive properties of the fluid;
as a function of an operator, we consider it as a Taylor series in
that operator. We require that $F\left(k\right)$ is an odd function,
so that $F\left(-i\partial_{x}\right)\phi$ contains only odd-degree
derivatives of $\phi$.

An extremum of the action is found by infinitesimally varying the
field $\phi$ and its derivatives $\partial_{t}\phi$, $\partial_{x}\phi$,
$\partial_{x}^{2}\phi$, etc., then setting the resulting variation
in $S$ to zero. The result is the following Euler-Lagrange equation:\begin{equation}
\frac{\partial L}{\partial\phi^{\star}}-\frac{\partial}{\partial t}\left(\frac{\partial L}{\partial\left(\partial_{t}\phi^{\star}\right)}\right)-\frac{\partial}{\partial x}\left(\frac{\partial L}{\partial\left(\partial_{x}\phi^{\star}\right)}\right)+\frac{\partial^{2}}{\partial x^{2}}\left(\frac{\partial L}{\partial\left(\partial_{x}^{2}\phi^{\star}\right)}\right)-\frac{\partial^{3}}{\partial x^{3}}\left(\frac{\partial L}{\partial\left(\partial_{x}^{3}\phi^{\star}\right)}\right)+\ldots=0\,.\label{eq:Euler_Lagrange_higher_derivatives}\end{equation}
Substituting $L$ from Eq. (\ref{eq:Lagrangian_with_dispersion})
- and using the fact that $F\left(k\right)$ is odd when performing
integration by parts - we find the acoustic wave equation \cite{Unruh-1995}:\begin{equation}
\left(\partial_{t}+\partial_{x}V\right)\left(\partial_{t}+V\partial_{x}\right)\phi+F^{2}\left(-i\partial_{x}\right)\phi=0\,.\label{eq:acoustic_wave_equation}\end{equation}
If $F^{2}\left(k\right)$ is a polynomial of finite degree, this is
a partial differential equation that can be solved numerically. However,
$F^{2}\left(k\right)$ need not be of finite degree, and it is useful
to express its operation on $\phi$ in terms of its Fourier transform \cite{Unruh-1995}: by definition, we have\begin{equation}
\phi\left(t,x\right)=\frac{1}{2\pi}\int_{-\infty}^{+\infty}dk\,\widetilde{\phi}\left(t,k\right)e^{ikx},\label{eq:inverse_Fourier_transform_x}\end{equation}
so that\begin{eqnarray}
F^{2}\left(-i\partial_{x}\right)\phi\left(t,x\right) & = & \frac{1}{2\pi}\int_{-\infty}^{+\infty}dk\, F^{2}\left(k\right)\widetilde{\phi}\left(t,k\right)e^{ikx}\nonumber \\
 & = & \frac{1}{2\pi}\int_{-\infty}^{+\infty}dk\,\int_{-\infty}^{+\infty}dx^{\prime}\, F^{2}\left(k\right)\phi\left(t,x^{\prime}\right)e^{ik\left(x-x^{\prime}\right)}\,.\label{eq:operation_of_F}\end{eqnarray}

\begin{figure}
\includegraphics[width=0.6\columnwidth]{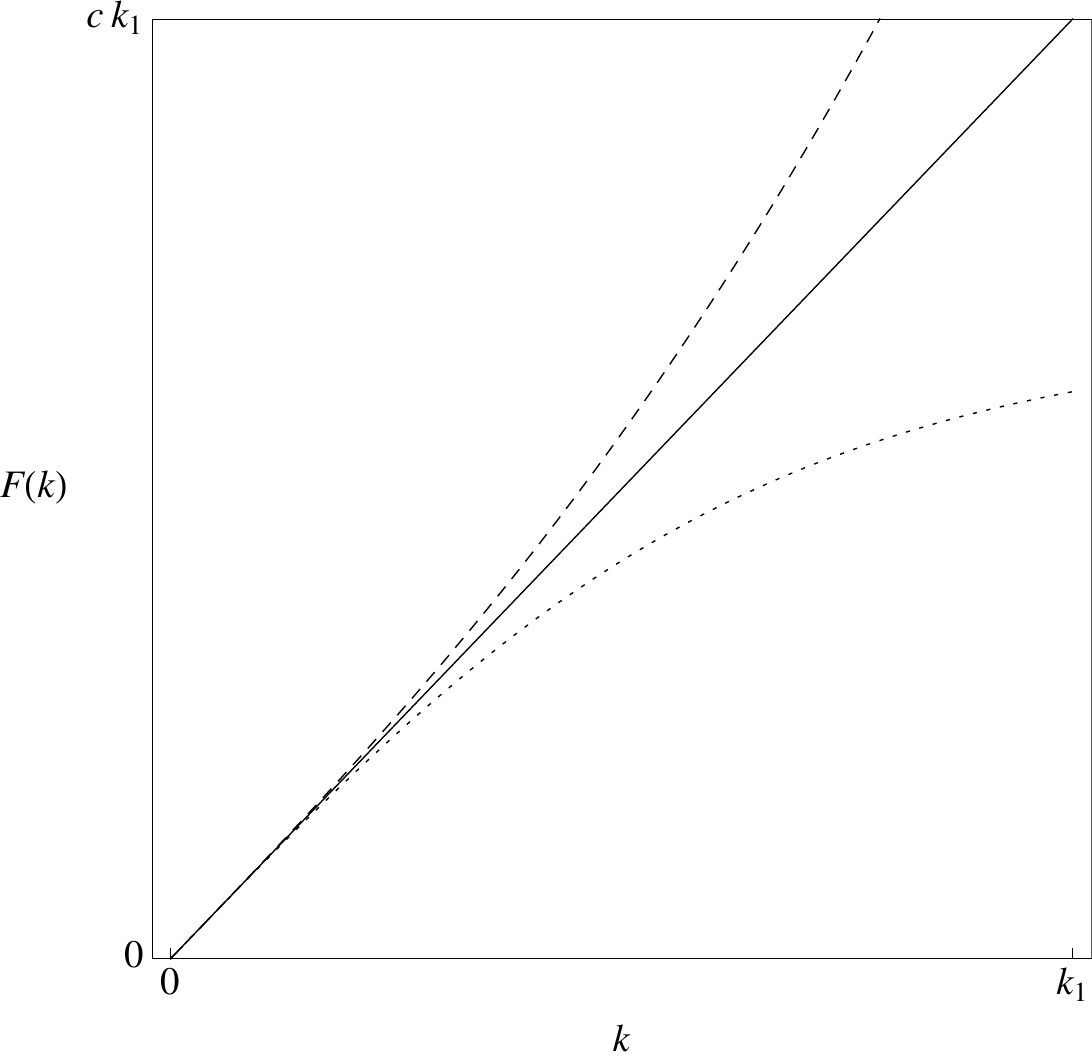}

\caption[\textsc{Dispersion profiles}]{\textsc{Dispersion profiles}: The solid line shows the dispersionless
curve, the dashed curve corresponds to superluminal dispersion, and
the dotted curve corresponds to subluminal dispersion.\label{fig:dispersion_sup_sub}}

\end{figure}

Let us assume for now that the flow velocity $V$ is constant in space.
Then, taking the Fourier transform of Eq. (\ref{eq:acoustic_wave_equation})
by making the substitutions $\phi\left(t,x\right)\rightarrow\widetilde{\phi}\left(\omega,k\right)$,
$\partial_{t}\rightarrow-i\omega$ and $\partial_{x}\rightarrow ik$,
we find \begin{equation}
\left[-\left(\omega-Vk\right)^{2}+F^{2}\left(k\right)\right]\widetilde{\phi}\left(\omega,k\right)=0\,,\label{eq:acoustic_dispersion_FT}\end{equation}
so that $\widetilde{\phi}\left(\omega,k\right)$ can only be non-zero
when $\omega$ and $k$ satisfy\begin{equation}
\left(\omega-Vk\right)^{2}=F^{2}\left(k\right)\,.\label{eq:acoustic_flow_dispersion}\end{equation}
This is the generalization to dispersive media of Eq. (\ref{eq:Doppler_formula}).
We see that the function $\pm F\left(k\right)$ simply gives the free-fall
frequency, $\omega_{\mathrm{ff}}$, as a function of the wavevector
$k$. (The sign is determined as in Eq. (\ref{eq:free-fall-freq_u-and-v}).)
In the absence of dispersion, the relation is one of direct proportionality:
$F\left(k\right)=ck$. Typically, waves exhibit very little dispersion
for low values of $k$, so that $F\left(k\right)\rightarrow ck$ as
$k\rightarrow0$. For higher values of $k$, the dispersion can take
two basic forms: superluminal and subluminal. These are defined according
to whether the magnitude of the phase velocity, $\left|\omega/k\right|$,
becomes higher or lower than $c$ as $k$ increases; examples are
shown in Figure \ref{fig:dispersion_sup_sub}. Our specification that
$F\left(k\right)$ be an odd function means that the fluid (in its
rest frame) is isotropic, for $-k$ is simply a spatial inversion
of $k$.

\begin{figure}
\subfloat{\includegraphics[width=0.45\columnwidth]{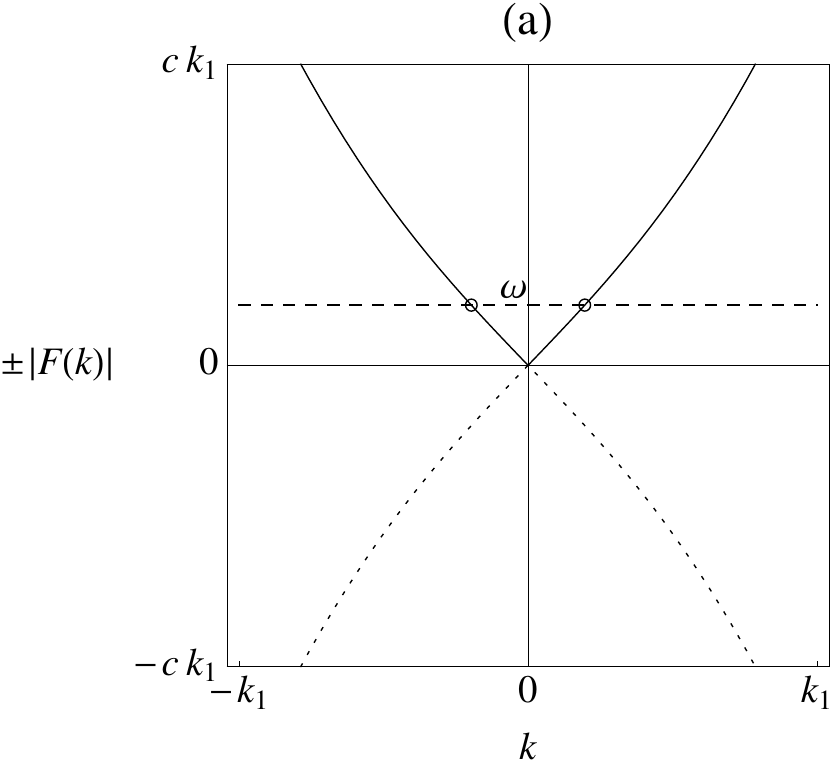}} \subfloat{\includegraphics[width=0.45\columnwidth]{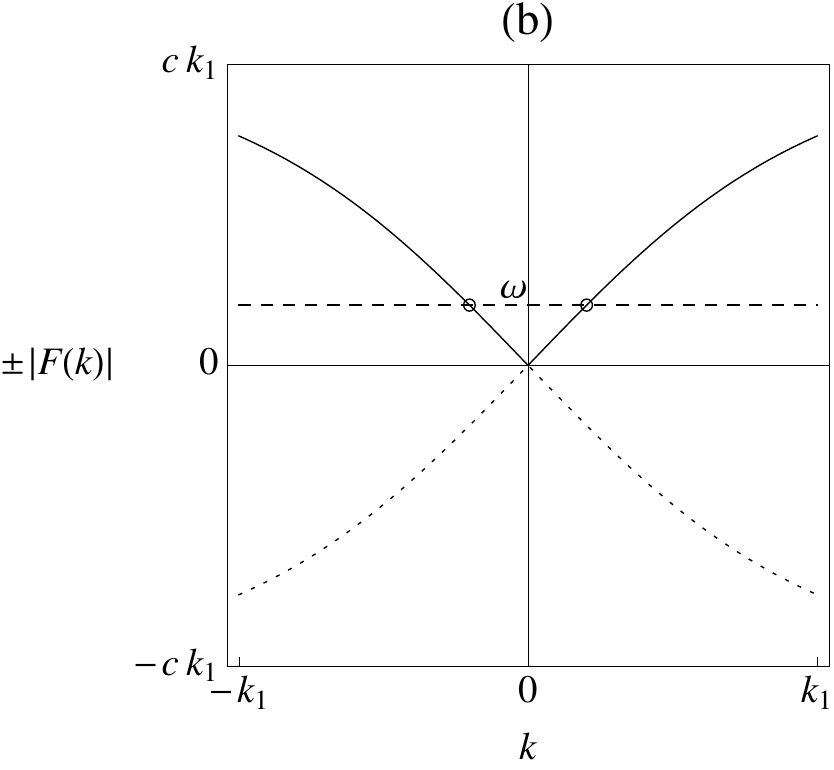}}

\subfloat{\includegraphics[width=0.45\columnwidth]{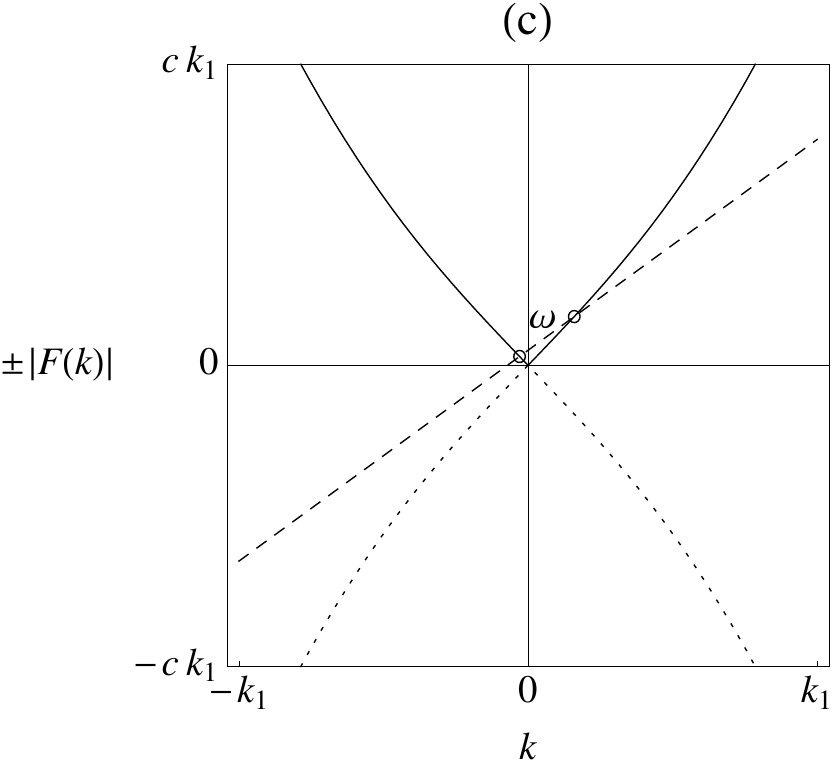}} \subfloat{\includegraphics[width=0.45\columnwidth]{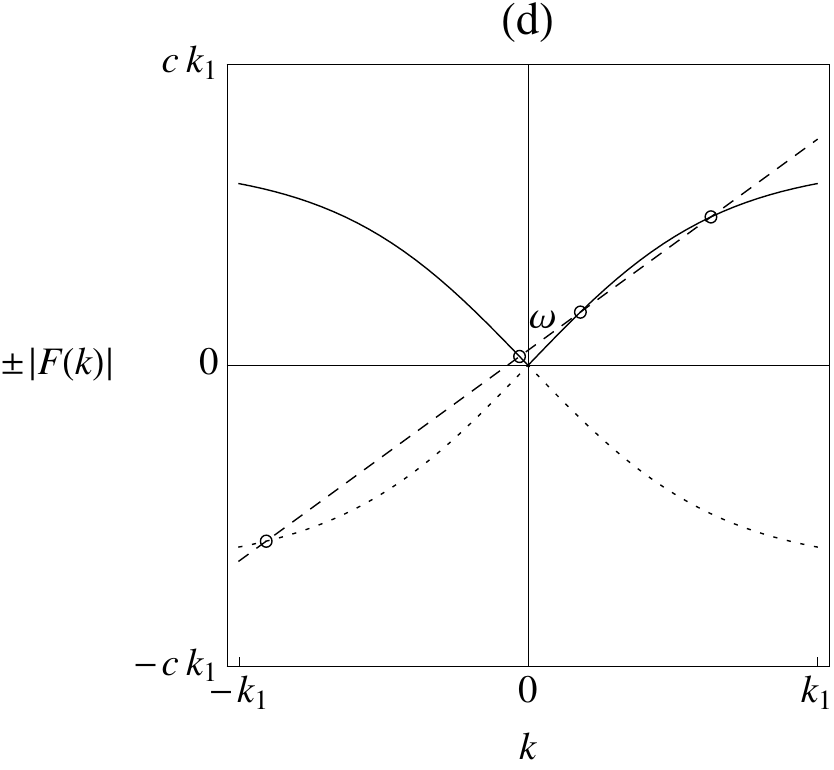}}

\caption[\textsc{Solutions at constant flow}]{\textsc{Solutions at constant flow}: For a given $\omega$ and $V$,
the possible values of $k$ are solutions of $\omega-Vk=\pm\left|F\left(k\right)\right|$.
Figures $\left(a\right)$ and $\left(b\right)$ show the solutions
when the fluid is static ($V=0$) for superluminal and subluminal
dispersion, respectively; each frequency has one $u$-mode and one
$v$-mode, as in the dispersionless case. Figures $\left(c\right)$
and $\left(d\right)$ show the solutions when $-c<V<0$; superluminal
dispersion still gives one $u$-mode and one $v$-mode, but subluminal
dispersion gives three possible $u$-modes, one of which has negative
free-fall frequency.\label{fig:Solutions-at-Constant-Flow}}

\end{figure}

For a given frequency, the possible (real) values of $k$ can by found
by plotting $\pm\left|F\left(k\right)\right|$ and the line $\omega-Vk$,
then reading off their points of intersection; examples are shown
in Figure \ref{fig:Solutions-at-Constant-Flow}. (The sign now determines
the sign of the free-fall frequency, and not whether the mode is on
the $u$- or $v$-branch.) When the fluid is static, the possible
wavevectors always occur in pairs of opposite sign and velocity: the
$u$- and $v$-branches, familiar from the dispersionless case. When
the fluid is flowing, the possible wavevectors still split into $u$-
and $v$-branches. But now, as in Fig. \ref{fig:Solutions-at-Constant-Flow}$\left(d\right)$,
something else can occur: a given $\omega$ can have several different
wavevectors on the same branch. It can even have wavevectors with
different signs of the free-fall frequency, and given remarks made
in §\ref{sub:Stationary_modes}, this suggests that a given $\omega$
may have solutions of oppositely-signed norm. We should bear this
possibility in mind, for it lies at the heart of the Hawking effect.
While the route may not be obvious at this stage, the fact that mixing
of modes of oppositely-signed norm leads to Hawking radiation in the
dispersionless case (see §\ref{sub:Spontaneous-creation}) would suggest
that the same is true here.

The case of a constant fluid flow is not particularly interesting,
for any wavepacket, centred around any of the possible solutions of
the dispersion relation (\ref{eq:acoustic_flow_dispersion}), simply
propagates from one infinity to the other. The analysis can be performed
in any Galilean frame, for the fluid flow velocity will always be
constant. The fact that certain values of $k$ share the same value
of $\omega$ in a certain reference frame is of no consequence, because
this will not be so in another reference frame; this mathematical
relationship is purely incidental. In particular, the relationship
between $k$-values with different signs of the free-fall frequency
is physically meaningless and purely a matter of definition; for,
we can always transform to the free-fall frame, where such a relationship
cannot (by definition) occur.

This picture changes radically when the fluid flow velocity is not
constant, but has some spatial dependence. In this situation, we cannot
transform into any Galilean frame without forcing the fluid velocity
profile to take on a time dependence as well as a spatial one. There
is now a naturally preferred reference frame: that in which the fluid
velocity profile is dependent only on position and not on time. The
\textit{Killing frequency} - that which is conserved - is the value
of $\omega$ measured in this preferred reference frame. During the
propagation of a wavepacket, $k$ may be forced to change, but it
can only change to a value of $k$ with the same value of $\omega$.
It is in this frame that such relationships - including that between
different signs of the free-fall frequency - are physically meaningful.

\section{Scalar product\label{sub:Scalar-product}}

As discussed in §\ref{sub:Acoustic_field_wave_equation}, invariance
of the Lagrangian under phase rotation, $\phi\rightarrow\phi\, e^{i\alpha}$,
leads to conservation of the scalar product:

\begin{eqnarray}
\left(\phi_{1},\phi_{2}\right) & = & i\int_{-\infty}^{+\infty}\left\{ \phi_{1}^{\star}\left(\partial_{t}+V\partial_{x}\right)\phi_{2}-\phi_{2}\left(\partial_{t}+V\partial_{x}\right)\phi_{1}^{\star}\right\} dx\nonumber \\
 & = & i\int_{-\infty}^{+\infty}\left\{ \phi_{1}^{\star}\pi_{2}-\phi_{2}\pi_{1}^{\star}\right\} dx\,,\label{eq:scalar_product_II}\end{eqnarray}
where the canonical momentum is given by\begin{equation}
\pi=\frac{\partial L}{\partial\left(\partial_{t}\phi^{\star}\right)}=\left(\partial_{t}+V\partial_{x}\right)\phi\,.\label{eq:canonical_momentum_II}\end{equation}
As before, the norm is defined as the scalar product of a wave with
itself.

If $V$ is constant, or if two wavepackets are localised in a region
where $V$ is constant, the scalar product takes a simple form. Here,
$\omega$ and $k$ are related via the dispersion relation in Eq.
(\ref{eq:acoustic_flow_dispersion}), so that $\omega=Vk\pm\left|F\left(k\right)\right|$,
where, as has already been mentioned, the plus or minus sign refers
to a positive or negative free-fall frequency. Rather than splitting
the dispersion relation into $u$- and $v$-branches, then, we may
split it into branches of positive and negative free-fall frequency.
A general solution of the wave equation can be expressed as a sum
of two Fourier integrals, one over each branch:\begin{equation}
\phi\left(x,t\right)=\frac{1}{2\pi}\int_{-\infty}^{+\infty}\widetilde{\phi}^{(+)}\left(k\right)e^{ikx-i\left(Vk+\left|F\left(k\right)\right|\right)t}dk+\frac{1}{2\pi}\int_{-\infty}^{+\infty}\widetilde{\phi}^{(-)}\left(k\right)e^{ikx-i\left(Vk-\left|F\left(k\right)\right|\right)t}dk\,.\label{eq:phi_general_soln}\end{equation}
The canonical momentum is given by Eq. (\ref{eq:canonical_momentum_II}):\begin{multline}
\pi\left(x,t\right)=\frac{-i}{2\pi}\int_{-\infty}^{+\infty}\left|F\left(k\right)\right|\widetilde{\phi}^{(+)}\left(k\right)e^{ikx-i\left(Vk+\left|F\left(k\right)\right|\right)t}dk\\
+\frac{i}{2\pi}\int_{-\infty}^{+\infty}\left|F\left(k\right)\right|\widetilde{\phi}^{(-)}\left(k\right)e^{ikx-i\left(Vk-\left|F\left(k\right)\right|\right)t}dk\,.\label{eq:pi_general_soln}\end{multline}
Then, applying Eq. (\ref{eq:scalar_product_II}), we find that the
scalar product is\begin{eqnarray}
\left(\phi_{1},\phi_{2}\right) & = & \frac{1}{\pi}\int_{-\infty}^{+\infty}\left|F\left(k\right)\right|\left\{ \widetilde{\phi}_{1}^{(+)\star}\left(k\right)\widetilde{\phi}_{2}^{(+)}\left(k\right)-\widetilde{\phi}_{1}^{(-)\star}\left(k\right)\widetilde{\phi}_{2}^{(-)}\left(k\right)\right\} dk\,.\label{eq:scalar_product_constant_v}\end{eqnarray}
That is, the two branches of the dispersion relation are decoupled,
contributing separately to the scalar product, and the integrand is
simply the product of the Fourier transforms multiplied by the free-fall
frequency. This formula is particularly useful when calculating the
norm of a wave, in which case it becomes\begin{equation}
\left(\phi,\phi\right)=\frac{1}{\pi}\int_{-\infty}^{+\infty}\left|F\left(k\right)\right|\left\{ \left|\widetilde{\phi}^{\left(+\right)}\left(k\right)\right|^{2}-\left|\widetilde{\phi}^{\left(-\right)}\left(k\right)\right|^{2}\right\} dk\,.\label{eq:norm_constant_v}\end{equation}

Conservation of the scalar product, and the norm in particular, is
a very important characteristic of the acoustic wave equation. Much
like conservation of energy and momentum in mechanics, it places a
restriction on the types of evolution that can occur. On quantizing
the acoustic field, normalized modes are multiplied by quantum amplitudes,
and the norm corresponds to the excitation or {}``particle'' number.
Taking this into account, Eq. (\ref{eq:norm_constant_v}) makes the
importance of mixing positive and negative free-fall frequencies more
apparent: these different spectral components contribute to the norm
\textit{with different signs} (thus confirming that Eq. (\ref{eq:norm_and_free-fall-freq})
continues to hold once dispersion is introduced). If a positive-norm
mode is partly converted into a negative-norm mode, this must be accompanied
by an increase in the norm from purely positive-norm modes; the positive-norm
modes are amplified. Thus, a positive-norm particle is not converted
into a negative-frequency particle; rather, two particles, one with
positive and one with negative norm, are created, so that the total
particle number increases. If the original state is vacuum, then these
pairs of particles are created out of vacuum fluctuations. This is
the process that gives rise to Hawking radiation.

\section{Decomposition into modes\label{sub:Decomposition-into-modes}}

The following analyis is inspired by, and closely follows, that of
Macher and Parentani \cite{Macher-Parentani-2008}.

Let us continue to assume that $V$ is constant. We saw in Eq. (\ref{eq:phi_general_soln})
that we can write $\phi$ as a sum over the plane waves $e^{ikx-i\omega t}$,
where $\omega=Vk\pm\left|F\left(k\right)\right|$. These two branches
of the dispersion relation have a very simple correspondence - a plane
wave on one branch is the complex conjugate of a plane wave on the
other:\[
e^{ikx-i\left(Vk+\left|F\left(k\right)\right|\right)t}=\left[e^{i\left(-k\right)x-i\left(V\left(-k\right)-\left|F\left(-k\right)\right|\right)t}\right]^{\star}\,.\]
Thus, we needn't worry about there being two branches to the dispersion
relation. For every value of $k$, we uniquely define $\omega=Vk+\left|F\left(k\right)\right|$,
and then we include the complex conjugates of the resulting plane
waves. This exhausts the possible modes; that is, they form a complete
set of solutions. (Note that we are presently working in the $k$-representation,
since we may write an exact expression for $\omega$ as a function
of $k$, but not vice versa.)

Before making use of the completeness of this set, we should first
normalise them with respect to the scalar product. Applying Eq. (\ref{eq:scalar_product_II}),
we find, for two positive-norm plane waves,\[
\left(e^{ik_{1}x-i\left(Vk_{1}+\left|F\left(k_{1}\right)\right|\right)t},e^{ik_{2}x-i\left(Vk_{2}+\left|F\left(k_{2}\right)\right|\right)t}\right)=4\pi\left|F\left(k_{1}\right)\right|\delta\left(k_{1}-k_{2}\right)\,,\]
while, for two plane waves of oppositely-signed norm,\[
\left(e^{ik_{1}x-i\left(Vk_{1}+\left|F\left(k_{1}\right)\right|\right)t},e^{ik_{2}x-i\left(Vk_{2}-\left|F\left(k_{2}\right)\right|\right)t}\right)=0\,.\]
This agrees with the corresponding result in the dispersionless case
(see §\ref{sub:Stationary_modes}), where an extra factor of $c$
appears because the delta function has its argument in $\omega$ rather
than $k$.

An inspection of Eq. (\ref{eq:scalar_product_II}) reveals the identities\begin{eqnarray}
\left(\phi_{1}^{\star},\phi_{2}^{\star}\right) & = & -\left(\phi_{1},\phi_{2}\right)^{\star}\,,\label{eq:scalar_product_conjugate}\\
\left(\phi_{2},\phi_{1}\right) & = & \left(\phi_{1},\phi_{2}\right)^{\star}\,.\label{eq:scalar_product_commutation}\end{eqnarray}
Defining the mode $\phi_{k}$ as follows:\begin{equation}
\phi_{k}\left(x,t\right)=\frac{1}{\sqrt{4\pi\left|F\left(k\right)\right|}}e^{ikx-i\left(Vk+\left|F\left(k\right)\right|\right)t}\,,\label{eq:k_mode_defn}\end{equation}
then the $\phi_{k}$ and their complex conjugates $\phi_{k}^{\star}$
form a complete orthonormal set, with\begin{alignat}{1}
\left(\phi_{k_{1}},\phi_{k_{2}}\right)=-\left(\phi_{k_{1}}^{\star},\phi_{k_{2}}^{\star}\right)=\delta\left(k_{1}-k_{2}\right)\,,\qquad & \left(\phi_{k_{1}},\phi_{k_{2}}^{\star}\right)=-\left(\phi_{k_{1}}^{\star},\phi_{k_{2}}\right)=0\,.\label{eq:k_mode_orthonormality}\end{alignat}
We refer to the $\phi_{k}$ as \textit{positive-norm modes} and to
the $\phi_{k}^{\star}$ as \textit{negative-norm modes}. In the free-fall
frame, positive-norm modes are those that:
\begin{itemize}
\item if $k>0$, are right-moving (i.e., $u$-modes);
\item if $k<0$, are left-moving (i.e., $v$-modes).
\end{itemize}
For negative-norm modes, this correspondence is reversed.

We may now write $\phi$ in terms of the modes defined in Eq. (\ref{eq:k_mode_defn}).
As before, we note that $\phi$ represents a real-valued quantity,
so that the coefficients of complex conjugate modes must themselves
be complex conjugates; we have\begin{equation}
\phi\left(x,t\right)=\int_{-\infty}^{+\infty}\left\{ A\left(k\right)\phi_{k}\left(x,t\right)+A^{\star}\left(k\right)\phi_{k}^{\star}\left(x,t\right)\right\} dk\,,\label{eq:phi_k_decomposition}\end{equation}
with canonical momentum given by\begin{equation}
\pi\left(x,t\right)=-i\int_{-\infty}^{+\infty}\left|F\left(k\right)\right|\left\{ A\left(k\right)\phi_{k}\left(x,t\right)-A^{\star}\left(k\right)\phi_{k}^{\star}\left(x,t\right)\right\} dk\,.\label{eq:pi_k_decomposition}\end{equation}
The orthonormality of the modes, characterised by Eq. (\ref{eq:k_mode_orthonormality}),
allows us to express the coefficients very simply:\begin{alignat}{1}
A\left(k\right)=\left(\phi_{k},\phi\right)\,,\qquad & A^{\star}\left(k\right)=-\left(\phi_{k}^{\star},\phi\right)\,.\label{eq:coefficients_k_decomposition}\end{alignat}

The mode decomposition of the field in Eq. (\ref{eq:phi_k_decomposition})
is readily quantized as before (see §\ref{sub:Quantization}).
We promote the real-valued quantities $\phi$ and $\pi$ to Hermitian
operators $\hat{\phi}$ and $\hat{\pi}$; $ $the coefficient of the
positive-norm modes, $A\left(k\right)$, to an annihilation operator,
$\hat{a}_{k}$; and the coefficient of the negative-norm modes, $A^{\star}\left(k\right)$,
to a creation operator, $\hat{a}_{k}^{\dagger}$, the Hermitian conjugate
of $\hat{a}_{k}$:\begin{equation}
\hat{\phi}\left(x,t\right)=\int_{-\infty}^{+\infty}\left\{ \hat{a}_{k}\phi_{k}\left(x,t\right)+\hat{a}_{k}^{\dagger}\phi_{k}^{\star}\left(x,t\right)\right\} dk\,,\label{eq:phi_k_decomposition_quantized}\end{equation}
\begin{equation}
\hat{\pi}\left(x,t\right)=-i\int_{-\infty}^{+\infty}\left|F\left(k\right)\right|\left\{ \hat{a}_{k}\phi_{k}\left(x,t\right)-\hat{a}_{k}^{\dagger}\phi_{k}^{\star}\left(x,t\right)\right\} dk\,.\label{eq:pi_k_decomposition_quantized}\end{equation}
Imposing the equal time commutation relation\begin{equation}
\left[\hat{\phi}\left(t,x\right),\hat{\pi}\left(t,x^{\prime}\right)\right]=i\,\delta\left(x-x^{\prime}\right)\,,\label{eq:phi_pi_commutation_relation}\end{equation}
the annihilation and creation operators are found to satisfy the Bose
commutation relations,\begin{alignat}{1}
\left[\hat{a}_{k_{1}},\hat{a}_{k_{2}}^{\dagger}\right]=\delta\left(k_{1}-k_{2}\right)\,,\qquad & \left[\hat{a}_{k_{1}},\hat{a}_{k_{2}}\right]=\left[\hat{a}_{k_{1}}^{\dagger},\hat{a}_{k_{2}}^{\dagger}\right]=0\,.\label{eq:Bose_commutation_relations_II}\end{alignat}

\section{Transforming to the $\omega$-representation\label{sub:Transforming-to-the-omega-repn}}

Eq. (\ref{eq:phi_k_decomposition_quantized}) describes the field
operator $\hat{\phi}$ as an integral over modes characterised by
the wavenumber $k$. However, since in the general case of an inhomogeneous
velocity profile it is the frequency $\omega$ which is conserved
rather than $k$, we would like to change this to an integral over
$\omega$, so that each mode is separately conserved. In the dispersionless
case, $\omega$ and $k$ are directly proportional, and the transformation
is straightforward. In the presence of dispersion, however, the transformation
depends on the dispersion profile, and one must be careful to include
all possible solutions corresponding to each value of $\omega$.

For the sake of example, let us suppose that the dispersion is subluminal.
We need only deal with the integral over the positive-norm modes,
i.e., those modes with $\omega=Vk+\left|F\left(k\right)\right|$,
keeping in mind that this does not necessarily mean that $\omega$
is positive (although the free-fall frequency $\omega_{\mathrm{ff}}$
certainly is). The cases of supersonic and subsonic flow are qualitatively
different, so let us examine them separately. (Note that, although
the following analysis has been kept fairly general, it may require
some modification given specific examples.)

\subsection{Supersonic flow}

\begin{figure}
\includegraphics[width=0.8\columnwidth]{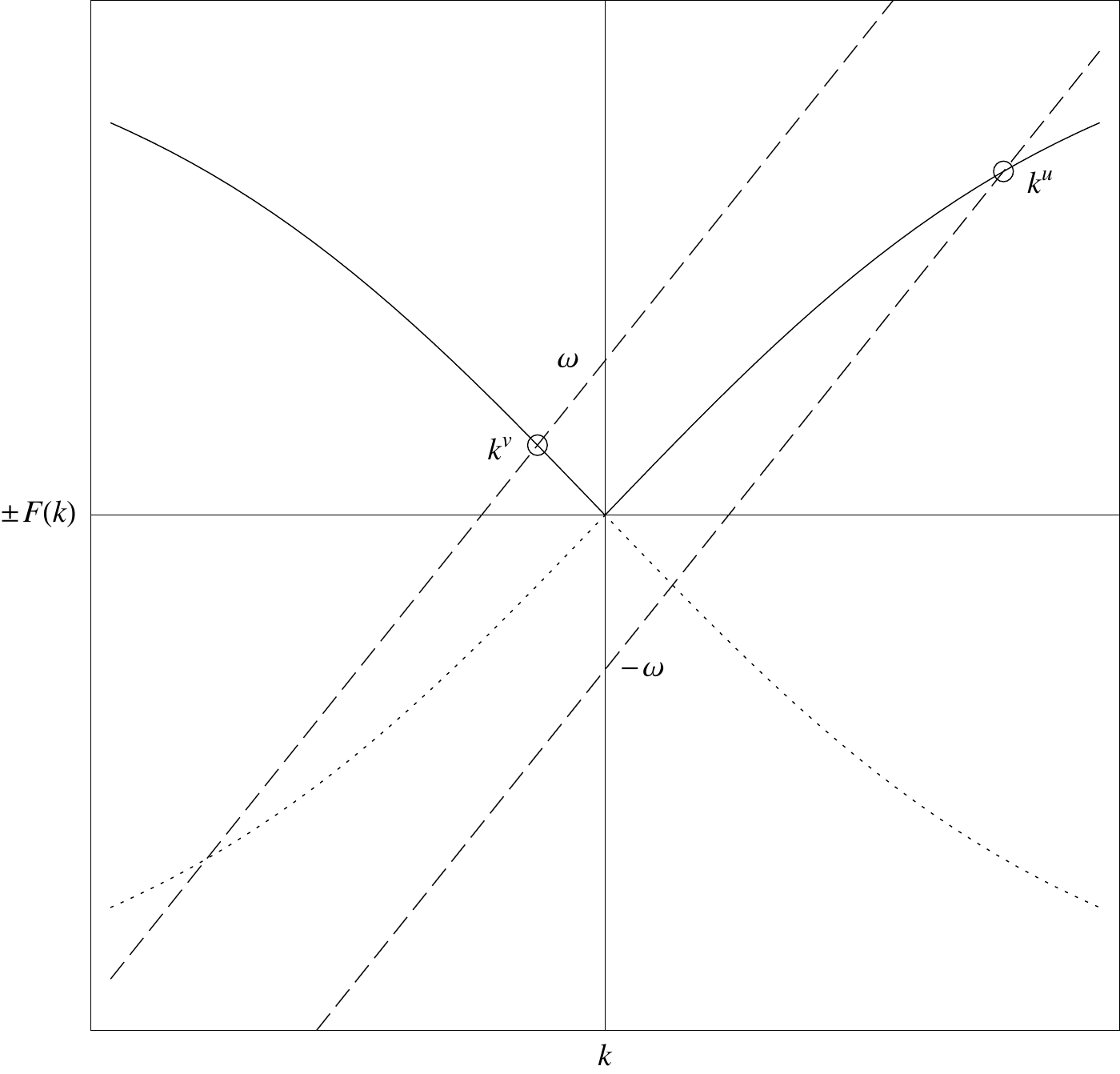}

\caption[\textsc{Solutions for supersonic flow and subluminal dispersion}]{\textsc{Solutions for supersonic flow and subluminal dispersion}:
Each value of $\omega$ corresponds to a single positive-norm mode.
If $\omega>0$, this is a $v$-mode; if $\omega<0$, it is a $u$-mode.
Each value of $\omega$ also corresponds to a single negative-norm
mode, which is the complex conjugate of the positive-norm mode corresponding
to $-\omega$.\label{fig:supersonic_k_values}}

\end{figure}

It is easily seen that, if $V<-c$ and the dispersion is subluminal,
then each value of $\omega$ corresponds to a single positive-norm
mode. (See Figure \ref{fig:supersonic_k_values}.) If $\omega$ is
positive, the positive-norm mode has a negative value of $k$, and
is therefore left-moving in the free-fall frame; these are the $v$-modes.
On the other hand, a negative $\omega$ has a positive-norm mode with
a positive value of $k$, which is right-moving in the free-fall frame;
these are the $u$-modes. Denoting the corresponding $k$-values by
$k^{v}$ and $k^{u}$, we have \begin{eqnarray}
\negthickspace\negthickspace\negthickspace\negthickspace\negthickspace\negthickspace\negthickspace\negthickspace\negthickspace\negthickspace\int_{-\infty}^{0}\left\{ \hat{a}_{k}\phi_{k}\left(x,t\right)+\hat{a}_{k}^{\dagger}\phi_{k}^{\star}\left(x,t\right)\right\} dk & = & \int_{0}^{\infty}\left\{ \hat{a}_{k^{v}\left(\omega\right)}\phi_{k^{v}\left(\omega\right)}\left(x,t\right)+\hat{a}_{k^{v}\left(\omega\right)}^{\dagger}\phi_{k^{v}\left(\omega\right)}^{\star}\left(x,t\right)\right\} \left|\frac{dk^{v}}{d\omega}\right|d\omega\,,\qquad\qquad\label{eq:v_branch_k2w}\\
\negthickspace\negthickspace\negthickspace\negthickspace\negthickspace\negthickspace\negthickspace\negthickspace\negthickspace\negthickspace\int_{0}^{+\infty}\left\{ \hat{a}_{k}\phi_{k}\left(x,t\right)+\hat{a}_{k}^{\dagger}\phi_{k}^{\star}\left(x,t\right)\right\} dk & = & \int_{-\infty}^{0}\left\{ \hat{a}_{k^{u}\left(\omega\right)}\phi_{k^{u}\left(\omega\right)}\left(x,t\right)+\hat{a}_{k^{u}\left(\omega\right)}^{\dagger}\phi_{k^{u}\left(\omega\right)}^{\star}\left(x,t\right)\right\} \left|\frac{dk^{u}}{d\omega}\right|d\omega\,.\qquad\qquad\label{eq:u_branch_k2w}\end{eqnarray}
We want to have integrals similar to those on the left-hand side,
parameterised by $\omega$ rather than $k$, without the extra factor
in the integrand:\begin{eqnarray}
\negthickspace\negthickspace\negthickspace\negthickspace\negthickspace\negthickspace\int_{-\infty}^{0}\left\{ \hat{a}_{k}\phi_{k}\left(x,t\right)+\hat{a}_{k}^{\dagger}\phi_{k}^{\star}\left(x,t\right)\right\} dk & = & \int_{0}^{\infty}\left\{ \hat{a}_{\omega}^{v}\phi_{\omega}^{v}\left(x,t\right)+\hat{a}_{\omega}^{v\dagger}\phi_{\omega}^{v\star}\left(x,t\right)\right\} d\omega\,,\label{eq:v_branch_k2w2}\\
\negthickspace\negthickspace\negthickspace\negthickspace\negthickspace\negthickspace\int_{0}^{+\infty}\left\{ \hat{a}_{k}\phi_{k}\left(x,t\right)+\hat{a}_{k}^{\dagger}\phi_{k}^{\star}\left(x,t\right)\right\} dk & = & \int_{-\infty}^{0}\left\{ \hat{a}_{\omega}^{u}\phi_{\omega}^{u}\left(x,t\right)+\hat{a}_{\omega}^{u\dagger}\phi_{\omega}^{u\star}\left(x,t\right)\right\} d\omega\,.\label{eq:u_branch_k2w2}\end{eqnarray}
We would also like the modes and operators which appear in the integrand
of the right-hand side to obey the usual normalisation and commutation
relations:\begin{equation}
\left(\phi_{\omega_{1}}^{u},\phi_{\omega_{2}}^{u}\right)=\delta\left(\omega_{1}-\omega_{2}\right)=\left|\frac{dk^{u}}{d\omega}\right|\delta\left(k_{1}^{u}-k_{2}^{u}\right)\,,\label{eq:mode_normalisation_k2w}\end{equation}
\begin{equation}
\left[\hat{a}_{\omega_{1}}^{u},\hat{a}_{\omega_{2}}^{u\dagger}\right]=\delta\left(\omega_{1}-\omega_{2}\right)=\left|\frac{dk^{u}}{d\omega}\right|\delta\left(k_{1}^{u}-k_{2}^{u}\right)\,,\label{eq:commutator_k2w}\end{equation}
with analogous relations for the $v$-branch. The conditions in Eqs.
(\ref{eq:v_branch_k2w2})-(\ref{eq:commutator_k2w}) are all satisfied
if the modes and operators in the $\omega$-representation are defined
as follows:\begin{alignat}{1}
\phi_{\omega}^{u}=\sqrt{\left|\frac{dk^{u}}{d\omega}\right|}\phi_{k^{u}}\,,\ \ \  & \phi_{\omega}^{v}=\sqrt{\left|\frac{dk^{v}}{d\omega}\right|}\phi_{k^{v}}\,,\label{eq:modes_k2w}\\
\hat{a}_{\omega}^{u}=\sqrt{\left|\frac{dk^{u}}{d\omega}\right|}\hat{a}_{k^{u}}\,,\ \ \  & \hat{a}_{\omega}^{v}=\sqrt{\left|\frac{dk^{v}}{d\omega}\right|}\hat{a}_{k^{v}}\,.\label{eq:operators_k2w}\end{alignat}
Then, writing explicitly the exponential time dependence of the modes,
the operator $\hat{\phi}$ is given by\begin{multline}
\hat{\phi}\left(x,t\right)=\int_{0}^{\infty}\left\{ \hat{a}_{\omega}^{v}\phi_{\omega}^{v}\left(x\right)e^{-i\omega t}+\hat{a}_{\omega}^{v\dagger}\phi_{\omega}^{v\star}\left(x\right)e^{i\omega t}\right\} d\omega\\
+\int_{-\infty}^{0}\left\{ \hat{a}_{\omega}^{u}\phi_{\omega}^{u}\left(x\right)e^{-i\omega t}+\hat{a}_{\omega}^{u\dagger}\phi_{\omega}^{u\star}\left(x\right)e^{i\omega t}\right\} d\omega\,,\label{eq:phi_operator_w}\end{multline}
where, substituting Eq. (\ref{eq:k_mode_defn}) in Eq. (\ref{eq:modes_k2w})
and recognising $d\omega/dk$ as the group velocity $v_{g}\left(k\right)$,
we have defined\begin{equation}
\phi_{\omega}^{u}\left(x\right)=\frac{1}{\sqrt{4\pi\left|F\left(k^{u}\left(\omega\right)\right)\right|\left|v_{g}\left(k^{u}\left(\omega\right)\right)\right|}}\exp\left(ik^{u}\left(\omega\right)x\right)\,,\label{eq:mode_w}\end{equation}
with an analogous identity for the $v$-branch.

Although Eq. (\ref{eq:phi_operator_w}) is correct, it is not yet
in its most useful form. Since $\omega$ is a conserved quantity,
any modes with the same value of $\omega$ can be scattered into each
other by an inhomogenous flow. What we have tried to do by writing
$\hat{\phi}$ as an integral over $\omega$ rather than $k$ is split
the modes into groups according to their frequency. Eq. (\ref{eq:phi_operator_w})
does not quite achieve this, because while we have separated the modes
into the complex conjugate pairs of positive and negative norm, we
have also included negative values of $\omega$ in the range of integration.
Each value of $\omega$, then, actually applies to \textit{two} different
terms of the integrand; this can be seen in Figure \ref{fig:supersonic_k_values},
where each value of $\omega$ clearly corresponds to one positive-norm
\textit{and} one negative-norm mode. These terms should be grouped
together, because mixing between them is possible. To do this, we
note simply that the integration over $\omega$ should be over positive
values only if the complex conjugate modes are included explicitly
in the integrand. Therefore, we need to rewrite the second integral
as follows:\[
\int_{-\infty}^{0}\left\{ \hat{a}_{\omega}^{u}\phi_{\omega}^{u}\left(x\right)e^{-i\omega t}+\hat{a}_{\omega}^{u\dagger}\phi_{\omega}^{u\star}\left(x\right)e^{i\omega t}\right\} d\omega=\int_{0}^{\infty}\left\{ \hat{a}_{-\omega}^{u}\phi_{-\omega}^{u}\left(x\right)e^{i\omega t}+\hat{a}_{-\omega}^{u\dagger}\phi_{-\omega}^{u\star}\left(x\right)e^{-i\omega t}\right\} d\omega\,,\]
so that Eq. (\ref{eq:phi_operator_w}) can be rewritten as\begin{eqnarray}
\hat{\phi}\left(x,t\right) & = & \int_{0}^{\infty}\left\{ \left[\hat{a}_{\omega}^{v}\phi_{\omega}^{v}\left(x\right)+\hat{a}_{-\omega}^{u\dagger}\phi_{-\omega}^{u\star}\left(x\right)\right]e^{-i\omega t}+\left[\hat{a}_{\omega}^{v\dagger}\phi_{\omega}^{v\star}\left(x\right)+\hat{a}_{-\omega}^{u}\phi_{-\omega}^{u}\left(x\right)\right]e^{i\omega t}\right\} d\omega\nonumber \\
 & = & \int_{0}^{\infty}\left\{ \hat{\phi}_{\omega}\left(x\right)e^{-i\omega t}+\hat{\phi}_{\omega}^{\dagger}\left(x\right)e^{i\omega t}\right\} \,,\end{eqnarray}
where we have defined\begin{equation}
\hat{\phi}_{\omega}\left(x\right)=\hat{a}_{\omega}^{v}\phi_{\omega}^{v}\left(x\right)+\hat{a}_{-\omega}^{u\dagger}\phi_{-\omega}^{u\star}\left(x\right)\,.\label{eq:mode_mixing}\end{equation}

Eq. (\ref{eq:mode_mixing}) shows that mixing between positive- and
negative-norm modes corresponds to mixing of annihilation and creation
operators. However, it must again be emphasised that, given the present
physical situation of a background flow which is both temporally and
spatially constant, such mixing cannot occur because $\omega$ and
$k$ are independently conserved. The grouping of $\phi_{-\omega}^{u\star}\left(x\right)$
with $\phi_{\omega}^{v}\left(x\right)$ in Eq. (\ref{eq:mode_mixing})
is, at present, merely a mathematical convenience with no physical
significance.

\subsection{Subsonic flow}

\begin{figure}
\includegraphics[width=0.8\columnwidth]{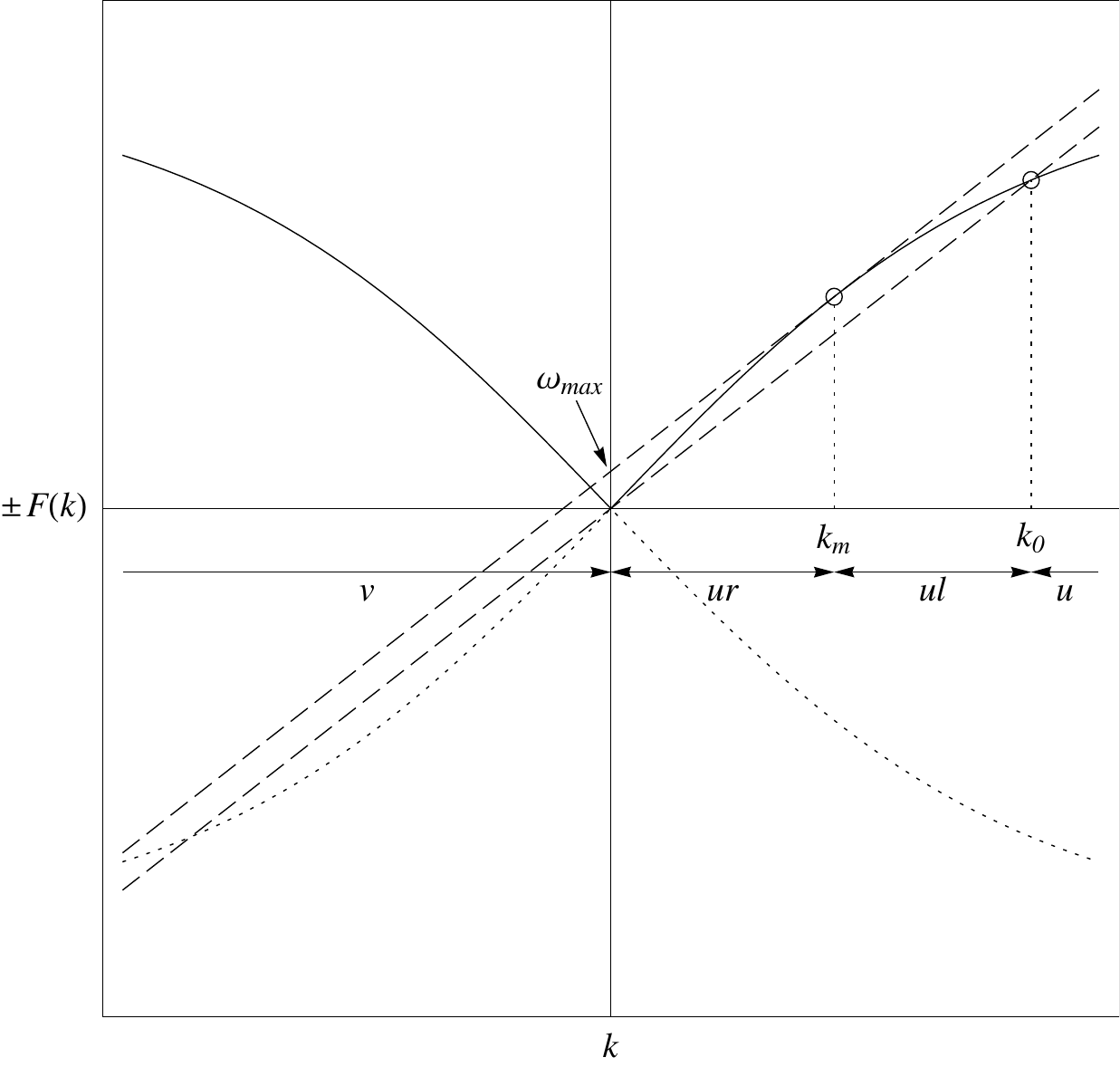}

\caption[\textsc{Solutions for subsonic flow and subluminal dispersion}]{\textsc{Solutions for subsonic flow and subluminal dispersion}: For
$0<\omega<\omega_{\mathrm{max}}$, each frequency corresponds to three
positive-norm modes - a $v$-mode, a $ur$-mode and a $ul$-mode -
and one negative-norm $u$-mode. For $\omega>\omega_{\mathrm{max}}$,
the situation is the same as for supersonic flow in Fig. \ref{fig:supersonic_k_values},
with one positive-norm $v$-mode and one negative-norm $u$-mode.\label{fig:subsonic_k_values}}

\end{figure}

If $-c<V<0$ so that the slope of the line $\omega-Vk$ is less than
$c$, then, as was remarked previously, there are some frequencies
with more than two possible values of $k$. The analysis performed
for supersonic flow is also applicable here. The transformation of
the $v$-branch carries over exactly, and the only change that occurs
for the $u$-branch is that it begins at some $k_{0}>0$ such that
$Vk_{0}+\left|F\left(k_{0}\right)\right|=0$, so that the lower limit
of the left-hand integral in Eqs. (\ref{eq:u_branch_k2w}) and (\ref{eq:u_branch_k2w2})
is $k_{0}$ rather than zero. Of course, this means that the operator
$\hat{\phi}$ in Eq. (\ref{eq:phi_operator_w}) is not equal to the
operator $\hat{\phi}$ in Eq. (\ref{eq:phi_k_decomposition_quantized}),
since we have not yet included those values of $k$ in the interval
from zero to $k_{0}$. This region also splits into two branches,
separated by a value $k_{m}$ for which the line $\omega-Vk$ is tangent
to $F\left(k\right)$. The frequency of this mode is the maximum frequency
with more than two solutions, and we denote it by $\omega_{max}=Vk_{m}+\left|F\left(k_{m}\right)\right|$.
This wavevector therefore has zero group velocity, since $d\omega/dk=0$
there. Let us call the region from zero to $k_{m}$ the $ur$-branch,
and that from $k_{m}$ to $k_{0}$ the $ul$-branch; these are illustrated
in Figure \ref{fig:subsonic_k_values}. These labels are chosen to
indicate that, while all these modes are right-moving in the free-fall
frame and therefore on the $u$-branch, those in the interval $\left(0,k_{m}\right)$
are right-moving in the observer's frame while those in the interval
$\left(k_{m},k_{0}\right)$ are left-moving in the observer's frame.
The remainder of the analysis proceeds exactly as before, and relations
analogous to those in Eqs. (\ref{eq:modes_k2w}) and (\ref{eq:operators_k2w}),
with an additional two relations for the $ur$- and $ul$-branches,
are found to implement the transformation to the $\omega$-representation.
The contribution to the field from the $ur$- and $ul$-branches is\begin{multline}
\int_{0}^{k_{0}}\left\{ \hat{a}_{k}\phi_{k}\left(x,t\right)+\hat{a}_{k}^{\dagger}\phi_{k}^{\star}\left(x,t\right)\right\} dk\\
=\int_{0}^{\omega_{\mathrm{max}}}\left\{ \left(\hat{a}_{\omega}^{ur}\phi_{\omega}^{ur}+\hat{a}_{\omega}^{ul}\phi_{\omega}^{ul}\right)e^{-i\omega t}+\left(\hat{a}_{\omega}^{ur\dagger}\phi_{\omega}^{ur\star}+\hat{a}_{\omega}^{ul\dagger}\phi_{\omega}^{ul\star}\right)e^{i\omega t}\right\} d\omega\,.\label{eq:ur_and_ul_modes}\end{multline}
The complete operator $\hat{\phi}$ of Eq. (\ref{eq:phi_k_decomposition_quantized})
becomes\begin{equation}
\hat{\phi}\left(x,t\right)=\int_{0}^{\infty}\left\{ \hat{\phi}_{\omega}\left(x\right)e^{-i\omega t}+\hat{\phi}_{\omega}^{\dagger}\left(x\right)e^{i\omega t}\right\} d\omega\label{eq:complete_phi_operator}\end{equation}
where\begin{equation}
\hat{\phi}_{\omega}\left(x\right)=\begin{cases}
\hat{a}_{\omega}^{v}\phi_{\omega}^{v}\left(x\right)+\hat{a}_{\omega}^{ur}\phi_{\omega}^{ur}\left(x\right)+\hat{a}_{\omega}^{ul}\phi_{\omega}^{ul}\left(x\right)+\hat{a}_{-\omega}^{u\dagger}\phi_{-\omega}^{u\star}\left(x\right)\, & \mathrm{for}\;0<\omega<\omega_{\mathrm{max}}\\
\hat{a}_{\omega}^{v}\phi_{\omega}^{v}\left(x\right)+\hat{a}_{-\omega}^{u\dagger}\phi_{-\omega}^{u\star}\left(x\right)\, & \mathrm{for}\;\omega>\omega_{\mathrm{max}}\end{cases}\,.\label{eq:phi_w_subsonic_flow}\end{equation}

\subsection{Inhomogeneous flow}

\begin{figure}
\includegraphics[width=0.8\columnwidth]{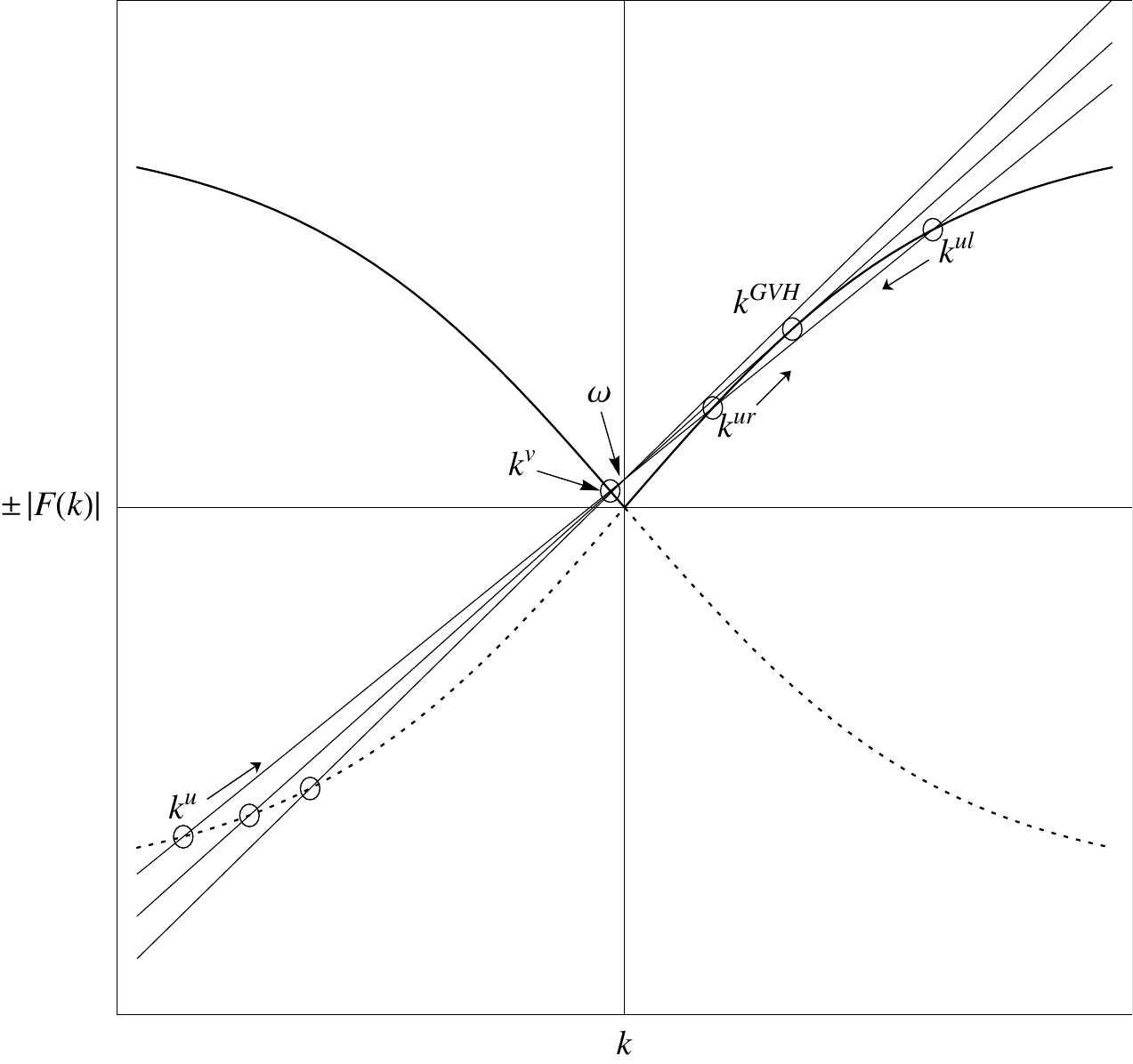}

\caption[\textsc{Variation of solutions with flow velocity}]{\textsc{Variation of solutions with flow velocity}: When $V$ varies
in space but not in time, $\omega$ is conserved but $k$ is not.
The variation in $k$ is seen by changing the slope of the line $\omega-Vk$
while $\omega$ is fixed. Thus, as $V$ becomes more negative, $k^{u}$
becomes less negative while $k^{v}$ remains almost fixed. $k^{ur}$
and $k^{ul}$, however, undergo a more significant change. These vary
smoothly towards each other, eventually merging at some value of $V$,
where a group-velocity horizon is established. As $V$ changes further,
there are no real solutions corresponding to the $ur$- and $ul$-branches,
and they have become complex.\label{fig:Variation-of-Solutions-with-V}}

\end{figure}

Let us now allow $V$ to be spatially dependent, although we shall
always assume that $V$ approaches constant limiting values as $x\rightarrow\pm\infty$.
The Lagrangian of Eq. (\ref{eq:Lagrangian_with_dispersion}) is no
longer invariant under spatial translations, and $k$ is no longer
conserved. However, since $V$ is time-independent, the Lagrangian
is still invariant under time translations, and frequency remains
a conserved quantity. In Figure \ref{fig:Variation-of-Solutions-with-V},
we represent pictorially the $x$-dependence of $k$ simply by varying
the slope of the line $\omega-Vk$ in Figs. \ref{fig:supersonic_k_values}
and \ref{fig:subsonic_k_values}, while its $y$-intercept, being
equal to $\omega$, is kept fixed.

\begin{figure}
\subfloat{\includegraphics[width=0.45\columnwidth]{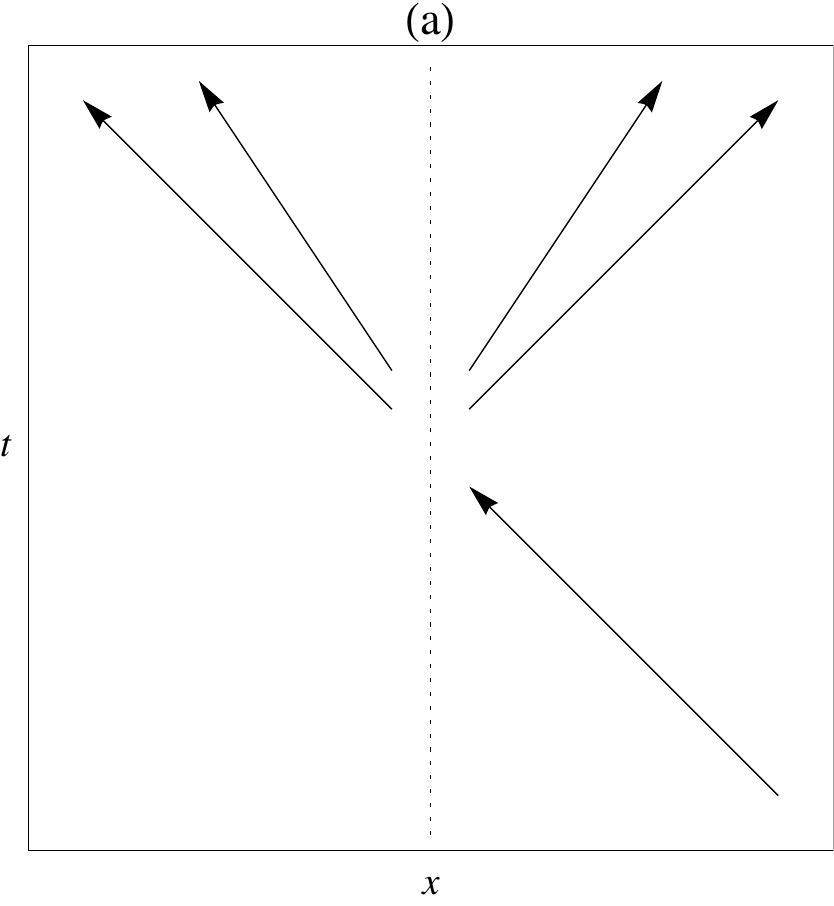}

} \subfloat{\includegraphics[width=0.45\columnwidth]{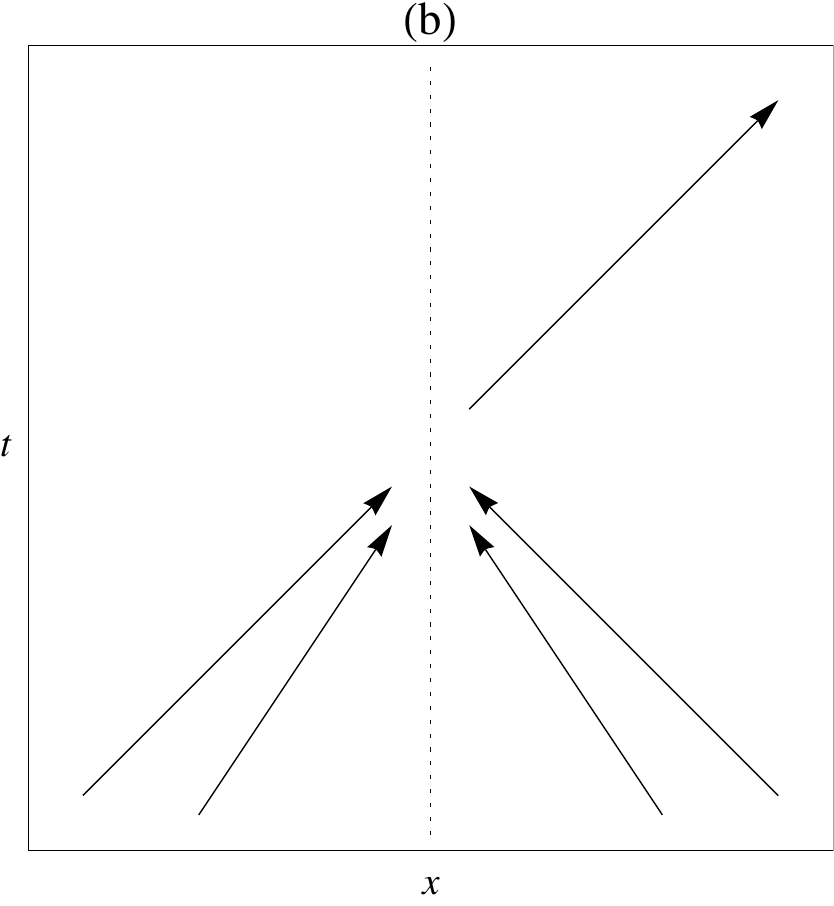}

}

\caption[\textsc{Space-time diagrams of in- and out-modes}]{\textsc{Space-time diagrams of in- and out-modes}: Figure $(a)$
represents an in-mode, containing a single ingoing wavepacket or wavenumber.
Figure $(b)$ is an out-mode, containing a single outgoing wavepacket
or wavenumber. The correspondence between wavepackets and wavenumbers
shows that the description of a wavenumber as ingoing or outgoing
depends on the direction of the group velocity.\label{fig:In_and_out_modes}}

\end{figure}

Firstly, let us remark that, in the context of an inhomogeneous flow,
the existence of the $ur$- and $ul$-branches in subsonic regions,
coupled with their absence in supersonic regions, is of very special
importance. If the flow varies from subsonic to supersonic, these
wavevectors experience a horizon: we see from Figure \ref{fig:Variation-of-Solutions-with-V}
that, as the flow speed increases, the two solutions vary continuously
towards each other, until they merge into a single value; as the flow
speed increases further, there are no real wavevectors corresponding
to these branches. Instead, the wavevector becomes complex, in such
a way that the solution is exponentially decreasing in the supersonic
region. The point at which the wavevectors merge is a \textit{group-velocity
horizon}: the line $\omega-Vk$ is tangent to $F\left(k\right)$ there,
so that the group velocity $d\omega/dk$ vanishes. Since the $ur$-
and $ul$-branches have oppositely-directed group velocities, reflection
of a wavepacket from a group-velocity horizon is possible if its wavevector
varies smoothly from one branch to the other. Numerical simulations
of wavepacket propagation - discussed in §\ref{sub:Wavepacket-propagation}
- show that this is indeed what happens. In particular, there is no
shifting to arbitrarily short wavelengths as the horizon is approached,
as in the dispersionless case (compare the out-modes of Fig. \ref{fig:Lemaitre-Null-Curves}$(b)$
with those of Fig. \ref{fig:In_and_out_modes}$(b)$). Dispersion
regularizes the wave behaviour at the horizon such that wavevectors
remain always finite; that is, there is no trans-Planckian problem.

Since $k$ is no longer a conserved quantity, the $\omega$-representation
of $\hat{\phi}$ developed above is now the only natural description
of the field. However, in this representation, we still have to assign
a different mode to each value of $k$, indicated by which branch
($u$, $v$, $ur$ or $ul$) it lies on. There arises an ambiguity:
if $k$ is not conserved, what do these separate modes mean? If a
wavepacket centred at $k^{u}\left(\omega\right)$ is scattered by
the inhomogeneous flow into other wavepackets with wavenumbers $k^{v}\left(\omega\right)$,
$k^{ul}\left(\omega\right)$ and $k^{ur}\left(\omega\right)$, on
what basis can we still designate this the $\phi_{\omega}^{u}$ mode?
There are two natural ways of assigning labels to the various modes:
\begin{itemize}
\item Modes may be determined by a single \textit{incoming} wavenumber,
which is then scattered into several \textit{outgoing} wavenumbers;
these are known as \textit{in-modes};
\item Modes may be determined by a single \textit{outgoing} wavenumber,
which results from the scattering of several \textit{ingoing} wavenumbers;
these are known as \textit{out-modes}.
\end{itemize}
This is the same way in which we categorized the modes in the dispersionless
case (see §\ref{sub:Out-modes-and-in-modes}).

Although the modes have a purely exponential time dependence, $e^{-i\omega t}$,
and therefore represent stationary solutions, their designation as
in- or out-modes is made more intuitive if we consider them as wavepackets
strongly peaked at the corresponding values of $\omega$ and $k$.
In that case, we can visualise the time evolution of the modes by
the scattering of wavepackets, and the in- and out-modes are determined
by whether we have a single incoming wavepacket at early times or
a single outgoing wavepacket at late times. (See Figure \ref{fig:In_and_out_modes}
for an illustration.) Wave\textit{packets} offer an intuitive way
of determining whether a particular wave\textit{number} is incoming
or outgoing, for this is clearly determined by the direction of the
group velocity $d\omega/dk$, rather than the phase velocity $\omega/k$.
(For an illustration of how the group velocity is determined, see
Figure \ref{fig:dispersion_group_velocities}.)

\begin{figure}
\subfloat{\includegraphics[width=0.45\columnwidth]{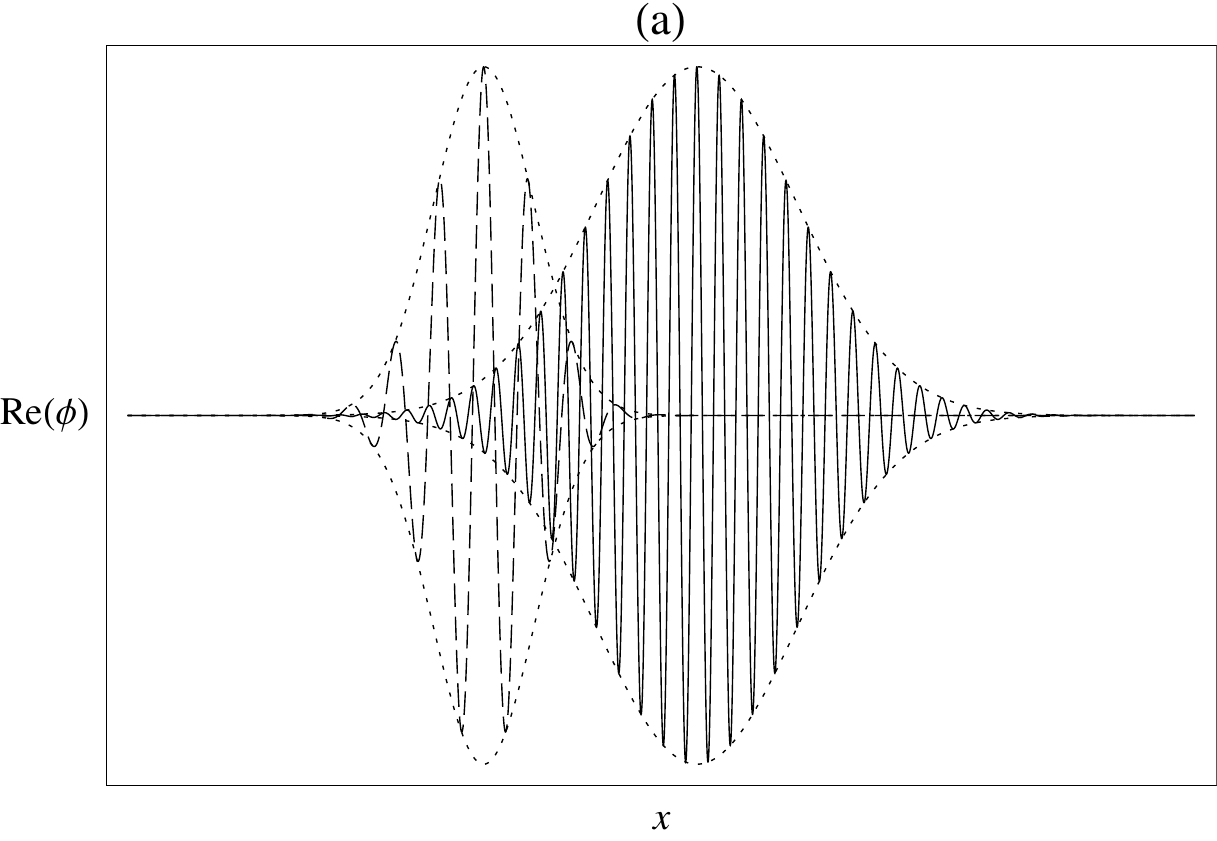}} \subfloat{\includegraphics[width=0.45\columnwidth]{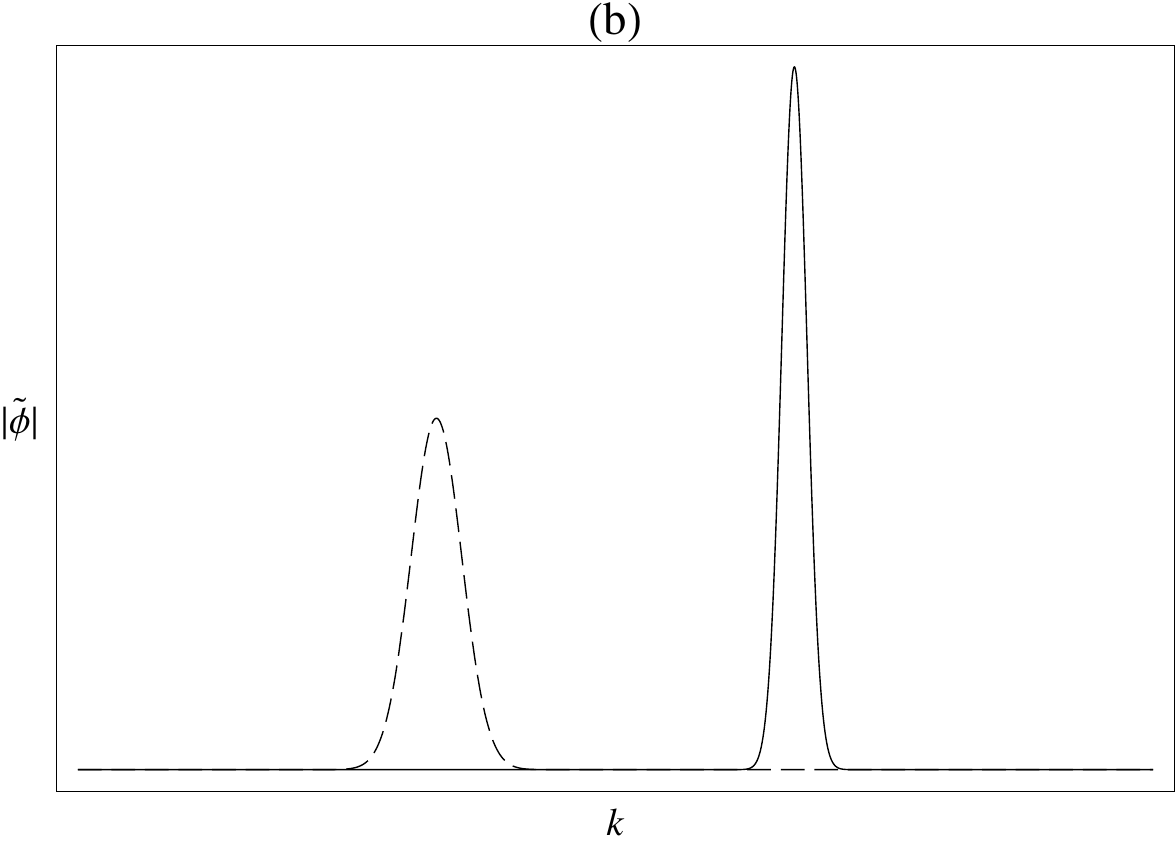}}

\caption[\textsc{Orthogonality of in- or out-modes}]{\textsc{Orthogonality of in- or out-modes}: Although two early- or
late-time wavepackets may have a significant overlap in real space
(Fig. $(a)$), they are well separated in Fourier space (Fig. $(b)$),
and their scalar product is vanishingly small. In the limit of a stationary
solution, the Fourier transforms approach zero thinness and the scalar
product vanishes completely.\label{fig:Scalar_product}}

\end{figure}

The wavepacket picture also provides a simple way of seeing that the
set of in-modes and the set of out-modes form complete orthonormal
sets of solutions to the wave equation. Consider, for example, a pair
of in-modes with different values of the initial incoming wavenumber.
At early times, each of these lives entirely in one of the asymptotic
constant velocity regions, and can be considered as wavepackets strongly
peaked around their respective values of $k$. Their scalar product,
as given in Eq. (\ref{eq:scalar_product_constant_v}), will nearly
vanish, and can be made arbitrarily close to zero by lengthening the
wavepackets or, equivalently, narrowing their Fourier transforms;
this is illustrated in Figure \ref{fig:Scalar_product}. Since the
scalar product is conserved, this will also be true at late times,
when the modes have scattered into several outgoing wavepackets. In
the limit of infinitely long initial wavepackets, corresponding to
the stationary modes, the scalar product vanishes. Thus, the in-modes
are orthogonal to each other. Moreover, they must be complete, for
they span all possible values of $k$, which form a complete set.
Therefore, Eqs. (\ref{eq:complete_phi_operator}) and (\ref{eq:phi_w_subsonic_flow})
are valid when the modes are interpreted as in-modes. Exactly the
same argument applies to the out-modes, the only difference being
that the scalar product between different modes is considered at late
times, where it is found to vanish in the same manner as the in-modes.
So, we may equally well interpret the modes as out-modes, which also
form a complete orthonormal set.

Henceforward, we shall assume that the asymptotic constant velocity
regions comprise one subsonic and one supersonic region. (This need
not be true, of course. In §\ref{sub:Transforming-to-the-omega-prime-repn},
we perform a similar analysis for light in an optical fibre, analogous
to the present case but with both asymptotic velocities subsonic.)
The existence of a group-velocity horizon for $ur$- and $ul$-modes,
discussed above, causes a slight change in the mode decomposition:
since the solution must be exponentially decreasing in the supersonic
region, only one linear combination of the $ur$- and $ul$-modes
is allowed, and they degenerate into a single mode. This is the mathematical
expression of the fact that one of these modes is ingoing while the
other is outgoing, and there is no sense in which these roles can
be swapped. The resulting mode shall be labelled the $url$-mode.
For $0<\omega<\omega_{\mathrm{max}}$, Eq. (\ref{eq:phi_w_subsonic_flow})
becomes\begin{eqnarray}
\hat{\phi}_{\omega}\left(x\right) & = & \hat{a}_{\omega}^{v,\mathrm{in}}\phi_{\omega}^{v,\mathrm{in}}\left(x\right)+\hat{a}_{\omega}^{url,\mathrm{in}}\phi_{\omega}^{url,\mathrm{in}}\left(x\right)+\left(\hat{a}_{-\omega}^{u,\mathrm{in}}\right)^{\dagger}\left[\phi_{-\omega}^{u,\mathrm{in}}\left(x\right)\right]^{\star}\nonumber \\
 & = & \hat{a}_{\omega}^{v,\mathrm{out}}\phi_{\omega}^{v,\mathrm{out}}\left(x\right)+\hat{a}_{\omega}^{url,\mathrm{out}}\phi_{\omega}^{url,\mathrm{out}}\left(x\right)+\left(\hat{a}_{-\omega}^{u,\mathrm{out}}\right)^{\dagger}\left[\phi_{-\omega}^{u,\mathrm{out}}\left(x\right)\right]^{\star}\,.\label{eq:modes_w_inhomogeneous_v}\end{eqnarray}

In the case, considered previously, of a velocity profile which is
constant in $x$, it still makes sense to define in- and out-modes
by considering the evolution of wavepackets. However, this case is
special in that the in-mode and out-mode corresponding to a certain
wavenumber are identical: an incoming wavepacket is not scattered,
and emerges as a single outgoing wavepacket with the same value of
$k$. So the modes and operators in each line of Eq. (\ref{eq:modes_w_inhomogeneous_v})
are identical, and the superscripts $in$ and $out$ superfluous.
Inhomogeneity of $V$ breaks this degeneracy.

Completeness of each set of modes ensures that we can express one
set as a linear transformation of the other. That is, we can write\begin{eqnarray}
\phi_{\omega}^{v,\mathrm{in}}\left(x\right) & = & C_{\omega}^{v,v}\phi_{\omega}^{v,\mathrm{out}}\left(x\right)+C_{\omega}^{v,url}\phi_{\omega}^{url,\mathrm{out}}\left(x\right)+C_{\omega}^{v,u}\left[\phi_{-\omega}^{u,\mathrm{out}}\left(x\right)\right]^{\star}\,,\nonumber \\
\phi_{\omega}^{url,\mathrm{in}}\left(x\right) & = & C_{\omega}^{url,v}\phi_{\omega}^{v,\mathrm{out}}\left(x\right)+C_{\omega}^{url,url}\phi_{\omega}^{url,\mathrm{out}}\left(x\right)+C_{\omega}^{url,u}\left[\phi_{-\omega}^{u,\mathrm{out}}\left(x\right)\right]^{\star}\,,\nonumber \\
\left[\phi_{-\omega}^{u,\mathrm{in}}\left(x\right)\right]^{\star} & = & C_{\omega}^{u,v}\phi_{\omega}^{v,\mathrm{out}}\left(x\right)+C_{\omega}^{u,url}\phi_{\omega}^{url,\mathrm{out}}\left(x\right)+C_{\omega}^{u,u}\left[\phi_{-\omega}^{u,\mathrm{out}}\left(x\right)\right]^{\star}\,.\label{eq:mode_in_out_3}\end{eqnarray}
The normalization of the modes, as expressed in Eq. (\ref{eq:mode_normalisation_k2w}),
allows the Bogoliubov coefficients to be written as scalar products
between in-modes and out-modes, e.g. $C_{\omega}^{v,v}\delta\left(\omega_{1}-\omega_{2}\right)=\left(\phi_{\omega_{1}}^{v,\mathrm{out}},\phi_{\omega_{2}}^{v,\mathrm{in}}\right)$
or $C_{\omega}^{v,u}\delta\left(\omega_{1}-\omega_{2}\right)=-\left(\left(\phi_{-\omega_{1}}^{u,\mathrm{out}}\right)^{\star},\phi_{\omega_{2}}^{v,\mathrm{in}}\right)$.
Using the identity in Eq. (\ref{eq:scalar_product_commutation}),
we can then express the out-modes in terms of in-modes, using the
same coefficients of Eqs. (\ref{eq:mode_in_out_3}).
The result is:\begin{eqnarray}
\phi_{\omega}^{v,\mathrm{out}}\left(x\right) & = & \left(C_{\omega}^{v,v}\right)^{\star}\phi_{\omega}^{v,\mathrm{in}}\left(x\right)+\left(C_{\omega}^{url,v}\right)^{\star}\phi_{\omega}^{url,\mathrm{in}}\left(x\right)-\left(C_{\omega}^{u,v}\right)^{\star}\left[\phi_{-\omega}^{u,\mathrm{in}}\left(x\right)\right]^{\star}\,,\nonumber \\
\phi_{\omega}^{url,\mathrm{out}}\left(x\right) & = & \left(C_{\omega}^{v,url}\right)^{\star}\phi_{\omega}^{v,\mathrm{in}}\left(x\right)+\left(C_{\omega}^{url,url}\right)^{\star}\phi_{\omega}^{url,\mathrm{in}}\left(x\right)-\left(C_{\omega}^{u,url}\right)^{\star}\left[\phi_{-\omega}^{u,\mathrm{in}}\left(x\right)\right]^{\star}\,,\nonumber \\
\left[\phi_{-\omega}^{u,\mathrm{out}}\left(x\right)\right]^{\star} & = & -\left(C_{\omega}^{v,u}\right)^{\star}\phi_{\omega}^{v,\mathrm{in}}\left(x\right)-\left(C_{\omega}^{url,u}\right)^{\star}\phi_{\omega}^{url,\mathrm{in}}\left(x\right)+\left(C_{\omega}^{u,u}\right)^{\star}\left[\phi_{-\omega}^{u,\mathrm{in}}\left(x\right)\right]^{\star}\,.\label{eq:mode_out_in_3}\end{eqnarray}
Substitution of Eqs. (\ref{eq:mode_in_out_3})
in Eq. (\ref{eq:modes_w_inhomogeneous_v}) yields the transformation
for the operators:\begin{eqnarray}
\hat{a}_{\omega}^{v,\mathrm{out}} & = & C_{\omega}^{v,v}\hat{a}_{\omega}^{v,\mathrm{in}}+C_{\omega}^{url,v}\hat{a}_{\omega}^{url,\mathrm{in}}+C_{\omega}^{u,v}\left(\hat{a}_{-\omega}^{u,\mathrm{in}}\right)^{\dagger}\,,\nonumber \\
\hat{a}_{\omega}^{url,\mathrm{out}} & = & C_{\omega}^{v,url}\hat{a}_{\omega}^{v,\mathrm{in}}+C_{\omega}^{url,url}\hat{a}_{\omega}^{url,\mathrm{in}}+C_{\omega}^{u,url}\left(\hat{a}_{-\omega}^{u,\mathrm{in}}\right)^{\dagger}\,,\nonumber \\
\left(\hat{a}_{-\omega}^{u,\mathrm{out}}\right)^{\dagger} & = & C_{\omega}^{v,u}\hat{a}_{\omega}^{v,\mathrm{in}}+C_{\omega}^{url,u}\hat{a}_{\omega}^{url,\mathrm{in}}+C_{\omega}^{u,u}\left(\hat{a}_{-\omega}^{u,\mathrm{in}}\right)^{\dagger}\,.\label{eq:operator_out_in_3}\end{eqnarray}
Performing the same substitution with Eqs. (\ref{eq:mode_out_in_3})
yields the inverse transformation:\begin{eqnarray}
\hat{a}_{\omega}^{v,\mathrm{in}} & = & \left(C_{\omega}^{v,v}\right)^{\star}\hat{a}_{\omega}^{v,\mathrm{out}}+\left(C_{\omega}^{v,url}\right)^{\star}\hat{a}_{\omega}^{url,\mathrm{out}}-\left(C_{\omega}^{v,u}\right)^{\star}\left(\hat{a}_{-\omega}^{u,\mathrm{out}}\right)^{\dagger}\,,\nonumber \\
\hat{a}_{\omega}^{url,\mathrm{in}} & = & \left(C_{\omega}^{url,v}\right)^{\star}\hat{a}_{\omega}^{v,\mathrm{out}}+\left(C_{\omega}^{url,url}\right)^{\star}\hat{a}_{\omega}^{url,\mathrm{out}}-\left(C_{\omega}^{url,u}\right)^{\star}\left(\hat{a}_{-\omega}^{u,\mathrm{out}}\right)^{\dagger}\,,\nonumber \\
\left(\hat{a}_{-\omega}^{u,\mathrm{in}}\right)^{\dagger} & = & -\left(C_{\omega}^{u,v}\right)^{\star}\hat{a}_{\omega}^{v,\mathrm{out}}-\left(C_{\omega}^{u,url}\right)^{\star}\hat{a}_{\omega}^{url,\mathrm{out}}+\left(C_{\omega}^{u,u}\right)^{\star}\left(\hat{a}_{-\omega}^{u,\mathrm{out}}\right)^{\dagger}\,.\label{eq:operator_in_out_3}\end{eqnarray}
The mixing of creation and annihilation operators is here made explicit:
a single annihilation or creation operator corresponding to an outgoing
particle can be equivalent to a sum of both annihilation and creation
operators for several incoming particles (and vice versa). We have
seen this before, in the dispersionless case (see §\ref{sub:Out-modes-and-in-modes});
and, as there, it leads to inequivalence of the in- and out-vacua.

\section{Spontaneous creation of phonons\label{sub:Spontaneous-creation-II}}

Making the reasonable assumption that the field is in the in-vacuum,
$\left|0_{\mathrm{in}}\right\rangle $ - that is, that there are no
incoming particles, and the quantum state is annihilated by all in-mode
annihilation operators:\begin{equation}
\hat{a}_{\omega}^{v,\mathrm{in}}\left|0_{\mathrm{in}}\right\rangle =\hat{a}_{\omega}^{url,\mathrm{in}}\left|0_{\mathrm{in}}\right\rangle =\hat{a}_{-\omega}^{u,\mathrm{in}}\left|0_{\mathrm{in}}\right\rangle =0\,\ \ \ \forall\ \omega\:\label{eq:in_vacuum_defn}\end{equation}
- we find that the number expectation value of outgoing particles
is non-zero. For example, in the outgoing $v$-mode, the number expectation
value is\begin{eqnarray}
\left\langle 0_{\mathrm{in}}\right|\left(\hat{a}_{\omega_{1}}^{v,\mathrm{out}}\right)^{\dagger}\hat{a}_{\omega_{2}}^{v,\mathrm{out}}\left|0_{\mathrm{in}}\right\rangle  & = & \left\langle 0_{\mathrm{in}}\right|\left\{ \left(C_{\omega_{1}}^{v,v}\right)^{\star}\left(\hat{a}_{\omega_{1}}^{v,\mathrm{in}}\right)^{\dagger}+\left(C_{\omega_{1}}^{url,v}\right)^{\star}\left(\hat{a}_{\omega_{1}}^{url,\mathrm{in}}\right)^{\dagger}+\left(C_{\omega_{1}}^{u,v}\right)^{\star}\hat{a}_{-\omega_{1}}^{u,\mathrm{in}}\right\} \nonumber \\
 &  & \qquad\qquad\qquad\times\left\{ C_{\omega_{2}}^{v,v}\hat{a}_{\omega_{2}}^{v,\mathrm{in}}+C_{\omega_{2}}^{url,v}\hat{a}_{\omega_{2}}^{url,\mathrm{in}}+C_{\omega_{2}}^{u,v}\left(\hat{a}_{-\omega_{2}}^{u,\mathrm{in}}\right)^{\dagger}\right\} \left|0_{\mathrm{in}}\right\rangle \nonumber \\
 & = & \left\langle 0_{\mathrm{in}}\right|\bigg\{\left(C_{\omega_{1}}^{u,v}\right)^{\star}\hat{a}_{-\omega_{1}}^{u,\mathrm{in}}\bigg\}\left\{ C_{\omega_{2}}^{u,v}\left(\hat{a}_{-\omega_{2}}^{u,\mathrm{in}}\right)^{\dagger}\right\} \left|0_{\mathrm{in}}\right\rangle \nonumber \\
 & = & \left(C_{\omega_{1}}^{u,v}\right)^{\star}C_{\omega_{2}}^{u,v}\left\langle 0_{\mathrm{in}}\right|\left\{ \left(\hat{a}_{-\omega_{1}}^{u,\mathrm{in}}\right)^{\dagger}\hat{a}_{-\omega_{2}}^{u,\mathrm{in}}+\delta\left(\omega_{1}-\omega_{2}\right)\right\} \left|0_{\mathrm{in}}\right\rangle \nonumber \\
 & = & \left|C_{\omega_{1}}^{u,v}\right|^{2}\delta\left(\omega_{1}-\omega_{2}\right)\,,\label{eq:v_out_number_expectation}\end{eqnarray}
where we used the commutator in Eq. (\ref{eq:commutator_k2w}) and
the normalization of the state $\left|0_{\mathrm{in}}\right\rangle $.
Similar calculations give the number expectation values for the other
outgoing modes:\begin{eqnarray}
\left\langle 0_{\mathrm{in}}\right|\left(\hat{a}_{\omega_{1}}^{url,\mathrm{out}}\right)^{\dagger}\hat{a}_{\omega_{2}}^{url,\mathrm{out}}\left|0_{\mathrm{in}}\right\rangle  & = & \left|C_{\omega_{1}}^{u,url}\right|^{2}\delta\left(\omega_{1}-\omega_{2}\right)\,,\label{eq:url_out_number_expectation}\\
\left\langle 0_{\mathrm{in}}\right|\left(\hat{a}_{-\omega_{1}}^{u,\mathrm{out}}\right)^{\dagger}\hat{a}_{-\omega_{2}}^{u,\mathrm{out}}\left|0_{\mathrm{in}}\right\rangle  & = & \left(\left|C_{\omega_{1}}^{v,u}\right|^{2}+\left|C_{\omega_{1}}^{url,u}\right|^{2}\right)\delta\left(\omega_{1}-\omega_{2}\right)\,.\label{eq:u_out_number_expectation}\end{eqnarray}

Before moving on, let us determine how these values are to be interpreted;
for this we follow an argument made in §IV B of Ref. \cite{Corley-Jacobson-1996}.
The appearance of the $\delta$ functions shows that the results of
Eqs. (\ref{eq:v_out_number_expectation})-(\ref{eq:u_out_number_expectation})
are densities rather than numbers. This is because the orthonormal
modes form a continuous spectrum, and are not normalized to unity
but to a delta function, as in Eq. (\ref{eq:mode_normalisation_k2w}).
We can overcome this problem by again turning to wavepackets. If $V$
is constant - as in the asmyptotic regions where phonons are detected
- then the wavepackets\begin{equation}
p_{jn}=\mathcal{N}\int_{j\epsilon}^{\left(j+1\right)\epsilon}dk\, e^{-i\omega\left(k\right)t+ikx+i2\pi nk/\epsilon}\label{eq:complete_set_of_wavepackets}\end{equation}
form a complete orthonormal set. This splits the entire space of modes
into wavepackets of reasonably well-defined wavevector (labelled by
$j$) and position (labelled by $n$). The wavepacket $p_{jn}$ is
localised at the point where the derivative of the phase in the integrand
vanishes, i.e., at the point\begin{equation}
-v_{g}\left(k\right)t+x+\frac{2\pi n}{\epsilon}=0\,,\end{equation}
and, at a fixed position $x$, the time between one wavepacket and
the next - or equivalently, the temporal width of a single wavepacket
- is $\Delta t=2\pi/\left(v_{g}\left(k\right)\epsilon\right)$. The
frequency width of a single wavepacket is related to its wavevector
width - and from the limits of the integral in Eq. (\ref{eq:complete_set_of_wavepackets})
we see that $\Delta k=\epsilon$ - by the formula $\Delta\omega=\left(d\omega/dk\right)\Delta k=v_{g}\left(k\right)\epsilon$.
Thus,\begin{equation}
\Delta t\,\Delta\omega=2\pi\,.\end{equation}
The number expectation value $N\left(\omega\right)$ of a single wavepacket
at frequency $\omega$ contains no $\delta$ function, and does indeed
describe the particle content of the wavepacket. In a time $\delta t\ll\Delta t$,
then, the rate of particles of frequency $\omega$ is approximately\begin{equation}
\frac{\delta N\left(\omega\right)}{\delta t}\approx\frac{N\left(\omega\right)}{\Delta t}=\frac{\Delta\omega}{2\pi}N\left(\omega\right)\,.\label{eq:particle_rate_single_wavepacket}\end{equation}
In the limit $\epsilon\rightarrow0$, the first equality of Eq. (\ref{eq:particle_rate_single_wavepacket})
becomes exact, and since $\Delta\omega$ becomes infinitesimal, we
find\begin{equation}
\frac{d^{2}N}{dt\, d\omega}=\frac{1}{2\pi}N\left(\omega\right)\,.\label{eq:spectral_flux_density}\end{equation}
The wavepackets themselves become plane waves in this limit, and $N\left(\omega\right)$
is simply the factor multiplying the $\delta$ functions in Eqs. (\ref{eq:v_out_number_expectation})-(\ref{eq:u_out_number_expectation}).
So, upon division by $2\pi$, these expressions give the \textit{spectral
flux density} of phonons: the number of phonons emitted per unit frequency
interval per unit time.

Particle creation arises from pairings of oppositely-signed modes
- in this case, the $v$-mode with the $u$-mode and the $url$-mode
with the $u$-mode. Indeed, as was earlier remarked, conservation
of the norm in the form of Eq. (\ref{eq:norm_constant_v}) suggests
that particles are created in precisely such pairs, and never singly.
We can show this to be true from the expectation values above: calculating
the norms of the modes $\left[\phi_{-\omega}^{u,\mathrm{in}}\right]^{\star}$
and $\Big[\phi_{-\omega}^{u,\mathrm{out}}\Big]^{\star}$ in Eqs. (\ref{eq:mode_in_out_3})
and (\ref{eq:mode_out_in_3}), we find\[
\left|C_{\omega}^{u,v}\right|^{2}+\left|C_{\omega}^{u,url}\right|^{2}-\left|C_{\omega}^{u,u}\right|^{2}=\left|C_{\omega}^{v,u}\right|^{2}+\left|C_{\omega}^{url,u}\right|^{2}-\left|C_{\omega}^{u,u}\right|^{2}=-1\,,\]
and, therefore,\begin{equation}
\left|C_{\omega}^{u,v}\right|^{2}+\left|C_{\omega}^{u,url}\right|^{2}=\left|C_{\omega}^{v,u}\right|^{2}+\left|C_{\omega}^{url,u}\right|^{2}=\left|C_{\omega}^{u,u}\right|^{2}-1\,.\label{eq:equal_total_rates}\end{equation}
Comparing with Eqs. (\ref{eq:v_out_number_expectation})-(\ref{eq:u_out_number_expectation}),
we see that the expectation value for the $u$-out mode is equal to
the sum of the expectation values for the $v$-out and $url$-out
modes, consistent with their being produced in pairs of oppositely-signed
norm.

That spontaneous creation occurs in such pairs can be shown explicity
by writing the in-vacuum $\left|0_{\mathrm{in}}\right\rangle $ in
terms of out-modes, similar to Eq. (\ref{eq:in-vacuum_out-Fock-basis}).
We define the outgoing Fock state $\left|n\right\rangle _{\omega}^{v,\mathrm{out}}$
as that state with $n_{\omega}^{v}$ quanta in the state $\phi_{\omega}^{v,\mathrm{out}}$,
and similarly for other modes. The out-mode creation and annihilation
operators act on these states according to the standard relations\begin{eqnarray}
\hat{a}_{\omega}^{v,\mathrm{out}}\left|n\right\rangle _{\omega}^{v} & = & \sqrt{n}\left|n-1\right\rangle _{\omega}^{v,\mathrm{out}}\,,\label{eq:annihilation_operation}\\
\left(\hat{a}_{\omega}^{v,\mathrm{out}}\right)^{\dagger}\left|n\right\rangle _{\omega}^{v} & = & \sqrt{n+1}\left|n+1\right\rangle _{\omega}^{v,\mathrm{out}}\,,\label{eq:creation_operation}\end{eqnarray}
with analogous relations for the other modes. Then
\begin{equation}
\left|0_{\mathrm{in}}\right\rangle =\prod_{\omega}\frac{1}{\left|C_{\omega}^{u,u}\right|}\sum_{m=0}^{\infty}\sum_{n=0}^{\infty}\left(\frac{C_{\omega}^{u,v}}{C_{\omega}^{u,u}}\right)^{m}\left(\frac{C_{\omega}^{u,url}}{C_{\omega}^{u,u}}\right)^{n}\sqrt{\frac{\left(m+n\right)!}{m!\, n!}}\left|m\right\rangle _{\omega}^{v,\mathrm{out}}\left|n\right\rangle _{\omega}^{url,\mathrm{out}}\left|m+n\right\rangle _{\omega}^{u,\mathrm{out}}\,,\label{eq:in_vacuum_out_Fock_states}\end{equation}
for this is the quantum state which is annihilated by all the in-mode annihilation operators of Eqs. (\ref{eq:operator_in_out_3}).
Note that the only states appearing in the expansion are those with pairs of
oppositely-signed norm - $v$ with $u$, or $url$ with $u$. Phonons
must be created only in such pairs.

Eq. (\ref{eq:in_vacuum_out_Fock_states}) is a generalization to a
dispersive fluid of the dispersionless in-vacuum of Eq. (\ref{eq:in-vacuum_out-Fock-basis}).
One noticeable difference is that it is now possible for $u$- and
$v$-modes to couple to each other, but in practice this is unimportant
since the strength of this coupling is typically very small. In what
follows, we shall normally neglect the Bogoliubov coefficients $C_{\omega}^{u,v}$
and $C_{\omega}^{v,u}$, so that Eq. (\ref{eq:equal_total_rates})
becomes\begin{equation}
\left|C_{\omega}^{u,url}\right|^{2}\approx\left|C_{\omega}^{url,u}\right|^{2}\approx\left|C_{\omega}^{u,u}\right|^{2}-1\,,\label{eq:rate_relations_neglecting_v}\end{equation}
and the creation rates of the $u$- and $url$-modes are approximately
equal. Eq. (\ref{eq:in_vacuum_out_Fock_states}) becomes\begin{equation}
\left|0_{\mathrm{in}}\right\rangle =\prod_{\omega}\frac{1}{\left|C_{\omega}^{u,u}\right|}\sum_{n=0}^{\infty}\left(\frac{C_{\omega}^{u,url}}{C_{\omega}^{u,u}}\right)^{n}\left|n\right\rangle _{\omega}^{url,\mathrm{out}}\left|n\right\rangle _{\omega}^{u,\mathrm{out}}\,.\label{eq:in-vacuum_neglecting-v}\end{equation}
There are two remaining differences with Eq. (\ref{eq:in-vacuum_out-Fock-basis}).
One is that, while the two out-modes of the dispersionless case are
definitely localized on opposite sides of the horizon, this is not
necessarily true in the dispersive case. The main part of the $url$-mode
is certainly localized in this way, but the $u$-mode is not, for,
as seen in Fig. \ref{fig:Variation-of-Solutions-with-V}, it behaves
perfectly regularly and experiences no group-velocity horizon. Thus
it can happen that, for a black hole horizon, the two modes are emitted
on opposite sides of the horizon, whilst for a white hole, they are
emitted on the \textit{same} side. The final difference is that, in
the dispersionless case, the coefficients in the expansion (\ref{eq:in-vacuum_neglecting-v})
are generally unknown, depending on the dispersion profile and the
flow velocity profile. However, their occurrence as Bogoliubov coefficients
in Eqs. (\ref{eq:mode_out_in_3}) provides
a way of calculating them numerically; the methods for this are described
in Chapter \ref{sec:Methods-of-Acoustic-Model}, with some results
given in Chapter \ref{sec:Results-for-Acoustic-Model}.

\section{Conclusion and discussion\label{sub:Conclusion-and-discussion}}

This chapter has shown how an acoustic field in a dispersive fluid
can be quantized via the usual methods of quantum field theory. The
process is very similar to that for a dispersionless fluid: a complete
set of orthonormal modes is found, into which the field is decomposed;
and the amplitudes of these modes are promoted to creation and annihilation
operators. If the fluid flow is not constant in space, there are two
natural sets of modes: those of a single ingoing wavevector, and those
of a single outgoing wavevector. Either of these sets can be used
to define a vacuum state, whose expectation value for any of the corresponding
{}``particles'' is zero. Under certain conditions - the mixing of
modes with norms of opposite sign - these two vacua are not equal:
the in-vacuum contains outgoing particles, and vice versa.

Dispersion \textit{does} alter the picture, however. The correspondence
between in- and out-modes is now one of scattering, and the Bogoliubov
coefficients are simply the elements of a scattering matrix, as seen
in Eqs. (\ref{eq:mode_in_out_3}) or Eqs.
(\ref{eq:mode_out_in_3}). In this sense,
the origin of mixing of positive- and negative-norm modes is simpler
to understand than in the dispersionless case. There is no need to
rely on there having been a formation, sometime in the past, of an
event horizon; whereas, in the dispersionless case, the assumption
of a past with no horizon allows us to determine how the in-modes
should be defined. This poses a problem for an eternal horizon; but
in the presence of dispersion, there is no such problem. This is closely
related to the trans-Planckian problem, which is also swept away by
dispersion. Tracing any outgoing wavepacket back into the past, we
no longer find it having arisen from arbitrarily short wavelengths;
rather, it scatters into several well-defined - and certainly finite
- ingoing wavevectors.

Of course, there are drawbacks. Although the Bogoliubov coefficients
have an intuitive interpretation as scattering amplitudes, they are
not so easy to calculate, and in general must be found numerically.
In Chapter \ref{sec:Methods-of-Acoustic-Model} we shall attempt an
analytic treatment by linearizing the velocity profile; but, as remarked
in §\ref{sub:Discussion}, since the wavepackets cannot now be traced
back in time arbitrarily close to the horizon, this linearized profile
can only give an approximate amplitude at best.

But perhaps the most surprising consequence of dispersion is that
there is no need to talk of an event horizon in the derivation of
spontaneous creation. Beginning with Eq. (\ref{eq:modes_w_inhomogeneous_v}),
we did assume the existence of one; but this only has the effect of
reducing the dimensionality of the modes because of the degeneracy
of the $ur$- and $ul$-modes, as discussed previously. We could have
assumed that both asymptotic regions are subsonic, so that Eq. (\ref{eq:phi_w_subsonic_flow})
continues to be valid, and it would have made no significant difference
to the conclusions. Only the scattering between positive- and negative-norm
modes is required for spontaneous creation to be possible. It may
be that the existence of an event horizon has a significant effect
on the calculated rate; we have yet to examine this. But even the
smallest of velocity changes, which comes nowhere near to establishing
an event horizon, can, in principle, induce particle creation.

\pagebreak{}

\chapter{Methods of Acoustic Model\label{sec:Methods-of-Acoustic-Model}}

Having derived the existence of Hawking radiation and a formula for
the radiation rate, we come now to the problem of how to calculate
this rate. Comparing Eqs. (\ref{eq:mode_out_in_3})
with Eqs. (\ref{eq:v_out_number_expectation})-(\ref{eq:u_out_number_expectation}),
we see that finding the rate is equivalent to finding the coefficients
of scattering into negative-norm modes. Neglecting the coupling into
$v$-modes so that Eqs. (\ref{eq:rate_relations_neglecting_v}) and
(\ref{eq:in-vacuum_neglecting-v}) are valid, and assuming that $V$
varies between a subsonic and a supersonic region so that the merging
of the $ul$- and $ur$-modes into a single $url$-mode occurs, the
only coefficient of interest is that for scattering between the $u$-
and $url$-modes: $\left|C_{\omega}^{u,url}\right|^{2}\approx\left|C_{\omega}^{url,u}\right|^{2}\equiv\left|\beta_{\omega}\right|^{2}$.
(We also use the standard notation for the elastic scattering, $\left|C_{\omega}^{u,u}\right|^{2}\equiv\left|\alpha_{\omega}\right|^{2}=1-\left|\beta_{\omega}\right|^{2}$.)
We may examine one of several different modes to find this coefficient.

In this chapter, we shall examine two numerical methods for finding
$ $$\left|\beta_{\omega}\right|^{2}$: in §\ref{sub:Wavepacket-propagation},
an FDTD (Finite Difference Time Domain) algorithm, which solves for
wavepacket propagation; and in §\ref{sub:Steady-state-solution},
an ordinary differential equation solver, for calculating steady-state
solutions. Steady-state solutions can be found algebraically for velocity
profiles constant everywhere except for step discontinuities, and
this is also examined. Finally, in §\ref{sub:WKB-approximation},
we examine the linearized velocity profile, in imitation of §\ref{sub:Event-horizon}
for the dispersionless case, to try to find an approximate analytic
expression for $\left|\beta_{\omega}\right|^{2}$.

\section{Wavepacket propagation\label{sub:Wavepacket-propagation}}

Here, we present an FDTD algorithm (due to Unruh \cite{Unruh-1995})
that solves directly for the propagation of a wavepacket. This is
not so useful for calculating $\left|\beta_{\omega}\right|^{2}$:
since a wavepacket is a linear superposition of solutions, the negative-norm
contribution is likewise such a superposition and will not be exactly
equal to $\left|\beta_{\omega}\right|^{2}$. Furthermore, $\left|\beta_{\omega}\right|^{2}$
is a spectral density, required to be integrated over a broad frequency
range, and the FDTD algorithm is too time-consuming to be practical
for repetition over a range of frequencies. However, it is useful
in visualising the solutions as scattering processes, as illustrated
in Figure \ref{fig:In_and_out_modes}; it shows clearly that the $ur$-
and $ul$-branches are shifted into each other by a group-velocity
horizon, and that the wavepacket does not simply come to a standstill.

We begin by expressing the second-order wave equation (\ref{eq:acoustic_wave_equation})
as two coupled first-order equations. This can be done using the canonical
momentum, $\pi$, as defined in Eq. (\ref{eq:canonical_momentum}).
We have:\begin{eqnarray}
\partial_{t}\phi+V\left(x\right)\partial_{x}\phi & = & \pi\,,\label{eq:acoustic_PDE_order1_1}\\
\partial_{t}\pi+\partial_{x}\left(V\left(x\right)\pi\right) & = & -F^{2}\left(-i\partial_{x}\right)\phi\,.\label{eq:acoustic_PDE_order1_2}\end{eqnarray}
Recall (see Eq. (\ref{eq:operation_of_F})) that the action of the
operator $F^{2}\left(-i\partial_{x}\right)$ on $\phi$ can be expressed
more explicitly using the Fourier transform $\widetilde{\phi}$:\begin{equation}
F^{2}\left(-i\partial_{x}\right)\phi=\frac{1}{2\pi}\int_{-\infty}^{+\infty}F^{2}\left(k\right)e^{ikx}\widetilde{\phi}\left(k\right)dk\,.\label{eq:F_squared_operator}\end{equation}
The action of $F^{2}\left(-i\partial_{x}\right)$ may be calculated
by repeated application of a Fast Fourier Transform algorithm. (See,
for example, Ref. \cite{NumericalRecipes} for a description.)

We now imagine a spatial grid, with separation $\Delta_{x}$ between
neighbouring points, on which we specify the values of $V$ and keep
track of the values of $\phi$ and $\pi$. Since $x$ is now replaced
by a discrete variable, which we label $i$, the continuous function
$V\left(x\right)$, $\phi\left(x\right)$ and $\pi\left(x\right)$
are replaced by their discretized versions $V_{i}=V\left(x_{i}\right)$,
$\phi_{i}=\phi\left(x_{i}\right)$ and $\pi_{i}=\pi\left(x_{i}\right)$.
The $x$-derivative at a point is approximated by the formula\begin{equation}
\partial_{x}\phi\left(x_{i}\right)=\frac{\phi_{i+1}-\phi_{i-1}}{2\Delta_{x}}\,.\end{equation}
A time grid is also introduced, with separation $\Delta_{t}$ between
neighbouring times. However, it is convenient to let the times on
which we specify $\phi$ and those on which we specify $\pi$ to be
offset by $\Delta_{t}/2$. Replacing $t$ by the discrete variable
$m$, the continuous functions $\phi\left(t,x\right)$ and $\pi\left(t,x\right)$
are replaced by their discretized versions $\phi_{m,i}=\phi\left(t_{m},x_{i}\right)$
and $\pi_{m,i}=\pi\left(t_{m}+\Delta_{t}/2,x_{i}\right)$. Since we
have to relate values of one to derivatives of the other, we form
time derivatives using the formula\begin{equation}
\partial_{t}\phi\left(t_{m}+\Delta_{t}/2,x_{i}\right)=\frac{\phi_{m+1,i}-\phi_{m,i}}{\Delta_{t}}\,,\label{eq:time_derivative}\end{equation}
while spatial derivatives are taken to be the average of the spatial
derivatives at times $t_{m}$ and $t_{m+1}$:\begin{eqnarray}
\partial_{x}\phi\left(t_{m}+\Delta_{t}/2,x_{i}\right) & = & \frac{1}{2}\left(\partial_{x}\phi\left(t_{m},x_{i}\right)+\partial_{x}\phi\left(t_{m+1},x_{i}\right)\right)\nonumber \\
 & = & \frac{1}{2}\left(\frac{\phi_{m,i+1}-\phi_{m,i-1}}{2\Delta_{x}}+\frac{\phi_{m+1,i+1}-\phi_{m+1,i-1}}{2\Delta_{x}}\right)\,.\label{eq:spatial_derivative}\end{eqnarray}
Applying Eqs. (\ref{eq:time_derivative}) and (\ref{eq:spatial_derivative})
to Eqs. (\ref{eq:acoustic_PDE_order1_1}) and (\ref{eq:acoustic_PDE_order1_2}),
we find the finite-difference equations\begin{eqnarray*}
\pi_{m,i} & = & \frac{\phi_{m+1,i}-\phi_{m,i}}{\Delta_{t}}+V_{i}\frac{1}{2}\left(\frac{\phi_{m,i+1}-\phi_{m,i-1}}{2\Delta_{x}}+\frac{\phi_{m+1,i+1}-\phi_{m+1,i-1}}{2\Delta_{x}}\right)\,,\\
-\left[F^{2}\left(-i\partial_{x}\right)\phi\right]_{m+1,i} & = & \frac{\pi_{m+1,i}-\pi_{m,i}}{\Delta_{t}}\\
 &  & +\frac{1}{2}\left(\frac{V_{i+1}\pi_{m,i+1}-V_{i-1}\pi_{m,i-1}}{2\Delta_{x}}+\frac{V_{i+1}\pi_{m+1,i+1}-V_{i-1}\pi_{m+1,i-1}}{2\Delta_{x}}\right)\,,\end{eqnarray*}
which can be written in the more suggestive form\begin{eqnarray}
\phi_{m+1,i}+\frac{\Delta_{t}}{4\Delta_{x}}V_{i}\left(\phi_{m+1,i+1}-\phi_{m+1,i-1}\right) & = & \phi_{m,i}-\frac{\Delta_{t}}{4\Delta_{x}}V_{i}\left(\phi_{m,i+1}-\phi_{m,i-1}\right)\nonumber \\
 &  & \qquad\qquad\qquad\qquad+\Delta_{t}\pi_{m,i}\,,\qquad\label{eq:acoustic_FDTD_1}\\
\pi_{m+1,i}+\frac{\Delta_{t}}{4\Delta_{x}}\left(V_{i+1}\pi_{m+1,i+1}-V_{i-1}\pi_{m+1,i-1}\right) & = & \pi_{m,i}-\frac{\Delta_{t}}{4\Delta_{x}}\left(V_{i+1}\pi_{m,i+1}-V_{i-1}\pi_{m,i-1}\right)\nonumber \\
 &  & \qquad\qquad\;-\Delta_{t}\left[F^{2}\left(-i\partial_{x}\right)\phi\right]_{m+1,i}\,.\qquad\label{eq:acoustic_FDTD_2}\end{eqnarray}

Equations (\ref{eq:acoustic_FDTD_1}) and (\ref{eq:acoustic_FDTD_2})
are almost closed, except at the boundaries, where the values of $\phi$
or $\pi$ at $i=1$ or $n$ depends on their values at $i=0$ or $n+1$.
The latter are not included in the grid, but this problem can be overcome
by imposing periodic boundary conditions; that is, we identify $i=0$
with $i=n$ and $i=n+1$ with $i=1$.

There remains the problem of specifying the initial profile for the
momentum $\pi$, once the initial form of $\phi$ has been specified.
In principle, there are no limitations on $\pi$, for there are two
possible dispersion branches and a general solution of the wave equation
can contain an arbitrary mixture of the two. However, we wish to have
a wavepacket centred at well-defined values of $k$ \textit{and} $\omega$;
this selects a single dispersion branch, and the momentum is uniquely
specified. Given that we use wavepackets which are right-moving in
the free-fall frame, and which are initially in a region of constant
flow velocity, we have $\omega=Vk+F\left(k\right)$ and the momentum
is given by\begin{equation}
\pi\left(x,t\right)=-\frac{i}{2\pi}\int_{-\infty}^{+\infty}F\left(k\right)\widetilde{\phi}\left(k\right)e^{ikx-i\left(Vk+F\left(k\right)\right)t}dk\,.\label{eq:pi_from_phi}\end{equation}
Thus, the momentum at time $t=0$ - which shall be labelled $\pi_{1/2,i}$,
since it comes half a timestep before $\pi_{1,i}$ - is easily calculated
using the Fast Fourier Transform algorithm. $\pi_{1,i}$ is then found
by propagating forward $\Delta_{t}/2$ using the modified difference
equation\begin{equation}
\pi_{1,i}=\pi_{1/2,i}-\frac{\Delta_{t}}{4\Delta_{x}}\left(V_{i+1}\pi_{1/2,i+1}-V_{i-1}\pi_{1/2,i-1}\right)-\frac{\Delta_{t}}{2}\left[F^{2}\left(-i\partial_{x}\right)\phi\right]_{1,i}\,.\label{eq:mod_diff_eqn_initial_pi}\end{equation}

With these conditions, Eqs. (\ref{eq:acoustic_FDTD_1}) and (\ref{eq:acoustic_FDTD_2})
form a closed linear system for each value of $m$: given $\phi_{m}$
and $\pi_{m}$, we can use Eq. (\ref{eq:acoustic_FDTD_1}) to find
$\phi_{m+1}$, and this in turn allows us to use Eq. (\ref{eq:acoustic_FDTD_2})
to find $\pi_{m+1}$. At each stage, the right-hand side is known.
The term $\left[F^{2}\left(-i\partial_{x}\right)\phi\right]_{m+1,i}$
is found via Eq. (\ref{eq:F_squared_operator}), using a Fast Fourier
Transform algorithm first to find $\widetilde{\phi}$, and then again
to take the inverse Fourier transform after multiplying by $F^{2}\left(k\right)$.
The knowns of the right-hand side form an $n$-dimensional vector;
the left-hand side contains $n$ unknowns in total. The whole linear
system can be written in matrix form as follows:\begin{eqnarray}
\negthickspace\negthickspace\negthickspace\negthickspace\negthickspace\negthickspace\negthickspace\negthickspace\negthickspace\negthickspace\negthickspace\negthickspace\negthickspace\negthickspace\negthickspace\negthickspace\left[\begin{array}{ccccccccc}
1 & aV_{1} &  &  &  &  &  &  & -aV_{1}\\
 & \ddots\\
 &  & \ddots\\
 &  & -aV_{i-1} & 1 & aV_{i-1}\\
 &  &  & -aV_{i} & 1 & aV_{i}\\
 &  &  &  & -aV_{i+1} & 1 & aV_{i+1}\\
 &  &  &  &  &  & \ddots\\
 &  &  &  &  &  &  & \ddots\\
aV_{n} &  &  &  &  &  &  & -aV_{n} & 1\end{array}\right]\left[\begin{array}{c}
\phi_{m+1,1}\\
\vdots\\
\vdots\\
\phi_{m+1,i-1}\\
\phi_{m+1,i}\\
\phi_{m+1,i+1}\\
\vdots\\
\vdots\\
\phi_{m+1,n}\end{array}\right] & = & \left[\begin{array}{c}
b_{m,1}\\
\vdots\\
\vdots\\
b_{m,i-1}\\
b_{m,i}\\
b_{m,i+1}\\
\vdots\\
\vdots\\
b_{m,n}\end{array}\right]\nonumber\\
\negthickspace\negthickspace\negthickspace\negthickspace\negthickspace\negthickspace\negthickspace\negthickspace\negthickspace\negthickspace\negthickspace\negthickspace\negthickspace\negthickspace\negthickspace\negthickspace\left[\begin{array}{ccccccccc}
1 & aV_{2} &  &  &  &  &  &  & -aV_{n}\\
 & \ddots\\
 &  & \ddots\\
 &  & -aV_{i-2} & 1 & aV_{i}\\
 &  &  & -aV_{i-1} & 1 & aV_{i+1}\\
 &  &  &  & -aV_{i} & 1 & aV_{i+2}\\
 &  &  &  &  &  & \ddots\\
 &  &  &  &  &  &  & \ddots\\
aV_{1} &  &  &  &  &  &  & -aV_{n-1} & 1\end{array}\right]\left[\begin{array}{c}
\pi_{m+1,1}\\
\vdots\\
\vdots\\
\pi_{m+1,i-1}\\
\pi_{m+1,i}\\
\pi_{m+1,i+1}\\
\vdots\\
\vdots\\
\pi_{m+1,n}\end{array}\right] & = & \left[\begin{array}{c}
c_{m,1}\\
\vdots\\
\vdots\\
c_{m,i-1}\\
c_{m,i}\\
c_{m,i+1}\\
\vdots\\
\vdots\\
c_{m,n}\end{array}\right] \nonumber \end{eqnarray}
where we have defined\begin{eqnarray}
a & = & \frac{\Delta_{t}}{4\Delta_{x}}\,,\label{eq:matrix_parameter_a}\\
b_{m,i} & = & \phi_{m,i}-\frac{\Delta_{t}}{4\Delta_{x}}V_{i}\left(\phi_{m,i+1}-\phi_{m,i-1}\right)+\Delta_{t}\pi_{m,i}\,,\label{eq:matrix_parameter_b}\\
c_{m,i} & = & \pi_{m,i}-\frac{\Delta_{t}}{4\Delta_{x}}\left(V_{i+1}\pi_{m,i+1}-V_{i-1}\pi_{m,i-1}\right)-\Delta_{t}\left[F^{2}\left(-i\partial_{x}\right)\phi\right]_{m+1,i}\,.\label{eq:matrix_parameter_c}\end{eqnarray}

The linear systems above are known as \textit{cyclic} equations,
given their periodic nature, and are easily solved using an appropriate
algorithm. (See Ref. \cite{NumericalRecipes}, §2.4.)

\begin{figure}
\subfloat{\includegraphics[width=0.45\columnwidth]{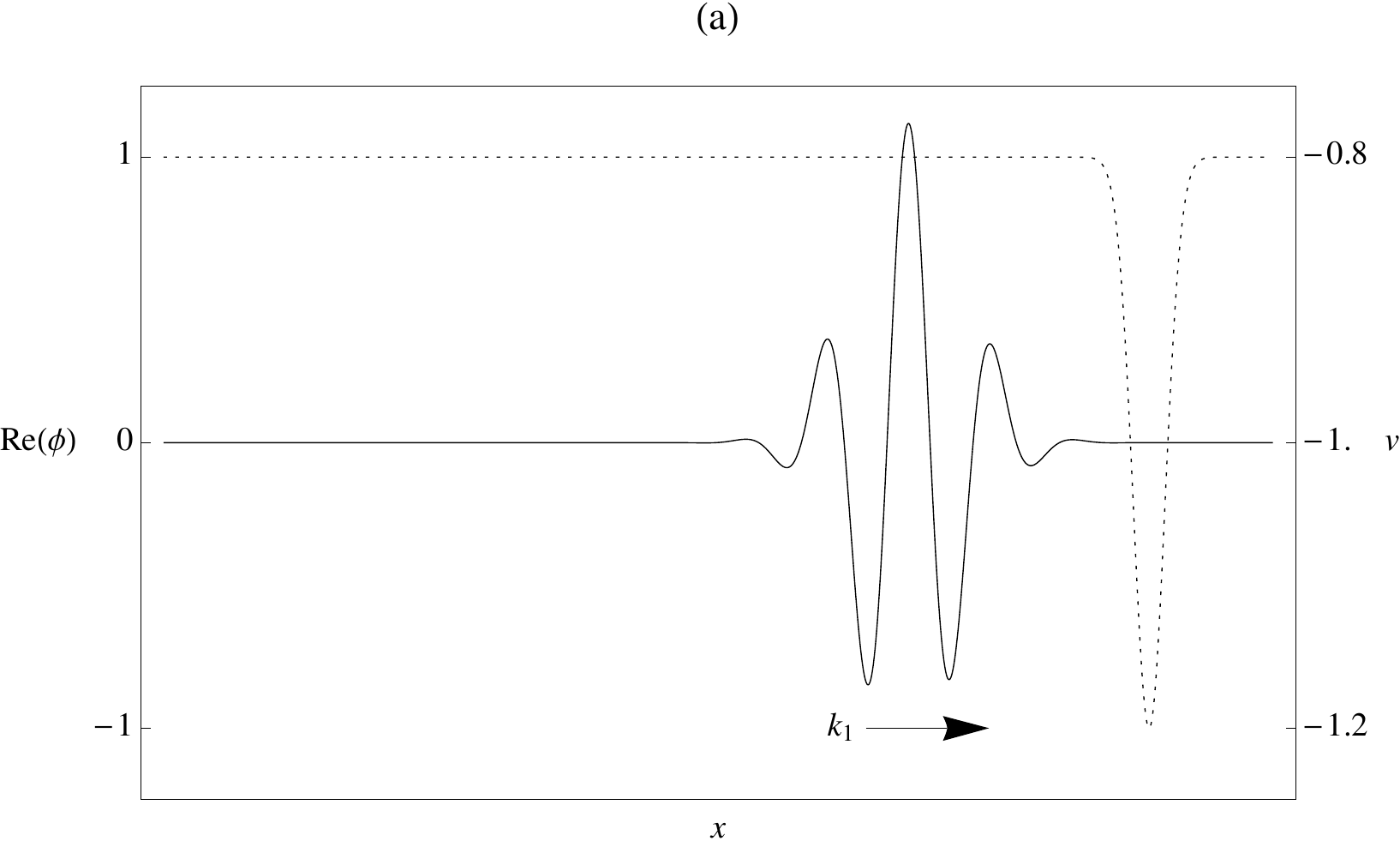}} \subfloat{\includegraphics[width=0.45\columnwidth]{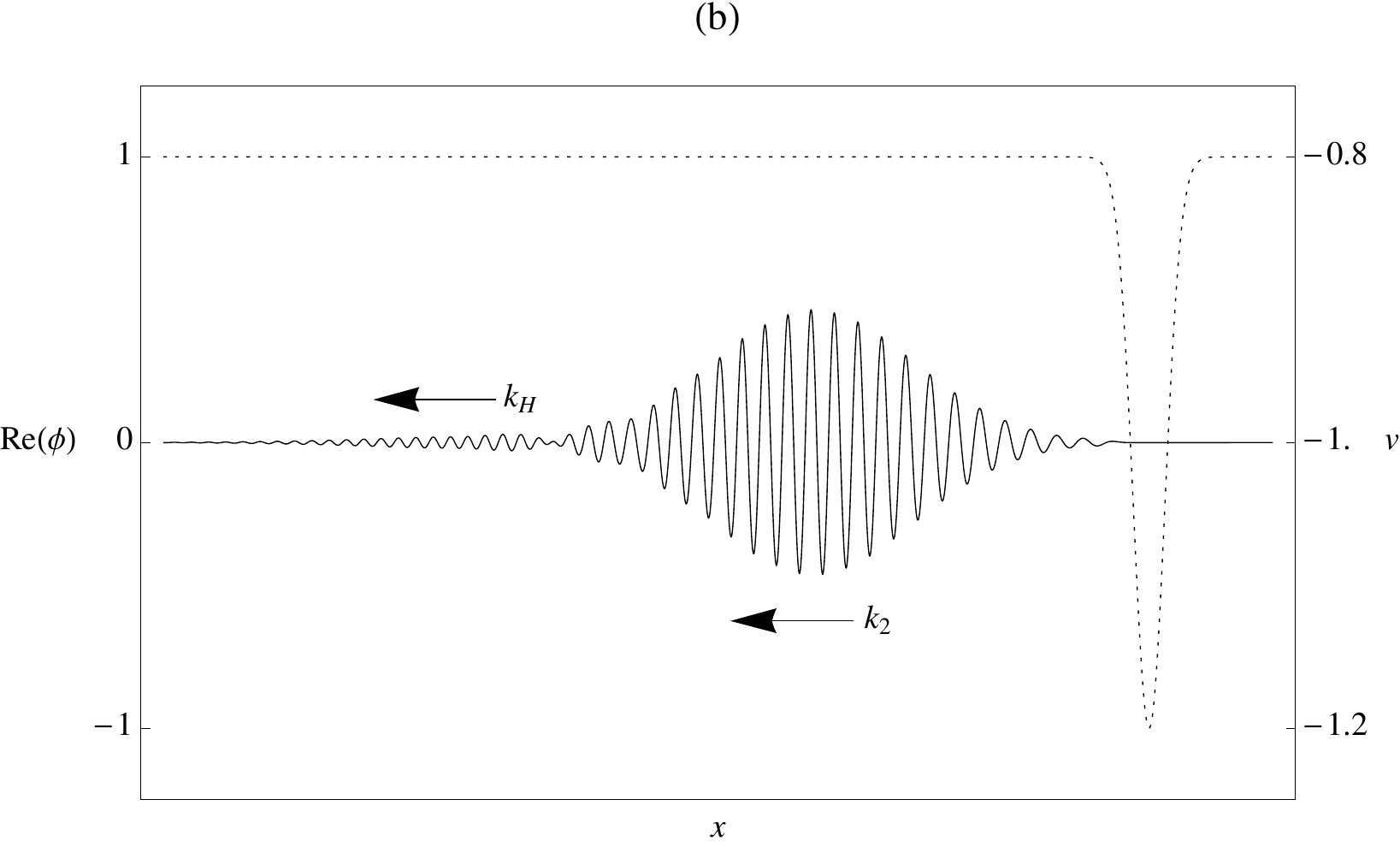}}

\subfloat{\includegraphics[width=0.45\columnwidth]{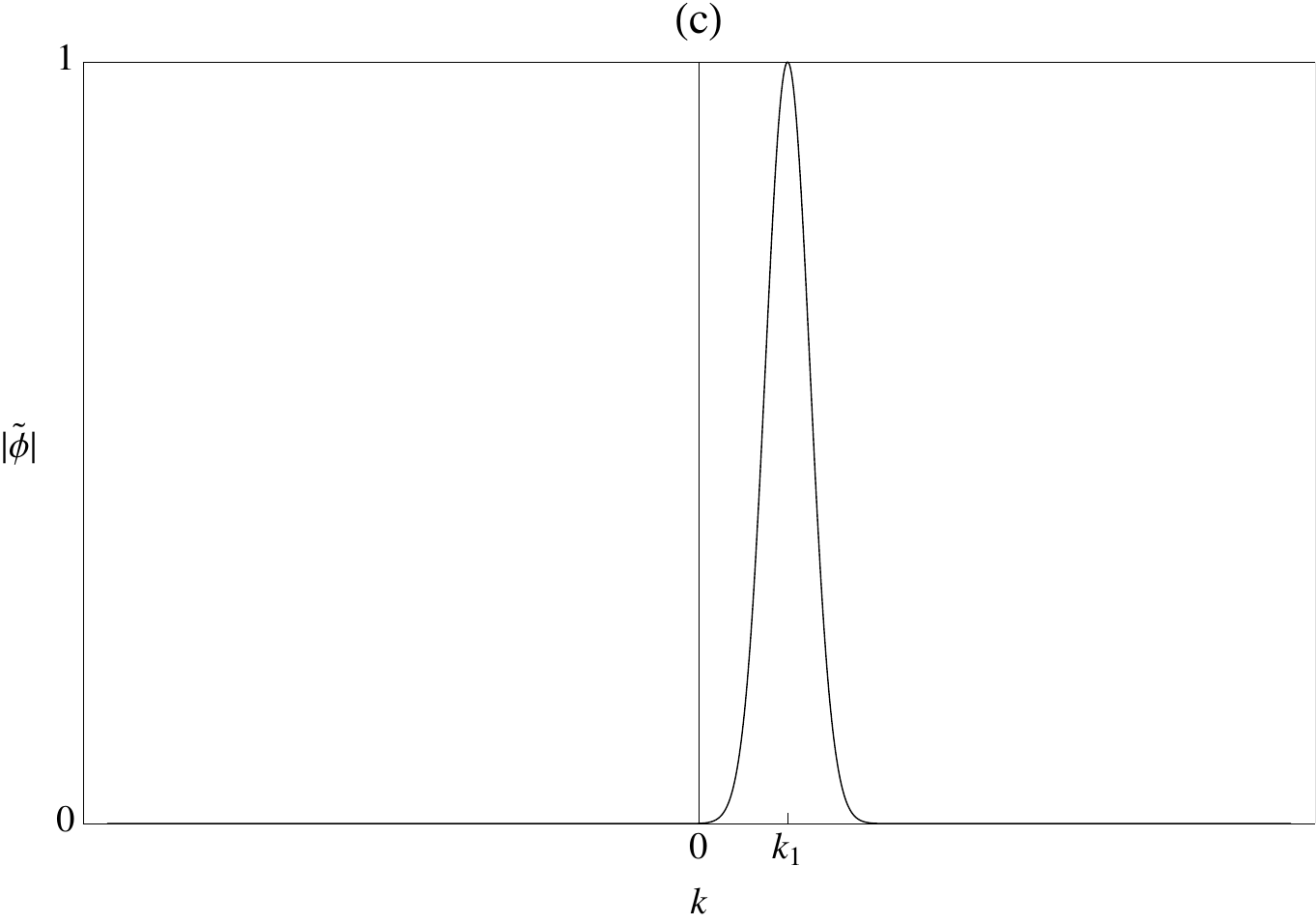}} \subfloat{\includegraphics[width=0.45\columnwidth]{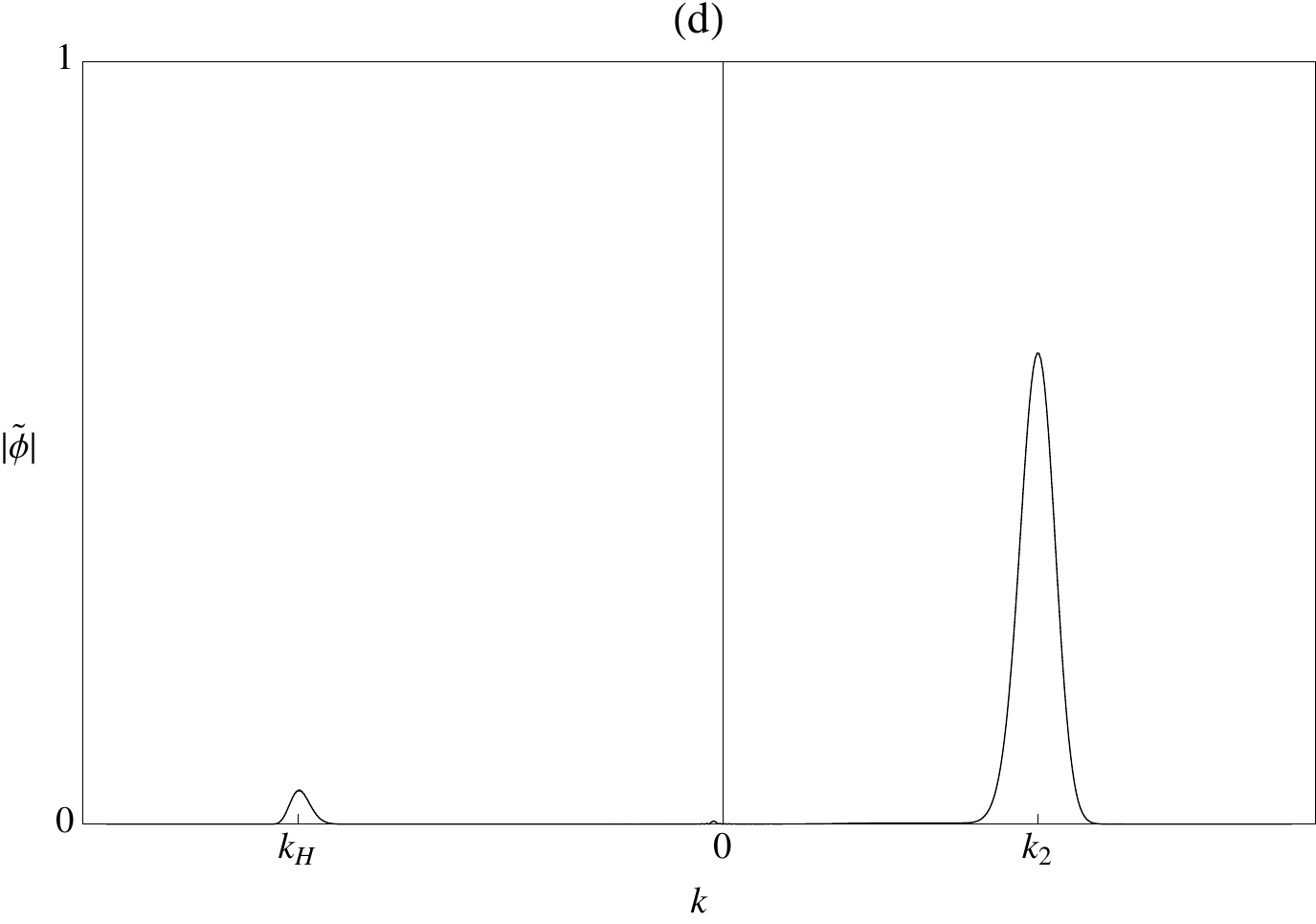}}

\caption[\textsc{Wavepacket propagation}]{\textsc{Wavepacket propagation}: These are results for the scattering
of a $ur$-mode wavepacket incident on a white-hole horizon. Figure
$\left(a\right)$ shows the incident wavepacket and velocity profile.
Figure $\left(b\right)$ shows the resulting outgoing wavepackets:
the $ur$-mode has been blueshifted into the $ul$-mode by the horizon,
and a negative-norm $u$-mode wavepacket has also been produced. Figures
$\left(c\right)$ and $\left(d\right)$ show the Fourier transforms
of the waveforms in Figs. $\left(a\right)$ and $\left(b\right)$,
respectively.\label{fig:wavepacket_propagation}}

\end{figure}

Calculation of the norm of the solution before and after the propagation
provides a check on the accuracy of the numerical procedure. Recall
(see Eqs. (\ref{eq:scalar_product}) and (\ref{eq:norm_constant_v}))
that the norm is given by Eq. (\ref{eq:scalar_product_II}) or, in
regions where $V$ is constant, by Eq. (\ref{eq:norm_constant_v}).
The negative-norm contribution, equivalent to the radiation rate,
is\begin{equation}
\left|\beta_{\omega}\right|^{2}=\frac{1}{\pi}\int\left|F\left(k\right)\right|\left|\widetilde{\phi}^{\left(-\right)}\left(k\right)\right|^{2}dk\,,\label{eq:negative_norm_contribution}\end{equation}
where $\widetilde{\phi}^{\left(-\right)}\left(k\right)$ is the Fourier
transform of the negative-norm branch. Therefore, if we begin with
a positive-norm $url$-mode wavepacket and allow it to scatter completely
into outgoing wavepackets, $\left|\beta_{\omega}\right|^{2}$ is found
by integrating Eq. (\ref{eq:negative_norm_contribution}) over the
$u$-mode part of the resulting Fourier transform.

See Figure \ref{fig:wavepacket_propagation} for an example which
shows frequency shifting between the $ur$- and $ul$-modes.

\section{Steady-state solution\label{sub:Steady-state-solution}}

The creation rate $\left|\beta_{\omega}\right|^{2}$ is exact only
for a stationary solution, the wavepackets of §\ref{sub:Wavepacket-propagation}
being superpositions of stationary solutions peaked at a certain value
of $\omega$. Solving for the stationary solutions directly is also
faster than performing the FDTD algorithm of §\ref{sub:Wavepacket-propagation},
which makes it the more practical method for calculating a spectrum
which extends over many frequencies. Here, we determine how to find
the stationary solutions. We begin with a general velocity profile,
which must be solved by numerical integration and linear algebra;
such a method is also described and used in Refs. \cite{Corley-Jacobson-1996,Corley-1997,Macher-Parentani-2008}.
We then examine the special case of a velocity profile that is constant
except for step discontinuities, which is solvable by linear algebra
only; see also Refs. \cite{Corley-1997,Recati-2009}.

\subsection{General velocity profile}

The time-independent acoustic wave equation is the ordinary differential
equation satisfied by $\phi_{\omega}\left(x\right)$, where\begin{equation}
\phi\left(x,t\right)=\phi_{\omega}\left(x\right)e^{-i\omega t}\,.\label{eq:steady_state_soln}\end{equation}
This is a stationary solution, whose time-dependence is purely oscillatory.
Plugging it into Eq. (\ref{eq:acoustic_wave_equation}), we find\begin{equation}
i\omega\left(i\omega-V^{\prime}\left(x\right)\right)\phi_{\omega}-2V\left(x\right)\left(i\omega-V^{\prime}\left(x\right)\right)\partial_{x}\phi_{\omega}+V^{2}\left(x\right)\partial_{x}^{2}\phi_{\omega}+F^{2}\left(-i\partial_{x}\right)\phi_{\omega}=0\,.\label{eq:acoustic_steady_state_eqn}\end{equation}
If $V$ is constant for all $x$, we may take the Fourier transform
of this equation to obtain Eq. (\ref{eq:acoustic_dispersion_FT}),
with $\omega$ now interpreted as a fixed parameter. The general solution
is simply a linear combination of plane waves,\begin{equation}
\phi_{\omega}\left(x\right)=\sum_{j}c_{j}\exp\left(ik_{j}x\right)\,,\label{eq:sum_of_exps}\end{equation}
where the index $j$ enumerates all possible solutions of the dispersion
relation (\ref{eq:acoustic_flow_dispersion}), and the $c_{j}$ are
complex constants. Note that some of the $k_{j}$ might be complex
(and if so, they will occur in complex conjugate pairs, since the
dispersion relation is real); however, given that they will increase
exponentially in one direction, these do not represent physical solutions
and can be discarded. Also note that there may be infinitely many
such plane waves, so that Eq. (\ref{eq:sum_of_exps}) is an infinite
series; we shall, however, assume $F^{2}\left(k\right)$ to be a finite
polynomial, so that $j$ has a maximum value.

Now let $V$ be spatially varying, yet still asymptotically constant:
$V\left(x\right)\rightarrow V_{\pm}$ as $x\rightarrow\pm\infty$.
Then, in the asymptotic regions, $\phi_{\omega}$ can still be represented
by the sum in Eq. (\ref{eq:sum_of_exps}). In general, the sums that
correspond to the negative and positive asymptotic regions will not
agree, even if $V_{+}=V_{-}$. This allows the appearance of complex
values of $k$ which are exponentially decreasing towards infinity,
for such a value of $k$ need not (indeed, \textit{will} not, for
physical solutions) appear in the opposite asymptotic region.

\begin{figure}
\includegraphics[width=0.8\columnwidth]{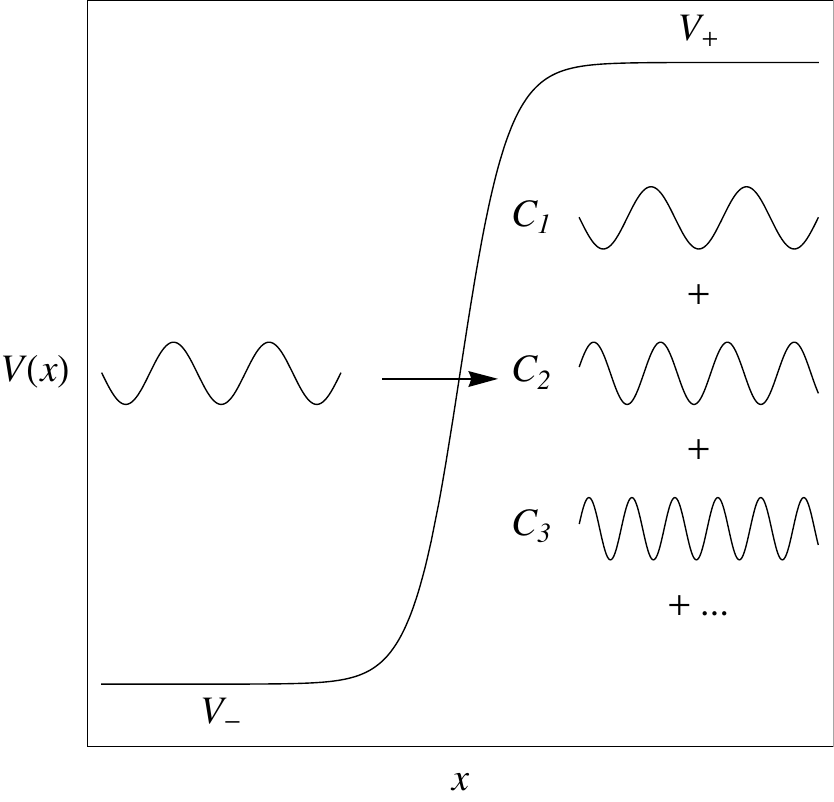}

\caption[\textsc{Integration of stationary solution}]{\textsc{Integration of stationary solution}: In one asymptotic region,
the solution is chosen to be a particular plane wave. Eq. (\ref{eq:acoustic_steady_state_eqn})
is then integrated through to the other region, where the solution
is a linear combination of plane waves. This is done for all possible
plane wave solutions in the initial region. The general solution is
a linear combination of these, thus giving rise to the linear transformation
of Eq. (\ref{eq:matrix_transformation}).\label{fig:Integration-of-Stationary-Solution}}

\end{figure}

The sums in the asymptotic regions are not arbitrary, but related
to each other via Eq. (\ref{eq:acoustic_steady_state_eqn}). Suppose
we specify that, in the negative asymptotic region, $\phi_{\omega}\left(x\right)$
should be given by the plane wave $\exp\left(ik_{1}^{-}x\right)$.
This is a boundary condition, and in a differential equation solver
- such as the Mathematica function NDSolve, used for this thesis - we can set $\phi_{\omega}$
and the necessary number of derivatives of $\phi_{\omega}$ equal
to those of $\exp\left(ik_{1}^{-}x\right)$ in the negative asymptotic
region, and solve Eq. (\ref{eq:acoustic_steady_state_eqn}) through
to the positive asymptotic region; this is illustrated in Figure \ref{fig:Integration-of-Stationary-Solution}.
Here, $\phi_{\omega}$ is again expressible as a sum of exponentials:
$\phi_{\omega}\left(x\right)\rightarrow\underset{j}{\sum}C_{1j}\exp\left(ik_{j}^{+}x\right)$.
This solution is wholly determined by the constants $C_{1j}$, which
form a vector: $\left(C_{11},C_{12},\ldots,C_{1n}\right)$. The same
integration can be performed for all other $k_{i}^{-}$ in the negative
asymptotic region, whose solutions in the positive asymptotic region
are wholly determined by vectors $\left(C_{i1},C_{i2},\ldots,C_{in}\right)$.
Moreover, since Eq. (\ref{eq:acoustic_steady_state_eqn}) is linear
in $\phi_{\omega}$, a linear superposition of solutions is itself
a solution. Thus, if we form a solution in the negative asymptotic
region of the form $\phi_{\omega}\rightarrow\underset{i}{\sum}a_{i}\exp\left(ik_{i}^{-}x\right)$
- which can be represented as the vector $\left(a_{1},a_{2},\ldots,a_{n}\right)$
- the solution in the positive asymptotic region, $\phi_{\omega}\rightarrow\underset{j}{\sum}b_{j}\exp\left(ik_{j}^{+}x\right)$,
will be represented by the vector $a_{1}\left(C_{1j}\right)+a_{2}\left(C_{2j}\right)+\ldots+a_{n}\left(C_{nj}\right)$.
It is clear that the vectors which specify the solutions in each of
the asymptotic regions are related via a linear transformation, which
can be written in matrix form:\begin{equation}
\left(\begin{array}{c}
b_{1}\\
b_{2}\\
\vdots\\
b_{n}\end{array}\right)=\left(\begin{array}{cccc}
C_{11} & C_{21} & \cdots & C_{n1}\\
C_{12} & C_{22} & \cdots & C_{n2}\\
\vdots & \vdots & \ddots & \vdots\\
C_{1n} & C_{2n} & \cdots & C_{nn}\end{array}\right)\left(\begin{array}{c}
a_{1}\\
a_{2}\\
\vdots\\
a_{n}\end{array}\right)\,.\label{eq:matrix_transformation}\end{equation}
The importance of Eq. (\ref{eq:matrix_transformation}) is that it
makes clear that there are only $n$ values which need be specified
to pick out a particular solution of $\phi_{\omega}$; and, moreover,
one of these is simply an overall multiplicative constant, which is
unimportant. The $n-1$ values remaining need not refer to coefficients
in a single asymptotic region. Indeed, when we solve for in- and out-modes,
it will be found that some of the coefficients on either side must
be set to zero, depending on whether they are ingoing or outgoing.
The remaining coefficients are found by solving Eq. (\ref{eq:matrix_transformation}),
with one of them set to an arbitrary constant which determines the
overall multiplicative factor; in practice, we usually set the coefficient
of the single ingoing or outgoing wave to unity.

\begin{figure}
\includegraphics[width=0.8\columnwidth]{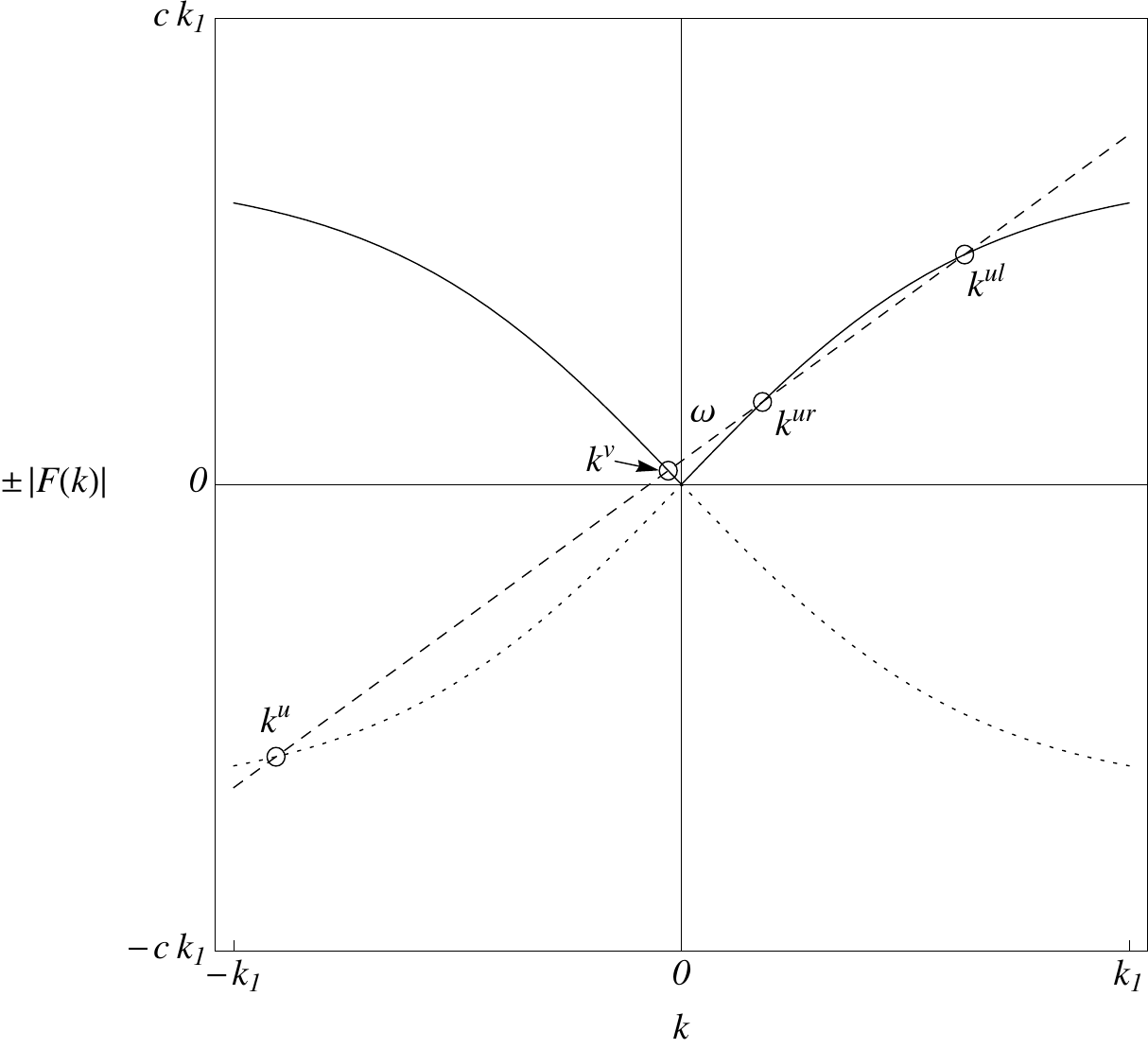}

\caption[\textsc{Solutions and their group velocities}]{\textsc{Solutions and their group velocities}: The group velocity
is $v_{g}=-\left(-V\right)\pm F^{\prime}\left(k\right)$. Therefore,
the direction of travel of a wave is determined by whether the slope
of $\pm F\left(k\right)$ is greater or less than the slope of $\omega-Vk$.
In this figure, $k^{u}$, $k^{v}$ and $k^{ul}$ all have a slope
$\pm F^{\prime}\left(k\right)$ which is less than the slope of $\omega-Vk$,
and are thus left-travelling wavepackets. However, $k^{ur}$ has a
slope $F^{\prime}\left(k\right)$ greater than that of $\omega-Vk$,
and is therefore a right-travelling wavepacket.\label{fig:dispersion_group_velocities}}

\end{figure}

If solving for an in- or out-mode, the $a_{i}$ and $b_{j}$ which
should be set to zero are determined by the direction of propagation
of the corresponding wavevectors. For example, if we are solving for
an in-mode as in Fig. \ref{fig:In_and_out_modes}$(a)$, then only
the initial wavevector can propagate \textit{towards} the horizon,
whilst any wavevectors propagating away from the horizon are allowed;
any ingoing wavevectors other than the initial one must have a coefficient
of zero. Whether a wave is ingoing or outgoing depends on the sign
of the group velocity,

\begin{equation}
v_{g}\left(k\right)=\frac{\partial\omega}{\partial k}=V\pm F^{\prime}\left(k\right)=-\left(-V\right)\pm F^{\prime}\left(k\right)\,.\end{equation}
Therefore, if the slope of $\pm F\left(k\right)$ is greater than
the slope of $\omega-Vk$, the group velocity is positive and a wavepacket
centred at the corresponding value of $k$ travels to the right; whereas,
if the slope of $\pm F\left(k\right)$ is less than the slope of $\omega-Vk$,
the group velocity is negative and such a wavepacket travels to the
left. In Fig. \ref{fig:dispersion_group_velocities}, $k^{u}$, $k^{v}$
and $k^{ul}$ have negative group velocities and travel to the left,
while $k^{ur}$ has a positive group velocity and travels to the right.
Therefore, for a given value of $V$, we can easily determine the
possible real values of $k$ and the direction of travel of their
wavepackets.

The Bogoliubov coefficients describe transformations between \textit{normalized}
solutions, which are given by Eq. (\ref{eq:mode_w}). Having calculated
the coefficient $C_{\omega,k}$ of the plane wave $\exp\left(ikx-i\omega t\right)$,
the norm of this wave - relative to the norm of the incoming or outgoing
wave - is given by \cite{Corley-1997}
\begin{equation}
\frac{\left(\phi_{\omega,k},\phi_{\omega,k}\right)}{\left(\phi_{\omega,k_{\mathrm{in/out}}},\phi_{\omega,k_{\mathrm{in/out}}}\right)}=\frac{\mathrm{sgn}\left(\omega-Vk\right)\left|F\left(k\right)\right|\left|v_{g}\left(k\right)\right|\left|C_{\omega,k}\right|^{2}}{\mathrm{sgn}\left(\omega-Vk_{\mathrm{in/out}}\right)\left|F\left(k_{\mathrm{in/out}}\right)\right|\left|v_{g}\left(k_{\mathrm{in/out}}\right)\right|\left|C_{\omega,k_{\mathrm{in/out}}}\right|^{2}}\,,\label{eq:norm_from_coefficient}\end{equation}
This equation allows us to determine the accuracy of the solution
by checking that the norm is conserved, i.e., that the sum of the
norms of all incoming waves is equal to the sum of the norms of all
outgoing waves. We can also find the creation rate $\left|\beta_{\omega}\right|^{2}$,
which is simply the relative norm between wavevectors of oppositely-signed
norm.

\subsection{Discontinuous velocity profile}

Given the complexity of Eq. (\ref{eq:acoustic_steady_state_eqn}),
purely analytic solutions are often impossible to obtain. An exception
to this, however, occurs when $V$ is constant everywhere except at
discrete points, where any changes are marked by a sudden discontinuity.
Such discontinuous velocity profiles form limiting cases, where the
derivative of $V$ has been allowed to increase without bound. Similar step-discontinuous
profiles have been examined in References \cite{Corley-Jacobson-1996} and \cite{Recati-2009}.

We have already observed that, in regions where the flow velocity
is constant, $\phi_{\omega}$ is simply a linear combination of plane
waves. Suppose that $F^{2}\left(k\right)$ is of order $2n$, giving
$2n$ solutions to the dispersion relation; then, in any such constant-velocity
region, the general solution is characterised by the $2n$ coefficients
corresponding to the different possible plane waves. Suppose also
that the velocity profile $V\left(x\right)$ has $r$ constant velocity
regions (and therefore $r-1$ points of discontinuity). Then there
are $2nr$ coefficients in total, and all must be specified to completely
characterise the solution. Of course, we still have only $2n$ in-
or out-modes, so the $2nr$ coefficients are not all independent:
they are restricted by $2n\left(r-1\right)$ conditions to be imposed
at the points of discontinuity, or $2n$ conditions at each such point.
These conditions relate the function $\phi_{\omega}$ and its first
$2n-1$ derivatives on either side of the point of discontinuity,
and are derived such that, as the velocity profile is gradually altered
towards the formation of the discontinuity, Eq. (\ref{eq:acoustic_steady_state_eqn})
is always satisfied.

Firstly, note that Eq. (\ref{eq:acoustic_steady_state_eqn}) contains
$V$ and its first derivative $V^{\prime}$ only. A step discontinuity
in $V$ results in $V^{\prime}$ being infinite (i.e. proportional
to the delta function) at the point of discontinuity. These must be
compensated for by corresponding discontinuities and delta functions
appearing in the derivatives of $\phi$. We can immediately conclude
that, since the highest derivative of $\phi$ appearing in Eq. (\ref{eq:acoustic_steady_state_eqn})
is of order $2n$, $\phi$ and its first $2n-2$ derivatives are continuous
everywhere; for, if this were not the case, $\partial_{x}^{2n}\phi$
would contain derivatives of the delta function, which would not be
compensated for by similar infinities in another quantity. Therefore,
at any point of discontinuity, the first $2n-1$ conditions imposed
are that $\phi,$ $\partial_{x}\phi$, $\ldots$, $\partial_{x}^{2n-2}\phi$
be continuous, their values on either side of the point being equal.
This leaves one condition remaining, which relates $\partial_{x}^{2n-1}\phi$
on either side of the point. There must be a discontinuity in $\partial_{x}^{2n-1}\phi$
in order to generate a delta function in $\partial_{x}^{2n}\phi$
to cancel that in $V^{\prime}$; but what is this discontinuity?

\begin{figure}
\subfloat{\includegraphics[width=0.45\columnwidth]{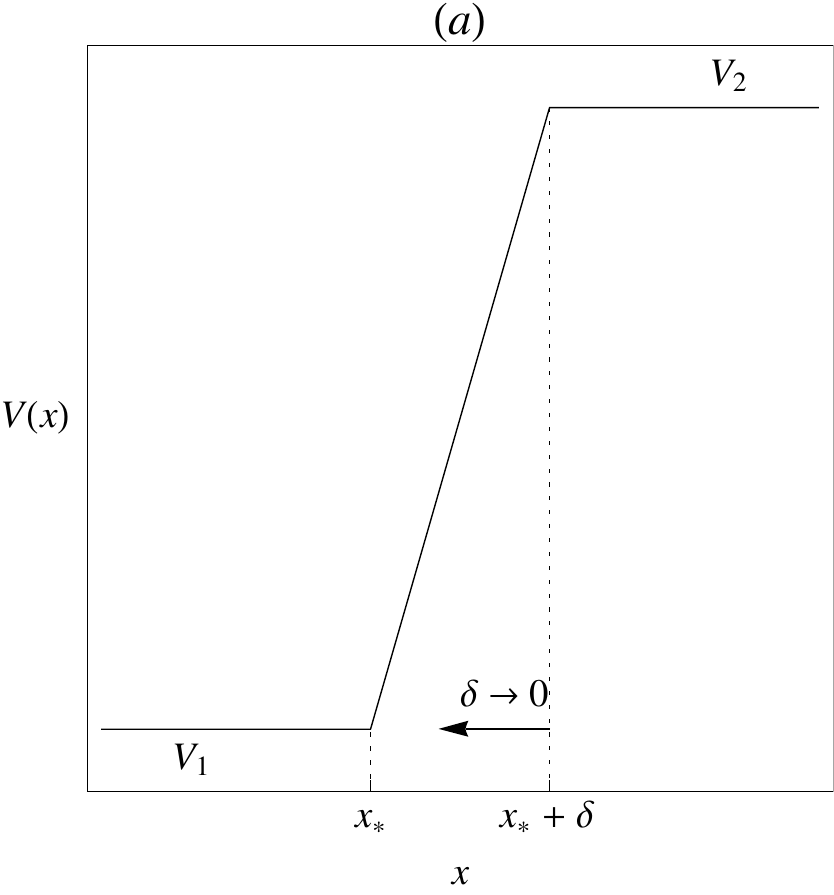}} \subfloat{\includegraphics[width=0.45\columnwidth]{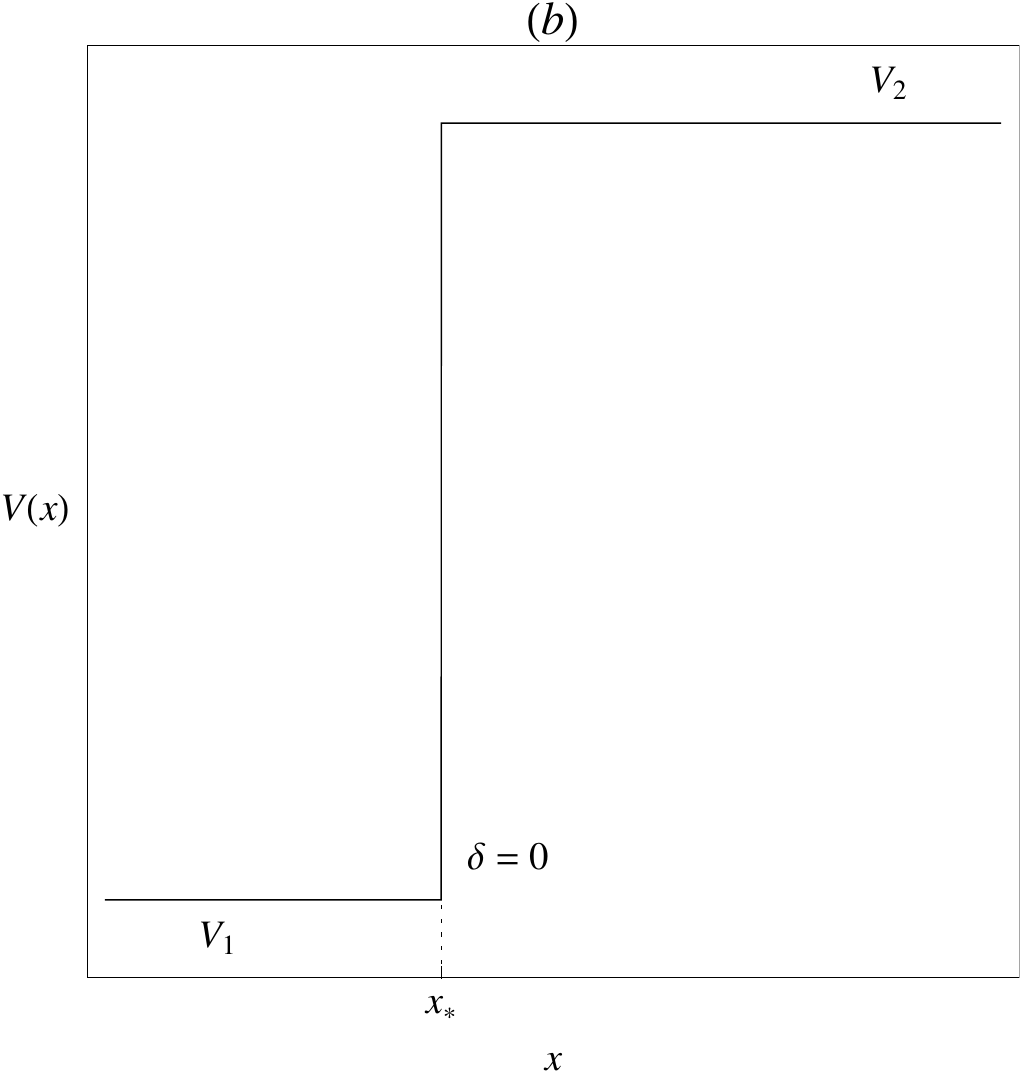}}

\caption[\textsc{Constructing a discontinuous velocity profile}]{\textsc{Constructing a discontinuous velocity profile}: In Figure
$\left(a\right)$, $V$ is continuous while $V^{\prime}$ is not.
As we let the length of the region of change go to zero, we obtain
Figure $\left(b\right)$, where $V$ is discontinuous.\label{fig:discontinuous_velocity_profile}}

\end{figure}

Let us proceed along lines suggested by Figure \ref{fig:discontinuous_velocity_profile}.
We consider a velocity profile which varies between two velocities,
which are constant in their respective regions. The variation between
them is assumed to be linear. We therefore have a discontinuity not
in $V$, but $V^{\prime}$, which is zero for $x<x_{\star}$ and $x>x_{\star}+\delta$,
and equal to $\left(V_{2}-V_{1}\right)/\delta$ for $x_{\star}<x<x_{\star}+\delta$.
Our task is to calculate the change in $\partial_{x}^{2n-1}\phi$
between $x=x_{\star}$ and $x=x_{\star}+\delta$, and then to take
the limit as $\delta\rightarrow0$.

With the velocity profile of Fig. \ref{fig:discontinuous_velocity_profile}(a),
in order that Eq. (\ref{eq:acoustic_steady_state_eqn}) be satisfied
on either side of the point $x=x_{\star}$ there must be a corresponding
discontinuity in another quantity to cancel that from $V^{\prime}$.
For reasons similar to those given above, only the highest-order derivative
of $\phi$ that appears in Eq. (\ref{eq:acoustic_steady_state_eqn})
can be discontinuous; for, if a lower-order derivative were discontinuous,
all higher-order derivatives would be infinite at $x=x_{\star}$,
and these infinities would not be cancelled. So, the discontinuity
in $V^{\prime}$ is compensated for by a corresponding discontinuity
in $\partial_{x}^{2n}\phi$. Writing $F^{2}\left(k\right)$ explicity
as a finite polynomial,

\begin{equation}
F^{2}\left(k\right)=\sum_{j=1}^{n}f_{j}k^{2j}\,,\label{eq:F_squared_polynomial}\end{equation}
Eq. (\ref{eq:acoustic_steady_state_eqn}) becomes\begin{equation}
-\omega^{2}\phi-2i\omega V\partial_{x}\phi-i\omega V^{\prime}\phi+2VV^{\prime}\partial_{x}\phi+V^{2}\partial_{x}^{2}\phi+\sum_{j=1}^{n}\left(-1\right)^{j}f_{j}\partial_{x}^{2j}\phi=0\,.\label{eq:acoustic_ODE_F_squared_polynomial}\end{equation}
In the limit as we approach the point of discontinuity - which we
label $x_{\star}$ - we must have\begin{equation}
\lim_{x\rightarrow x_{\star}^{-}}\left[\left(-i\omega\phi+2V\partial_{x}\phi\right)V^{\prime}+\left(-1\right)^{n}f_{n}\partial_{x}^{2n}\phi\right]=\lim_{x\rightarrow x_{\star}^{+}}\left[\left(-i\omega\phi+2V\partial_{x}\phi\right)V^{\prime}+\left(-1\right)^{n}f_{n}\partial_{x}^{2n}\phi\right]\,.\label{eq:continuity_condition_v_prime}\end{equation}
Therefore, defining $\Delta V^{\prime}=\lim_{x\rightarrow x_{\star}^{+}}V^{\prime}-\lim_{x\rightarrow x_{\star}^{-}}V^{\prime}$,
and similarly for $\Delta\left(\partial_{x}^{2n}\phi\right)$, we
have\begin{eqnarray}
\Delta\left(\partial_{x}^{2n}\phi\right) & = & \frac{\left(-1\right)^{n}}{f_{n}}\left(i\omega\phi-2V\partial_{x}\phi\right)\Delta V^{\prime}\nonumber \\
 & = & \frac{\left(-1\right)^{n}}{f_{n}}\left(i\omega\phi-2V_{1}\partial_{x}\phi\right)\frac{\left(V_{2}-V_{1}\right)}{\delta}\,,\label{eq:discontinuity_in_2n_derivative}\end{eqnarray}
where it is understood that $\phi$ and $\partial_{x}\phi$ are evaluated
at $x=x_{\star}$ (since they are continuous, there is no ambiguity).
Although this step discontinuity in $\partial_{x}^{2n}\phi$ will
cause $\partial_{x}^{2n+1}\phi$ to become infinite (i.e. proportional
to the delta function) precisely at the point of discontinuity, there
may also be a discontinuity in the limiting values of $\partial_{x}^{2n+1}\phi$.
This can be found exactly as above, but to find the differential equation
containing $\partial_{x}^{2n+1}\phi$, we must differentiate Eq. (\ref{eq:acoustic_ODE_F_squared_polynomial}).
The result is\begin{multline}
-\omega^{2}\partial_{x}\phi-3i\omega V^{\prime}\partial_{x}\phi-2i\omega V\partial_{x}^{2}\phi-i\omega V^{\prime\prime}\phi+2\left(V^{\prime}\right)^{2}\partial_{x}\phi+2VV^{\prime\prime}\partial_{x}\phi+4VV^{\prime}\partial_{x}^{2}\phi+V^{2}\partial_{x}^{3}\phi\\
+\sum_{j=1}^{n}\left(-1\right)^{j}f_{j}\partial_{x}^{2j+1}\phi=0\,.\label{eq:acoustic_ODE_F_squared_polynomial_differentiated}\end{multline}
We have already argued that all derivatives up to and including $\partial_{x}^{2n-1}\phi$
must be continuous at $x_{\star}$; therefore, the only possible discontinuous
derivative appearing in Eq. (\ref{eq:acoustic_ODE_F_squared_polynomial_differentiated})
is $\partial_{x}^{2n+1}\phi$. Moreover, note that $V^{\prime\prime}$,
though infinite precisely at $x=x_{\star}$ (as is $\partial_{x}^{2n+1}\phi$),
is zero on either side of this point. The discontinuity in $V^{\prime}$
must be balanced by a corresponding discontinuity (or, rather, a difference
in limiting values) in $\partial_{x}^{2n+1}\phi$; indeed, we have\begin{multline}
\lim_{x\rightarrow x_{\star}^{-}}\left[\left(-3i\omega\partial_{x}\phi+4V\partial_{x}^{2}\phi\right)v^{\prime}+2\left(\partial_{x}\phi\right)\left(V^{\prime}\right)^{2}+\left(-1\right)^{n}f_{n}\partial_{x}^{2n+1}\phi\right]\\
=\lim_{x\rightarrow x_{\star}^{+}}\left[\left(-3i\omega\partial_{x}\phi+4V\partial_{x}^{2}\phi\right)v^{\prime}+2\left(\partial_{x}\phi\right)\left(V^{\prime}\right)^{2}+\left(-1\right)^{n}f_{n}\partial_{x}^{2n+1}\phi\right]\label{eq:continuity_condition_2n+1_derivative}\end{multline}
which implies\begin{eqnarray}
\negthickspace\negthickspace\negthickspace\negthickspace\negthickspace\negthickspace\negthickspace\negthickspace\negthickspace\negthickspace\negthickspace\negthickspace\Delta\left(\partial_{x}^{2n+1}\phi\right) & = & \frac{\left(-1\right)^{n}}{f_{n}}\left[\left(3i\omega\partial_{x}\phi-4V\partial_{x}^{2}\phi\right)\Delta V^{\prime}-2\left(\partial_{x}\phi\right)\Delta\left(V^{\prime}\right)^{2}\right]\nonumber \\
 & = & \frac{\left(-1\right)^{n}}{f_{n}}\left[\left(3i\omega\partial_{x}\phi-4V_{1}\partial_{x}^{2}\phi\right)\frac{\left(V_{2}-V_{1}\right)}{\delta}-2\left(\partial_{x}\phi\right)\frac{\left(V_{2}-V_{1}\right)^{2}}{\delta^{2}}\right]\,,\label{eq:discontinuity_in_2n+1_derivative}\end{eqnarray}
where, again, $\partial_{x}\phi$ and $\partial_{x}^{2}\phi$ are
understood to be evaluated at $x=x_{\star}$.

Note that higher derivatives of $\phi$ will also have discontinuities,
which can be found by differentiating Eq. (\ref{eq:acoustic_ODE_F_squared_polynomial_differentiated})
further. In the resulting differential equations, however, this will
never yield a higher power of $V^{\prime}$ than $\left(V^{\prime}\right)^{2}$.
Therefore, the discontinuities in higher derivatives can, at most,
vary as $1/\delta^{2}$.

The change in $\partial_{x}^{2n-1}\phi$ over the region of change
in $V$ can be found using a Taylor expansion:\begin{eqnarray*}
\negthickspace\negthickspace\negthickspace\left.\partial_{x}^{2n-1}\phi\right|_{x=x_{\star}+\delta}-\left.\partial_{x}^{2n-1}\phi\right|_{x=x_{\star}} & = & \left.\partial_{x}^{2n}\phi\right|_{x\rightarrow x_{\star}^{+}}\delta+\frac{1}{2}\left.\partial_{x}^{2n+1}\phi\right|_{x\rightarrow x_{\star}^{+}}\delta^{2}+\frac{1}{6}\left.\partial_{x}^{2n+2}\phi\right|_{x\rightarrow x_{\star}^{+}}\delta^{3}\ldots\\
 & = & \left.\partial_{x}^{2n}\phi\right|_{x\rightarrow x_{\star}^{-}}\delta+\frac{1}{2}\left.\partial_{x}^{2n+1}\phi\right|_{x\rightarrow x_{\star}^{-}}\delta^{2}+\frac{1}{6}\left.\partial_{x}^{2n+2}\phi\right|_{x\rightarrow x_{\star}^{-}}\delta^{3}\\
 &  & \negthickspace\negthickspace\negthickspace+\Delta\left(\partial_{x}^{2n}\phi\right)\delta+\frac{1}{2}\Delta\left(\partial_{x}^{2n+1}\phi\right)\delta^{2}+\frac{1}{6}\Delta\left(\partial_{x}^{2n+2}\phi\right)\delta^{3}+\ldots\end{eqnarray*}
Letting $\delta\rightarrow0$, we find that the first three terms
of the second line vanish, since they are unaffected by the discontinuity.
Moreover, since the discontinuities can have at most a $1/\delta^{2}$
dependence, only the terms in $\Delta\left(\partial_{x}^{2n}\phi\right)$
and $\Delta\left(\partial_{x}^{2n+1}\phi\right)$ can yield terms
that are independent of $\delta$. Consequently, we find that there
is a discontinuity in $\partial_{x}^{2n-1}\phi$ which is given by\begin{eqnarray}
\Delta\left(\partial_{x}^{2n-1}\phi\right) & = & \frac{\left(-1\right)^{n}}{f_{n}}\left[\left(i\omega\phi-2V_{1}\partial_{x}\phi\right)\left(V_{2}-V_{1}\right)-\partial_{x}\phi\left(V_{2}-V_{1}\right)^{2}\right]\nonumber \\
 & = & \frac{\left(-1\right)^{n}}{f_{n}}\left[i\omega\phi-\left(V_{1}+V_{2}\right)\partial_{x}\phi\right]\left(V_{2}-V_{1}\right)\,.\label{eq:discontinuity_2n-1_derivative}\end{eqnarray}

Eq. (\ref{eq:discontinuity_2n-1_derivative}) is the final condition
to be imposed at each point of discontinuity. (To the author's knowledge,
this general expression for the discontinuity in $\partial^{2n-1}_{x}\phi$ has
not been presented previously in the literature.) Having begun with $2nr$
coefficients, we have seen that the differential equation imposes
$2n\left(r-1\right)$ linear conditions on them, which leaves only
$2n$ degrees of freedom: precisely the number required to uniquely
specify a solution of the $2n$-order differential equation (\ref{eq:acoustic_steady_state_eqn}).

\section{Linearized velocity profile\label{sub:WKB-approximation}}

In our quest to find $\left|\beta_{\omega}\right|^{2}$, the amount
of mode conversion into negative-frequency modes, we would like to
have even an approximate analytic formula for the process. This allows
us to see which parameters are important, and the derivation of such
a formula can offer insight into the fundamental nature of the process.

\begin{figure}
\includegraphics[width=0.8\columnwidth]{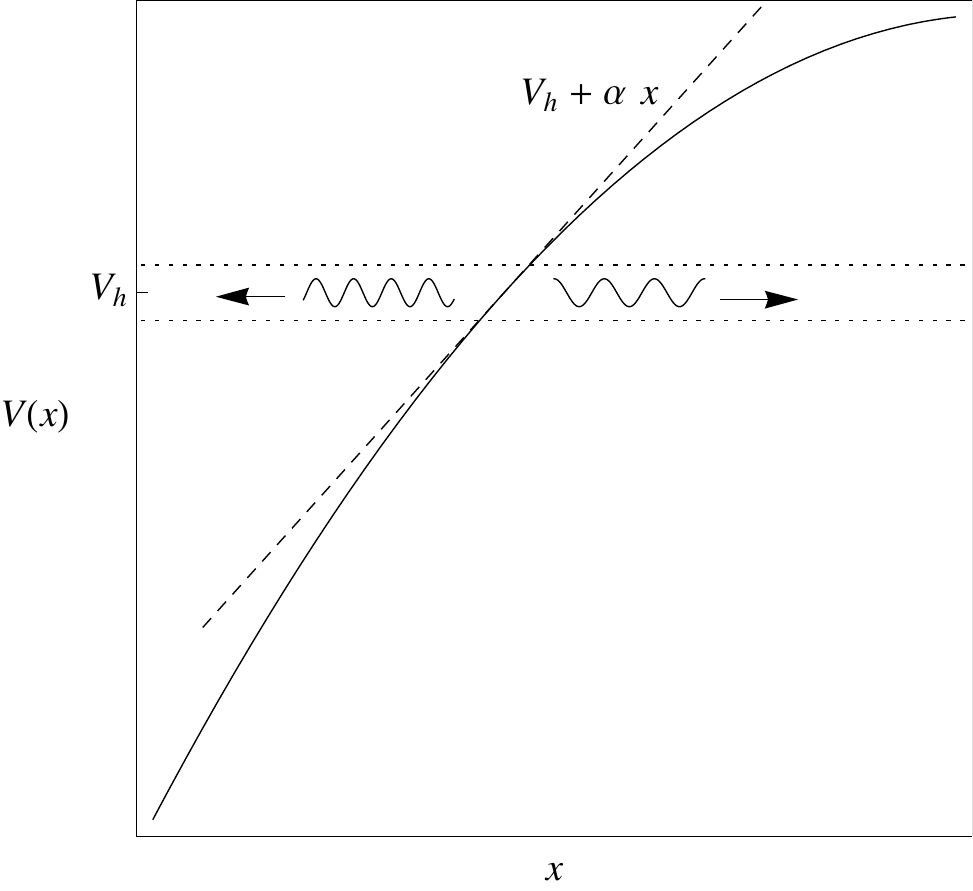}

\caption[\textsc{Linearizing the velocity profile}]{\textsc{Linearizing the velocity profile}: Assuming that mode conversion
originates primarily in a small region around the horizon, where $V=V_{h}$,
we may develop an approximate analytic description of the process
by using a linearized form of the velocity profile, $V_{h}+\alpha x$.\label{fig:Linearizing-the-Velocity-Profile}}

\end{figure}

Let us continue to deal with stationary solutions of the form (\ref{eq:steady_state_soln}),
and which satisfy the differential equation (\ref{eq:acoustic_steady_state_eqn}).
Dispersion adds higher derivatives to this equation through the operator
$F^{2}\left(-i\partial_{x}\right)$, making it difficult to solve
in position space and generally requiring numerical integration to
do so, as in §\ref{sub:Steady-state-solution}. However, in Fourier
space, the operator simply becomes a multiplicative function $F^{2}\left(k\right)$,
and does not determine the order of the differential equation. Instead,
the velocity profile $V\left(x\right)$ is replaced by the operator
$V\left(i\partial_{k}\right)$, so it is $V\left(x\right)$ which
determines the order of the differential equation. Generally, the
resulting equation is also very complicated; but, if we make the simplifying
assumption that the origins of mode conversion are mainly in the vicinity
of the horizon, we may approximate $V\left(x\right)$ by its linearized
form\begin{equation}
V\left(x\right)\approx V_{h}+\alpha x\,\label{eq:linearised_velocity_profile}\end{equation}
(having taken the zero of $x$ to be at the horizon, and assuming
that $V^{\prime}$ is non-zero there). This is illustrated pictorially
in Figure \ref{fig:Linearizing-the-Velocity-Profile}. The mathematical
result is the reduction of Eq. (\ref{eq:acoustic_steady_state_eqn})
to a second-order differential equation, making it much more tractable.
(Such a transformation is also performed in Refs. \cite{Corley-1998} and \cite{Unruh-Schutzhold-2005}.)

\subsection{Solution in Fourier space}

Performing the transformation just described - that is, replacing
$V\left(x\right)$ with the approximation in Eq. (\ref{eq:linearised_velocity_profile}),
then transforming to the Fourier representation via the substitutions
$\phi_{\omega}\left(x\right)\rightarrow\widetilde{\phi}_{\omega}\left(k\right)$,
$x\rightarrow i\partial_{k}$ and $\partial_{x}\rightarrow ik$ -
the differential equation (\ref{eq:acoustic_steady_state_eqn}) becomes\begin{multline}
\alpha^{2}k^{2}\partial_{k}^{2}\widetilde{\phi}_{\omega}+\left(2\alpha^{2}k+2i\alpha k\left(\omega-V_{h}k\right)\right)\partial_{k}\widetilde{\phi}_{\omega}\\
+\left(F^{2}\left(k\right)-\left(\omega-V_{h}k\right)^{2}+i\alpha\left(\omega-2V_{h}k\right)\right)\widetilde{\phi}_{\omega}=0\,.\label{eq:2nd_order_in_k_space}\end{multline}
This can be reduced to a standard form second-order differential equation
by defining another field $\widetilde{\psi}_{\omega}$ such that\begin{eqnarray}
\widetilde{\phi}_{\omega} & = & \widetilde{\psi}_{\omega}\exp\left(-\frac{1}{2}\int\frac{2\alpha^{2}k+2i\alpha k\left(\omega-V_{h}k\right)}{\alpha^{2}k^{2}}dk\right)\nonumber \\
 & = & \widetilde{\psi}_{\omega}\exp\left(\left(-1-i\frac{\omega}{\alpha}\right)\ln k+i\frac{V_{h}}{\alpha}k\right)\,,\label{eq:extracting_phase}\end{eqnarray}
for Eq. (\ref{eq:2nd_order_in_k_space}) then becomes the remarkably
simple equation\begin{equation}
\partial_{k}^{2}\widetilde{\psi}_{\omega}+\frac{F^{2}\left(k\right)}{\alpha^{2}k^{2}}\widetilde{\psi}_{\omega}=0\,.\label{eq:2nd_order_in_k_space_psi}\end{equation}
This is identical in form to the time-independent Schrödinger equation.
By direct analogy, we can apply the WKB method \cite{Landau-Lifshitz-QM}
(see also \cite{Brout-et-al-1995,Corley-1998,Unruh-Schutzhold-2005} for applications to Hawking radiation)
to find the approximate solution

\begin{equation}
\widetilde{\psi}_{\omega}\left(k\right)\approx C^{u}\sqrt{\frac{k}{F\left(k\right)}}\exp\left(\frac{i}{\alpha}\int^{k}\frac{F\left(k^{\prime}\right)}{k^{\prime}}dk^{\prime}\right)+C^{v}\sqrt{\frac{k}{F\left(k\right)}}\exp\left(-\frac{i}{\alpha}\int^{k}\frac{F\left(k^{\prime}\right)}{k^{\prime}}dk^{\prime}\right)\,,\label{eq:WKB_approx_in_k_space}\end{equation}
where $C^{u}$ and $C^{v}$ are arbitrary constants. (The meaning
of the labels shall soon become clear.) Eq. (\ref{eq:WKB_approx_in_k_space})
is valid so long as the {}``de Broglie wavelength'', $\lambda\left(k\right)=2\pi\left|\alpha\right|k/F\left(k\right)$,
varies negligibly over a single cycle; more precisely \cite{Landau-Lifshitz-QM},\begin{equation}
\left|\frac{\lambda^{\prime}\left(k\right)}{2\pi}\right|=\frac{\left|\alpha\right|}{F^{2}\left(k\right)}\left|F\left(k\right)-kF^{\prime}\left(k\right)\right|\ll1\,.\label{eq:WKB_condition}\end{equation}
The full Fourier transform $\widetilde{\phi}_{\omega}\left(k\right)$
is found via substitution in Eq. (\ref{eq:extracting_phase}):\begin{eqnarray}
\!\!\!\!\!\!\!\!\!\!\widetilde{\phi}_{\omega}\left(k\right) & \approx & C^{u}\sqrt{\frac{k}{F\left(k\right)}}\exp\left(\frac{i}{\alpha}\int^{k}\frac{F\left(k^{\prime}\right)}{k^{\prime}}dk^{\prime}+\left(-1-i\frac{\omega}{\alpha}\right)\ln k+i\frac{V_{h}}{\alpha}k\right)\nonumber \\
 &  & \qquad+C^{v}\sqrt{\frac{k}{F\left(k\right)}}\exp\left(-\frac{i}{\alpha}\int^{k}\frac{F\left(k^{\prime}\right)}{k^{\prime}}dk^{\prime}+\left(-1-i\frac{\omega}{\alpha}\right)\ln k+i\frac{V_{h}}{\alpha}k\right)\nonumber \\
 & = & \frac{C^{u}}{k}\sqrt{\frac{k}{F\left(k\right)}}\exp\left(-\frac{i}{\alpha}\int^{k}\left(\frac{\omega-F\left(k^{\prime}\right)}{k^{\prime}}-V_{h}\right)dk^{\prime}\right)\nonumber \\
 &  & \qquad\qquad\qquad+\frac{C^{v}}{k}\sqrt{\frac{k}{F\left(k\right)}}\exp\left(-\frac{i}{\alpha}\int^{k}\left(\frac{\omega+F\left(k^{\prime}\right)}{k^{\prime}}-V_{h}\right)dk^{\prime}\right)\,.\label{eq:WKB_approx_phase_reinserted}\end{eqnarray}

\subsection{Transforming back to position space}

$\phi_{\omega}\left(x\right)$ is, of course, the inverse Fourier
transform of $\widetilde{\phi}_{\omega}\left(k\right)$:\begin{eqnarray}
\phi_{\omega}\left(x\right) & = & \int_{-\infty}^{+\infty}\widetilde{\phi}_{\omega}\left(k\right)\exp\left(ikx\right)dk\nonumber \\
 & \approx & \int_{-\infty}^{+\infty}\frac{1}{k}\sqrt{\frac{k}{F\left(k\right)}}\left[C^{u}\exp\left(i\varphi^{u}\left(k\right)\right)+C^{v}\exp\left(i\varphi^{v}\left(k\right)\right)\right]dk\,,\label{eq:inverse_Fourier_transform}\end{eqnarray}
where we have defined\begin{eqnarray*}
\varphi^{u}\left(k\right) & = & -\frac{1}{\alpha}\int^{k}\left(\frac{\omega-F\left(k^{\prime}\right)}{k^{\prime}}-V_{h}-\alpha x\right)dk^{\prime}\,,\\
\varphi^{v}\left(k\right) & = & -\frac{1}{\alpha}\int^{k}\left(\frac{\omega+F\left(k^{\prime}\right)}{k^{\prime}}-V_{h}-\alpha x\right)dk^{\prime}\,.\end{eqnarray*}
We can find an approximate solution to Eq. (\ref{eq:inverse_Fourier_transform})
using the saddle-point approximation. This states that the main contribution
to the integral in this equation comes from those points where the
first derivatives of the phases $\varphi^{u}\left(k\right)$ and $\varphi^{v}\left(k\right)$
vanish. Then, to lowest order, the contribution from each of these
points behaves like an overall constant multiplied by a Gaussian integral,
so long as the integration contour may be taken along the direction
of steepest descent. Given that the phases are integrals themselves,
the saddle points are those where the integrands vanish, i.e.,\begin{eqnarray*}
\frac{d\varphi^{u}}{dk}=0 & \Rightarrow & -\frac{1}{\alpha}\left(\frac{\omega-F\left(k\right)}{k}-V_{h}-\alpha x\right)=0\,,\\
\frac{d\varphi^{v}}{dk}=0 & \Rightarrow & -\frac{1}{\alpha}\left(\frac{\omega+F\left(k\right)}{k}-V_{h}-\alpha x\right)=0\,.\end{eqnarray*}
Recall that, since we have linearized the velocity around the horizon,
$V_{h}+\alpha x$ is simply the flow velocity $V\left(x\right)$.
Therefore, we can rewrite the above conditions in the form\begin{eqnarray}
\frac{d\varphi^{u}}{dk}=0 & \Rightarrow & \omega=V\left(x\right)k+F\left(k\right)\,,\label{eq:saddle_points}\\
\frac{d\varphi^{v}}{dk}=0 & \Rightarrow & \omega=V\left(x\right)k-F\left(k\right)\,.\end{eqnarray}
At any position $x$, the saddle points of the phases in Eq. (\ref{eq:inverse_Fourier_transform})
are simply those values of $k$ that solve the dispersion relation
with the local value of the flow velocity. The phase $\varphi^{u}\left(k\right)$
has saddle points at those solutions which are right-moving with respect
to the fluid, and therefore lie on the $u$-branch; the saddle points
of the phase $\varphi^{v}\left(k\right)$ are solutions that are left-moving
with respect to the fluid, and lie on the $v$-branch. For the sake
of simplicity, we shall continue to assume that there is negligible
coupling between the $u$- and $v$-branches, dealing exclusively
with the $u$-branch; that is, we set $C^{v}=0$.

In order to perform the integration by approximation with Gaussian
integrals, we also need the second derivative of the phase at its
saddle points:\begin{alignat}{3}
\frac{d^{2}\varphi^{u}}{dk^{2}} & = & \,\frac{1}{\alpha}\frac{\left(\omega-F\left(k\right)+kF^{\prime}\left(k\right)\right)}{k^{2}} & = & \,\frac{1}{\alpha}\frac{\left(V\left(x\right)+F^{\prime}\left(k\right)\right)}{k}\,,\end{alignat}
where, in the second equality, we have evaluated $\omega-F\left(k\right)$
at the saddle points using Eq. (\ref{eq:saddle_points}). Therefore,
in the vicinity of a solution $k_{j}$ of Eq. (\ref{eq:saddle_points}),
the phase $\varphi^{u}\left(k\right)$ is approximately given by\begin{eqnarray}
\varphi^{u}\left(k\right) & \approx & \varphi^{u}\left(k_{j}\right)+\frac{1}{2\alpha}\frac{\left(V\left(x\right)+F^{\prime}\left(k_{j}\right)\right)}{k_{j}}\left(k-k_{j}\right)^{2}\nonumber \\
 & = & \varphi^{u}\left(k_{j}\right)+\frac{v_{g}\left(k_{j}\right)}{2\alpha k_{j}}\left(k-k_{j}\right)^{2}\,,\label{eq:phase_quadratic_approximation}\end{eqnarray}
where from Eq. (\ref{eq:saddle_points}) we have identified $V\left(x\right)+F^{\prime}\left(k_{j}\right)=d\omega/dk\left(k_{j}\right)=v_{g}\left(k_{j}\right)$,
the local group velocity. (Note that, as a solution of Eq. (\ref{eq:saddle_points}),
$k_{j}$ is itself $x$-dependent.)

We may now perform the integral in Eq. (\ref{eq:inverse_Fourier_transform}).
Firstly, note that the integral is properly Gaussian if $\varphi^{u}\left(k\right)$
increases in the $+i$ direction; therefore, the direction of steepest
descent is proportional to $\exp\left(i\frac{\pi}{4}\mathrm{sgn}\left(v_{g}\left(k_{j}\right)/\alpha k_{j}\right)\right)$.
Assuming that the integration contour may be deformed so as to pass
through each saddle point in its direction of steepest descent, the
position-space solution becomes

\begin{equation}
\phi_{\omega}\left(x\right)\approx C^{u}\,\sqrt{2\pi\left|\alpha\right|}\sum_{j}\frac{\mathrm{sgn}\left(k_{j}\right)}{\sqrt{\left|v_{g}\left(k_{j}\right)\right|\left|F\left(k_{j}\right)\right|}}\exp\left(i\frac{\pi}{4}\mathrm{sgn}\left[\frac{v_{g}\left(k_{j}\right)}{\alpha k_{j}}\right]+i\varphi^{u}\left(k_{j}\right)\right)\,.\label{eq:saddle_point_approx_sum}\end{equation}
Notice that the factors of $\left|v_{g}\left(k_{j}\right)F\left(k_{j}\right)\right|^{-1/2}$
are exactly those required to normalize the various plane waves in
the $\omega$-representation according to Eq. (\ref{eq:mode_w}).
Therefore, the relative norms of the modes themselves are entirely
determined by the phases $\varphi^{u}\left(k_{j}\right)$.

\subsection{Generalization to nonlinear velocity profiles}

Within the limits of the saddle-point approximation, Eq. (\ref{eq:saddle_point_approx_sum})
is strictly valid only in the case of the linear velocity profile
(\ref{eq:linearised_velocity_profile}). However, this fact has been
made implicit by writing the solution in its present form, and it
is easy to extrapolate the solution to nonlinear velocity profiles.
To see this, note that the prefactor is unimportant, and its dependence
on $\alpha$ can be removed simply by making $C^{u}$ proportional
to $\left|\alpha\right|^{-1/2}$. $\alpha$ appears in the sign of
the $\frac{\pi}{4}$ part of the phase; however, it is only the sign
of $\alpha$ that matters here, and so long as $V$ is monotonic in
$x$, this sign is uniquely determined. We thus restrict our attention
to monotonic velocity profiles. Now the only part of Eq. (\ref{eq:saddle_point_approx_sum})
still dependent on the linear form of Eq. (\ref{eq:linearised_velocity_profile})
is the phase $\varphi^{u}\left(k_{j}\right)$:\begin{eqnarray}
\varphi^{u}\left(k_{j}\right) & = & -\frac{1}{\alpha}\int^{k_{j}}\left(\frac{\omega-F\left(k^{\prime}\right)}{k^{\prime}}-V_{h}-\alpha x\right)dk^{\prime}\nonumber \\
 & = & \int^{k_{j}}\left(x-x\left(k^{\prime}\right)\right)dk^{\prime}\,,\label{eq:generalized_phase}\end{eqnarray}
where the function $x\left(k\right)$ is defined according to Eq.
(\ref{eq:saddle_points}) with the linearized velocity profile (\ref{eq:linearised_velocity_profile})
substituted for $V\left(x\right)$. $x\left(k\right)$, then, is the
position $x$ at which the wavevector $k$ is a solution of (the $u$-branch
of) the dispersion relation. Although this has been derived for the
special case of a linear velocity profile, it is easily extendable
to any monotonic velocity profile. When expressed in this general form,
we also note that $\alpha$, the derivative of $V$ at the horizon, no longer
has a privileged status. We may not interpret Eq. (\ref{eq:saddle_point_approx_sum})
as being applicable to any monotonic velocity profile, and we may write\begin{equation}
\phi_{\omega}\left(x\right)\approx C^{\prime}\sum_{j}\frac{\mathrm{sgn}\left(k_{j}\right)}{\sqrt{\left|v_{g}\left(k_{j}\right)\right|\left|F\left(k_{j}\right)\right|}}\exp\left(i\frac{\pi}{4}\mathrm{sgn}\left[\frac{v_{g}\left(k_{j}\right)}{V^{\prime}k_{j}}\right]+i\int^{k_{j}}\left(x-x\left(k^{\prime}\right)\right)dk^{\prime}\right)\,.\label{eq:generalized_solution}\end{equation}
Note that, while Eq. (\ref{eq:generalized_solution}) is not an original result, its application to the entire velocity profile -
not just the linearized form - is a novel approach, and leads to original results.

Not only are the velocity profiles we are interested in monotonic,
but they also approach asymptotic limiting values as $x\rightarrow\pm\infty$.
Thus, at those values of $k$ which solve the dispersion relation
in the asymptotic regions, $x\left(k\right)$ rapidly diverges. The
$x$-dependence of the phase is not obvious in the form of Eq. (\ref{eq:generalized_phase});
however, differentiating with respect to $x$, and recalling that
$k_{j}$ is itself a function of $x$, we find $d\varphi^{u}/dx=k_{j}$,
and so

\begin{equation}
\varphi^{u}\left(k_{j}\right)=\int^{k_{j}}\left(x-x\left(k^{\prime}\right)\right)dk^{\prime}=\int^{x}k_{j}\left(x^{\prime}\right)dx^{\prime}\label{eq:phase_integral_equivalence}\end{equation}
up to an unimportant additive constant. (This is simply the formula
for integration by parts.) The right side of Eq. (\ref{eq:phase_integral_equivalence})
is the more usual form of the phase generated by the WKB approximaton
in position space, and it clearly shows that the individual components
become plane waves in the asymptotic regions.

The first equality of Eq. (\ref{eq:phase_integral_equivalence}) is
the most useful for relating different \textit{wavevectors} (as opposed
to different \textit{positions}) to each other. Their relative phase
is simply the integral of $x-x\left(k\right)$ between them; their
relative norm, or squared amplitude, is determined by the imaginary
part of this integral. Since $x$ is always real, it is the imaginary
part of the function $x\left(k\right)$ which is important. Thus the
norm of one wavevector $k_{1}$ relative to that of another
wavevector $k_{2}$ is given by\begin{alignat}{1}
\frac{\left(\phi_{\omega,k_{1}},\phi_{\omega,k_{1}}\right)}{\left(\phi_{\omega,k_{2}},\phi_{\omega,k_{2}}\right)}\approx\frac{\mathrm{sgn}\left(\omega-Vk_{1}\right)}{\mathrm{sgn}\left(\omega-Vk_{2}\right)}\exp\left(2\,\mathrm{Im}\left[J\right]\right)\,,\qquad & J=\int_{k_{2}}^{k_{1}}x\left(k^{\prime}\right)dk^{\prime}\,.\label{eq:norm_phase_integral}\end{alignat}
In particular, for particle creation, the relative weight between the Bogoliubov coefficients is given by
\begin{equation}
\left|\frac{\beta}{\alpha}\right|^{2}=\exp\left(2\,\mathrm{Im}\left[\int_{k_{\alpha}}^{k_{\beta}}x\left(k^{\prime}\right)dk^{\prime}\right]\right)\,.\label{eq:creation_amp_integral}
\end{equation}

Eqs. (\ref{eq:norm_phase_integral}) and (\ref{eq:creation_amp_integral}) are new results, derived simply from the extension
of the applicability of Eq. (\ref{eq:generalized_solution}) to the entire velocity profile. Thus
the validity of the former depends on the validity of the latter.
Certainly, they cannot be applied when the WKB approximation breaks down and the condition
(\ref{eq:WKB_condition}) is not satisfied - that is, in the non-adiabatic limit of large $|\alpha|$.
The question of when Eqs. (\ref{eq:norm_phase_integral}) and (\ref{eq:creation_amp_integral}) \textit{are} applicable
is rather more difficult to address, for, strictly speaking, Eq. (\ref{eq:generalized_solution}) is
\textit{in}valid for any nonlinear velocity profile!  Since the limits of the integrals in Eqs. (\ref{eq:norm_phase_integral})
and (\ref{eq:creation_amp_integral}) correspond to asymptotic regions where $V$ approaches constant values, they implicity imply that
$V$ is nonlinear, and we cannot expect \textit{a priori} that are valid in any regime.
That they do possess a regime of validity will be shown in the following chapter by comparison with
numerical results, but the reason for the existence of this regime remains a mystery.

It is important to note that, apart from the considerations just discussed,
Eqs. (\ref{eq:norm_phase_integral}) and (\ref{eq:creation_amp_integral}) are valid only
for a wave solution which can be expressed in terms of a single contour
integral of the form in Eq. (\ref{eq:inverse_Fourier_transform}).
In general, owing to the multivaluedness of the imaginary part of
$x\left(k\right)$, there are several possible such integrals, with
different results. This is to be expected: there are generally several
solutions to the dispersion relation, each one giving rise to an independent
solution of the wave equation. It is also generally true that, as
described in §\ref{sub:Steady-state-solution}, a particular in- or
out-mode is formed via linear combination of these solutions. In this
case, the relative norm of a particular wavevector will be given by
some linear combination of its various norms of the form (\ref{eq:norm_phase_integral}).

\section{Summary}

This chapter has demonstrated three ways of determining the amount
of mode conversion induced by a given velocity profile:
\begin{itemize}
\item an FDTD algorithm to solve for the evolution of a wavepacket directly;
\item numerical integration of an ordinary differential equation to find
the steady-state solution for a given Killing frequency; this is done
both generally and for the case of a profile with step discontinuities;
and
\item an analytic treatment involving linearizing the velocity profile around
the horizon and extrapolating the result to more general profiles.
\end{itemize}
\pagebreak{}

\chapter{Results for Acoustic Model\label{sec:Results-for-Acoustic-Model}}

In chapters \ref{sec:Theoretical-Origins-of-Hawking-Radiation} and
\ref{sec:The-Acoustic-Model}, we demonstrated, from a purely theoretical
standpoint, that mode conversion between positive- and negative-norm
modes in both nondispersive and dispersive fluids leads to spontaneous
creation of phonons. Chapter \ref{sec:Methods-of-Acoustic-Model}
turned to the practicalities of calculating the amount of mode conversion,
and consequently the phonon creation rate. The descriptions and derivations
of these chapters have been kept as general as possible, so as to
be applicable to many different velocity and dispersion profiles.

In the present chapter, we turn our attention to specific examples,
so as to provide a quantitative account of the Hawking phonon radiation
expected in fluids. The velocity and dispersion profiles used are
kept as simple as possible, and we expect many of the qualitative
features of the results still to be applicable in the general case.

\section{Dispersion profile}

Firstly, let us define the dispersion relation to be used. Recall
that the dispersionless case (see Eqs. (\ref{eq:Doppler_formula})
and (\ref{eq:acoustic_flow_dispersion})) has $F^{2}\left(k\right)=c^{2}k^{2}$;
and that, since the fluid is isotropic in its rest frame, $F^{2}\left(k\right)$
should contain only even powers of $k$. Since the complexities of
dispersion arise through the introduction of higher-order derivatives
in the wave equation, the simplest deviation from the dispersionless
model is through the inclusion of a quartic term in $F^{2}\left(k\right)$:\begin{equation}
F^{2}\left(k\right)=c^{2}k^{2}\left(1-\frac{k^{2}}{k_{d}^{2}}\right)\,.\label{eq:quartic_dispersion}\end{equation}

\begin{figure}
\includegraphics[width=0.8\columnwidth]{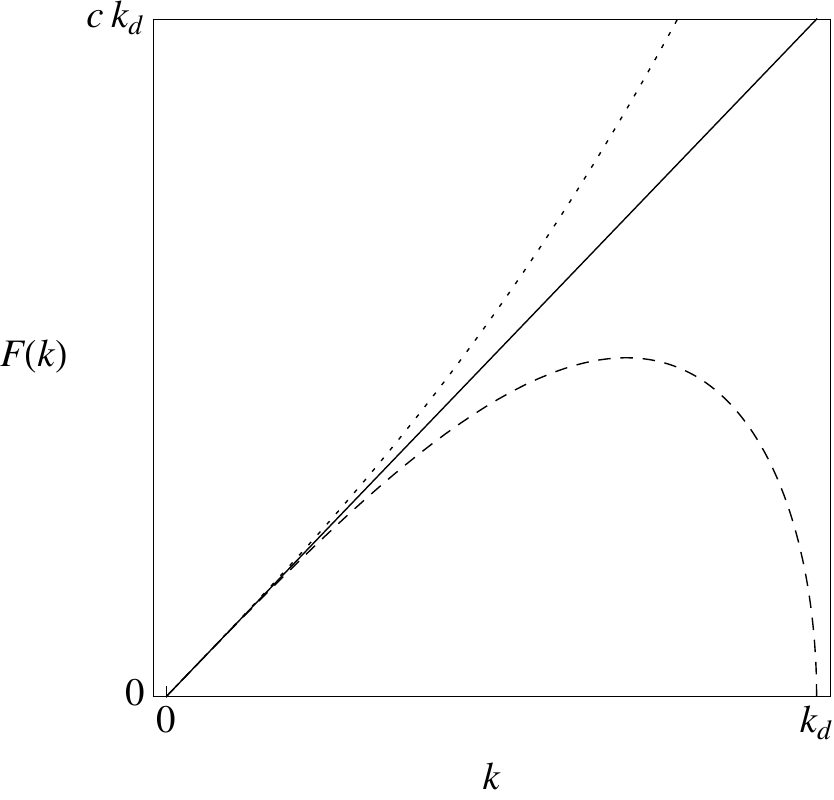}

\caption[\textsc{Quartic dispersion}]{\textsc{Quartic dispersion}: The dashed line plots Eq. (\ref{eq:quartic_dispersion}),
which describes subluminal dispersion; the dotted line corresponds
to superluminal dispersion, where the minus sign in Eq. (\ref{eq:quartic_dispersion})
is replaced with a plus sign.\label{fig:quartic_dispersion}}

\end{figure}

It is usually found that $F^{2}\left(k\right)$ approaches a dispersionless
form for low values of $k$, and this feature has been incorporated
in Eq. (\ref{eq:quartic_dispersion}); $c$, then, is the limiting
wave velocity (both phase and group velocity) for small wavevectors.
These are quantitatively defined as being much smaller than the parameter
$k_{d}$, i.e., there is no or little dispersion for wavevectors $k$
such that $k^{2}/k_{d}^{2}\ll1$. The parameter $k_{d}$, then, determines
the point at which significant deviations from nondispersive behaviour
first appear, and can be said to signify the strength of the dispersion:
the larger is $k_{d}$, the larger is the value of $k$ at which dispersive
effects begin to take hold, and so the weaker is the dispersion. Finally,
the minus attached to the deviation $k^{2}/k_{d}^{2}$ corresponds
to subluminal dispersion (phase velocity becoming less than $c$);
had this been a plus sign, it would describe superluminal dispersion
(phase velocity becoming greater than $c$). We have restricted ourselves
to subluminal dispersion, since superluminal dispersion gives results
which are qualitatively very similar, the main difference occurring
in the coupling to $v$-modes \cite{Macher-Parentani-2008} which
is neglected here. Both Eq. (\ref{eq:quartic_dispersion}) and its
superluminal counterpart are plotted in Figure \ref{fig:quartic_dispersion}.

\section{Velocity profile}

\begin{figure}
\includegraphics[width=0.8\columnwidth]{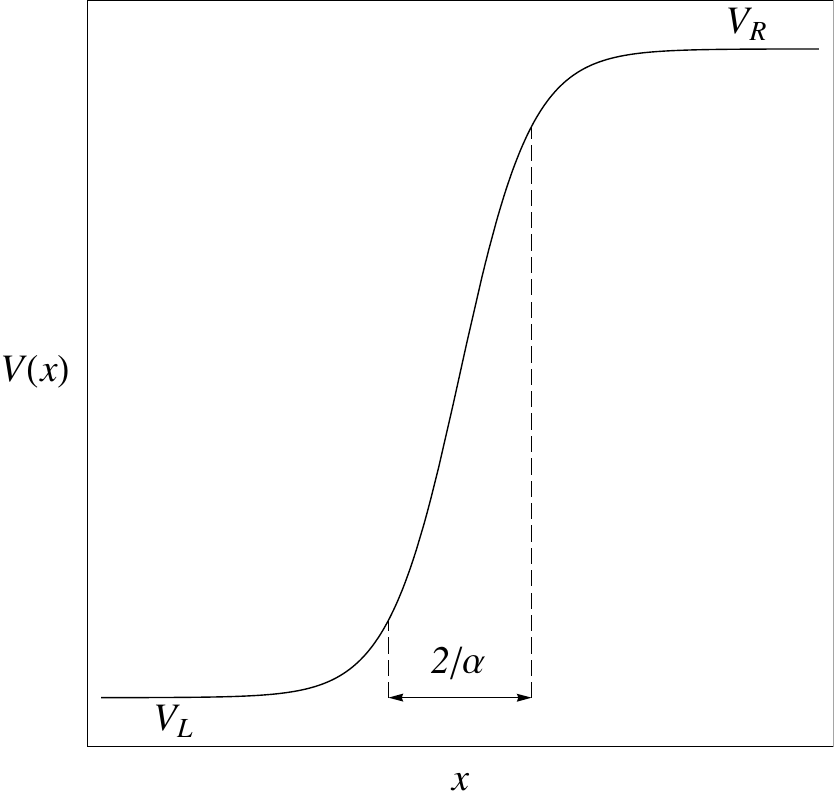}

\caption[\textsc{Hyperbolic tangent velocity profile}]{\textsc{Hyperbolic tangent velocity profile}: This is the simplest
monotonic profile with asymptotically constant values, and is expressed
by Eq. (\ref{eq:hyperbolic_tangent_velocity_profile}). $\alpha$
is inversely related to the length of the transition region.\label{fig:tanh_velocity_profile}}

\end{figure}

Endeavouring to keep the physical system as simple as possible, we
shall use a velocity profile that varies monotonically between two
asymptotic values; such a profile shall then be invertible, which
is useful for application of the analytic results in Eqs. (\ref{eq:generalized_solution})
and (\ref{eq:norm_phase_integral}). The simplest candidate is the
hyperbolic tangent function, and accordingly we restrict ourselves
to velocity profiles of the form\begin{equation}
V\left(x\right)=\frac{1}{2}\left(V_{R}+V_{L}\right)+\frac{1}{2}\left(V_{R}-V_{L}\right)\tanh\left(\alpha x\right)\,.\label{eq:hyperbolic_tangent_velocity_profile}\end{equation}
$V_{R}$ and $V_{L}$ are the asymptotic values of $V\left(x\right)$
as $x$ tends to $+\infty$ or $-\infty$, respectively. The parameter
$\alpha$ characterizes the steepness of the velocity profile, but
this also depends on the difference $V_{R}-V_{L}$; a more direct
interpretation of $\alpha$ is that it governs the length of the region
of transition between $V_{L}$ and $V_{R}$, which is of order $2/\alpha$
as shown in Figure \ref{fig:tanh_velocity_profile}.

It may be argued that a velocity profile with equal asymptotic velocities
would be a more desirable option. Certainly, such a profile would
have greater practical applicability: the change in flow velocity
is normally caused by a local disturbance, such as an obstacle that
alters the depth of fluid and hence the speed of its flow, or an intense
pulse of light that alters the refractive index of an optical fibre.
In these practical scenarios, the change is local, and the physical
conditions in a region far from the change are the same whether we
are ahead of the change or behind it. But the corresponding velocity
profile is not invertible, so that the analytic treatment of §\ref{sub:WKB-approximation}
is inapplicable; and a horizon, should one occur, must be accompanied
by a partner horizon on the other side of the change. Since the derivatives
of the flow velocity at these horizons must have opposite signs, one
of them will be a black hole horizon, the other a white hole horizon.
The Hawking spectrum emitted by such a system will be some superposition
of the spectra of each horizon, and might exhibit interference. Given
these complications, it is wise to limit ourselves to the investigation
of a single horizon, or at least a monotonic variation, for which
we must pay the reasonable price of physical equivalence of the asymptotic
regions.

\section{Normalizing the wave equation}

With the dispersion and velocity profiles given by Eqs. (\ref{eq:quartic_dispersion})
and (\ref{eq:hyperbolic_tangent_velocity_profile}), the acoustic
wave equation (\ref{eq:acoustic_wave_equation}) becomes\begin{multline}
\partial_{t}^{2}\phi+V^{\prime}\left(x\right)\partial_{t}\phi+2V\left(x\right)\partial_{t}\partial_{x}\phi+2V\left(x\right)V^{\prime}\left(x\right)\partial_{x}\phi+V^{2}\left(x\right)\partial_{x}^{2}\phi\\
-c^{2}\partial_{x}^{2}\phi-\frac{c^{2}}{k_{d}^{2}}\partial_{x}^{4}\phi=0\,.\label{eq:acoustic_wave_eqn_quartic_dispersion}\end{multline}
This equation contains five parameters: $c$ and $k_{d}$ from the
dispersion profile; and $V_{L}$, $V_{R}$ and $\alpha$ from the
velocity profile. However, two of these correspond to the scaling
of space and time; in normalizing the variables to make them dimensionless,
we can reduce the number of independent parameters to three.

Firstly, let us normalize the velocity profile with respect to $c$,
the low-$k$ limit of the wave velocity. We define the dimensionless
velocity $U\left(x\right)=V\left(x\right)/c$, and rewrite the wave
equation in the form\begin{multline*}
\frac{1}{c^{2}}\partial_{t}^{2}\phi+\frac{1}{c}U^{\prime}\left(x\right)\partial_{t}\phi+2\frac{1}{c}U\left(x\right)\partial_{t}\partial_{x}\phi+2U\left(x\right)U^{\prime}\left(x\right)\partial_{x}\phi+U^{2}\left(x\right)\partial_{x}^{2}\phi\\
-\partial_{x}^{2}\phi-\frac{1}{k_{d}^{2}}\partial_{x}^{4}\phi=0\,.\end{multline*}
Next, let us normalize the position variable $x$ (and, consequently,
the wavenumber $k$) by requiring that the dispersion parameter $k_{d}$
be normalized to unity. This is done by defining the dimensionless
position variable $X=k_{d}x$; then the partial derivative $\partial_{x}=k_{d}\partial_{X}$,
and the dimensionless wavenumber $K=k/k_{d}$. With this substitution,
the wave equation becomes\begin{multline*}
\frac{1}{c^{2}k_{d}^{2}}\partial_{t}^{2}\phi+\frac{1}{ck_{d}}U^{\prime}\left(X\right)\partial_{t}\phi+2\frac{1}{ck_{d}}U\left(X\right)\partial_{t}\partial_{X}\phi+2U\left(X\right)U^{\prime}\left(X\right)\partial_{X}\phi\\
+U^{2}\left(X\right)\partial_{X}^{2}\phi-\partial_{X}^{2}\phi-\partial_{X}^{4}\phi=0\,,\end{multline*}
where\begin{equation}
U\left(X\right)=\frac{1}{2}\left(U_{R}+U_{L}\right)+\frac{1}{2}\left(U_{R}-U_{L}\right)\tanh\left(aX\right)\,\label{eq:normalized_velocity_profile}\end{equation}
and $a=\alpha/k_{d}$. Finally, we shall normalize the time variable
$t$ (and the frequency $\omega$). The present form of the wave equation
shows that the tidiest way to do this is to define the dimensionless
time variable $T=ck_{d}t$; consequently, the partial derivative $\partial_{t}=ck_{d}\partial_{T}$,
the dimensionless frequency $\Omega$ is given by the relation $\Omega=\omega/\left(ck_{d}\right)$,
and the wave equation takes the form\begin{multline}
\partial_{T}^{2}\phi+U^{\prime}\left(X\right)\partial_{T}\phi+2U\left(X\right)\partial_{T}\partial_{X}\phi+2U\left(X\right)U^{\prime}\left(X\right)\partial_{X}\phi+U^{2}\left(X\right)\partial_{X}^{2}\phi\\
-\partial_{X}^{2}\phi-\partial_{X}^{4}\phi=0\,.\end{multline}
This can be factorized to give\begin{equation}
\left(\partial_{T}+\partial_{X}U\right)\left(\partial_{T}+U\partial_{X}\right)\phi+f^{2}\left(-i\partial_{X}\right)\phi=0\,,\end{equation}
where the rescaled dispersion relation is\begin{equation}
f^{2}\left(K\right)=K^{2}\left(1-K^{2}\right)\,.\label{eq:normalized_dispersion}\end{equation}

The wave equation now contains only three parameters: 
\begin{itemize}
\item $U_{R}$ and $U_{L}$, the asymptotic values of the flow velocity
relative to the low-frequency wave speed; and
\item $a$, a parameter that combines steepness of the flow velocity profile
with dispersion strength. A higher value of $a$ indicates greater
steepness / stronger dispersion.
\end{itemize}
Often, the central value of the velocity is taken to be $-1$, so
that this point would be the event horizon in the dispersionless case.
Then, $U_{R}+U_{L}$ is fixed (at $-2$), and the number of parameters
is reduced to two: one is still $a$; the other is taken to be $h=\left(U_{R}-U_{L}\right)/2$,
the {}``height'' of the velocity profile, where $U_{R}=-1+h$ and
$U_{L}=-1-h$. We shall examine this case; but we will also go on
to examine more general velocity profiles where $-1$ is not the central
value, including those where $V$ is nowhere equal to $-1$. Although,
in the latter case, there is no horizon for low wavevectors, the discussion
in §\ref{sub:Conclusion-and-discussion} suggests that spontaneous
phonon creation is still a possibility.

\section{Thermal behaviour\label{sub:Thermal-behaviour}}

\begin{figure}
\includegraphics[width=0.8\columnwidth]{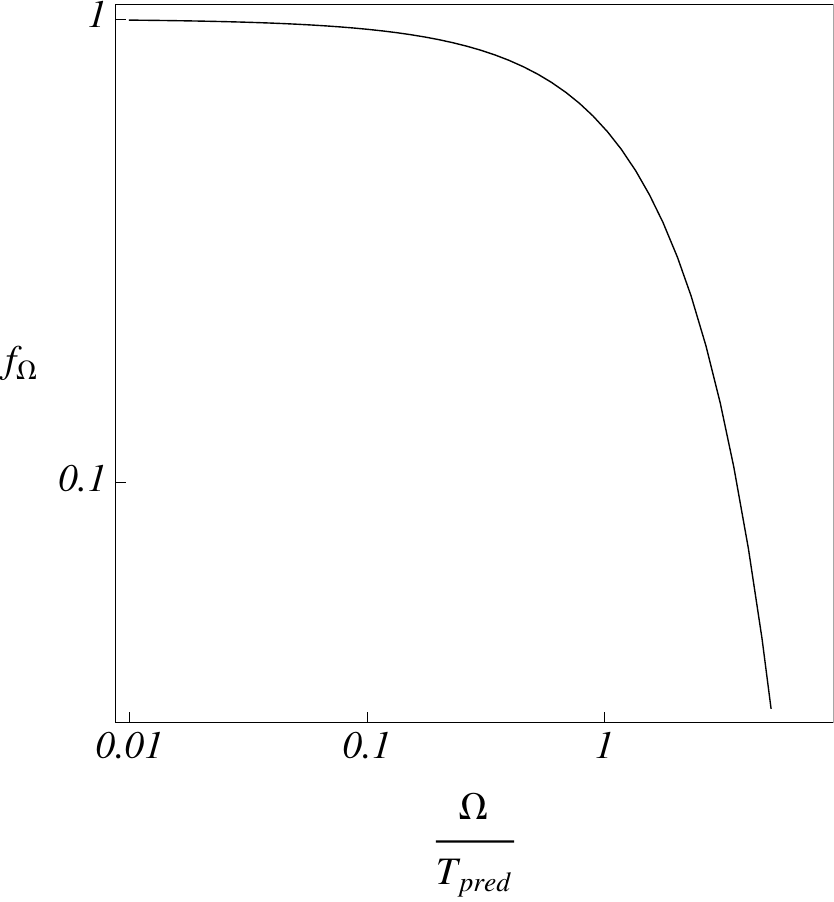}

\caption[\textsc{The thermal spectrum}]{\textsc{The thermal spectrum}: This shows the rescaled energy spectrum
(\ref{eq:normalized_spectrum_defn}) which is purely thermal. Both
axes use a logarithmic scale.\label{fig:thermal_spectrum}}

\end{figure}

Purely thermal behaviour, as in Eq. (\ref{eq:thermal_spectrum}),
is characterized by the Planck spectrum

\begin{equation}
\left|\beta_{\Omega}\right|_{T}^{2}=\frac{1}{\exp\left(\Omega/T\right)-1}\,.\label{eq:thermal_prediction}\end{equation}
Clearly, $\left|\beta_{\Omega}\right|_{T}^{2}\rightarrow T/\Omega$
as $\Omega\rightarrow0$, and the low-frequency temperature of any
spectrum which behaves in a similar way can be defined accordingly;
that is, we say that the spectrum has low-frequency temperature $T$
if\begin{equation}
\left|\beta_{\Omega}\right|^{2}\rightarrow\frac{T}{\Omega}\qquad\mathrm{as}\qquad\Omega\rightarrow0\,.\label{eq:low_frequency_temperature}\end{equation}
In order to regularize the spectrum, it is useful to define a new
quantity \cite{Macher-Parentani-2008}\begin{equation}
f_{\Omega}=\frac{\Omega}{T_{\mathrm{pred}}}\left|\beta_{\Omega}\right|^{2}\,.\label{eq:normalized_spectrum_defn}\end{equation}
where $T_{\mathrm{pred}}$ is some predicted value for the temperature.
Since $\left|\beta_{\Omega}\right|^{2}$ is a number spectrum, $f_{\Omega}$
is proportional to the energy spectrum. Its behaviour as $\Omega\rightarrow0$
is very instructive, for if $\left|\beta_{\Omega}\right|^{2}$ has
a low-frequency temperature $T$ in the sense defined above, we have\begin{equation}
f_{\Omega}\rightarrow\frac{T}{T_{\mathrm{pred}}}\qquad\mathrm{as}\qquad\Omega\rightarrow0\,.\label{eq:low_freq_temp}\end{equation}
$f_{\Omega}$, then, measures the low-frequency temperature relative
to $T_{\mathrm{pred}}$; and in the special case when $T_{\mathrm{pred}}=T$,
the spectral form of $f_{\Omega}$ provides a reference thermal curve
(shown in Figure \ref{fig:thermal_spectrum}) against which other
spectra may be compared.

Thermal effects, then, are dominant at low frequencies. Generally,
the Hawking spectra are not purely thermal (see §\ref{sub:Critical-values-of-Omega}
for further discussion). However, when there is a point at which $U=-1$,
there is an event horizon for all wavevectors in the limit $K\rightarrow0$,
since the dispersion profile approaches the dispersionless limit in
this regime. We might reasonably expect that low-frequency thermal
behaviour is preserved when $U=-1$ at some point; the numerical results
of §\ref{sub:Numerical-results} show that this is the case. To determine
the value of the low-frequency temperature, we must look at two separate
regimes.

\subsection{Validity of linearized velocity profile}

If the linearized velocity profile of Eq. (\ref{eq:linearised_velocity_profile})
is valid over some length scale, we expect the phase integral form
of the creation rate in Eq. (\ref{eq:norm_phase_integral}) to hold;
and, moreover, it should agree with Eq. (\ref{eq:Hawking_temperature})
for the temperature in the dispersionless case,\begin{equation}
T=\frac{1}{2\pi}\left|\left.\frac{dU}{dX}\right|_{U=-1}\right|\,.\label{eq:thermal_prediction_temp}\end{equation}
With the velocity profile given in Eq. (\ref{eq:normalized_velocity_profile}),
the event horizon occurs at the point $X_{h}$ that satisfies\[
-1=\frac{1}{2}\left(U_{R}+U_{L}\right)+\frac{1}{2}\left(U_{R}-U_{L}\right)\tanh\left(aX_{h}\right)\,,\]
or\[
\tanh\left(aX_{h}\right)=-\frac{\left(2+U_{R}+U_{L}\right)}{\left(U_{R}-U_{L}\right)}\,.\]
The derivative of the velocity profile at the event horizon is\begin{eqnarray*}
U^{\prime}\left(X_{h}\right) & = & \frac{1}{2}\left(U_{R}-U_{L}\right)a\,\mathrm{sech}^{2}\left(aX_{h}\right)\\
 & = & \frac{1}{2}\left(U_{R}-U_{L}\right)a\,\left(1-\tanh^{2}\left(aX_{h}\right)\right)\\
 & = & \frac{1}{2}\left(U_{R}-U_{L}\right)a\,\left(\frac{\left(U_{R}-U_{L}\right)^{2}-\left(2+U_{R}+U_{L}\right)^{2}}{\left(U_{R}-U_{L}\right)^{2}}\right)\\
 & = & -2a\frac{\left(1+U_{L}\right)\left(1+U_{R}\right)}{\left(U_{R}-U_{L}\right)}\,.\end{eqnarray*}
Thus, the temperature predicted by the dispersionless model is\begin{equation}
T_{\mathrm{pred}}=\frac{a}{\pi}\left|\frac{\left(1+U_{L}\right)\left(1+U_{R}\right)}{\left(U_{R}-U_{L}\right)}\right|\,,\label{eq:thermal_prediction-1}\end{equation}
which reduces to $T_{\mathrm{pred}}=ha/\left(2\pi\right)$ when $-1$
is the central value of $U$, i.e. when $U_{R}=-1+h$ and $U_{L}=-1-h$.

For the purpose of consistency, when we form the scaled energy spectrum
of Eq. (\ref{eq:normalized_spectrum_defn}) it is the predicted temperature
of Eq. (\ref{eq:thermal_prediction-1}) that shall be used.

\subsection{Validity of discontinous velocity profile}

If the change in velocity occurs over a much shorter length than some
characteristic length scale, the linearized velocity profile is invalid
and the system is better approximated by a step-discontinuous profile.
Here, the method and analysis of §\ref{sub:Steady-state-solution}
come into play. It was seen there that the mode conversion - and,
consequently, the temperature - approaches a finite limiting value
as the slope $a$ increases without bound. The prediction of Eq. (\ref{eq:thermal_prediction-1})
cannot hold; the temperature should approach some limiting value,
$T_{\infty}$, which is independent of $a$.

The calculation of $T_{\infty}$ is not conceptually difficult, but
it is quite lengthy, and has been relegated to Appendix \ref{sec:Appendix_Acoustic}.
The result is\begin{equation}
T_{\infty}=\left(1-U_{R}^{2}\right)^{1/2}\left|\frac{\left(1+U_{R}\right)\left(1+U_{L}\right)\left(U_{R}+U_{L}\right)}{\left(1-U_{R}\right)\left(1-U_{L}\right)\left(U_{R}-U_{L}\right)}\right|\,.\label{eq:limiting_temperature}\end{equation}
(This was presented previously in Ref. \cite{Corley-1997} and is very similar to a result given more recently in Ref. \cite{Recati-2009}.)
Note that Eq. (\ref{eq:thermal_prediction-1}) predicts a temperature
of exactly $T_{\infty}$ when $a$ is given by\[
a=\pi\left(1-U_{R}^{2}\right)^{1/2}\left|\frac{\left(U_{R}+U_{L}\right)}{\left(1-U_{R}\right)\left(1-U_{L}\right)}\right|\,.\]
It is when $a$ is on the order of this value that the transition
between the limiting forms of the temperature, given by Eqs. (\ref{eq:thermal_prediction-1})
and (\ref{eq:limiting_temperature}), takes place.

\section{Nonthermal behaviour and critical frequencies\label{sub:Critical-values-of-Omega}}

\begin{figure}
\subfloat{\includegraphics[width=0.45\columnwidth]{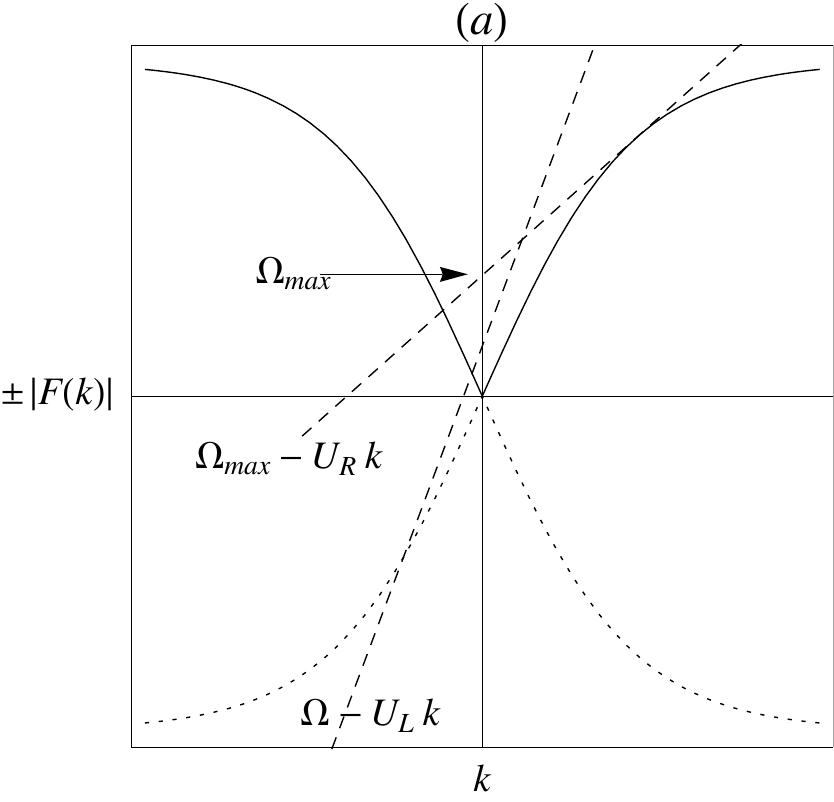}} \subfloat{\includegraphics[width=0.45\columnwidth]{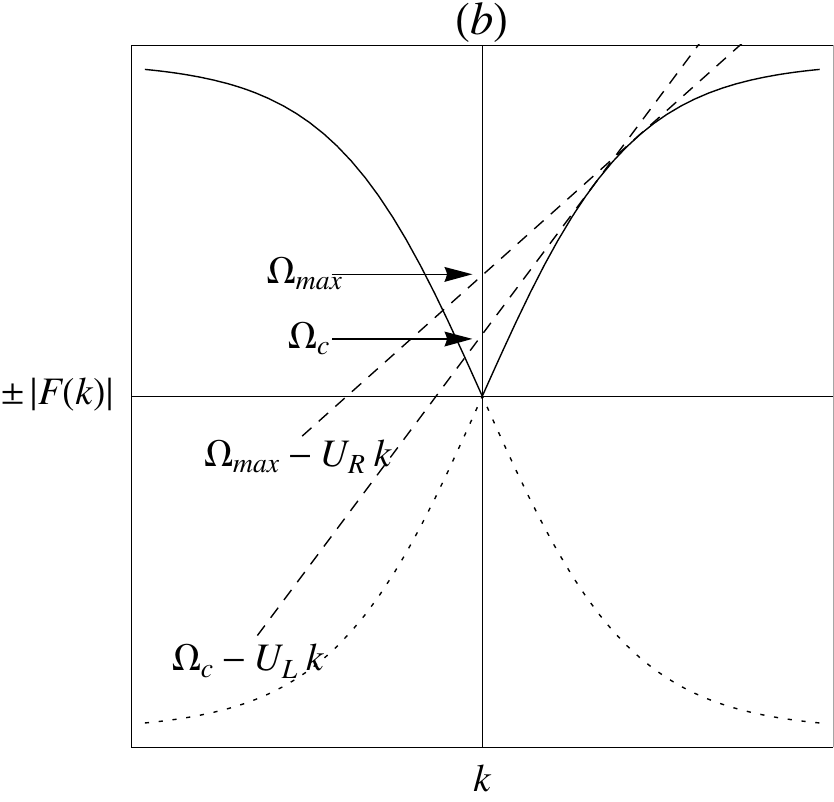}}

\caption[\textsc{Critical frequencies}]{\textsc{Critical frequencies}: In $(a)$, there is a point at which
$U=-1$, so that there is a region of subsonic flow and one of supersonic
flow. The dispersion relation in the region of supersonic flow has
a maximum frequency at which $ur$- and $ul$-modes appear; but in
the supersonic region, there is no such cut-off point. Figure $(b)$
shows the altered picture when $U$ is nowhere equal to $-1$ and
the flow is everywhere subsonic. Then both regions have a cut-off
frequency; frequencies between the two cut-offs experience a group-velocity
horizon, while those below both cut-offs, including very low frequencies,
do not.\label{fig:Critical-Frequencies}}

\end{figure}

When $U=-1$ at some point, it might be expected that the phonon spectrum
is thermal at low frequencies, as discussed in §\ref{sub:Thermal-behaviour}.
However, this spectrum cannot be purely thermal, for an examination
of the dispersion relation (see Figure \ref{fig:Critical-Frequencies}$\left(a\right)$)
shows that there is a cut-off frequency above which mixing of positive-
and negative-norm modes cannot occur. (It may occur between $u$-
and $v$-modes, but this coupling is typically negligibly small.)

However, nonthermality must also come into play for low frequencies
when $U$ is nowhere equal to $-1$. Since group velocity is frequency-dependent,
there may still be group-velocity horizons for certain frequencies,
provided the velocity profile conforms to the relations given above.
There will still be a cut-off frequency $\Omega_{\mathrm{max}}$,
above which no group-velocity horizons exist and no mixing between
positive and negative norms occurs. But there will also be no group-velocity
horizons for very low frequencies, since, as $\Omega\rightarrow0$,
the group velocity approaches unity. Therefore, there exists a frequency,
$\Omega_{c}$, which is the minimum frequency for which there exists
a group-velocity horizon (see Figure \ref{fig:Critical-Frequencies}$\left(b\right)$).
The horizon begins, when $\Omega=\Omega_{c}$, at one infinity (in
one of the asymptotic regions), and as $\Omega$ increases to $\Omega_{\mathrm{max}}$,
it moves smoothly to the other infinity (the other asymptotic region).

What happens for $\Omega<\Omega_{c}$? Although there is no group-velocity
horizon, mixing between positive and negative norms \textit{is} possible;
this is discussed in §\ref{sub:Conclusion-and-discussion}. What is
more, the dimension of the space of solutions is greater than when
a group-velocity horizon exists, because there is no merging of the $ur$-
and $ul$-modes into a single $url$-mode. Therefore, $\Omega_{c}$,
far from being a minimum frequency at which norm-mixing can occur,
simply marks a boundary between two different types of behaviour,
which must be treated separately in calculations. It may be noticed
that, taking the thermal prediction at face value, it is to be concluded
that virtually no radiation should be present since $U$ is never
equal to $-1$; or, at the very least, radiation should only be omitted
in the frequency range $\Omega_{c}<\Omega<\Omega_{\mathrm{max}}$,
when there is a group-velocity horizon. As we shall see, these predictions
are very far from the truth, and such horizons are unnecessary for
the emission of radiation.

\section{Phase integral analysis}

\begin{figure}
\includegraphics[width=0.8\columnwidth]{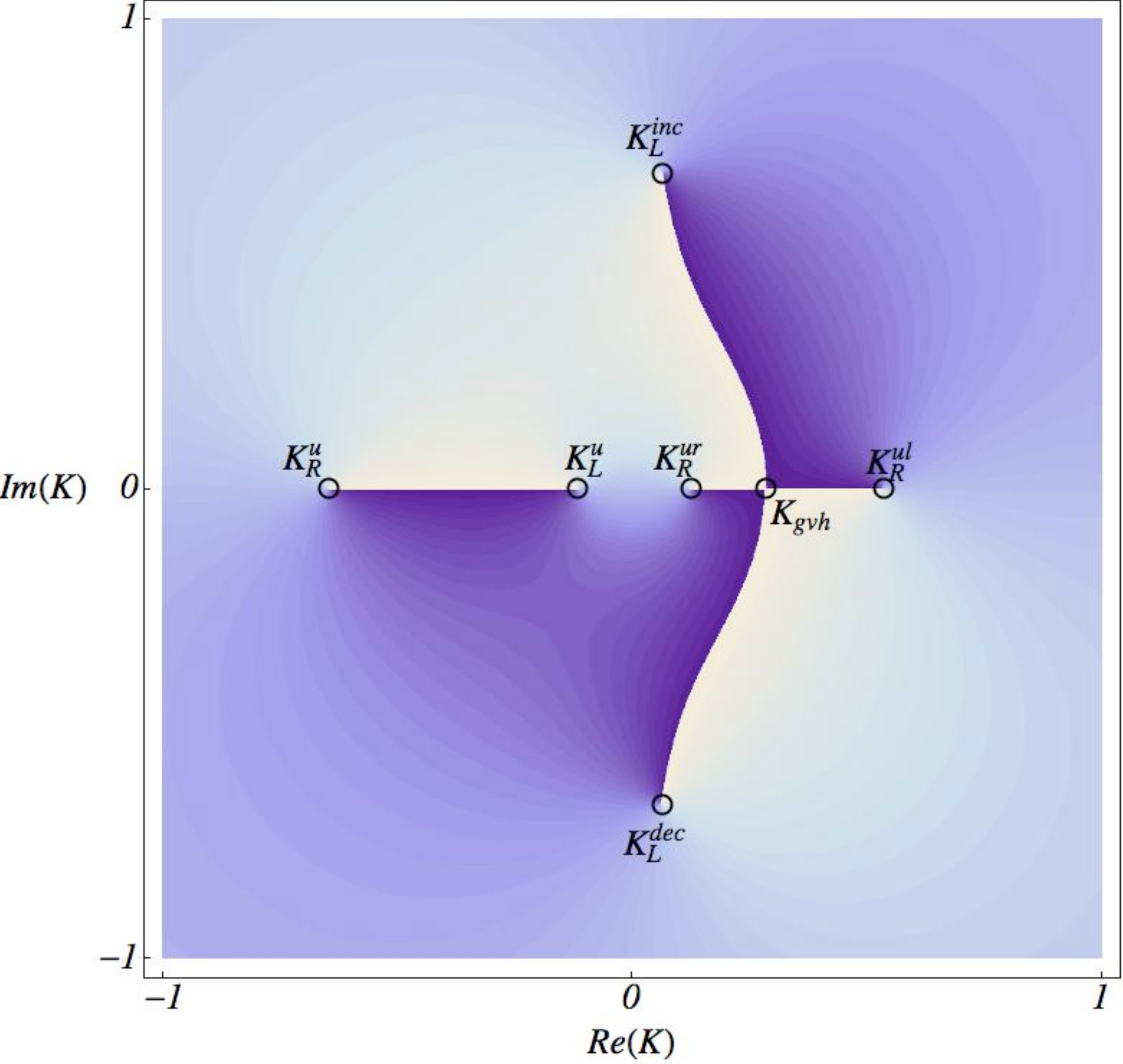}

\caption[\textsc{Phase integrals with a group-velocity horizon}]{\textsc{Phase integrals with a group-velocity horizon}: The degree
of shading shows the imaginary part of $X\left(K\right)$, which is
restricted to the interval between $0$ and $\pi/a$. The branch cuts
show where it is equal to these boundary values; thus they show where
$X\left(K\right)$ is real, tracing out the values of $K$ that are
realised somewhere in the velocity profile.\label{fig:WKB_contour_plot}}

\end{figure}

Before we explore the results of numerical calculations, let us pause
to examine the results of the one analytic tool at our disposal: the
phase integral approximation of Eq. (\ref{eq:creation_amp_integral}).
Substituting the appropriate wavevectors, this predicts a Hawking
emission rate given by\begin{equation}
\left|\frac{\beta_{\Omega}}{\alpha_{\Omega}}\right|^{2}=\exp\left(2\,\mathrm{Im}\left\{ J\right\} \right)\label{eq:WKB_radiation_rate}\end{equation}
where\begin{equation}
J=\int_{K^{ur}_{R}}^{K^{u}_{R}}X\left(K\right)dK\label{eq:WKB_phase_integral}\end{equation}
and $X\left(K\right)$ is defined implicitly by the conservation of
$\Omega$ according to the dispersion relation $\Omega=U\left(X\right)K+f\left(K\right)$.
Beginning with the $X$-dependent dispersion relation\[
\Omega=\left[\frac{1}{2}\left(U_{R}+U_{L}\right)+\frac{1}{2}\left(U_{R}-U_{L}\right)\tanh\left(aX\right)\right]K+f\left(K\right)\,,\]
we can find $X$ as a function of $K$:\[
\tanh\left(aX\right)=\frac{2}{U_{R}-U_{L}}\left(\frac{\Omega-f\left(K\right)}{K}-\frac{1}{2}\left(U_{R}+U_{L}\right)\right)\,,\]
or\begin{equation}
X\left(K\right)=\frac{1}{a}\,\mathrm{arctanh}\left[\frac{2}{U_{R}-U_{L}}\left(\frac{\Omega-f\left(K\right)}{K}-\frac{1}{2}\left(U_{R}+U_{L}\right)\right)\right]\,.\label{eq:X_as_function_of_K}\end{equation}
The only ambiguity is that $\tanh\left(x+i\pi\right)=\tanh\left(x\right)$,
i.e., the hyperbolic tangent is periodic in the imaginary part of
its argument, with period $\pi$. $X\left(K\right)$, therefore, is
uniquely defined up to an integer multiple of $i\pi/a$. We restrict
$0\le\mathrm{Im}\left\{ X\left(K\right)\right\} <\pi/a$. A typical
contour plot of the imaginary part of $X\left(K\right)$ is shown
in Fig. \ref{fig:WKB_contour_plot}.

The branch cuts of Fig. \ref{fig:WKB_contour_plot} show clearly those
values of $K$ where $\mathrm{Im}\left\{ X\left(K\right)\right\} =0$,
or where $X\left(K\right)$ is real. These lines trace out the instantaneous
values of $K$ (for a particular value of $\Omega$) as one varies
from one asymptotic region to the other. Their endpoints, which are
labelled in the diagram, are those values of $K$ which correspond
to plane waves in the asymptotic regions themselves, i.e., they are
solutions of either $\Omega-U_{R}K=f\left(K\right)$ or $\Omega-U_{L}K=f\left(K\right)$.
The cross-bow shaped feature, which joins together four different
values of $K$ - two of them real, two of them complex conjugates
- is indicative of the presence of a group-velocity horizon. The solutions
merge at the horizon itself, where they are equal to $K_{\mathrm{gvh}}$;
moving in one direction, they split into two real values, reaching
$K_{1}$ and $K_{2}$ in the asymptotic region; moving in the other
direction, they split into two complex conjugate values, one of which
is exponentially increasing while the other is decreasing. The other
feature corresponds to the Hawking wave, which experiences no group-velocity
horizon and simply varies from one real value to another.

\begin{figure}
\subfloat{\includegraphics[width=0.45\columnwidth]{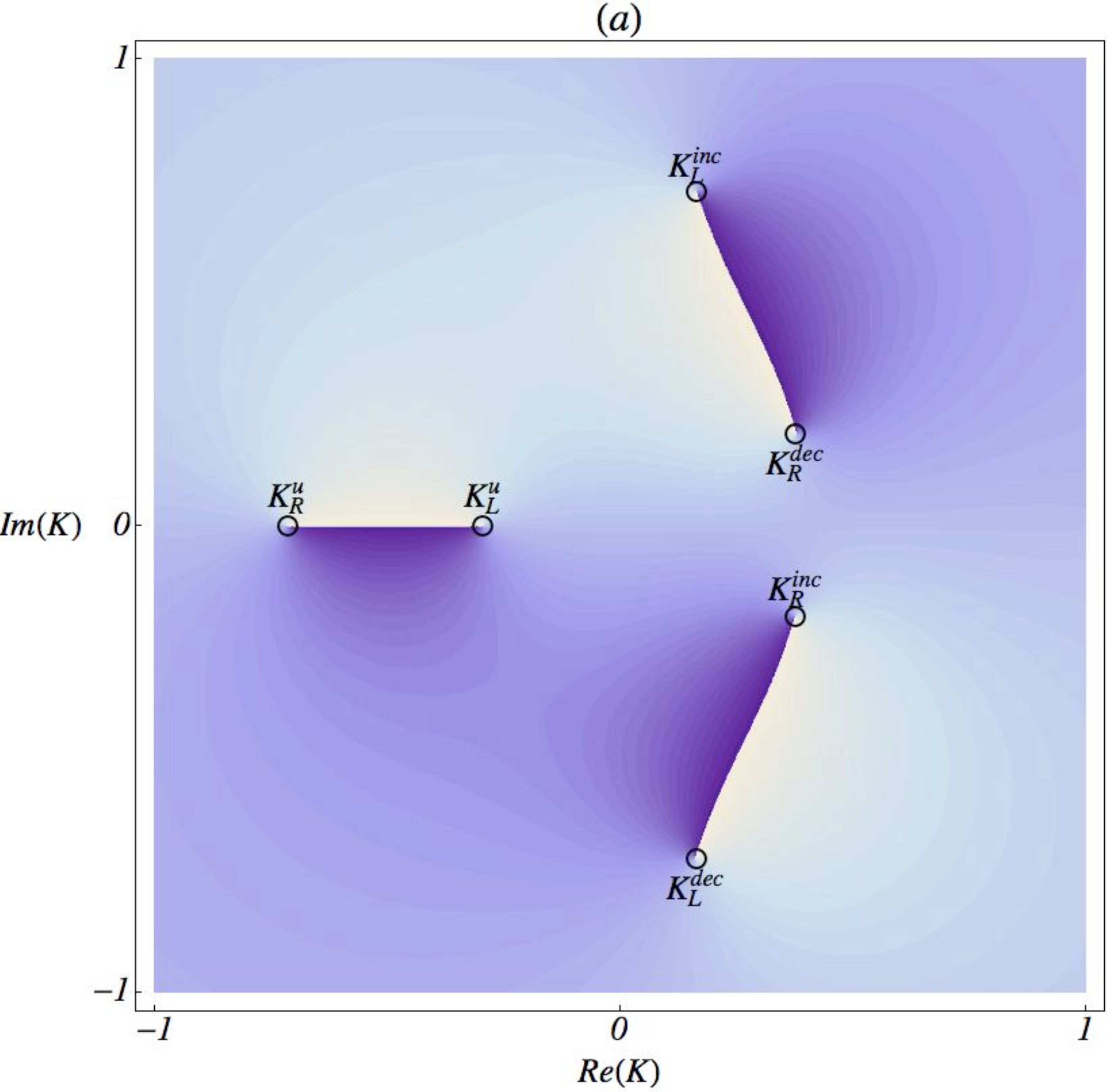}}\subfloat{\includegraphics[width=0.45\columnwidth]{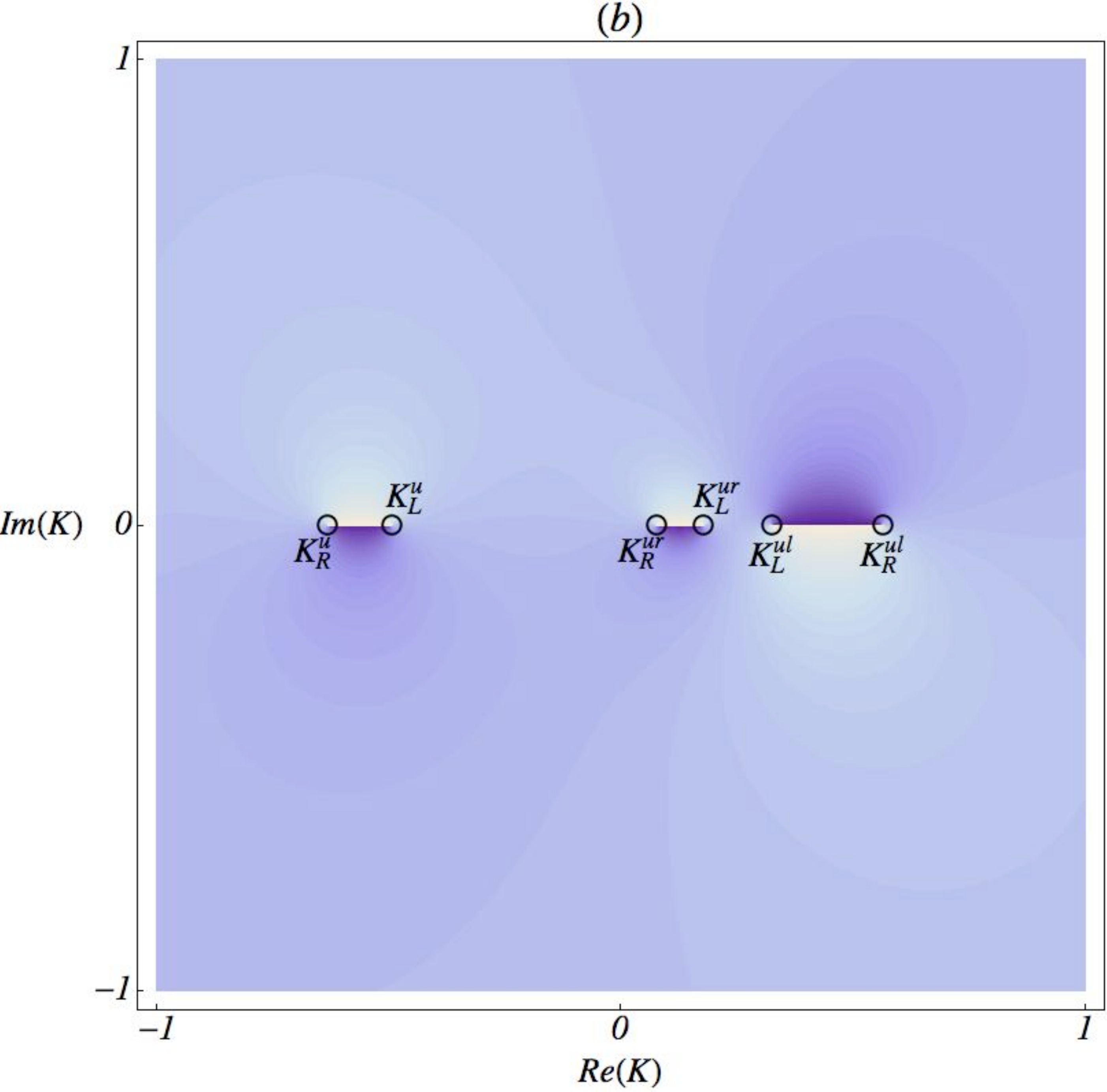}}

\caption[\textsc{Phase integrals with no group-velocity horizon}]{\textsc{Phase integrals with no group-velocity horizon}: The degree
of shading shows the imaginary part of $X\left(K\right)$, restricted
to the interval between $0$ and $\pi/a$. The branch cuts show where
$X\left(K\right)$ is real. In $\left(a\right)$, $\Omega>\Omega_{\mathrm{max}}$;
in $\left(b\right)$, $\Omega<\Omega_{c}$.\label{fig:WKB_no_gvh}}

\end{figure}

Figure \ref{fig:WKB_no_gvh} shows the other possibilities mentioned
in §\ref{sub:Critical-values-of-Omega}. Figure \ref{fig:WKB_no_gvh}$\left(a\right)$
shows the imaginary part of $X\left(K\right)$ in the case where $\Omega>\Omega_{\mathrm{max}}$.
The cross-bow pattern has disappeared, in accordance with the fact
that there is no longer a group-velocity horizon. Instead, only the
complex branches remain, so that the only real solution, for all values
of $X$, is the negative value of $K$. No positive-norm modes exist,
and hence no mode mixing is possible. Figure \ref{fig:WKB_no_gvh}$\left(b\right)$
corresponds to the case where $\Omega<\Omega_{c}$. (This requires
that $U$ is nowhere equal to $-1$.) We again have no group-velocity
horizon, but this time it is the real branches that remain, so that
we have three real solutions everywhere. Mixing of positive and negative
norms is therefore possible in both asymptotic regions.

According to Eqs. (\ref{eq:WKB_radiation_rate}) and (\ref{eq:WKB_phase_integral}),
we can find an approximation to the Hawking radiation rate by taking
the imaginary part of the integral of $X\left(K\right)$ from the
initial value of $K$ to the Hawking value. If we restrict the integration
to real values of $K$, this is simply the integral of the imaginary
part of $X\left(K\right)$. Fig (\ref{fig:WKB_contour_plot}) shows
that we need integrate only between $K^{ur}_{R}$ and $K^{u}_{L}$, since
beyond these points, $X\left(K\right)$ is real. What is more, for
any real value of $K$, the argument of the inverse hyperbolic tangent
in Eq. (\ref{eq:X_as_function_of_K}) is real; if the imaginary part
of $X\left(K\right)$ is not zero, then it must be equal to $\pi/\left(2a\right)$.
Thus we have\[
\mathrm{Im}\left\{ J\right\} =\frac{\pi}{2a}\left(K_{R}^{ur}-K_{L}^{u}\right)\,,\]
so that\begin{equation}
\left|\frac{\beta}{\alpha}\right|^{2}=\exp\left(-\frac{\pi}{a}\left(K_{R}^{ur}-K_{L}^{u}\right)\right)\,.\label{eq:WKB_beta-alpha_prediction}\end{equation}
Using the relation $\left|\alpha\right|^{2}-\left|\beta\right|^{2}=1$, which corresponds
to conservation of norm, this predicts a Hawking emission rate of
\begin{equation}
\left|\beta\right|^{2}=\frac{1}{\exp\left(\frac{\pi}{a}\left(K_{R}^{ur}-K_{L}^{u}\right)\right)-1}\,.\label{eq:WKB_prediction}
\end{equation}

Eq. (\ref{eq:WKB_prediction}) is clearly very similar to a Planck spectrum of the form (\ref{eq:thermal_prediction}),
and since thermal behaviour is essentially characterised by a $1/\Omega$ pole near $\Omega=0$, we expect
that Eq. (\ref{eq:WKB_prediction}) will reduce to such a form in the limit of low frequency. Indeed, as shown in
Appendix \ref{sec:Appendix_Acoustic}, for small $\Omega$ the wavevectors $K_{R}^{ur}$ and $K_{L}^{u}$
are approximately given by
\begin{alignat}{1}
K_{R}^{ur}\approx\frac{1}{1+U_{R}}\Omega\,,\,\,\,\,\, & K_{L}^{u}\approx\frac{1}{1+U_{L}}\Omega\,,
\end{alignat}
which means that
\begin{equation}
\frac{\pi}{a}\left(K_{R}^{ur}-K_{L}^{u}\right)\approx\frac{\pi}{a}\frac{\left(U_{L}-U_{R}\right)}{\left(1+U_{R}\right)\left(1+U_{L}\right)}\Omega\,.
\end{equation}
Thus, Eq. (\ref{eq:WKB_prediction}) corresponds exactly to the Hawking prediction, with temperature given by Eq. (\ref{eq:thermal_prediction-1}),
in the limit of low frequency. For higher frequencies, we expect deviations from thermality, and it remains to be seen
whether Eq. (\ref{eq:WKB_prediction}) will correspond well with numerical results.

\section{Numerical results\label{sub:Numerical-results}}

\subsection{Central horizon}

The typical example is one in which $-1$ is the central value of
the flow velocity, so that the event horizon occurs exactly at the
centre of the variation. The asymptotic velocities $U_{R}$ and $U_{L}$
can be replaced by the single parameter $h$, the height of the step
from $-1$ to either of the asymptotic velocities. As we have seen
from Eq. (\ref{eq:thermal_prediction-1}), the Hawking prediction
gives a temperature\begin{equation}
T_{\mathrm{pred}}=\frac{ha}{2\pi}\,.\label{eq:thermal_prediction_central_horizon}\end{equation}

\begin{figure}
\subfloat{\includegraphics[width=0.45\columnwidth]{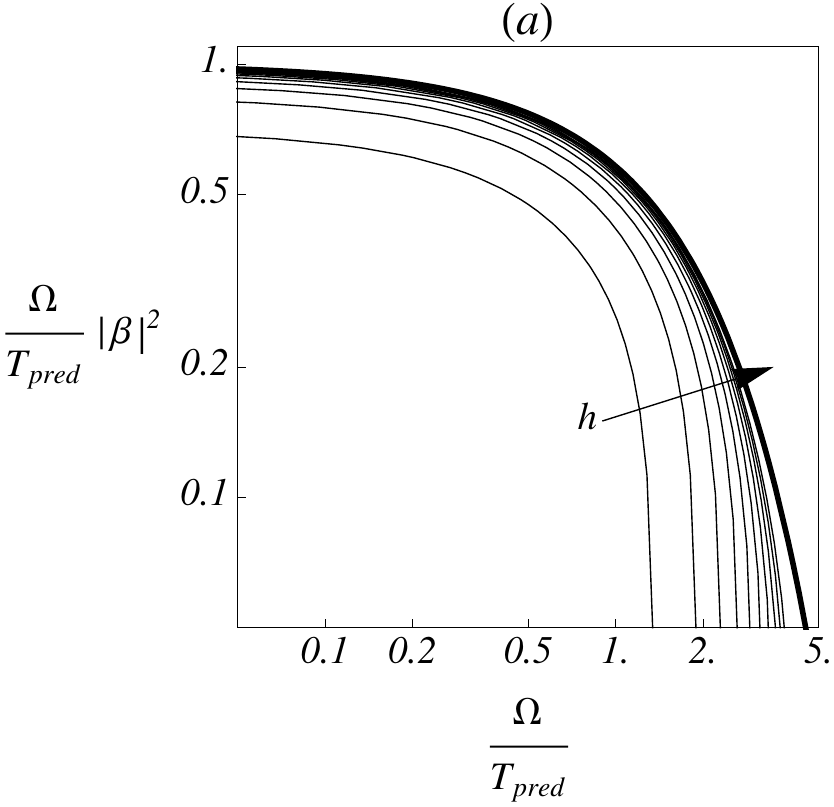}} \subfloat{\includegraphics[width=0.45\columnwidth]{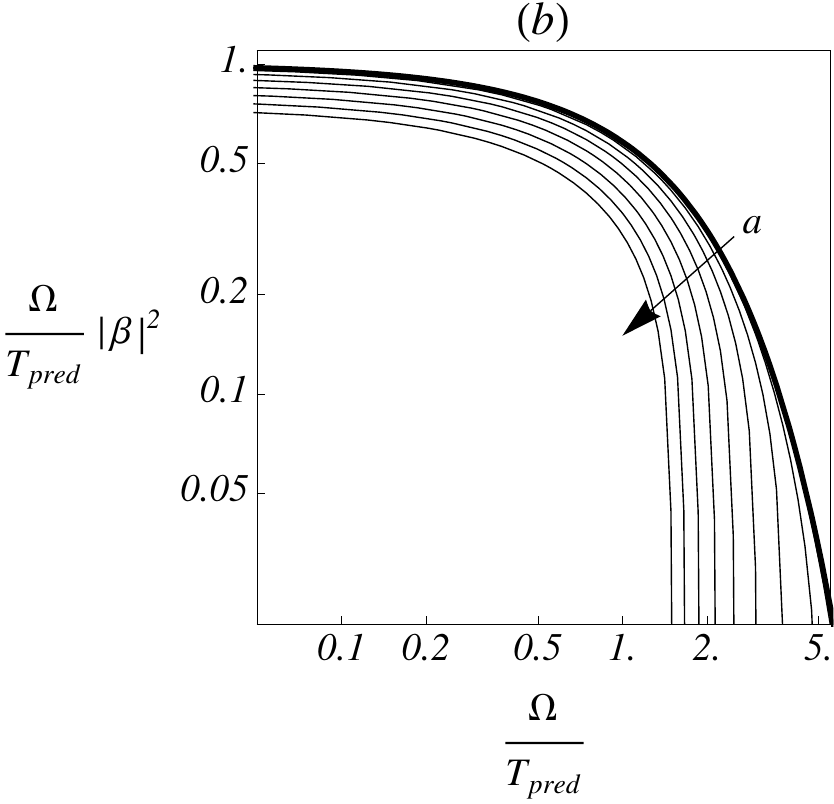}}

\subfloat{\includegraphics[width=0.45\columnwidth]{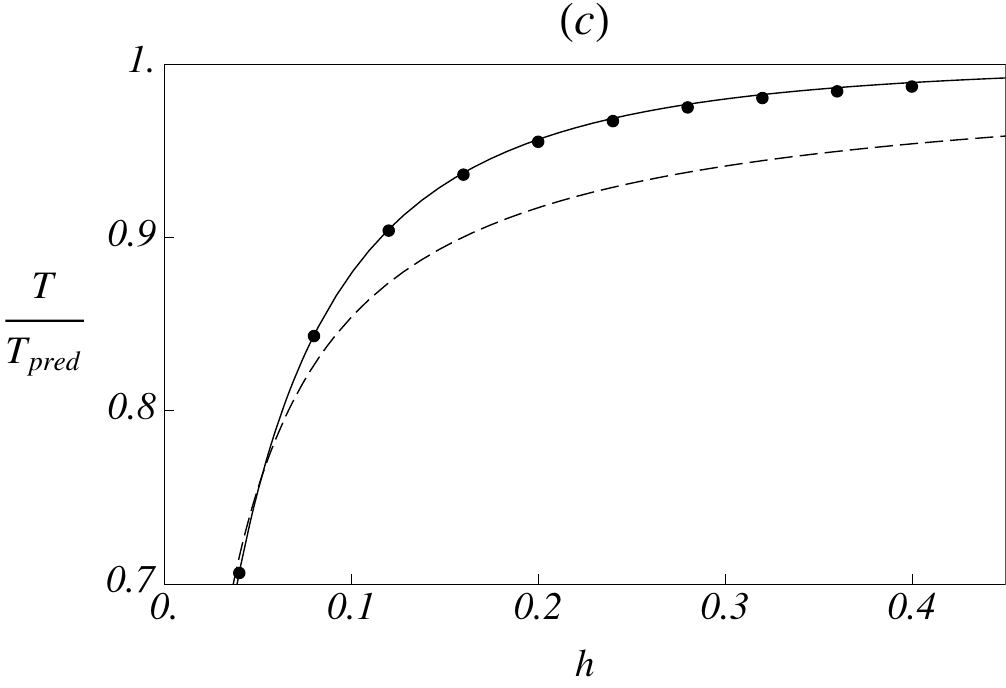}} \subfloat{\includegraphics[width=0.45\columnwidth]{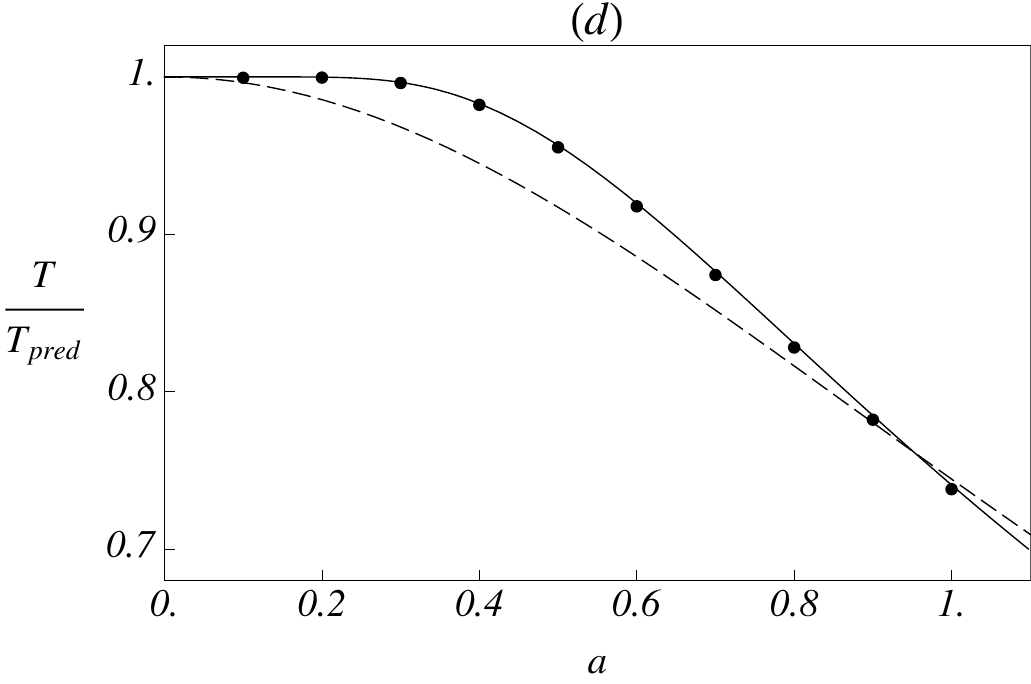}}

\caption[\textsc{Hawking spectra with a central horizon, 1}]{\textsc{Hawking spectra with a central horizon, 1}: In Figure $\left(a\right)$,
the parameter $a$ is fixed at $0.5$ while $h$ is varied between
$0.04$ and $0.40$; in Figure $\left(b\right)$, $h$ is fixed at
$0.20$ while $a$ is varied between $0.1$ and $1.0$. The thick
curve represents a thermal spectrum, and the arrows show the direction
in which the relevant parameter increases. Figures $\left(c\right)$
and $\left(d\right)$ show $T/T_{\mathrm{pred}}$, the low-frequency
limit of the spectra in Figures $\left(a\right)$ and $\left(b\right)$,
respectively. The functional form of Eq. (\ref{eq:guess_T_curve_1})
is shown as a dashed curve, while the solid curve corresponds to Eq.
(\ref{eq:guess_T_curve_2}).\label{fig:aco_Hawking-spectra_central-horizon_1}}

\end{figure}

Figure \ref{fig:aco_Hawking-spectra_central-horizon_1} shows a variety
of spectra, normalized according to Eq. (\ref{eq:normalized_spectrum_defn})
with the temperature given by Eq. (\ref{eq:thermal_prediction_central_horizon}).
Figure \ref{fig:aco_Hawking-spectra_central-horizon_1}$\left(a\right)$
shows several spectra with the same value of $a$ and different values
of $h$; Figure \ref{fig:aco_Hawking-spectra_central-horizon_1}$\left(b\right)$
shows spectra with the same value of $h$ but various values of $a$.
In each plot, a pure thermal spectrum is also shown.

\begin{figure}
\subfloat{\includegraphics[width=0.45\columnwidth]{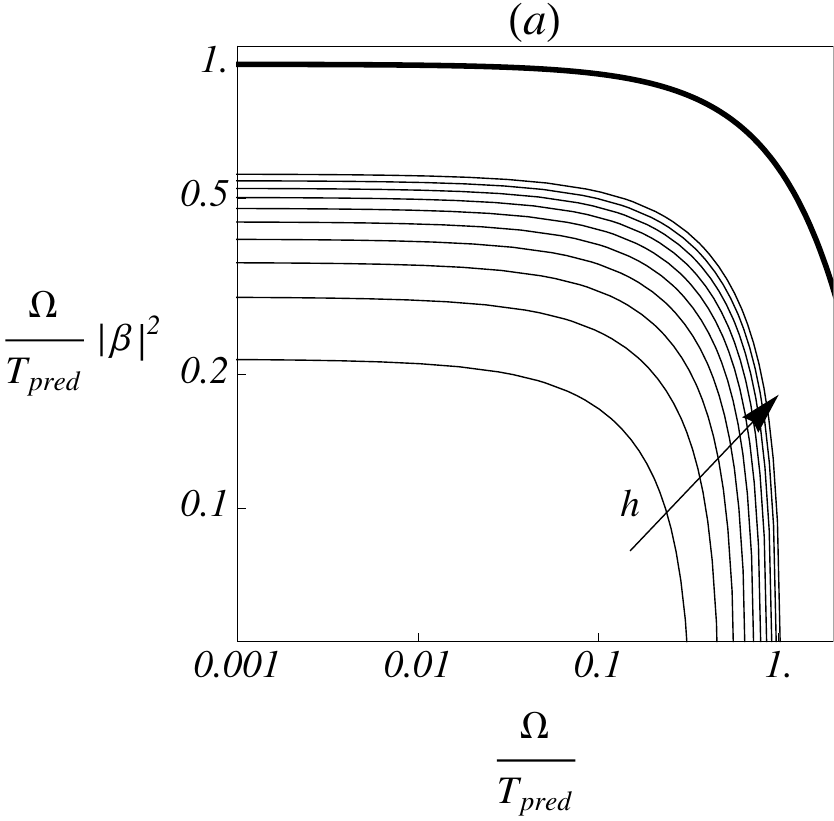}} \subfloat{\includegraphics[width=0.45\columnwidth]{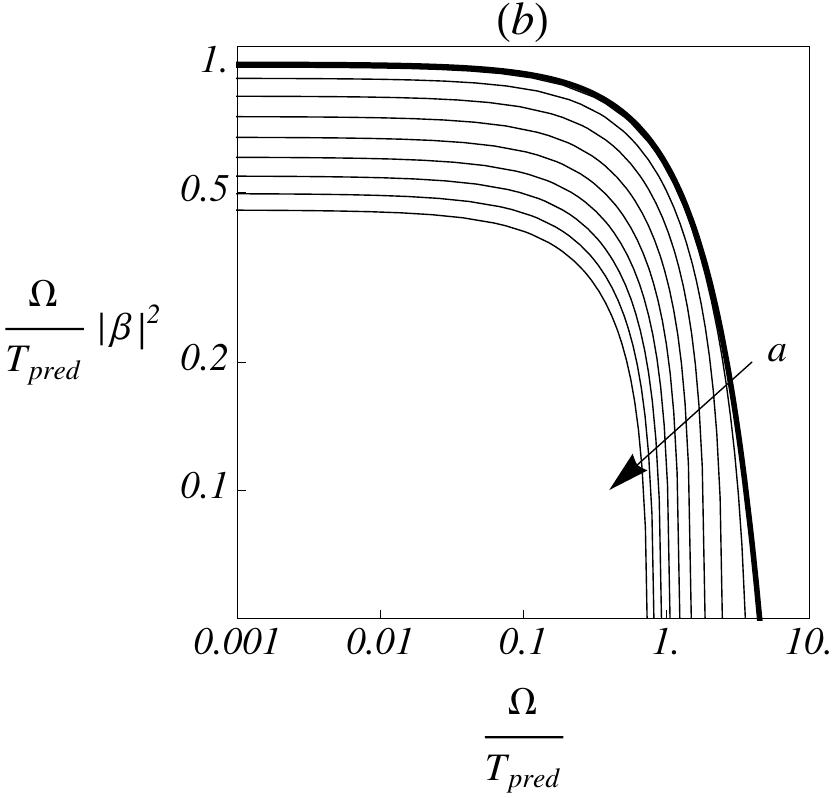}}

\subfloat{\includegraphics[width=0.45\columnwidth]{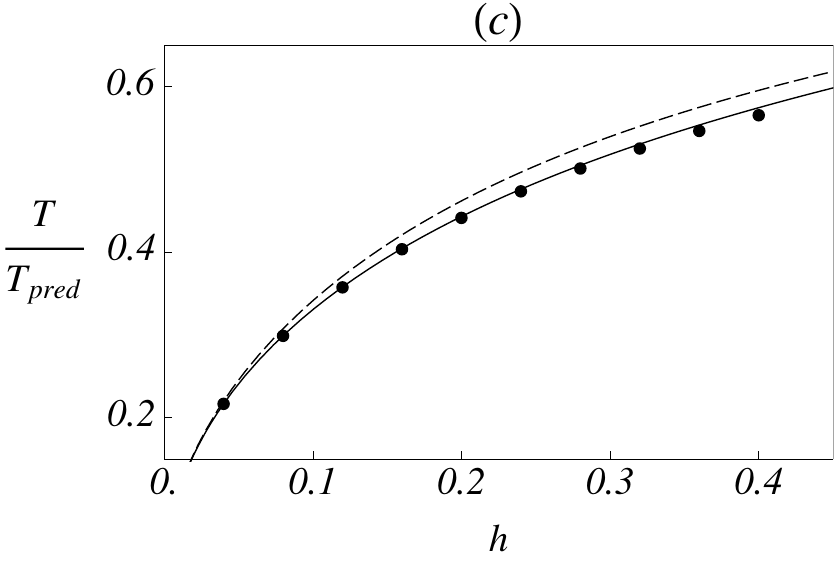}} \subfloat{\includegraphics[width=0.45\columnwidth]{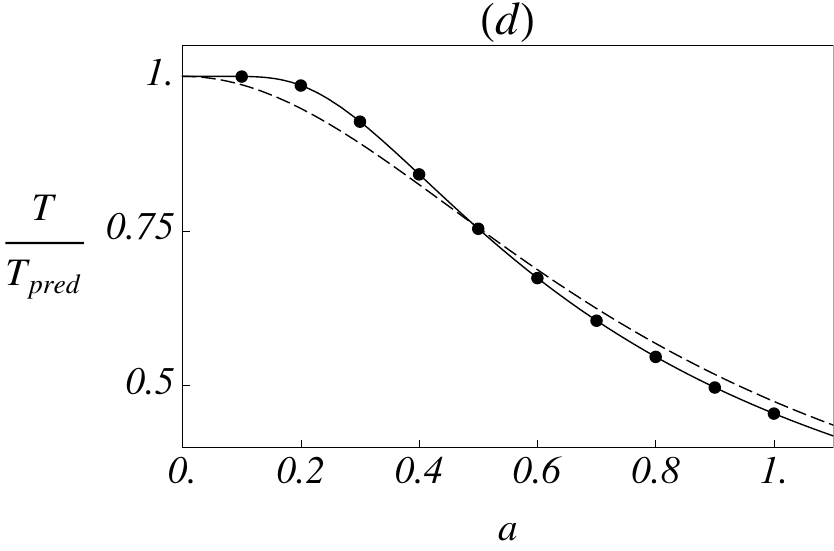}}

\caption[\textsc{Hawking spectra with a central horizon, 2}]{\textsc{Hawking spectra with a central horizon, 2}: In Figure $\left(a\right)$,
the parameter $a$ is fixed at $2.0$ while $h$ is varied between
$0.04$ and $0.40$; in Figure $\left(b\right)$, $h$ is fixed at
$0.05$ while $a$ is varied between $0.1$ and $1.0$. The thick
curve represents a thermal spectrum, and the arrows show the direction
in which the relevant parameter increases. Figures $\left(c\right)$
and $\left(d\right)$ plot $T/T_{\mathrm{pred}}$, the low-frequency
limit of the spectra in Figures $\left(a\right)$ and $\left(b\right)$.
The functional form given in Eq. (\ref{eq:guess_T_curve_1}) is plotted
as a dashed line, while the solid line corresponds to Eq. (\ref{eq:guess_T_curve_2}).\label{fig:aco_Hawking-spectra_central-horizon_2}}

\end{figure}

These spectra (of Figs. \ref{fig:aco_Hawking-spectra_central-horizon_1}$\left(a\right)$
and $\left(b\right)$) all have a faster fall to zero than the thermal
spectrum, as expected: they must go to zero at the maximum frequency
$\Omega_{\mathrm{max}}$, while the thermal spectrum is non-zero for
all frequencies. Also, the measured spectra all approach limiting
values for low frequencies, indicating that the creation rate $\left|\beta_{\Omega}\right|^{2}/\left(2\pi\right)$
always has the $\frac{1}{\Omega}$ pole that is characteristic of
a thermal spectrum. However, since they do not all approach a limiting
value of $1$, it is clear that they do not all accord with the
prediction of Eq. (\ref{eq:thermal_prediction_central_horizon}),
which seems to be true only in the limit of high $h$ and low $a$.
(These trends have been noted previously \cite{Macher-Parentani-2008}.)
Recalling Eq. (\ref{eq:low_freq_temp}), the limiting values of the
spectra as $\Omega\rightarrow0$ are precisely the values of $T/T_{\mathrm{pred}}$,
where $T$ is the true low-frequency temperature. Some spectra, then,
have a value of $T$ which is lower than $T_{\mathrm{pred}}$. Again,
this is to be expected: the Hawking prediction yields a temperature
which can increase without bound, whereas we have seen that there
is a limiting temperature as $a\rightarrow\infty$ (given by Eq. (\ref{eq:limiting_temperature})).

Figures \ref{fig:aco_Hawking-spectra_central-horizon_1}$\left(c\right)$
and $\left(d\right)$ plot the values of $T/T_{\mathrm{pred}}$ for
the spectra of $\left(a\right)$ and $\left(b\right)$, respectively.
They also show two analytic, though heuristically derived, curves
- one of which gives excellent agreement with the measured values.
To find this, let us consider the dependence of $T$ on $a$, i.e.,
we treat $h$ as fixed. It is clear that, at low $a$, the temperature
is given by $T_{\mathrm{pred}}=ha/\left(2\pi\right)$, and is thus
linear in $a$. The temperature increases monotonically with $a$,
approaching the limiting value $T_{\mathrm{\infty}}$, which is independent
of $a$. There are many functional forms that would behave as such,
but the hyperbolic tangent is a natural guess, given that the velocity
profile has precisely this form. Even so, there are two possible guesses
for the form of the temperature which behave in the required way:\begin{alignat}{1}
T=T_{\mathrm{\infty}}\tanh\left(\frac{ha}{2\pi\, T_{\mathrm{\infty}}}\right)\,,\qquad & \mathrm{or}\qquad T=\frac{ha}{2\pi}\tanh\left(\frac{2\pi\, T_{\mathrm{\infty}}}{ha}\right)\,.\label{eq:guess_T_curves}\end{alignat}
Substituting in Eq. (\ref{eq:limiting_temperature}) the values $U_{R}=-1+h$
and $U_{L}=-1-h$, we find\begin{equation}
T_{\infty}=\frac{h^{3/2}}{\left(2-h\right)^{1/2}\left(2+h\right)}\,.\label{eq:limiting_T_central_horizon}\end{equation}
Plugging this into the above forms for the temperature and dividing
by $T_{\mathrm{pred}}=ha/\left(2\pi\right)$, the guessed temperature
curves are\begin{eqnarray}
\frac{T}{T_{\mathrm{pred}}} & = & \frac{2\pi h^{1/2}}{a\left(2-h\right)^{1/2}\left(2+h\right)}\tanh\left(\frac{a\left(2-h\right)^{1/2}\left(2+h\right)}{2\pi h^{1/2}}\right)\,,\label{eq:guess_T_curve_1}\\
\mathrm{or}\quad\frac{T}{T_{\mathrm{pred}}} & = & \tanh\left(\frac{2\pi h^{1/2}}{a\left(2-h\right)^{1/2}\left(2+h\right)}\right)\,.\label{eq:guess_T_curve_2}\end{eqnarray}
As shown in Fig. \ref{fig:aco_Hawking-spectra_central-horizon_1},
Eq. (\ref{eq:guess_T_curve_2}) agrees remarkably well with numerical
results, over the entire parameter space.  This is the first demonstration of such
a universally valid expression for the low-frequency temperature.

More spectra are shown in Fig. \ref{fig:aco_Hawking-spectra_central-horizon_2},
with fixed values of $a=2.0$ and $h=0.05$ chosen so as to increase
the deviation from the thermal prediction of Eq. (\ref{eq:thermal_prediction_central_horizon}).
The values of $T/T_{\mathrm{max}}$ are also plotted, and once again
it is found that Eq. (\ref{eq:guess_T_curve_2}) agrees very well
with numerical results.

\subsection{Non-central horizon}

We have seen that, when the central value of the velocity profile
is equal to $-1$, then the ratio of the measured temperature to the
predicted temperature,\[
\frac{T}{T_{\mathrm{pred}}}=\tanh\left(\frac{T_{\infty}}{T_{\mathrm{pred}}}\right)\,.\]
It is reasonable to generalise this to the case of an arbitrary velocity
profile, so long as it is equal to $-1$ somewhere (for, if it is
not, then the Hawking prediction of Eq. (\ref{eq:thermal_prediction_temp})
is inapplicable). Plugging in the expressions for $T_{\mathrm{pred}}$
and $T_{\infty}$ as given in Eqs. (\ref{eq:thermal_prediction-1})
and (\ref{eq:limiting_temperature}), respectively, then we find\begin{equation}
\frac{T}{T_{\mathrm{pred}}}=\tanh\left(\frac{\pi}{a}\left(1-U_{R}^{2}\right)^{1/2}\frac{\left(U_{L}+U_{R}\right)}{\left(1-U_{R}\right)\left(U_{L}-1\right)}\right)\,,\label{eq:T-over-Tpred_general_U}\end{equation}
or\begin{equation}
T=\frac{a}{\pi}\frac{\left(1+U_{R}\right)\left(1+U_{L}\right)}{\left(U_{L}-U_{R}\right)}\tanh\left(\frac{\pi}{a}\left(1-U_{R}^{2}\right)^{1/2}\frac{\left(U_{L}+U_{R}\right)}{\left(1-U_{R}\right)\left(U_{L}-1\right)}\right)\,.\label{eq:modified_temp_general_U}\end{equation}

\begin{figure}
\includegraphics[width=0.8\columnwidth]{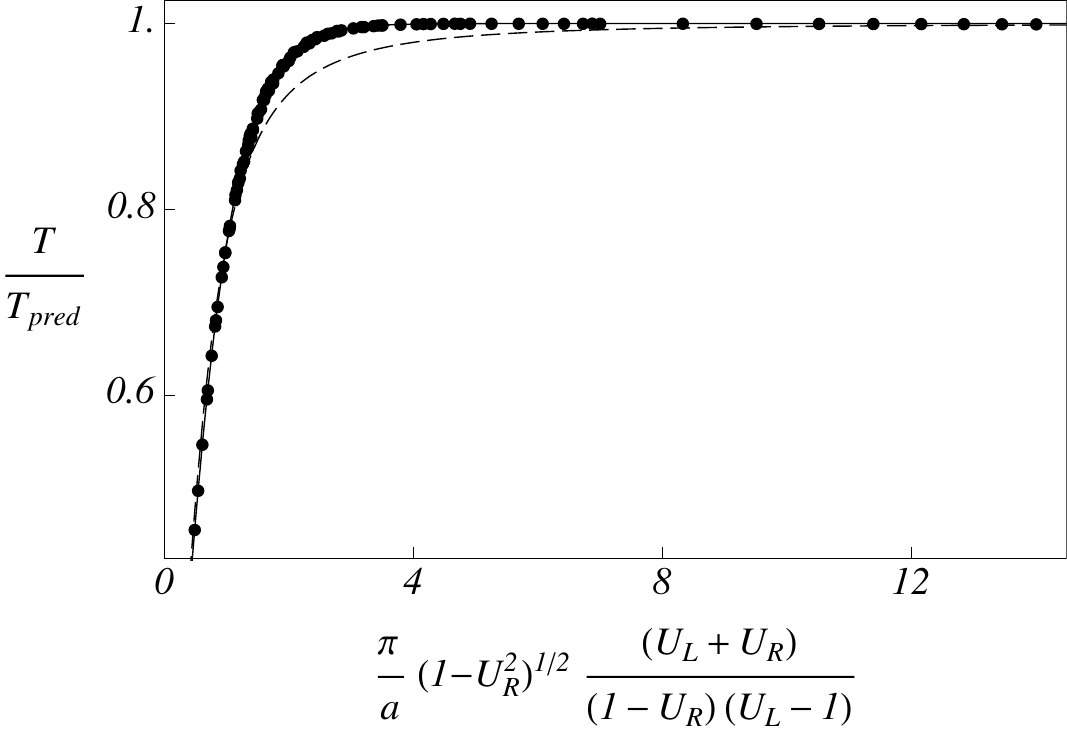}

\caption[\textsc{Low-frequency temperature}]{\textsc{Low-frequency temperature}: This figure plots $T/T_{\mathrm{pred}}$,
or the low-frequency limit of $\Omega\left|\beta\right|^{2}/T_{\mathrm{pred}}$,
for various values of $U_{R}$, $U_{L}$ and $a$. Writing the abscissa
as simply $x$, the solid curve plots $\tanh\left(x\right)$, while
the dashed curve plots $x\,\tanh\left(1/x\right)$, which is the generalised
form of Eq. (\ref{eq:guess_T_curve_1}). The low-frequency temperature
is thus seen to agree very well with Eq. (\ref{eq:T-over-Tpred_general_U}).\label{fig:Hawking_aco_low-freq-temp_tanh}}

\end{figure}

Figure \ref{fig:Hawking_aco_low-freq-temp_tanh} plots the values
of $T/T_{\mathrm{pred}}$ for various values of the parameters $U_{R}$,
$U_{L}$ and $a$. The independent variable in the plot is taken to
be the combination of these parameters that appears in the argument
of the hyperbolic tangent function in Eqs. (\ref{eq:T-over-Tpred_general_U}).
This equation predicts that the resulting plot should be simply the
hyperbolic tangent function; that is indeed found to be the case.

\begin{figure}
\includegraphics[width=0.8\columnwidth]{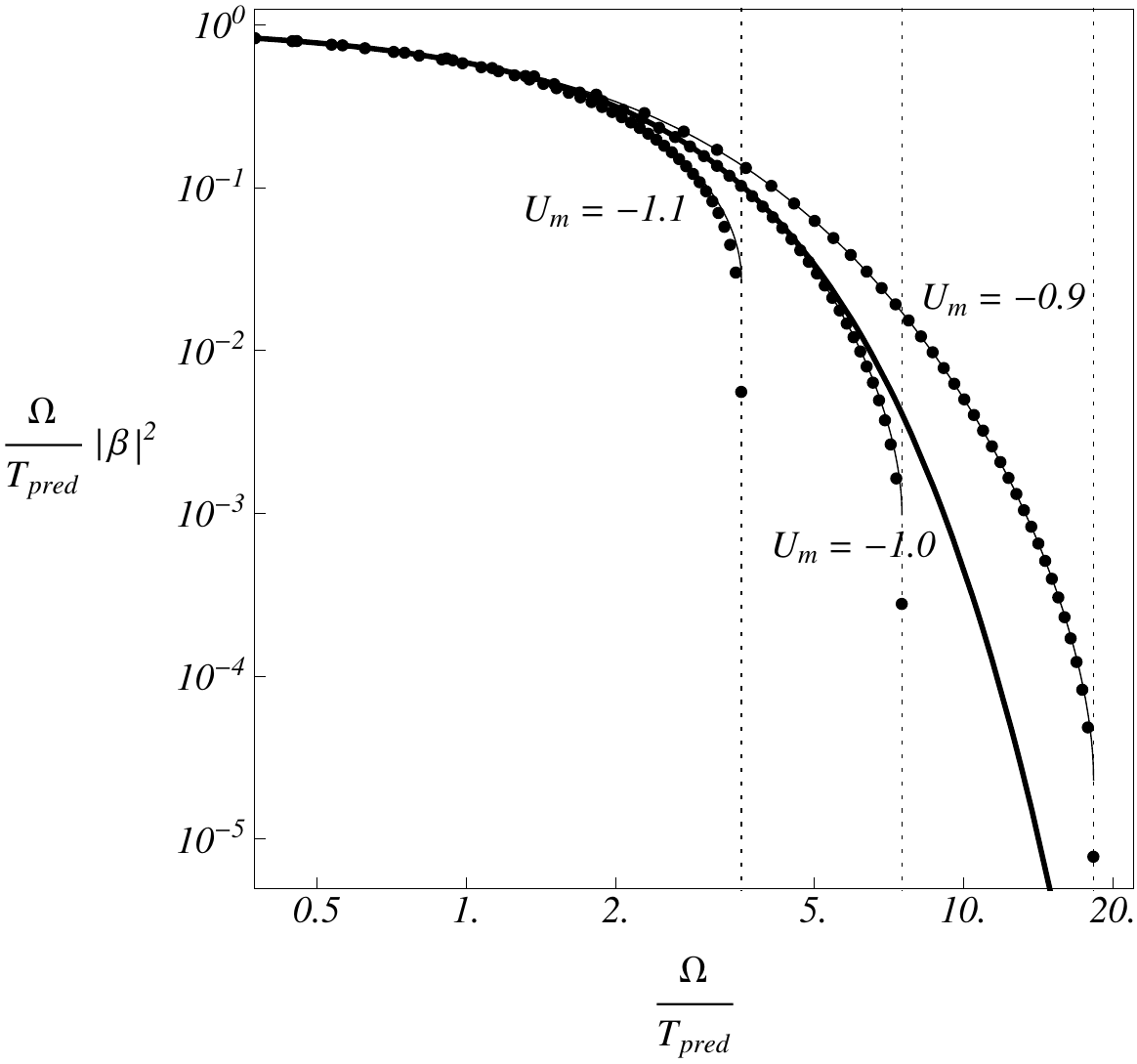}

\caption[\textsc{Hawking spectra with non-central horizon}]{\textsc{Hawking spectra with non-central horizon}: The thick curve
represents a thermal spectrum, while the thinner curves are the spectra
obtained from Eq. (\ref{eq:WKB_prediction}) with various
values of $U_{m}$. (We have fixed $h=0.2$ and $a=0.2$, so that
the low-frequency temperature (\ref{eq:modified_temp_general_U})
agrees very well with Eq. (\ref{eq:thermal_prediction-1}).) The dots
are data points obtained by solving for the spectra numerically. The
dotted vertical lines show the positions of $\Omega_{\mathrm{max}}/T_{\mathrm{pred}}$
for each curve.\label{fig:aco_Hawking-spectra_WKB}}

\end{figure}

When the central value of $U$ - hereafter referred to as $U_{m}$
- is equal to $-1$, the deviation of the spectra from the thermal
spectrum occurs as a result of $ $the cut-off frequency $\Omega_{\mathrm{max}}$,
so that they fall sharply to zero as this frequency is approached.
However, when $U_{m}$ is different from $-1$, the deviations from
the thermal spectrum are much more significant. Figure \ref{fig:aco_Hawking-spectra_WKB}
shows examples for different values of $U_{m}$; the values of $h$
and $a$ are fixed at values such that $T/T_{\mathrm{pred}}\approx1$.
The most striking difference occurs when $U_{m}>-1$, for then the
observed spectrum is \textit{greater} than the thermal spectrum (at
the high-frequency end).

Also of note in Fig. \ref{fig:aco_Hawking-spectra_WKB} is how accurate
the phase integral prediction of Eq. (\ref{eq:WKB_prediction})
is, clearly in accordance with the deviations just mentioned. This is the first
demonstration of an analytic expression valid over the entire spectrum. The
only significant deviation occurs very close to $\Omega_{\mathrm{max}}$,
when again the measured spectrum falls off more sharply than the model
predicts. It is also worth noting that the phase-integral approximation
is only accurate when the low-frequency temperature agrees very well
with the Hawking prediction, i.e., when $T/T_{\mathrm{pred}}\approx1$.
Indeed, taking the low-frequency limit of Eq. (\ref{eq:WKB_prediction})
yields precisely the temperature of Eq. (\ref{eq:thermal_prediction-1}).
It is not clear whether the phase integral approximation can be modified
to accord with the modified temperature of Eq. (\ref{eq:modified_temp_general_U}).

\subsection{No low-frequency horizon}

What happens if there is no point at which $U=-1$? Such a regime is never
examined in the literature, for it does not correspond to a black hole in the
dispersionless limit. The Hawking prediction of Eqs. (\ref{eq:thermal_prediction}) and
(\ref{eq:thermal_prediction_temp}) cannot be applied, so it might be
imagined that the spectrum vanishes. However, we have already seen
that the Hawking prediction does not always apply, the relation between
the temperature and the derivative at the horizon being rather less
straightforward; and, moreover, that the spectra are not purely thermal
anyway, exhibiting large deviations from thermality when $U_{m}>-1$.
(This would, of course, be the case if $U_{L}$ were increased to
be greater than $-1$.) Recalling the observation made in §\ref{sub:Conclusion-and-discussion}
that an event horizon is not a necessary ingredient for Hawking radiation
to be possible, we conclude that there is no reason to suppose that
the Hawking spectrum will vanish when $U$ is everywhere greater than
$-1$ - except, perhaps, that it is likely to be far from thermal.

Recall that, in this case, there are two types of frequency, separated
by a critical frequency $\Omega_{c}$: those frequencies above $\Omega_{c}$
(and, of course, below $\Omega_{\mathrm{max}}$) experience a group-velocity
horizon, and are treated in the same way as before; frequencies below
$\Omega_{c}$, however, do not have a group-velocity horizon, and
can occur in pairs of $u-ul$ or $u-ur$ modes. (We continue to ignore
the backward-travelling $v$-modes.) We must take account of both
of these pairings, and their spectra will in general be different.

\begin{figure}
\subfloat{\includegraphics[width=0.45\columnwidth]{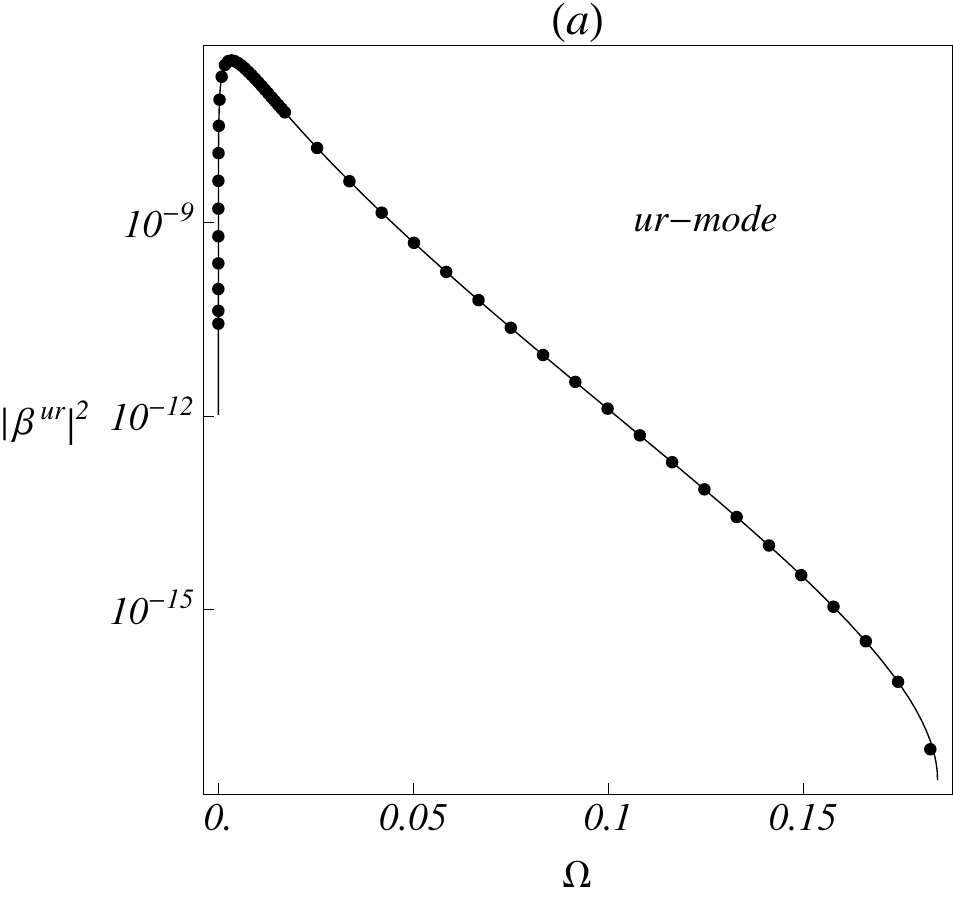}} \subfloat{\includegraphics[width=0.45\columnwidth]{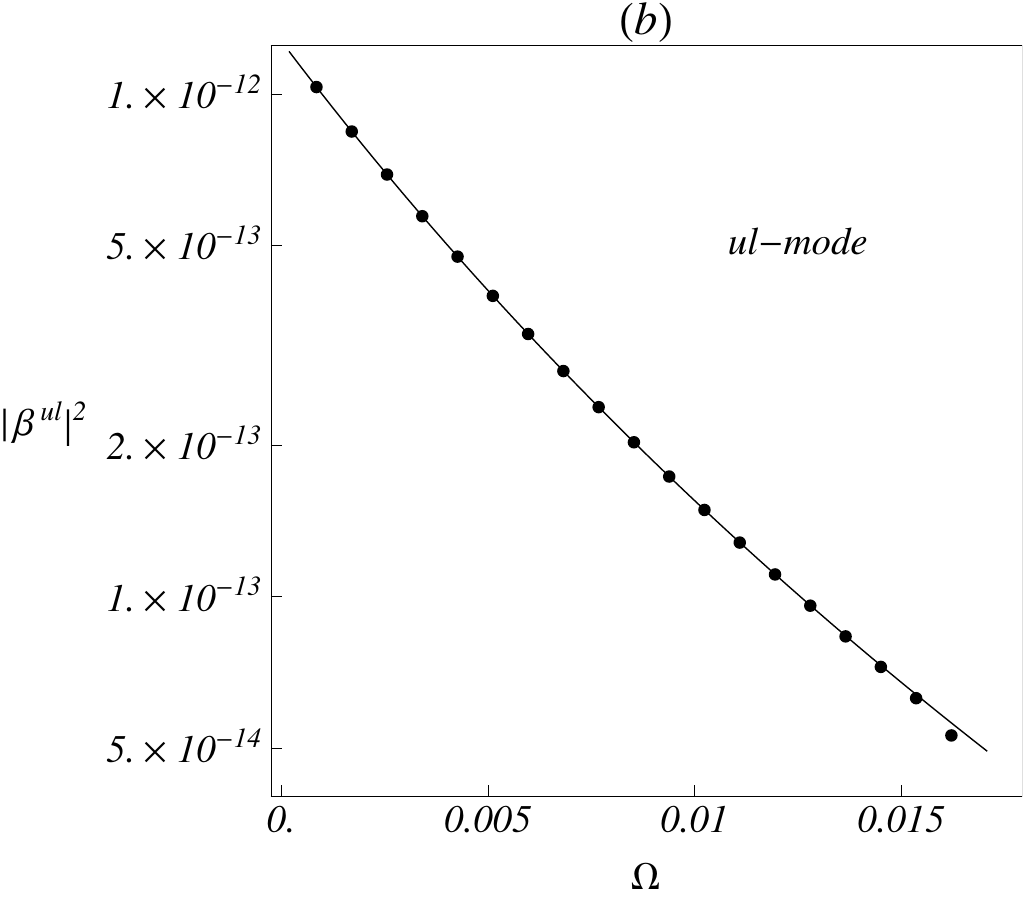}}

\caption[\textsc{Hawking spectrum with no low-frequency horizon, 1}]{\textsc{Hawking spectrum with no low-frequency horizon, 1}: Here,
we have taken $U_{R}=-0.5$, $U_{L}=-0.9$ and $a=0.1$, so that $U$
is nowhere equal to $-1$. Figure $\left(a\right)$ shows the spectrum
for the $ur$-$u$ mode pairing, while Figure $\left(b\right)$ corresponds
to the $ul$-$u$ mode pairing. The dots show the results of numerical
calculations, while the curves plot the modified phase-integral predictions
of Eqs. (\ref{eq:WKB_no-horizon_gvh-rate})-(\ref{eq:WKB_no-horizon_ur-rate}).
Note that the $ul$-$u$ pairing can only occur for $\Omega<\Omega_{c}$,
while the $ur$-$u$ pairing is possible right up to $\Omega_{\mathrm{max}}$.
The two spectra approach zero at the upper limit of their frequency ranges,
while the $ur$-$u$ spectra has a maximum value at the frequency $\Omega_{c}$.\label{fig:Hawking_aco_no-horizon_WKB-good}}

\end{figure}

Figure \ref{fig:Hawking_aco_no-horizon_WKB-good} shows an example
of such a spectrum, with $U_{R}=-0.5$, $U_{L}=-0.9$ and $a=0.1$.
It is clearly non-zero, and is also clearly non-thermal. The two possible channels
for phonon creation give very different spectra, and, for the $ur$-$u$ pair,
there is a change of behaviour at the frequency $\Omega_{c}$, the spectrum
reaching a maximum at this point and approaching zero at both zero frequency and $\Omega_{\mathrm{max}}$.

The numerically calculated spectra are also in good agreement with the phase-integral approximation, although
this claim requires some further explanation. When $\Omega_{c}<\Omega<\Omega_{\mathrm{max}}$,
so that we have a group-velocity horizon, the phase-integral approximation
gives the same expression as Eq. (\ref{eq:WKB_prediction}),\begin{equation}
\left|\beta_{\Omega}^{ur}\right|^{2}\approx\exp\left(-\frac{\pi}{a}\left(K_{R}^{ur}-K_{L}^{u}\right)\right)\,.\label{eq:WKB_no-horizon_gvh-rate}\end{equation}
(These values are small enough that we needn't make the modification
of Eq. (\ref{eq:WKB_prediction}).) When $\Omega<\Omega_{c}$,
this requires some modification. The spectrum is calculated using
the $u$-in mode, the creation rates given by the norms of the out-components
for the $ul$- and $ur$-modes. The phase integral of Eq. (\ref{eq:norm_phase_integral})
is then taken between the $K^{u}$-branch and either the $K^{ur}$-
or $K^{ul}$-branch. Taking the $ul$-$u$ pairing first: the imaginary
part of $X\left(K\right)$ is $\pi/\left(2a\right)$ in the regions
between the branches; however, on traversing the $K^{ur}$-branch,
$X\left(K\right)$ is not purely real but has an imaginary part of
$\pi/a$. The total integral, then, has an imaginary part of $\pi\left(K_{L}^{ul}-K_{L}^{u}+K_{L}^{ur}-K_{R}^{ur}\right)/\left(2a\right)$,
and the predicted creation rate is\begin{equation}
\left|\beta_{\Omega}^{ul}\right|^{2}\approx\exp\left(-\frac{\pi}{a}\left(K_{L}^{ul}-K_{L}^{u}+K_{L}^{ur}-K_{R}^{ur}\right)\right)\,.\label{eq:WKB_no-horizon_ul-rate}\end{equation}
The creation rate of the $ur$-$u$ pairing would seem to have the
same expression as in Eq. (\ref{eq:WKB_no-horizon_gvh-rate}). However,
numerics show that the spectrum becomes linear in $\Omega$ as $\Omega\rightarrow0$,
a form not given by this exponential factor. It is found heuristically,
using the fact that both $K_{R}^{ur}$ and $K_{L}^{ur}$ approach
zero as $\Omega\rightarrow0$, that the $ur$-$u$ spectrum conforms
to the expression\begin{equation}
\left|\beta_{\Omega}^{ur}\right|^{2}\approx\exp\left(-\frac{\pi}{a}\left(K_{R}^{ur}-K_{L}^{u}\right)\right)-\exp\left(-\frac{\pi}{a}\left(K_{L}^{ur}-K_{L}^{u}\right)\right)\,.\label{eq:WKB_no-horizon_ur-rate}\end{equation}
Although the discontinuous ($a\rightarrow\infty$) velocity profile
can be solved to show that $\left|\beta_{\Omega}^{ur}\right|^{2}$
is indeed linear in $\Omega$ at low frequencies, it is unclear why
the spectrum should obey Eq. (\ref{eq:WKB_no-horizon_ur-rate}) so
well. (This may require a more sophisticated phase-integral method,
mentioned in §\ref{sub:WKB-approximation}, in which various solutions
are linearly combined to give the required mode; the corresponding
norms might then exhibit linear combination of various phase integrals,
as suggested by Eq. (\ref{eq:WKB_no-horizon_ur-rate}).)  Note that
Eqs. (\ref{eq:WKB_no-horizon_ur-rate}) and (\ref{eq:WKB_no-horizon_gvh-rate})
together indicate that the $ur$-$u$ spectrum reaches a maximum at $\Omega_{c}$,
the boundary between the existence and non-existence of a group-velocity horizon.

\begin{figure}
\subfloat{\includegraphics[width=0.45\columnwidth]{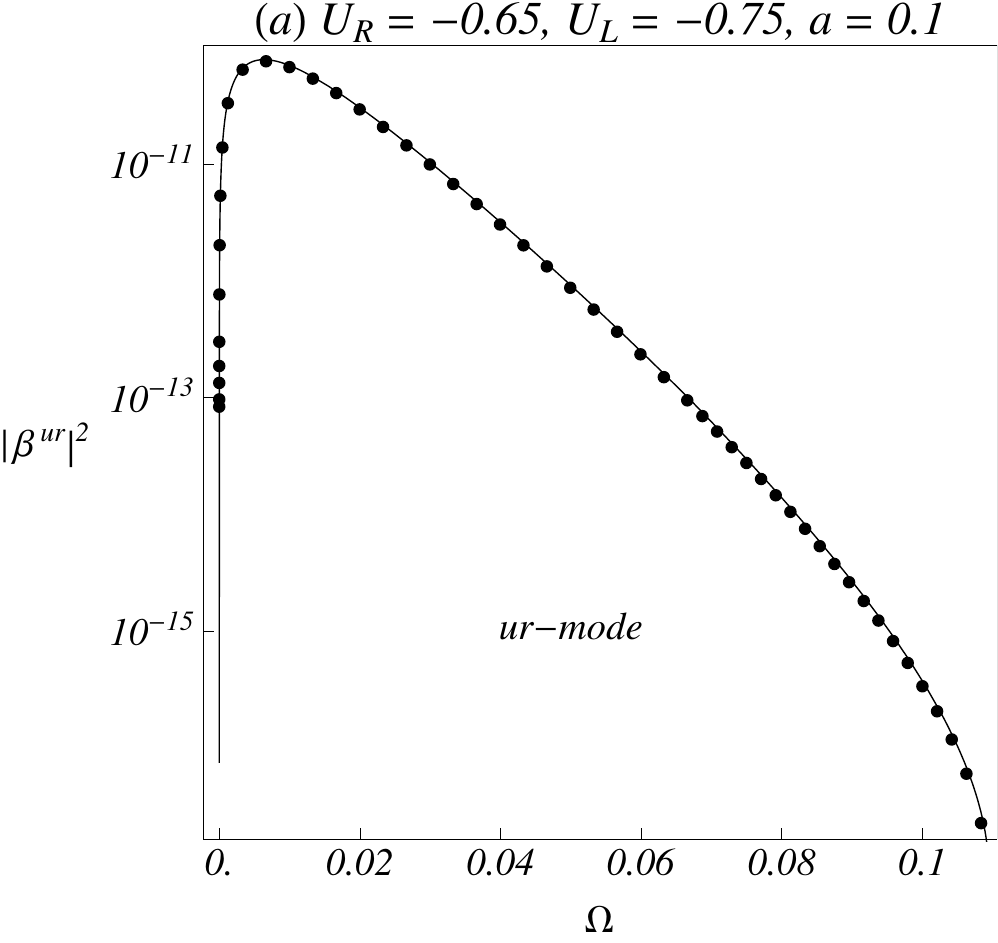}} \subfloat{\includegraphics[width=0.45\columnwidth]{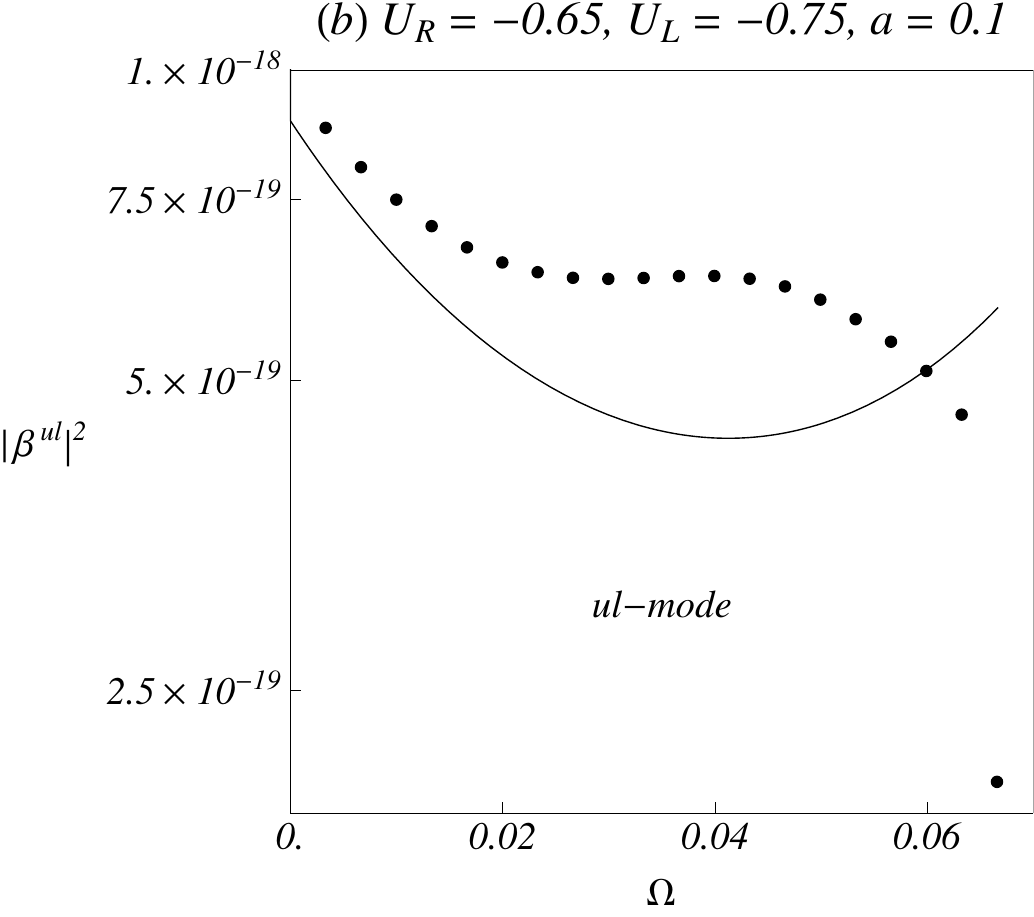}}

\subfloat{\includegraphics[width=0.45\columnwidth]{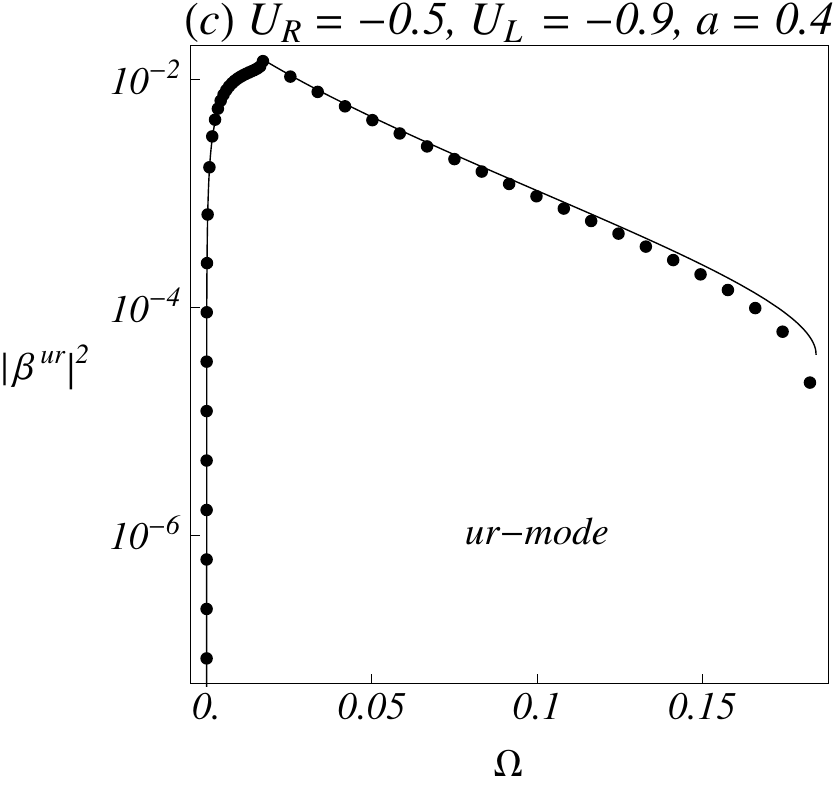}} \subfloat{\includegraphics[width=0.45\columnwidth]{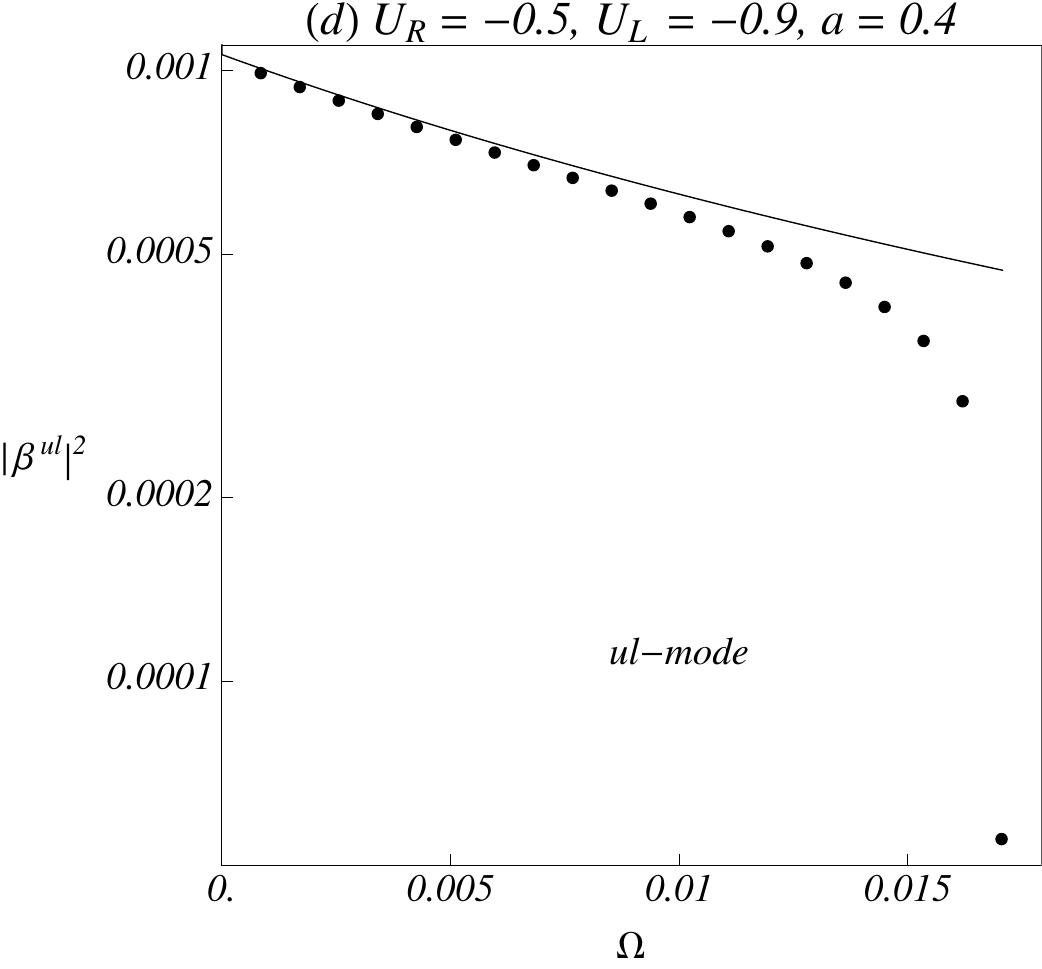}}

\caption[\textsc{Hawking spectra with no low-frequency horizon, 2}]{\textsc{Hawking spectra with no low-frequency horizon, 2}: Figures
$\left(a\right)$ and $\left(b\right)$ show the $ur$- and $ul$-mode
spectra for $a=0.1$, as in Fig. \ref{fig:Hawking_aco_no-horizon_WKB-good},
but with a decreased step height of $0.05$; similarly, Figures $\left(c\right)$
and $\left(d\right)$ show the $ur$- and $ul$-mode spectra for $U_{R}=-0.5$
and $U_{L}=-0.9$, as in Fig. \ref{fig:Hawking_aco_no-horizon_WKB-good},
but with an increased steepness of $a=0.4$.\label{fig:Hawking_aco_no-horizon_WKB-bad}}

\end{figure}

However, we cannot expect the phase-integral approximation to hold
universally, and indeed it does not. If the parameters are altered
in such a way that invalidates the phase-integral approximation in
the presence of a horizon - that is, if the steepness $a$ is increased
or the step height $h=\left(U_{R}-U_{L}\right)/2$ is decreased -
then the phase-integral approximation for the case of no low-frequency
horizon also becomes less accurate. Figure \ref{fig:Hawking_aco_no-horizon_WKB-bad}
shows such spectra: the upper plots have the same value of $a$ as
Fig. \ref{fig:Hawking_aco_no-horizon_WKB-good}, but the step height
has been decreased to $0.05$; the lower plots have the same asymptotic
velocities as Fig. \ref{fig:Hawking_aco_no-horizon_WKB-good}, but
$a$ has been increased to $0.4$. It is perhaps worthy of note that
the $ul$-mode spectrum deviates more from the phase-integral prediction
than does the $ur$-mode spectrum.

\section{Concluding remarks}

The question of whether Hawking radiation is robust with respect
to dispersive effects can be answered affirmatively. Hawking
radiation is still predicted once dispersion is introduced, though
the resulting spectra are no longer purely thermal. When we have a
low-frequency event horizon, the low-frequency end of the spectrum
is thermal, with a temperature given by Eq. (\ref{eq:modified_temp_general_U}).
This has been derived heuristically, though it agrees with Hawking's
original result when the hyperbolic tangent is approximately $1$.
At higher frequencies, the spectrum deviates from thermality, though
it is still well-approximated by the phase-integral prediction of
Eq. (\ref{eq:WKB_prediction}).

We have also found that, when dispersion is included, Hawking radiation
does not require the presence of a low-frequency event horizon, an observation
that does not seem to have been appreciated before. Though
$U$ might nowhere equal $-1$, a spectrum of spontaneously created
phonons is predicted, even for frequencies which do not experience
a group-velocity horizon. These spectra are non-thermal, though they
can be well-approximated by a modified phase-integral prediction,
as given in Eqs. (\ref{eq:WKB_no-horizon_gvh-rate})-(\ref{eq:WKB_no-horizon_ur-rate}).

This raises the question of the precise origins of the created quanta.
In the dispersionless case, which requires an event horizon for Hawking
radiation, one can imagine the creation of a virtual pair at the event
horizon, one on either side. The horizon prevents them from recombining,
so that they are rendered in reality. But when we have no event horizon
- when the created quanta may actually be emitted in the same direction,
like the $u$-$ul$ pair - how are they produced? We might say that
there is a \textit{phase}-velocity horizon, for the created quanta,
having oppositely signed (free-fall) frequencies, must have oppositely
directed phase velocities. This, however, is not a well-defined concept,
for the phase velocity $\omega/k$ can never become zero if $\omega$
is conserved. Moreover, if the existence of a phase-velocity horizon
truly determined the Hawking spectrum, we would expect that a greater
height in the velocity profile would increase the range of frequencies
for which a phase-velocity horizon occurs; yet, an increase in the
width of this part of the spectrum is not observed under an increase
in the step height. The origin of the radiation thus remains a mystery.

\pagebreak{}

\part{Hawking Radiation in Optical Fibres\label{par:Horizons-in-Optical-Fibres}}

\chapter{Nonlinear Fibre Optics\label{sec:Nonlinear-Fibre-Optics}}

The fluid model is difficult to implement in practice. The velocity
gradients and the Hawking temperatures they induce are typically so
small (around 10 nK \cite{Giovanazzi-et-al-2004}) that a fluid with temperature close to absolute zero must be
used. Thus, the proposed systems of Helium-3 \cite{Jacobson-Volovik-1998,Volovik-1999,Fischer-Volovik-2001,Volovik2001195,HeliumDroplet}
and Bose-Einstein condensate \cite{Garay-et-al-2000,Garay-et-al-2001,Barcelo-Liberati-Visser-2001-arXiv,Barcelo-Liberati-Visser-2001}
represent the most feasible means of observing spontaneous particle creation
in fluids, but require sophisticated technology to do so.

From a practical viewpoint, light in moving media provides an attractive
experimental setup for the observation of Hawking radiation. Light
it the simplest of quantum objects, exhibiting quantum properties
without the need for very low temperatures. Indeed, there is evidence
that Hawking radiation has already been observed in the optical arena
\cite{Faccio-2010}. In the remainder of this thesis, we shall examine
the simple one-dimensional system of light in optical fibres, a system
routinely used in the laboratory setting, looking particularly at
its capacity for event horizon analogies and the generation of Hawking
radiation.

This chapter examines some of the essential physics of light in optical
fibres. After reviewing the basic linear response in §\ref{sub:Linear-effects},
in §\ref{sub:Nonlinear-effects} we look at the lowest-order nonlinear
response and some of its effects, culminating in the interaction of
a soliton with a weak continuous probe wave. Finally, in §\ref{sub:Co-moving-frame},
we introduce a coordinate frame in which this system resembles wave
propagation in a dispersive moving medium, before concluding the chapter
in §\ref{sub:FIBRES-Conclusion-and-discussion}.

\section{Linear effects\label{sub:Linear-effects}}

Electromagnetic waves (such as light) are propagating disturbances
of electric field $\mathbf{E}$ and magnetic field $\mathbf{H}$,
which are governed by Maxwell's equations \cite{Jackson}:\begin{eqnarray}
\nabla\times\mathbf{E} & = & -\frac{\partial\mathbf{B}}{\partial t}\,,\label{eq:Maxwell_equation_1}\\
\nabla\times\mathbf{H} & = & \mathbf{J}+\frac{\partial\mathbf{D}}{\partial t}\,,\label{eq:Maxwell_equation_2}\\
\nabla\cdot\mathbf{D} & = & \rho_{f}\,,\label{eq:Maxwell_equation_3}\\
\nabla\cdot\mathbf{B} & = & 0\,,\label{eq:Maxwell_equation_4}\end{eqnarray}
where $\mathbf{J}$ is the free current density, $\rho_{f}$ is the
free charge density, and $\mathbf{B}$ and $\mathbf{D}$ are related
to $\mathbf{E}$ and $\mathbf{H}$ via the constitutive relations:\begin{eqnarray}
\mathbf{D} & = & \epsilon_{0}\mathbf{E}+\mathbf{P}\,,\label{eq:constitutive_relation_D}\\
\mathbf{B} & = & \mu_{0}\mathbf{H}+\mathbf{M}\,.\label{eq:constitutive_relation_B}\end{eqnarray}
$\mathbf{P}$ and $\mathbf{M}$ are the polarization and magnetization,
respectively. Restricting our attention to optical fibres \cite{Okoshi,Agrawal},
which are non-magnetic ($\mathbf{M}=\mathbf{0}$), non-conducting
($\mathbf{J}=\mathbf{0}$) and electrically neutral ($\rho_{f}=0$),
we can substitute the constitutive relations into Maxwell's equations
and eliminate $\mathbf{D}$ and $\mathbf{B}$, yielding\begin{eqnarray}
\nabla\times\mathbf{E} & = & -\mu_{0}\frac{\partial\mathbf{H}}{\partial t}\,,\label{eq:Maxwell_fibre_1}\\
\nabla\times\mathbf{H} & = & \epsilon_{0}\frac{\partial\mathbf{E}}{\partial t}+\frac{\partial\mathbf{P}}{\partial t}\,,\label{eq:Maxwell_fibre_2}\\
\nabla\cdot\mathbf{E} & = & -\frac{1}{\epsilon_{0}}\nabla\cdot\mathbf{P}\,,\label{eq:Maxwell_fibre_3}\\
\nabla\cdot\mathbf{H} & = & 0\,.\label{eq:Maxwell_fibre_4}\end{eqnarray}
$\mathbf{H}$ may be eliminated from these equations by taking the
curl of Eq. (\ref{eq:Maxwell_fibre_1}) and substituting Eq. (\ref{eq:Maxwell_fibre_2}),
resulting in the wave equation\begin{equation}
\nabla\times\left(\nabla\times\mathbf{E}\right)=-\mu_{0}\epsilon_{0}\frac{\partial^{2}\mathbf{E}}{\partial t^{2}}-\mu_{0}\frac{\partial^{2}\mathbf{P}}{\partial t^{2}}\,.\label{eq:general_wave_equation_for_E}\end{equation}
The product $\mu_{0}\epsilon_{0}=1/c^{2}$, where $c$ is the speed
of light in vacuum; for vacuum, which contains no bound charges, cannot
be polarized, and has $\mathbf{P}=\mathbf{0}$. Optical fibres, however,
are made of dielectric material, so that $\mathbf{P}$ is non-zero
and must be accounted for.

In order to solve Eq. (\ref{eq:general_wave_equation_for_E}), we
must determine the relationship between the polarization $\mathbf{P}$
and the electric field $\mathbf{E}$. Since $\mathbf{P}$ results
from the response of bound charges within the medium to the electric
field $\mathbf{E}$, and this response is unlikely to be instantaneous,
our first guess at this relationship might be \cite{Agrawal}
\begin{equation}
\mathbf{P}=\epsilon_{0}\int_{-\infty}^{t}\chi\left(t-t^{\prime}\right)\mathbf{E}\left(t^{\prime}\right)dt^{\prime}\,.\label{eq:linear_polarization}\end{equation}
This describes the polarization as a linear response to the electric
field, containing terms in only the first power of $\mathbf{E}$;
but, to incorporate the non-instantaneous nature of the response,
it is built up of contributions from $\mathbf{E}$ at all earlier
times. The weight of the contribution from $\mathbf{E}$ at time $t$
before the measurement of $\mathbf{P}$ is given by the response function,
$\chi\left(t\right)$. Substituting in Eq. (\ref{eq:general_wave_equation_for_E}),
the wave equation becomes\begin{equation}
\nabla\times\left(\nabla\times\mathbf{E}\right)+\frac{1}{c^{2}}\frac{\partial^{2}}{\partial t^{2}}\left(\mathbf{E}+\int_{-\infty}^{t}\chi\left(t-t^{\prime}\right)\mathbf{E}\left(t^{\prime}\right)dt^{\prime}\right)=0\,.\label{wave_equation_linear_polarization}\end{equation}
This is more readily solved by taking the Fourier transform with respect
to time \cite{Okoshi,Agrawal}, which gives\begin{equation}
\nabla\times\left(\nabla\times\widetilde{\mathbf{E}}\right)-\frac{\omega^{2}}{c^{2}}\left(1+\widetilde{\chi}\left(\omega\right)\right)\widetilde{\mathbf{E}}=0\,,\label{eq:Fourier_wave_eqn_linear_polarization}\end{equation}
where we have defined\begin{equation}
\mathbf{\widetilde{E}}\left(\omega,\mathbf{r}\right)=\int_{-\infty}^{+\infty}e^{i\omega t}\,\mathbf{E}\left(t,\mathbf{r}\right)dt\label{eq:Fourier_E_defn}\end{equation}
and\begin{equation}
\widetilde{\chi}\left(\omega\right)=\int_{-\infty}^{+\infty}e^{i\omega t}\,\chi\left(t\right)dt\,.\label{eq:susceptibility_defn}\end{equation}
In regions where $\widetilde{\chi}\left(\omega\right)$ is independent
of position, i.e., where the medium is homogeneous, Eq. (\ref{eq:Fourier_wave_eqn_linear_polarization})
can be simplified. For, taking the Fourier transform with respect
to time of Eq. (\ref{eq:Maxwell_fibre_3}), we find\begin{equation}
\nabla\cdot\left(\widetilde{\mathbf{E}}+\frac{1}{\epsilon_{0}}\widetilde{\mathbf{P}}\right)=\nabla\cdot\left(\left(1+\widetilde{\chi}\left(\omega\right)\right)\widetilde{\mathbf{E}}\right)=\left(1+\widetilde{\chi}\left(\omega\right)\right)\nabla\cdot\widetilde{\mathbf{E}}=0\,,\label{eq:divergence_of_Fourier_E_1}\end{equation}
which implies\begin{equation}
\nabla\cdot\mathbf{\widetilde{E}}=0\,.\label{eq:divergence_of_Fourier_E_2}\end{equation}
We also use the vector identity\begin{equation}
\nabla\times\left(\nabla\times\widetilde{\mathbf{E}}\right)=\nabla\left(\nabla\cdot\widetilde{\mathbf{E}}\right)-\nabla^{2}\widetilde{\mathbf{E}}=-\nabla^{2}\mathbf{\widetilde{E}}\,.\label{eq:curl_of_curl_of_E}\end{equation}
Substituting into Eq. (\ref{eq:Fourier_wave_eqn_linear_polarization}),
we find the Helmholtz equation:\begin{equation}
\nabla^{2}\widetilde{\mathbf{E}}+\frac{\omega^{2}}{c^{2}}\left(1+\widetilde{\chi}\left(\omega\right)\right)\widetilde{\mathbf{E}}=0\,.\label{eq:Helmholtz_eqn}\end{equation}

Eq. (\ref{eq:Helmholtz_eqn}) describes wave propagation in media.
We see that the effect of any particular medium on wave propagation
depends mainly on the \textit{linear susceptibility} $\widetilde{\chi}\left(\omega\right)$,
and is clearly frequency-dependent. We define the \textit{permittivity}
$\epsilon\left(\omega\right)=\left(n\left(\omega\right)+ic\alpha\left(\omega\right)/2\omega\right)^{2}\equiv1+\widetilde{\chi}\left(\omega\right)$;
if the imaginary part of the susceptibility is small, as is often
the case, then $\epsilon\left(\omega\right)$ is approximately equal
to $n^{2}\left(\omega\right)$. $n\left(\omega\right)$ is the \textit{refractive
index}, and Eq. (\ref{eq:Helmholtz_eqn}) can be written\begin{equation}
\nabla^{2}\widetilde{\mathbf{E}}+\frac{\omega^{2}}{c^{2}}n^{2}\left(\omega\right)\widetilde{\mathbf{E}}=0\,,\label{eq:Helmholtz_eqn_with_n}\end{equation}
whence it is clear that the refractive index describes a frequency-dependent
wave speed: assuming a monochromatic wave, and comparing Eq. (\ref{eq:Helmholtz_eqn_with_n})
with the vacuum case $n\left(\omega\right)=1$, we see that the vacuum
wave speed $c$ is simply replaced by $c/n\left(\omega\right)$, and
it admits plane wave solutions of the form $\exp\left[-i\omega\left(t-n\left(\omega\right)z/c\right)\right]$.
So the refractive index describes the dispersion of the fibre. Looking
retrospectively at Eqs. (\ref{eq:linear_polarization})-(\ref{eq:Fourier_wave_eqn_linear_polarization}),
we see that, in optical fibres, dispersion is a consequence of the
non-instantaneous linear response of the material polarization to
the electric field. It is therefore associated with a nonlocality
in \textit{time}; this should be compared to the acoustic case of
Chapter \ref{sec:The-Acoustic-Model}, where the dispersion was assumed
to arise from nonlocality in \textit{space}.

Confining light to an optical fibre introduces further dispersive
effects, for a fibre acts as a waveguide, forcing the waves to travel
in a single direction. Using cylindrical coordinates $\rho$, $\phi$
and $z$, Eq. (\ref{eq:Helmholtz_eqn_with_n}) becomes \cite{Okoshi,Agrawal}
\begin{equation}
\partial_{\rho}^{2}\widetilde{\mathbf{E}}+\frac{1}{\rho}\partial_{\rho}\widetilde{\mathbf{E}}+\frac{1}{\rho^{2}}\partial_{\phi}^{2}\widetilde{\mathbf{E}}+\partial_{z}^{2}\widetilde{\mathbf{E}}+\frac{\omega^{2}}{c^{2}}n^{2}\left(\omega\right)\widetilde{\mathbf{E}}=0\,.\label{eq:Helmholtz_eqn_cylindrical_coords}\end{equation}
Focusing for clarity on the component $\widetilde{E}_{z}$ and separating
the variables as follows,\begin{equation}
\widetilde{E}_{z}\left(\mathbf{r},\omega\right)=A\left(\omega\right)R\left(\rho\right)\exp\left(im\phi\right)\exp\left(i\beta z\right)\,,\label{eq:Ez_separation_of_variables}\end{equation}
we find that Eq. (\ref{eq:Helmholtz_eqn_cylindrical_coords}) is equivalent
to the following differential equation for $R\left(\rho\right)$:\begin{equation}
\partial_{\rho}^{2}R+\frac{1}{\rho}\partial_{\rho}R+\left(\frac{\omega^{2}}{c^{2}}n^{2}\left(\omega\right)-\beta^{2}-\frac{m^{2}}{\rho^{2}}\right)R=0\,.\label{eq:diff_eqn_for_R}\end{equation}
This is the equation for Bessel functions of order $m$ \cite{Abramowitz-Stegun}. In order
that the solution for the radial dependence should be free of singularities
within the fibre core and exponentially decreasing far from the fibre
core, only a certain number of discrete solutions are allowed, each
corresponding to a particular value of $\beta$. Under certain conditions
(small core radius and small refractive index difference between core
and cladding \cite{Agrawal}), only a single value of $\beta$ corresponds
to a given frequency.

For every mode, there are in fact two orthogonal solutions. (When
$\beta$ is single-valued, these correspond approximately to polarization
of the electric field along the $x$- or the $y$-axis \cite{Okoshi}.)
Depending on the transverse structure of the fibre, these two orthogonal
modes can have different values of $ $$\beta$, so that both frequency
\textit{and} polarization can affect wave speed. This phenomenon is
known as \textit{birefringence}. The two polarizations are often denoted
the \textit{fast} and \textit{slow }axes, depending on their relative
speed: hence the slow axis has a higher refractive index than the
fast axis.

Factoring out the transverse form of the guided modes (i.e., their
$\rho$- and $\phi$-dependence), the wave equation (\ref{eq:Helmholtz_eqn_with_n})
is reduced to the one-dimensional form \cite{Philbin-et-al}\begin{equation}
\left(\partial_{z}^{2}\widetilde{E}+\beta^{2}\left(\omega\right)\widetilde{E}\right)=0\,,\label{eq:linear_wave_equation}\end{equation}
and we may define a modified refractive index to satisfy\begin{equation}
\beta\left(\omega\right)=\frac{n^{\prime}\left(\omega\right)\omega}{c}\,.\label{eq:modified_refractive_index}\end{equation}
$n^{\prime}\left(\omega\right)$ differs from $n\left(\omega\right)$
in Eq. (\ref{eq:Helmholtz_eqn_with_n}) in that it includes the effects
of waveguide dispersion, arising from the restriction of light propagation
to a single spatial dimension. In what follows, we always assume a
one-dimensional optical fibre, and for convenience we shall simply
use $n\left(\omega\right)$ to mean the modified refractive index
defined in Eq. (\ref{eq:modified_refractive_index}).

Eq. (\ref{eq:linear_wave_equation}) may be transformed back to real
space via the substitutions $\widetilde{E}\rightarrow E$ and $\omega\rightarrow i\partial_{t}$.
At this point, however, it is convenient to return to Eq. (\ref{eq:general_wave_equation_for_E}),
and to recall that we have only included the linear polarization in
Eq. (\ref{eq:linear_wave_equation}). Reintroducing the nonlinear
part of the polarization $P_{\mathrm{}}^{\mathrm{NL}}$,
the full wave equation in real space becomes \cite{Philbin-et-al}\begin{equation}
c^{2}\left(\partial_{z}^{2}+\beta^{2}\left(i\partial_{t}\right)\right)E=\frac{1}{\epsilon_{0}}\partial_{t}^{2}P_{\mathrm{}}^{\mathrm{NL}}\,.\label{eq:full_wave_equation}\end{equation}
The operator $\beta^{2}\left(i\partial_{t}\right)$ is formed by inserting
the operator $i\partial_{t}$ into the Taylor expansion for $\beta^{2}\left(\omega\right)$,
analogously to the operator $F^{2}\left(-i\partial_{x}\right)$ in
Eq. (\ref{eq:acoustic_wave_equation}). When the polarization is assumed
linear and $P_{\mathrm{}}^{\mathrm{NL}}=0$, the general solution
is simply a linear superposition of plane waves with frequency and
propagation constant related by $\beta\left(\omega\right)$. These
plane waves evolve independently of each other. When $P_{\mathrm{}}^{\mathrm{NL}}$
is non-zero, the various plane waves may couple to each other, resulting
in frequency shifting and Hawking radiation \cite{Philbin-et-al}.

\section{Nonlinear effects\label{sub:Nonlinear-effects}}

We have seen how the linear polarization induced in an optical fibre
leads to dispersion and birefringence, which are frequency- and polarization-dependencies
of the phase velocity of plane waves; or, equivalently, of the propagation
constant $\beta$. Both of these effects are highly manipulable in
photonic crystal fibres \cite{Agrawal,Russell}, and can be made relatively
strong. We shall now consider nonlinear effects in fibres.

\subsection{Nonlinear polarization}

Although Eq. (\ref{eq:linear_polarization}) yields accurate predictions
of wave propagation if the optical power is low, a more general form
of the polarization can be written in the manner of a Taylor series,
as follows:\begin{equation}
\mathbf{P}=\mathbf{P}^{(1)}+\mathbf{P}^{(2)}+\mathbf{P}^{(3)}+\ldots\label{eq:total_polarization}\end{equation}
where $\mathbf{P}^{(n)}$ is of order $n$ in the electric field:\begin{equation}
P_{i}^{\left(n\right)}=\epsilon_{0}\int_{-\infty}^{t}dt_{1}\,\ldots\int_{-\infty}^{t}dt_{n}\;\chi_{ij_{1}j_{2}\ldots j_{n}}^{\left(n\right)}\left(t-t_{1},\ldots,t-t_{n}\right)\, E_{j_{1}}\left(t_{1}\right)\,\ldots\, E_{j_{n}}\left(t_{n}\right)\:.\label{eq:nth_order_polarization}\end{equation}
The $n^{\mathrm{th}}$-order susceptibility $\chi^{\left(n\right)}$
is a tensor of rank $n+1$, and we adopt the convention that repeated
indices ($j_{1}$, $j_{2}$, etc.) are summed over. Eq. (\ref{eq:linear_polarization})
is a special case of Eqs. (\ref{eq:total_polarization}) and (\ref{eq:nth_order_polarization})
in which it is assumed that only $\chi^{\left(1\right)}$ is non-zero,
and that $\chi^{\left(1\right)}\left(t\right)$ is simply a function
of time multiplied by the identity operator. This must be true in
an isotropic medium, so $\chi^{\left(1\right)}$ will continue to
have this form. The next-highest order susceptibility, $\chi^{\left(2\right)}$,
is found to be zero when the molecules of the dielectric are symmetric
\cite{Agrawal,Loudon}. This is true of silica; therefore, the lowest-order
nonlinear effects are derived from $\chi^{\left(3\right)}$ and the
third-order polarization. This is generated from the nonlinear response
of electronic vibrations in the fibre medium, and works on a timescale
comparable to the transit time of light over the diameter of an atom
\cite{Agrawal}. We can, therefore, assume this nonlinear response
to be instantaneous, and the time dependence of $\chi^{\left(3\right)}$
to be simply the product of three delta functions:\begin{equation}
\chi_{ij_{1}j_{2}j_{3}}^{\left(3\right)}\left(t-t_{1},t-t_{2},t-t_{3}\right)=\chi_{ij_{1}j_{2}j_{3}}^{\left(3\right)}\delta\left(t-t_{1}\right)\delta\left(t-t_{2}\right)\delta\left(t-t_{3}\right)\,.\label{eq:chi_3_instantaneous}\end{equation}

We have seen that there are two orthogonal polarizations of the electric
field, which we shall denote by $E_{\pm}$, where the subscripts $+$
and $-$ represent the fast and slow axes of the fibre, respectively.
There will also be two corresponding nonlinear polarizations $P_{\pm}$,
which, combining Eqs. (\ref{eq:nth_order_polarization}) and (\ref{eq:chi_3_instantaneous}),
are related to $E_{\pm}$ by\begin{equation}
P_{\pm}^{\mathrm{NL}}=\epsilon_{0}\chi_{\pm j_{1}j_{2}j_{3}}^{\left(3\right)}E_{j_{1}}E_{j_{2}}E_{j_{3}}\,,\label{eq:3rd_order_polarization_instantaneous}\end{equation}
where $j_{1}$, $j_{2}$ and $j_{3}$ are summed over $+$ and $-$.
Some physical considerations help to simplify Eq. (\ref{eq:3rd_order_polarization_instantaneous}).
Firstly, note that $P_{\pm}^{\mathrm{NL}}$ cannot depend on odd powers
of $E_{\mp}$: for, if it did, the direction of $P_{\pm}^{\mathrm{NL}}$
would be determined by the direction of $E_{\mp}$, in violation of
isotropy. Secondly, and also a consequence of isotropy, we have the
relation \cite{Agrawal}\begin{equation}
\chi_{\pm\pm\pm\pm}^{\left(3\right)}=\chi_{\pm\pm\mp\mp}^{\left(3\right)}+\chi_{\pm\mp\pm\mp}^{\left(3\right)}+\chi_{\pm\mp\mp\pm}^{\left(3\right)}\,.\label{eq:chi3_isotropy_relation}\end{equation}
Thus, Eq. (\ref{eq:3rd_order_polarization_instantaneous}) becomes\begin{equation}
P_{\pm}^{\mathrm{NL}}=\epsilon_{0}\chi_{\pm\pm\pm\pm}^{\left(3\right)}\left(E_{\pm}^{3}+E_{\mp}^{2}E_{\pm}\right)\,.\label{eq:real_3rd_order_polarization}\end{equation}
Simple though Eq. (\ref{eq:real_3rd_order_polarization}) may be,
it is useful to modify it further. So far, we have taken $E$ and
$P^{\mathrm{NL}}$ to be real quantities; however, given their oscillatory
nature, it is easier for the sake of calculation to replace them with
complex quantities by making the substitutions\begin{eqnarray}
P_{\pm}^{\mathrm{NL}} & \rightarrow & \frac{1}{2}\left(P_{\pm}^{\mathrm{NL}}+P_{\pm}^{\mathrm{NL}\star}\right)\,,\label{eq:P_real_to_complex}\\
E_{\pm} & \rightarrow & \frac{1}{2}\left(E_{\pm}+E_{\pm}^{\star}\right)\,.\label{eq:E_real_to_complex}\end{eqnarray}
Then Eq. (\ref{eq:real_3rd_order_polarization}) becomes\begin{equation}
P_{\pm}^{\mathrm{NL}}=\frac{1}{4}\epsilon_{0}\chi_{\pm\pm\pm\pm}^{\left(3\right)}\left(E_{\pm}^{3}+3\, E_{\pm}^{2}E_{\pm}^{\star}+E_{\mp}^{2}E_{\pm}+E_{\mp}^{2}E_{\pm}^{\star}+2\, E_{\pm}E_{\mp}E_{\mp}^{\star}\right)\,.\label{eq:real_polarization_to_complex}\end{equation}
Finally, we note that, in general, only that component of the polarization
$\mathbf{P}^{\mathrm{NL}}$ which has a similar frequency to the electric
field $\mathbf{E}$ will have a significant effect on wave propagation.
(Different frequency components require phase matching, which usually
occurs only by design.) Taking $E_{\pm}$ to have a positive frequency,
only those terms of $P_{\pm}^{\mathrm{NL}}$ which include a complex
conjugate can satisfy the frequency-matching condition. Finally, then,
Eq. (\ref{eq:real_polarization_to_complex}) becomes \cite{Philbin-et-al}
\begin{equation}
P_{\pm}^{\mathrm{NL}}=\frac{3}{4}\epsilon_{0}\chi_{\pm\pm\pm\pm}^{\left(3\right)}\left(\left|E_{\pm}\right|^{2}E_{\pm}+\frac{2}{3}\left|E_{\mp}\right|^{2}E_{\pm}+\frac{1}{3}E_{\mp}^{2}E_{\pm}^{\star}\right)\,.\label{eq:complex_3rd_order_polarization}\end{equation}

Although Eq. (\ref{eq:complex_3rd_order_polarization}) is wholly
derived from the assumption of a third-order polarization, its various
terms are associated with different qualitative effects. The first
and second terms describe, respectively, the phenomena of \textit{self-phase
modulation} and \textit{cross-phase modulation}; that is, they lead
to phase shifts in the electric field as a result of either the nonlinear
effect of a polarized field on itself, or of one polarized field on
the orthogonally polarized field \cite{Agrawal}. The third term describes \textit{four-wave
mixing} between the two polarizations \cite{Agrawal}.

The self-phase modulation term can be decomposed further. If the electric
field is composed of two distinct frequency bands, it is useful to
separate these components by writing\begin{equation}
E_{\pm}=E_{1}e^{i\beta_{1}z-i\omega_{1}t}+E_{2}e^{i\beta_{2}z-i\omega_{2}t}\,,\label{eq:separation_of_frequencies}\end{equation}
where $E_{1}$ and $E_{2}$ are assumed to be slowly-varying compared
with the plane waves they multiply. Then the first term of Eq. (\ref{eq:complex_3rd_order_polarization})
becomes\begin{equation}
\left|E_{\pm}\right|^{2}E_{\pm}=\left(\left|E_{1}\right|^{2}+2\left|E_{2}\right|^{2}\right)E_{1}e^{i\beta_{1}z-i\omega_{1}t}+\left(\left|E_{2}\right|^{2}+2\left|E_{1}\right|^{2}\right)E_{2}e^{i\beta_{2}z-i\omega_{2}t}\,,\label{eq:SPM_XPM_across_frequencies}\end{equation}
where we have neglected all terms that do not oscillate at $\omega_{1}$
or $\omega_{2}$ (and therefore require phase matching to have a significant
effect). Thus, we see that, within a single polarization, we can separate
the nonlinear effects of a certain frequency band on itself and on
other frequency bands; these effects are also commonly referred to
as self- and cross-phase modulation, where the descriptive terms are
now in reference to the frequency rather than the direction of polarization.

\subsection{Optical solitons}

One of the most notable consequences of third-order polarization is
the existence of stable localised solutions of the wave equation;
that is, pulses whose shapes are not distorted during propagation
through the fibre \cite{Agrawal}. In these specific cases, the nonlinear effects
contrive to cancel the linear (dispersive) effects, which acting alone
would tend to stretch and smear the pulse. The possibility of stable
pulses is useful in forming an analogy with a moving fluid: in the
frame co-moving with a stable pulse (introduced in §\ref{sub:Co-moving-frame}),
the background is time-independent.

Let us derive these stable pulse solutions. Firstly, since we are
considering a single electric field with well-defined frequency, the
nonlinear polarization includes only the self-phase modulation term;
that is, $P^{\mathrm{NL}}=\frac{3}{4}\epsilon_{0}\chi^{\left(3\right)}\left|E\right|^{2}E$.
(The subscripts $\pm$ have been emitted for clarity, though it should
be remembered that $\chi^{\left(3\right)}$ may be polarization-dependent.)
Secondly, we assume the validity of the \textit{slowly-varying envelope
approximation}, i.e., we write the total electric field as the product
of an envelope and a carrier wave,\begin{equation}
E\left(z,t\right)=\mathcal{E}\left(z,t\right)\exp\left(i\beta_{s}z-i\omega_{s}t\right)\,,\label{eq:envelope_and_carrier}\end{equation}
(where $\beta_{s}\equiv\beta\left(\omega_{s}\right)$) and assume
that the envelope varies on scales much longer than an optical cycle.
Specifically, this amounts to the assumption of the validity of the
following relations:\begin{alignat}{1}
\left|\partial_{z}^{2}\mathcal{E}\right|\ll\left|\beta_{s}\partial_{z}\mathcal{E}\right|\,,\qquad & \left|\partial_{t}^{2}\mathcal{E}\right|\ll\left|\omega_{s}\partial_{t}\mathcal{E}\right|\,.\label{eq:SVEA_conditions}\end{alignat}
Under the slowly-varying envelope approximation, the two sides of
the wave equation (\ref{eq:full_wave_equation}) can be well-approximated
by terms containing only first-order derivatives:\begin{eqnarray}
\!\!\!\!\!\!\!\!\!\!\!\!\!\!c^{2}\left(\partial_{z}^{2}+\beta^{2}\left(i\partial_{t}\right)\right)E & \approx & \exp\left(i\beta_{s}z-i\omega_{s}t\right)2\beta_{s}c^{2}\left(i\partial_{z}+\beta-\beta_{s}\right)\mathcal{E}\,,\label{eq:SVEA_SPM_wave_eqn_LHS}\\
\frac{1}{\epsilon_{0}}\partial_{t}^{2}P^{\mathrm{NL}} & \approx & -\exp\left(i\beta_{s}z-i\omega_{s}t\right)\frac{3}{4}i\chi^{\left(3\right)}\omega_{s}\mathcal{E}\left(-i\omega_{s}\left|\mathcal{E}\right|^{2}+3\partial_{t}\left|\mathcal{E}\right|^{2}\right)\,.\label{eq:SVEA_SPM_wave_eqn_RHS}\end{eqnarray}
In Eq. (\ref{eq:SVEA_SPM_wave_eqn_LHS}), when the operator $\beta$
acts on the envelope $\mathcal{E}$ rather than the full electric
field $E$, it should be remembered that it is the difference frequency
$\omega-\omega_{s}$ which is replaced by the operator $i\partial_{t}$,
and not $\omega$ itself. Forming a second-order Taylor expansion
of $\beta$ at the pulse frequency $\omega_{s}$,\begin{equation}
\beta\approx\beta_{s}+\beta_{1,s}\left(\omega-\omega_{s}\right)+\frac{1}{2}\beta_{2,s}\left(\omega-\omega_{s}\right)^{2}\,,\label{eq:2nd_order_expansion_of_beta}\end{equation}
we can write Eq. (\ref{eq:SVEA_SPM_wave_eqn_LHS}) in the form\begin{equation}
c^{2}\left(\partial_{z}^{2}+\beta^{2}\left(i\partial_{t}\right)\right)E\approx\exp\left(i\beta_{s}z-i\omega_{s}t\right)2\beta_{s}c^{2}\left(i\partial_{z}+i\beta_{1,s}\partial_{t}-\frac{1}{2}\beta_{2,s}\partial_{t}^{2}\right)\mathcal{E}\,.\label{eq:SVEA_SPM_wave_eqn_LHS_beta_expanded}\end{equation}
In Eq. (\ref{eq:SVEA_SPM_wave_eqn_RHS}), we assume that the intensity
$\left|\mathcal{E}\right|^{2}$ also varies slowly in comparison with
the frequency $\omega_{s}$, so that we can neglect the derivative
term. Making these approximations and setting the two sides of the
wave equation equal, we obtain\begin{equation}
i\left(\partial_{t}+\frac{1}{\beta_{1,s}}\partial_{z}\right)\mathcal{E}-\frac{\beta_{2,s}}{2\beta_{1,s}}\partial_{t}^{2}\mathcal{E}+\frac{3}{8}\chi^{\left(3\right)}\frac{\omega_{s}^{2}}{c^{2}\beta_{s}\beta_{1,s}}\left|\mathcal{E}\right|^{2}\mathcal{E}=0\,.\label{eq:NLS_eqn_z_and_t}\end{equation}
The combination of first-order derivatives in Eq. (\ref{eq:NLS_eqn_z_and_t})
strongly suggests the usefulness of the following coordinate transformation:\begin{eqnarray}
z^{\prime} & = & z\,,\label{eq:comoving_frame_z_prime_defn}\\
\tau & = & t-\beta_{1,s}z\,.\label{eq:comoving_frame_tau_defn}\end{eqnarray}
Since $\beta_{1,s}$ is simply the inverse of the group velocity of
the pulse, this is a transformation into a frame of reference that
moves with the pulse; we shall have more to say about it later. The
partial derivatives of the various coordinates are related as follows:\begin{eqnarray}
\partial_{z} & = & \partial_{z^{\prime}}-\beta_{1,s}\partial_{\tau}\,,\label{eq:z_derivative}\\
\partial_{t} & = & \partial_{\tau}\,,\label{eq:t_derivative}\end{eqnarray}
and substituting these in Eq. (\ref{eq:NLS_eqn_z_and_t}) yields\begin{equation}
i\partial_{z^{\prime}}\mathcal{E}-\frac{1}{2}\beta_{2,s}\partial_{\tau}^{2}\mathcal{E}+\frac{3\chi^{\left(3\right)}\omega_{s}^{2}}{8c^{2}\beta_{s}}\left|\mathcal{E}\right|^{2}\mathcal{E}=0\,.\label{eq:NLS_eqn_z_prime_and_tau}\end{equation}
Eq. (\ref{eq:NLS_eqn_z_prime_and_tau}) is very similar in form to
the Schrödinger equation; the only mathematical difference is that the {}``potential''
depends on the squared magnitude of the {}``wavefunction'' $\mathcal{E}$.
(Of course, there is also the physical difference that space and time coordinates
have been interchanged, a consequence of the transformation (\ref{eq:comoving_frame_z_prime_defn})-(\ref{eq:comoving_frame_tau_defn}).
We shall return to this point in \S\ref{sub:Co-moving-frame}.)
For this reason, it is often referred to as the \textit{nonlinear
}Schrödinger equation. It describes the propagation of a pulse, taking
into account dispersive effects via the $\partial_{\tau}^{2}\mathcal{E}$
term, and nonlinear effects via the $\left|\mathcal{E}\right|^{2}\mathcal{E}$
term. As has been stated, one of its most important properties is
its admittance of stable solutions whose shapes do not change during
their propagation, thus achieving a perfect balance between the dispersive
and nonlinear effects. These solutions are only possible when $\beta_{2,s}<0$,
i.e., when the group velocity dispersion is anomalous. This makes
intuitive sense, for the nonlinearity cannot be negative, and so one
sign of the dispersion should help to increase the nonlinear effects
while the other should contrive to cancel it out.

The stable solution is \cite{Agrawal,Zakharov-Shabat-1972}
\begin{eqnarray}
\mathcal{E} & = & \mathcal{E}_{0}\,\mathrm{sech}\left(\frac{\tau}{T_{s}}\right)\exp\left(i\frac{\left|\beta_{2,s}\right|z^{\prime}}{2T_{s}^{2}}\right)\label{eq:soliton_z_prime_and_tau}\\
 & = & \mathcal{E}_{0}\,\mathrm{sech}\left(\frac{t-\beta_{1,s}z}{T_{s}}\right)\exp\left(i\frac{\left|\beta_{2,s}\right|z}{2T_{s}^{2}}\right)\label{eq:soliton_z_and_t}\end{eqnarray}
where\begin{equation}
\mathcal{E}_{0}^{2}=\frac{8c^{2}\beta_{s}\left|\beta_{2,s}\right|}{3\chi^{\left(3\right)}\omega_{s}^{2}T_{s}^{2}}\,.\label{eq:intensity_of_soliton}\end{equation}
A pulse of the form of Eq. (\ref{eq:soliton_z_and_t}) is called a
\textit{fundamental soliton} \cite{Agrawal}. (Higher-order solitons
exist, with shapes which are not constant but periodic in $z$ \cite{Agrawal,Zakharov-Shabat-1972}.
These, however, are not relevant to our discussions.) Note that, at
a given frequency $\omega_{s}$ (which must obey the condition $\beta_{2,s}=\partial^{2}\beta/\partial\omega^{2}\left(\omega_{s}\right)<0$),
the fundamental solitons form a one-parameter family of solutions
of the nonlinear Schrödinger equation. The parameter that characterises
them is $T_{s}$, the {}``width'' of the soliton. (For the intensity
profile $\left|\mathcal{E}\right|^{2}$, the full-width half-maximum
is approximately $1.76\: T_{s}$). Note that, as a consequence of
its nonlinear nature, the peak intensity of the soliton is entirely
determined by $T_{s}$, and becomes larger as $T_{s}$ is made smaller.
Thus, solitons vary between long, weak pulses and short, intense pulses.

\begin{figure}
\includegraphics[width=0.8\columnwidth]{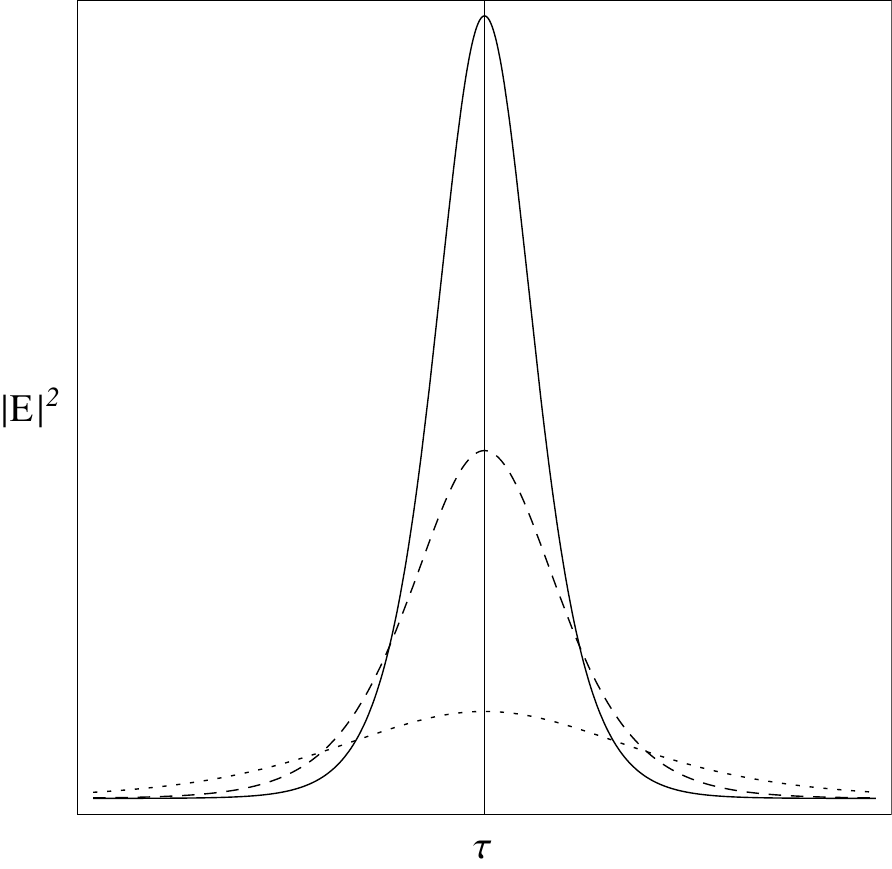}

\caption[\textsc{Optical solitons}]{\textsc{Optical solitons}: For a frequency at which $\partial^{2}\beta/\partial\omega^{2}<0$,
the dispersive and nonlinear effects of wave propagation contrive
to cancel each other out, and stable solutions called solitons are
possible. Here, the optical intensity of several solitons is plotted;
and, in accordance with Eq. (\ref{eq:intensity_of_soliton}), the
peak power is proportional to $1/T_{s}^{2}$, where $T_{s}$ is the
soliton width.\label{fig:Optical-Solitons}}

\end{figure}

\subsection{Pulse and probe}

For our purposes, we shall mostly be interested in cross-phase modulation.
We shall consider two fields: one an optical pulse, which we usually
take to be a fundamental soliton; the other a continuous-wave probe,
whose intensity is much less than that of the pulse. These are well-separated
in frequency, and we denote their central frequencies by $\omega_{s}$
and $\omega_{p}$, respectively. The nonlinear effects acting between
these two fields will be essentially unidirectional, with the pulse
acting on the probe, since the low intensity of the latter means that,
to a very good approximation, we can neglect any back-reaction on
the pulse. From Eqs. (\ref{eq:complex_3rd_order_polarization}) and
(\ref{eq:SPM_XPM_across_frequencies}), we find that the appropriate
nonlinear polarization is\begin{equation}
P^{\mathrm{NL}}=\frac{r}{2}\epsilon_{0}\chi^{\left(3\right)}\left|E_{s}\right|^{2}E_{p}\,,\label{eq:pulse_probe_polarization}\end{equation}
where\begin{equation}
r=\begin{cases}
3 & \textrm{fields similarly polarized}\\
1 & \textrm{fields orthogonally polarized}\end{cases}\,.\label{eq:r_defn}\end{equation}
This nonlinear polarization is in the same direction as the probe
field $E_{p}$. Substituting in the wave equation (\ref{eq:full_wave_equation}),
we find\begin{equation}
c^{2}\left[\partial_{z}^{2}+\beta^{2}\left(i\partial_{t}\right)\right]E_{p}=\partial_{t}^{2}\left(\chi E_{p}\right)\,,\label{eq:wave_eqn_for_probe}\end{equation}
where the nonlinear susceptibility $\chi$ induced by the pulse is
given by\begin{equation}
\chi=\frac{r}{2}\chi^{\left(3\right)}\left|E_{s}\right|^{2}\,.\label{eq:nonlinearity_of_pulse}\end{equation}
Since the propagation constant $\beta=\omega n\left(\omega\right)/c$,
Eq. (\ref{eq:wave_eqn_for_probe}) can be written in the form\begin{equation}
c^{2}\partial_{z}^{2}E_{p}-\partial_{t}^{2}n^{2}\left(i\partial_{t}\right)E_{p}-\partial_{t}^{2}\left(\chi E_{p}\right)=0\,,\label{eq:wave_eqn_for_probe_with_n}\end{equation}
whence we see that the influence of $\chi$ on the probe is as a local
change of refractive index,\begin{equation}
n_{\mathrm{eff}}^{2}=n^{2}+\chi\,.\label{eq:effective_n_including_chi}\end{equation}
This increase in refractive index is often referred to as the \textit{Kerr
effect}.

Let us briefly consider the implications of Eq. (\ref{eq:effective_n_including_chi}).
The behaviour of waves is usually described by a refractive index
profile, and where we restrict our attention to waves of relatively
low intensity, the refractive index, while it depends on frequency,
is constant along the fibre. Thus, a certain frequency will propagate
with the same phase and group velocity at all times, and the complexity
of propagation of a wavepacket is due to the fact that the phase and
group velocities differ for different frequencies. This is the dispersion
phenomenon, which is entirely described by the linear polarization,
all higher-order polarizations being negligible.

With the introduction of a relatively intense pulse and the third-order
polarization induced by it, a new phenomenon comes to light: while
we can still describe the behaviour of waves by a refractive index
profile, this profile now obtains a dependence on position (and time),
the change being proportional to the intensity of the pulse and thus
propagating along the fibre as the pulse does. (Recall that we have
assumed an instantaneous third-order polarization; in reality the
refractive index change will lag very slightly behind the pulse, but
in most circumstances we can neglect this lag.) If a probe wave also
propagates down the fibre, it will experience a transient increase
in refractive index and, hence, a decrease in its phase velocity.
Dispersion will also cause a small shift in the frequency of the wave.
Usually, this transient effect will be almost negligible; but, if
either the phase or group velocity of the probe is initially very
close to the group velocity of the pulse, so that their relative speed
is very small, the effect of the pulse may be very significant.

\section{Co-moving frame\label{sub:Co-moving-frame}}

\subsection{Coordinate transformation}

\begin{figure}
\subfloat{\includegraphics[width=0.45\columnwidth]{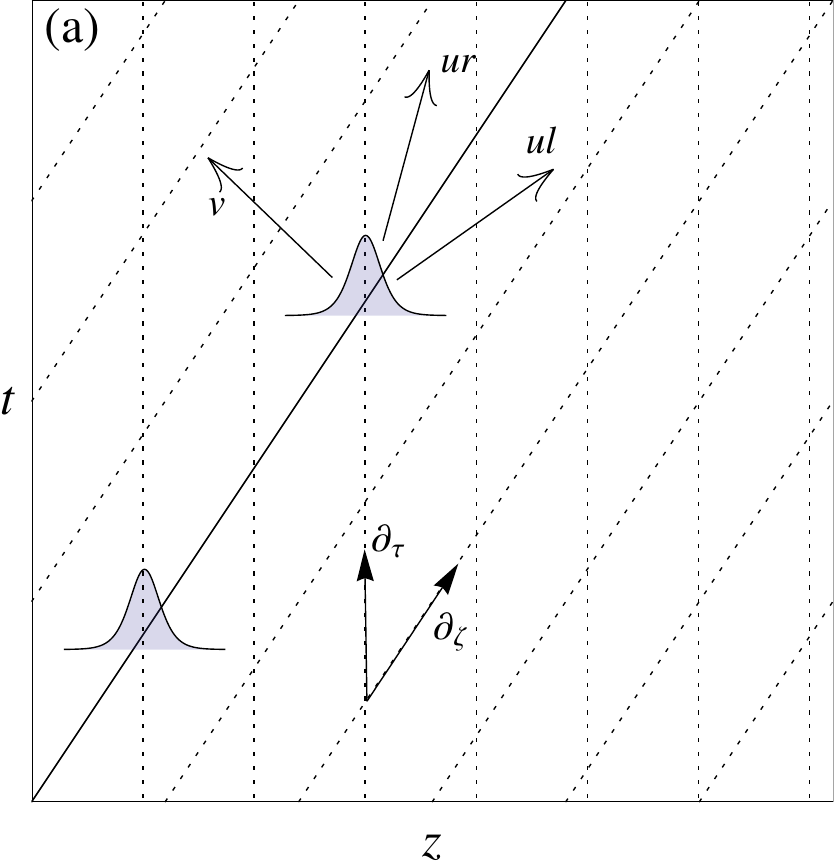}}\subfloat{\includegraphics[width=0.45\columnwidth]{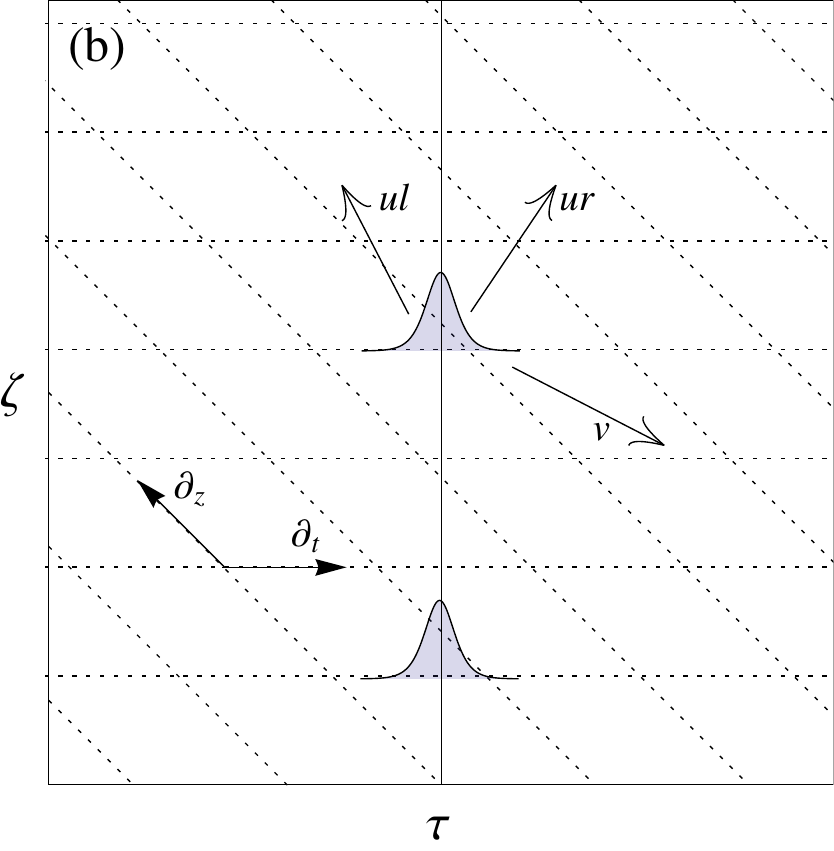}}

\caption[\textsc{Co-moving frame}]{\textsc{Co-moving frame}: The system of a pulse in an optical fibre, as seen from the lab frame in Figure $(a)$, and from the co-moving frame in Figure $(b)$; these are related via the coordinate transformation in Eqs. (\ref{eq:comoving_frame_z_prime_defn}) and (\ref{eq:comoving_frame_tau_defn}). Note the behaviour of waves in the two frames: the co-propagating $u$-modes are split between those moving faster or slower than the pulse, and these have different directions in the co-moving frame; the counter-propagating $v$-modes, travelling towards negative $z$, appear to move backwards in \textit{time} in the co-moving frame. Also note the apparent flip between left and right: in the co-moving frame, points with a negative value of $\tau$ (retarded time) are \textit{further along the fibre} than the pulse.\label{fig:co-moving-frame}}

\end{figure}

Let us examine further the setup of an intense pulse propagating in
an optical fibre, and its effect on weak probe waves. This is most
easily done in a frame of reference in which the pulse is stationary.
Such a frame is constructed via the following coordinate transformation:\begin{eqnarray}
\zeta & = & \frac{z}{u}\,,\label{eq:transformation_zeta}\\
\tau & = & t-\frac{z}{u}\,,\label{eq:transformation_tau}\end{eqnarray}
where $u$ is the group velocity of the pulse; this transformation is illustrated in Figure \ref{fig:co-moving-frame}. Note that this is essentially
the same as the transformation made in Eqs. (\ref{eq:comoving_frame_z_prime_defn})
and (\ref{eq:comoving_frame_tau_defn}), the only difference being
that, instead of keeping the propagation distance, we have now normalised
it with respect to the velocity to get the propagation time. It is
very similar to a Galilean transformation, differing in that the roles
of space and time have been reversed. (This causes the counter-propagating $v$-modes to appear to travel backwards in time, as shown in Fig. \ref{fig:co-moving-frame}.) It is clear that the pulse remains
centered at $\tau=0$ and is independent of $\zeta$, so that the
nonlinear susceptibility is simply a function of $\tau$: $\chi=\chi\left(\tau\right)$.
Note that the transformation defined by Eqs. (\ref{eq:transformation_zeta})
and (\ref{eq:transformation_tau}) is not a Lorentz transformation,
an inequality made more pronounced by the fact that $u$, the speed
of a pulse in an optical fibre, is comparable with $c$, the speed
of light in vacuum. The coordinates $\zeta$ and $\tau$, then, do
not correspond to space and time coordinates in the co-moving frame.
They do, however, define a perfectly valid (non-inertial) coordinate
system; we can solve the wave equation in this coordinate system and
invert the transformation to find the solution in the original laboratory
frame.

The partial derivatives in the co-moving and the lab frame are related
via\begin{eqnarray}
\partial_{z} & = & \frac{1}{u}\left(\partial_{\zeta}-\partial_{\tau}\right)\,,\label{eq:transformation_derivative_z}\\
\partial_{t} & = & \partial_{\tau}\,.\label{eq:transformation_derivative_t}\end{eqnarray}
Substituting into the wave equation (\ref{eq:wave_eqn_for_probe})
for the probe/pulse system, we find\begin{equation}
\left(\partial_{\zeta}-\partial_{\tau}\right)^{2}E_{p}+u^{2}\beta^{2}\left(i\partial_{\tau}\right)E_{p}-\frac{u^{2}}{c^{2}}\partial_{\tau}^{2}\left(\chi E_{p}\right)=0\,.\label{eq:wave_eqn_for_probe_comoving_frame}\end{equation}
This is the desired wave equation for weak probe waves in the presence
of an intense pulse. It is now clear why the roles of space and time
have been switched from the usual Galilean transformation in Eqs.
(\ref{eq:transformation_zeta}) and (\ref{eq:transformation_tau}):
this leaves the time derivative unchanged, and hence the operator
$\beta^{2}\left(i\partial_{t}\right)$ maintains its simple form.
(This should be compared with Eq. (\ref{eq:acoustic_wave_equation}),
where the dispersion operator contains derivatives with respect to
space rather than time.)

It is more useful to express the optical wave equation in terms of
the vector potential $A_{p}$, which is related to the electric and
magnetic fields via\begin{alignat}{1}
E_{p}=-\partial_{t}A_{p}\,,\qquad & B_{p}=\partial_{z}A_{p}\,,\label{eq:vector_potential_defn}\end{alignat}
for then Eq. (\ref{eq:wave_eqn_for_probe_comoving_frame}) becomes\begin{equation}
\left(\partial_{\zeta}-\partial_{\tau}\right)^{2}A_{p}+u^{2}\beta^{2}\left(i\partial_{\tau}\right)A_{p}-\frac{u^{2}}{c^{2}}\partial_{\tau}\left(\chi\partial_{\tau}A_{p}\right)=0\,.\label{eq:wave_eqn_for_vector_potential}\end{equation}
Although this difference may seem trivial, Eq. (\ref{eq:wave_eqn_for_vector_potential})
is preferable to Eq. (\ref{eq:wave_eqn_for_probe_comoving_frame})
because it can be derived from a relatively simple Lagrangian, analogous
to that of Eq. (\ref{eq:Lagrangian_with_dispersion}) for the acoustic
model. We defer discussion of this until §\ref{sub:FIBRES-Action-and-scalar-product}.

\subsection{Dispersion relation}

The wave equation is generally solved by decomposing a solution into
its plane wave components. It is useful, therefore, to know the dispersion
relation in the co-moving frame, to see how these plane waves behave.
First, we must define appropriate variables. In the usual laboratory
frame, in which we have coordinates $\left(z,t\right)$, we also have
wavenumber $k$ and frequency $\omega$. These are defined as derivatives
of the phase: we assume a solution of the form

\begin{equation}
A=\mathcal{A}\:\exp\left(i\varphi\right)\label{eq:amplitude_and_phase}\end{equation}
where $\mathcal{A}$ and $\varphi$ are real quantities. $k$ and
$\omega$ are then defined as follows:\begin{alignat}{1}
k=\partial_{z}\varphi\,,\qquad & \omega=-\partial_{t}\varphi\,.\label{eq:k_and_omega_defns}\end{alignat}
In the co-moving frame defined by the coordinate transformation Eqs.
(\ref{eq:transformation_zeta}) and (\ref{eq:transformation_tau}),
we have coordinates $\left(\tau,\zeta\right)$, both of which have
dimensions of time. We define two frequencies, $\omega$ and $\omega^{\prime}$,
as follows:\begin{alignat}{1}
\omega=-\partial_{\tau}\varphi\,,\qquad & \omega^{\prime}=-\partial_{\zeta}\varphi\,.\label{eq:omega_and_omega_prime_defns}\end{alignat}
The definitions of Eqs. (\ref{eq:k_and_omega_defns}) and (\ref{eq:omega_and_omega_prime_defns})
are related via the transformation of the partial derivatives, Eqs.
(\ref{eq:transformation_derivative_z}) and (\ref{eq:transformation_derivative_t}).
Eq. (\ref{eq:transformation_derivative_t}) shows that the two definitions
of $\omega$ are equivalent, hence the use of the same variable. Eq.
(\ref{eq:transformation_derivative_z}) yields the relation\begin{alignat}{1}
k=\frac{1}{u}\left(\omega-\omega^{\prime}\right)\,,\qquad & \mathrm{or}\quad\omega^{\prime}=\omega-uk\,.\label{eq:Doppler_formula_FIBRES}\end{alignat}
$\omega^{\prime}$, then, is simply the Doppler-shifted frequency
measured in the co-moving frame. For this reason, we shall refer to
it as the \textit{co-moving} frequency; in contrast, $\omega$ will
be referred to as the \textit{lab} frequency.

We saw in Eq. (\ref{eq:effective_n_including_chi}) that the nonlinearity
$\chi$ behaves as an increase in the square of the refractive index:
$n_{\mathrm{eff}}^{2}=n^{2}+\chi$. Assuming $\chi$ is constant in
both space and time, this can be incorporated into the dispersion
relation. Defining\begin{equation}
\beta_{\chi}\left(\omega\right)=\frac{n_{\mathrm{eff}}\left(\omega\right)\omega}{c}\approx\frac{\omega}{c}\left(n\left(\omega\right)+\frac{\chi}{2n\left(\omega\right)}\right)\,,\label{eq:beta_with_chi}\end{equation}
where the second step is valid if $\chi\ll n^{2}\left(\omega\right)$,
then we have\begin{equation}
k=\frac{1}{u}\left(\omega-\omega^{\prime}\right)=\pm\beta_{\chi}\left(\omega\right)\,,\label{eq:dispersion_with_chi-1}\end{equation}
or, rearranging,\begin{eqnarray}
\omega^{\prime} & = & \omega\mp u\beta_{\chi}\left(\omega\right)\nonumber \\
 & = & \omega\left(1\mp\frac{u}{c}n_{\mathrm{eff}}\left(\omega\right)\right)\nonumber \\
 & \approx & \omega\left(1\mp\frac{u}{c}\left(n\left(\omega\right)+\frac{\chi}{2n\left(\omega\right)}\right)\right)\,.\label{eq:dispersion_with_chi-2}\end{eqnarray}
This is the dispersion relation in the co-moving frame. The effect
of the nonlinearity is to alter the form of the dispersion profile
$\beta_{\chi}\left(\omega\right)$; it does not affect the velocity
$u$, which is assumed constant. There are two branches of the dispersion
relation, corresponding to probe waves which propagate either with
(\textit{co-propagating}, or $u$-modes) or against (\textit{counter-propagating},
or $v$-modes) the pulse. We shall deal mainly with co-propagating
waves, which take the plus sign in Eq. (\ref{eq:dispersion_with_chi-1})
and the minus sign in Eq. (\ref{eq:dispersion_with_chi-2}).

\section{Conclusion and discussion\label{sub:FIBRES-Conclusion-and-discussion}}

The interaction between electromagnetic waves and matter is entirely
determined by the induced polarization, $\mathbf{P}$, of the matter.
We have seen that the linear response determines the refractive index
$n\left(\omega\right)$, but this is not sufficient to allow different
wave components to interact. For this, we must introduce the nonlinear
component of the polarization, the lowest order of which is the third
order. We have seen that third-order polarization gives rise to several
different effects, and we paid particular attention to the existence
of optical solitons and the interaction of an intense pulse with a
low-intensity probe wave. The coordinate transformation of Eqs. (\ref{eq:transformation_zeta})
and (\ref{eq:transformation_tau}) defines a reference frame in which
the pulse is stationary. The interaction of this pulse with a probe
wave is described by a wave equation, Eq. (\ref{eq:wave_eqn_for_vector_potential}),
which in turn yields the dispersion relation of Eq. (\ref{eq:dispersion_with_chi-2}).

It is this dispersion relation that is most useful in pinning down
the analogy with waves in moving fluids. Since it is independent of
$\zeta$, $\omega^{\prime}$ is a conserved quantity, much as $\omega$
is conserved in the acoustic model. $\omega^{\prime}$ depends on
both $\omega$ and $\chi$, where $\chi$ is in general $\tau$-dependent;
in the acoustic model, $\omega$ is a function of $k$ and $V$, where
$V$ is in general $x$-dependent. The essential difference between
the two models is the following: in the acoustic model, the inhomogeneity
of wave speed is due to a change in the flow velocity, the dispersion
profile remaining unchanged; in the optical model, the flow velocity
is constant, and the change in wave speed is brought about by an effective
change in the dispersion profile, and hence a change in the wave speed
with respect to the medium. These behave differently when the effects
on co-propagating and counter-propagating wave are compared: for an
inhomogeneous flow velocity, the change in speed is in the same direction
for both $u$- and $v$-modes; for an inhomogeneous dispersion profile,
the change in speed is opposite in direction for $u$- and $v$-modes.
But, for a single branch, the two situations are indistinguishable,
the total wave velocity being the important quantity. We thus expect
the interaction of co-propagating probe waves with a light pulse to
exhibit the same features of horizons already demonstrated by the
acoustic waves in a moving medium. In the next Chapter, we shall encounter
frequency shifting, like that seen in wavepacket propagation simulations
for acoustic waves, showing that group-velocity horizons can exist
in optical fibres. In Chapter \ref{sec:The-Fibre-Optical-Model},
we shall develop the theory of spontaneous photon creation in optical
fibres, closely following the derivation given in Chapter \ref{sec:The-Acoustic-Model}
for the acoustic model.

\pagebreak{}

\chapter{Frequency Shifting in Optical Fibres\label{sec:Frequency-Shifting-in-Optical-Fibres}}

The existence of event horizons in optical fibres can be demonstrated
by the classical effect of frequency shifting \cite{Philbin-et-al,Robertson-Leonhardt-2010},
and it is to this phenomenon that we first direct our attention. We
begin by expounding the black and white hole analogies, and using
straightforward geometrical optics we explain why we expect these
to be induced by a light pulse in an optical fibre. We then solve
the full wave equation, and compare our theoretical predictions with
experimental observations. There are some discrepancies. Another nonlinear
effect - the \textit{Raman effect} - is invoked in an attempt to explain
these discrepancies, but ultimately it fails to do so.

The main analysis and results of this chapter are presented in Reference
\cite{Robertson-Leonhardt-2010}.

\section{The fiber-optical analogy and geometrical optics}

As discussed in Chapter \ref{sec:Nonlinear-Fibre-Optics}, pulses
in fibres simulate a non-trivial flow velocity profile that moves
with respect to the medium. Through the Kerr effect, the pulse effectively
increases the local refractive index of the fiber. Since the speed
of light waves is determined by the refractive index, this behaves
like a perturbation in the flow velocity of the medium. In the co-moving
frame in which the pulse is stationary, the fiber appears as a moving
medium, with constant background flow equal in magnitude - but opposite
in direction - to the velocity of the pulse. A co-propagating wave
attempts to move against this flow, and the decrease in velocity induced
by interaction with the pulse can be described as a dip in the local
flow velocity; that is, to complete the analogy with a moving fluid,
the medium must be considered as moving faster (in the backward direction)
in the vicinity of the pulse. If the incident wave already has a low
speed in the co-moving frame, then it is possible that this dip forms
an event horizon. In fact, if a horizon is formed at all, there must
be a pair of horizons: a white hole at the pulse's trailing edge,
where the medium's speed decreases in the direction of flow, and a
black hole horizon at the leading edge, where the opposite occurs.

\begin{figure}
\includegraphics[width=0.8\columnwidth]{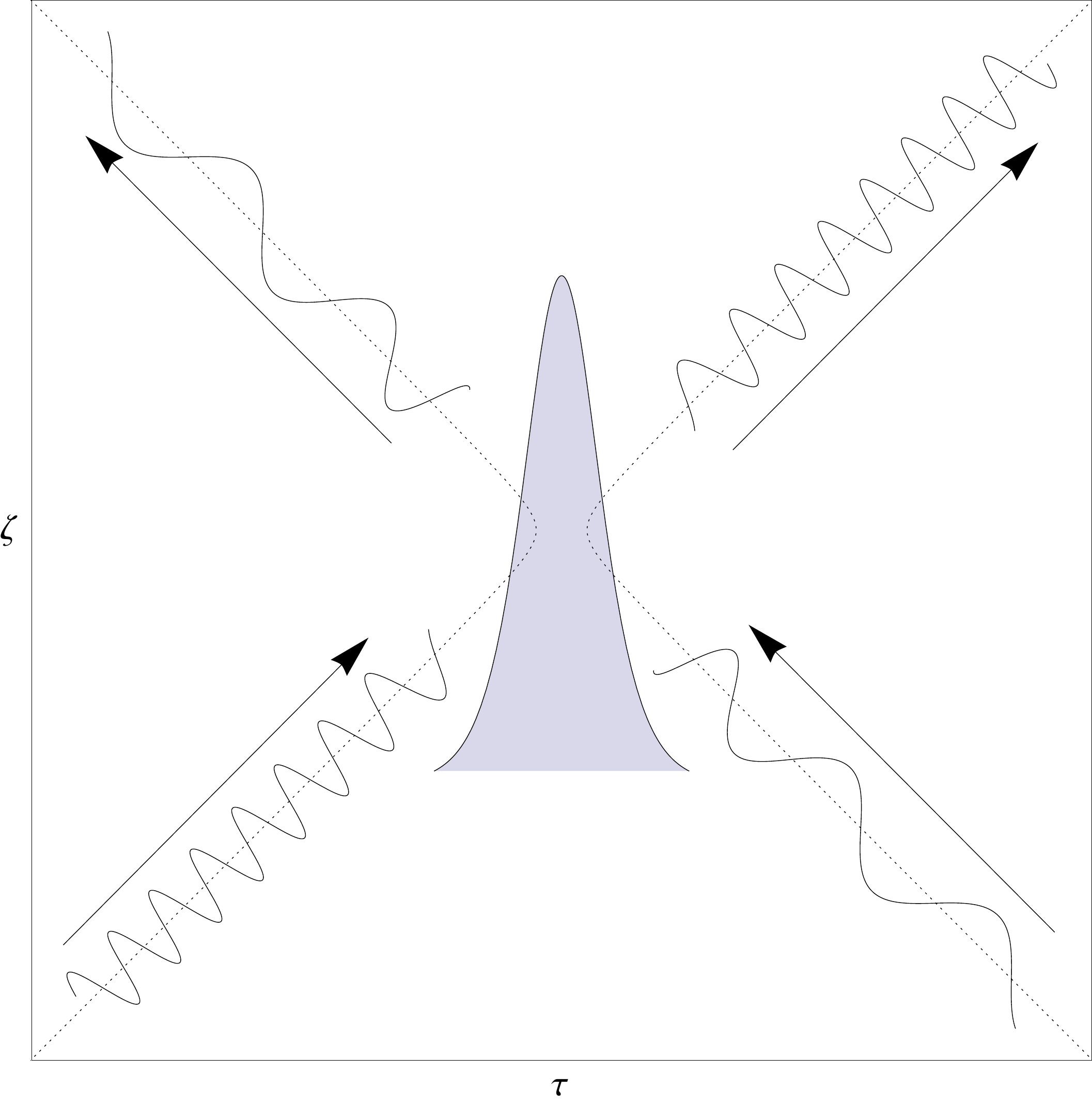}

\caption[\textsc{Frequency shifting by a pulse}]{\textsc{Frequency shifting by a pulse}: Nonlinear effects effectively
increase the refractive index in the vicinity of the pulse. This causes
a slight decrease in the phase velocity of incident waves, changing
their wavelength; in turn, the dispersion of the fibre causes a corresponding
change in the group velocity. For some frequencies, this results in
a reflection from the pulse. The white hole horizon at the trailing
edge of the pulse cause blueshifting, while the black hole horizon
at the leading edge causes redshifting.\label{fig:Frequency-shifting-by-a-Pulse}}

\end{figure}

Suppose there is a frequency at which this pair of horizons is formed.
Let a wavepacket at this frequency be incident on the pulse. How does
this wavepacket behave? Numerical simulations show that it is reflected
from the pulse with a different frequency, a process called \textit{frequency
shifting}. This is illustrated in Figure \ref{fig:Frequency-shifting-by-a-Pulse}
- it is also analogous to the frequency shifting seen in Figure \ref{fig:wavepacket_propagation}
- and can be understood conceptually as follows. Let us focus initially
on the white hole horizon on the pulse's trailing edge. An incident
wave, propagating against the flow of the medium, is slowed by the
increase in flow velocity. The front of the wavepacket interacts with
the pulse first, and will be slowed down before the back of the wavepacket;
this results in a compression of its wavefronts, and an increase in
its frequency. It might be considered that this continues indefinitely,
the wavepacket becoming squashed at the horizon with ever-increasing
frequency. Dispersion, however, limits this shifting, for an increase
in frequency, as experienced by the incident wavepacket, is also accompanied
by a decrease in speed with respect to the medium (assuming normal
dispersion). At some point, the wave's speed becomes less than the
flow velocity of the medium, and the wave is dragged back away from
the horizon. The overall result is reflection of the wavepacket at
a higher frequency: a white hole causes \textit{blueshifting}.

A similar process can occur at the black hole horizon on the pulse's
leading edge, but it is more difficult to describe in real time. Thankfully,
the time-reversal symmetry of black holes and white holes can be invoked
here: the process that occurs at the black hole horizon is simply
the time-reverse of a corresponding process at the white hole horizon.
If a certain frequency is blueshifted by the white hole horizon, then
the resulting frequency will be shifted back to the original frequency
by the black hole horizon. A black hole, then, causes \textit{redshifting}.

The value of the frequency shift is easily determined, for the frequency
shifting process must conform to the conservation of co-moving frequency,
$\omega^{\prime}$. This allows a diagrammatic interpretation of the
process. Let us examine the dispersion relation, including the $\tau$-dependent
change induced by the pulse, so that $\beta$ becomes a function of
both $\omega$ and $\tau$:\begin{equation}
\beta\left(\omega,\tau\right)=\frac{\omega}{c}\left(n\left(\omega\right)+\delta n\left(\omega,\tau\right)\right)\,.\end{equation}
(Recall that $\delta n$ is approximately equal to $\chi/2n$.) The
co-moving frequency is given by\begin{equation}
\omega^{\prime}=\omega-u\beta\left(\omega,\tau\right)=\omega\left(1-\frac{u}{c}\left(n\left(\omega\right)+\delta n\left(\omega,\tau\right)\right)\right)\,.\label{eq:dispersion_with_pulse_perturbation}\end{equation}

\begin{figure}
\includegraphics[width=0.8\columnwidth]{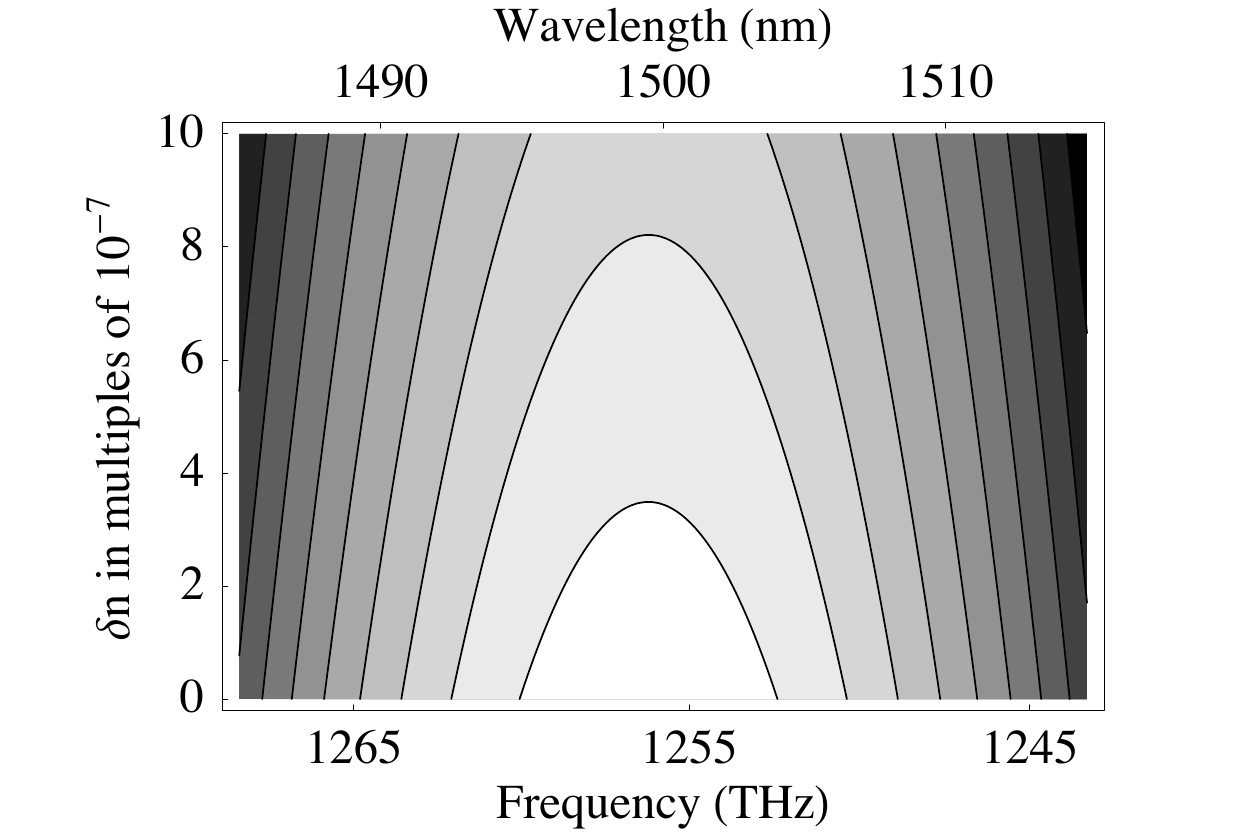}

\caption[\textsc{Frequency shifting contours}]{\textsc{Frequency shifting contours}: Each contour corresponds to
a single co-moving frequency, $\omega^{\prime}$, in the $\omega$-$\delta n$
plane. As described by Eq. (\ref{eq:parabolic_contour}), these contours
form parabolas centred at the group-velocity-matching frequency, $\omega_{m}$.\label{fig:Frequency_shifting_contours}}

\end{figure}

Assuming $u/c$ and $n\left(\omega\right)$ are given, the condition
that $\omega^{\prime}$ is constant determines a family of contours
in the $\omega$-$\delta n$ plane. An example is shown in Fig. \ref{fig:Frequency_shifting_contours}.
The incoming lab frequency $\omega$ determines the value of $\omega^{\prime}$
($\delta n=0$ far from the pulse), and hence on which of these contours
the wave lies. During the interaction, it must always lie on this
contour. Frequency shifting, which causes reflection in the co-moving
frame, takes place in a frequency window around the frequency $\omega_{m}$
at which the group velocity in the co-moving frame vanishes; $\omega_{m}$
is the group-velocity-matching frequency which has the same group
velocity as the pulse. Taylor expanding $\beta\left(\omega\right)$
around this frequency, we have\begin{equation}
\beta\left(\omega\right)=\beta_{0,m}+\beta_{1,m}\left(\omega-\omega_{m}\right)+\frac{1}{2}\beta_{2,m}\left(\omega-\omega_{m}\right)^{2}\end{equation}
where $\beta_{j,m}\equiv\partial^{j}\beta/\partial\omega^{j}\left(\omega_{m}\right)$.
Using the fact that $\beta_{1}$ is the inverse of the group velocity,
and so $\beta_{1,m}=1/u$, plugging this into Eq. (\ref{eq:dispersion_with_pulse_perturbation})
gives\begin{equation}
\omega^{\prime}\approx\omega_{m}^{\prime}-\frac{1}{2}\beta_{2,m}u\left(\omega-\omega_{m}\right)^{2}-\frac{u}{c}\omega\,\delta n\,,\end{equation}
or\begin{equation}
\delta n\approx\frac{c}{u}\frac{1}{\omega_{m}}\left[\omega_{m}^{\prime}-\omega^{\prime}-\frac{1}{2}\beta_{2,m}u\left(\omega-\omega_{m}\right)^{2}\right]\,,\label{eq:parabolic_contour}\end{equation}
where $\omega_{m}^{\prime}\equiv\omega_{m}-u\beta_{0,m}$. Around
the group-velocity-matching frequency, the contours of constant $\omega^{\prime}$
form parabolas centred at $\omega_{m}$. When a wave first encounters
the pulse, it moves to a higher value of $\delta n$, and its lab
frequency $\omega$ must also change accordingly. The end result depends
on the maximum value of $\delta n$. It is seen from Eq. (\ref{eq:parabolic_contour})
that the height of the contour is $\left(\omega_{m}^{\prime}-\omega^{\prime}\right)c/u\omega_{m}$.
If this is greater than $\delta n_{max}$, then the wave cannot pass
over the peak of the contour. It sees the pulse as a transient change
in velocity, moving to a slightly different frequency while traversing
it, then returning to its original frequency on clearing the pulse;
it experiences no event horizon. On the other hand, if the height
of the contour is \textit{less} than $\delta n_{max}$, the situation
is very different: now, the wave cannot pass over the peak of the
\textit{pulse}, and must pass over the peak of the contour instead.
Thus it lands at the shifted frequency $2\omega_{m}-\omega$ (the
second zero of the parabola), whose group velocity is opposite to
that of the incident frequency. The position of the event horizon
is well-defined: it is that point at which the value of $\delta n$
is equal to the height of the contour, and beyond which the wave cannot
penetrate. What is more, the contour whose height is equal to $\delta n_{max}$
marks the edge of the frequency window for which frequency shifting
is possible: those frequencies enclosed within this contour will experience
an event horizon and be frequency-shifted, while those outside of
this contour will be able to pass through the pulse.

\section{Wave solutions of the pulse-probe interaction}

So far, we have used geometrical optics - in which waves travel on
well-defined ray trajectories - to discuss the frequency shifting
process. This led to the conclusion that a given wave will either
pass through the pulse or be reflected from it at a shifted frequency.
Real waves, however, are spatially extended, and cannot have well-defined
values of both $\omega$ and $\tau$ as required by Eq. (\ref{eq:dispersion_with_pulse_perturbation}).
Frequency shifting can be described more accurately by a full wave
treatment of the optical field.

Firstly, let us define the difference frequencies\begin{alignat*}{1}
\bar{\omega}=\omega-\omega_{m}\,,\qquad & \bar{\omega}^{\prime}=\omega^{\prime}-\omega_{m}^{\prime}\,,\end{alignat*}
so that the dispersion relation (not including any local perturbations
due to the pulse) is simply\begin{equation}
\bar{\omega}^{\prime}=-\frac{1}{2}\beta_{2,m}u\bar{\omega}^{2}\,.\label{eq:dispersion_difference_frequencies}\end{equation}
Defining the envelope $\mathcal{A}$ such that the probe field $A_{p}=\mathcal{A}\,\exp\left(-i\omega_{m}^{\prime}\zeta-i\omega_{m}\tau\right)$,
the pulse-probe interaction is described by the wave equation \cite{Agrawal}\begin{equation}
-i\partial_{\zeta}\mathcal{A}+\frac{1}{2}\beta_{2,m}u\partial_{\tau}^{2}\mathcal{A}=\frac{2}{3}ru\gamma_{m}P\mathcal{A}\,,\label{eq:pulse_probe_interaction}\end{equation}
where $P$ is the optical power of the pulse, $\gamma_{m}$ is the
nonlinear coefficient at the group-velocity-matching frequency, and
$r$ is a dimensionless constant that depends on the relative polarization
of the pulse and probe ($r=3$ for co-polarized, $r=1$ for cross-polarized).
We have assumed that the intensity of the probe is very weak compared
to that of the pulse, so that any back-reaction on the pulse and any
self-interaction of the probe is negligible.

Assuming the pulse power $P$ is independent of $\zeta$, Eq. (\ref{eq:pulse_probe_interaction})
is equivalent in form to the time-dependent Schrödinger equation.
Factoring out a $\zeta$-dependent phase,\begin{equation}
\mathcal{A}=\psi_{\bar{\omega}}\,\exp\left(-i\bar{\omega}^{\prime}\zeta\right)\,,\label{eq:factoring_out_phase}\end{equation}
leads to the time-independent form\begin{equation}
-\frac{2}{\beta_{2,m}u}\bar{\omega}^{\prime}\psi_{\bar{\omega}}=\bar{\omega}^{2}\psi_{\bar{\omega}}=-\partial_{\tau}^{2}\psi_{\bar{\omega}}+\frac{2r\gamma_{m}}{\beta_{2,m}}P\psi_{\bar{\omega}}\,,\label{eq:pulse_probe_time_indep_Schrodinger}\end{equation}
where we used Eq. (\ref{eq:dispersion_difference_frequencies}) to
simplify the left-hand side.

If the pulse power $P$ is truly independent of $\zeta$, then the
pulse must be a fundamental soliton. Therefore, the optical power
has the form\begin{equation}
P\left(\tau\right)=\frac{\left|\beta_{2,s}\right|}{T_{s}^{2}\gamma_{s}}\mathrm{sech}^{2}\left(\frac{\tau}{T_{s}}\right)\,,\label{eq:soliton_power}\end{equation}
where $T_{s}$ is the soliton width (full width at half maximum $\approx1.76\, T_{s}$),
$\gamma_{s}$ is the nonlinear coefficient at the soliton frequency
$\omega_{s}$, and $\beta_{2,s}=\partial^{2}\beta/\partial\omega^{2}\left(\omega_{s}\right)<0$
(since solitons can only exist in regions of anomalous dispersion).
Eq. (\ref{eq:pulse_probe_time_indep_Schrodinger}) now becomes\begin{equation}
\bar{\omega}^{2}T_{s}^{2}\psi_{\bar{\omega}}=-T_{s}^{2}\partial_{\tau}^{2}\psi_{\bar{\omega}}+2r\frac{\gamma_{m}}{\gamma_{s}}\frac{\left|\beta_{2,s}\right|}{\beta_{2,m}}\mathrm{sech}^{2}\left(\frac{\tau}{T_{s}}\right)\psi_{\bar{\omega}}\,.\label{eq:pulse_probe_soliton}\end{equation}
This equation is exactly solvable (see Problem 4, §25 of Ref. \cite{Landau-Lifshitz-QM});
the solution is\begin{equation}
\psi_{\bar{\omega}}=\left(\frac{1}{4}\left(1-\xi^{2}\right)\right)^{i\bar{\omega}T_{s}/2}{}_{2}F_{1}\left[i\bar{\omega}T_{s}-\sigma,\, i\bar{\omega}T_{s}+\sigma+1;\: i\bar{\omega}T_{s}+1;\:\frac{1}{2}\left(1-\xi\right)\right]\,,\label{eq:exact_solution_2F1}\end{equation}
where\[
\xi=\tanh\left(\frac{\tau}{T_{s}}\right)\]
and\begin{equation}
\sigma=\frac{1}{2}\left(-1+\sqrt{1-8r\frac{\gamma_{m}}{\gamma_{s}}\frac{\left|\beta_{2,s}\right|}{\beta_{2,m}}}\right)\,.\label{eq:sigma_defn}\end{equation}
$_{2}F_{1}\left[a,\, b;\: c;\: z\right]$ is a hypergeometric function:\begin{alignat*}{1}
_{2}F_{1}\left[a,\, b;\: c;\: z\right]=\sum_{n=0}^{\infty}\frac{\left(a\right)_{n}\left(b\right)_{n}}{\left(c\right)_{n}}\frac{z^{n}}{n!}\,,\qquad & \left(a\right)_{n}=a\left(a+1\right)\ldots\left(a+n-1\right)=\frac{\Gamma\left(a+n\right)}{\Gamma\left(a\right)}\,.\end{alignat*}
Although Eq. (\ref{eq:exact_solution_2F1}) is exact, it is not especially
helpful in this form. Its complicated nature makes it very difficult
to use in calculations, and it is not obvious what the solution actually
looks like. Fortunately, it is relatively easy to simplify it. Since
the {}``potential'' of Eq. (\ref{eq:pulse_probe_time_indep_Schrodinger})
is proportional to the pulse power, it is non-zero only in a relatively
short region; outside this region, the general solution is simply
a sum of the plane waves $\exp\left(-i\bar{\omega}\tau\right)$ and
$\exp\left(i\bar{\omega}\tau\right)$. These are precisely the plane
waves converted into each other by frequency shifting (they share
the same value of $\omega^{\prime}$, and are equidistant from the
group-velocity-matching frequency $\omega_{m}$, which of course has
a difference frequency $\bar{\omega}=0$). So, in the limit $\tau\rightarrow\pm\infty$
(or $\xi\rightarrow\pm1$), the exact solution Eq. (\ref{eq:exact_solution_2F1})
should degenerate into sums of these plane waves. Indeed, the solution
is normalized such that, as $\tau\rightarrow\infty$, $\psi_{\bar{\omega}}\rightarrow\exp\left(-i\bar{\omega}\tau\right)$.
In the other asymptotic region, as $\tau\rightarrow-\infty$, the
form of $\psi_{\bar{\omega}}$ is found to be\[
\psi_{\bar{\omega}}\rightarrow\frac{\Gamma\left(i\bar{\omega}T_{s}+1\right)\Gamma\left(i\bar{\omega}T_{s}\right)}{\Gamma\left(i\bar{\omega}T_{s}+\sigma+1\right)\Gamma\left(i\bar{\omega}T_{s}-\sigma\right)}\exp\left(-i\bar{\omega}\tau\right)+\frac{\Gamma\left(i\bar{\omega}T_{s}+1\right)\Gamma\left(-i\bar{\omega}T_{s}\right)}{\Gamma\left(\sigma+1\right)\Gamma\left(-\sigma\right)}\exp\left(i\bar{\omega}\tau\right)\,.\]
Multiplying through by a constant, we can write the asymptotic form
of the solution\begin{equation}
\psi_{\bar{\omega}}\left(\tau\right)\rightarrow\begin{cases}
\exp\left(-i\bar{\omega}\tau\right)+R_{\bar{\omega}}\exp\left(i\bar{\omega}\tau\right)\,, & \tau\rightarrow-\infty\\
T_{\bar{\omega}}\exp\left(-i\bar{\omega}\tau\right)\,, & \tau\rightarrow+\infty\end{cases}\,,\label{eq:asymptotic_solns}\end{equation}
where we have defined\begin{equation}
R_{\bar{\omega}}=\frac{\Gamma\left(-i\bar{\omega}T_{s}\right)\Gamma\left(i\bar{\omega}T_{s}+\sigma+1\right)\Gamma\left(i\bar{\omega}T_{s}-\sigma\right)}{\Gamma\left(i\bar{\omega}T_{s}\right)\Gamma\left(\sigma+1\right)\Gamma\left(-\sigma\right)}\,\label{eq:reflection_coefficient}\end{equation}
and\begin{equation}
T_{\bar{\omega}}=\frac{\Gamma\left(i\bar{\omega}T_{s}+\sigma+1\right)\Gamma\left(i\bar{\omega}T_{s}-\sigma\right)}{\Gamma\left(i\bar{\omega}T_{s}+1\right)\Gamma\left(i\bar{\omega}T_{s}\right)}\,.\label{eq:transmission_coefficient}\end{equation}

\begin{figure}
\subfloat{\includegraphics[width=0.45\columnwidth]{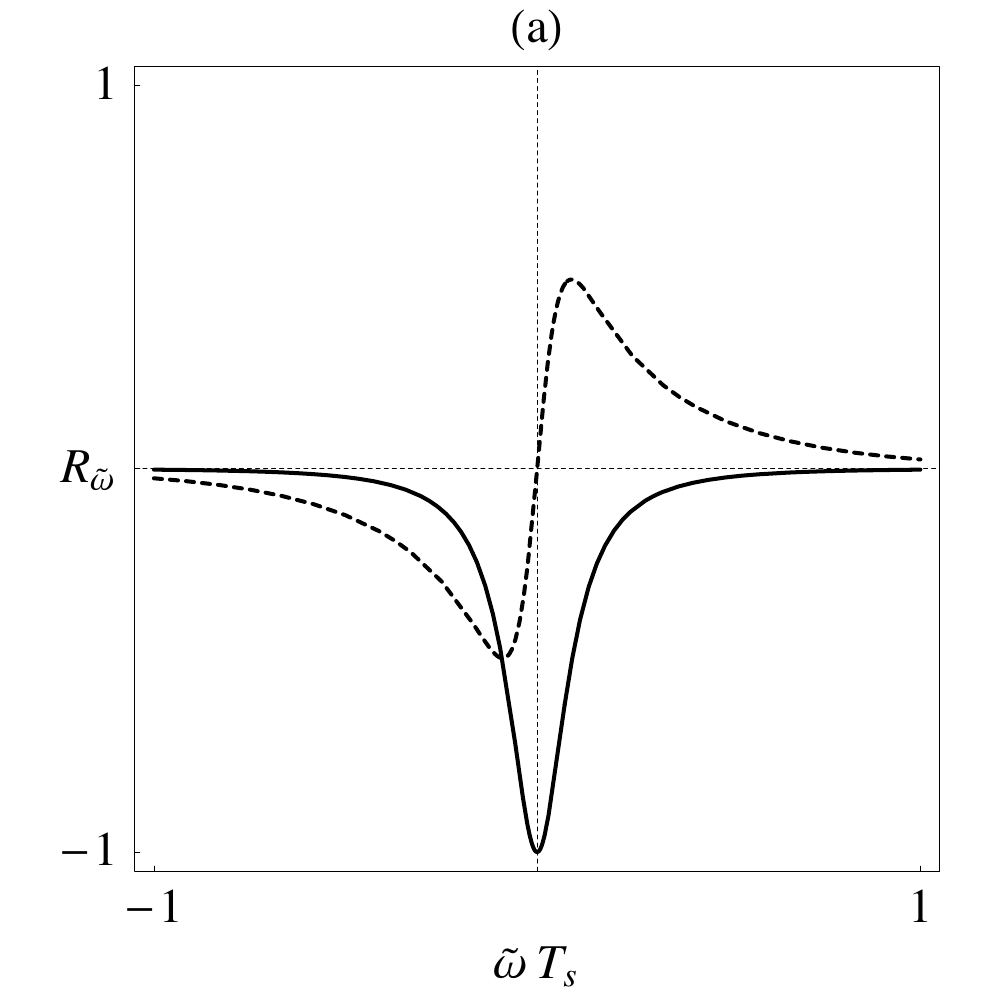}}\subfloat{\includegraphics[width=0.45\columnwidth]{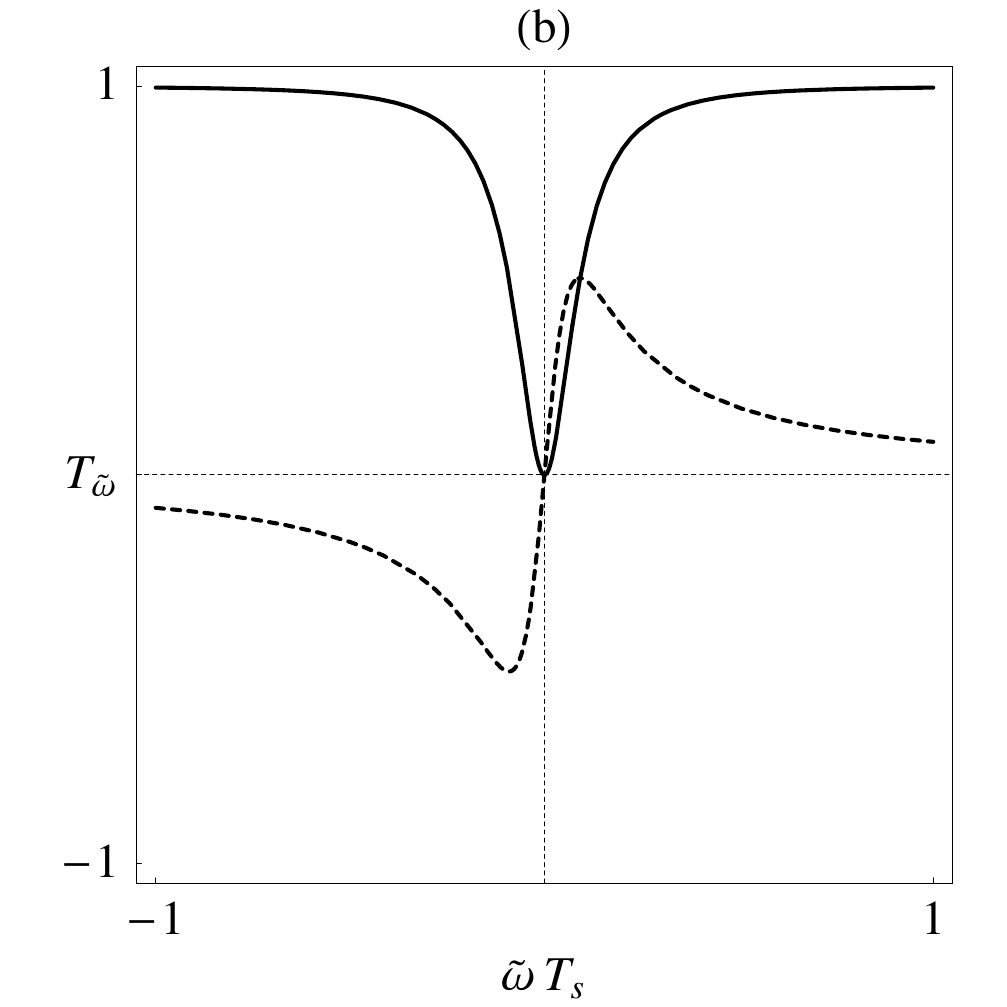}}

\subfloat{\includegraphics[width=0.45\columnwidth]{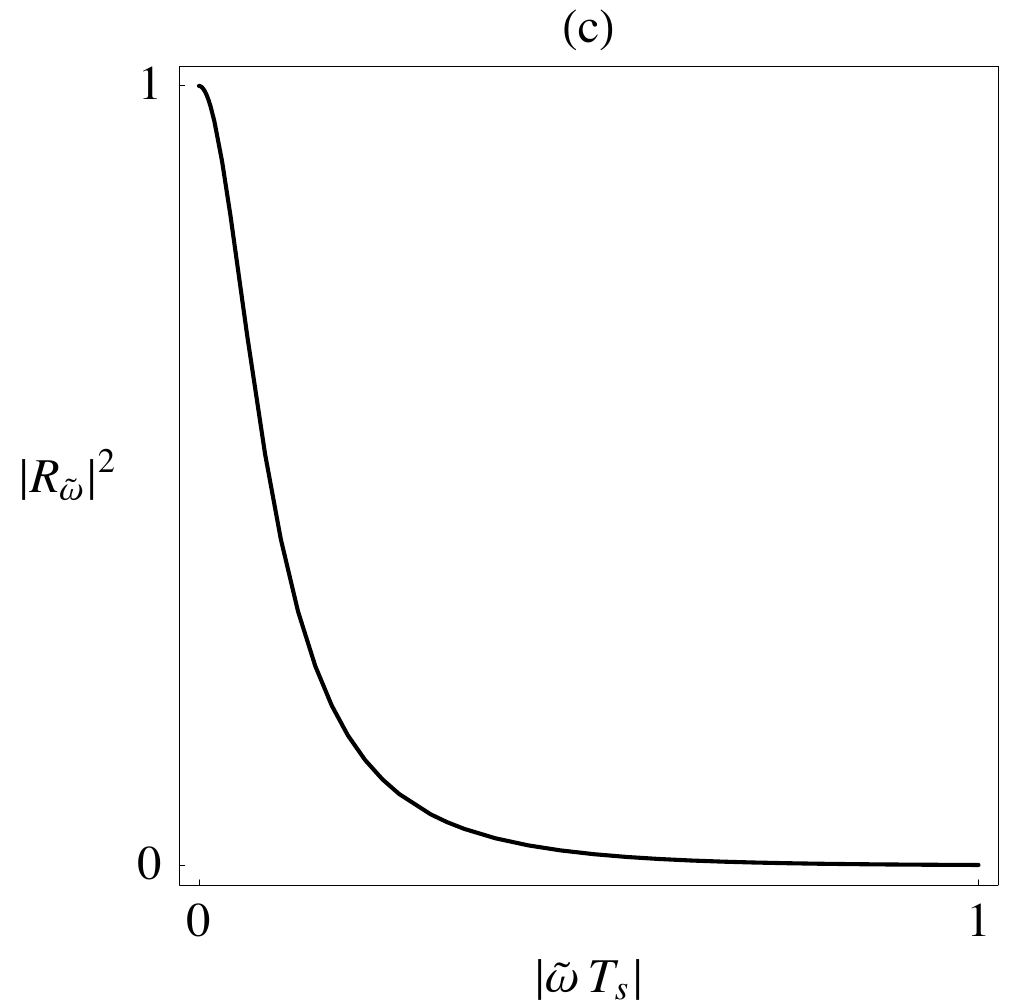}}\subfloat{\includegraphics[width=0.45\columnwidth]{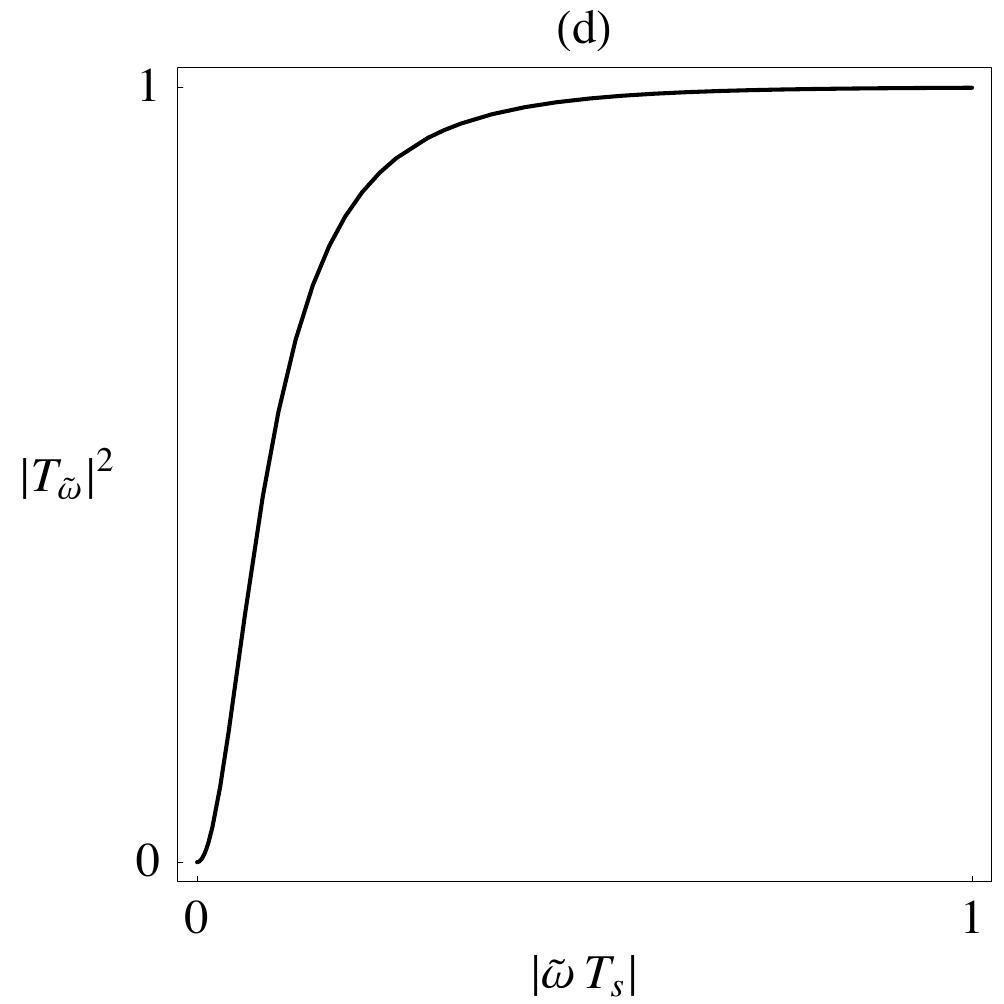}}

\caption[\textsc{Reflection and transmission amplitudes}]{\textsc{Reflection and transmission amplitudes}: Figures $\left(a\right)$
and $\left(b\right)$ show the real (solid line) and imaginary (dotted
line) parts of $R_{\widetilde{\omega}}$ and $T_{\widetilde{\omega}}$,
given by Eqs. (\ref{eq:reflection_coefficient}) and (\ref{eq:transmission_coefficient}).
Figures $\left(c\right)$ and $\left(d\right)$ show the squared magnitudes
$\left|R_{\widetilde{\omega}}\right|^{2}=1-\left|T_{\widetilde{\omega}}\right|^{2}$
and $\left|T_{\widetilde{\omega}}\right|^{2}$, given by Eq. (\ref{eq:transmission_squared_magnitude}).
Here, we have taken $\sigma=-0.092$, in accordance with experiment.\label{fig:Reflection_and_transmission}}

\end{figure}

Assuming for the moment that $\bar{\omega}$ is positive - that is,
$\omega$ is greater than the group-velocity-matching frequency $\omega_{m}$
- the asymptotic solutions of Eq. (\ref{eq:asymptotic_solns}) describe
a plane wave, $\exp\left(-i\bar{\omega}\tau\right)$, incident on
the pulse from negative $\tau$. This implies, in the lab frame, that
the wave of frequency $\omega$ has a group velocity less than that
of the pulse, and that the pulse is able to catch up with it. It interacts
with the pulse's leading edge, where there may be a black hole event
horizon. There is thus a reflected component, $R_{\bar{\omega}}\exp\left(i\bar{\omega}\tau\right)$,
at a lower (redshifted) frequency and travelling back towards negative
$\tau$ (or with group velocity faster than that of the pulse in the
lab frame). There is also a transmitted component, $T_{\bar{\omega}}\exp\left(-i\bar{\omega}\tau\right)$,
at the original frequency and continuing on towards positive $\tau$,
having managed to traverse the entire pulse. Unlike geometrical optics,
in which a wave is either reflected or transmitted, the wave picture
shows that these two effects are always present, albeit in varying
degrees. The relative amounts of wave energy which are reflected or
transmitted are given by the squared moduli $\left|R_{\bar{\omega}}\right|^{2}$
and $\left|T_{\bar{\omega}}\right|^{2}$, with conservation of energy
encoded in the identity $\left|R_{\bar{\omega}}\right|^{2}+\left|T_{\bar{\omega}}\right|^{2}=1$.
We find\begin{equation}
\left|T_{\bar{\omega}}\right|^{2}=\begin{cases}
\sinh^{2}\left(\pi\bar{\omega}T_{s}\right)/\left[\sinh^{2}\left(\pi\bar{\omega}T_{s}\right)+\cos^{2}\left(\frac{\pi}{2}\sqrt{1-8r\frac{\gamma_{m}}{\gamma_{s}}\frac{\left|\beta_{2,s}\right|}{\beta_{2,m}}}\right)\right]\,, & 8r\frac{\gamma_{m}}{\gamma_{s}}\frac{\left|\beta_{2,s}\right|}{\beta_{2,m}}<1\\
\sinh^{2}\left(\pi\bar{\omega}T_{s}\right)/\left[\sinh^{2}\left(\pi\bar{\omega}T_{s}\right)+\cosh^{2}\left(\frac{\pi}{2}\sqrt{8r\frac{\gamma_{m}}{\gamma_{s}}\frac{\left|\beta_{2,s}\right|}{\beta_{2,m}}-1}\right)\right]\,, & 8r\frac{\gamma_{m}}{\gamma_{s}}\frac{\left|\beta_{2,s}\right|}{\beta_{2,m}}>1\end{cases}\,.\label{eq:transmission_squared_magnitude}\end{equation}
The reflection and transmission coefficients are plotted in Fig. \ref{fig:Reflection_and_transmission}.
Note that we have total reflection at $\bar{\omega}=0$, and that
the amount of reflection decreases as we move away from the group-velocity-matching
frequency. This is in accordance with our expectation that frequency
shifting should occur most strongly for frequencies close to $\omega_{m}$,
so that, in the co-moving frame, they are significantly slowed by
the pulse.

We have seen that the solution Eq. (\ref{eq:exact_solution_2F1})
corresponds to redshifting at the black hole horizon when $\bar{\omega}$
is positive. Blueshifting at the white hole horizon must also correspond
to some solutions of Eq. (\ref{eq:pulse_probe_soliton}), but these
are not found simply by allowing $\bar{\omega}$ to be negative. The
reason is that the blueshifting process occurs at the trailing edge
of the pulse, so that the incoming and reflected waves must be present
for $\tau\rightarrow+\infty$, while only the transmitted wave should
be present for $\tau\rightarrow-\infty$. The solution of Eq. (\ref{eq:exact_solution_2F1})
clearly does not share this basic property of the blueshifting process.
Rather, we must make use of the reflection symmetry of the wave equation
(\ref{eq:pulse_probe_soliton}), i.e., its invariance under the transformation
$\tau\rightarrow-\tau$. This means that, if $\psi_{\bar{\omega}}\left(\tau\right)$
is a solution, then $\psi_{\bar{\omega}}\left(-\tau\right)$ is also
a solution. This transformation simply changes the sign of $\xi$
in Eq. (\ref{eq:exact_solution_2F1}), while the asymptotic solutions
become\[
\psi_{\bar{\omega}}\left(-\tau\right)\rightarrow\begin{cases}
T_{\bar{\omega}}\exp\left(i\bar{\omega}\tau\right)\,, & \tau\rightarrow-\infty\\
\exp\left(i\bar{\omega}\tau\right)+R_{\bar{\omega}}\exp\left(-i\bar{\omega}\tau\right)\,, & \tau\rightarrow+\infty\end{cases}\,.\]
This corresonds precisely to blueshifting at the white hole horizon:
the plane wave $\exp\left(i\bar{\omega}\tau\right)$, with frequency
less than $\omega_{m}$, is incident from positive values of $\tau$
(i.e., in the lab frame, it has a group velocity greater than that
of the pulse, and catches up with it); this results in a reflected
blueshifted wave, $R_{\bar{\omega}}\exp\left(-i\bar{\omega}\tau\right)$,
which travels back toward positive $\tau$; and a transmitted wave,
$T_{\bar{\omega}}\exp\left(i\bar{\omega}\tau\right)$, a part of the
original frequency which has managed to traverse the whole pulse.
Note that the reflection and transmission amplitudes are exactly the
same as for the redshifting case.

\section{General solution}

A general solution of Eq. (\ref{eq:pulse_probe_interaction}) can
be expressed as a linear combination of eigenfunctions (those solutions
with a specified value of $\bar{\omega}^{\prime}$). In order to form
such an expression, the coefficients of the various eigenfunctions
must be obtainable. This is possible through product integration if
the eigenfunctions themselves are orthonormal. We first ensure that
we have a complete, orthonormal set of eigenfunctions, and then we
use them to form a general solution of the pulse-probe interaction.

The functions $\psi_{\bar{\omega}}\left(\tau\right)$ and $\psi_{\bar{\omega}}\left(-\tau\right)$
- where $\bar{\omega}$ varies from zero to infinity - form a complete
set of solutions of the wave equation (\ref{eq:pulse_probe_soliton}).
Those with different values of $\bar{\omega}$ have different eigenvalues,
and are therefore orthogonal. However, $\psi_{\bar{\omega}}\left(\tau\right)$
has the same eigenvalue as $\psi_{\bar{\omega}}\left(-\tau\right)$;
these solutions are degenerate, and need not be orthogonal. In order
to have a complete set of orthonormal solutions, we define the following
sets of functions:\begin{eqnarray*}
\psi_{\bar{\omega}}^{+}\left(\tau\right) & = & \frac{1}{\sqrt{4\pi}}\left[\psi_{\bar{\omega}}\left(\tau\right)+\psi_{\bar{\omega}}\left(-\tau\right)\right]\,,\\
\psi_{\bar{\omega}}^{-}\left(\tau\right) & = & \frac{1}{\sqrt{4\pi}}\left[\psi_{\bar{\omega}}\left(\tau\right)-\psi_{\bar{\omega}}\left(-\tau\right)\right]\,.\end{eqnarray*}
$\psi_{\bar{\omega}}^{+}$ and $\psi_{\bar{\omega}}^{-}$ are even
and odd functions of $\tau$, respectively. Their product is therefore
odd, and integrates to zero, so by virtue of their definitions $\psi_{\bar{\omega}}^{+}$
and $\psi_{\bar{\omega}}^{-}$ are orthogonal to each other. The overall
factor of $\left(4\pi\right)^{-1/2}$ ensures that they are ortho\textit{normal},
i.e., that their products integrate to give $\delta\left(\bar{\omega}_{1}-\bar{\omega}_{2}\right)$.
Since the integral over any finite region must be finite, the contribution
to the $\delta$ function comes from integration over the asymptotic
regions. Let us, then, write the asymptotic expressions explicitly:\begin{equation}
\psi_{\widetilde{\omega}}^{+}\left(\tau\right)\rightarrow\frac{1}{\sqrt{4\pi}}\begin{cases}
\exp\left(-i\bar{\omega}\tau\right)+\left(R_{\bar{\omega}}+T_{\bar{\omega}}\right)\exp\left(i\bar{\omega}\tau\right)\,, & \tau\rightarrow-\infty\\
\exp\left(i\bar{\omega}\tau\right)+\left(R_{\bar{\omega}}+T_{\bar{\omega}}\right)\exp\left(-i\bar{\omega}\tau\right)\,, & \tau\rightarrow+\infty\end{cases}\,,\label{eq:even_asymptotics}\end{equation}
\begin{equation}
\psi_{\widetilde{\omega}}^{-}\left(\tau\right)\rightarrow\frac{1}{\sqrt{4\pi}}\begin{cases}
\exp\left(-i\bar{\omega}\tau\right)+\left(R_{\bar{\omega}}-T_{\bar{\omega}}\right)\exp\left(i\bar{\omega}\tau\right)\,, & \tau\rightarrow-\infty\\
-\exp\left(i\bar{\omega}\tau\right)-\left(R_{\bar{\omega}}-T_{\bar{\omega}}\right)\exp\left(-i\bar{\omega}\tau\right)\,, & \tau\rightarrow+\infty\end{cases}\,.\label{eq:odd_asymptotics}\end{equation}
Then the product of $\psi_{\bar{\omega}}^{+}$ with itself gives\begin{eqnarray*}
\int_{-\infty}^{+\infty}\left[\psi_{\bar{\omega}_{1}}^{+}\left(\tau\right)\right]^{\star}\psi_{\bar{\omega}_{2}}^{+}\left(\tau\right)\, d\tau & \approx & \frac{1}{2\pi}\int_{0}^{\infty}\left[e^{-i\bar{\omega}_{1}\tau}+\left(R_{\bar{\omega}_{1}}^{\star}+T_{\bar{\omega}_{1}}^{\star}\right)e^{i\bar{\omega}_{1}\tau}\right]\left[e^{i\bar{\omega}_{2}\tau}+\left(R_{\bar{\omega}_{2}}+T_{\bar{\omega}_{2}}\right)e^{-i\bar{\omega}_{2}\tau}\right]d\tau\\
 & = & \frac{1}{2\pi}\int_{0}^{\infty}\left[e^{-i\left(\bar{\omega}_{1}-\bar{\omega}_{2}\right)\tau}+\left(R_{\bar{\omega}_{1}}^{\star}+T_{\bar{\omega}_{1}}^{\star}\right)e^{i\left(\bar{\omega}_{1}+\bar{\omega}_{2}\right)\tau}\right.\\
 &  & \qquad\qquad\left.+\left(R_{\bar{\omega}_{2}}+T_{\bar{\omega}_{2}}\right)e^{-i\left(\bar{\omega}_{1}+\bar{\omega}_{2}\right)\tau}+\left(R_{\bar{\omega}_{1}}^{\star}+T_{\bar{\omega}_{1}}^{\star}\right)\left(R_{\bar{\omega}_{2}}+T_{\bar{\omega}_{2}}\right)e^{i\left(\bar{\omega}_{1}-\bar{\omega}_{2}\right)\tau}\right]d\tau\\
 & = & \frac{1}{2\pi}\left[\pi\,\delta\left(\bar{\omega}_{1}-\bar{\omega}_{2}\right)+\left(R_{\bar{\omega}_{1}}^{\star}+T_{\bar{\omega}_{1}}^{\star}\right)\left(R_{\bar{\omega}_{2}}+T_{\bar{\omega}_{2}}\right)\pi\,\delta\left(\bar{\omega}_{1}-\bar{\omega}_{2}\right)\right]\end{eqnarray*}
where, in the last line, we have included only terms in $\delta\left(\bar{\omega}_{1}-\bar{\omega}_{2}\right)$,
since the positivity of $\bar{\omega}$ means that $\delta\left(\bar{\omega}_{1}+\bar{\omega}_{2}\right)$
is always equal to zero, and any finite contributions can be ignored
on the grounds that the exact solution must be proportional to a $\delta$
function (since it is a solution of the Schrödinger equation). The
coefficient of the second term can be rewritten with $\bar{\omega}_{2}$
set equal to $\bar{\omega}_{1}$:\[
\left|R_{\bar{\omega}_{1}}+T_{\bar{\omega}_{1}}\right|^{2}=\left|R_{\bar{\omega}_{1}}\right|^{2}+R_{\bar{\omega}_{1}}^{\star}T_{\bar{\omega}_{1}}+R_{\bar{\omega}_{1}}T_{\bar{\omega}_{1}}^{\star}+\left|T_{\bar{\omega}_{1}}\right|^{2}\,.\]
Conservation of energy, as mentioned previously, implies $\left|R_{\bar{\omega}_{1}}\right|^{2}+\left|T_{\bar{\omega}_{1}}\right|^{2}=1$.
The product $R_{\bar{\omega}_{1}}^{\star}T_{\bar{\omega}_{1}}$ can
be calculated directly from Eqs. (\ref{eq:reflection_coefficient})
and (\ref{eq:transmission_coefficient}):\begin{eqnarray*}
R_{\bar{\omega}_{1}}^{\star}T_{\bar{\omega}_{1}} & = & \frac{\Gamma\left(i\bar{\omega}_{1}T_{s}\right)\Gamma\left(-i\bar{\omega}_{1}T_{s}+\sigma^{\star}+1\right)\Gamma\left(-i\bar{\omega}_{1}T_{s}-\sigma^{\star}\right)}{\Gamma\left(-i\bar{\omega}_{1}T_{s}\right)\Gamma\left(\sigma^{\star}+1\right)\Gamma\left(-\sigma^{\star}\right)}\frac{\Gamma\left(i\bar{\omega}_{1}T_{s}+\sigma+1\right)\Gamma\left(i\bar{\omega}_{1}T_{s}-\sigma\right)}{\Gamma\left(i\bar{\omega}_{1}T_{s}+1\right)\Gamma\left(i\bar{\omega}_{1}T_{s}\right)}\\
 & = & \frac{1}{i\bar{\omega}_{1}T_{s}}\frac{\left|\Gamma\left(i\bar{\omega}_{1}T_{s}+\sigma+1\right)\right|^{2}\left|\Gamma\left(i\bar{\omega}_{1}T_{s}-\sigma\right)\right|^{2}}{\left|\Gamma\left(i\bar{\omega}_{1}T_{s}\right)\right|^{2}\Gamma\left(\frac{1}{2}+\left(\sigma^{\star}+\frac{1}{2}\right)\right)\Gamma\left(\frac{1}{2}-\left(\sigma^{\star}+\frac{1}{2}\right)\right)}\,.\end{eqnarray*}
Given the definition of $\sigma$ in Eq. (\ref{eq:sigma_defn}), $\frac{1}{2}+\left(\sigma^{\star}+\frac{1}{2}\right)$
and $\frac{1}{2}-\left(\sigma^{\star}+\frac{1}{2}\right)$ are either
purely real or complex conjugates of each other. $R_{\bar{\omega}_{1}}^{\star}T_{\bar{\omega}_{1}}$,
then, is purely imaginary, and $R_{\bar{\omega}_{1}}^{\star}T_{\bar{\omega}_{1}}+R_{\bar{\omega}_{1}}T_{\bar{\omega}_{1}}^{\star}=0$.
Therefore, $\left|R_{\bar{\omega}_{1}}+T_{\bar{\omega}_{1}}\right|^{2}=1$,
and we have\[
\int_{-\infty}^{+\infty}\left[\psi_{\bar{\omega}_{1}}^{+}\left(\tau\right)\right]^{\star}\psi_{\bar{\omega}_{2}}^{+}\left(\tau\right)\, d\tau=\delta\left(\bar{\omega}_{1}-\bar{\omega}_{2}\right)\,,\]
as desired. An entirely analogous calculation shows that $\left|R_{\bar{\omega}_{1}}-T_{\bar{\omega}_{1}}\right|^{2}=1$
and, consequently, that\[
\int_{-\infty}^{+\infty}\left[\psi_{\bar{\omega}_{1}}^{-}\left(\tau\right)\right]^{\star}\psi_{\bar{\omega}_{2}}^{-}\left(\tau\right)\, d\tau=\delta\left(\bar{\omega}_{1}-\bar{\omega}_{2}\right)\,.\]

The general solution of Eq. (\ref{eq:pulse_probe_interaction}), where
$P\left(\tau\right)$ is given by Eq. (\ref{eq:soliton_power}), is\begin{equation}
\mathcal{A}\left(\tau,\zeta\right)=\int_{0}^{\infty}\left\{ c_{+}\left(\bar{\omega}\right)\psi_{\bar{\omega}}^{+}\left(\tau\right)+c_{-}\left(\bar{\omega}\right)\psi_{\bar{\omega}}^{-}\left(\tau\right)\right\} \exp\left(i\frac{1}{2}\beta_{2,m}u\zeta\bar{\omega}^{2}\right)d\widetilde{\omega}\,,\label{eq:general_soln}\end{equation}
where we have replaced the phase factor that was removed in the transition
from the time-dependent to the time-independent Schrödinger equation
(see Eq. (\ref{eq:factoring_out_phase})). The coefficients $c_{+}\left(\bar{\omega}\right)$
and $c_{-}\left(\bar{\omega}\right)$ are arbitrary, but they are
completely determined by the inital form of $\mathcal{A}$. The orthonormality
condition on the eigenfunctions allows their calculation:\begin{eqnarray}
\int_{-\infty}^{+\infty}\left[\psi_{\bar{\omega}}^{\pm}\left(\tau\right)\right]^{\star}\mathcal{A}\left(\tau,0\right)d\tau & = & \int_{0}^{\infty}\left\{ c_{+}\left(w\right)\int_{-\infty}^{+\infty}\left[\psi_{\bar{\omega}}^{\pm}\left(\tau\right)\right]^{\star}\psi_{w}^{+}\left(\tau\right)\, d\tau\right.\nonumber \\
 &  & \qquad\left.+c_{-}\left(w\right)\int_{-\infty}^{+\infty}\left[\psi_{\bar{\omega}}^{\pm}\left(\tau\right)\right]^{\star}\psi_{w}^{-}\left(\tau\right)\, d\tau\right\} dw\nonumber \\
 & = & \int_{0}^{\infty}c_{\pm}\left(w\right)\delta\left(\bar{\omega}-w\right)dw\nonumber \\
 & = & c_{\pm}\left(\bar{\omega}\right)\,.\label{eq:calculating_coefficients}\end{eqnarray}

\section{Solution for a continuous wave}

In experiment \cite{Philbin-et-al}, a soliton is sent into the fiber
alongside a continuous probe wave, whose frequency we denote $\omega_{p}$
($\bar{\omega}_{p}=\omega_{p}-\omega_{m}$). The initial form of the
probe wave, $\mathcal{A}\left(\tau,0\right)$, is its value at the
entrance to the fiber ($z=0$) as a function of time, since here $\tau=t$.
Thus,\begin{equation}
\mathcal{A}\left(\tau,0\right)=\exp\left(-i\bar{\omega}_{p}\tau\right)\,,\label{eq:initial_CW_wave}\end{equation}
and the coefficients $c_{\pm}\left(\bar{\omega}\right)$ are just
Fourier transforms of (complex conjugates of) eigenfunctions:\begin{equation}
c_{\pm}\left(\bar{\omega}\right)=\left[\widetilde{\psi}_{\bar{\omega}}^{\pm}\left(\bar{\omega}_{p}\right)\right]^{\star}\label{eq:coefficients_of_CW_wave}\end{equation}
where we have defined\begin{equation}
\widetilde{\psi}_{\bar{\omega}}^{\pm}\left(\bar{w}\right)=\int_{-\infty}^{+\infty}\psi_{\bar{\omega}}^{\pm}\left(\tau\right)\,\exp\left(i\bar{w}\tau\right)\, d\tau\,.\label{eq:Fourier_transform_eigenfunctions}\end{equation}
Plugging Eq. (\ref{eq:coefficients_of_CW_wave}) into Eq. (\ref{eq:general_soln}),
then taking its Fourier transform, we find:\begin{equation}
\mathcal{\widetilde{A}}\left(\bar{w},\zeta\right)=\int_{0}^{\infty}\left\{ \left[\widetilde{\psi}_{\bar{\omega}}^{+}\left(\bar{\omega}_{p}\right)\right]^{\star}\widetilde{\psi}_{\bar{\omega}}^{+}\left(\bar{w}\right)+\left[\widetilde{\psi}_{\bar{\omega}}^{-}\left(\bar{\omega}_{p}\right)\right]^{\star}\widetilde{\psi}_{\bar{\omega}}^{-}\left(\bar{w}\right)\right\} \exp\left(i\frac{1}{2}\beta_{2,m}u\zeta\bar{\omega}^{2}\right)d\bar{\omega}\,.\label{eq:Fourier_transformed_soln}\end{equation}
These Fourier transforms can be calculated to a very good approximation
if we perform the Fourier integration using the asymptotic forms of
Eqs. (\ref{eq:even_asymptotics}) and (\ref{eq:odd_asymptotics}).
These give\begin{multline}
\widetilde{\psi}_{\bar{\omega}}^{+}\left(\bar{w}\right)\approx\frac{1}{\sqrt{4\pi}}\left[\pi\left(1+R_{\bar{\omega}}+T_{\bar{\omega}}\right)\left(\delta\left(\bar{\omega}-\bar{w}\right)+\delta\left(\bar{\omega}+\bar{w}\right)\right)\right.\\
\left.+i\left(1-R_{\bar{\omega}}-T_{\bar{\omega}}\right)\left(\frac{1}{\bar{\omega}-\bar{w}}+\frac{1}{\bar{\omega}+\bar{w}}\right)\right]\,,\label{eq:even_Fourier_transform}\end{multline}
\begin{multline}
\widetilde{\psi}_{\bar{\omega}}^{-}\left(\bar{w}\right)\approx\frac{1}{\sqrt{4\pi}}\left[\pi\left(1-R_{\bar{\omega}}+T_{\bar{\omega}}\right)\left(\delta\left(\bar{\omega}-\bar{w}\right)-\delta\left(\bar{\omega}+\bar{w}\right)\right)\right.\\
\left.+i\left(1+R_{\bar{\omega}}-T_{\bar{\omega}}\right)\left(\frac{1}{\bar{\omega}-\bar{w}}-\frac{1}{\bar{\omega}+\bar{w}}\right)\right]\,.\label{eq:odd_Fourier_transform}\end{multline}
Plugging Eqs. (\ref{eq:even_Fourier_transform}) and (\ref{eq:odd_Fourier_transform})
into Eq. (\ref{eq:Fourier_transformed_soln}), we find that we can
write $\widetilde{\mathcal{A}}\left(\bar{w},\zeta\right)$ as a sum
of three distinct terms, corresponding to different terms of the integrand
when Eqs. (\ref{eq:even_Fourier_transform}) and (\ref{eq:odd_Fourier_transform})
are substituted in Eq. (\ref{eq:Fourier_transformed_soln}).
The products of $\delta$ functions with other $\delta$ functions
indicate the appearance of a $\delta$ function in the result; cross
terms of $\delta$ functions with poles can be evaluated by simple
substitution; and only terms involving products of poles remain to
be integrated. Separating these terms explicitly,\begin{equation}
\mathcal{\widetilde{A}}\left(\bar{w},\zeta\right)\approx C_{\delta}\left(\zeta\right)\,2\pi\,\delta\left(\bar{w}-\bar{\omega}_{p}\right)+\widetilde{\mathcal{A}}_{\mathrm{cross}}\left(\bar{w},\zeta\right)+\widetilde{\mathcal{A}}_{\mathrm{poles}}\left(\bar{w},\zeta\right)\,.\label{eq:separating_cross_and_poles}\end{equation}

The $\delta$ function arises because, far from the pulse, the initial
probe wave evolves as if it were purely a continuous wave; in other
words, the presence of the pulse affects only a finite portion of
the probe wave. This term, then, is the same as would be expected
if there were no pulse present at all. Since, in that case, plane
waves have the form $\exp\left(-i\bar{\omega}\tau+i\frac{1}{2}\beta_{2,m}u\zeta\bar{\omega}^{2}\right)$
(see Eq. (\ref{eq:dispersion_difference_frequencies})), we have\[
C_{\delta}\left(\zeta\right)=\exp\left(i\frac{1}{2}\beta_{2,m}u\zeta\bar{\omega}_{p}^{2}\right)\,.\]

$\widetilde{\mathcal{A}}_{\mathrm{cross}}\left(\bar{w},\zeta\right)$
is fairly straightforward to evaluate, although some care must be
taken with regard to the signs of $\bar{u}$ and $\bar{\omega}_{p}$
since the integral is only half infinite. The result is\[
\widetilde{\mathcal{A}}_{\mathrm{cross}}\left(\bar{w},\zeta\right)=\frac{f_{\mathrm{cross}}\left(\bar{w}\right)-f_{\mathrm{cross}}\left(\bar{\omega}_{p}\right)}{\bar{w}-\bar{\omega}_{p}}+\frac{f_{\mathrm{cross}}^{\prime}\left(\bar{w}\right)-f_{\mathrm{cross}}^{\prime}\left(-\bar{\omega}_{p}\right)}{\bar{w}+\bar{\omega}_{p}}\,,\]
where we have defined\begin{eqnarray*}
f_{\mathrm{cross}}\left(\bar{w}\right) & = & \mathrm{Im}\left[T_{\bar{w}}\right]\exp\left(i\frac{1}{2}\beta_{2,m}u\zeta\bar{w}^{2}\right)\,,\\
f_{\mathrm{cross}}^{\prime}\left(\bar{w}\right) & = & \mathrm{Im}\left[R_{\bar{w}}\right]\exp\left(i\frac{1}{2}\beta_{2,m}u\zeta\bar{w}^{2}\right)\,.\end{eqnarray*}

The remaining term, $\widetilde{\mathcal{A}}_{\mathrm{poles}}\left(\bar{w},\zeta\right)$,
can, after some manipulation, be brought to the simplified form\[
\widetilde{A}_{\mathrm{poles}}\left(\bar{w},\zeta\right)=\frac{f_{\mathrm{poles}}\left(\bar{w}\right)-f_{\mathrm{poles}}\left(\bar{\omega}_{p}\right)}{\bar{w}-\bar{\omega}_{p}}+\frac{f_{\mathrm{poles}}^{\prime}\left(\bar{w}\right)-f_{\mathrm{poles}}^{\prime}\left(-\bar{\omega}_{p}\right)}{\bar{w}+\bar{\omega}_{p}}\,,\]
where we have defined\begin{eqnarray*}
f_{\mathrm{poles}}\left(\bar{w}\right) & = & \frac{1}{\pi}\int_{-\infty}^{+\infty}\mathrm{Re}\left[1-T_{\bar{\omega}}\right]\frac{\exp\left(i\frac{1}{2}\beta_{2,m}u\zeta\bar{\omega}^{2}\right)}{\bar{\omega}-\bar{w}}d\bar{\omega}\,,\\
f_{\mathrm{poles}}^{\prime}\left(\bar{w}\right) & = & \frac{1}{\pi}\int_{-\infty}^{+\infty}\mathrm{Re}\left[-R_{\bar{\omega}}\right]\frac{\exp\left(i\frac{1}{2}\beta_{2,m}u\zeta\bar{\omega}^{2}\right)}{\bar{\omega}-\bar{w}}d\bar{\omega}\,.\end{eqnarray*}
Combining these, we find a more illuminating form of the spectrum
than Eq. (\ref{eq:separating_cross_and_poles}); that is, we can write\begin{equation}
\widetilde{\mathcal{A}}\left(\bar{w},\zeta\right)\approx\exp\left(i\frac{1}{2}\beta_{2,m}u\zeta\bar{\omega}_{p}^{2}\right)\,2\pi\,\delta\left(\bar{w}-\bar{\omega}_{p}\right)+\frac{F\left(\bar{w}\right)-F\left(\bar{\omega}_{p}\right)}{\bar{w}-\bar{\omega}_{p}}+\frac{F^{\prime}\left(\bar{w}\right)-F^{\prime}\left(-\bar{\omega}_{p}\right)}{\bar{w}+\bar{\omega}_{p}}\,,\label{eq:separating_two_peaks}\end{equation}
where we have defined\begin{alignat}{1}
F\left(\bar{w}\right)=f_{\mathrm{cross}}\left(\bar{w}\right)+f_{\mathrm{poles}}\left(\bar{w}\right)\,,\qquad & F^{\prime}\left(\bar{w}\right)=f_{\mathrm{cross}}^{\prime}\left(\bar{w}\right)+f_{\mathrm{poles}}^{\prime}\left(\bar{w}\right)\,.\end{alignat}
Rather than splitting into three parts for mathematical convenience,
as in Eq. (\ref{eq:separating_cross_and_poles}), we see in Eq. (\ref{eq:separating_two_peaks})
that the spectrum splits into three parts of physical significance:
a $\delta$ function at the input frequency $\bar{\omega}_{p}$, a
shoulder surrounding this $\delta$ function, and a peak at the shifted
frequency $-\bar{\omega}_{p}$.

\begin{figure}
\subfloat{\includegraphics[width=0.45\columnwidth]{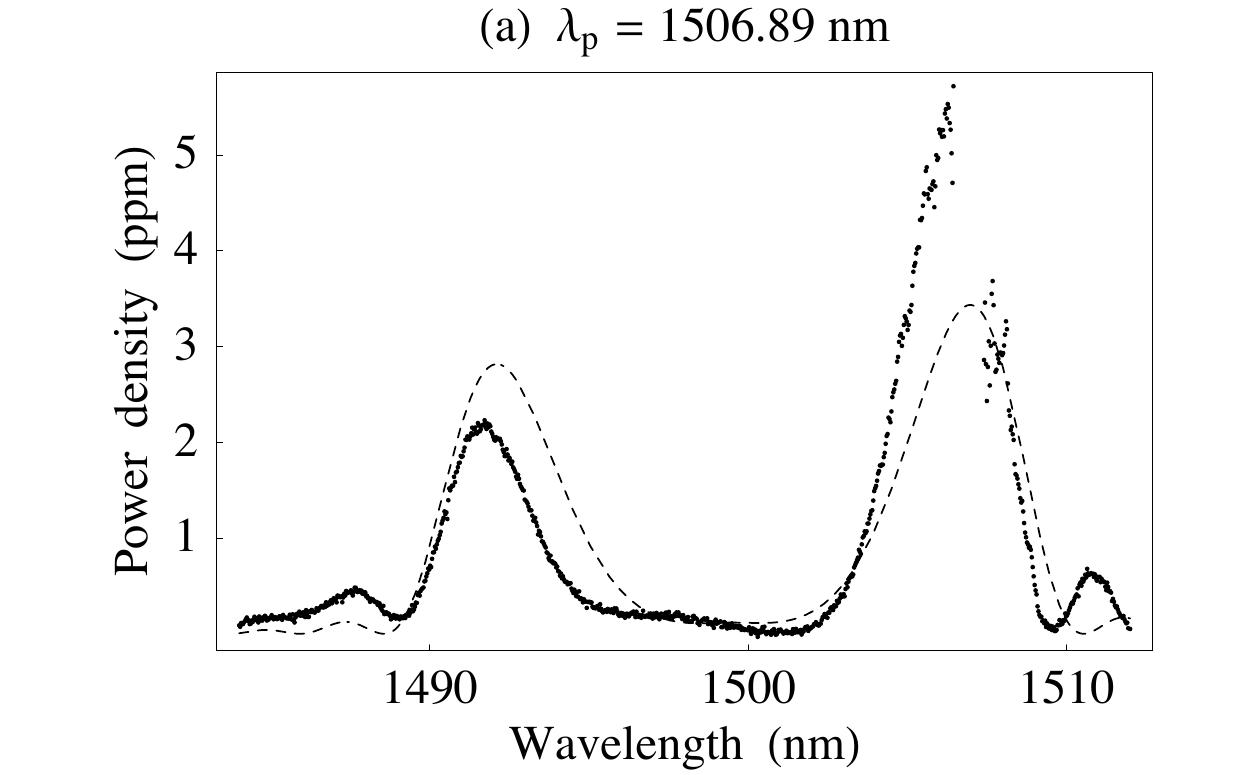}}\subfloat{\includegraphics[width=0.45\columnwidth]{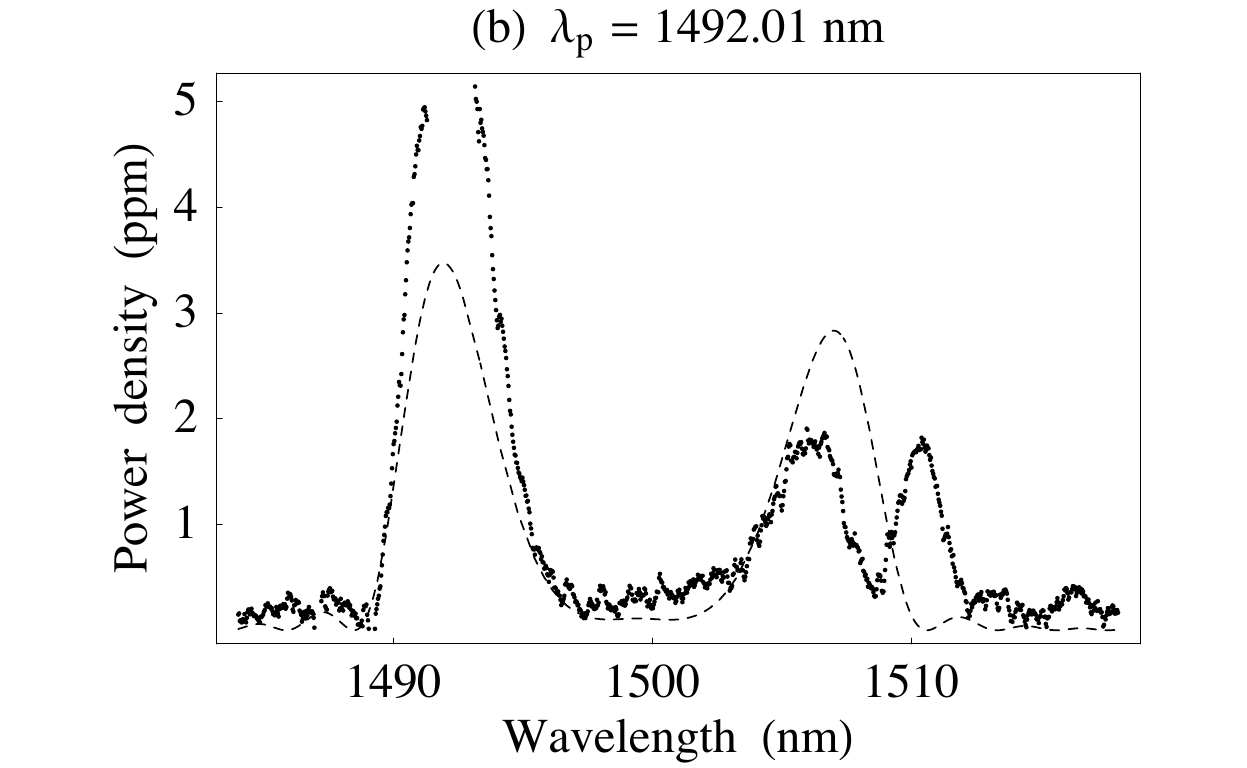}}

\caption[\textsc{Frequency shifting spectra for a constant-velocity pulse}]{\textsc{Frequency shifting spectra for a constant-velocity pulse}:
The group-velocity-matching wavelength is $\lambda_{m}=1499.47\,\mathrm{nm}$.
Figure $\left(a\right)$ shows a blueshifting spectrum, while Figure
$\left(b\right)$ is a redshifting spectrum. Data points represent
experimental observations, while dashed curves are theoretical predictions
(from Eq. (\ref{eq:separating_two_peaks})). The experimentally obtained
spectra are normalized with respect to the probe power and are measured
in parts per million (ppm). The heights of the theoretical curves
were fitted using the least-$\chi^{2}$ method. (Data courtesy
of Chris Kuklewicz and Friedrich K\"onig.) \label{fig:Comparison_non_decelerating}}

\end{figure}

Figure \ref{fig:Comparison_non_decelerating} shows plots of the predicted
spectrum of Eq. (\ref{eq:separating_two_peaks}), together with experimentally
observed frequency-shifting spectra.

\section{The Raman effect}

Despite the careful analysis given above, there is a discrepancy between
the theoretical predictions it generates and the experimental results.
Qualitatively, the appearance of the $\delta$ function and the peaks
at the initial and shifted frequencies has been explained. However,
there is disagreement over the relative heights of these peaks, and
there are noticeable differences with regard to the side-bands, which
are relatively small according to theory but quite prominent in experiment.

A possible explanation for these discrepancies is our neglect of the
Raman effect \cite{Agrawal} in the previous calculation. Whereas the Kerr effect
- which is responsible for the interaction between the pulse and the
probe - is caused by the nonlinear response of the electrons of the
fiber medium to the electric field, the Raman effect is caused by
the nonlinear response of the nuclei (i.e., molecular vibrations)
of the fibre medium. This nuclear response is much slower than the
electronic response, and in silica takes place on a time scale of
around $60-70\:\mathrm{fs}$ \cite{Stolen-1989,Blow-Wood-1989,Hollenbeck-Cantrell-2002,Lin-Agrawal-2006}. For pulses much longer
than this, the nuclear response is essentially instantaneous and cannot
be distinguished from the electronic response, so that there is no
discernible Raman effect. However, for pulses with lengths on the
same order as this time scale, the nuclear response cannot be regarded
as instantaneous, and new effects emerge. Energy is transferred to
vibrational modes within the fiber, at the cost of energy of the pulse.
This causes a gradual redshift of the pulse's carrier frequency; and,
since the pulse is a soliton and exists in a region of anomalous dispersion,
this redshift is accompanied by a decrease in its group velocity.
As far as frequency shifting is concerned, this deceleration is the
Raman effect's most important consequence. As we have seen, energy
is shifted between frequencies that are equidistant from the group-velocity-matching
frequency, $\omega_{m}$, but this depends crucially on the group
velocity of the pulse. Since the dispersion at $\omega_{m}$ is normal,
$\omega_{m}$ must increase as $\omega_{s}$ decreases, and an incident
probe wave will be reflected at a frequency which is continuously
increasing.

To account for a decelerating pulse, the coordinate transformation
that moves into a co-moving frame must be different from that which
we have used. Suppose that, in the lab frame, the pulse propagates
with velocity\begin{equation}
u\left(z\right)=u_{0}-\frac{az}{u_{0}}\,,\label{eq:decreasing_soliton_velocity}\end{equation}
where $a$ has units of acceleration. The time taken for the pulse
to propagate a distance $z$ is\begin{multline*}
T\left(z\right)=\int_{0}^{z}\frac{dz^{\prime}}{u\left(z^{\prime}\right)}=\frac{1}{u_{0}}\int_{0}^{z}\frac{dz^{\prime}}{1-az^{\prime}/u_{0}^{2}}\approx\frac{1}{u_{0}}\int_{0}^{z}\left(1+\frac{az^{\prime}}{u_{0}^{2}}\right)dz^{\prime}=\frac{z}{u_{0}}+\frac{az^{2}}{2u_{0}^{3}}\,.\end{multline*}
Note that this is valid only when $\left|az/u_{0}^{2}\right|\ll1$;
this is true if the relative change in the velocity of the pulse is
much less than unity. The co-moving frame - in which the pulse appears
stationary - can be described by the coordinates\begin{alignat}{1}
\tau=t-\frac{z}{u_{0}}-\frac{az^{2}}{2u_{0}^{3}}\,,\qquad & \zeta=\frac{z}{u_{0}}\,.\label{eq:coordinates_decelerating_frame}\end{alignat}
These coordinates reduce to those of a reference frame with uniform
velocity $u_{0}$ as $a\rightarrow0$. Partial derivatives transform
according to the chain rule:\begin{alignat}{1}
\partial_{z}=\frac{1}{u_{0}}\left[\partial_{\zeta}-\left(1+\frac{a\zeta}{u_{0}}\right)\partial_{\tau}\right]\,,\qquad & \partial_{t}=\partial_{\tau}\,.\label{eq:derivatives_decelerating_frame}\end{alignat}
Just as we define wavenumber and frequency in the lab frame as derivatives
of the phase,\begin{alignat*}{1}
k=\frac{\partial\varphi}{\partial z}\,,\qquad & \omega=-\frac{\partial\varphi}{\partial t}\,,\end{alignat*}
so we define two frequencies in the co-moving frame as derivatives
of $\tau$ and $\zeta$,\begin{alignat*}{1}
\omega=-\frac{\partial\varphi}{\partial\tau}\,,\qquad & \omega^{\prime}=-\frac{\partial\varphi}{\partial\zeta}\,,\end{alignat*}
and these are related via the transformation of partial derivatives
given above. As before, we find $\omega$ is the same in both frames
(hence the use of the same label). On the other hand,\[
\omega^{\prime}=\omega-u_{0}k+\frac{a\zeta}{u_{0}}\omega=\omega-u_{0}\beta\left(\omega\right)+\frac{a\zeta}{u_{0}}\omega\,,\]
where, in the second step, we have imposed the condition that $k$
be positive, and hence that all waves be forward-propagating. The
only difference from the non-decelerating case is the appearance of
a $\zeta$-dependent term.

To find the optical wave equation in the decelerating frame, we return
to the full wave equation,\begin{equation}
\partial_{z}^{2}E+\beta^{2}\left(i\partial_{t}\right)E=\frac{1}{\epsilon_{0}c^{2}}\partial_{t}^{2}P\,.\label{eq:electric_field_wave_equation_FULL}\end{equation}
Since $\partial_{t}=\partial_{\tau}$, an identity that holds true
even in the non-decelerating case, the terms $\beta^{2}\left(i\partial_{t}\right)E$
and $\partial_{t}^{2}P/\left(\epsilon_{0}c^{2}\right)$ are unaltered
from the previous analysis, while $\partial_{z}^{2}E$ is computed
below. As before, we define the slowly-varying envelope $\mathcal{E}$
such that\[
E=\mathcal{E}\,\exp\left(i\beta_{m}z-i\omega_{m}t\right)=\mathcal{E}\,\exp\left(-i\omega_{m}\tau-i\left(\omega_{m}-u_{0}\beta_{m}\right)\zeta-i\omega_{m}\frac{a\zeta^{2}}{2u_{0}}\right)\,,\]
where we have replaced the lab coordinates $z$ and $t$ with the
new coordinates of the decelerating frame, as defined in Eq. (\ref{eq:coordinates_decelerating_frame}).
Then, from previous calculations, we have\begin{eqnarray}
\!\!\!\!\!\!\!\!\!\!\!\!\!\!\!\!\!\!\!\!\!\!\!\!\!\beta^{2}\left(i\partial_{t}\right)E & \approx & \left[\beta_{m}^{2}\mathcal{E}+2i\beta_{m}\frac{1}{u_{0}}\partial_{\tau}\mathcal{E}-\left(\beta_{m}\beta_{2,m}+\frac{1}{u_{0}^{2}}\right)\partial_{\tau}^{2}\mathcal{E}\right]\exp\left(i\beta_{m}z-i\omega_{m}t\right)\,\,\,\,\,\,\,\,\,\,\label{eq:decelerating_frame_betasq}\\
\frac{1}{\epsilon_{0}c^{2}}\partial_{t}^{2}P & \approx & -2\kappa\frac{\omega_{m}^{2}}{c^{2}}I\mathcal{E}\exp\left(i\beta_{m}z-i\omega_{m}t\right)\,.\label{eq:decelerating_frame_dt2_polarization}\end{eqnarray}
For $\partial_{z}^{2}E$, we substitute the partial derivatives given
in Eq. (\ref{eq:derivatives_decelerating_frame}):\begin{eqnarray}
\partial_{z}^{2}E & = & \frac{1}{u_{0}^{2}}\left[\partial_{\zeta}-\left(1+\frac{a\zeta}{u_{0}}\right)\partial_{\tau}\right]^{2}\mathcal{E}\,\exp\left(-i\omega_{m}\tau-i\left(\omega_{m}-u_{0}\beta_{m}\right)\zeta-i\omega_{m}\frac{a\zeta^{2}}{2u_{0}}\right)\nonumber \\
 & = & \frac{1}{u_{0}^{2}}\left[\partial_{\zeta}-\left(1+\frac{a\zeta}{u_{0}}\right)\partial_{\tau}\right]\left[\partial_{\zeta}\mathcal{E}-\left(1+\frac{a\zeta}{u_{0}}\right)\partial_{\tau}\mathcal{E}+iu_{0}\beta_{m}\mathcal{E}\right]\nonumber \\
 &  & \qquad\qquad\qquad\qquad\qquad\qquad\qquad\times\exp\left(-i\omega_{m}\tau-i\left(\omega_{m}-u_{0}\beta_{m}\right)\zeta-i\omega_{m}\frac{a\zeta^{2}}{2u_{0}}\right)\nonumber \\
 & = & \frac{1}{u_{0}^{2}}\left[\partial_{\zeta}^{2}\mathcal{E}-2\left(1+\frac{a\zeta}{u_{0}}\right)\partial_{\zeta}\partial_{\tau}\mathcal{E}+2iu_{0}\beta_{m}\partial_{\zeta}\mathcal{E}-\left(2iu_{0}\beta_{m}\left(1+\frac{a\zeta}{u_{0}}\right)+\frac{a}{u_{0}}\right)\partial_{\tau}\mathcal{E}\right.\nonumber \\
 &  & \qquad\qquad\left.+\left(1+\frac{a\zeta}{u_{0}}\right)^{2}\partial_{\tau}^{2}\mathcal{E}-u_{0}^{2}\beta_{m}^{2}\mathcal{E}\right]\exp\left(-i\omega_{m}\tau-i\left(\omega_{m}-u_{0}\beta_{m}\right)\zeta-i\omega_{m}\frac{a\zeta^{2}}{2u_{0}}\right)\nonumber \\
 & \approx & \frac{1}{u_{0}^{2}}\left[2iu_{0}\beta_{m}\partial_{\zeta}\mathcal{E}-\left(2iu_{0}\beta_{m}\left(1+\frac{a\zeta}{u_{0}}\right)+\frac{a}{u_{0}}\right)\partial_{\tau}\mathcal{E}+\left(1+\frac{a\zeta}{u_{0}}\right)^{2}\partial_{\tau}^{2}\mathcal{E}-u_{0}^{2}\beta_{m}^{2}\mathcal{E}\right]\nonumber \\
 &  & \qquad\qquad\qquad\qquad\qquad\qquad\qquad\qquad\qquad\qquad\times\exp\left(i\beta_{m}z-i\omega_{m}t\right)\,,\label{eq:decelerating_frame_dz2_electric_field}\end{eqnarray}
where, in the last step, we used the slowly-varying envelope approximation
to neglect the second-order derivatives $\partial_{\zeta}^{2}\mathcal{E}$
and $\partial_{\zeta}\partial_{\tau}\mathcal{E}$. Substituting Eqs.
(\ref{eq:decelerating_frame_betasq})-(\ref{eq:decelerating_frame_dz2_electric_field})
into Eq. (\ref{eq:electric_field_wave_equation_FULL}), we find the
wave equation for the envelope $\mathcal{E}$:\[
i\partial_{\zeta}\mathcal{E}=\left(i\frac{a\zeta}{u_{0}}+\frac{a}{2\beta_{m}u_{0}^{2}}\right)\partial_{\tau}\mathcal{E}+\left(\frac{\beta_{2,m}u_{0}}{2}-\frac{a\zeta}{\beta_{m}u_{0}^{2}}\left(1+\frac{a\zeta}{2u_{0}}\right)\right)\partial_{\tau}^{2}\mathcal{E}-2\kappa\frac{\omega_{m}^{2}}{c^{2}}I\mathcal{E}\,.\]
It can be shown that the second term of each bracket is negligibly
small in comparison to the first, so that this simplifies to\begin{equation}
i\partial_{\zeta}\mathcal{E}=i\frac{a\zeta}{u_{0}}\partial_{\tau}\mathcal{E}+\frac{\beta_{2,m}u_{0}}{2}\partial_{\tau}^{2}\mathcal{E}-2\kappa\frac{\omega_{m}^{2}}{c^{2}}I\mathcal{E}\,.\label{eq:envelope_wave_eqn}\end{equation}
Let us make the substitution\begin{alignat}{1}
\mathcal{E}=\psi\,\exp\left(-i\phi\right)\,,\qquad & \phi=\frac{a\zeta\tau}{\beta_{2,m}u_{0}^{2}}+\frac{a^{2}\zeta^{3}}{6\beta_{2,m}u_{0}^{3}}\,.\label{eq:extracting_phase_from_E}\end{alignat}
Then Eq. (\ref{eq:envelope_wave_eqn}) becomes\begin{multline*}
i\left(\partial_{\zeta}\psi-i\frac{a\tau}{\beta_{2,m}u_{0}^{2}}\psi-i\frac{a^{2}\zeta^{2}}{2\beta_{2,m}u_{0}^{3}}\psi\right)e^{-i\phi}=i\frac{a\zeta}{u_{0}}\left(\partial_{\tau}\psi-i\frac{a\zeta}{\beta_{2,m}u_{0}^{2}}\psi\right)e^{-i\phi}\\
+\frac{\beta_{2,m}u_{0}}{2}\left(\partial_{\tau}^{2}\psi-2i\frac{a\zeta}{\beta_{2,m}u_{0}^{2}}\partial_{\tau}\psi-\frac{a^{2}\zeta^{2}}{\beta_{2,m}^{2}u_{0}^{4}}\psi\right)e^{-i\phi}-2\kappa\frac{\omega_{m}^{2}}{c^{2}}I\psi e^{-i\phi}\,,\end{multline*}
and, rearranging, we have\begin{equation}
-i\partial_{\zeta}\psi=-\frac{\beta_{2,m}u_{0}}{2}\partial_{\tau}^{2}\psi+\left(\frac{a\tau}{\beta_{2,m}u_{0}^{2}}+2\kappa\frac{\omega_{m}^{2}}{c^{2}}I\right)\psi\,.\label{eq:Schrodinger_eqn}\end{equation}
If the pulse is a fundamental soliton, this becomes\begin{equation}
-i\partial_{\zeta}\psi=-\frac{\beta_{2,m}u_{0}}{2}\partial_{\tau}^{2}\psi+\left(\frac{a\tau}{\beta_{2,m}u_{0}^{2}}+ru_{0}\frac{\gamma_{m}}{\gamma_{s}}\frac{\left|\beta_{2,s}\right|}{T_{s}^{2}}\mathrm{sech}^{2}\left(\frac{\tau}{T_{s}}\right)\right)\psi\,.\label{eq:Schrodinger_eqn_soliton}\end{equation}
As in the non-decelerating case, we have arrived at a wave equation
equivalent in form to the time-dependent Schrödinger equation. The
mathematical effect of the deceleration is now made apparent: the
Schrödinger {}``potential'' now includes not only the pulse intensity,
but also a term linear in $\tau$ and proportional to the magnitude
of the deceleration.

\section{Accelerated waves in the absence of a pulse}

Given the typically short duration of the pulse, we are, as before,
more interested in asymptotic solutions of Eq. (\ref{eq:Schrodinger_eqn})
which hold in the limit $\tau\rightarrow\pm\infty$. In these regions,
the intensity of the pulse is zero, and the wave equation is simply\begin{equation}
-i\partial_{\zeta}\psi=-\frac{\beta_{2,m}u_{0}}{2}\partial_{\tau}^{2}\psi+\frac{a\tau}{\beta_{2,m}u_{0}^{2}}\psi\,.\label{eq:time_dep_Schrodinger_linear_potential}\end{equation}
Factoring out a $\zeta$-dependent phase,\begin{equation}
\psi=\psi_{\tau_{0}}\exp\left(i\frac{a\zeta\tau_{0}}{\beta_{2,m}u_{0}^{2}}\right)\,,\label{eq:extracting_phase_(Airy)}\end{equation}
where $\tau_{0}$ is a free parameter (analogous to energy in the
usual Schrödinger equation), the wave equation is transformed to the
form of the time-independent Schrödinger equation,\begin{equation}
-\frac{\beta_{2,m}u_{0}}{2}\partial_{\tau}^{2}\psi+\frac{a\left(\tau-\tau_{0}\right)}{\beta_{2,m}u_{0}^{2}}\psi=0\,.\label{eq:Schrodinger_linear_potential}\end{equation}
This can be simplified further by defining the dimensionless variables\begin{alignat}{1}
\theta=\frac{\tau}{\tau_{1}}\,,\qquad & \theta_{0}=\frac{\tau_{0}}{\tau_{1}}\,,\label{eq:dimensionless_variables}\end{alignat}
where the characteristic time scale $\tau_{1}$ is given by\begin{equation}
\tau_{1}^{3}=\frac{\beta_{2,m}^{2}u_{0}^{3}}{2a}\,.\label{eq:tau1_defn}\end{equation}
Then Eq. (\ref{eq:Schrodinger_linear_potential}) becomes\begin{equation}
\partial_{\theta}^{2}\psi_{\theta_{0}}=\left(\theta-\theta_{0}\right)\psi_{\theta_{0}}\,.\end{equation}
This is the well-known Airy equation; its solutions are the Airy functions
$\mathrm{Ai}\,\left(\theta-\theta_{0}\right)$ and $\mathrm{Bi}\,\left(\theta-\theta_{0}\right)$.
Since $\mathrm{Bi}\,\left(\theta-\theta_{0}\right)$ increases exponentially
as $\theta\rightarrow+\infty$, in the absence of a pulse $\mathrm{Ai}\,\left(\theta-\theta_{0}\right)$
represents the only physical solution.

\begin{figure}
\includegraphics[width=0.8\columnwidth]{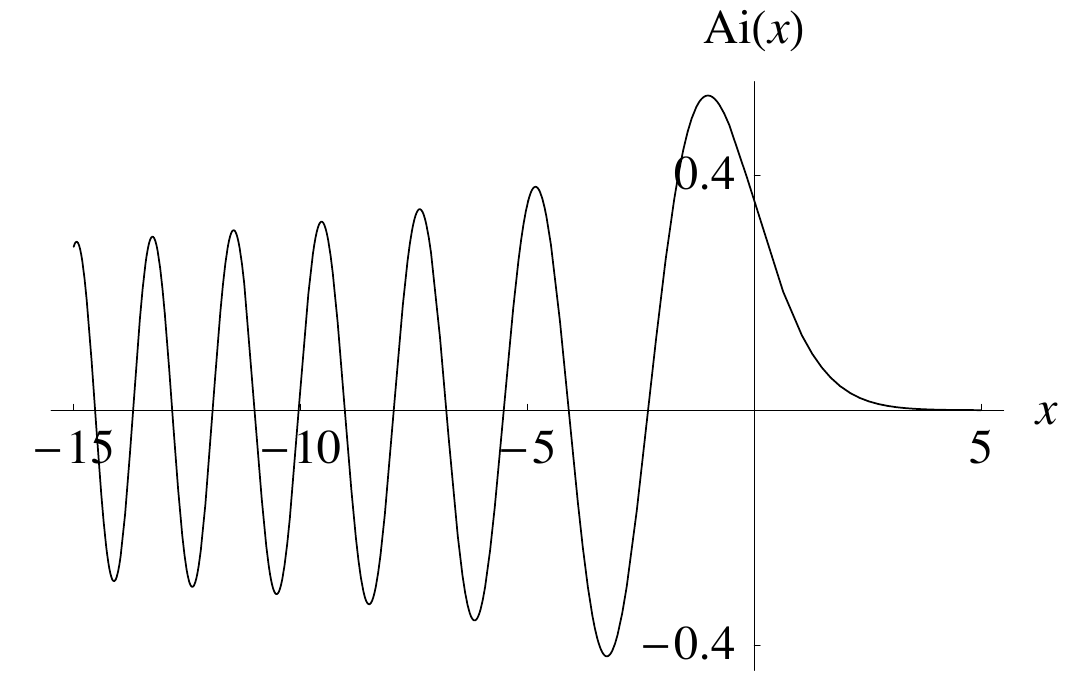}

\caption[\textsc{The Airy function}]{\textsc{The Airy function}: $\mathrm{Ai}\,\left(x\right)$ is exponentially
decreasing for positive $x$ and oscillatory for negative $x$.\label{fig:Airy_function}}

\end{figure}

Let us pause to consider the meanings of the parameters $\tau_{0}$
and $\tau_{1}$. Consider the Airy function $\mathrm{Ai}\,\left(\theta-\theta_{0}\right)$,
plotted in Fig. (\ref{fig:Airy_function}): for positive $\theta-\theta_{0}$,
it decays exponentially; for negative $\theta-\theta_{0}$, its amplitude
decays as $\left(\theta_{0}-\theta\right)^{-1/4}$, while it oscillates
with continuously increasing frequency $\omega\propto\left(\theta_{0}-\theta\right)^{1/2}$.

$\tau_{1}$ is a wave scale; it determines the characteristic oscillation
period of the envelope. As we would expect, these oscillations become
more prominent for larger acceleration $a$, since $\tau_{1}$ - and
hence the oscillation period - decreases.

$\tau_{0}$ represents the {}``turning point'' which marks the boundary
between oscillatory behaviour and exponential decay; it is a point
beyond which the wave barely penetrates. This boundedness of the wave
is closely related to pulse trapping \cite{Nishizawa-Goto-1,Nishizawa-Goto-2,Nishizawa-Goto-3,Hill-2009},
which was explained by Gorbach and Skryabin \cite{Gorbach-Skryabin}
using a linear {}``gravity-like'' potential like that in Eq. (\ref{eq:time_dep_Schrodinger_linear_potential}).

The general solution of Eq. (\ref{eq:time_dep_Schrodinger_linear_potential})
is found by replacing the phase of Eq. (\ref{eq:extracting_phase_(Airy)})
and forming a linear superposition of the eigenfunctions $\psi_{\theta_{0}}=\mathrm{Ai}\,\left(\theta-\theta_{0}\right)$:\begin{equation}
\psi\left(\theta,\zeta\right)=\int_{-\infty}^{+\infty}c\left(\theta_{0}\right)\mathrm{Ai}\,\left(\theta-\theta_{0}\right)\exp\left(i\frac{a\zeta\tau_{1}}{\beta_{2,m}u_{0}^{2}}\theta_{0}\right)\, d\theta_{0}\,.\end{equation}

\section{Including the pulse-probe interaction}

Complete solutions of the pulse-probe interaction, where the pulse
experiences a Raman deceleration, must obey Eq. (\ref{eq:Schrodinger_eqn}).
However, there is no analytic solution for a potential which includes
both a linear term and the soliton ($\mathrm{sech}^{2}$) term. We
require a further approximation in order to find analytic expressions
for the eigenfunctions. We shall assume that the Raman deceleration
can be neglected on time scales of the order of the soliton width
$T_{s}$; that is, we assume that there is a region outside, but close
to, the pulse, in which the solutions found for a non-decelerating
pulse (Eqs. (\ref{eq:exact_solution_2F1}) and (\ref{eq:asymptotic_solns}))
are valid. (We can include the asymptotic solutions because these
are valid wherever the pulse intensity is essentially zero.) The solutions
here can be matched to Airy functions $\mathrm{Ai}\left(\theta-\theta_{0}\right)$
and $\mathrm{Bi}\left(\theta-\theta_{0}\right)$ in the asymptotic
regions.

We have already made one important conclusion about the solutions
in the decelerating case: due to the exponentially increasing behaviour
of $\mathrm{Bi}\left(\theta-\theta_{0}\right)$, the solution as $\theta\rightarrow+\infty$
must contain $\mathrm{Ai}\left(\theta-\theta_{0}\right)$ exclusively.
The pulse-probe interaction, then, must manifest itself through the
appearance of a term including $\mathrm{Bi}\left(\theta-\theta_{0}\right)$
in the asymptotic region $\theta\rightarrow-\infty$. The general
solution for $\psi_{\theta_{0}}$ is then\begin{equation}
\psi_{\theta_{0}}\left(\theta\right)\,=\,\mathcal{N}_{\theta_{0}}\,\begin{cases}
c_{\theta_{0}}^{A}\mathrm{Ai}\left(\theta-\theta_{0}\right)+c_{\theta_{0}}^{B}\mathrm{Bi}\left(\theta-\theta_{0}\right)\,, & \theta\rightarrow-\infty\\
\mathrm{Ai}\left(\theta-\theta_{0}\right)\,, & \theta\rightarrow+\infty\end{cases}\,,\label{eq:general_asymptotics_decelerating_case}\end{equation}
where $\mathcal{N}_{\theta_{0}}$ is a prefactor that ensures the
eigenfunctions are normalized according to the condition\begin{equation}
\int_{-\infty}^{+\infty}\psi_{\theta_{0}^{\prime}}\left(\theta\right)\,\psi_{\theta_{0}}\left(\theta\right)\, d\theta\,=\,\delta\left(\theta_{0}-\theta_{0}^{\prime}\right)\,.\label{eq:normalization_decelerating_case}\end{equation}
At least for the exact solutions, it must be possible to choose $\mathcal{N}_{\theta_{0}}$
such that this normalization condition holds, since they are simply
solutions of the Schrödinger equation. The $\delta$ function, however,
is determined by the asymptotics of the integrand, so we may use the
asymptotic forms of Eq. (\ref{eq:general_asymptotics_decelerating_case})
to find $\mathcal{N}_{\theta_{0}}$. The integral over positive $\theta$
involves the product of two $\mathrm{Ai}$ functions; since $\mathrm{Ai}\left(x\right)$
decays exponentially for positive $x$ ($\mathrm{Ai}\left(x\right)\sim\exp\left(-2x^{3/2}/3\right)$
\cite{Vallee-Soares,Abramowitz-Stegun}), this integral is finite,
and cannot contribute to the $\delta$ function. We are then left
with the integral over negative $\theta$, where the Airy functions
are oscillatory \cite{Vallee-Soares,Abramowitz-Stegun}:\begin{eqnarray*}
\mathrm{Ai}\left(\theta-\theta_{0}\right) & \rightarrow & \frac{1}{\sqrt{\pi}}\left(\theta_{0}-\theta\right)^{-1/4}\sin\left(\frac{2}{3}\left(\theta_{0}-\theta\right)^{3/2}+\frac{\pi}{4}\right)\\
 & \rightarrow & \frac{1}{\sqrt{\pi}}\left(-\theta\right)^{-1/4}\sin\left(\frac{2}{3}\left(-\theta\right)^{3/2}+\theta_{0}\left(-\theta\right)^{1/2}+\frac{\pi}{4}\right)\,,\end{eqnarray*}
and similarly for $\mathrm{Bi}\left(\theta-\theta_{0}\right)$ with
$\sin$ replaced by $\cos$. The products $\mathrm{Ai}\left(\theta-\theta_{0}^{\prime}\right)\mathrm{Ai}\left(\theta-\theta_{0}\right)$
and $\mathrm{Bi}\left(\theta-\theta_{0}^{\prime}\right)\mathrm{Bi}\left(\theta-\theta_{0}\right)$
result in a cosine term whose argument is the difference of the arguments
of the individual terms; more explicitly, we have\begin{eqnarray*}
\int_{-\infty}^{0}\mathrm{Ai}\left(\theta-\theta_{0}^{\prime}\right)\mathrm{Ai}\left(\theta-\theta_{0}\right) & \approx & \int_{-\infty}^{0}\frac{1}{2\pi}\left(-\theta\right)^{1/2}\cos\left(\left(\theta_{0}-\theta_{0}^{\prime}\right)\left(-\theta\right)^{1/2}\right)d\theta\\
 & = & \frac{1}{\pi}\int_{0}^{\infty}\cos\left(\left(\theta_{0}-\theta_{0}^{\prime}\right)x\right)dx\\
 & = & \frac{1}{2\pi}\int_{0}^{\infty}\left(\exp\left(i\left(\theta_{0}-\theta_{0}^{\prime}\right)x\right)+\exp\left(-i\left(\theta_{0}-\theta_{0}^{\prime}\right)x\right)\right)dx\\
 & \approx & \delta\left(\theta_{0}-\theta_{0}^{\prime}\right)\,,\end{eqnarray*}
where, in the last line, we have included only the $\delta$ function.
This calculation also holds for the integral of $\mathrm{Bi}\left(\theta-\theta_{0}^{\prime}\right)\mathrm{Bi}\left(\theta-\theta_{0}\right)$.
For the mixed product $\mathrm{Ai}\left(\theta-\theta_{0}^{\prime}\right)\mathrm{Bi}\left(\theta-\theta_{0}\right)$,
the relevant term is a sine rather than a cosine, so we have\begin{eqnarray*}
\int_{-\infty}^{0}\mathrm{Ai}\left(\theta-\theta_{0}^{\prime}\right)\mathrm{Bi}\left(\theta-\theta_{0}\right)d\theta & \approx & \int_{-\infty}^{0}\frac{1}{2\pi}\left(-\theta\right)^{1/2}\sin\left(\left(\theta_{0}-\theta_{0}^{\prime}\right)\left(-\theta\right)^{1/2}\right)d\theta\\
 & = & \frac{1}{\pi}\int_{0}^{\infty}\sin\left(\left(\theta_{0}-\theta_{0}^{\prime}\right)x\right)dx\\
 & = & \frac{1}{2i\pi}\int_{0}^{\infty}\left(\exp\left(i\left(\theta_{0}-\theta_{0}^{\prime}\right)x\right)-\exp\left(-i\left(\theta_{0}-\theta_{0}^{\prime}\right)x\right)\right)dx\\
 & \approx & 0\,.\end{eqnarray*}
The last line includes only the $\delta$ function term, which vanishes.
Therefore, on substitution of Eq. (\ref{eq:general_asymptotics_decelerating_case})
into Eq. (\ref{eq:normalization_decelerating_case}), we find\[
\mathcal{N}_{\theta_{0}^{\prime}}\mathcal{N}_{\theta_{0}}\left(c_{\theta_{0}^{\prime}}^{A}c_{\theta_{0}}^{A}+c_{\theta_{0}^{\prime}}^{B}c_{\theta_{0}}^{B}\right)\delta\left(\theta_{0}-\theta_{0}^{\prime}\right)=\delta\left(\theta_{0}-\theta_{0}^{\prime}\right),\]
which holds if the normalizing prefactor is defined to be\begin{equation}
\mathcal{N}_{\theta_{0}}=\frac{1}{\sqrt{\left(c_{\theta_{0}}^{A}\right)^{2}+\left(c_{\theta_{0}}^{B}\right)^{2}}}\,.\end{equation}

All that remains is to find the coefficients $c_{\theta_{0}}^{A}$
and $c_{\theta_{0}}^{B}$. For this purpose, it is useful to distinguish
three cases: $\theta_{0}\lesssim-1$, $-1\lesssim\theta_{0}\lesssim1$
and $\theta_{0}\gtrsim1$.

In the case $\theta_{0}\lesssim-1$, the turning point is far on the
negative side. We can argue, as before, that due to its exponentially
increasing behaviour for positive argument, $\mathrm{Bi}\left(\theta-\theta_{0}\right)$
cannot appear here, and the solution is purely $\mathrm{Ai}\left(\theta-\theta_{0}\right)$.
In terms of the coefficients, this implies $c_{\theta_{0}}^{A}=1$
and $c_{\theta_{0}}^{B}=0$. From the perspective of geometrical optics,
the probe wave turns around before it has a chance to interact with
the pulse, so the solution is essentially the same as if no pulse
were present.

In the case $-1\lesssim\theta_{0}\lesssim1$, the behaviour of the
Airy functions near $\theta=0$ is not well-approximated by either
of their asymptotic forms, so there are no approximations that can
be made. Instead, we must directly solve Eq. (\ref{eq:Schrodinger_eqn_soliton}),
or, with the $\zeta$-dependent phase of Eq. (\ref{eq:extracting_phase_(Airy)})
and the dimensionless variables defined in Eqs. (\ref{eq:dimensionless_variables})
and (\ref{eq:tau1_defn}), the following equation:\begin{equation}
\partial_{\theta}^{2}\psi_{\theta_{0}}=\left[\theta-\theta_{0}+2r\frac{\gamma_{m}}{\gamma_{s}}\frac{\left|\beta_{2,s}\right|}{\beta_{2,m}}\frac{\tau_{1}^{2}}{T_{s}^{2}}\mathrm{sech}^{2}\left(\theta\frac{\tau_{1}}{T_{s}}\right)\right]\psi_{\theta_{0}}\,.\label{eq:Schrodinger_eqn_soliton_dimensionless}\end{equation}
This is done by picking a point $\theta_{\star}$ situated well outside
the soliton (typically, $\theta_{\star}\gtrsim5T_{s}/\tau_{1}$),
setting the value and first derivative of $\psi_{\theta_{0}}$ at
$\theta_{\star}$ to be $\mathrm{Ai}\left(\theta_{\star}-\theta_{0}\right)$
and $\mathrm{Ai}^{\prime}\left(\theta_{\star}-\theta_{0}\right)$,
respectively, and solving Eq. (\ref{eq:Schrodinger_eqn_soliton_dimensionless})
back to, say, $-2\theta_{\star}$. The solution in the region between
$-2\theta_{\star}$ and $-\theta_{\star}$ is then fitted to the linear
combination $c_{\theta_{0}}^{A}\mathrm{Ai}\left(\theta-\theta_{0}\right)+c_{\theta_{0}}^{B}\mathrm{Bi}\left(\theta-\theta_{0}\right)$.

In the case $\theta_{0}\gtrsim1$, the turning point is far on the
positive side, and the behaviour of the Airy functions near $\theta=0$
is oscillatory. We can write asymptotic expressions as previously,
this time treating $\left|\theta/\theta_{0}\right|\ll1$:\begin{eqnarray}
\mathrm{Ai}\left(\theta-\theta_{0}\right) & \approx & \frac{1}{\sqrt{\pi}}\left(\theta_{0}-\theta\right)^{-1/4}\sin\left(\frac{2}{3}\left(\theta_{0}-\theta\right)^{3/2}+\frac{\pi}{4}\right)\nonumber \\
 & \approx & \frac{1}{\sqrt{\pi}}\theta_{0}^{-1/4}\sin\left(\frac{2}{3}\theta_{0}^{3/2}-\theta_{0}^{1/2}\theta+\frac{\pi}{4}\right)\nonumber \\
 & = & \frac{1}{\sqrt{\pi}}\theta_{0}^{-1/4}\frac{1}{2i}\left[\exp\left(i\left(\frac{2}{3}\theta_{0}^{3/2}-\theta_{0}^{1/2}\theta+\frac{\pi}{4}\right)\right)\right.\nonumber \\
 &  & \qquad\qquad\qquad\left.-\exp\left(-i\left(\frac{2}{3}\theta_{0}^{3/2}-\theta_{0}^{1/2}\theta+\frac{\pi}{4}\right)\right)\right]\,,\label{eq:Ai_near_zero}\end{eqnarray}
\begin{eqnarray}
\mathrm{Bi}\left(\theta-\theta_{0}\right) & \approx & \frac{1}{\sqrt{\pi}}\left(\theta_{0}-\theta\right)^{-1/4}\cos\left(\frac{2}{3}\left(\theta_{0}-\theta\right)^{3/2}+\frac{\pi}{4}\right)\nonumber \\
 & \approx & \frac{1}{\sqrt{\pi}}\theta_{0}^{-1/4}\cos\left(\frac{2}{3}\theta_{0}^{3/2}-\theta_{0}^{1/2}\theta+\frac{\pi}{4}\right)\nonumber \\
 & = & \frac{1}{\sqrt{\pi}}\theta_{0}^{-1/4}\frac{1}{2}\left[\exp\left(i\left(\frac{2}{3}\theta_{0}^{3/2}-\theta_{0}^{1/2}\theta+\frac{\pi}{4}\right)\right)\right.\nonumber \\
 &  & \qquad\qquad\qquad\left.+\exp\left(-i\left(\frac{2}{3}\theta_{0}^{3/2}-\theta_{0}^{1/2}\theta+\frac{\pi}{4}\right)\right)\right]\,.\label{eq:Bi_near_zero}\end{eqnarray}
Combining these expressions leads to the expression for a plane wave
in terms of Airy functions, valid near $\theta=0$:\begin{equation}
\exp\left(-i\theta_{0}^{1/2}\theta\right)\approx\sqrt{\pi}\theta_{0}^{1/4}\exp\left(-i\frac{2}{3}\theta_{0}^{3/2}\right)\left[e^{i\pi/4}\mathrm{Ai}\left(\theta-\theta_{0}\right)+e^{-i\pi/4}\mathrm{Bi}\left(\theta-\theta_{0}\right)\right]\,.\label{eq:plane_wave_to_Airy}\end{equation}
$\exp\left(i\theta_{0}^{1/2}\theta\right)$ is, of course, given simply
by the complex conjugate of this. We may now use the non-decelerating
solutions of Eq. (\ref{eq:asymptotic_solns}) to find the decelerating
solutions. Starting from the fact that the solution for positive $\theta$
must be $\mathrm{Ai}\left(\theta-\theta_{0}\right)$ (we ignore the
normalization for now), we expand this using Eq. (\ref{eq:Ai_near_zero})
above to find the form of the solution, for positive $\theta$, near
the soliton:\[
\psi_{\theta_{0}}\sim\frac{1}{\sqrt{\pi}}\theta_{0}^{-1/4}\frac{1}{2i}\exp\left(i\left(\frac{2}{3}\theta_{0}^{3/2}+\frac{\pi}{4}\right)\right)\exp\left(-i\theta_{0}^{1/2}\theta\right)+\mathrm{c.c.}\]
Eq. (\ref{eq:asymptotic_solns}) transforms this into a solution,
still near the soliton, but for \textit{negative} $\theta$:\[
\psi_{\theta_{0}}\sim\frac{1}{\sqrt{\pi}}\theta_{0}^{-1/4}\frac{1}{2i}\exp\left(i\left(\frac{2}{3}\theta_{0}^{3/2}+\frac{\pi}{4}\right)\right)\frac{1}{T_{\theta_{0}^{1/2}/\tau_{1}}}\left[\exp\left(-i\theta_{0}^{1/2}\theta\right)+R_{\theta_{0}^{1/2}/\tau_{1}}\exp\left(i\theta_{0}^{1/2}\theta\right)\right]\,+\,\mathrm{c.c.}\,,\]
where $R_{\bar{\omega}}$ and $T_{\bar{\omega}}$ are the reflection
and transmission amplitudes defined in Eqs. (\ref{eq:reflection_coefficient})
and (\ref{eq:transmission_coefficient}). We may now use Eq. (\ref{eq:plane_wave_to_Airy})
and its complex conjugate to express this solution in terms of Airy
functions, allowing it to hold for all negative $\theta$:\begin{eqnarray*}
\psi_{\theta_{0}} & \sim & \frac{1}{2T_{\theta_{0}^{1/2}/\tau_{1}}}\left[\exp\left(i\frac{4}{3}\theta_{0}^{3/2}\right)R_{\theta_{0}^{1/2}/\tau_{1}}\left(\mathrm{Bi}\left(\theta-\theta_{0}\right)-i\,\mathrm{Ai}\left(\theta-\theta_{0}\right)\right)+\mathrm{Ai}\left(\theta-\theta_{0}\right)-i\,\mathrm{Bi}\left(\theta-\theta_{0}\right)\right]+\mathrm{c.c.}\\
 & = & \mathrm{Re}\left\{ \frac{1}{T_{\theta_{0}^{1/2}/\tau_{1}}}\left[1-i\,\exp\left(i\frac{4}{3}\theta_{0}^{3/2}\right)R_{\theta_{0}^{1/2}/\tau_{1}}\right]\right\} \mathrm{Ai}\left(\theta-\theta_{0}\right)\\
 &  & \qquad\qquad+\,\mathrm{Im}\left\{ \frac{1}{T_{\theta_{0}^{1/2}/\tau_{1}}}\left[1+i\,\exp\left(i\frac{4}{3}\theta_{0}^{3/2}\right)R_{\theta_{0}^{1/2}/\tau_{1}}\right]\right\} \mathrm{Bi}\left(\theta-\theta_{0}\right)\,.\end{eqnarray*}
We can now read off the expressions for the coefficients $c_{\theta_{0}}^{A}$
and $c_{\theta_{0}}^{B}$:\begin{eqnarray}
c_{\theta_{0}}^{A} & = & \mathrm{Re}\left\{ \frac{1}{T_{\theta_{0}^{1/2}/\tau_{1}}}\left[1-i\,\exp\left(i\frac{4}{3}\theta_{0}^{3/2}\right)R_{\theta_{0}^{1/2}/\tau_{1}}\right]\right\} \,,\\
c_{\theta_{0}}^{B} & = & \mathrm{Im}\left\{ \frac{1}{T_{\theta_{0}^{1/2}/\tau_{1}}}\left[1+i\,\exp\left(i\frac{4}{3}\theta_{0}^{3/2}\right)R_{\theta_{0}^{1/2}/\tau_{1}}\right]\right\} \,.\end{eqnarray}

\section{Solution in the decelerating case}

Having derived the form of the eigenfunctions when a constant deceleration
of the pulse is included, we now use these in the calculation of the
spectrum due to frequency shifting. The general solution of Eq. (\ref{eq:Schrodinger_eqn_soliton_dimensionless})
is a linear superposition of the eigenfunctions $\psi_{\theta_{0}}$
with the $\zeta$-dependent phase of Eq. (\ref{eq:extracting_phase_(Airy)})
replaced:\begin{equation}
\psi\left(\theta,\zeta\right)=\int_{-\infty}^{+\infty}c\left(\theta_{0}\right)\psi_{\theta_{0}}\left(\theta\right)\exp\left(i\frac{a\zeta\tau_{1}}{\beta_{2,m}u_{0}^{2}}\theta_{0}\right)d\theta_{0}\,.\label{eq:general_soln_decelerating}\end{equation}
The coefficients $c\left(\theta_{0}\right)$ are completely determined
by the value of $\psi$ at $\zeta=0$; for, using the orthonormality
condition of Eq. (\ref{eq:normalization_decelerating_case}), we have\begin{eqnarray*}
\int_{-\infty}^{+\infty}\psi\left(\theta,0\right)\psi_{\theta_{0}}\left(\theta\right)d\theta & = & \int_{-\infty}^{+\infty}\int_{-\infty}^{+\infty}c\left(\theta_{0}^{\prime}\right)\psi_{\theta_{0}^{\prime}}\left(\theta\right)\psi_{\theta_{0}}\left(\theta\right)d\theta_{0}^{\prime}d\theta_{0}\\
 & = & \int_{-\infty}^{+\infty}c\left(\theta_{0}^{\prime}\right)\delta\left(\theta_{0}-\theta_{0}^{\prime}\right)d\theta_{0}^{\prime}\\
 & = & c\left(\theta_{0}\right)\,.\end{eqnarray*}
The general solution for the envelope $\mathcal{E}$, where the electric
field $E=\mathcal{E}\exp\left(i\beta_{m}z-i\omega_{m}t\right)$, is
found by replacing the phases extracted in Eq. (\ref{eq:extracting_phase_from_E}):\begin{equation}
\mathcal{E}\left(\theta,\zeta\right)=\exp\left(-i\frac{a\zeta\tau_{1}}{\beta_{2,m}u_{0}^{2}}\theta-i\frac{a^{2}\zeta^{3}}{6\beta_{2,m}u_{0}^{3}}\right)\psi\left(\theta,\zeta\right)\,.\end{equation}

As before, our initial wave is the value of the continuous-wave probe
at the input of the fibre ($\zeta=0$); that is,\begin{equation}
\psi\left(\theta,0\right)=\exp\left(-i\bar{\omega}_{p}\tau_{1}\theta\right)\,,\label{eq:initial_value_decelerating}\end{equation}
where $\bar{\omega}_{p}=\omega_{p}-\omega_{m}$. The coefficients
$c\left(\theta_{0}\right)$, then, are given by\begin{eqnarray*}
c\left(\theta_{0}\right) & = & \int_{-\infty}^{+\infty}\exp\left(-i\bar{\omega}_{p}\tau_{1}\theta\right)\psi_{\theta_{0}}\left(\theta\right)d\theta\\
 & = & \widetilde{\psi}_{\theta_{0}}\left(-\bar{\omega}_{p}\tau_{1}\right)\,,\end{eqnarray*}
where we have defined the Fourier-transformed eigenfunction\begin{equation}
\widetilde{\psi}_{\theta_{0}}\left(w\right)=\int_{-\infty}^{+\infty}\exp\left(iw\theta\right)\psi_{\theta_{0}}\left(\theta\right)d\theta\,.\label{eq:Fourier_transformed_eigenfunction}\end{equation}
The continuous wave of Eq. (\ref{eq:initial_value_decelerating})
therefore evolves according to\[
\psi\left(\theta,\zeta\right)=\int_{-\infty}^{+\infty}\widetilde{\psi}_{\theta_{0}}\left(-\bar{\omega}_{p}\tau_{1}\right)\psi_{\theta_{0}}\left(\theta\right)\exp\left(i\frac{a\zeta\tau_{1}}{\beta_{2,m}u_{0}^{2}}\theta_{0}\right)d\theta_{0}\,,\]
and the probe envelope is given by\[
\mathcal{E}\left(\theta,\zeta\right)=\exp\left(-i\frac{a^{2}\zeta^{3}}{6\beta_{2,m}u_{0}^{3}}\right)\int_{-\infty}^{+\infty}\widetilde{\psi}_{\theta_{0}}\left(-\bar{\omega}_{p}\tau_{1}\right)\psi_{\theta_{0}}\left(\theta\right)\exp\left(-i\frac{a\zeta\tau_{1}}{\beta_{2,m}u_{0}^{2}}\left(\theta-\theta_{0}\right)\right)d\theta_{0}\,.\]

In order to find the power spectrum, which is the measured quantity
in experiment, we must take the Fourier transform of the probe envelope,
which gives\begin{equation}
\widetilde{\mathcal{E}}\left(w,\zeta\right)=\exp\left(-i\frac{a^{2}\zeta^{3}}{6\beta_{2,m}u_{0}^{3}}\right)\int_{-\infty}^{+\infty}\widetilde{\psi}_{\theta_{0}}\left(-\bar{\omega}_{p}\tau_{1}\right)\widetilde{\psi}_{\theta_{0}}\left(w-\frac{a\zeta\tau_{1}}{\beta_{2,m}u_{0}^{2}}\right)\exp\left(i\frac{a\zeta\tau_{1}}{\beta_{2,m}u_{0}^{2}}\theta_{0}\right)d\theta_{0}\,.\label{eq:Fourier_transformed_spectrum}\end{equation}
This is not yet in a form suitable for numerical integration, because
the spectrum contains a $\delta$ function, for the same reason given
in the non-decelerating case: a finite portion of the probe wave interacts
with the pulse, while far from the pulse it evolves just as a continuous
wave in the absence of a pulse. We therefore expect Eq. (\ref{eq:Fourier_transformed_spectrum})
to contain a term $\delta\left(w-\bar{\omega}_{p}\tau_{1}\right)$,
and this must first be extracted in order to make the numerical integration
stable. To this end, we introduce some further definitions. Consider
the identity\begin{equation}
\int_{-\infty}^{+\infty}\exp\left(iw\theta\right)\mathrm{Ai}\left(\theta-\theta_{0}\right)d\theta=\exp\left(-i\frac{w^{3}}{3}+iw\theta_{0}\right)\label{eq:Ai_Fourier}\end{equation}
as motivation for the following definitions:\begin{eqnarray}
\int_{-\infty}^{0}\exp\left(iw\theta\right)\mathrm{Ai}\left(\theta-\theta_{0}\right)d\theta & = & \exp\left(-i\frac{w^{3}}{3}+iw\theta_{0}\right)a_{\theta_{0}}\left(w\right)\,,\label{eq:Ai_half_Fourier}\\
\int_{-\infty}^{0}\exp\left(iw\theta\right)\mathrm{Bi}\left(\theta-\theta_{0}\right)d\theta & = & \exp\left(-i\frac{w^{3}}{3}+iw\theta_{0}\right)b_{\theta_{0}}\left(w\right)\,.\label{eq:Bi_half_Fourier}\end{eqnarray}
Both $a_{\theta_{0}}\left(w\right)$ and $b_{\theta_{0}}\left(w\right)$
must approach zero as $\theta_{0}\rightarrow-\infty$, while as $\theta_{0}\rightarrow+\infty$,
$a_{\theta_{0}}\left(w\right)$ approaches unity while $b_{\theta_{0}}\left(w\right)$
diverges exponentially. Note that Eqs. (\ref{eq:Ai_Fourier}) and
(\ref{eq:Ai_half_Fourier}) together imply the following:\begin{equation}
\int_{0}^{+\infty}\exp\left(iw\theta\right)\mathrm{Ai}\left(\theta-\theta_{0}\right)d\theta=\exp\left(-i\frac{w^{3}}{3}+iw\theta_{0}\right)\left[1-a_{\theta_{0}}\left(w\right)\right]\,.\label{eq:Ai_other_half_Fourier}\end{equation}
Substitution of Eqs. (\ref{eq:Ai_half_Fourier})-(\ref{eq:Ai_other_half_Fourier})
in Eq. (\ref{eq:Fourier_transformed_eigenfunction}) gives, for the
Fourier-transformed eigenfunction,\begin{equation}
\widetilde{\psi}_{\theta_{0}}\left(w\right)=\exp\left(-i\frac{w^{3}}{3}+iw\theta_{0}\right)\xi_{\theta_{0}}\left(w\right)\,,\label{eq:Fourier_eigenfunction}\end{equation}
where we have defined\begin{equation}
\xi_{\theta_{0}}\left(w\right)=\mathcal{N}_{\theta_{0}}\left[1+\left(c_{\theta_{0}}^{A}-1\right)a_{\theta_{0}}\left(w\right)+c_{\theta_{0}}^{B}b_{\theta_{0}}\left(w\right)\right]\,.\end{equation}
The function $\xi_{\theta_{0}}\left(w\right)$ approaches unity both
as $\theta_{0}\rightarrow-\infty$ and as $\theta_{0}\rightarrow+\infty$.
It is this asymptotic behaviour that generates the $\delta$ function.
Plugging Eq. (\ref{eq:Fourier_eigenfunction}) into Eq. (\ref{eq:Fourier_transformed_spectrum}),
we have\begin{eqnarray}
\!\!\!\!\!\!\!\!\!\!\!\!\!\!\!\!\!\!\!\!\widetilde{\mathcal{E}}\left(w,\zeta\right) & = & \exp\left(-i\frac{a^{2}\zeta^{3}}{6\beta_{2,m}u_{0}^{3}}+i\frac{1}{3}\left(\bar{\omega}_{p}\tau_{1}\right)^{3}-i\frac{1}{3}\left(w-\frac{a\zeta\tau_{1}}{\beta_{2,m}u_{0}^{2}}\right)^{3}\right)\nonumber \\
 &  & \qquad\int_{-\infty}^{+\infty}\exp\left(i\left(w-\bar{\omega}_{p}\tau_{1}\right)\theta_{0}\right)\xi_{\theta_{0}}\left(-\bar{\omega}_{p}\tau_{1}\right)\xi_{\theta_{0}}\left(w-\frac{a\zeta\tau_{1}}{\beta_{2,m}u_{0}^{2}}\right)d\theta_{0}\nonumber \\
 & = & \exp\left(-i\frac{a^{2}\zeta^{3}}{6\beta_{2,m}u_{0}^{3}}+i\frac{1}{3}\left(\bar{\omega}_{p}\tau_{1}\right)^{3}-i\frac{1}{3}\left(w-\frac{a\zeta\tau_{1}}{\beta_{2,m}u_{0}^{2}}\right)^{3}\right)\nonumber \\
 &  & \Bigg\{2\pi\,\delta\left(w-\bar{\omega}_{p}\tau_{1}\right)\nonumber \\
 &  & +\int_{-\infty}^{+\infty}\exp\left(i\left(w-\bar{\omega}_{p}\tau_{1}\right)\theta_{0}\right)\left[\xi_{\theta_{0}}\left(-\bar{\omega}_{p}\tau_{1}\right)\xi_{\theta_{0}}\left(w-\frac{a\zeta\tau_{1}}{\beta_{2,m}u_{0}^{2}}\right)-1\right]d\theta_{0}\Bigg\}\,\,\,\,\,\,\,\,\,\,\label{eq:decelerating_spectrum}\end{eqnarray}

\section{Comparison with experimental results}

In order to take the Raman deceleration into account, we must have
some idea of its value, or, equivalently, of the value of $\tau_{1}$.
This can be estimated from measurable quantities as follows. Dispersion
is normally measured via the group-velocity dispersion $D=\partial^{2}\beta/\partial\omega\partial\lambda$,
or, since $\partial\beta/\partial\omega$ is simply the inverse of
the group velocity, $D=\partial\left(1/v_{g}\right)/\partial\lambda$.
If the wavelength shift of the soliton, $\Delta\lambda_{s}$, is measured,
then we have\begin{equation}
\frac{1}{v_{f}}-\frac{1}{v_{i}}\approx D_{s}\Delta\lambda_{s}\,,\end{equation}
where $v_{i}$ and $v_{f}$ are the initial and final values of the
soliton's group velocity, and $D_{s}$ is the value of $D$ at the
soliton wavelength. Then\begin{equation}
v_{f}-v_{i}=v_{f}v_{i}\left(\frac{v_{f}-v_{i}}{v_{f}v_{i}}\right)=v_{f}v_{i}\left(\frac{1}{v_{i}}-\frac{1}{v_{f}}\right)\approx-u_{0}^{2}D_{s}\Delta\lambda_{s}\,,\end{equation}
where $u_{0}$ is the initial velocity of the soliton, and we have
assumed that its fractional change over the course of the fibre is
much less than unity. According to Eq. (\ref{eq:decreasing_soliton_velocity}),
we have $v_{f}-v_{i}=-aL/u_{0}$, where $L$ is the length of the
fibre. Therefore,\begin{equation}
\frac{a}{u_{0}^{3}}\approx\frac{D_{s}\Delta\lambda_{s}}{L}\,.\end{equation}
Plugging this into the definition of $\tau_{1}$ (Eq. (\ref{eq:tau1_defn})),
and using the identity $\beta_{2}=\partial^{2}\beta/\partial\omega^{2}=-\left(\lambda^{2}/2\pi c\right)D$,
we find\begin{equation}
\tau_{1}\approx\left(\frac{D_{m}^{2}\lambda_{m}^{4}L}{8\pi^{2}c^{2}D_{s}\Delta\lambda_{s}}\right)^{1/3}\,.\label{eq:estimating_tau1}\end{equation}
In our experimental setup, this gives $\tau_{1}\approx650\,\mathrm{fs}$.

The above derivation of the value of $\tau_{1}$ took an experimentally
observed value of the soliton wavelength shift, $\Delta\lambda_{s}$.
We may also reinforce the conclusion by using an analytic expression
for this shift (as given in Reference \cite{Agrawal}):\begin{equation}
\Delta\omega_{s}=\omega_{f}-\omega_{i}=-\frac{8T_{R}\left|\beta_{2,s}\right|L}{15T_{s}^{4}}=-\frac{4T_{R}\lambda_{s}^{2}D_{s}L}{15\pi cT_{s}^{4}}\,,\end{equation}
where $T_{s}$ is the soliton width (as appears in Eq. (\ref{eq:soliton_power}))
and $T_{R}$ is a Raman parameter that is found to be around $3\,\mathrm{fs}$
in silica. The change in wavelength is therefore\begin{equation}
\Delta\lambda_{s}\approx\frac{\partial\lambda}{\partial\omega}\Delta\omega_{s}=-\frac{\lambda_{s}^{2}}{2\pi c}\Delta\omega_{s}=\frac{2T_{R}\lambda_{s}^{4}D_{s}L}{15\pi^{2}c^{2}T_{0}^{4}}\,.\end{equation}
Plugging this into Eq. (\ref{eq:estimating_tau1}), we find\begin{equation}
\tau_{1}\approx\left(\frac{15\lambda_{m}^{4}D_{m}^{2}T_{s}}{16\lambda_{s}^{4}D_{s}^{2}T_{R}}\right)^{1/3}T_{s}\,.\end{equation}
Again, we find $\tau_{1}\approx650\,\mathrm{fs}$.

\begin{figure}
\subfloat{\includegraphics[width=0.45\columnwidth]{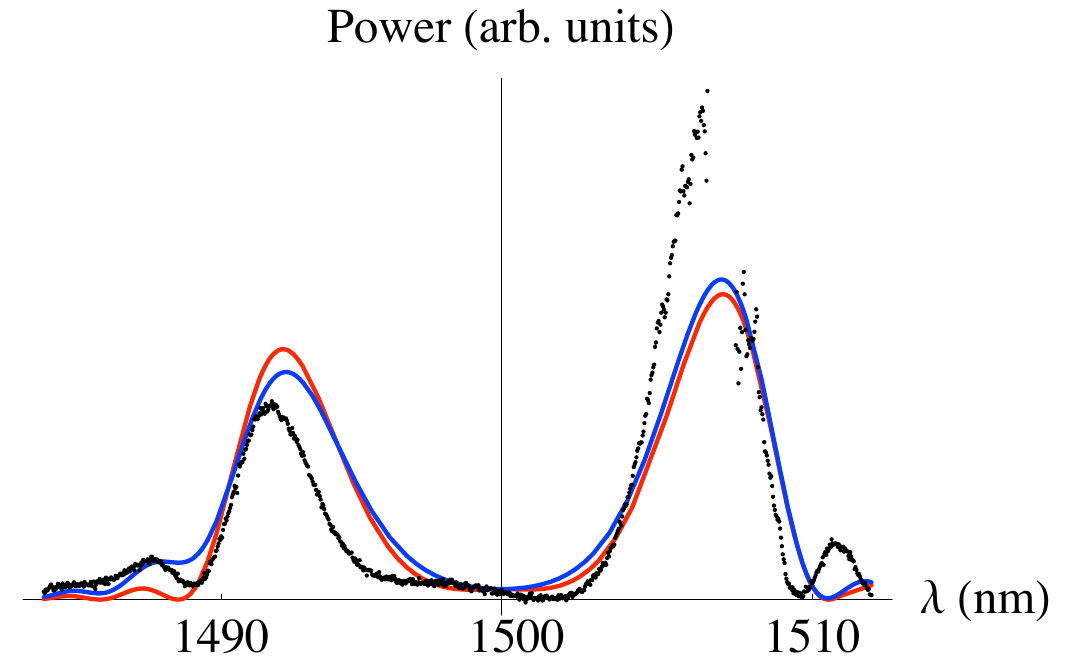}}\subfloat{\includegraphics[width=0.45\columnwidth]{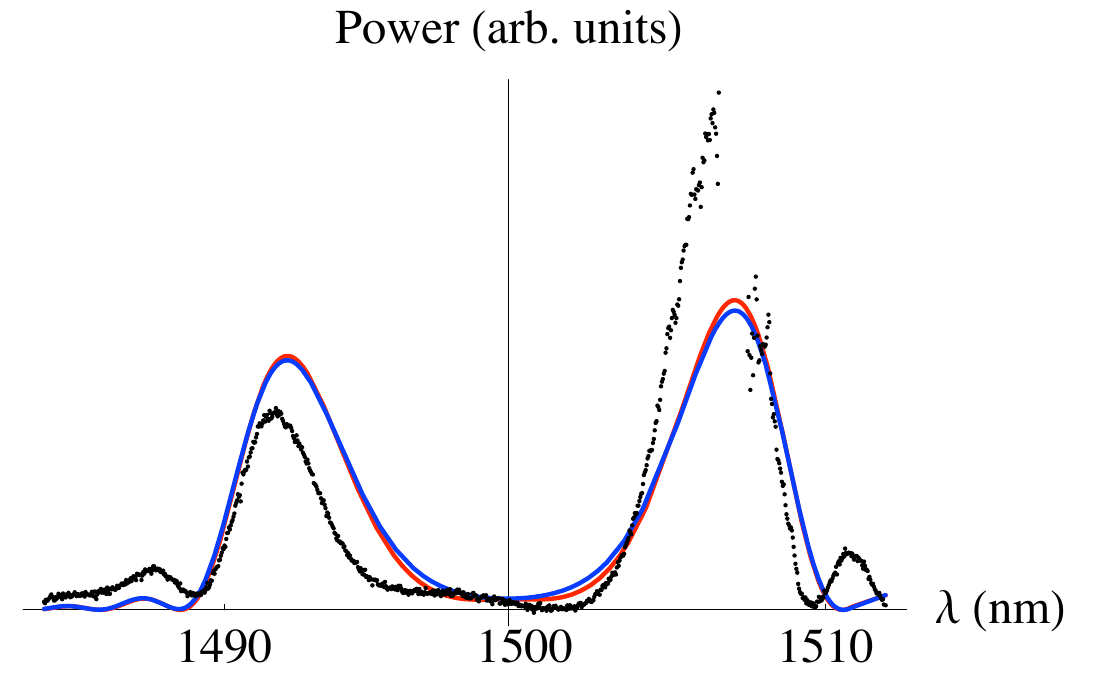}}

\subfloat{\includegraphics[width=0.45\columnwidth]{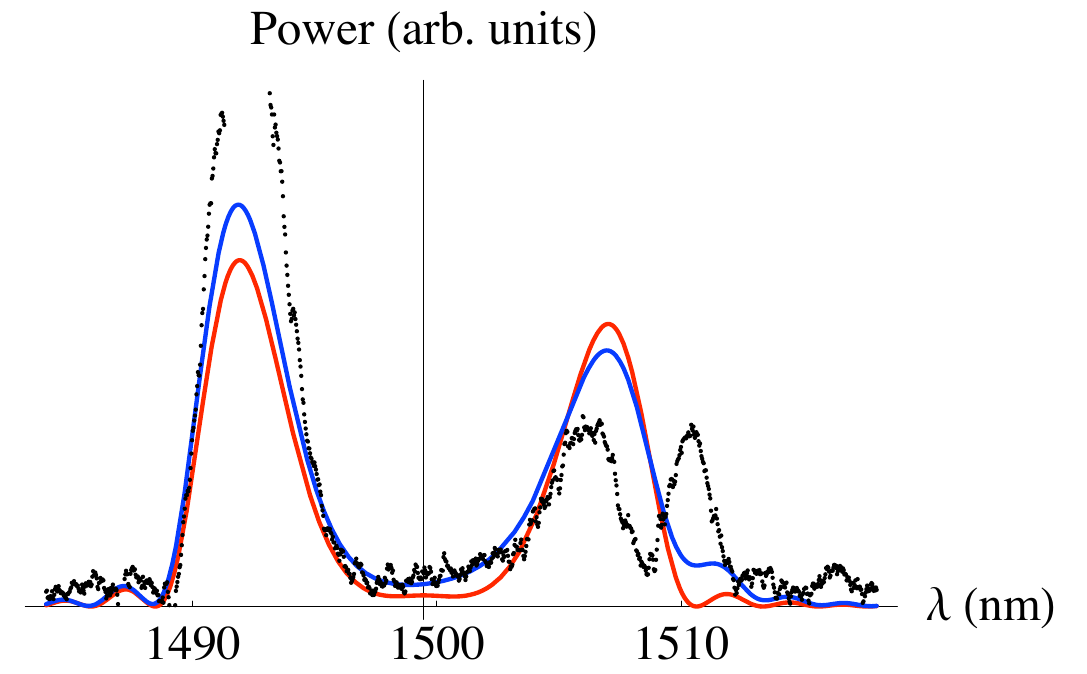}}\subfloat{\includegraphics[width=0.45\columnwidth]{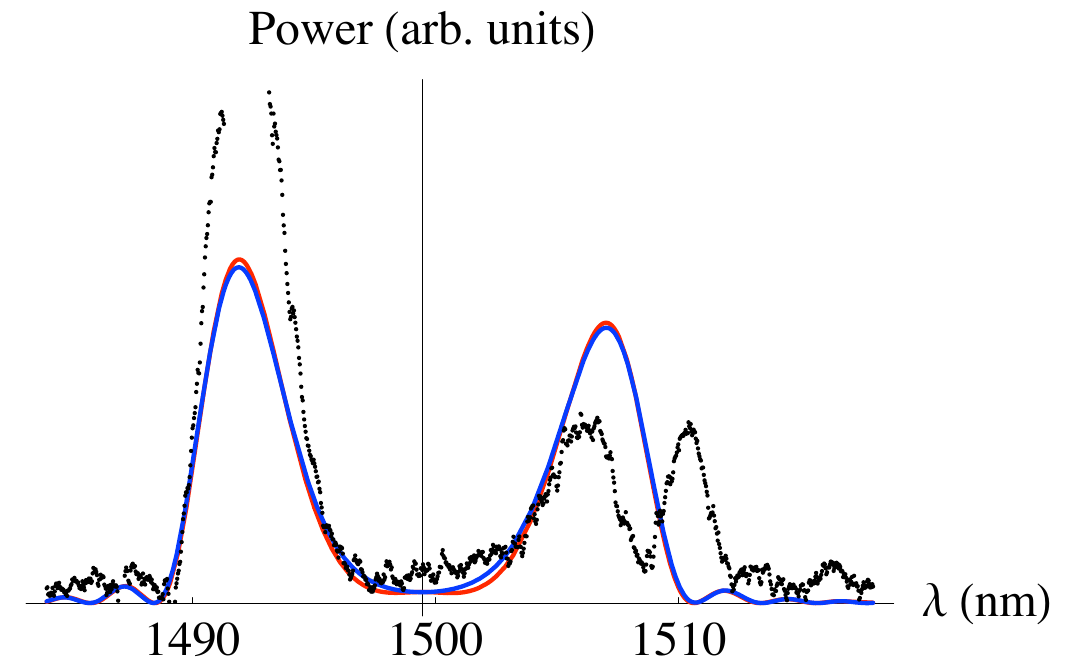}}

\caption[\textsc{Frequency shifting spectra including Raman deceleration}]{\textsc{Frequency shifting spectra including Raman deceleration}:
The dots are experimental data, while the curves are theoretical predictions,
without (dashed line) and with (solid line) the Raman effect. The
experimentally obtained spectra are normalized with respect to the
probe power and are measured in parts per million (ppm). The heights
of the theoretical curves were fitted using the least-$\chi^{2}$
method. (Data courtesy
of Chris Kuklewicz and Friedrich K\"onig.) \label{fig:Comparing_theory_and_experiment-1}}

\end{figure}

Figure \ref{fig:Comparing_theory_and_experiment-1} shows plots of
the theoretical predictions for the spectra - both without (Eq. (\ref{eq:separating_two_peaks}))
and with (Eq. (\ref{eq:decelerating_spectrum})) the Raman effect
- and the experimentally observed spectra. We see that, if $\tau_{1}$
is taken to be $650\,\mathrm{fs}$ as deduced above, then the Raman
deceleration of the soliton has very little effect: the spectra are
almost the same as those calculated when the soliton propagates at
a constant speed. Plots are also shown when $\tau_{1}$ takes a value
of $325\,\mathrm{fs}$, for it is only at around this value that the
Raman effect makes appreciable differences to the frequency-shifting
spectra. Since $a\propto\tau_{1}^{-3}$, this corresponds to a deceleration
- and a wavelength shift - higher than the deduced values by a factor
of $8$. It thus appears that the discrepancies between theory and
experiment are \textit{not} due to the Raman effect.

\section{Conclusion and discussion}

The Kerr interaction between pulse and probe does, indeed, induce
the equivalent of a group-velocity horizon for some probe frequencies.
Frequency shifting occurs, and has been experimentally observed. The
theoretical prediction varies negligibly when the Raman effect is
taken into account - the Raman deceleration must be increased by an
order of magnitude before it has any noticeable effect. However, a
discrepancy between theory and experiment remains, and is as yet unresolved.

\pagebreak{}

\chapter{The Fibre-Optical Model for Hawking Radiation\label{sec:The-Fibre-Optical-Model}}

The previous chapter has shown that pulses in optical fibres do exhibit
features of black- and white-hole horizons, and that the effects of
these horizons are experimentally detectable. This raises expectations
that Hawking radiation might also be among these features. In this
chapter, we present an analysis almost exactly analogous to that of
Chapter \ref{sec:The-Acoustic-Model}, but using instead the probe/pulse
system described in §§\ref{sub:Nonlinear-effects} and \ref{sub:Co-moving-frame}.
This shall lead us to conclude that Hawking radiation is indeed possible
in this system.

\section{Action and scalar product\label{sub:FIBRES-Action-and-scalar-product}}

We begin, as before, with an action integral:\begin{equation}
S=\int\int d\tau\, d\zeta\, u\, L\left(A,\partial_{\zeta}A,\partial_{\tau}A,\partial_{\tau}^{2}A,\ldots\right)\,.\label{eq:action_integral_FIBRES}\end{equation}
The Lagrangian for electromagnetic fields is given by \cite{Landau-Lifshitz-Fields}
\begin{eqnarray}
L & = & \frac{1}{2}\epsilon_{0}\left(n^2|E|^2-c^2|B|^2\right) \nonumber \\
& = & \frac{1}{2}\epsilon_{0}\left(n^2|\partial_{t}A|^2-c^2|\partial_{z}A|^2\right)\,.
\end{eqnarray}
The refractive index depends on both the dispersion of the fibre and, as we have seen
(see Eq. (\ref{eq:effective_n_including_chi})), on the nonlinearity induced by a pulse.
Given that the dispersion is normally described via the propagation constant
$\beta=n\omega/c$, the Lagrangian is naturally modified to include the dispersive and
nonlinear effects as follows:
\begin{eqnarray}
L & = & \frac{1}{2}\epsilon_{0}\left[c^2|\beta\left(i\partial_{t}\right)A|^{2}+\chi|\partial_{t}A|^{2}-c^2|\partial_{z}A|^{2}\right] \nonumber \\
& = & \frac{1}{2}\epsilon_{0}\left[c^{2}\left|\beta\left(i\partial_{\tau}\right)A\right|^{2}+\chi\left|\partial_{\tau}A\right|^{2}-\frac{c^{2}}{u^{2}}\left|\left(\partial_{\zeta}-\partial_{\tau}\right)A\right|^{2}\right]\,,\label{eq:Lagrangian_FIBRES}\end{eqnarray}
where, in the second line, we have transformed to the coordinates $\tau$ and $\zeta$ via the transformation of
Eqs. (\ref{eq:transformation_derivative_z}) and (\ref{eq:transformation_derivative_t}).
Then, plugging into the Euler-Lagrange equation (\ref{eq:Euler_Lagrange_higher_derivatives})
(with appropriate re-labelling), we find the wave equation\begin{equation}
\frac{c^{2}}{u^{2}}\left(\partial_{\zeta}-\partial_{\tau}\right)^{2}A+c^{2}\beta^{2}\left(i\partial_{\tau}\right)A-\partial_{\tau}\left(\chi\partial_{\tau}A\right)=0\,.\label{eq:wave_equation_FIBRES}\end{equation}

In the case where the fibre is dispersionless, with a refractive index
$n$ that is independent of frequency, the Lagrangian reduces to\begin{equation}
L=\frac{1}{2}\epsilon_{0}\left[\left(n^{2}+\chi\right)\left|\partial_{\tau}A\right|^{2}-\frac{c^{2}}{u^{2}}\left|\left(\partial_{\zeta}-\partial_{\tau}\right)A\right|^{2}\right]\,,\label{eq:Lagrangian_dispersionless_FIBRES}\end{equation}
and we see that the effect of the nonlinearity $\chi$ is to increase
the refractive index in accordance with Eq. (\ref{eq:effective_n_including_chi}).
Unfortunately, Eq. (\ref{eq:Lagrangian_dispersionless_FIBRES}) cannot
be written in the form for a massless scalar field, i.e., there is
no metric such that $L=\frac{1}{2}\sqrt{-g}g^{\mu\nu}\partial_{\mu}A^{\star}\partial_{\nu}A$.
For, in order to find such a metric, we must multiply Eq. (\ref{eq:Lagrangian_dispersionless_FIBRES})
by a factor of $u/c/\sqrt{n^{2}+\chi}$, and since $\chi$ is $\tau$-dependent,
this alters the wave equation. However, we can gain some insight by
continuing down this road, the following results being exact if $\chi$
is constant. The resulting Lagrangian corresponds to a massless scalar
field in a spacetime with metric\begin{equation}
ds^{2}=\frac{u^{2}}{c^{2}}\left(n^{2}+\chi\right)d\zeta^{2}-\left(d\tau+d\zeta\right)^{2}\,.\label{eq:approx_dispersionless_metric}\end{equation}
Comparing with the general form of Eq. (\ref{eq:generalized_Lemaitre_metric}),
we see that Eq. (\ref{eq:approx_dispersionless_metric}) describes
a moving medium in which the flow velocity is constant ($V=-1$),
but the velocity of null trajectories (i.e., the velocity of waves)
with respect to the medium has the $\tau$-dependent value $\frac{u^{2}}{c^{2}}\left(n^{2}+\chi\right)$.
This is in accordance with the discussion in §\ref{sub:FIBRES-Conclusion-and-discussion}:
the pulse affects not the velocity of the fibre, but of the waves
with respect to the fibre. An event horizon is established wherever
$\chi=c^{2}/u^{2}-n^{2}$, and the Hawking temperature is given, as
in Eq. (\ref{eq:Hawking_temperature}), by the derivative of the wave
velocity at the horizon:\begin{equation}
T=\frac{1}{2\pi}\frac{d}{d\tau}\left[\frac{u}{c}\sqrt{n^{2}+\chi}-1\right]_{\chi=c^{2}/u^{2}-n^{2}}=\frac{1}{4\pi}\frac{u^{2}}{c^{2}}\chi^{\prime}\left(\tau_{h}\right)\,.\label{eq:Hawking_temp_approx_metric}\end{equation}
This is inexact, but we shall see in §\ref{sub:Predicted-behaviour}
that, if $\chi=c^{2}/u^{2}-n^{2}$ at some point and the variation
in $\chi$ is small, the resulting Hawking temperature does indeed
agree with Eq. (\ref{eq:Hawking_temp_approx_metric}).

Returning to the exact Lagrangian of Eq. (\ref{eq:Lagrangian_FIBRES}),
we remark that, like its acoustic counterpart (see §\ref{sub:The-wave-equation}),
it is invariant with respect to both translation in $\zeta$ ({}``time''
translation) and phase rotation of $A$. The former leads to the existence
of stationary modes, discussed in the next section; the latter leads
to conservation of the scalar product\begin{eqnarray}
\left(A_{1},A_{2}\right) & = & i\,\epsilon_{0}\frac{c^{2}}{u^{2}}\int d\tau\left[A_{1}^{\star}\left(\partial_{\zeta}-\partial_{\tau}\right)A_{2}-A_{2}\left(\partial_{\zeta}-\partial_{\tau}\right)A_{1}^{\star}\right]\nonumber \\
 & = & i\,\int d\tau\left[A_{2}\pi_{1}^{\star}-A_{1}^{\star}\pi_{2}\right]\,,\label{eq:scalar_product_FIBRES}\end{eqnarray}
where $\pi$ is the canonical momentum,\begin{equation}
\pi=\frac{\partial L}{\partial\left(\partial_{\zeta}A^{\star}\right)}=-\epsilon_{0}\frac{c^{2}}{u^{2}}\left(\partial_{\zeta}-\partial_{\tau}\right)A\,.\label{eq:optical_canonical_momentum_density}\end{equation}
In regions where $\chi$ is constant in $\tau$, the formula for the
scalar product can be written in a more instructive form. We separate
the branches of the dispersion relation into those with positive and
negative lab frame wavenumbers, i.e., we define the two branches via
the dispersion relations $\left(\omega-\omega^{\prime}\right)/u=\left|\beta_{\chi}\left(\omega\right)\right|$
and $\left(\omega-\omega^{\prime}\right)/u=-\left|\beta_{\chi}\left(\omega\right)\right|$,
or\begin{equation}
\omega^{\prime}=\omega\mp u\left|\beta_{\chi}\left(\omega\right)\right|\,,\label{eq:two_dispersion_branches}\end{equation}
where $\beta_{\chi}\left(\omega\right)=\beta\left(\omega\right)\sqrt{1+\chi/n^{2}\left(\omega\right)}$.
(Note that the \textit{minus} sign in Eq. (\ref{eq:two_dispersion_branches})
corresponds to a \textit{positive} lab-frame wavenumber, and vice
versa.) Any solution of the optical wave equation can be expressed
as a sum of two Fourier integrals, covering the two branches of the
dispersion relation:\begin{multline}
A\left(\tau,\zeta\right)=\frac{1}{2\pi}\int_{-\infty}^{+\infty}\widetilde{A}^{\left(+\right)}\left(\omega\right)e^{-i\omega\tau-i\left(\omega-u\left|\beta_{\chi}\left(\omega\right)\right|\right)\zeta}d\omega\\
+\frac{1}{2\pi}\int_{-\infty}^{+\infty}\widetilde{A}^{\left(-\right)}\left(\omega\right)e^{-i\omega\tau-i\left(\omega+u\left|\beta_{\chi}\left(\omega\right)\right|\right)\zeta}d\omega\,.\end{multline}
The canonical momentum, defined in Eq. (\ref{eq:optical_canonical_momentum_density}),
is\begin{multline}
\pi\left(\tau,\zeta\right)=\frac{-i}{2\pi}\frac{\epsilon_{0}c^{2}}{u}\int_{-\infty}^{+\infty}\left|\beta_{\chi}\left(\omega\right)\right|\widetilde{A}^{\left(+\right)}\left(\omega\right)e^{-i\omega\tau-i\left(\omega-u\left|\beta_{\chi}\left(\omega\right)\right|\right)\zeta}d\omega\\
+\frac{i}{2\pi}\frac{\epsilon_{0}c^{2}}{u}\int_{-\infty}^{+\infty}\left|\beta_{\chi}\left(\omega\right)\right|\widetilde{A}^{\left(-\right)}\left(\omega\right)e^{-i\omega\tau-i\left(\omega+u\left|\beta_{\chi}\left(\omega\right)\right|\right)\zeta}d\omega\,.\end{multline}
Then, applying Eq. (\ref{eq:scalar_product_FIBRES}) for the scalar
product, we find\begin{equation}
\left(A_{1,}A_{2}\right)=\frac{1}{\pi}\frac{\epsilon_{0}c^{2}}{u}\int_{-\infty}^{+\infty}\left|\beta_{\chi}\left(\omega\right)\right|\left\{ \widetilde{A}_{1}^{\left(+\right)\star}\left(\omega\right)\widetilde{A}_{2}^{\left(+\right)}\left(\omega\right)-\widetilde{A}_{1}^{\left(-\right)\star}\left(\omega\right)\widetilde{A}_{2}^{\left(-\right)}\left(\omega\right)\right\} d\omega\,.\end{equation}
In the case of the norm of a wave - the scalar product of the wave
with itself - this becomes\begin{equation}
\left(A,A\right)=\frac{1}{\pi}\frac{\epsilon_{0}c^{2}}{u}\int_{-\infty}^{+\infty}\left|\beta_{\chi}\left(\omega\right)\right|\left\{ \left|\widetilde{A}^{\left(+\right)}\left(\omega\right)\right|^{2}-\left|\widetilde{A}^{\left(-\right)}\left(\omega\right)\right|^{2}\right\} d\omega\,.\end{equation}
This is completely analogous to the acoustic case, where we also found
(see §\ref{sub:Scalar-product}) that the norm was determined by the
difference of the squared magnitudes of the Fourier components corresponding
to the two different branches of the dispersion relation. That these
contribute to the norm with different signs means that it is possible
to increase the norm of each separately while maintaining the value
of the overall norm; this corresponds to pair creation. Our switching
of the roles of space and time has caused the integration to be over
$\omega$ (rather than $k$), and the factor multiplying the Fourier
components is the lab-frame wavenumber (rather than the free-fall
frequency).

\section{Decomposition into modes}

Just as for the acoustic field, we can express the general optical
field $A\left(\tau,\zeta\right)$ as a linear superposition of orthonormal
modes. For now, we shall continue to assume that $\chi$ is constant
in $\tau$ (this condition is relaxed in §\ref{sub:Transforming-to-the-omega-prime-repn}).
We first note that the branches of the dispersion relation characterised
by the sign of the norm are related via complex conjugation:\[
e^{-i\omega\tau-i\left(\omega-\left|\beta_{\chi}\left(\omega\right)\right|\right)\zeta}=\left[e^{-i\left(-\omega\right)\tau-i\left(-\omega+\left|\beta_{\chi}\left(-\omega\right)\right|\right)\zeta}\right]^{\star}\,.\]
We therefore define $\omega^{\prime}=\omega-u\left|\beta_{\chi}\left(\omega\right)\right|$
(taking the norm to be positive) and include the complex conjugates
of the resulting plane waves, forming a complete set of solutions.
The normalized plane wave solutions are defined as\begin{equation}
A_{\omega}\left(\tau,\zeta\right)=\sqrt{\frac{u}{4\pi\epsilon_{0}c^{2}\left|\beta_{\chi}\left(\omega\right)\right|}}e^{-i\omega\tau-i\left(\omega-u\left|\beta_{\chi}\left(\omega\right)\right|\right)\zeta}\,.\label{eq:optical_mode_normalization}\end{equation}
Given this definition, the scalar products between modes are given
by\begin{alignat}{1}
\left(A_{\omega_{1}},A_{\omega_{2}}\right)=-\left(A_{\omega_{1}}^{\star},A_{\omega_{2}}^{\star}\right)=\delta\left(\omega_{1}-\omega_{2}\right),\qquad & \left(A_{\omega_{1}},A_{\omega_{2}}^{\star}\right)=\left(A_{\omega_{1}}^{\star},A_{\omega_{2}}\right)=0\,.\label{eq:orthonormality_of_optical_modes}\end{alignat}
If we transform back to the lab coordinates $z$ and $t$ (using Eqs.
(\ref{eq:transformation_zeta}) and (\ref{eq:transformation_tau})),
the phase of the mode $A_{\omega}$ transforms as\begin{eqnarray*}
\varphi=-\omega\tau-\left(\omega-u\left|\beta_{\chi}\left(\omega\right)\right|\right)\zeta & = & -\omega\left(t-\frac{z}{u}\right)-\left(\omega-u\left|\beta_{\chi}\left(\omega\right)\right|\right)\frac{z}{u}\\
 & = & \left|\beta_{\chi}\left(\omega\right)\right|z-\omega t\,,\end{eqnarray*}
so positive-norm modes are those that
\begin{itemize}
\item if $\omega>0$, propagate along the fiber in the forward (co-propagating)
direction, i.e., they are $u$-modes;
\item if $\omega<0$, propagate along the fiber in the backward (counter-propagating)
direction, i.e., they are $v$-modes.
\end{itemize}
For negative-norm modes, the sign in front of $\left|\beta_{\chi}\left(\omega\right)\right|$ is changed, and the reverse is true.

It would do well to re-emphasise that $\omega$ is the frequency
\textit{as measured in the laboratory}, and it is generally \textit{not} conserved.
It should not be confused with the frequency $\omega$ as defined in the context
of the acoustic model, which is a genuine Killing frequency. Rather, the co-moving
frequency $\omega^{\prime}$ is a Killing frequency here, while $\omega$ is analogous
to the wavevector $k$ in the acoustic model.
So that statement that $v$-modes with $\omega<0$ have positive norm, despite first
appearances, is actually in accordance with the similar statement made in \S\ref{sub:Decomposition-into-modes}.

Expressing the total optical field $A$ as a sum over the individual
modes, and constraining $A$ to be real since it describes a physical
quantity, we have\begin{equation}
A\left(\tau,\zeta\right)=\int_{-\infty}^{+\infty}\left\{ a\left(\omega\right)A_{\omega}\left(\tau,\zeta\right)+a^{\star}\left(\omega\right)A_{\omega}^{\star}\left(\tau,\zeta\right)\right\} d\omega\,,\end{equation}
with canonical momentum\begin{equation}
\pi\left(\tau,\zeta\right)=-i\frac{\epsilon_{0}c^{2}}{u}\int_{-\infty}^{+\infty}\left|\beta_{\chi}\left(\omega\right)\right|\left\{ a\left(\omega\right)A_{\omega}\left(\tau,\zeta\right)-a^{\star}\left(\omega\right)A_{\omega}^{\star}\left(\tau,\zeta\right)\right\} d\omega\,,\end{equation}
where the coefficient functions can be expressed very simply using
the orthonormality conditions of Eqs. (\ref{eq:orthonormality_of_optical_modes}):\begin{alignat}{1}
a\left(\omega\right)=\left(A_{\omega},A\right)\,,\qquad & a^{\star}\left(\omega\right)=-\left(A_{\omega}^{\star},A\right)\,.\end{alignat}
The optical field is readily quantized: the real-valued quantities
$A$ and $\pi$ are promoted to the Hermitian operators $\hat{A}$
and $\hat{\pi}$, and the coefficients $a\left(\omega\right)$ and
$a^{\star}\left(\omega\right)$ to the Hermitian conjugate operators
$\hat{a}_{\omega}$ and $\hat{a}_{\omega}^{\dagger}$:\begin{equation}
\hat{A}\left(\tau,\zeta\right)=\int_{-\infty}^{+\infty}\left\{ \hat{a}_{\omega}A_{\omega}\left(\tau,\zeta\right)+\hat{a}_{\omega}^{\dagger}A_{\omega}^{\star}\left(\tau,\zeta\right)\right\} d\omega\,,\label{eq:optical_field_operator}\end{equation}
\begin{equation}
\hat{\pi}\left(\tau,\zeta\right)=-i\frac{\epsilon_{0}c^{2}}{u}\int_{-\infty}^{+\infty}\left|\beta_{\chi}\left(\omega\right)\right|\left\{ \hat{a}_{\omega}A_{\omega}\left(\tau,\zeta\right)-\hat{a}_{\omega}^{\dagger}A_{\omega}^{\star}\left(\tau,\zeta\right)\right\} d\omega\,.\label{eq:optical_momentum_operator}\end{equation}
Imposing the equal time (i.e., equal propagation time, $\zeta$) commutation
relation\begin{equation}
\left[\hat{A}\left(\tau_{1},\zeta\right),\hat{\pi}\left(\tau_{2},\zeta\right)\right]=i\,\delta\left(\tau_{1}-\tau_{2}\right)\,,\label{eq:field_momentum_commutation_relation}\end{equation}
we find that $\hat{a}_{\omega}$ and $\hat{a}_{\omega}^{\dagger}$
satisfy the Bose commutation relations\begin{alignat}{1}
\left[\hat{a}_{\omega_{1}},\hat{a}_{\omega_{2}}^{\dagger}\right]=\delta\left(\omega_{1}-\omega_{2}\right)\,,\qquad & \big[\hat{a}_{\omega_{1}},\hat{a}_{\omega_{2}}\big]=\left[\hat{a}_{\omega_{1}}^{\dagger},\hat{a}_{\omega_{2}}^{\dagger}\right]=0\,.\label{eq:optical_Bose_commutation_relations}\end{alignat}

\section{Transforming to the $\omega^{\prime}$-representation\label{sub:Transforming-to-the-omega-prime-repn}}

Eqs. (\ref{eq:optical_field_operator})-(\ref{eq:optical_Bose_commutation_relations})
describe the field in the $\omega$-representation, i.e., the variable
used to specify the various field modes is the lab frequency $\omega$.
This is fine when $\chi$ is constant. However, if $\chi$ is a function
of $\tau$, then $\omega$ is not conserved, and the field modes defined
above are only valid in regions where $\chi$ is approximately constant.
Even then, these modes do not evolve independently, but can be transformed
into each other by the inhomogeneous nonlinearity profile. To express
the field in terms of modes which do evolve independently, we must
use a representation based on the conserved co-moving frequency $\omega^{\prime}$.
(This is analogous to the transformation from the $k$- to the $\omega$-representation
performed in §\ref{sub:Transforming-to-the-omega-repn}.)

\subsection{Constant $\chi$}

\begin{figure}
\includegraphics[width=0.8\columnwidth]{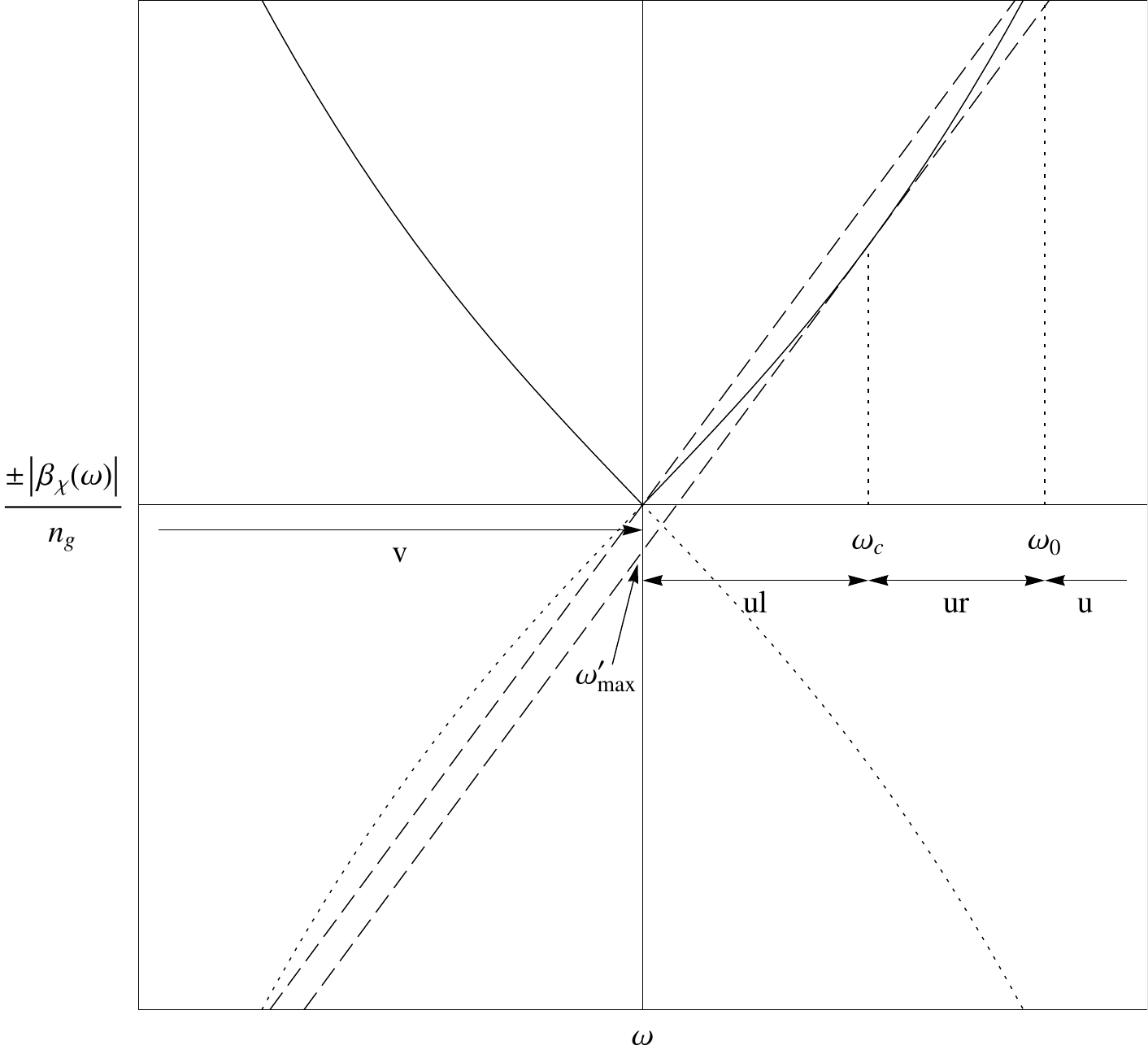}

\caption[\textsc{Solutions of fibre-optical dispersion relation}]{\textsc{Solutions of fibre-optical dispersion relation}: For given
values of $\chi$ and $\omega^{\prime}$, the possible lab frequencies
$\omega$ are the points of intersection of the straight line $\omega-\omega^{\prime}$
and the dispersion curve $\pm u\left|\beta_{\chi}\left(\omega\right)\right|$.
If $\chi<c^{2}/u^{2}-1$, the real $\omega$-axis splits into the
four branches $v$, $ul$, $ur$ and $u$; these are delimited by
$\omega^{\prime}=0$ and $\omega^{\prime}=\omega_{\mathrm{max}}^{\prime}$,
whose corresponding curves are plotted here as dashed lines. If $\omega^{\prime}>0$,
there are two positive-norm solutions, which lie on the $v$- and
$u$-branches; if $-\omega_{\mathrm{max}}^{\prime}<\omega^{\prime}<0$,
there are still two positive-norm solutions, but they lie on the $ul$-
and $ur$-branches.\label{fig:optical_omega_values}}

\end{figure}

We begin by assuming a constant value of $\chi$. The dispersion relation
is given by the positive-norm branch of Eq. (\ref{eq:two_dispersion_branches}),\begin{equation}
\omega-\omega^{\prime}=u\left|\beta_{\chi}\left(\omega\right)\right|=\frac{u}{c}\left|\omega\right|n\left(\omega\right)\sqrt{1+\frac{\chi}{n^{2}\left(\omega\right)}}\,,\label{eq:dispersion_relation_postitive_norm}\end{equation}
and the possible values of $\omega$ corresponding to a single value
of $\omega^{\prime}$ are found by plotting these two functions and
noting their points of intersection, as illustrated in Figure \ref{fig:optical_omega_values}.

Since the function $\omega\left(\omega^{\prime}\right)$ is multi-valued,
the transformation from an integral over $\omega$ to an integral
over $\omega^{\prime}$ must be split into several terms, each of
which covers a branch over which $\omega\left(\omega^{\prime}\right)$
is single-valued. An examination of Fig. \ref{fig:optical_omega_values}
shows that there are four such branches:
\begin{itemize}
\item $-\infty<\omega<0$, which covers all plane waves that are backward-propagating
in the lab frame; this is the $v$-branch;
\item $0<\omega<\omega_{c}$, where $\omega_{c}$ has the most negative
value of the co-moving frequency, which we shall call $-\omega_{\mathrm{max}}^{\prime}$;
these waves are forward-propagating and have a negative-directed (left-moving)
group velocity in the co-moving frame, and thus make up the $ul$-branch;
\item $\omega_{c}<\omega<\omega_{0}$, where $\omega_{0}$ has zero co-moving
frequency; these forward-propagating waves have a positive-directed
(right-moving) group velocity in the co-moving frame, and make up
the $ur$-branch;
\item $\omega_{0}<\omega<\infty$, which covers the remainder of the waves
that are forward-propagating in the lab frame; this is the remaining
$u$-branch.
\end{itemize}
The integral over all $\omega$ is a sum of the integrals over each
of these branches:\begin{align}
\int_{-\infty}^{0}\hat{a}_{\omega}A_{\omega}\left(\tau,\zeta\right)d\omega & = & \int_{0}^{\infty}\hat{a}_{\omega^{v}\left(\omega^{\prime}\right)}A_{\omega^{v}\left(\omega^{\prime}\right)}\left(\tau,\zeta\right)\left|\frac{d\omega^{v}}{d\omega^{\prime}}\right|d\omega^{\prime} & = & \int_{0}^{\infty}\hat{a}_{\omega^{\prime}}^{v}A_{\omega^{\prime}}^{v}\left(\tau\right)e^{-i\omega^{\prime}\zeta}d\omega^{\prime}\,,\\
\int_{0}^{\omega_{c}}\hat{a}_{\omega}A_{\omega}\left(\tau,\zeta\right)d\omega & = & \int_{-\omega_{max}^{\prime}}^{0}\hat{a}_{\omega^{ul}\left(\omega^{\prime}\right)}A_{\omega^{ul}\left(\omega^{\prime}\right)}\left(\tau,\zeta\right)\left|\frac{d\omega^{ul}}{d\omega^{\prime}}\right|d\omega^{\prime} & = & \int_{-\omega_{max}^{\prime}}^{0}\hat{a}_{\omega^{\prime}}^{ul}A_{\omega^{\prime}}^{ul}\left(\tau\right)e^{-i\omega^{\prime}\zeta}d\omega^{\prime}\,,\\
\int_{\omega_{c}}^{\omega_{0}}\hat{a}_{\omega}A_{\omega}\left(\tau,\zeta\right)d\omega & = & \int_{-\omega_{max}^{\prime}}^{0}\hat{a}_{\omega^{ur}\left(\omega^{\prime}\right)}A_{\omega^{ur}\left(\omega^{\prime}\right)}\left(\tau,\zeta\right)\left|\frac{d\omega^{ur}}{d\omega^{\prime}}\right|d\omega^{\prime} & = & \int_{-\omega_{max}^{\prime}}^{0}\hat{a}_{\omega^{\prime}}^{ur}A_{\omega^{\prime}}^{ur}\left(\tau\right)e^{-i\omega^{\prime}\zeta}d\omega^{\prime}\,,\\
\int_{\omega_{0}}^{\infty}\hat{a}_{\omega}A_{\omega}\left(\tau,\zeta\right)d\omega & = & \int_{0}^{\infty}\hat{a}_{\omega^{u}\left(\omega^{\prime}\right)}A_{\omega^{u}\left(\omega^{\prime}\right)}\left(\tau,\zeta\right)\left|\frac{d\omega^{u}}{d\omega^{\prime}}\right|d\omega^{\prime} & = & \int_{0}^{\infty}\hat{a}_{\omega^{\prime}}^{u}A_{\omega^{\prime}}^{u}\left(\tau\right)e^{-i\omega^{\prime}\zeta}d\omega^{\prime}\,,\end{align}
where we have defined\begin{alignat}{1}
\hat{a}_{\omega^{\prime}}^{v}=\sqrt{\left|\frac{d\omega^{v}}{d\omega^{\prime}}\right|}\hat{a}_{\omega^{v}\left(\omega^{\prime}\right)}\,,\qquad & A_{\omega^{\prime}}^{v}\left(\tau\right)e^{-i\omega^{\prime}\zeta}=\sqrt{\left|\frac{d\omega^{v}}{d\omega^{\prime}}\right|}A_{\omega^{v}\left(\omega^{\prime}\right)}\left(\tau,\zeta\right)\,,\label{eq:transformation_to_omega_prime_representation}\end{alignat}
and similarly for the other branches. These definitions ensure that
the newly-defined modes are orthonormal with respect to the co-moving
frequency $\omega^{\prime}$,\begin{eqnarray}
\left(A_{\omega_{1}^{\prime}}^{v}\left(\tau\right)e^{-i\omega_{1}^{\prime}\zeta},A_{\omega_{2}^{\prime}}^{v}\left(\tau\right)e^{-i\omega_{2}^{\prime}\zeta}\right) & = & \left|\frac{d\omega^{v}}{d\omega^{\prime}}\right|\left(A_{\omega^{v}\left(\omega_{1}^{\prime}\right)}\left(\tau,\zeta\right),A_{\omega^{v}\left(\omega_{2}^{\prime}\right)}\left(\tau,\zeta\right)\right)\nonumber \\
 & = & \left|\frac{d\omega^{v}}{d\omega^{\prime}}\right|\delta\left(\omega^{v}\left(\omega_{1}^{\prime}\right)-\omega^{v}\left(\omega_{2}^{\prime}\right)\right)\nonumber \\
 & = & \delta\left(\omega_{1}^{\prime}-\omega_{2}^{\prime}\right)\,,\label{eq:scalar_product_omega_prime_representation}\end{eqnarray}
while the operators obey commutation relations that are now defined
with respect to $\omega^{\prime}$,\begin{eqnarray}
\left[\hat{a}_{\omega_{1}^{\prime}}^{v},\hat{a}_{\omega_{2}^{\prime}}^{v\dagger}\right] & = & \left|\frac{d\omega^{v}}{d\omega^{\prime}}\right|\left[\hat{a}_{\omega^{v}\left(\omega_{1}^{\prime}\right)},\hat{a}_{\omega^{v}\left(\omega_{2}^{\prime}\right)}^{\dagger}\right]\nonumber \\
 & = & \left|\frac{d\omega^{v}}{d\omega^{\prime}}\right|\delta\left(\omega^{v}\left(\omega_{1}^{\prime}\right)-\omega^{v}\left(\omega_{2}^{\prime}\right)\right)\nonumber \\
 & = & \delta\left(\omega_{1}^{\prime}-\omega_{2}^{\prime}\right)\,.\label{eq:commutator_omega_prime_representation}\end{eqnarray}
Similar relations hold for the other branches.

Replacing the Hermitian conjugates, the total optical field is given
by\begin{multline}
\!\!\!\!\!\!\!\!\!\!\!\!\!\!\!\!\!\!\!\!\hat{A}\left(\tau,\zeta\right)=\int_{0}^{\infty}\left\{ \left(\hat{a}_{\omega^{\prime}}^{v}A_{\omega^{\prime}}^{v}\left(\tau\right)+\hat{a}_{\omega^{\prime}}^{u}A_{\omega^{\prime}}^{u}\left(\tau\right)\right)e^{-i\omega^{\prime}\zeta}+\left(\left(\hat{a}_{\omega^{\prime}}^{v}\right)^{\dagger}\left[A_{\omega^{\prime}}^{v}\left(\tau\right)\right]^{\star}+\left(\hat{a}_{\omega^{\prime}}^{u}\right)^{\dagger}\left[A_{\omega^{\prime}}^{u}\left(\tau\right)\right]^{\star}\right)e^{i\omega^{\prime}\zeta}\right\} d\omega^{\prime}\\
+\int_{-\omega_{max}^{\prime}}^{0}\left\{ \left(\hat{a}_{\omega^{\prime}}^{ul}A_{\omega^{\prime}}^{ul}\left(\tau\right)+\hat{a}_{\omega^{\prime}}^{ur}A_{\omega^{\prime}}^{ur}\left(\tau\right)\right)e^{-i\omega^{\prime}\zeta}+\left(\left(\hat{a}_{\omega^{\prime}}^{ul}\right)^{\dagger}\left[A_{\omega^{\prime}}^{ul}\left(\tau\right)\right]^{\star}+\left(\hat{a}_{\omega^{\prime}}^{ur}\right)^{\dagger}\left[A_{\omega^{\prime}}^{ur}\left(\tau\right)\right]^{\star}\right)e^{i\omega^{\prime}\zeta}\right\} d\omega^{\prime}\,.\end{multline}
A mode with negative $\omega^{\prime}$ is the Hermitian conjugate
of a mode with positive $\omega^{\prime}$, and vice versa. Since
any modes with the same co-moving frequency can be mixed, our final
task it to rewrite the optical field operator such that all modes
with the same $\omega^{\prime}$ are grouped together, and each value
of $\omega^{\prime}$ appears only once. As discussed in §\ref{sub:Transforming-to-the-omega-repn},
this amounts to rewriting the second term above - an integral over
negative frequencies - as an integral over positive frequencies:\begin{multline}
\int_{-\omega_{max}^{\prime}}^{0}\left\{ \left(\hat{a}_{\omega^{\prime}}^{ul}A_{\omega^{\prime}}^{ul}\left(\tau\right)+\hat{a}_{\omega^{\prime}}^{ur}A_{\omega^{\prime}}^{ur}\left(\tau\right)\right)e^{-i\omega^{\prime}\zeta}+\left(\left(\hat{a}_{\omega^{\prime}}^{ul}\right)^{\dagger}\left[A_{\omega^{\prime}}^{ul}\left(\tau\right)\right]^{\star}+\left(\hat{a}_{\omega^{\prime}}^{ur}\right)^{\dagger}\left[A_{\omega^{\prime}}^{ur}\left(\tau\right)\right]^{\star}\right)e^{i\omega^{\prime}\zeta}\right\} d\omega^{\prime}\\
\shoveleft{=\int_{0}^{\omega_{max}^{\prime}}\left\{ \left(\hat{a}_{-\omega^{\prime}}^{ul}A_{-\omega^{\prime}}^{ul}\left(\tau\right)+\hat{a}_{-\omega^{\prime}}^{ur}A_{-\omega^{\prime}}^{ur}\left(\tau\right)\right)e^{i\omega^{\prime}\zeta}\right.}\\
\left.+\left(\left(\hat{a}_{-\omega^{\prime}}^{ul}\right)^{\dagger}\left[A_{-\omega^{\prime}}^{ul}\left(\tau\right)\right]^{\star}+\left(\hat{a}_{-\omega^{\prime}}^{ur}\right)^{\dagger}\left[A_{-\omega^{\prime}}^{ur}\left(\tau\right)\right]^{\star}\right)e^{-i\omega^{\prime}\zeta}\right\} d\omega^{\prime}\,.\end{multline}
Finally, the optical field operator can be expressed in the simple
form\begin{equation}
\hat{A}\left(\tau,\zeta\right)=\int_{0}^{\infty}\left\{ \hat{A}_{\omega^{\prime}}\left(\tau\right)e^{-i\omega^{\prime}\zeta}+\hat{A}_{\omega^{\prime}}^{\dagger}\left(\tau\right)e^{i\omega^{\prime}\zeta}\right\} d\omega^{\prime}\,,\label{eq:optical_field_omega_prime_representation}\end{equation}
where we have defined\begin{equation}
\hat{A}_{\omega^{\prime}}\left(\tau\right)=\begin{cases}
\hat{a}_{\omega^{\prime}}^{v}A_{\omega^{\prime}}^{v}\left(\tau\right)+\hat{a}_{\omega^{\prime}}^{u}A_{\omega^{\prime}}^{u}\left(\tau\right)+\left(\hat{a}_{-\omega^{\prime}}^{ul}\right)^{\dagger}\left[A_{-\omega^{\prime}}^{ul}\left(\tau\right)\right]^{\star}+\left(\hat{a}_{-\omega^{\prime}}^{ur}\right)^{\dagger}\left[A_{-\omega^{\prime}}^{ur}\left(\tau\right)\right]^{\star}\,, & \omega^{\prime}<\omega_{\mathrm{max}}^{\prime}\\
\hat{a}_{\omega^{\prime}}^{v}A_{\omega^{\prime}}^{v}\left(\tau\right)+\hat{a}_{\omega^{\prime}}^{u}A_{\omega^{\prime}}^{u}\left(\tau\right)\,, & \omega^{\prime}>\omega_{\mathrm{max}}^{\prime}\end{cases}\,.\label{eq:optical_field_omega_prime_representation_2}\end{equation}
Just as in the acoustic case, the mixing of positive- and negative-norm
modes corresponds to a mixing of annihilation and creation operators.
This mixing, however, is still purely incidental if $\chi$ is independent
of $\tau$, for then $\omega$ is separately conserved.

\subsection{Nonconstant $\chi$}

Now let $\chi$ be a nonconstant function of $\tau$. As usual, we
assume that $\chi$ approaches limiting asymptotic values as $\tau\rightarrow\pm\infty$;
in practice, since light pulses are localised, these asymptotic values
will both be zero, but we needn't make this assumption. Since $\omega$
is no longer conserved, the description of the field in the $\omega^{\prime}$-representation
given by Eqs. (\ref{eq:optical_field_omega_prime_representation})
and (\ref{eq:optical_field_omega_prime_representation_2}) is now
its natural form. The assignment of branch labels $\left(v,u,ul,ur\right)$
follows the example set by the acoustic model in §\ref{sub:Transforming-to-the-omega-repn}:
we specify two types of modes, labelled in- and out-modes, which are
characterised either by a single incident frequency scattered into
many outgoing frequencies, or many incident frequencies scattered
into a single outgoing frequency; the branch label applies to this
single incident or outgoing frequency. For a graphical illustration,
refer back to Fig. \ref{fig:In_and_out_modes}.

\begin{figure}
\includegraphics[width=0.8\columnwidth]{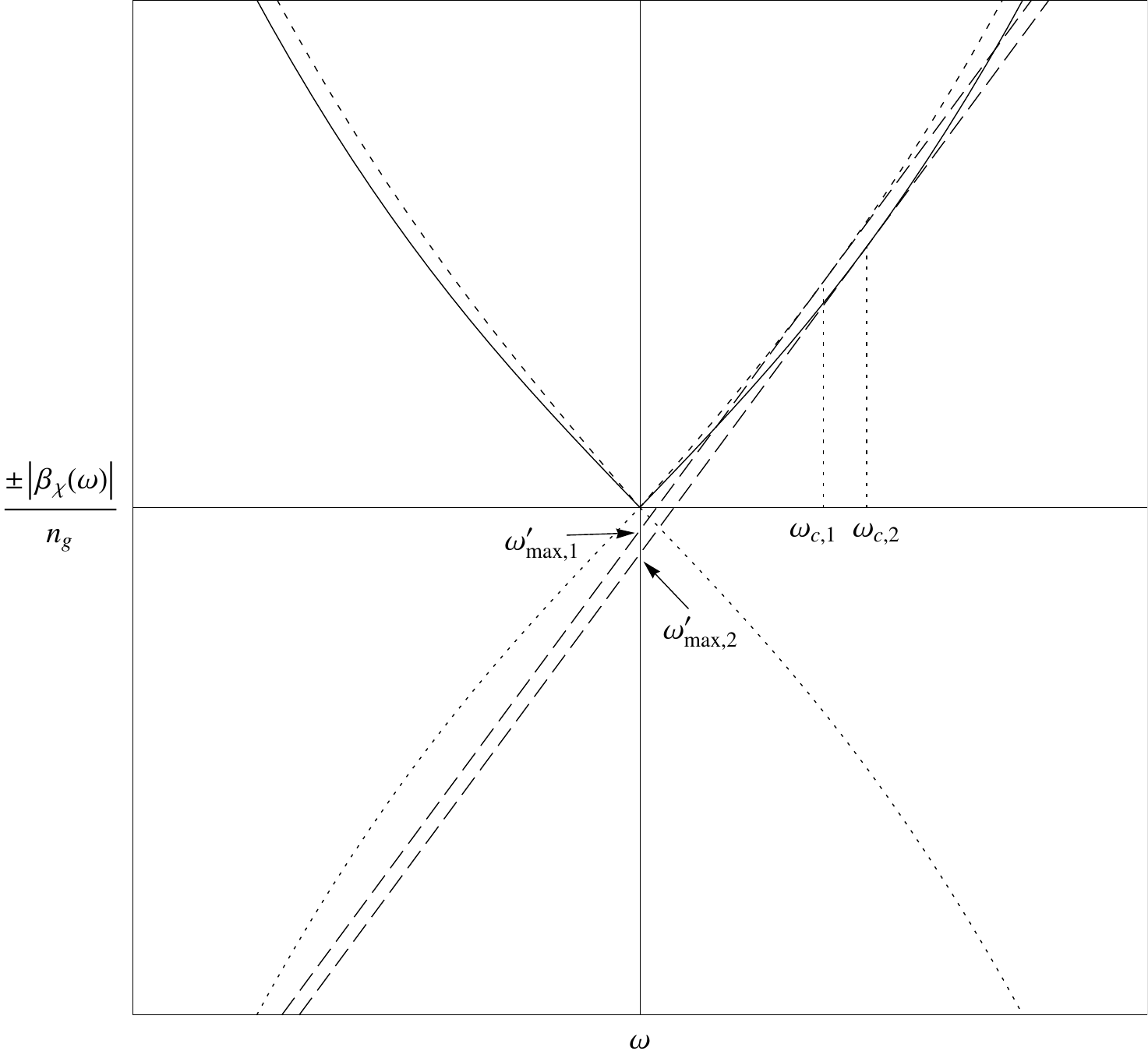}

\caption[\textsc{Effect of nonlinearity on solutions of dispersion relation}]{\textsc{Effect of nonlinearity on solutions of dispersion relation}:
Varying $\chi$ varies the dispersion profile $\beta_{\chi}\left(\omega\right)$,
so having different asymptotic values of $\chi$ results in different
asymptotic values of $\omega_{\mathrm{max}}^{\prime}$. Those values
of $\omega^{\prime}$ that lie between these two values experience
a group-velocity horizon.\label{fig:optical_omega_values_2}}

\end{figure}

Another difference arises if the asymptotic values of $\chi$ are
distinct. Then the dispersion relation on the right-hand side of Eq.
(\ref{eq:dispersion_relation_postitive_norm}) is slightly different
in each asymptotic region; in particular, $\omega_{max}^{\prime}$
differs, as shown in Figure \ref{fig:optical_omega_values_2}. Let
us label these different values as $\omega_{\mathrm{max},1}^{\prime}$
and $\omega_{\mathrm{max},2}^{\prime}$, with $\omega_{\mathrm{max},2}^{\prime}>\omega_{\mathrm{max},1}^{\prime}$.
Then there are three distinct cases:
\begin{itemize}
\item $0<\omega^{\prime}<\omega_{\mathrm{max},1}^{\prime}$: Both values
of $\chi$ have four real solutions for $\omega$, and these solutions
vary smoothly as $\chi$ varies. The mode is decomposed according
to the first line of Eq. (\ref{eq:optical_field_omega_prime_representation_2}).
\item $\omega_{\mathrm{max},1}^{\prime}<\omega^{\prime}<\omega_{\mathrm{max},2}^{\prime}$:
One value of $\chi$ has four real solutions for $\omega$, while
the other has only two. The two that exist for both values of $\chi$
correspond to the $v$- and $u$- branches, and these vary smoothly
as $\chi$ is varied. The other two solutions correspond to the $ur$-
and $ul$-branches, and these exist only for one asymptotic value
of $\chi$. As $\chi$ varies, these solutions merge, becoming complex
conjugates in the other asymptotic region; they thus experience a
group-velocity horizon, and, as discussed in §\ref{sub:Transforming-to-the-omega-repn},
are combined into a single branch, which will be labelled the $url$-branch.
The field mode is decomposed similarly to the first line of Eq. (\ref{eq:modes_w_inhomogeneous_v}).
\item $\omega^{\prime}>\omega_{\mathrm{max},2}^{\prime}$: Each value of
$\chi$ has only two real solutions for $\omega$, which vary smoothly
as $\chi$ varies. These correspond to the $v$- and $u$-branches,
and the mode is expanded in accordance with the second line of Eq.
(\ref{eq:optical_field_omega_prime_representation_2}).
\end{itemize}
To summarise, the total field remains in the form of Eq. (\ref{eq:optical_field_omega_prime_representation}),
and the individual modes are now decomposed as follows:\[
\!\!\!\!\!\!\!\!\!\!\hat{A}_{\omega^{\prime}}\left(\tau\right)=\begin{cases}
\hat{a}_{\omega^{\prime}}^{v}A_{\omega^{\prime}}^{v}\left(\tau\right)+\hat{a}_{\omega^{\prime}}^{u}A_{\omega^{\prime}}^{u}\left(\tau\right)+\left(\hat{a}_{-\omega^{\prime}}^{ul}\right)^{\dagger}\left[A_{-\omega^{\prime}}^{ul}\left(\tau\right)\right]^{\star}+\left(\hat{a}_{-\omega^{\prime}}^{ur}\right)^{\dagger}\left[A_{-\omega^{\prime}}^{ur}\left(\tau\right)\right]^{\star}\,, & 0<\omega^{\prime}<\omega_{\mathrm{max},1}^{\prime}\\
\hat{a}_{\omega^{\prime}}^{v}A_{\omega^{\prime}}^{v}\left(\tau\right)+\hat{a}_{\omega^{\prime}}^{u}A_{\omega^{\prime}}^{u}\left(\tau\right)+\left(\hat{a}_{-\omega^{\prime}}^{url}\right)^{\dagger}\left[A_{-\omega^{\prime}}^{url}\left(\tau\right)\right]^{\star}\,, & \!\!\!\!\!\!\!\!\!\!\!\!\!\! \omega_{\mathrm{max},1}^{\prime}<\omega^{\prime}<\omega_{\mathrm{max},2}^{\prime}\\
\hat{a}_{\omega^{\prime}}^{v}A_{\omega^{\prime}}^{v}\left(\tau\right)+\hat{a}_{\omega^{\prime}}^{u}A_{\omega^{\prime}}^{u}\left(\tau\right)\,. & \omega^{\prime}>\omega_{\mathrm{max},2}^{\prime}\end{cases}\,.\]
The modes appearing in this decomposition may be in- or out-modes,
each of these forming a complete orthonormal set. (Again, see §\ref{sub:Transforming-to-the-omega-repn}.)
Therefore, we have, for $0<\omega^{\prime}<\omega_{\mathrm{max},1}^{\prime}$,
the identity\begin{multline}
\hat{A}_{\omega^{\prime}}\left(\tau\right)=\hat{a}_{\omega^{\prime}}^{v,\mathrm{in}}A_{\omega^{\prime}}^{v,\mathrm{in}}\left(\tau\right)+\hat{a}_{\omega^{\prime}}^{u,\mathrm{in}}A_{\omega^{\prime}}^{u,\mathbf{\mathrm{in}}}\left(\tau\right)+\left(\hat{a}_{-\omega^{\prime}}^{ur,\mathrm{in}}\right)^{\dagger}\left[A_{-\omega^{\prime}}^{ur,\mathrm{in}}\left(\tau\right)\right]^{\star}+\left(\hat{a}_{-\omega^{\prime}}^{ul,\mathrm{in}}\right)^{\dagger}\left[A_{-\omega^{\prime}}^{ul,\mathrm{in}}\left(\tau\right)\right]^{\star}\\
=\hat{a}_{\omega^{\prime}}^{v,\mathrm{out}}A_{\omega^{\prime}}^{v,\mathrm{out}}\left(\tau\right)+\hat{a}_{\omega^{\prime}}^{u,\mathrm{out}}A_{\omega^{\prime}}^{u,\mathrm{out}}\left(\tau\right)+\left(\hat{a}_{-\omega^{\prime}}^{ur,\mathrm{out}}\right)^{\dagger}\left[A_{-\omega^{\prime}}^{ur,\mathrm{out}}\left(\tau\right)\right]^{\star}+\left(\hat{a}_{-\omega^{\prime}}^{ul,\mathrm{out}}\right)^{\dagger}\left[A_{-\omega^{\prime}}^{ul,\mathrm{out}}\left(\tau\right)\right]^{\star}\,,\label{eq:correspondence_between_in_and_out_modes}\end{multline}
with similar identities for other values of $\omega^{\prime}$.

Since typically $\omega_{\mathrm{max},1}^{\prime}$ and $\omega_{\mathrm{max},2}^{\prime}$
are very close in value, we shall assume in what follows $0<\omega^{\prime}<\omega_{\mathrm{max},1}^{\prime}$,
so that there is no group-velocity horizon; for $\omega_{\mathrm{max},1}^{\prime}<\omega^{\prime}<\omega_{\mathrm{max},2}^{\prime}$,
the analysis is akin to that of §\ref{sub:Transforming-to-the-omega-repn}.
Completeness of the modes ensures that we can express one type of
mode as a linear combination of the other, so we may write \begin{eqnarray*}
\!\!\!\!\!\!\!\!\!\!A_{\omega^{\prime}}^{v,\mathrm{in}}\left(\tau\right) & = & C_{\omega^{\prime}}^{v,v}A_{\omega^{\prime}}^{v,\mathrm{out}}\left(\tau\right)+C_{\omega^{\prime}}^{v,u}A_{\omega^{\prime}}^{u,\mathrm{out}}\left(\tau\right)+C_{\omega^{\prime}}^{v,ur}\left[A_{-\omega^{\prime}}^{ur,\mathrm{out}}\left(\tau\right)\right]^{\star}+C_{\omega^{\prime}}^{v,ul}\left[A_{-\omega^{\prime}}^{ul,\mathrm{out}}\left(\tau\right)\right]^{\star}\,,\\
A_{\omega^{\prime}}^{u,\mathrm{in}}\left(\tau\right) & = & C_{\omega^{\prime}}^{u,v}A_{\omega^{\prime}}^{v,\mathrm{out}}\left(\tau\right)+C_{\omega^{\prime}}^{u,u}A_{\omega^{\prime}}^{u,\mathrm{out}}\left(\tau\right)+C_{\omega^{\prime}}^{u,ur}\left[A_{-\omega^{\prime}}^{ur,\mathrm{out}}\left(\tau\right)\right]^{\star}+C_{\omega^{\prime}}^{u,ul}\left[A_{-\omega^{\prime}}^{ul,\mathrm{out}}\left(\tau\right)\right]^{\star}\,,\\
\left[A_{-\omega^{\prime}}^{ur,\mathrm{in}}\left(\tau\right)\right]^{\star} & = & C_{\omega^{\prime}}^{ur,v}A_{\omega^{\prime}}^{v,\mathrm{out}}\left(\tau\right)+C_{\omega^{\prime}}^{ur,u}A_{\omega^{\prime}}^{u,\mathrm{out}}\left(\tau\right)+C_{\omega^{\prime}}^{ur,ur}\left[A_{-\omega^{\prime}}^{ur,\mathrm{out}}\left(\tau\right)\right]^{\star}+C_{\omega^{\prime}}^{ur,ul}\left[A_{-\omega^{\prime}}^{ul,\mathrm{out}}\left(\tau\right)\right]^{\star}\,,\\
\left[A_{-\omega^{\prime}}^{ul,\mathrm{in}}\left(\tau\right)\right]^{\star} & = & C_{\omega^{\prime}}^{ul,v}A_{\omega^{\prime}}^{v,\mathrm{out}}\left(\tau\right)+C_{\omega^{\prime}}^{ul,u}A_{\omega^{\prime}}^{u,\mathrm{out}}\left(\tau\right)+C_{\omega^{\prime}}^{ul,ur}\left[A_{-\omega^{\prime}}^{ur,\mathrm{out}}\left(\tau\right)\right]^{\star}+C_{\omega^{\prime}}^{ul,ul}\left[A_{-\omega^{\prime}}^{ul,\mathrm{out}}\left(\tau\right)\right]^{\star}\,.\end{eqnarray*}
 The coefficients are equal to scalar products between modes, e.g.,
$C_{\omega^{\prime}}^{v,v}=\left(A_{\omega^{\prime}}^{v,\mathrm{out}},A_{\omega^{\prime}}^{v,\mathrm{in}}\right)$
and $C_{\omega^{\prime}}^{v,ur}=-\left(\left[A_{-\omega^{\prime}}^{ur,\mathrm{out}}\right]^{\star},A_{\omega^{\prime}}^{v,\mathrm{in}}\right)\,$.
Applying the identity $\left(A_{2},A_{1}\right)=\left(A_{1},A_{2}\right)^{\star}$
- which follows from the definition of the scalar product in Eq. (\ref{eq:scalar_product_FIBRES})
- we can express the out-modes as combinations of in-modes using the
same coefficients:\begin{eqnarray*}
A_{\omega^{\prime}}^{v,\mathrm{out}}\left(\tau\right) & = & \left(C_{\omega^{\prime}}^{v,v}\right)^{\star}A_{\omega^{\prime}}^{v,\mathrm{in}}\left(\tau\right)+\left(C_{\omega^{\prime}}^{u,v}\right)^{\star}A_{\omega^{\prime}}^{u,\mathrm{in}}\left(\tau\right)-\left(C_{\omega^{\prime}}^{ur,v}\right)^{\star}\left[A_{-\omega^{\prime}}^{ur,\mathrm{in}}\left(\tau\right)\right]^{\star}-\left(C_{\omega^{\prime}}^{ul,v}\right)^{\star}\left[A_{-\omega^{\prime}}^{ul,\mathrm{in}}\left(\tau\right)\right]^{\star}\,,\\
A_{\omega^{\prime}}^{u,\mathrm{out}}\left(\tau\right) & = & \left(C_{\omega^{\prime}}^{v,u}\right)^{\star}A_{\omega^{\prime}}^{v,\mathrm{in}}\left(\tau\right)+\left(C_{\omega^{\prime}}^{u,u}\right)^{\star}A_{\omega^{\prime}}^{u,\mathrm{in}}\left(\tau\right)-\left(C_{\omega^{\prime}}^{ur,u}\right)^{\star}\left[A_{-\omega^{\prime}}^{ur,\mathrm{in}}\left(\tau\right)\right]^{\star}-\left(C_{\omega^{\prime}}^{ul,u}\right)^{\star}\left[A_{-\omega^{\prime}}^{ul,\mathrm{in}}\left(\tau\right)\right]^{\star}\,,\\
\left[A_{-\omega^{\prime}}^{ur,\mathrm{out}}\left(\tau\right)\right]^{\star} & = & -\left(C_{\omega^{\prime}}^{v,ur}\right)^{\star}A_{\omega^{\prime}}^{v,\mathrm{in}}\left(\tau\right)-\left(C_{\omega^{\prime}}^{u,ur}\right)^{\star}A_{\omega^{\prime}}^{u,\mathrm{in}}\left(\tau\right)+\left(C_{\omega^{\prime}}^{ur,ur}\right)^{\star}\left[A_{-\omega^{\prime}}^{ur,\mathrm{in}}\left(\tau\right)\right]^{\star}\\
 &  & \qquad\qquad\qquad\qquad\qquad\qquad\qquad\qquad\qquad\qquad+\left(C_{\omega^{\prime}}^{ul,ur}\right)^{\star}\left[A_{-\omega^{\prime}}^{ul,\mathrm{in}}\left(\tau\right)\right]^{\star}\,,\\
\left[A_{-\omega^{\prime}}^{ul,\mathrm{out}}\left(\tau\right)\right]^{\star} & = & -\left(C_{\omega^{\prime}}^{v,ul}\right)^{\star}A_{\omega^{\prime}}^{v,\mathrm{in}}\left(\tau\right)-\left(C_{\omega^{\prime}}^{u,ul}\right)^{\star}A_{\omega^{\prime}}^{u,\mathrm{in}}\left(\tau\right)+\left(C_{\omega^{\prime}}^{ur,ul}\right)^{\star}\left[A_{-\omega^{\prime}}^{ur,\mathrm{in}}\left(\tau\right)\right]^{\star}\\
 &  & \qquad\qquad\qquad\qquad\qquad\qquad\qquad\qquad\qquad\qquad+\left(C_{\omega^{\prime}}^{ul,ul}\right)^{\star}\left[A_{-\omega^{\prime}}^{ul,\mathrm{in}}\left(\tau\right)\right]^{\star}\,.\end{eqnarray*}
Substitution in Eq. (\ref{eq:correspondence_between_in_and_out_modes})
gives the corresponding transformation of the annihilation and creation
operators:\begin{eqnarray*}
\hat{a}_{\omega^{\prime}}^{v,\mathrm{out}} & = & C_{\omega^{\prime}}^{v,v}\hat{a}_{\omega^{\prime}}^{v,\mathrm{in}}+C_{\omega^{\prime}}^{u,v}\hat{a}_{\omega^{\prime}}^{u,\mathrm{in}}+C_{\omega^{\prime}}^{ur,v}\left(\hat{a}_{-\omega^{\prime}}^{ur,\mathrm{in}}\right)^{\dagger}+C_{\omega^{\prime}}^{ul,v}\left(\hat{a}_{-\omega^{\prime}}^{ul,\mathrm{in}}\right)^{\dagger}\,,\\
\hat{a}_{\omega^{\prime}}^{u,\mathrm{out}} & = & C_{\omega^{\prime}}^{v,u}\hat{a}_{\omega^{\prime}}^{v,\mathrm{in}}+C_{\omega^{\prime}}^{u,u}\hat{a}_{\omega^{\prime}}^{u,\mathrm{in}}+C_{\omega^{\prime}}^{ur,u}\left(\hat{a}_{-\omega^{\prime}}^{ur,\mathrm{in}}\right)^{\dagger}+C_{\omega^{\prime}}^{ul,u}\left(\hat{a}_{-\omega^{\prime}}^{ul,\mathrm{in}}\right)^{\dagger}\,,\\
\left(\hat{a}_{-\omega^{\prime}}^{ur,\mathrm{out}}\right)^{\dagger} & = & C_{\omega^{\prime}}^{v,ur}\hat{a}_{\omega^{\prime}}^{v,\mathrm{in}}+C_{\omega^{\prime}}^{u,ur}\hat{a}_{\omega^{\prime}}^{u,\mathrm{in}}+C_{\omega^{\prime}}^{ur,ur}\left(\hat{a}_{-\omega^{\prime}}^{ur,\mathrm{in}}\right)^{\dagger}+C_{\omega^{\prime}}^{ul,ur}\left(\hat{a}_{-\omega^{\prime}}^{ul,\mathrm{in}}\right)^{\dagger}\,,\\
\left(\hat{a}_{-\omega^{\prime}}^{ul,\mathrm{out}}\right)^{\dagger} & = & C_{\omega^{\prime}}^{v,ul}\hat{a}_{\omega^{\prime}}^{v,\mathrm{in}}+C_{\omega^{\prime}}^{u,ul}\hat{a}_{\omega^{\prime}}^{u,\mathrm{in}}+C_{\omega^{\prime}}^{ur,ul}\left(\hat{a}_{-\omega^{\prime}}^{ur,\mathrm{in}}\right)^{\dagger}+C_{\omega^{\prime}}^{ul,ul}\left(\hat{a}_{-\omega^{\prime}}^{ul,\mathrm{in}}\right)^{\dagger}\,.\end{eqnarray*}
Similarly, the inverse transformation is found:\begin{eqnarray*}
\hat{a}_{\omega^{\prime}}^{v,\mathrm{in}} & = & \left(C_{\omega^{\prime}}^{v,v}\right)^{\star}\hat{a}_{\omega^{\prime}}^{v,\mathrm{out}}+\left(C_{\omega^{\prime}}^{v,u}\right)^{\star}\hat{a}_{\omega^{\prime}}^{u,\mathrm{out}}-\left(C_{\omega^{\prime}}^{v,ur}\right)^{\star}\left(\hat{a}_{-\omega^{\prime}}^{ur,\mathrm{out}}\right)^{\dagger}-\left(C_{\omega^{\prime}}^{v,ul}\right)^{\star}\left(\hat{a}_{-\omega^{\prime}}^{ul,\mathrm{out}}\right)^{\dagger}\,,\\
\hat{a}_{\omega^{\prime}}^{u,\mathrm{in}} & = & \left(C_{\omega^{\prime}}^{u,v}\right)^{\star}\hat{a}_{\omega^{\prime}}^{v,\mathrm{out}}+\left(C_{\omega^{\prime}}^{u,u}\right)^{\star}\hat{a}_{\omega^{\prime}}^{u,\mathrm{out}}-\left(C_{\omega^{\prime}}^{u,ur}\right)^{\star}\left(\hat{a}_{-\omega^{\prime}}^{ur,\mathrm{out}}\right)^{\dagger}-\left(C_{\omega^{\prime}}^{u,ul}\right)^{\star}\left(\hat{a}_{-\omega^{\prime}}^{ul,\mathrm{out}}\right)^{\dagger}\,,\\
\left(\hat{a}_{-\omega^{\prime}}^{ur,\mathrm{in}}\right)^{\dagger} & = & -\left(C_{\omega^{\prime}}^{ur,v}\right)^{\star}\hat{a}_{\omega^{\prime}}^{v,\mathrm{out}}-\left(C_{\omega^{\prime}}^{ur,u}\right)^{\star}\hat{a}_{\omega^{\prime}}^{u,\mathrm{out}}+\left(C_{\omega^{\prime}}^{ur,ur}\right)^{\star}\left(\hat{a}_{-\omega^{\prime}}^{ur,\mathrm{out}}\right)^{\dagger}+\left(C_{\omega^{\prime}}^{ur,ul}\right)^{\star}\left(\hat{a}_{-\omega^{\prime}}^{ul,\mathrm{out}}\right)^{\dagger}\,,\\
\left(\hat{a}_{-\omega^{\prime}}^{ul,\mathrm{in}}\right)^{\dagger} & = & -\left(C_{\omega^{\prime}}^{ul,v}\right)^{\star}\hat{a}_{\omega^{\prime}}^{v,\mathrm{out}}-\left(C_{\omega^{\prime}}^{ul,u}\right)^{\star}\hat{a}_{\omega^{\prime}}^{u,\mathrm{out}}+\left(C_{\omega^{\prime}}^{ul,ur}\right)^{\star}\left(\hat{a}_{-\omega^{\prime}}^{ur,\mathrm{out}}\right)^{\dagger}+\left(C_{\omega^{\prime}}^{ul,ul}\right)^{\star}\left(\hat{a}_{-\omega^{\prime}}^{ul,\mathrm{out}}\right)^{\dagger}\,.\end{eqnarray*}
As in the acoustic case, this is where the significance of mixing
between positive- and negative-norm modes becomes manifest: a single
operator corresponding to an in- or out-mode, when written in terms
of those operators for the other type of mode, can contain both annihilation
and creation operators.

\section{Spontaneous creation of photons}

Assuming that the quantum state is the in-vacuum, $\left|0_{\mathrm{in}}\right\rangle $,
which contains no incoming particles and is annihilated by all in-mode
annihilation operators, i.e.,\begin{equation}
\hat{a}_{\omega^{\prime}}^{v,\mathrm{in}}\left|0_{\mathrm{in}}\right\rangle =\hat{a}_{\omega^{\prime}}^{u,\mathrm{in}}\left|0_{\mathrm{in}}\right\rangle =\hat{a}_{-\omega^{\prime}}^{ur,\mathrm{in}}\left|0_{\mathrm{in}}\right\rangle =\hat{a}_{-\omega^{\prime}}^{ul,\mathrm{in}}\left|0_{\mathrm{in}}\right\rangle =0\,,\end{equation}
we find that the mixing of different types of operator still allows
the presence of outgoing particles. For example, the expectation number
for the outgoing $u$-mode is\begin{eqnarray}
\!\!\!\!\!\!\!\!\!\!\!\!\!\!\left\langle 0_{\mathrm{in}}\right|\left(\hat{a}_{\omega_{1}^{\prime}}^{u,\mathrm{out}}\right)^{\dagger}\hat{a}_{\omega_{2}^{\prime}}^{u,\mathrm{out}}\left|0_{\mathrm{in}}\right\rangle  & = & \left\langle 0_{\mathrm{in}}\right|\left\{ \left(C_{\omega_{1}^{\prime}}^{ur,u}\right)^{\star}\hat{a}_{-\omega_{1}^{\prime}}^{ur,\mathrm{in}}+\left(C_{\omega_{1}^{\prime}}^{ul,u}\right)^{\star}\hat{a}_{-\omega_{1}^{\prime}}^{ul,\mathrm{in}}\right\} \nonumber \\
 &  & \qquad\qquad\qquad\qquad\times\left\{ C_{\omega_{2}^{\prime}}^{ur,u}\left(\hat{a}_{-\omega_{2}^{\prime}}^{ur,\mathrm{in}}\right)^{\dagger}+C_{\omega_{2}^{\prime}}^{ul,u}\left(\hat{a}_{-\omega_{2}^{\prime}}^{ul,\mathrm{in}}\right)^{\dagger}\right\} \left|0_{\mathrm{in}}\right\rangle \nonumber \\
 & = & \left\langle 0_{\mathrm{in}}\right|\left\{ \left(C_{\omega_{1}^{\prime}}^{ur,u}\right)^{\star}C_{\omega_{2}^{\prime}}^{ur,u}\hat{a}_{-\omega_{1}^{\prime}}^{ur,\mathrm{in}}\left(\hat{a}_{-\omega_{2}^{\prime}}^{ur,\mathrm{in}}\right)^{\dagger}+\left(C_{\omega_{1}^{\prime}}^{ul,u}\right)^{\star}C_{\omega_{2}^{\prime}}^{ul,u}\hat{a}_{-\omega_{1}^{\prime}}^{ul,\mathrm{in}}\left(\hat{a}_{-\omega_{2}^{\prime}}^{ul,\mathrm{in}}\right)^{\dagger}\right\} \left|0_{\mathrm{in}}\right\rangle \nonumber \\
 & = & \left\langle 0_{\mathrm{in}}\right|\left\{ \left(C_{\omega_{1}^{\prime}}^{ur,u}\right)^{\star}C_{\omega_{2}^{\prime}}^{ur,u}\left(\delta\left(\omega_{1}^{\prime}-\omega_{2}^{\prime}\right)+\left(\hat{a}_{-\omega_{2}^{\prime}}^{ur,\mathrm{in}}\right)^{\dagger}\hat{a}_{-\omega_{1}^{\prime}}^{ur,\mathrm{in}}\right)\right.\nonumber \\
 &  & \qquad\qquad\left.+\left(C_{\omega_{1}^{\prime}}^{ul,u}\right)^{\star}C_{\omega_{2}^{\prime}}^{ul,u}\left(\delta\left(\omega_{1}^{\prime}-\omega_{2}^{\prime}\right)+\left(\hat{a}_{-\omega_{2}^{\prime}}^{ul,\mathrm{in}}\right)^{\dagger}\hat{a}_{-\omega_{1}^{\prime}}^{ul,\mathrm{in}}\right)\right\} \left|0_{\mathrm{in}}\right\rangle \nonumber \\
 & = & \left(\left|C_{\omega_{1}^{\prime}}^{ur,u}\right|^{2}+\left|C_{\omega_{1}^{\prime}}^{ul,u}\right|^{2}\right)\delta\left(\omega_{1}^{\prime}-\omega_{2}^{\prime}\right)\,,\end{eqnarray}
where the normalization of the state $\left|0_{\mathrm{in}}\right\rangle $
and the commutator of Eq. (\ref{eq:commutator_omega_prime_representation})
were used. Similar expressions hold for the expectation values of
the other modes:\begin{eqnarray}
\left\langle 0_{\mathrm{in}}\right|\left(\hat{a}_{\omega_{1}^{\prime}}^{v,\mathrm{out}}\right)^{\dagger}\hat{a}_{\omega_{2}^{\prime}}^{v,\mathrm{out}}\left|0_{\mathrm{in}}\right\rangle  & = & \left(\left|C_{\omega_{1}^{\prime}}^{ur,v}\right|^{2}+\left|C_{\omega_{1}^{\prime}}^{ul,v}\right|^{2}\right)\delta\left(\omega_{1}^{\prime}-\omega_{2}^{\prime}\right)\,,\\
\left\langle 0_{\mathrm{in}}\right|\left(\hat{a}_{\omega_{1}^{\prime}}^{ur,\mathrm{out}}\right)^{\dagger}\hat{a}_{\omega_{2}^{\prime}}^{ur,\mathrm{out}}\left|0_{\mathrm{in}}\right\rangle  & = & \left(\left|C_{\omega_{1}^{\prime}}^{v,ur}\right|^{2}+\left|C_{\omega_{1}^{\prime}}^{u,ur}\right|^{2}\right)\delta\left(\omega_{1}^{\prime}-\omega_{2}^{\prime}\right)\,,\\
\left\langle 0_{\mathrm{in}}\right|\left(\hat{a}_{\omega_{1}^{\prime}}^{ul,\mathrm{out}}\right)^{\dagger}\hat{a}_{\omega_{2}^{\prime}}^{ul,\mathrm{out}}\left|0_{\mathrm{in}}\right\rangle  & = & \left(\left|C_{\omega_{1}^{\prime}}^{v,ul}\right|^{2}+\left|C_{\omega_{1}^{\prime}}^{u,ul}\right|^{2}\right)\delta\left(\omega_{1}^{\prime}-\omega_{2}^{\prime}\right)\,,\end{eqnarray}
and also for those modes with co-moving frequency such that $\omega_{\mathrm{max,1}}^{\prime}<\omega^{\prime}<\omega_{\mathrm{max,2}}^{\prime}$:\begin{eqnarray}
\left\langle 0_{\mathrm{in}}\right|\left(\hat{a}_{\omega_{1}^{\prime}}^{v,\mathrm{out}}\right)^{\dagger}\hat{a}_{\omega_{2}^{\prime}}^{v,\mathrm{out}}\left|0_{\mathrm{in}}\right\rangle  & = & \left|C_{\omega_{1}^{\prime}}^{url,v}\right|^{2}\delta\left(\omega_{1}^{\prime}-\omega_{2}^{\prime}\right)\,,\\
\left\langle 0_{\mathrm{in}}\right|\left(\hat{a}_{\omega_{1}^{\prime}}^{u,\mathrm{out}}\right)^{\dagger}\hat{a}_{\omega_{2}^{\prime}}^{u,\mathrm{out}}\left|0_{\mathrm{in}}\right\rangle  & = & \left|C_{\omega_{1}^{\prime}}^{url,u}\right|^{2}\delta\left(\omega_{1}^{\prime}-\omega_{2}^{\prime}\right)\,,\\
\left\langle 0_{\mathrm{in}}\right|\left(\hat{a}_{\omega_{1}^{\prime}}^{url,\mathrm{out}}\right)^{\dagger}\hat{a}_{\omega_{2}^{\prime}}^{url,\mathrm{out}}\left|0_{\mathrm{in}}\right\rangle  & = & \left(\left|C_{\omega_{1}^{\prime}}^{url,v}\right|^{2}+\left|C_{\omega_{1}^{\prime}}^{url,u}\right|^{2}\right)\delta\left(\omega_{1}^{\prime}-\omega_{2}^{\prime}\right)\,.\end{eqnarray}
However, for those modes with co-moving frequency $\omega^{\prime}>\omega_{\mathrm{max,2}}^{\prime}$,
there is no mixing of positive- and negative-norm modes and hence
no particle creation.

The $\delta$ functions appearing in the expectation values above
are due to the continuous nature of the stationary modes into which
the optical field is decomposed. Arguments entirely analogous to those
given in §\ref{sub:Spontaneous-creation-II} show that the expectation
values should be interpreted, upon division by $2\pi$, as spectral
flux densities:\begin{equation}
\frac{d^{2}N}{d\zeta\, d\omega^{\prime}}=\frac{1}{2\pi}\left|\beta_{\omega^{\prime}}\right|^{2}\,,\label{eq:FIBRES_spectral_flux_density_zeta-omegaprime}\end{equation}
where $\left|\beta_{\omega^{\prime}}\right|^{2}$ is the factor multiplying
the delta funtions in the expectation values above. However, this
is defined in the co-moving frame, and requires some modification
to be turned into a more useful quantity, defined with respect to
the lab frame. Instead of the variables $\zeta$ and $\omega^{\prime}$,
we would like a spectral flux density in terms of $t$ and $\omega$.
Firstly, it should be noted that Eq. (\ref{eq:FIBRES_spectral_flux_density_zeta-omegaprime})
refers to a single pulse; in practice, a stream of pulses are coupled
into the fibre with repetition rate $\nu_{\mathrm{rep}}$. So, in
order to find the number of photons emitted per unit time, we first
find the number of photons emitted per pulse, and multiply by $\nu_{\mathrm{rep}}$.
The number of photons per pulse is found by integrating with respect
to $\zeta$ over the entire length of the fibre, and assuming that
the pulse is a soliton so that it remains stationary in the co-moving
frame, so amounts to multiplying by $L/u$, where $L$ is the length
of the fibre and $u$ is the group velocity of the pulse. Therefore,
Eq. (\ref{eq:FIBRES_spectral_flux_density_zeta-omegaprime}) becomes\begin{equation}
\frac{d^{2}N}{dt\, d\omega^{\prime}}=\frac{1}{2\pi}\frac{L\,\nu_{\mathrm{rep}}}{u}\left|\beta_{\omega^{\prime}}\right|^{2}\,,\label{eq:FIBRES_spectral_flux_density_t-omegaprime}\end{equation}
or, multiplying by $\left|d\omega^{\prime}/d\omega\right|$ to effect
the transformation from variable $\omega^{\prime}$ to $\omega$,\begin{equation}
\frac{d^{2}N}{dt\, d\omega}=\frac{1}{2\pi}\frac{L_{\mathrm{}}\,\nu_{\mathrm{rep}}}{u}\left|\frac{d\omega^{\prime}}{d\omega}\right|\left|\beta_{\omega^{\prime}}\right|^{2}\,.\label{eq:FIBRES_spectral_flux_density_t-omega}\end{equation}
Taking the values from the experiment in Chapter \ref{sec:Frequency-Shifting-in-Optical-Fibres},
$L=1.5\,\mathrm{m}$, $\nu_{\mathrm{rep}}=80\,\mathrm{MHz}$ and $u=c/n_{g}$
where $n_{g}\approx1.5$, we find that $L\,\nu_{\mathrm{rep}}/u\approx0.6$;
loosening these restrictions slightly, we can say that the transformation
from the co-moving frame to the lab frame roughly preserves the order
of magnitude of the spectral flux density. However, it should be borne
in mind that the actual value depends on the factor $L\,\nu_{\mathrm{rep}}/u$.
Of the quantities appearing there, the most important is $L$, for
in the derivation we assumed that the pulse maintains its shape throughout
the fibre. However, if instead we make use of self-steepening mechanisms,
so that the pulse shape changes with its propagation, then $L$ should
be replaced by an effective length $L_{\mathrm{eff}}$, over which
the pulse emits photons most efficiently.

\section{Concluding remarks}

Having quantized the optical probe field in the presence of a strong
pulse, we have shown, as in the acoustic case, that there are two
natural sets of modes into which the field may be decomposed: in-modes
and out-modes. These sets define two vacuum states, and it is found
that, due to the mixing of modes of positive and negative norm, these
vacuum states are not equal. Thus, in the in-vacuum, which has no
incoming photons, there are outgoing photons, which are emitted spontaneously
by the pulse.

\pagebreak{}

\chapter{Methods of Fibre-Optical Model\label{sec:Methods-of-Fibre-Optical-Model}}

Having developed the theoretical foundations of our fibre-optical
model of the event horizon, we come now to the methods employed to
calculate the expected spectrum of Hawking radiation. These bear many
similarities to the methods of Chapter \ref{sec:Methods-of-Acoustic-Model}
used in the acoustic model. As there, we examine an FDTD algorithm
for wavepacket propagation; an ordinary differential equation solver
and algebraic technique for finding steady-state solutions; and an
analytic method derived from linearization of the nonlinearity profile.

\section{Wavepacket propagation\label{sub:FDTD-wavepacket-propagation-FIBRES}}

When discussing methods for solving the acoustic case in §\ref{sub:Wavepacket-propagation},
we began by following in the spirit of Unruh \cite{Unruh-1995}, calculating
negative-frequency contributions by solving for the propagation of
a wavepacket directly using an FDTD (Finite Difference Time Domain)
algorithm. The starting point is, of course, the fibre-optical wave
equation (\ref{eq:wave_equation_FIBRES}), which can be decomposed
into two PDEs of the first order in $\zeta$ using (a quantity proportional
to) the canonical momentum $\pi$ of Eq. (\ref{eq:optical_canonical_momentum_density}):\begin{eqnarray}
\partial_{\zeta}A-\partial_{\tau}A & = & \Pi\,,\label{eq:optical_PDE_order1_1}\\
\partial_{\zeta}\Pi-\partial_{\tau}\Pi & = & -u^{2}\beta^{2}\left(i\partial_{\tau}\right)A+\frac{u^{2}}{c^{2}}\left(\partial_{\tau}\chi\right)\partial_{\tau}A+\frac{u^{2}}{c^{2}}\chi\partial_{\tau}^{2}A\,.\label{eq:optical_PDE_order1_2}\end{eqnarray}
Notice that Eq. (\ref{eq:optical_PDE_order1_2}) contains a term in
$\partial_{\tau}^{2}A$, and is therefore not purely first-order.
The operation of $\beta^{2}$ on $A$ is analogous to that of $F^{2}$
on $\phi$ in the acoustic model: if $\widetilde{A}\left(\omega\right)$
is the Fourier transform of $A$ with respect to $\tau$,\begin{equation}
\widetilde{A}\left(\omega\right)=\int_{-\infty}^{+\infty}e^{i\omega\tau}A\left(\tau\right)d\tau\,,\label{eq:A_Fourier}\end{equation}
then the operation of $\beta^{2}$ on $A$ is given by\begin{equation}
\beta^{2}\left(i\partial_{\tau}\right)A\left(\tau\right)=\frac{1}{2\pi}\int_{-\infty}^{+\infty}\beta^{2}\left(\omega\right)e^{-i\omega\tau}\widetilde{A}\left(\omega\right)d\omega\,.\label{eq:beta_squared_operator}\end{equation}
The Fourier transform is implemented numerically using the FFT (Fast
Fourier Transform) algorithm.

As in §\ref{sub:Wavepacket-propagation}, we convert Eqs. (\ref{eq:optical_PDE_order1_1})
and (\ref{eq:optical_PDE_order1_2}) into finite difference equations
by discretizing the $\zeta$- and $\tau$-axes into grids, with separations
$\Delta_{\zeta}$ and $\Delta_{\tau}$ respectively. We also allow
the grids representing values of $A$ and $\Pi$ to be offset by $\Delta_{\zeta}/2$.
Then Eqs. (\ref{eq:optical_PDE_order1_1}) and (\ref{eq:optical_PDE_order1_2})
become\begin{multline*}
\frac{A_{m+1,i}-A_{m,i}}{\Delta_{\zeta}}-\frac{1}{2}\left[\frac{A_{m,i+1}-A_{m,i-1}}{2\Delta_{\tau}}+\frac{A_{m+1,i+1}-A_{m+1,i-1}}{2\Delta_{\tau}}\right]=\Pi_{m,i}\,,\\
\frac{\Pi_{m+1,i}-\Pi_{m,i}}{\Delta_{\zeta}}-\frac{1}{2}\left[\frac{\Pi_{m,i+1}-\Pi_{m,i-1}}{2\Delta_{\tau}}+\frac{\Pi_{m+1,i+1}-\Pi_{m+1,i-1}}{2\Delta_{\tau}}\right]\\
=-u^{2}\left[\beta^{2}\left(i\partial_{\tau}\right)\right]_{m+1,i}+\frac{u^{2}}{c^{2}}\left[\partial_{\tau}\chi\right]_{i}\frac{A_{m+1,i+1}-A_{m+1,i-1}}{2\Delta_{\tau}}+\frac{u^{2}}{c^{2}}\chi_{i}\frac{A_{m+1,i+1}-2A_{m+1,i}+A_{m+1,i-1}}{\Delta_{\tau}^{2}}\,.\end{multline*}
Rearranging, we have\begin{equation}
A_{m+1,i}-\frac{\Delta_{\zeta}}{4\Delta_{\tau}}\left(A_{m+1,i+1}-A_{m+1,i-1}\right)=A_{m,i}+\frac{\Delta_{\zeta}}{4\Delta_{\tau}}\left(A_{m,i+1}-A_{m,i-1}\right)+\Delta_{\zeta}\Pi_{m,i}\,,\end{equation}
\begin{multline}
\Pi_{m+1,i}-\frac{\Delta_{\zeta}}{4\Delta_{\tau}}\left(\Pi_{m+1,i+1}-\Pi_{m+1,i-1}\right)=\Pi_{m,i}+\frac{\Delta_{\zeta}}{4\Delta_{\tau}}\left(\Pi_{m,i+1}-\Pi_{m,i-1}\right)-\Delta_{\zeta}u^{2}\left[\beta^{2}\left(i\partial_{\tau}\right)A\right]_{m+1,i}\\
+\frac{\Delta_{\zeta}}{2\Delta_{\tau}}\frac{u^{2}}{c^{2}}\left[\partial_{\tau}\chi\right]_{i}\left(A_{m+1,i+1}-A_{m+1,i-1}\right)+\frac{\Delta_{\zeta}}{\Delta_{\tau}^{2}}\frac{u^{2}}{c^{2}}\chi_{i}\left(A_{m+1,i+1}-2A_{m+1,i}+A_{m+1,i-1}\right)\,,\end{multline}
or, in matrix form,\begin{eqnarray}
\negthickspace\negthickspace\negthickspace\negthickspace\left[\begin{array}{ccccccc}
1 & -a &  &  &  &  & a\\
 & \ddots\\
 &  & \ddots\\
 &  & a & 1 & -a\\
 &  &  &  & \ddots\\
 &  &  &  &  & \ddots\\
-a &  &  &  &  & a & 1\end{array}\right]\left[\begin{array}{c}
A_{m+1,1}\\
\vdots\\
\vdots\\
A_{m+1,i}\\
\vdots\\
\vdots\\
A_{m+1,n}\end{array}\right] & = & \left[\begin{array}{c}
b_{m,1}\\
\vdots\\
\vdots\\
b_{m,i}\\
\vdots\\
\vdots\\
b_{m,n}\end{array}\right]\,,\qquad\qquad\label{eq:optical_FDTD_matrix_1}\\
\negthickspace\negthickspace\negthickspace\negthickspace\left[\begin{array}{ccccccc}
1 & -a &  &  &  &  & a\\
 & \ddots\\
 &  & \ddots\\
 &  & a & 1 & -a\\
 &  &  &  & \ddots\\
 &  &  &  &  & \ddots\\
-a &  &  &  &  & a & 1\end{array}\right]\left[\begin{array}{c}
\Pi_{m+1,1}\\
\vdots\\
\vdots\\
\Pi_{m+1,i}\\
\vdots\\
\vdots\\
\Pi_{m+1,n}\end{array}\right] & = & \left[\begin{array}{c}
c_{m,1}\\
\vdots\\
\vdots\\
c_{m,i}\\
\vdots\\
\vdots\\
c_{m,n}\end{array}\right]\,,\qquad\qquad\label{eq:optical_FDTD_matrix_2}\end{eqnarray}
where we have assumed periodic boundary conditions, identifying $i=n+1$
with $i=1$ and $i=0$ with $i=n$, and where we have defined\begin{eqnarray}
a & = & \frac{\Delta_{\zeta}}{4\Delta_{\tau}}\,,\\
b_{m,i} & = & A_{m,i}+\frac{\Delta_{\zeta}}{4\Delta_{\tau}}\left(A_{m,i+1}-A_{m,i-1}\right)+\Delta_{\zeta}\Pi_{m,i}\,,\\
c_{m,i} & = & \Pi_{m,i}+\frac{\Delta_{\zeta}}{4\Delta_{\tau}}\left(\Pi_{m,i+1}-\Pi_{m,i-1}\right)-\Delta_{\zeta}u^{2}\left[\beta^{2}\left(i\partial_{\tau}\right)A\right]_{m+1,i}+\frac{\Delta_{\zeta}}{2\Delta_{\tau}}\frac{u^{2}}{c^{2}}\left[\partial_{\tau}\chi\right]_{i}\left(A_{m+1,i+1}-A_{m+1,i-1}\right)\nonumber \\
 &  & \qquad\qquad+\frac{\Delta_{\zeta}}{\Delta_{\tau}^{2}}\frac{u^{2}}{c^{2}}\chi_{i}\left(A_{m+1,i+1}-2A_{m+1,i}+A_{m+1,i-1}\right)\,.\end{eqnarray}
The solution follows as before. The vector $\chi_{i}$ is assumed
fixed in $\zeta$ (hence its independence of $m$), and determines
the nonlinearity profile; $\left[\partial_{\tau}\chi\right]_{i}$
is its $\tau$-derivative, and can be stored as a separate vector
for convenience. Given the vectors $A_{m}$ and $\Pi_{m}$, we can
calculate $b_{m}$ and solve Eq. (\ref{eq:optical_FDTD_matrix_1})
for $A_{m+1}$, using the algorithm for cyclic equations from Ref.
\cite{NumericalRecipes}; then, given the vectors $\Pi_{m}$ and $A_{m+1}$,
we can calculate $c_{m}$ and solve Eq. (\ref{eq:optical_FDTD_matrix_2})
for $\Pi_{m+1}$. The only remaining unknowns are the initial values
$A_{1}$ and $\Pi_{1}$, which are related via the Fourier Transform
as in Eq. (\ref{eq:pi_from_phi}) and a half-step finite difference
equation similar to Eq. (\ref{eq:mod_diff_eqn_initial_pi}).

The FDTD algorithm does not run so smoothly for the optical model
as for the acoustic model. One problem is with the dispersion profile,
which naturally increases with frequency, in principle without bound.
Allowing such an unbounded form for $\beta\left(\omega\right)$ can
make the algorthim unstable; however, this problem is easily solved
by applying an appropriate cut-off frequency, beyond which $\beta$
is set to zero. A less tractable problem is that the degree of backscattering,
or scattering from $u$-modes into $v$-modes, is greater in the optical
model than in the acoustic model. Although the amount of backscattering
in terms of norm is still small, the amplitude of the wave, which
is proportional to $\beta\left(\omega\right)^{-1/2}$, is fairly large,
since the $v$-modes occur at very low frequencies. This makes the
above FDTD algorithm of limited value, especially visually, since
this unphysical low-frequency mode quickly fills up the domain.

\section{Steady-state solution\label{sub:Steady-state-solution-FIBRES}}

In practice, the spectrum of Hawking radiation is found by finding
steady-state solutions for various values of $\omega^{\prime}$, and
finding their negative-norm components. The general nonlinearity profile
is solved in an entirely analogous fashion to the acoustic model (see
§\ref{sub:Steady-state-solution}), and will be discussed only briefly.
For the special case of a nonlinearity profile which in constant apart
from step discontinuities, the method is again analogous to the acoustic
model, but the resulting equations are slightly different and will
be discussed in full.

\subsection{General nonlinearity profile}

Assuming a stationary solution of the form

\[
A\left(\zeta,\tau\right)=A_{\omega^{\prime}}\left(\tau\right)e^{-i\omega^{\prime}\zeta}\,,\]
the optical wave equation (\ref{eq:wave_equation_FIBRES}) is transformed
into the ordinary differential equation

\begin{equation}
-\omega^{\prime2}A_{\omega^{\prime}}+2i\omega^{\prime}\partial_{\tau}A_{\omega^{\prime}}+\partial_{\tau}^{2}A_{\omega^{\prime}}+u^{2}\beta^{2}\left(i\partial_{\tau}\right)A_{\omega^{\prime}}-\frac{u^{2}}{c^{2}}\partial_{\tau}\left(\chi\left(\tau\right)\partial_{\tau}A_{\omega^{\prime}}\right)=0\,.\label{eq:optical_steady_state_ODE}\end{equation}
The order of this equation is the same as the order of $\beta^{2}\left(\omega\right)$.
Assuming a constant value of $\chi$ and taking the Fourier transform
via the substitutions $A_{\omega^{\prime}}\left(\tau\right)\rightarrow\widetilde{A}_{\omega^{\prime}}\left(\omega\right)$
and $\partial_{\tau}\rightarrow-i\omega$, Eq. (\ref{eq:optical_steady_state_ODE})
becomes\[
\left\{ -\left(\omega^{\prime}-\omega\right)^{2}+u^{2}\beta^{2}\left(\omega\right)+\frac{u^{2}}{c^{2}}\chi\omega^{2}\right\} \widetilde{A}_{\omega^{\prime}}=0\,,\]
and again we find the dispersion relation of Eqs. (\ref{eq:beta_with_chi})-(\ref{eq:dispersion_with_chi-2}),\begin{equation}
\left(\omega^{\prime}-\omega\right)^{2}=u^{2}\beta^{2}\left(\omega\right)+\frac{u^{2}}{c^{2}}\chi\omega^{2}=u^{2}\beta_{\chi}^{2}\left(\omega\right)\,.\label{eq:optical_dispersion_relation}\end{equation}
If $\chi$ approaches asymptotic values for $\tau\rightarrow\pm\infty$,
then it can be approximated as constant in the asymptotic regions,
and the dispersion relation of Eq. (\ref{eq:optical_dispersion_relation})
will apply. The general solution in each asymptotic region is a sum
of plane waves with frequencies determined by the dispersion relation,
and the plane wave coefficients in each region are related via a scattering
matrix as in Eq. (\ref{eq:matrix_transformation}). (Refer to Fig.
\ref{fig:Integration-of-Stationary-Solution} for an illustration.)
The in- or out-mode that is to be solved determines which coefficients
to set to zero, and the remaining coefficients are solved via linear
algebra. The modes are normalized in the $\omega^{\prime}$-representation
according to Eqs. (\ref{eq:optical_mode_normalization}) and (\ref{eq:transformation_to_omega_prime_representation}):\begin{equation}
A_{\omega^{\prime}}\left(\tau\right)=\sqrt{\frac{u}{4\pi\epsilon_{0}c^{2}\left|\beta_{\chi}\left(\omega\right)\right|\left|\partial\omega^{\prime}/\partial\omega\right|}}e^{-i\omega\tau-i\omega^{\prime}\zeta}\,.\end{equation}
Therefore, if we have the plane wave coefficients, the norm of a particular
plane wave in relation to the single ingoing or outgoing wave is\begin{equation}
\frac{\left(A_{\omega^{\prime},\omega},A_{\omega^{\prime},\omega}\right)}{\left(A_{\omega^{\prime},\omega_{\mathrm{in/out}}},A_{\omega^{\prime},\omega_{\mathrm{in/out}}}\right)}=\frac{\mathrm{sgn}\left(\omega-\omega^{\prime}\right)\left|\beta_{\chi}\left(\omega\right)\right|\left|\partial\omega^{\prime}/\partial\omega\left(\omega\right)\right|\left|C_{\omega^{\prime},\omega}\right|^{2}}{\mathrm{sgn}\left(\omega_{\mathrm{in/out}}-\omega^{\prime}\right)\left|\beta_{\chi}\left(\omega_{\mathrm{in/out}}\right)\right|\left|\partial\omega^{\prime}/\partial\omega\left(\omega_{\mathrm{in/out}}\right)\right|\left|C_{\omega^{\prime},\omega_{\mathrm{in/out}}}\right|^{2}}\,.\end{equation}

\subsection{Discontinuous nonlinearity profile}

Eq. (\ref{eq:optical_steady_state_ODE}) may be solved analytically
in the special case where $\chi$ is constant everywhere except at
discrete points. This might correspond, for example, to a square pulse,
whose intensity is constant in a certain region and zero elsewhere.
It is of course an idealised case, formed by taking the limit as the
derivative of $\chi$ increases without bound.

We proceed exactly as in §\ref{sub:Steady-state-solution}. We assume
that $\beta^{2}\left(\omega\right)$ is a polynomial of order $2n$,\[
\beta^{2}\left(\omega\right)=\sum_{j=1}^{n}b_{j}\omega^{2j}\,,\]
which would result in $2n$ solutions of the dispersion relation.
In any region of constant $\chi$, the solution is specified by the
$2n$ coefficients of the possible plane waves; if there are $r$
regions, or $r-1$ points of discontinuity, there are $2nr$ coefficients
in total. However, there can only be $2n$ linearly independent solutions,
which means there must be $2n\left(r-1\right)$ conditions on these
coefficients, or $2n$ conditions for each point of discontinuity.
These relate the limiting values of $A_{\omega^{\prime}}$ and its
first $2n-1$ derivatives on each side of these points.

Expanding the last term of Eq. (\ref{eq:optical_steady_state_ODE}),
and replacing the operator $\beta^{2}\left(i\partial_{\tau}\right)$
with its polynomial expansion, we have\begin{multline}
-\omega^{\prime2}A_{\omega^{\prime}}+2i\omega^{\prime}\partial_{\tau}A_{\omega^{\prime}}+\partial_{\tau}^{2}A_{\omega^{\prime}}+\frac{u^{2}}{c^{2}}\chi^{\prime}\left(\tau\right)\partial_{\tau}A_{\omega^{\prime}}+\frac{u^{2}}{c^{2}}\chi\left(\tau\right)\partial_{\tau}^{2}A_{\omega^{\prime}}\\
+u^{2}\sum_{j=1}^{n}\left(-1\right)^{j}b_{j}\partial^{2j}A_{\omega^{\prime}}=0\,.\label{eq:steady_state_ODE_beta_polynomial}\end{multline}
We see that the steady-state equation contains $\chi$ and its first
derivative $\chi^{\prime}$ only, and is therefore \textit{almost}
equivalent to the acoustic case, which contains the flow velocity
$V$ and its first derivative $V^{\prime}$ only. (The difference
here is that $\chi$ and $\chi^{\prime}$ are not multiplied, and
the discontinuities in the derivatives of $A_{\omega^{\prime}}$ can
have at most a $1/\delta$ dependence, in contrast to the $1/\delta^{2}$
dependence in the acoustic model.) We note that a discontinuity in
$\chi$ must be accompanied by a delta function in $\chi^{\prime}$,
and that these can only be cancelled by a discontinuity in $\partial_{\tau}^{2n-1}A_{\omega^{\prime}}$
accompanied by a delta function in $\partial_{\tau}^{2n}A_{\omega^{\prime}}$.
All other quantities must be continuous. Therefore, the first $2n-1$
of the $2n$ conditions to be imposed at a point of discontinuity
are the continuity of $A_{\omega^{\prime}}$ and its first $2n-2$
derivatives; the final condition involves imposing the correct discontinuity
on $\partial_{\tau}^{2n-1}A_{\omega^{\prime}}$.

Proceeding as in the acoustic case, we imagine a nonlinearity profile
that varies linearly over a region of length $\delta$ between two
constant-$\chi$ regions, and is therefore continuous in $\chi$ but
discontinuous in $\chi^{\prime}$; the case of discontinuous $\chi$
is found by taking the limit $\delta\rightarrow0$. (See Fig. \ref{fig:discontinuous_velocity_profile},
replacing $V$ with $\chi$.) A discontinuity in $\chi^{\prime}$
at the point $\tau=\tau_{\star}$ must be accompanied by a corresponding
discontinuity in $\partial_{\tau}^{2n}A_{\omega^{\prime}}$ at $\tau_{\star}$,
so that Eq. (\ref{eq:steady_state_ODE_beta_polynomial}) is always
satisfied. We have\[
\lim_{\tau\rightarrow\tau_{\star}^{-}}\left[\frac{u^{2}}{c^{2}}\chi^{\prime}\partial_{\tau}A_{\omega^{\prime}}-u^{2}\left(-1\right)^{n}b_{n}\partial_{\tau}^{2n}A_{\omega^{\prime}}\right]=\lim_{\tau\rightarrow\tau_{\star}^{+}}\left[\frac{u^{2}}{c^{2}}\chi^{\prime}\partial_{\tau}A_{\omega^{\prime}}-u^{2}\left(-1\right)^{n}b_{n}\partial_{\tau}^{2n}A_{\omega^{\prime}}\right]\,,\]
or, defining $\Delta\chi^{\prime}\equiv\lim_{\tau\rightarrow\tau_{\star}^{+}}\chi^{\prime}-\lim_{\tau\rightarrow\tau_{\star}^{-}}\chi^{\prime}$
and similarly for $\Delta\left(\partial_{\tau}^{2n}A_{\omega^{\prime}}\right)$,\begin{eqnarray}
\Delta\left(\partial_{\tau}^{2n}A_{\omega^{\prime}}\right) & = & \frac{\left(-1\right)^{n}}{b_{n}c^{2}}\left(\partial_{\tau}A_{\omega^{\prime}}\right)\Delta\chi^{\prime}\nonumber \\
 & = & \frac{\left(-1\right)^{n}}{b_{n}c^{2}}\left(\partial_{\tau}A_{\omega^{\prime}}\right)\frac{\left(\chi_{2}-\chi_{1}\right)}{\delta}\,,\label{eq:discontinuity_in_2n_A_derivative}\end{eqnarray}
where $\partial_{\tau}A_{\omega^{\prime}}$ is to be evaluated at
$\tau=\tau_{\star}$ (this is continuous, so there is no ambiguity).

To find the change in $\partial_{\tau}^{2n-1}A_{\omega^{\prime}}$
over the change in $\chi$, we use the Taylor expansion\begin{eqnarray*}
\left.\partial_{\tau}^{2n-1}A_{\omega^{\prime}}\right|_{\tau=\tau_{\star}+h}-\left.\partial_{\tau}^{2n-1}A_{\omega^{\prime}}\right|_{\tau=\tau_{\star}} & = & \left.\partial_{\tau}^{2n}A_{\omega^{\prime}}\right|_{\tau\rightarrow\tau_{\star}^{+}}\delta+\frac{1}{2}\left.\partial_{\tau}^{2n+1}A_{\omega^{\prime}}\right|_{\tau\rightarrow\tau_{\star}^{+}}\delta^{2}+\ldots\\
 & = & \left.\partial_{\tau}^{2n}A_{\omega^{\prime}}\right|_{\tau\rightarrow\tau_{\star}^{-}}\delta+\frac{1}{2}\left.\partial_{\tau}^{2n+1}A_{\omega^{\prime}}\right|_{\tau\rightarrow\tau_{\star}^{-}}\delta^{2}+\ldots\\
 &  & +\Delta\left(\partial_{\tau}^{2n}A_{\omega^{\prime}}\right)\delta+\frac{1}{2}\Delta\left(\partial_{\tau}^{2n+1}A_{\omega^{\prime}}\right)\delta^{2}+\ldots\end{eqnarray*}
Letting $h\rightarrow0$, the only non-vanishing term is $\Delta\left(\partial_{\tau}^{2n}A_{\omega^{\prime}}\right)\delta$,
and we find that the discontinuity in $\partial_{\tau}^{2n-1}A_{\omega^{\prime}}$
is\begin{equation}
\Delta\left(\partial_{\tau}^{2n-1}A_{\omega^{\prime}}\right)=\frac{\left(-1\right)^{n}}{b_{n}c^{2}}\left(\partial_{\tau}A_{\omega^{\prime}}\right)\left(\chi_{2}-\chi_{1}\right)\,.\label{eq:discontinuity_in_2n-1_A_derivative}\end{equation}
This is the final condition to be imposed at each point of discontinuity.
We are thus left with $2n$ degrees of freedom, corresponding to the
$2n$-dimensional space of solutions, for which we have the bases
of $2n$ in-modes or $2n$ out-modes.

\section{Linearized intensity profile\label{sub:Linearized-intensity-profile}}

Using an approach very similar to that used in §\ref{sub:WKB-approximation},
we may derive an analytic expression for the creation rate. The essential
assumption is that the creation process occurs in a very narrow region
around the horizon, and therefore that we may approximate the intensity
profile as a linear one - just as, in the acoustic case, we approximated
the velocity profile as linear.

Assume a steady-state solution $A_{\omega^{\prime}}$, which satisfies
Eq. (\ref{eq:optical_steady_state_ODE}). Let the nonlinearity $\chi\left(\tau\right)$
take the form\begin{equation}
\chi\left(\tau\right)=\chi_{h}-\alpha\tau\,.\label{eq:linearized_intensity_profile}\end{equation}
The optical wave equation now becomes first-order in $\tau$. Making
the substitutions $A_{\omega^{\prime}}\left(\tau\right)\rightarrow\widetilde{A}_{\omega^{\prime}}\left(\omega\right)$,
$\tau\rightarrow-i\partial_{\omega}$ and $\partial_{\tau}\rightarrow-i\omega$,
we find a first-order differential equation for the Fourier transform:\begin{equation}
-\left(\omega^{\prime}-\omega\right)^{2}\widetilde{A}_{\omega^{\prime}}+u^{2}\beta^{2}\left(\omega\right)\widetilde{A}_{\omega^{\prime}}+\frac{u^{2}}{c^{2}}\chi_{h}\omega^{2}\widetilde{A}_{\omega^{\prime}}-i\frac{u^{2}}{c^{2}}\alpha\omega\widetilde{A}_{\omega^{\prime}}+i\frac{u^{2}}{c^{2}}\alpha\partial_{\omega}\left(\omega^{2}\widetilde{A}_{\omega^{\prime}}\right)=0\,.\label{eq:Fourier_transform_1st_order}\end{equation}
Rearranging, this becomes\begin{equation}
\partial_{\omega}\left(\omega^{2}\widetilde{A}_{\omega^{\prime}}\right)=\left\{ \frac{1}{\omega}+\frac{i}{\alpha}\left[\chi_{h}+c^{2}\frac{\beta^{2}\left(\omega\right)}{\omega^{2}}-\frac{c^{2}}{u^{2}}\left(1-\frac{\omega^{\prime}}{\omega}\right)^{2}\right]\right\} \omega^{2}\widetilde{A}_{\omega^{\prime}}\,.\end{equation}
The solution is found to be\begin{equation}
\omega^{2}\widetilde{A}_{\omega^{\prime}}=C\,\exp\left(\log\,\omega+\frac{i}{\alpha}\int^{\omega}\left[\chi_{h}+c^{2}\frac{\beta^{2}\left(\bar{\omega}\right)}{\bar{\omega}^{2}}-\frac{c^{2}}{u^{2}}\left(1-\frac{\omega^{\prime}}{\bar{\omega}}\right)^{2}\right]d\bar{\omega}\right)\end{equation}
or\begin{equation}
\widetilde{A}_{\omega^{\prime}}=\frac{C}{\omega}\exp\left(\frac{i}{\alpha}\int^{\omega}\left[\chi_{h}+c^{2}\frac{\beta^{2}\left(\bar{\omega}\right)}{\bar{\omega}^{2}}-\frac{c^{2}}{u^{2}}\left(1-\frac{\omega^{\prime}}{\bar{\omega}}\right)^{2}\right]d\bar{\omega}\right)\,,\end{equation}
where $C$ is an arbitrary constant. This solution for the Fourier
transform is exact if $\chi\left(\tau\right)$ is exactly linear.
This differs from the acoustic case, where the WKB approximation was
invoked to provide an approximate solution for the Fourier transform.

In real space, the solution is given by the inverse Fourier transform,\begin{eqnarray}
A_{\omega^{\prime}}\left(\tau\right) & = & \int_{-\infty}^{+\infty}\widetilde{A}_{\omega^{\prime}}\left(\omega\right)\exp\left(-i\omega\tau\right)d\omega\nonumber \\
 & = & \int_{-\infty}^{+\infty}\frac{C}{\omega}\exp\left(i\varphi\left(\omega\right)\right)d\omega\label{eq:real_space_solution}\end{eqnarray}
where we have defined\begin{eqnarray}
\varphi\left(\omega\right) & = & \frac{1}{\alpha}\int^{\omega}\left[\chi_{h}-\alpha\tau+c^{2}\frac{\beta^{2}\left(\bar{\omega}\right)}{\bar{\omega}^{2}}-\frac{c^{2}}{u^{2}}\left(1-\frac{\omega^{\prime}}{\bar{\omega}}\right)^{2}\right]d\bar{\omega}\nonumber \\
 & = & \frac{1}{\alpha}\int^{\omega}\left[\chi\left(\tau\right)+c^{2}\frac{\beta^{2}\left(\bar{\omega}\right)}{\bar{\omega}^{2}}-\frac{c^{2}}{u^{2}}\left(1-\frac{\omega^{\prime}}{\bar{\omega}}\right)^{2}\right]d\bar{\omega}\,.\label{eq:phase_in_inverse_FT}\end{eqnarray}
As before, this is easily solved using the method of stationary phase:
the main contributions to the integral come from near those points
where $\varphi^{\prime}\left(\omega\right)=0$, where we approximate
the integral as Gaussian. The stationary points $\omega_{j}$ are
easily found, for they satisfy\[
\varphi^{\prime}\left(\omega_{j}\right)=\frac{1}{\alpha}\left[\chi\left(\tau\right)+c^{2}\frac{\beta^{2}\left(\omega_{j}\right)}{\omega_{j}^{2}}-\frac{c^{2}}{u^{2}}\left(1-\frac{\omega^{\prime}}{\omega_{j}}\right)^{2}\right]=0\,,\]
or, rearranging,\begin{equation}
\left(\omega_{j}-\omega^{\prime}\right)^{2}=u^{2}\beta^{2}\left(\omega_{j}\right)+\frac{u^{2}}{c^{2}}\chi\left(\tau\right)\omega_{j}^{2}\,.\label{eq:local_dispersion_relation}\end{equation}
That is, the stationary points are exactly those values of $\omega$
that solve the dispersion relation, with $\chi$ given by its value
at $\tau$. Around these points, the phase $\varphi\left(\omega\right)$
can be approximated as quadratic in $\omega$, and for this we need
the second derivative of $\varphi\left(\omega\right)$:\[
\varphi^{\prime\prime}\left(\omega\right)=\frac{1}{\alpha}\frac{c^{2}\left(\omega^{2}\left(2\beta\left(\omega\right)\beta^{\prime}\left(\omega\right)-2\left(\omega-\omega^{\prime}\right)/u^{2}\right)-2\omega\left(\beta^{2}\left(\omega\right)-\left(\omega-\omega^{\prime}\right)^{2}/u^{2}\right)\right)}{\omega^{4}}\,.\]
If $\omega$ solves the dispersion relation of Eq. (\ref{eq:local_dispersion_relation}),
then clearly\[
\frac{c^{2}}{u^{2}}\left(u^{2}\beta^{2}\left(\omega\right)-\left(\omega-\omega^{\prime}\right)^{2}\right)=-\chi\left(\tau\right)\omega^{2}\,.\]
Differentiating, we have\[
\frac{c^{2}}{u^{2}}\left(u^{2}2\beta\left(\omega\right)\beta^{\prime}\left(\omega\right)-2\left(\omega-\omega^{\prime}\right)\left(1-\frac{\partial\omega^{\prime}}{\partial\omega}\right)\right)=-\chi\left(\tau\right)2\omega,\]
or, rearranging,\[
\frac{c^{2}}{u^{2}}\left(2u^{2}\beta\left(\omega\right)\beta^{\prime}\left(\omega\right)-2\left(\omega-\omega^{\prime}\right)\right)=-2\frac{c^{2}}{u^{2}}\left(\omega-\omega^{\prime}\right)\frac{\partial\omega^{\prime}}{\partial\omega}-2\chi\left(\tau\right)\omega\,.\]
Substituting into the expression for $\varphi^{\prime\prime}\left(\omega\right)$
above, we find the much simpler expression\begin{eqnarray}
\varphi^{\prime\prime}\left(\omega_{j}\right) & = & \frac{1}{\alpha\omega_{j}^{4}}\left(-2\frac{c^{2}}{u^{2}}\omega_{j}^{2}\left(\omega_{j}-\omega^{\prime}\right)\frac{\partial\omega^{\prime}}{\partial\omega}\left(\omega_{j}\right)-2\chi\left(\tau\right)\omega_{j}^{3}+2\chi\left(\tau\right)\omega_{j}^{3}\right)\nonumber \\
 & = & \frac{2}{\alpha}\frac{c^{2}}{u^{2}}\frac{1}{\omega_{j}^{2}}\left(\omega_{j}-\omega^{\prime}\right)v_{g}\left(\omega_{j}\right)\,,\end{eqnarray}
where $v_{g}\left(\omega\right)=-\partial\omega^{\prime}/\partial\omega$
is the group velocity in the co-moving frame. Around the frequency
$\omega_{j}$, then, $\varphi\left(\omega\right)$ can be approximated
by\begin{eqnarray}
\varphi\left(\omega\right) & \approx & \varphi\left(\omega_{j}\right)+\frac{1}{2}\varphi^{\prime\prime}\left(\omega_{j}\right)\left(\omega-\omega_{j}\right)^{2}\nonumber \\
 & = & \varphi\left(\omega_{j}\right)+\frac{1}{\alpha}\frac{c^{2}}{u^{2}}\frac{1}{\omega_{j}^{2}}\left(\omega_{j}-\omega^{\prime}\right)v_{g}\left(\omega_{j}\right)\left(\omega-\omega_{j}\right)^{2}\,.\label{eq:quadratic_form_of_phase}\end{eqnarray}
The integral in Eq. (\ref{eq:real_space_solution}) can now be approximated
as\begin{eqnarray}
A_{\omega^{\prime}}\left(\tau\right) & \approx & \sum_{j}\frac{C}{\omega_{j}}\int_{-\infty}^{+\infty}\exp\left(i\varphi\left(\omega_{j}\right)+\frac{i}{\alpha}\frac{c^{2}}{u^{2}}\frac{1}{\omega_{j}^{2}}\left(\omega_{j}-\omega^{\prime}\right)v_{g}\left(\omega_{j}\right)\left(\omega-\omega_{j}\right)^{2}\right)d\omega\nonumber \\
 & = & C\sum_{j}\frac{1}{\omega_{j}}\exp\left(i\varphi\left(\omega_{j}\right)\right)\sqrt{\pi\left|\alpha\right|\omega_{j}^{2}\frac{u^{2}}{c^{2}}\frac{1}{\left|\left(\omega_{j}-\omega^{\prime}\right)v_{g}\left(\omega_{j}\right)\right|}}\exp\left(i\frac{\pi}{4}\mathrm{sgn}\,\left[\left(\omega_{j}-\omega^{\prime}\right)v_{g}\left(\omega_{j}\right)\right]\right)\nonumber \\
 & = & C\,\sqrt{\pi\,\left|\alpha\right|}\,\frac{u}{c}\,\sum_{j}\frac{\mathrm{sgn}\,\left[\omega_{j}\right]}{\sqrt{\left|\left(\omega_{j}-\omega^{\prime}\right)v_{g}\left(\omega_{j}\right)\right|}}\exp\left(i\frac{\pi}{4}\mathrm{sgn}\,\left[\left(\omega_{j}-\omega^{\prime}\right)v_{g}\left(\omega_{j}\right)\right]+i\varphi\left(\omega_{j}\right)\right)\,.\nonumber \end{eqnarray}
Note that the factor $\left|\left(\omega_{j}-\omega^{\prime}\right)v_{g}\left(\omega_{j}\right)\right|^{-1/2}$
is precisely that factor needed to normalize the plane wave of frequency
$\omega_{j}$. It seems reasonable that the remaining exponential
term, $\exp\left(i\varphi\left(\omega_{j}\right)\right)$, becomes
just such a plane wave in the asymptotic region, where $\chi$ becomes
constant. Let us examine this term; from Eq. (\ref{eq:phase_in_inverse_FT}),\begin{equation}
\varphi\left(\omega_{j}\right)=\frac{1}{\alpha}\int^{\omega_{j}}\left[\chi_{h}-\alpha\tau+\frac{c^{2}}{u^{2}}\frac{u^{2}\beta^{2}\left(\omega\right)-\left(\omega-\omega^{\prime}\right)^{2}}{\omega^{2}}\right]d\omega\,,\end{equation}
valid when $\chi\left(\tau\right)=\chi_{h}-\alpha\tau$. However,
if we plug this form of $\chi\left(\tau\right)$ into the dispersion
relation Eq. (\ref{eq:local_dispersion_relation}), we find that we
can express $\tau$ as a function of $\omega$:\begin{equation}
\tau\left(\omega\right)=\frac{1}{\alpha}\left[\chi_{h}+\frac{c^{2}}{u^{2}}\frac{u^{2}\beta^{2}\left(\omega\right)-\left(\omega-\omega^{\prime}\right)^{2}}{\omega^{2}}\right]\,.\end{equation}
The phase $\varphi\left(\omega_{j}\right)$, then, is simply given
by\begin{equation}
\varphi\left(\omega_{j}\right)=-\omega_{j}\tau+\int^{\omega_{j}}\tau\left(\omega\right)d\omega\,,\end{equation}
and we generalize the result by allowing $\tau\left(\omega\right)$
to take an arbitrary form.

In the form of an integral over $\omega$, the simple nature of the
phase is obscured; for, differentiating with respect to $\tau$, and
bearing in mind that $\omega_{j}$ itself is a function of $\tau$,
we find $d\varphi/d\tau=-\omega_{j}$. Therefore, $\varphi$ can also
be written\begin{equation}
\varphi=-\int^{\tau}\omega_{j}\left(\tau^{\prime}\right)d\tau^{\prime}\,.\end{equation}
Clearly, in the asymptotic regions, $\varphi\rightarrow-\omega_{j}\tau$,
and the solution becomes a sum over the normalized modes. The only
way, in this approximation, that different modes can have different
amplitudes is if $\varphi\left(\omega\right)$ has an imaginary part.
The norm, then, of one frequency $\omega_{1}$ relative to another $\omega_{2}$ is
\begin{alignat}{1}
\frac{\left(A_{\omega^{\prime},\omega_{1}},A_{\omega^{\prime},\omega_{1}}\right)}{\left(A_{\omega^{\prime},\omega_{2}},A_{\omega^{\prime},\omega_{2}}\right)}\approx\frac{\mathrm{sgn}\left(\omega_{1}-\omega^{\prime}\right)}{\mathrm{sgn}\left(\omega_{2}-\omega^{\prime}\right)}\exp\left(-2\,\mathrm{Im}\left\{ J\right\} \right)\,,\qquad & J=\int_{\omega_{2}}^{\omega_{1}}\tau\left(\bar{\omega}\right)d\bar{\omega}\,.\end{alignat}
In particular, for photon creation, the relative weight between the Bogoliubov coefficients is
\begin{equation}
\left|\frac{\beta}{\alpha}\right|^{2}=\exp\left(-2\,\mathrm{Im}\left[\int_{\omega_{\alpha}}^{\omega_{\beta}}\tau\left(\bar{\omega}\right)\,d\bar{\omega}\right]\right)\,.\label{eq:creat_phase_FIBRES}
\end{equation}

\section{Summary}

The methods described in this chapter are entirely analogous to those
developed in the context of the acoustic model. They are:
\begin{itemize}
\item an FDTD algorithm, for solving wavepacket propagation;
\item numerical integration to find stationary modes; and
\item an analytic approximation for a linearized intensity profile.
\end{itemize}
Of these, as before, numerical integration of stationary modes is
the most convenient for calculating the Hawking spectra, while the
analytic approximation is useful in pointing the way towards a better
analytic expression.

\pagebreak{}

\chapter{Results for the Fibre-Optical Model\label{sec:Results-for-the-Fibre-Optical-Model}}

We come now to the presentation of the results obtained from solutions
to the fibre-optical wave equation. This is the first such presentation for the fibre-optical model,
although the main results - both old and new - of Chapter \ref{sec:Results-for-Acoustic-Model} in
the context of the acoustic model are replicated here. However, when using realistic profiles
for the pulse-induced nonlinearity, we find ourselves in a new regime which not only contains
no low-frequency horizon, but whose spectrum is dominated by a frequencies which experience
no group-velocity horizon. The qualitative results in this regime are rather different; in particular, we shall
see that the width of the spectrum remains constant in $h$, the height of the nonlinearity, but
that its value is proportional to $h^{2}$.

\section{Dispersion relation\label{sub:Dispersion-relation}}

The dispersion is treated as subluminal, so that $\beta^{\prime}\left(\omega\right)$
increases with $\omega$. (Recall that $\beta\left(\omega\right)$
is the wavenumber as a function of frequency; it is essentially the
inverse of the function $F\left(k\right)$ that was used in the acoustic
model. Subluminal dispersion, then, is expressed as a decrease in
$F^{\prime}\left(k\right)$ with $k$ or an increase in $\beta^{\prime}\left(\omega\right)$
with $\omega$.) As before, the simplest deviation from the dispersionless
case is the introduction of a quartic term in $\beta^{2}\left(\omega\right)$:\begin{equation}
\beta^{2}\left(\omega\right)=\frac{\omega^{2}}{c^{2}}\left(1+\frac{\omega^{2}}{\omega_{d}^{2}}\right)\,.\label{eq:optical_quartic_dispersion}\end{equation}
This is the first form of the dispersion relation we shall use. It
cannot, of course, capture all the details of any realistic dispersion
relation; indeed, it is characterised by a single parameter, $\omega_{d}$,
which determines the scale at which deviations from the dispersionless
case first become appreciable. We could, if we wished, more closely
approximate a certain dispersion profile by adding higher even powers
of $\omega$; but such adjustments increase the complexity of the
calculations - the order of the differential equation is increased,
and usually some of the possible frequencies are complex, which can
affect the accuracy of the numerics - and do not return their worth
with significantly different results. Therefore, we shall not consider
them.

It is convenient to express $\beta$ not in terms of $\omega_{d}$,
but of the lab frequency with zero co-moving frequency, which we shall
call $\omega_{0}$. This is defined via the relation\[
\omega^{\prime}\left(\omega_{0}\right)=\omega_{0}-\frac{u}{c}\omega_{0}\sqrt{1+\frac{\omega_{0}^{2}}{\omega_{d}^{2}}}=0\Longrightarrow\omega_{0}^{2}=\omega_{d}^{2}\left(\frac{c^{2}}{u^{2}}-1\right)\,.\]
Eq. (\ref{eq:optical_quartic_dispersion}) may now be written:\begin{equation}
\beta^{2}\left(\omega\right)=\frac{\omega^{2}}{c^{2}}\left(1+\left(\frac{c^{2}}{u^{2}}-1\right)\frac{\omega^{2}}{\omega_{0}^{2}}\right)\,.\label{eq:optical_quartic_dispersion_omega_zero}\end{equation}

However, even leaving the dispersion relation in this quartic form,
there is still some room for manoeuvre. We shall be especially interested
in the behaviour of high frequencies, around the frequency $\omega_{0}$.
To find a form of $\beta$ which models the dispersion around this
frequency well, we wish both that $\omega^{\prime}$ vanish there
and that the derivative $d\omega^{\prime}/d\omega$ is equal to a
specified value (which may then be measured). Let $\xi=-d\omega^{\prime}/d\omega\mid_{\omega=\omega_{0}}$.
Matching these two conditions requires two variable parameters for
$\beta$, but Eq. (\ref{eq:optical_quartic_dispersion}) has only
one: the parameter $\omega_{d}$. This is because we have chosen $\beta$
such that the phase velocity approaches $c$ for low frequencies.
If we require a better match at high frequencies, we must sacrifice
this low-frequency condition. (There should be no qualms about doing this,
for realistic fibres have a non-trivial refractive index as $\omega\rightarrow0$,
so the dispersionless limit is unrealistic anyway.) Doing so, we may write $\beta^{2}$
in the more general form\[
\beta^{2}\left(\omega\right)=\frac{\omega^{2}}{c^{2}}\left(b_{1}+b_{2}\omega^{2}\right)\,,\]
then apply the conditions we want satisfied:\begin{alignat*}{3}
\omega^{\prime}\left(\omega_{0}\right) & = & \,\omega_{0}-\frac{u}{c}\omega_{0}\sqrt{b_{1}+b_{2}\omega_{0}^{2}} & = & 0\,,\\
\frac{d\omega^{\prime}}{d\omega}\left(\omega_{0}\right) & = & 1-\frac{u}{c}\frac{b_{1}+2\omega_{0}^{2}}{\sqrt{b_{1}+b_{2}\omega_{0}^{2}}} & = & \,-\xi\,.\end{alignat*}
These are easily solved to give\begin{alignat*}{1}
b_{1}=\frac{c^{2}}{u^{2}}\left(1-\xi\right)\,,\qquad & b_{2}=\frac{c^{2}}{u^{2}}\frac{\xi}{\omega_{0}^{2}}\,.\end{alignat*}
Finally, we have\begin{equation}
\beta^{2}\left(\omega\right)=\frac{\omega^{2}}{u^{2}}\left(1-\xi+\xi\frac{\omega^{2}}{\omega_{0}^{2}}\right)\,.\label{eq:optical_quartic_dispersion_with_xi}\end{equation}

Comparing Eqs. (\ref{eq:optical_quartic_dispersion_omega_zero}) and
(\ref{eq:optical_quartic_dispersion_with_xi}), we see that the second
can be made equal to the first via the substitution $\xi\rightarrow1-u^{2}/c^{2}$.
Therefore, in what follows, we shall use the dispersion relation in
Eq. (\ref{eq:optical_quartic_dispersion_with_xi}) exclusively, and
without loss of generality. For the purpose of comparison with the acoustic model,
we shall include the case where $n\rightarrow1$ as $\omega\rightarrow0$, bearing
in mind that this is unrealistic. We shall also look at a more realistic model,
whose value of $\xi$ is determined by the high-frequency dispersion, near $\omega_{0}$, of a real fibre.
Therefore, the two cases we shall examine are:
\begin{itemize}
\item $\xi\approx0.56$ gives $\beta^{2}\left(\omega\right)\rightarrow\frac{\omega}{c}$ at
low frequencies (near $\omega=0$); and
\item $\xi\approx0.036$ gives a more realistic $\beta^{2}\left(\omega\right)$
at high frequencies (near $\omega=\omega_{0}$).
\end{itemize}

Let us also note that a purely subluminal dispersion relation, like those given above, is incompatible with the alleged
existence of optical solitons in the fibre: as noted in \S\ref{sub:Nonlinear-effects}, solitons are only possible in regions where
$\partial^{2}\beta/\partial\omega^{2}<0$, which is clearly not the case for the above forms of $\beta$.  However, the purpose of
the soliton - if, indeed, the pulse is a soliton - is to provide the nonlinear change in refractive index; its frequency is irrelevant
for the purposes of the mode conversion.  It is the dispersion at the modes that are scattered into each other which is relevant
for calculating the Hawking spectrum.  Even then, as already noted, the quartic dispersion cannot be expected to apply to
all frequencies, being only the simplest analytic deviation from the dispersionless case.

\section{Intensity profile\label{sub:Intensity-profile}}

The simplest intensity profile, by direct analogy with the acoustic
model, is the hyperbolic tangent,\begin{equation}
\chi\left(\tau\right)=\frac{1}{2}\left(\chi_{R}+\chi_{L}\right)+\frac{1}{2}\left(\chi_{R}-\chi_{L}\right)\tanh\left(\alpha\tau\right)\,,\label{eq:chi_tanh}\end{equation}
which varies monotonically between two asymptotic values: $\chi_{L}$
on the left ($\tau\rightarrow-\infty$) and $\chi_{R}$ on the right
($\tau\rightarrow+\infty$). Another possibility is the hyperbolic
secant profile that would be associated with a fundamental soliton:\begin{equation}
\chi\left(\tau\right)=\chi_{0}\,\mathrm{sech}^{2}\left(\frac{\tau}{T_{0}}\right)\,.\label{eq:chi_sech2}\end{equation}
Although Eq. (\ref{eq:chi_sech2}) describes an intensity profile
which is exact for a fundamental soliton, and in that sense more realistic
than Eq. (\ref{eq:chi_tanh}), there are instances when this relationship
might change. In particular, the phenomenon of self-steepening perturbs
a soliton in such a way that its trailing edge becomes much steeper
than its leading edge. Since radiation rates are strongly dependent
on the steepness of the change in intensity, this single steep edge
dominates the process, so that Eq. (\ref{eq:chi_tanh}) becomes more
physically relevant than Eq. (\ref{eq:chi_sech2}), whose leading
and trailing edges are equally steep.

\section{Normalizing the wave equation\label{sub:Normalizing-the-wave-eqn-FIBRES}}

We now transform the optical wave equation into a form with as few
parameters as possible. Beginning with the general fiber-optical wave
equation (\ref{eq:wave_eqn_for_vector_potential}),\begin{equation}
\frac{c^{2}}{u^{2}}\left(\partial_{\zeta}-\partial_{\tau}\right)^{2}A+c^{2}\beta^{2}\left(i\partial_{\tau}\right)A-\partial_{\tau}\left(\chi\partial_{\tau}A\right)=0\,,\label{eq:optical_wave_equation}\end{equation}
we substitute the quartic form of $\beta\left(\omega\right)$ given
in Eq. (\ref{eq:optical_quartic_dispersion_with_xi}) to get\begin{equation}
\frac{c^{2}}{u^{2}}\left(\partial_{\zeta}-\partial_{\tau}\right)^{2}A-\frac{c^{2}}{u^{2}}\left(1-\xi\right)\partial_{\tau}^{2}A+\frac{c^{2}}{u^{2}}\frac{\xi}{\omega_{0}^{2}}\partial_{\tau}^{4}A-\partial_{\tau}\left(\chi\partial_{\tau}A\right)=0\,.\end{equation}
Transforming to dimensionless variables $T=\omega_{0}\tau$ and $Z=\omega_{0}\zeta$,
this becomes\begin{equation}
\partial_{Z}^{2}A-2\partial_{Z}\partial_{T}A+\xi\left(\partial_{T}^{2}A+\partial_{T}^{4}A\right)-\partial_{T}\left(\frac{u^{2}}{c^{2}}\chi\,\partial_{T}A\right)=0\,,\label{eq:optical_wave_eqn_norm_with_xi}\end{equation}
or\begin{alignat}{1}
\left(\partial_{Z}-\partial_{T}\right)^{2}A+B^{2}\left(i\partial_{T}\right)A-\partial_{T}\left(X\partial_{T}A\right)=0\,,\qquad & \mathrm{where}\quad B^{2}\left(\Omega\right)=\left(1-\xi\right)\Omega^{2}+\xi\Omega^{4}\nonumber \\
 & \mathrm{and}\qquad X\left(T\right)=\frac{u^{2}}{c^{2}}\chi\left(T\right)\,.\label{eq:optical_wave_eqn_norm_with_xi_B_X}\end{alignat}
If $\chi$ takes the hyperbolic tangent form of Eq. (\ref{eq:chi_tanh}),
then $X$ may be written\[
X\left(T\right)=\frac{1}{2}\left(X_{R}+X_{L}\right)+\frac{1}{2}\left(X_{R}-X_{L}\right)\tanh\left(aT\right)\,,\]
and we see that the normalized optical wave equation contains four
parameters:
\begin{itemize}
\item $X_{R}$ and $X_{L}$, the asymptotic values of the intensity-dependent
nonlinearity;
\item $a=\alpha/\omega_{0}$, a parameter that combines the steepness of
the intensity profile with the strength of the dispersion; and
\item $\xi$, which determines the group velocity in the co-moving frame,
or, in the lab frame, the speed of the pulse relative to the frequency
regime in which we are interested.
\end{itemize}

\section{Predicted behaviour for a low-frequency event horizon\label{sub:Predicted-behaviour}}

In the absence of dispersion, the wave equation (\ref{eq:optical_wave_equation})
takes the form\begin{equation}
n_{g}^{2}\partial_{\zeta}^{2}A-2n_{g}^{2}\partial_{\zeta}\partial_{\tau}A-\left(\partial_{\tau}\chi\right)\partial_{\tau}A-\left(\chi-\left(n_{g}^{2}-1\right)\right)\partial_{\tau}^{2}A=0\,,\end{equation}
where $n_{g}=c/u$ is the group index at the pulse frequency. It is
clear that there is an event horizon for all frequencies at $\chi=n_{g}^{2}-1$;
this is equivalent to the point $U=-1$ in the acoustic case. Even
in the dispersive case, if such a point exists then all forward-propagating,
positive-norm frequencies experience a group-velocity horizon. We
expect the emission of Hawking radiation, which is thermal at low
frequencies.

What is the low-frequency temperature of the emitted radiation? In
the non-dispersive acoustic case, by direct analogy with gravitational
black holes, the temperature was found to be proportional to the derivative
of the velocity profile at the horizon. It was then found, by both
numerical simulations and analytic approximations, that this form
of the temperature does not change if the slope is small and the dispersion
weak. Since the mathematical analogy between the optical model and
the acoustic model is not exact, it is not so clear what form the
low-frequency temperature should take, although it should be approximately
proportional to the derivative of $\chi$ at the event horizon. Instead,
let us make use of the fact that linearizing the velocity profile
in the vicinity of the horizon led in the acoustic model to an expression
for the temperature that is valid in the low-slope, weak-dispersion
regime. We also have such an expression for the optical model:
according to Eq. (\ref{eq:creat_phase_FIBRES}), the predicted Hawking creation rate is given by
\begin{equation}
\left|\frac{\beta_{\Omega^{\prime}}}{\alpha_{\Omega^{\prime}}}\right|^{2}=\exp\left(-2\,\mathrm{Im}\left\{ J\right\} \right)\end{equation}
where\begin{equation}
J=\int_{\Omega_{L}^{ul}}^{\Omega_{L}^{u}}T\left(\Omega\right)d\Omega\,.\end{equation}
This requires that $T$ be given as a function of $\Omega$, which
is done in the same way as for the acoustic case; the result is\begin{equation}
T\left(\Omega\right)=\frac{1}{a}\,\mathrm{arctanh}\left[\frac{2}{\chi_{R}-\chi_{L}}\left\{ \frac{n_{g}^{2}\left(\Omega-\Omega^{\prime}\right)^{2}-B^{2}\left(\Omega\right)}{\Omega^{2}}-\frac{1}{2}\left(\chi_{R}+\chi_{L}\right)\right\} \right]\,.\end{equation}
Restricting our attention to real values of $\Omega$, the only contribution
to $\mathrm{Im}\left\{ J\right\} $ comes from those values of $\Omega$
that do not correspond to real values of $T$. For these frequencies,
$T\left(\Omega\right)=i\pi/\left(2a\right)$, up to an integer multiple
of $i\pi/a$. Therefore,\begin{equation}
\left|\frac{\beta_{\Omega^{\prime}}}{\alpha_{\Omega^{\prime}}}\right|^{2}=\exp\left(-\frac{\pi}{a}\left(\Omega_{L}^{ul}-\Omega_{R}^{u}\right)\right)\,,\label{eq:predicted_Hawking_rate_exp}\end{equation}
where $\Omega_{L}^{ul}$ and $\Omega_{R}^{u}$ are frequencies in
the asymptotic regions, and mark the extremities of the real branches
of $T\left(\Omega\right)$. (We ignore the counter-propagating $v$-modes,
whose contribution to the phase integral is negligible.)
Using the relation $\left|\alpha_{\Omega^{\prime}}\right|^{2}-\left|\beta_{\Omega^{\prime}}\right|^{2}=1$,
this becomes the thermal-like spectrum
\begin{equation}
\left|\beta_{\Omega^{\prime}}\right|^{2}=\frac{1}{\exp\left(\frac{\pi}{a}\left(\Omega_{L}^{ul}-\Omega_{R}^{u}\right)\right)-1}\,.\label{eq:predicted_Hawking_rate}\end{equation}
To first order in $\Omega^{\prime}$ we find\begin{alignat*}{1}
\Omega_{L}^{ul}=\frac{n_{g}}{n_{g}-\sqrt{\chi_{L}+1}}\,\Omega^{\prime}\,,\qquad & \Omega_{R}^{u}=-\frac{n_{g}}{\sqrt{\chi_{R}+1}-n_{g}}\,\Omega^{\prime}\,,\end{alignat*}
so that\[
\Omega_{L}^{ul}-\Omega_{R}^{u}=\frac{n_{g}\left(\sqrt{\chi_{R}+1}-\sqrt{\chi_{L}+1}\right)}{\left(n_{g}-\sqrt{\chi_{L}+1}\right)\left(\sqrt{\chi_{R}+1}-n_{g}\right)}\,\Omega^{\prime}\,.\]
For low co-moving frequencies, then, we have\[
\left|\beta_{\Omega^{\prime}}\right|^{2}=\frac{1}{\exp\left(\Omega^{\prime}/T_{\mathrm{pred}}\right)-1}\]
where the predicted temperature $T_{\mathrm{pred}}$ is given by\begin{equation}
T_{\mathrm{pred}}=\frac{a\left(n_{g}-\sqrt{\chi_{L}+1}\right)\left(\sqrt{\chi_{R}+1}-n_{g}\right)}{\pi n_{g}\left(\sqrt{\chi_{R}+1}-\sqrt{\chi_{L}+1}\right)}\,.\label{eq:predicted_temperature}\end{equation}
Analogous with Eq. (\ref{eq:normalized_spectrum_defn}) for the acoustic
model, we define a rescaled energy spectrum\begin{equation}
f_{\Omega^{\prime}}=\frac{\Omega^{\prime}}{T_{\mathrm{pred}}}\left|\beta_{\Omega^{\prime}}\right|^{2}\,,\end{equation}
so that, if the spectrum is thermal at low frequencies with temperature
$T$, then\begin{equation}
f_{\Omega^{\prime}}\rightarrow\frac{T}{T_{\mathrm{pred}}}\qquad\mathrm{as}\qquad\Omega^{\prime}\rightarrow0\,.\end{equation}

To see how this is related to the slope of the nonlinearity, let us
find $\chi^{\prime}\left(T_{h}\right)$, where $\chi\left(T_{h}\right)=n_{g}^{2}-1$.
We have\[
n_{g}^{2}-1=\frac{1}{2}\left(\chi_{R}+\chi_{L}\right)+\frac{1}{2}\left(\chi_{R}-\chi_{L}\right)\tanh\left(aT_{h}\right)\,,\]
and so\[
\tanh\left(aT_{h}\right)=\frac{2\left(n_{g}^{2}-1\right)-\left(\chi_{R}+\chi_{L}\right)}{\left(\chi_{R}-\chi_{L}\right)}=-\frac{\left(\chi_{R}+1-n_{g}^{2}\right)-\left(n_{g}^{2}-1-\chi_{L}\right)}{\left(\chi_{R}+1-n_{g}^{2}\right)+\left(n_{g}^{2}-1-\chi_{L}\right)}\,.\]
The derivative is given by\begin{eqnarray*}
\chi^{\prime}\left(T_{h}\right) & = & \frac{a}{2}\left(\chi_{R}-\chi_{L}\right)\left(1-\tanh^{2}\left(aT_{h}\right)\right)\\
 & = & \frac{a}{2}\left(\chi_{R}-\chi_{L}\right)\frac{\left(\left(\chi_{R}+1-n_{g}^{2}\right)+\left(n_{g}^{2}-1-\chi_{L}\right)\right)^{2}-\left(\left(\chi_{R}+1-n_{g}^{2}\right)-\left(n_{g}^{2}-1-\chi_{L}\right)\right)^{2}}{\left(\left(\chi_{R}+1-n_{g}^{2}\right)+\left(n_{g}^{2}-1-\chi_{L}\right)\right)^{2}}\\
 & = & \frac{a}{2}\frac{2\left(\chi_{R}+1-n_{g}^{2}\right)2\left(n_{g}^{2}-1-\chi_{L}\right)}{\left(\chi_{R}-\chi_{L}\right)}\\
 & = & 2a\frac{\left(\sqrt{\chi_{R}+1}-n_{g}\right)\left(n_{g}-\sqrt{\chi_{L}+1}\right)}{\left(\sqrt{\chi_{R}+1}-\sqrt{\chi_{L}+1}\right)}\frac{\left(\sqrt{\chi_{R}+1}+n_{g}\right)\left(n_{g}+\sqrt{\chi_{L}+1}\right)}{\left(\sqrt{\chi_{R}+1}+\sqrt{\chi_{L}+1}\right)}\,.\end{eqnarray*}
The temperature of Eq. (\ref{eq:predicted_temperature}), then, is
related to the derivative at the horizon by\begin{equation}
T_{\mathrm{pred}}=\frac{\chi^{\prime}\left(T_{h}\right)}{2\pi n_{g}}\frac{\left(\sqrt{\chi_{R}+1}+\sqrt{\chi_{L}+1}\right)}{\left(\sqrt{\chi_{R}+1}+n_{g}\right)\left(n_{g}+\sqrt{\chi_{L}+1}\right)}\,.\label{eq:low_a_temperature_optical}\end{equation}
If $\chi_{R}$ and $\chi_{L}$ are close to the horizon, i.e., if
$\chi_{R}\approx\chi_{L}\approx n_{g}^{2}-1$, this reduces to the
simple formula\begin{equation}
T_{\mathrm{pred}}\approx\frac{\chi^{\prime}\left(T_{h}\right)}{4\pi n_{g}^{2}}\,.\end{equation}
This prediction agrees with that of Eq. (\ref{eq:Hawking_temp_approx_metric}),
which was derived from the approximate description of the pulse/probe
system as a spacetime geometry.

Eq. (\ref{eq:low_a_temperature_optical}) suggests that the temperature
increases without bound as the derivative of the nonlinearity does.
However, we have seen that, in the presence of dispersion, the solutions
of the wave equation are well-behaved in the limit $a\rightarrow\infty$.
Eqs. (\ref{eq:predicted_temperature}) and (\ref{eq:low_a_temperature_optical})
are valid only in the low-$a$ limit; in the high-$a$ limit, the
temperature is determined by solving the discontinuous nonlinearity
profile. Therefore, as $a$ increases, the temperature becomes independent
of $a$ by approaching a limit. This limiting case is solved in Appendix
\ref{sec:Appendix_Optical}; the resulting low-frequency temperature
is\begin{equation}
T_{\infty}=\left(n_{g}^{2}-1-\chi_{L}\right)^{1/2}\frac{\sqrt{1+\chi_{L}}\left(\chi_{R}+1-n_{g}^{2}\right)\left(n_{g}-\sqrt{1+\chi_{L}}\right)}{n_{g}\left(\chi_{R}-\chi_{L}\right)\left(n_{g}+\sqrt{1+\chi_{L}}\right)}\,.\label{eq:high_a_temperature_optical}\end{equation}
Comparing Eqs. (\ref{eq:predicted_temperature}) and (\ref{eq:high_a_temperature_optical}),
we find that the low-$a$ and high-$a$ forms of the temperature agree
if we take\[
a=\pi\left(n_{g}^{2}-1-\chi_{L}\right)^{1/2}\frac{\sqrt{1+\chi_{L}}\left(\sqrt{1+\chi_{R}}+n_{g}\right)}{\left(\sqrt{1+\chi_{R}}+\sqrt{1+\chi_{L}}\right)\left(n_{g}+\sqrt{1+\chi_{L}}\right)}\,,\]
and it is expected that the transition from one regime to the other
occurs when $a$ is comparable to this value.

\section{Numerical results\label{sub:Numerical-results-FIBRES}}

\subsection{With a low-frequency horizon}

We begin, as in the acoustic case, with the case where an event horizon
for low frequencies is present. In the optical case, this occurs at
$\chi=n_{g}^{2}-1$.

\begin{figure}
\includegraphics[width=0.8\columnwidth]{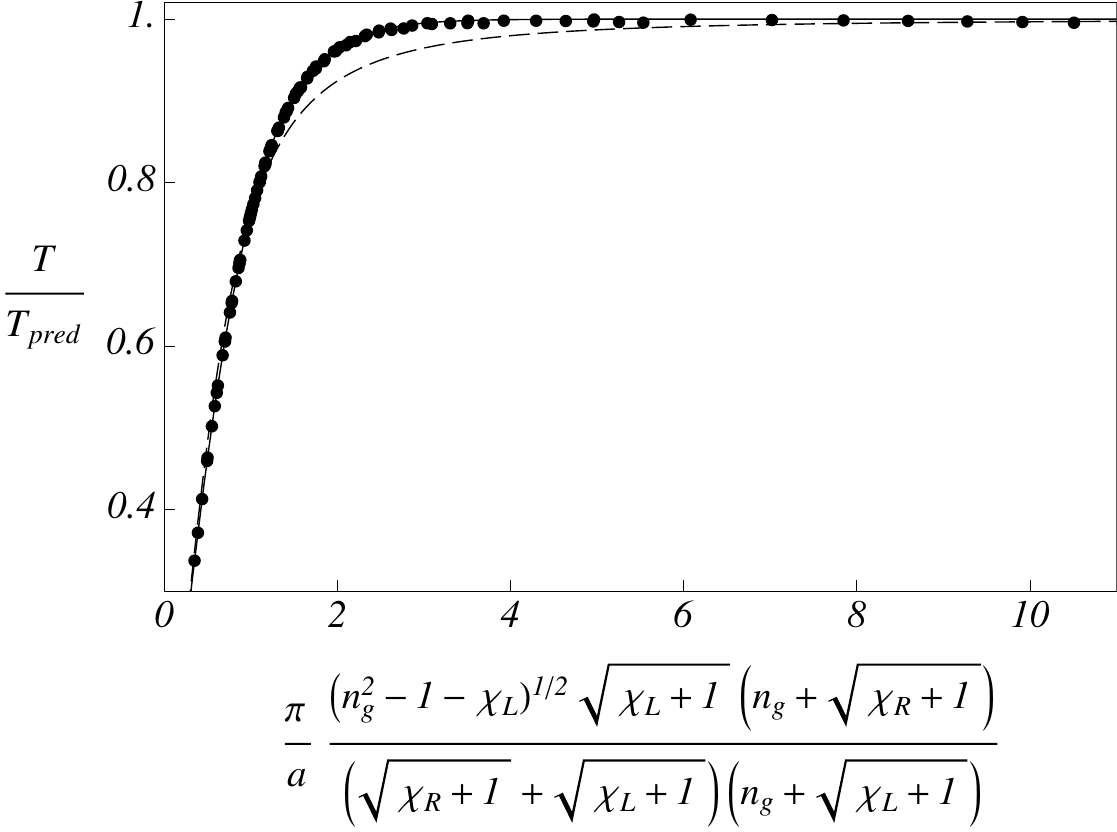}

\caption[\textsc{Low-frequency temperature}]{\textsc{Low-frequency temperature}: This plots the limiting low-$\Omega^{\prime}$
temperature, or $\Omega^{\prime}\left|\beta\right|^{2}/T_{\mathrm{pred}}$,
for various values of $\chi_{L}$, $\chi_{R}$ and $a$. Writing the
abscissa simply as $x$, the solid line plots $\tanh\left(x\right)$,
while the dashed line plots $x\,\tanh\left(1/x\right)$.\label{fig:Hawking_opt_low-freq-temp_tanh}}

\end{figure}

We have seen in Eq. (\ref{eq:predicted_temperature}) that the low-frequency
temperature is predicted to be proportional to $a$ when $a$ is sufficiently
small; we have also seen, in Eq. (\ref{eq:high_a_temperature_optical}),
that this temperature approaches a limiting value as $a\rightarrow\infty$.
This is exactly as was found in the acoustic model, where we were
able to fit the resulting temperatures to a hyperbolic tangent curve
(see Fig. \ref{fig:Hawking_aco_low-freq-temp_tanh}). Since the nonlinearity
profile we use here is also a hyperbolic tangent, it is natural to
guess that the temperature assumes a similar form. We have already
seen, in §\ref{sub:Numerical-results}, that there are two
possibilities:\begin{align}
\frac{T}{T_{\mathrm{pred}}}=\frac{T_{\infty}}{T_{\mathrm{pred}}}\tanh\left(\frac{T_{\mathrm{pred}}}{T_{\infty}}\right)\,,\qquad & \mathrm{or}\qquad\frac{T}{T_{\mathrm{pred}}}=\tanh\left(\frac{T_{\infty}}{T_{\mathrm{pred}}}\right)\,.\label{eq:two_temperature_forms}\end{align}
Both of these are plotted in Figure \ref{fig:Hawking_opt_low-freq-temp_tanh},
along with results of the optical case with various values of the
parameters $\chi_{L}$, $\chi_{R}$ and $a$. Once again, we find
that the second of Eqs. (\ref{eq:two_temperature_forms}) agrees remarkably
well with the numerical results, clearly much more so than the first
equation. Substituting Eqs. (\ref{eq:predicted_temperature}) and
(\ref{eq:high_a_temperature_optical}), we have\begin{equation}
T=\frac{a\left(n_{g}-\sqrt{1+\chi_{L}}\right)\left(\sqrt{1+\chi_{R}}-n_{g}\right)}{\pi n_{g}\left(\sqrt{1+\chi_{R}}-\sqrt{1+\chi_{L}}\right)}\tanh\left(\frac{\pi\left(n_{g}^{2}-1-\chi_{L}\right)^{1/2}\sqrt{1+\chi_{L}}\left(\sqrt{1+\chi_{R}}+n_{g}\right)}{a\left(\sqrt{1+\chi_{R}}+\sqrt{1+\chi_{L}}\right)\left(n_{g}+\sqrt{1+\chi_{L}}\right)}\right)\,.\label{eq:measured_temperature}\end{equation}

\begin{figure}
\includegraphics[width=0.8\columnwidth]{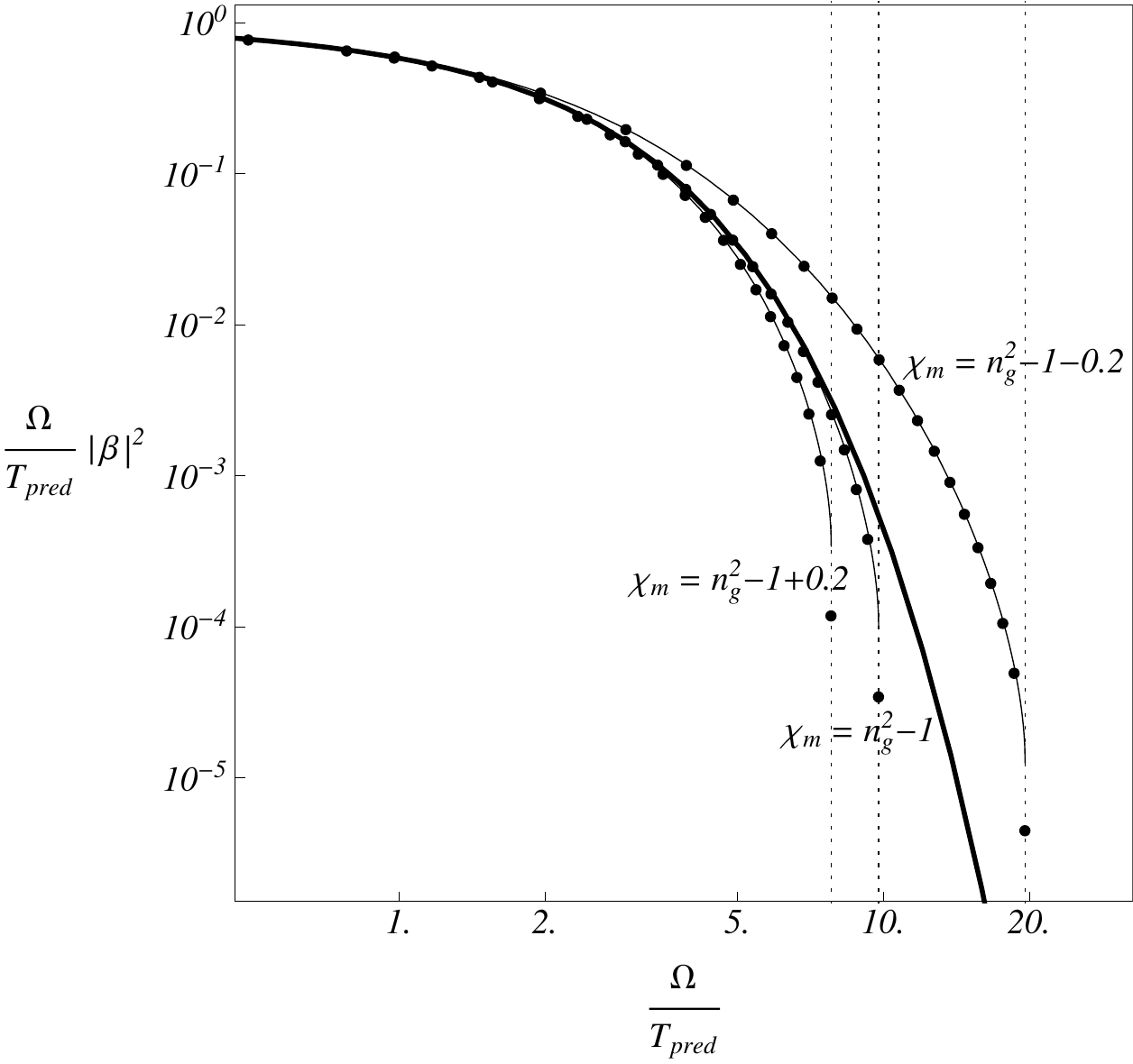}

\caption[\textsc{Hawking spectra with a low-frequency horizon}]{\textsc{Hawking spectra with a low-frequency horizon}: The nonlinearity
profile is such that $\chi=n_{g}^{2}-1$ at some point. The dots plot
the numerical results, the thin black curves plot the predictions
of Eq. (\ref{eq:predicted_Hawking_rate}), and the thick curve shows
the thermal spectrum. We have taken $n_{g}=1.6$, and fixed $\chi_{L}=n_{g}^{2}-1-0.6$
and $a=0.2$. These values ensure that $T/T_{\mathrm{pred}}\approx1$,
and since $T/T_{\mathrm{pred}}$ has a very weak dependence on $\chi_{R}$,
we can vary the latter to achieve various values of $\chi_{m}=\left(\chi_{L}+\chi_{R}\right)/2$.\label{fig:opt_Hawking-spectra_WKB}}

\end{figure}

It is also found that, if the parameters $\chi_{L}$, $\chi_{R}$
and $a$ are chosen such that $T/T_{\mathrm{pred}}\approx1$, then
the resulting spectra agree very well with the phase integral prediction
of Eq. (\ref{eq:predicted_Hawking_rate}). Figure \ref{fig:opt_Hawking-spectra_WKB}
shows several such spectra, together with the predicted spectra and
a purely thermal spectrum. Again we find, as in the acoustic case,
that the resulting spectrum is closest to thermal when the horizon
- the point where $\chi=n_{g}^{2}-1$ - is at the midpoint of the
variation. Strikingly, if the middle value $\chi_{m}=\left(\chi_{L}+\chi_{R}\right)/2$
is decreased, so that the event horizon is moved towards the far side
of the variation, then the high-frequency end of the spectrum is enhanced
in comparison to thermality.

\subsection{With realistic values of $\chi$}

Thermal emission from a horizon where $\chi=n_{g}^{2}-1$, though
interesting and useful in constructing the analogy between the fibre-optical
model and the more conventional acoustic model, is very difficult
to set up in practice. A typical group index in glass is $n_{g}\simeq1.5$,
so that this horizon occurs at $\chi\simeq1.25$. However, for femtosecond
pulses, the peak value of $\chi$ is typically less than $10^{-3}$.
Can we still expect the emission of a measurable Hawking spectrum
for such modest values of $\chi$?

\begin{figure}
\includegraphics[width=0.8\columnwidth]{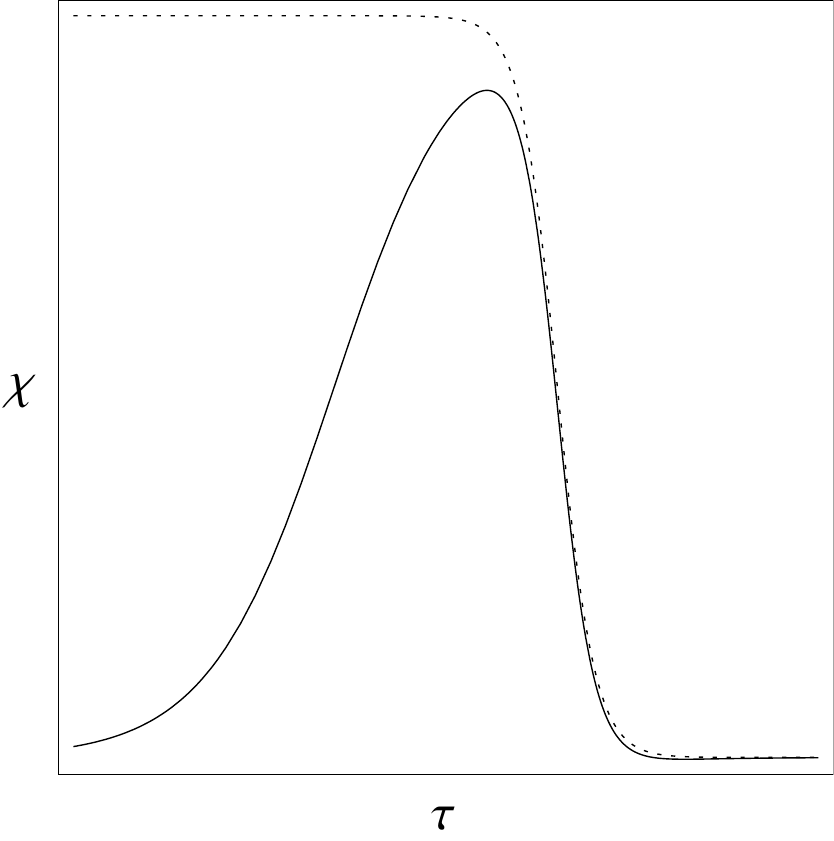}

\caption[\textsc{Self-steepening of a pulse}]{\textsc{Self-steepening of a pulse}: Higher-order nonlinear effects cause the optical intensity to become concentrated near the trailing edge of the pulse.  Since steepness is the main ingredient of Hawking radiation, we may approximate this as a single horizon with a hyperbolic tangent profile, as represented by the dotted curve. \label{fig:Self-steepening}}

\end{figure}

In a real fibre, of course, $\chi=0$ everywhere except in the vicinity
of the light pulse. This is realisable mathematically if we assume
a fundamental soliton, so that $\chi$ is of the form\begin{equation}
\chi\left(T\right)=h\,\mathrm{sech}^{2}\left(0.65\, a\, T\right)\,,\label{eq:chi_sech^2_form}\end{equation}
where $h$ and $a$ are parameters determining the height and width,
respectively, of the soliton. However, self-steepening \cite{Agrawal}
perturbs the pulse such that its trailing edge is much steeper than
its leading edge, so that the trailing edge dominates (see Figure \ref{fig:Self-steepening}). In this case,
we can focus solely on the trailing edge of the pulse by using the
hyperbolic tangent profile and setting $\chi_{R}=0$ and $\chi_{L}=h$:\begin{equation}
\chi\left(T\right)=\frac{h}{2}\left(1-\tanh\left(a\, T\right)\right)\,.\label{eq:chi_tanh_form}\end{equation}
Again, $h$ determines the height of the profile, while $a$ determines
the {}``width'' or length of the transition region. In both Eqs.
(\ref{eq:chi_sech^2_form}) and (\ref{eq:chi_tanh_form}), the length
of the transition region, where $\chi$ varies between $0$ and $h$,
is given approximately by $2/a$; due to the normalization of §\ref{sub:Normalizing-the-wave-eqn-FIBRES},
$a=1/\pi\approx0.3$ means that the transition occurs over about one
cycle of the frequency $\omega_{0}$, a steepness in intensity which is
achievable using few-cycle pulses \cite{Kartner,Brabec-Krausz-2000}. (This correspondence between
the two profiles explains the appearance of the factor of $0.65$
in Eq. (\ref{eq:chi_sech^2_form}).) It should be borne in mind that
the soliton form of Eq. (\ref{eq:chi_sech^2_form}) has two such variations,
one increasing and one decreasing.

Although the evolution of a pulse undergoing self-steepening is complicated,
it is clear that the asymptotic values of $\chi$ are always zero.
This is clearly not the case for Eq. (\ref{eq:chi_tanh_form}), where
$\chi\rightarrow h$ as $T\rightarrow-\infty$. We can compensate
for this by adding another term to $\chi$ as follows:\begin{eqnarray}
\chi\left(T\right) & = & \frac{h}{2}\left(\left(1-\tanh\left(a\, T\right)\right)-\left(1-\tanh\left(a^{\prime}\left(T+T^{\prime}\right)\right)\right)\right)\nonumber \\
 & = & \frac{h}{2}\left(\tanh\left(a^{\prime}\left(T+T^{\prime}\right)\right)-\tanh\left(a\, T\right)\right)\,,\label{eq:chi_asymmetric_tanh_form}\end{eqnarray}
where $a^{\prime}<a$ and $T^{\prime}>0$. This supplies a drop in
$\chi$ from $h$ to zero as we approach negative $T$, and the low
value of $a^{\prime}$ ensures that the steepness of this leading
edge is less than the steepness of the trailing edge, still characterised
by $a$. The main purpose of this form of $\chi$ is to examine the
effects of setting both asymptotic values of $\chi$ to zero, and
hence to check the validity of the results obtained using the form
of Eq. (\ref{eq:chi_tanh_form}).

\begin{figure}
\subfloat{\includegraphics[width=0.45\columnwidth]{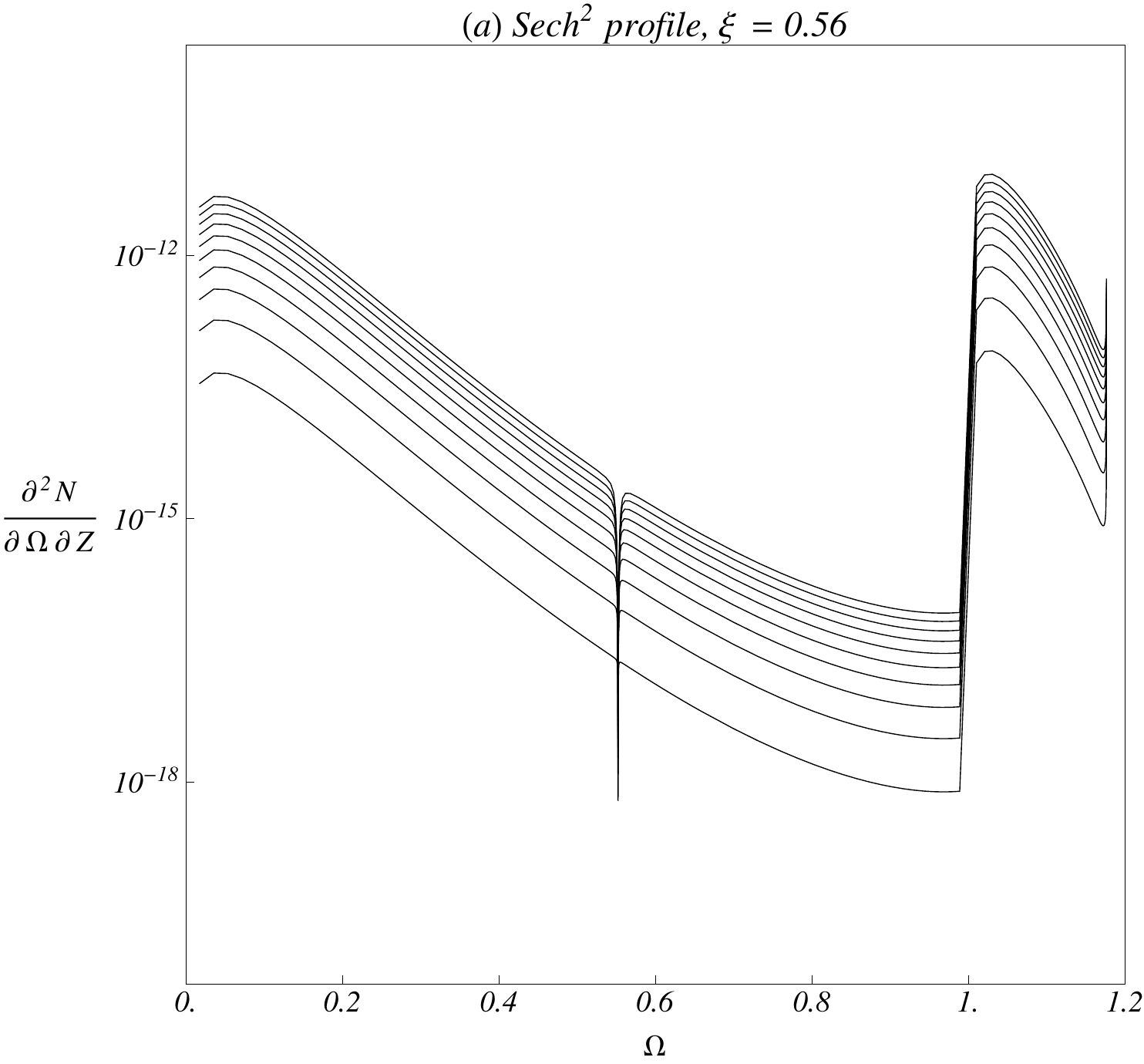}} \subfloat{\includegraphics[width=0.45\columnwidth]{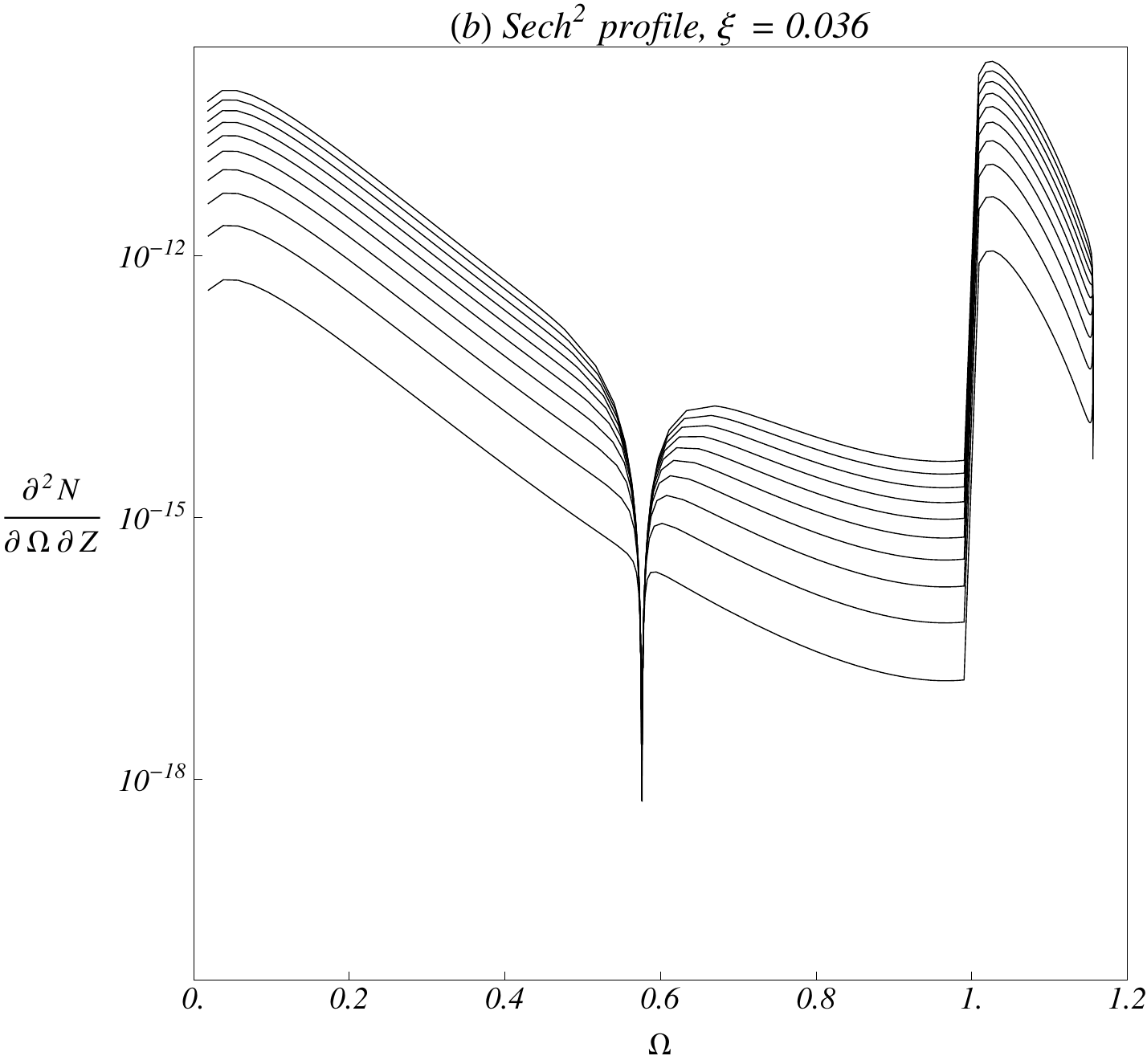}}

\subfloat{\includegraphics[width=0.45\columnwidth]{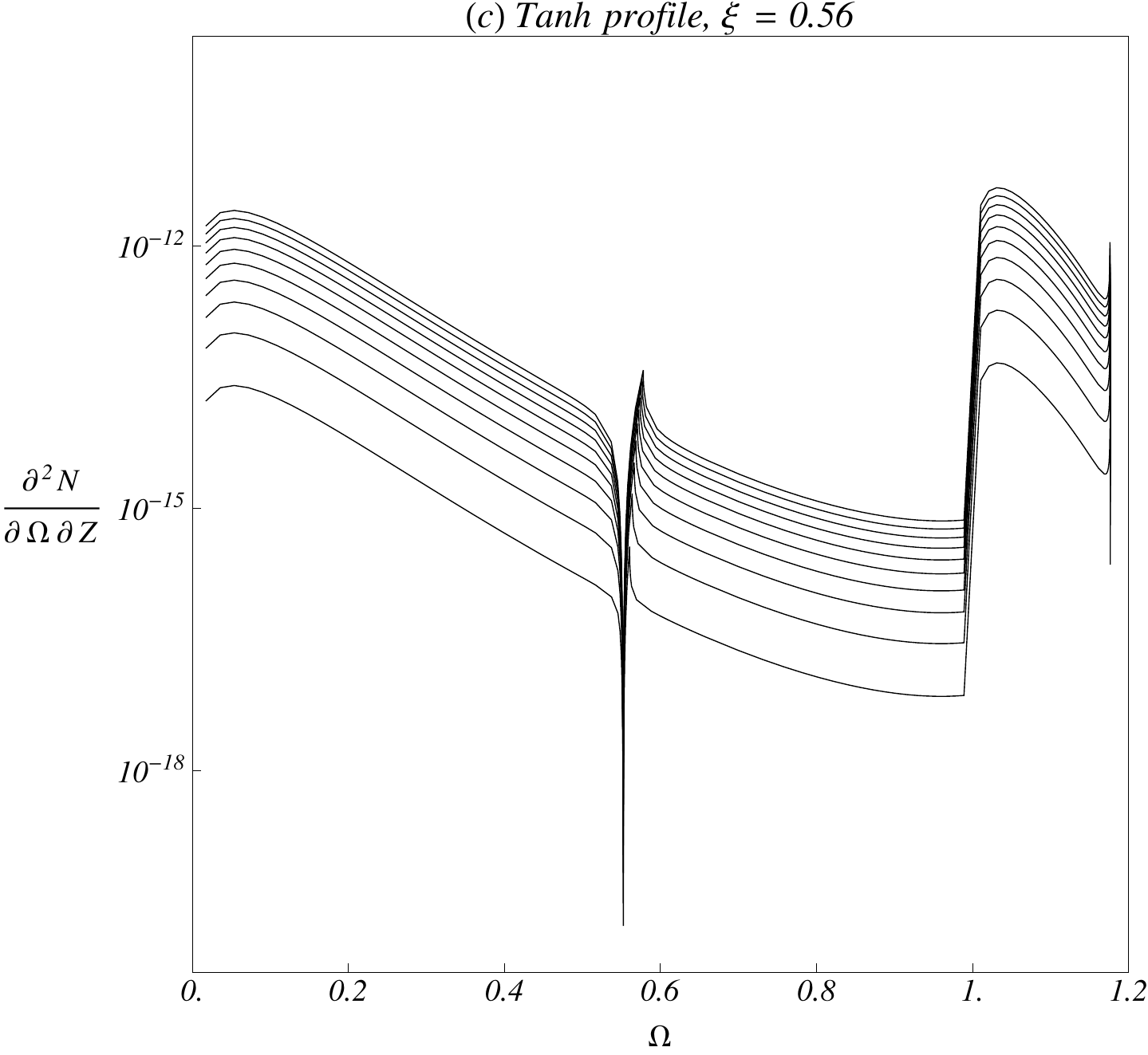}} \subfloat{\includegraphics[width=0.45\columnwidth]{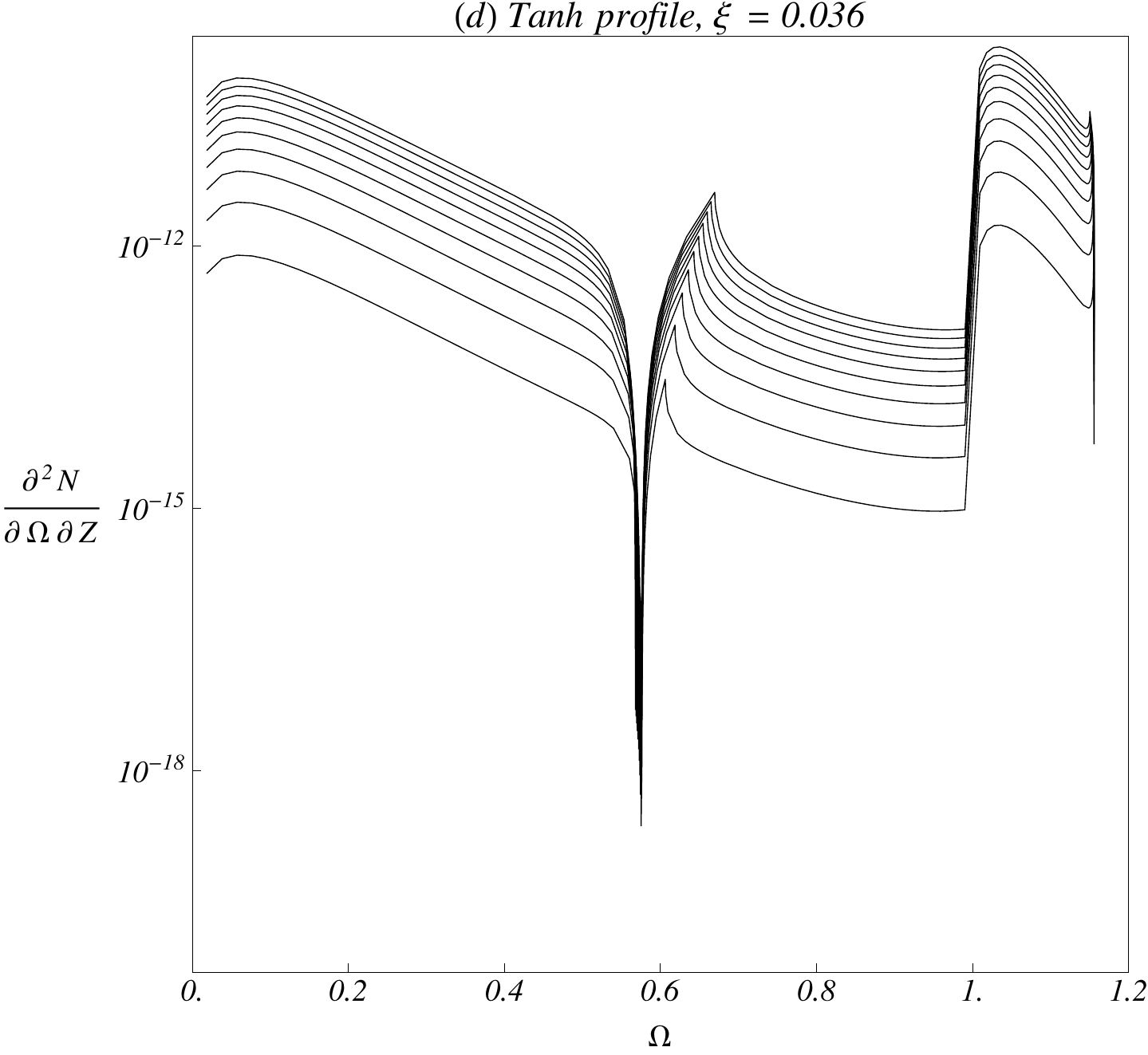}}

\caption[\textsc{Hawking spectra for realistic values of $\chi$, 1}]{\textsc{Hawking spectra for realistic values of $\chi$, 1}: These
plot photon flux spectra as functions of the lab frequency. All plots
use the same logarithmic scale for the creation rate, and a linear
scale for the (dimensionless) frequency $\Omega=\omega/\omega_{0}$.
The top plots, Figs. $\left(a\right)$ and $\left(b\right)$, are
obtained using a soliton-type nonlinearity, given in Eq. (\ref{eq:chi_sech^2_form});
the bottom plots, Figs. $\left(c\right)$ and $\left(d\right)$, are
for the hyperbolic tangent profile of Eq. (\ref{eq:chi_tanh_form}).
In the left-hand plots $\left(a\right)$ and $\left(c\right)$, $n\rightarrow1$
at low frequencies and $\xi=0.56$ in
Eqs. (\ref{eq:optical_wave_eqn_norm_with_xi}) and (\ref{eq:optical_wave_eqn_norm_with_xi_B_X});
in the right-hand plots $\left(b\right)$ and $\left(d\right)$, the
dispersion is matched to a realistic fibre near $\omega_{0}$ ($\Omega=1$), and $\xi=0.036$. Throughout,
$a=0.3$ while $h$ varies between $10^{-4}$ and $10^{-3}$.\label{fig:Hawking_no-horizon_varying-h}}

\end{figure}

\begin{figure}
\subfloat{\includegraphics[width=0.45\columnwidth]{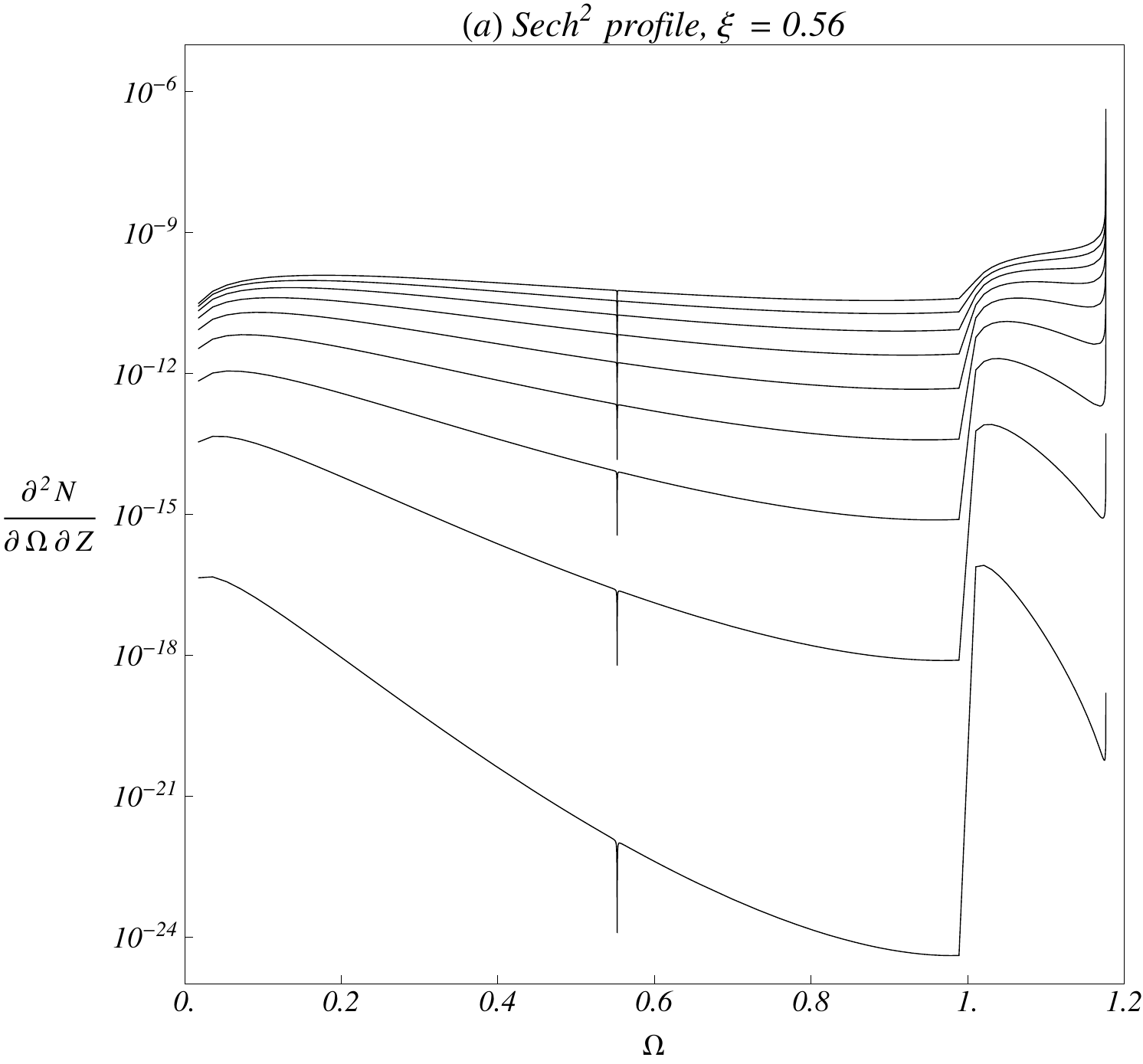}} \subfloat{\includegraphics[width=0.45\columnwidth]{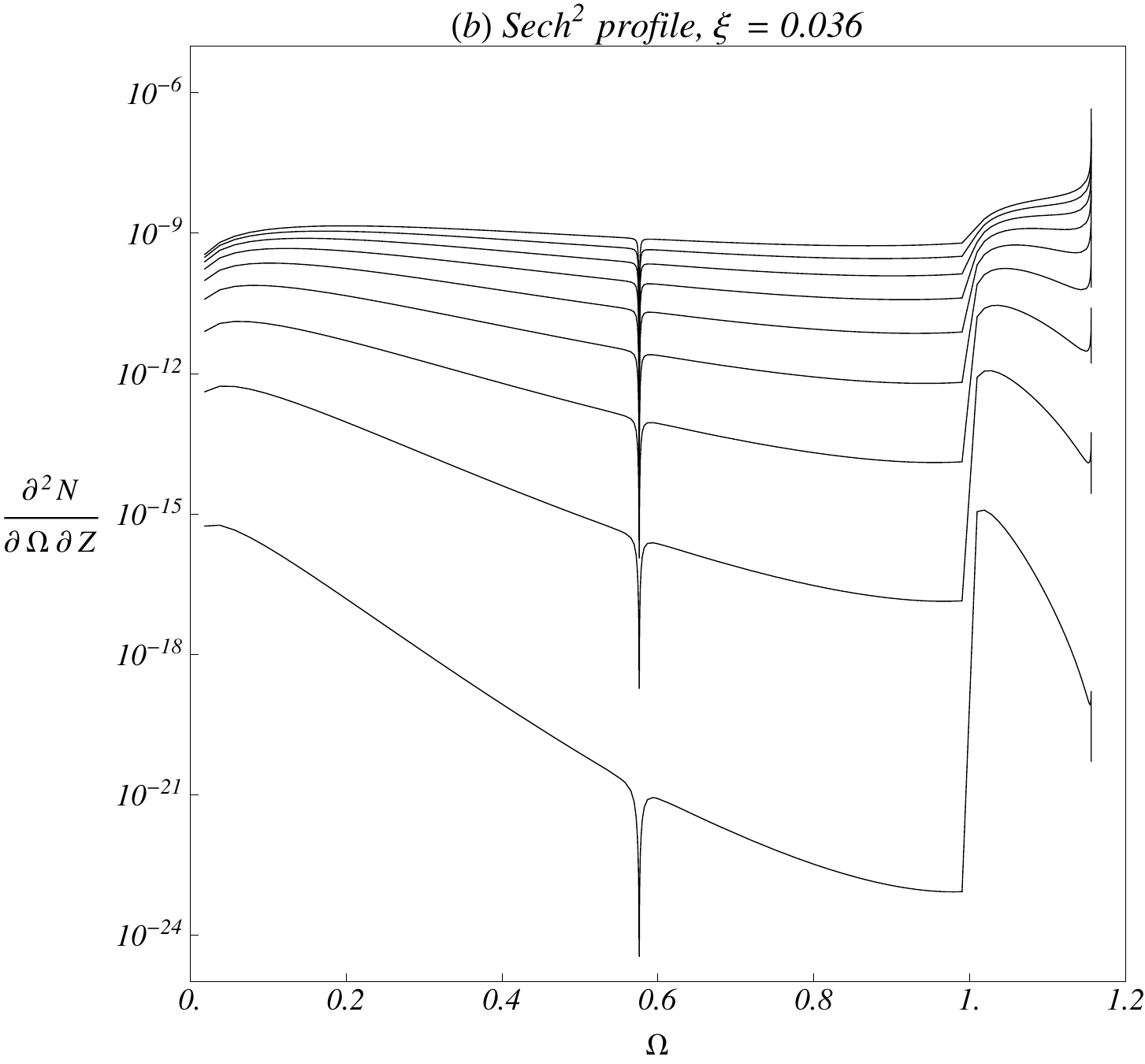}}

\subfloat{\includegraphics[width=0.45\columnwidth]{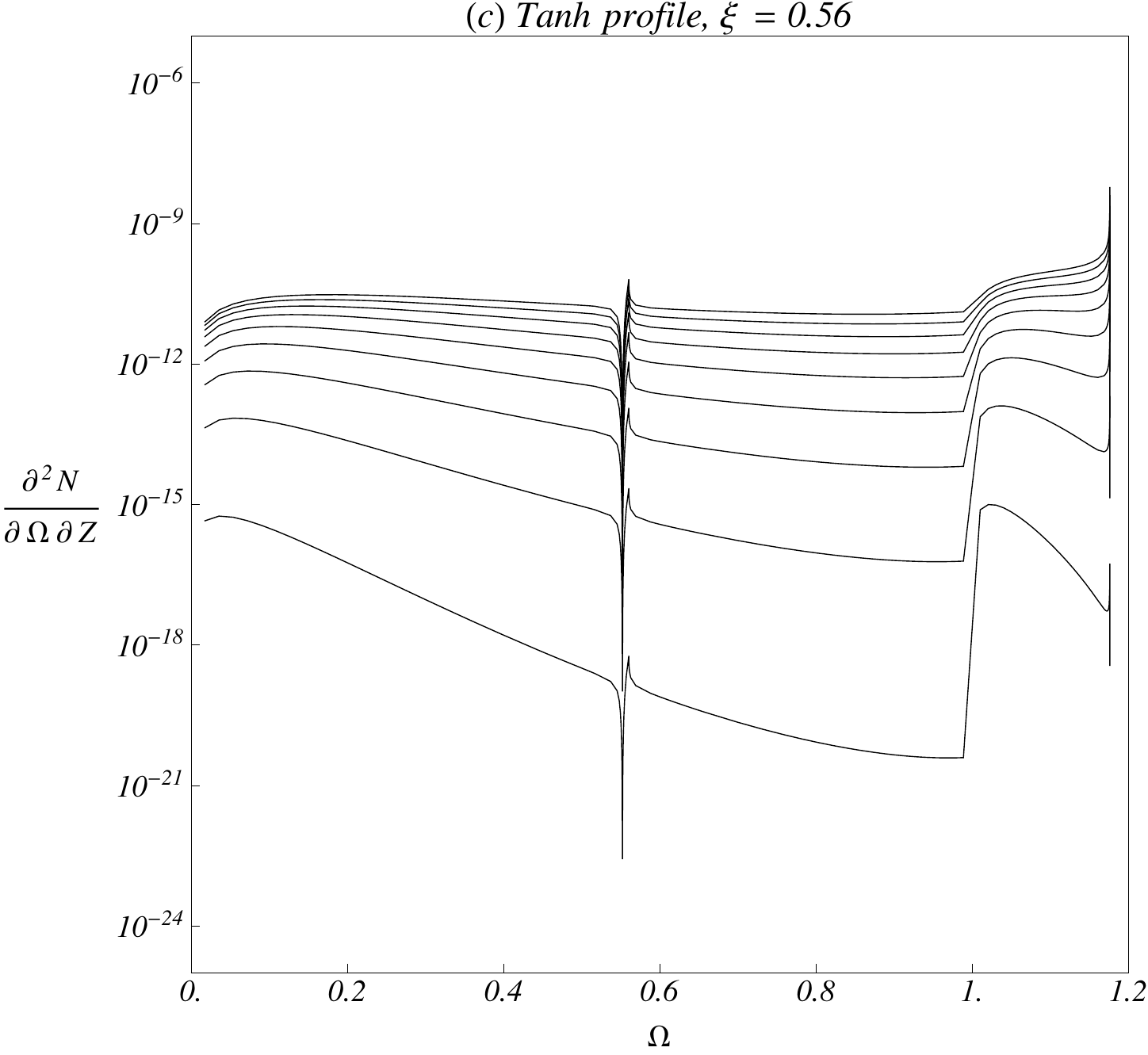}} \subfloat{\includegraphics[width=0.45\columnwidth]{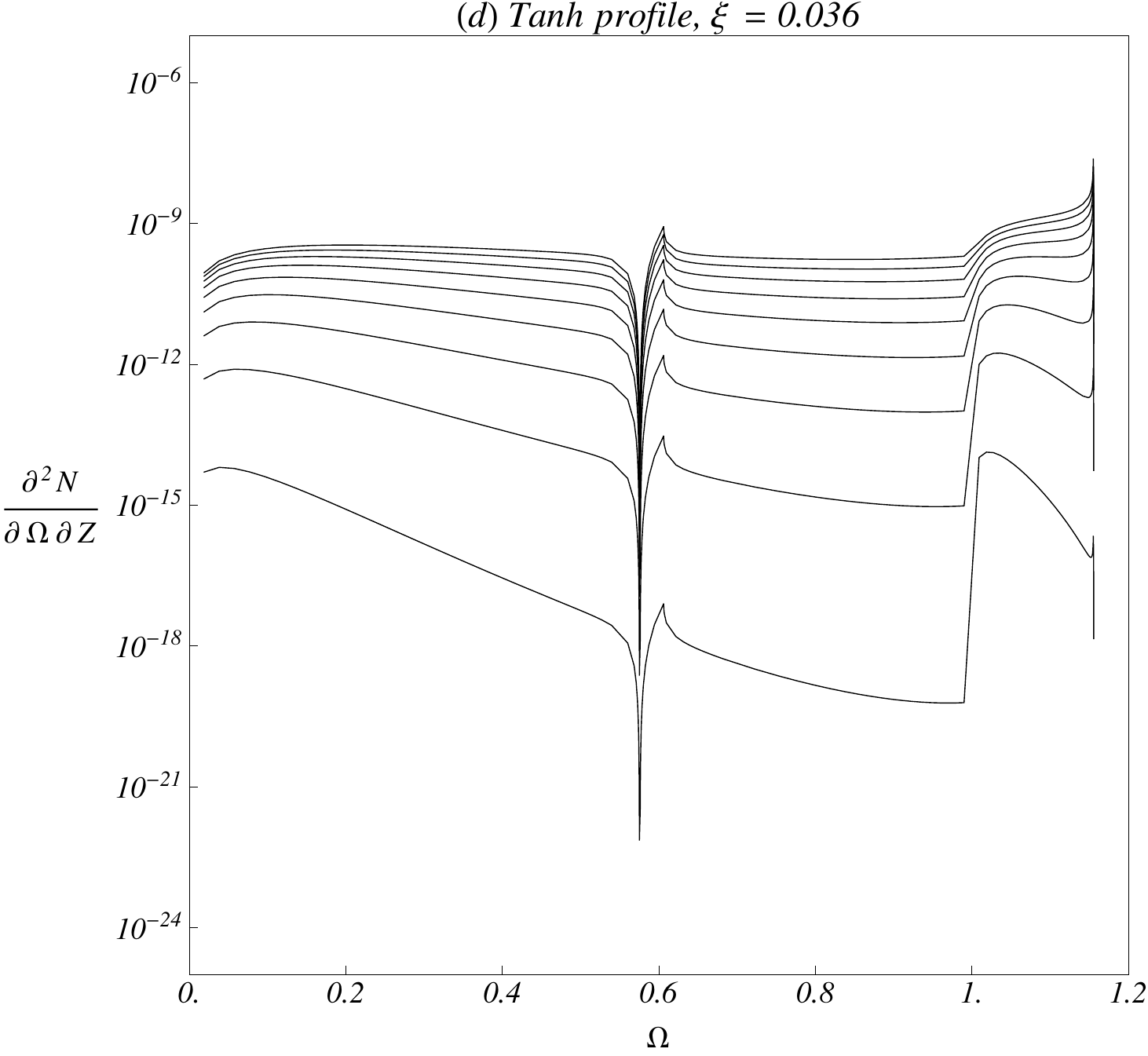}}

\caption[\textsc{Hawking spectra for realistic values of $\chi$, 2}]{\textsc{Hawking spectra for realistic values of $\chi$, 2}: These
plot photon flux spectra as functions of the lab frequency. The plots
are arranged similarly to Figure \ref{fig:Hawking_no-horizon_varying-h}.
The only difference is that, here, $h=10^{-4}$ while $a$ varies
between $0.2$ and $1.0$.\label{fig:Hawking_no-horizon_varying-a}}

\end{figure}

Figures \ref{fig:Hawking_no-horizon_varying-h} and \ref{fig:Hawking_no-horizon_varying-a}
show some results obtained using these more reasonable values of $\chi$.
These compare both the use of the soliton (Eq. (\ref{eq:chi_sech^2_form}))
and hyperbolic tangent (Eq. (\ref{eq:chi_tanh_form})) forms of the
nonlinearity, and the use of the different values of $\xi$ given in \S\ref{sub:Dispersion-relation},
which correspond to $n\rightarrow1$ at low frequencies ($\xi=0.56$)
and the more realistic case which is accurate near $\Omega=1$ ($\xi=0.036$).
In Fig. \ref{fig:Hawking_no-horizon_varying-h}, $a$
is fixed while spectra for different values of $h$ are plotted; in
Fig. \ref{fig:Hawking_no-horizon_varying-a}, $h$ is fixed while
$a$ is varied.

There are several points to note about these results:
\begin{itemize}
\item There is no $1/\Omega$ singularity at low frequencies, so the spectra
are not thermal even in the low-frequency regime.
\item There is a sharp, narrow peak near the maximum lab frequency, whose
co-moving frequency is $\Omega_{\mathrm{max}}^{\prime}$. At the same
time, there is a sharp, narrow dip at the group-velocity-matched frequency,
whose co-moving frequency is also $\Omega_{\mathrm{max}}^{\prime}$.
This is because, while the photon creation rate increases at these
co-moving frequencies, they are spread over a relatively large frequency
band in the lab frame.
\item The hyperbolic tangent nonlinearity profile creates a sharp peak on
the high-frequency side of the sharp dip. This is smoothed out if
a soliton-type nonlinearity is used.
\item The soliton profile gives lower spectra than the hyperbolic tangent
profile for low values of $a$, but climbs to higher spectra for higher
values of $a$.
\item Setting the parameter $\xi$ to $0.036$
gives a higher creation rate than is obtained when $\xi=0.56$.
\item For the most part, the spectra are very nearly proportional to $h^{2}$.
The only regions where this proportionality does not hold are where
the co-moving frequency is at its maximum, i.e., near the group-velocity-matching
frequency and its negative-norm counterpart. This makes sense: as
$h$ increases, so the range of frequencies which experience a group-velocity
horizon (the \textit{capture range}) widens, and the peak and dip
at these frequencies will also widen.
\item For large $a$ the spectra approach limiting curves, as expected.
For low $a$, they are highly sensitive to its value. If the change
in $\chi$ takes place over more than about one cycle of the frequency
$\omega_{0}$, then it is by far the \textit{low} co-moving frequencies
that dominate the spectra, reaching heights greater than the narrow
peak at the maximum frequency. What is more, the coupling between
the low-frequency $ul$-modes and the high-frequency $u$-modes vastly
outweighs the coupling between the $ur$- and $u$-modes.$ $ As $a$
increases, however, the whole spectrum increases exponentially and
the low-frequency peak is smoothed out.
\end{itemize}
The asymmetric hyperbolic tangent profile produces Hawking spectra
very similar to those for the pure hyperbolic tangent. The main difference
lies in the region around the group-velocity-matching frequency: the
sharp peak on the high-frequency side is smoothed, but unlike for
the soliton profile, this peak clearly still exists.

\begin{figure}
\subfloat{\includegraphics[width=0.45\columnwidth]{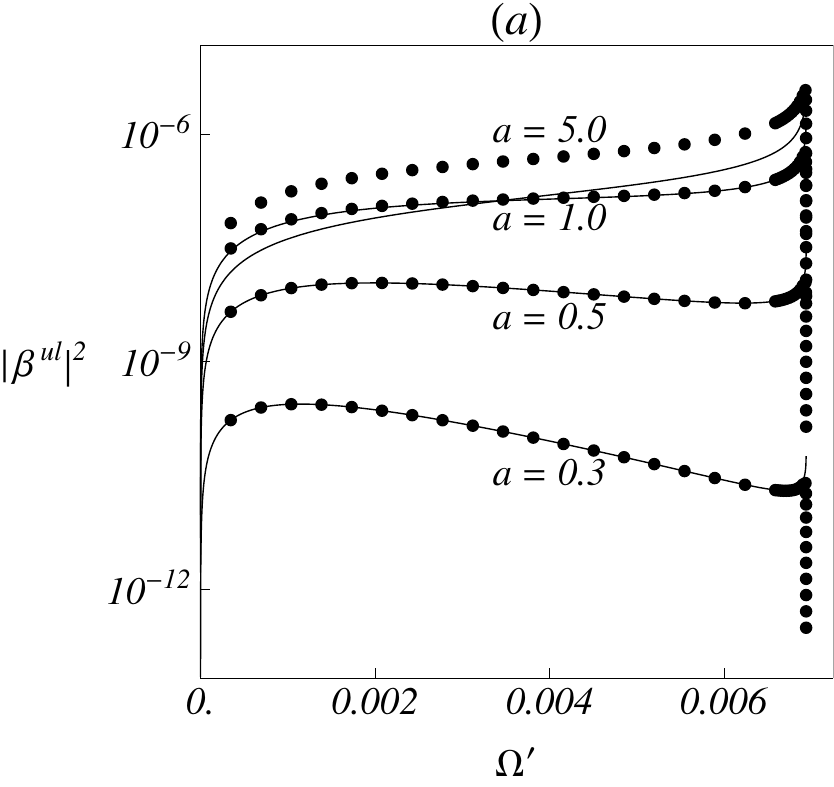}} \subfloat{\includegraphics[width=0.45\columnwidth]{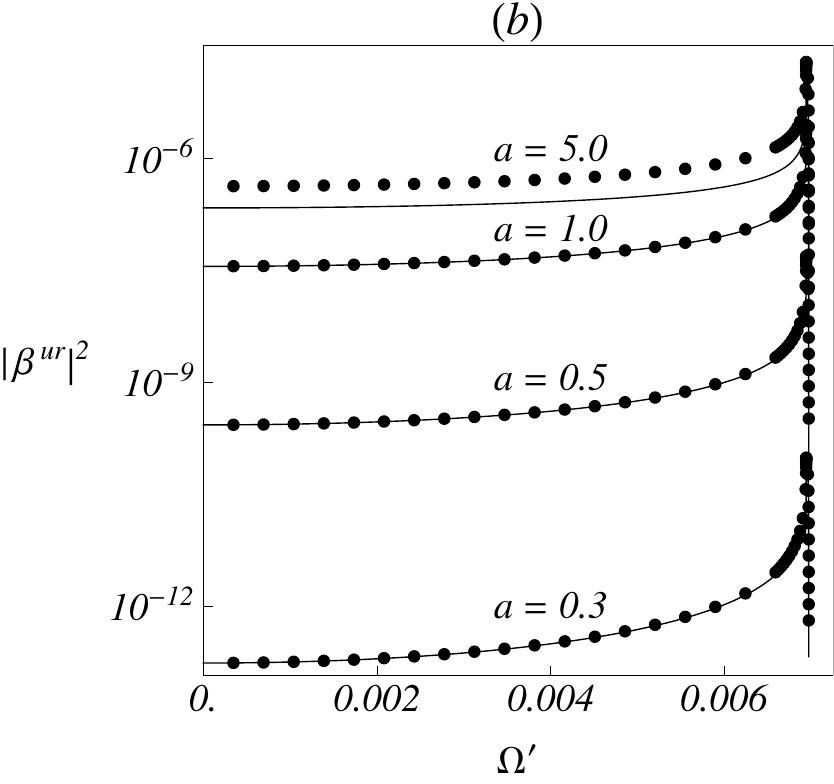}}

\caption[\textsc{Hawking spectra and the phase-integral approximation}]{\textsc{Hawking spectra and the phase-integral approximation}: These
show numerically calculated Hawking spectra (dots) as functions of
the co-moving frequency (\textit{not} the lab frequency, for the sake
of convenience). The purpose is to compare this with the modified
phase-integral predictions of Eqs. (\ref{eq:Hawking_opt_no-horizon_gvh})-(\ref{eq:Hawking_opt_no-horizon_ur})
(curves), with which there is seen to be excellent agreement for $a\lesssim1$.
Throughout, $h=10^{-4}$.\label{fig:Hawking_no-horizon_WKB}}

\end{figure}

It should be borne in mind that the calculated spectra, which are photon flux densities according to Eq. (\ref{eq:FIBRES_spectral_flux_density_zeta-omegaprime}), are dimensionless quantities. In order to yield a measurable quantity, we must replace the factor of $\omega_{0}$ that was factored out in \S \ref{sub:Normalizing-the-wave-eqn-FIBRES}. Typically, $\omega_{0}$ lies in the ultraviolet regime at about $300\,\mathrm{nm}$ or $6\times10^{15}\,\mathrm{Hz}$. Integrated over the range of $\Omega$, a constant value of $10^{-15}$ for $\partial^{2}N/\partial\Omega/\partial Z$ corresponds roughly to one emitted photon per second. This must then be modified by the factor $L\nu_{\mathrm{rep}}/u$ appearing in Eq. (\ref{eq:FIBRES_spectral_flux_density_t-omega}), which takes account of repeated pulses and propagation length. This can vary from about unity for a soliton in a metre-long fibre down to about $10^{-3}$ for short pulses or effective propagation lengths. So Figure \ref{fig:Hawking_no-horizon_varying-h}, where the intensity change occurs over a single cycle of the frequency $\omega_{0}$ and the height of the nonlinearity is on the order of $10^{-3}$, yields results ranging from less than a photon per minute to thousands of photons per second.

Remarkably, these spectra are still fairly well-approximated by analytic
expressions analogous to those found in the acoustic model. As was
the case there, some modification is required: it is found that, rather
than simply a single exponential, the spectrum is formed from a linear
combination of exponentials. For instance, when we have a group-velocity
horizon (which occurs in the tail of the $ur$-$u$ spectrum, near
$\Omega_{\mathrm{max}}^{\prime}$), the spectrum is not given by Eq.
(\ref{eq:predicted_Hawking_rate_exp}) but is much better approximated
by\begin{multline}
\left|\beta^{ur}\right|^{2}\approx\exp\left(-\frac{\pi}{a}\left(\Omega_{R}^{ul}-\Omega_{L}^{u}\right)\right)-\exp\left(-\frac{\pi}{a}\left(\Omega_{R}^{ur}-\Omega_{L}^{u}\right)\right)\\
-\exp\left(-\frac{\pi}{a}\left(\Omega_{R}^{ul}-\Omega_{R}^{u}\right)\right)+\exp\left(-\frac{\pi}{a}\left(\Omega_{R}^{ur}-\Omega_{R}^{u}\right)\right)\,.\label{eq:Hawking_opt_no-horizon_gvh}\end{multline}
Similarly, we find, when there is no group-velocity horizon, that
the $ul$-$u$ and $ur$-$u$ spectra are approximately given by the
following:\begin{multline}
\left|\beta^{ul}\right|^{2}\approx\exp\left(-\frac{\pi}{a}\left(\Omega_{R}^{ul}-\Omega_{L}^{u}\right)\right)-\exp\left(-\frac{\pi}{a}\left(\Omega_{L}^{ul}-\Omega_{L}^{u}\right)\right)\\
-\exp\left(-\frac{\pi}{a}\left(\Omega_{R}^{ul}-\Omega_{R}^{u}\right)\right)+\exp\left(-\frac{\pi}{a}\left(\Omega_{L}^{ul}-\Omega_{R}^{u}\right)\right)\,,\label{eq:Hawking_opt_no-horizon_ul}\end{multline}
\begin{multline}
\left|\beta^{ur}\right|^{2}\approx\exp\left(-\frac{\pi}{a}\left(\Omega_{L}^{ul}-\Omega_{R}^{ul}\right)\right)\left\{ \exp\left(-\frac{\pi}{a}\left(\Omega_{L}^{ur}-\Omega_{L}^{u}\right)\right)-\exp\left(-\frac{\pi}{a}\left(\Omega_{R}^{ur}-\Omega_{L}^{u}\right)\right)\right.\\
\left.-\exp\left(-\frac{\pi}{a}\left(\Omega_{L}^{ur}-\Omega_{R}^{u}\right)\right)+\exp\left(-\frac{\pi}{a}\left(\Omega_{R}^{ur}-\Omega_{R}^{u}\right)\right)\right\} \,.\label{eq:Hawking_opt_no-horizon_ur}\end{multline}
These expressions are compared with numerical results in Figure \ref{fig:Hawking_no-horizon_WKB}.
Note that agreement is good for lower values of $a$ ($\lesssim1$),
but not for higher values, where the linearization of $\chi$ performed
in §\ref{sub:Linearized-intensity-profile} becomes invalid and the
spectrum approaches that obtained from a discontinuous nonlinearity
profile.

\section{Concluding remarks\label{sub:Concluding-remarks-RESULTS-FIBRES}}

We have found, perhaps surprisingly, that Hawking radiation is not
only predicted by the optical model, but it is also subject to the
same observations made in the context of the acoustic model. When
an event horizon is present, the spectrum is thermal at low frequencies,
and the temperature is given by Eq. (\ref{eq:measured_temperature})
- a direct analogy of Eq. (\ref{eq:modified_temp_general_U}) found
for the acoustic case. The spectra deviate from thermality at higher
frequencies, but are well-approximated by the phase-integral approximation.
When there is no low-frequency horizon - the more realistic case -
the spectra are no longer thermal but are well-approximated by a modified
version of the phase-integral approximation which linearly combines
different values of the phase integral; these are given in Eqs. (\ref{eq:Hawking_opt_no-horizon_gvh})-(\ref{eq:Hawking_opt_no-horizon_ur}).
In particular, with small values of the step size $h$, the spectra (or those parts
of the spectra not subject to a group-velocity horizon) are proportional to $h^{2}$.

While these results look promising, it is worth examining a serious
criticism of the optical model. This has to do with the validity of
the low-frequency $ul$-modes, and is the subject of the next chapter.

\pagebreak{}

\chapter{The Low-Frequency Problem\label{sec:The-Low-Frequency-Problem}}

In solving the complete optical wave equation, we have seen in Chapter
\ref{sec:Results-for-the-Fibre-Optical-Model} that, upon quantization
of the optical field, spontaneous creation of photon pairs of vastly
different frequencies occurs. Numerical simulations robustly conclude
that the dominant pairing is between the very high frequencies of
the $u$-mode and the very low frequencies of the $ul$-mode. This
poses a problem unique to the optical model, to do with the strength
of the nonlinear coupling between probe light and the pulse. In the
acoustic model, the spontaneous creation of phonon pairs with wavenumbers
of different orders of magnitude is perfectly plausible, provided
the dispersion relation is valid in both regimes. These wavenumbers,
however different, will {}``see'' the same flow velocity profile,
which is externally fixed - a property of the medium rather than the
waves. In the optical model, the refractive index change induced by
a pulse can be written as $\delta n=2n_{2}I$ \cite{Agrawal}. (The
factor of $2$ appears because the pulse and probe are assumed to
have different frequencies, so that they interact through cross-phase
modulation; see Eq. (\ref{eq:SPM_XPM_across_frequencies}).) $I$
is the intensity of the pulse, and can be considered as externally
fixed since it is independent of the probe waves. $n_{2}$, however,
is a frequency-dependent quantity: $n_{2}=n_{2}\left(\omega\right)$,
where $\omega$ is the frequency of the probe. Frequencies of a similar
order of magnitude - such as the $u$- and $ur$-modes around the
point where $\omega^{\prime}=0$ - will have similar values of $n_{2}$,
and the refractive index change can be assumed identical for such
frequencies. This assumption is invalid for frequencies of different
orders of magnitude. Indeed, there should be virtually no nonlinear
coupling at very low frequencies - precisely those frequencies which,
as seen in Fig. \ref{fig:Hawking_no-horizon_varying-h}, turn out
to dominate the Hawking radiation!

How can we modify our model in order to suppress the low frequencies,
leaving the $u$- and $ur$-modes as the dominant Hawking pairs?

\section{Frequency-dependent $\chi$\label{sub:Frequency-dependent-chi}}

A first modification is to introduce an explicit frequency-dependence
into the form of $\chi$.

\subsection{Quadratic dependence on $\omega$}

We assume that $\chi$ can be written as the product\begin{equation}
\chi\left(\omega,\tau\right)=\kappa\left(\omega\right)\chi\left(\tau\right)\,,\label{eq:frequency_dependent_chi}\end{equation}
so that the frequency $\omega$ {}``sees'' the nonlinearity profile
$\kappa\left(\omega\right)\chi\left(\tau\right)$. $\chi\left(\tau\right)$
is assumed to be the valid form of the nonlinearity when $\omega$
is near $\omega_{0}$, the frequency with vanishing co-moving frequency.;
therefore, $\kappa\left(\omega_{0}\right)=1$. To make the nonlinearity
vanish at low frequencies, we take $\kappa\left(0\right)=0$. Since
the frequencies $\omega$ and $-\omega$ must {}``see'' identical
nonlinearities, $\kappa\left(\omega\right)$ must be an even function,
and hence we find that the simplest form of $\kappa$ is\begin{equation}
\kappa\left(\omega\right)=\frac{\omega^{2}}{\omega_{0}^{2}}\,.\label{eq:quadratic_frequency_dependence}\end{equation}
This is a highly artificial construction, but it should help to eliminate
the unrealistic frequencies and give some indication of what the Hawking
spectrum might look like in practice.

To implement Eqs. (\ref{eq:frequency_dependent_chi}) and (\ref{eq:quadratic_frequency_dependence}),
we replace $\omega$ by the operator $i\partial_{\tau}$ and substitute
into the optical wave equation. However, the arrangement of the two
new differential operators must be chosen with care. This is done
in such a way that the scalar product of Eq. (\ref{eq:scalar_product_FIBRES})
remains constant with respect to $\zeta$; upon the taking of this
derivative, and the substitution of the wave equation in the resulting
integrand, we may use integration by parts to obtain cancelling terms
only when the derivative operators $i\partial_{\tau}$ are placed
symmetrically around $\chi$. Therefore, the new wave equation is\begin{equation}
n_{g}^{2}\left(\partial_{\zeta}-\partial_{\tau}\right)^{2}A+c^{2}\beta^{2}\left(i\partial_{\tau}\right)A+\frac{1}{\omega_{0}^{2}}\partial_{\tau}^{2}\left(\chi\partial_{\tau}^{2}A\right)=0\,.\label{eq:optical_wave_eqn_freq_dep_chi}\end{equation}
Plugging in the quartic form of the dispersion in Eq. (\ref{eq:optical_quartic_dispersion_with_xi})
and normalizing the wave equation according to Eq. (\ref{eq:optical_wave_eqn_norm_with_xi}),
this becomes

\begin{equation}
\partial_{Z}^{2}A-2\partial_{Z}\partial_{T}A+\xi\left(\partial_{T}^{2}A+\partial_{T}^{4}A\right)+\partial_{T}^{2}\left(X\partial_{T}^{2}A\right)=0\,.\label{eq:optical_wave_eqn_high_freq_disp_freq_dep_chi_normalized}\end{equation}

\subsection{Results}

Eq. (\ref{eq:optical_wave_eqn_high_freq_disp_freq_dep_chi_normalized})
can be solved numerically as before. We assume a hyperbolic tangent
nonlinearity profile:\begin{equation}
X\left(T\right)=\frac{h}{2}\left(1-\tanh\left(aT\right)\right)\,.\label{eq:hyperbolic_tangent_chi}\end{equation}
We also assume that the contributions from the $v$-mode are negligible,
so that the $u$-mode pairs only with the $ul$- and $ur$-modes.
$ $Then, we may use the method of §\ref{sub:Steady-state-solution-FIBRES}
to numerically solve for the $u$-in mode, and the norms of the $ul$-
and $ur$-out modes are interpreted as the spectral amplitudes $\left|\beta^{ul}\right|^{2}$
and $\left|\beta^{ur}\right|^{2}$; the spectrum of the $u$-mode
is $\left|\beta^{u}\right|^{2}=\left|\beta^{ul}\right|^{2}+\left|\beta^{ur}\right|^{2}$.

\begin{figure}
\subfloat{\includegraphics[width=0.45\columnwidth]{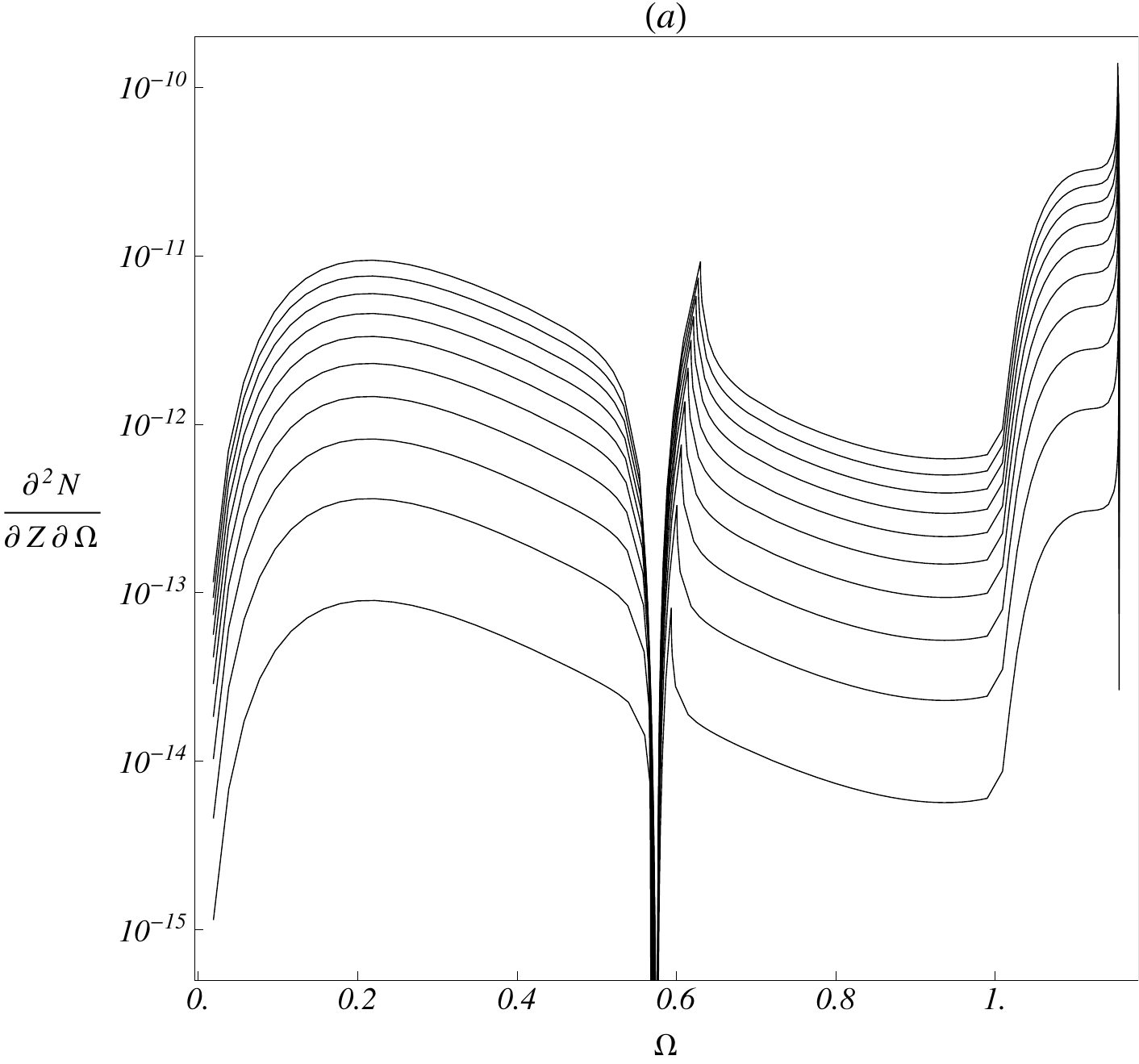}} \subfloat{\includegraphics[width=0.45\columnwidth]{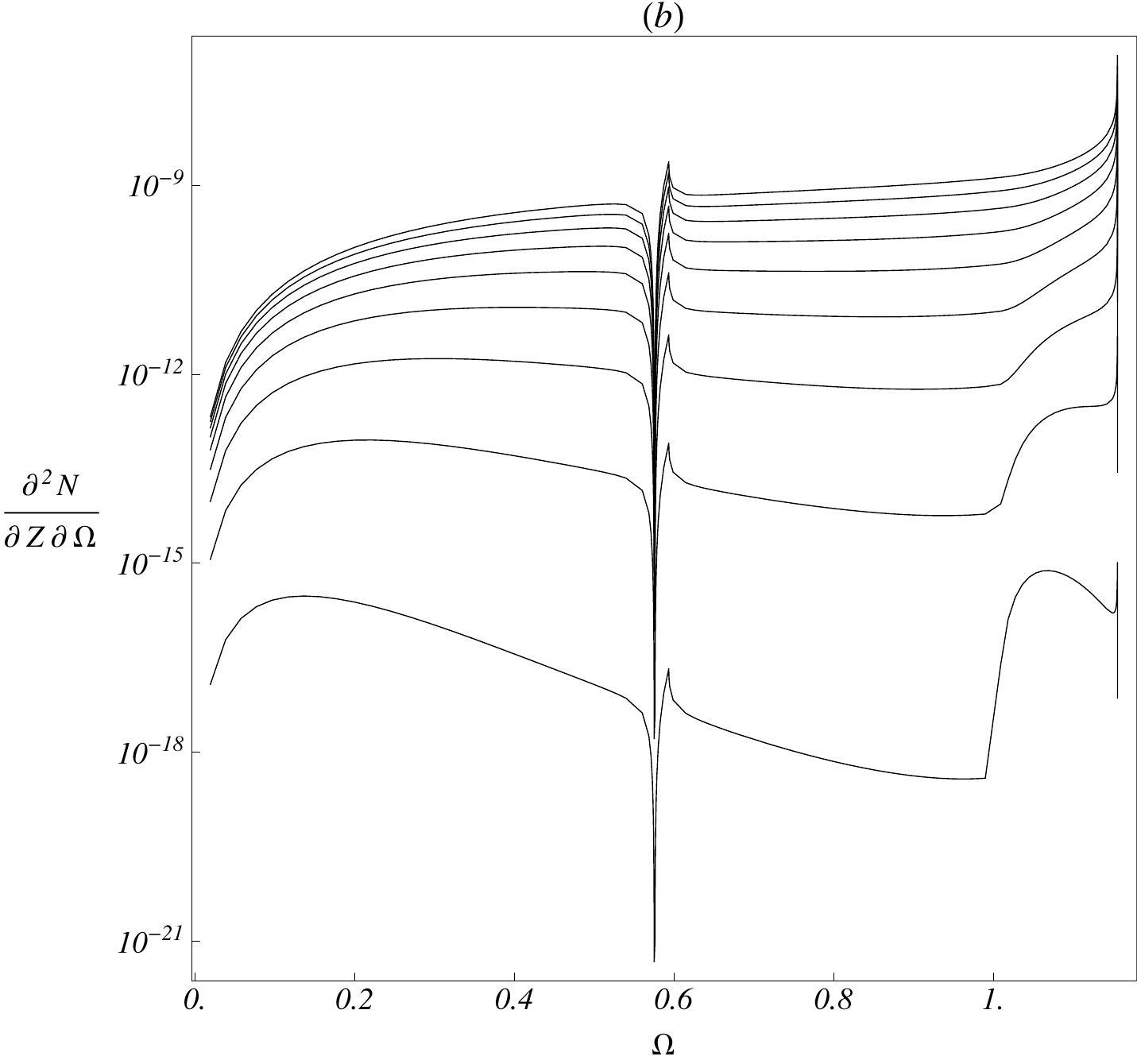}}

\caption[\textsc{Hawking spectra with frequency-dependent $\chi$}]{\textsc{Hawking spectra with frequency-dependent $\chi$}: The spectral
flux density is plotted on a logarithmic scale against the (normalized)
lab frequency on a linear scale. $\chi$ has the quadratic frequency
dependence of Eq. (\ref{eq:quadratic_frequency_dependence}) and the
hyperbolic tangent spatial dependence of Eq. (\ref{eq:hyperbolic_tangent_chi}).
In Figure $\left(a\right)$, $a$ is fixed at $0.3$ while $h$ varies
between $10^{-4}$ and $10^{-3}$; in Figure $\left(b\right)$, $h=10^{-4}$
while $a$ varies between $0.2$ and $1.0$.
In both plots, $\xi=0.036$, the more realistic value.\label{fig:spectra_freq-dep-chi}}

\end{figure}

Figure \ref{fig:spectra_freq-dep-chi} shows some spectra obtained
with this model. We see that the contribution from the low-frequency
modes has indeed been greatly diminished, although there is still
a significant contribution from the $ul$-mode when the frequency
is not too close to zero. Other than this reduction, the spectra are
almost identical with their counterparts when $\chi$ is independent
of frequency, and if $a$ is low then the low-frequency component
of the radiation remains the dominant one.

\begin{figure}
\subfloat{\includegraphics[width=0.45\columnwidth]{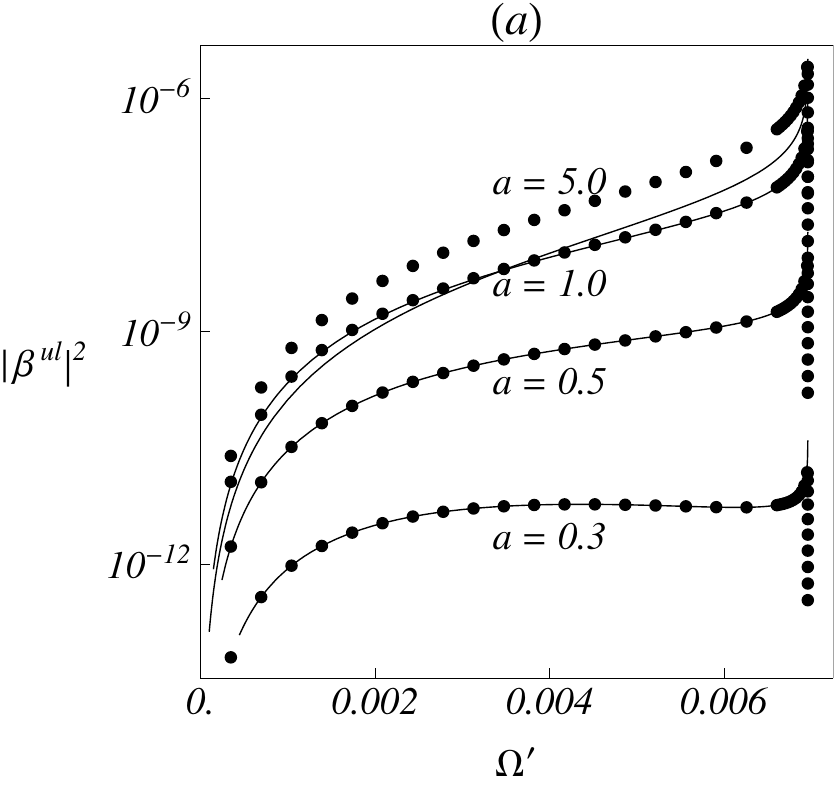}} \subfloat{\includegraphics[width=0.45\columnwidth]{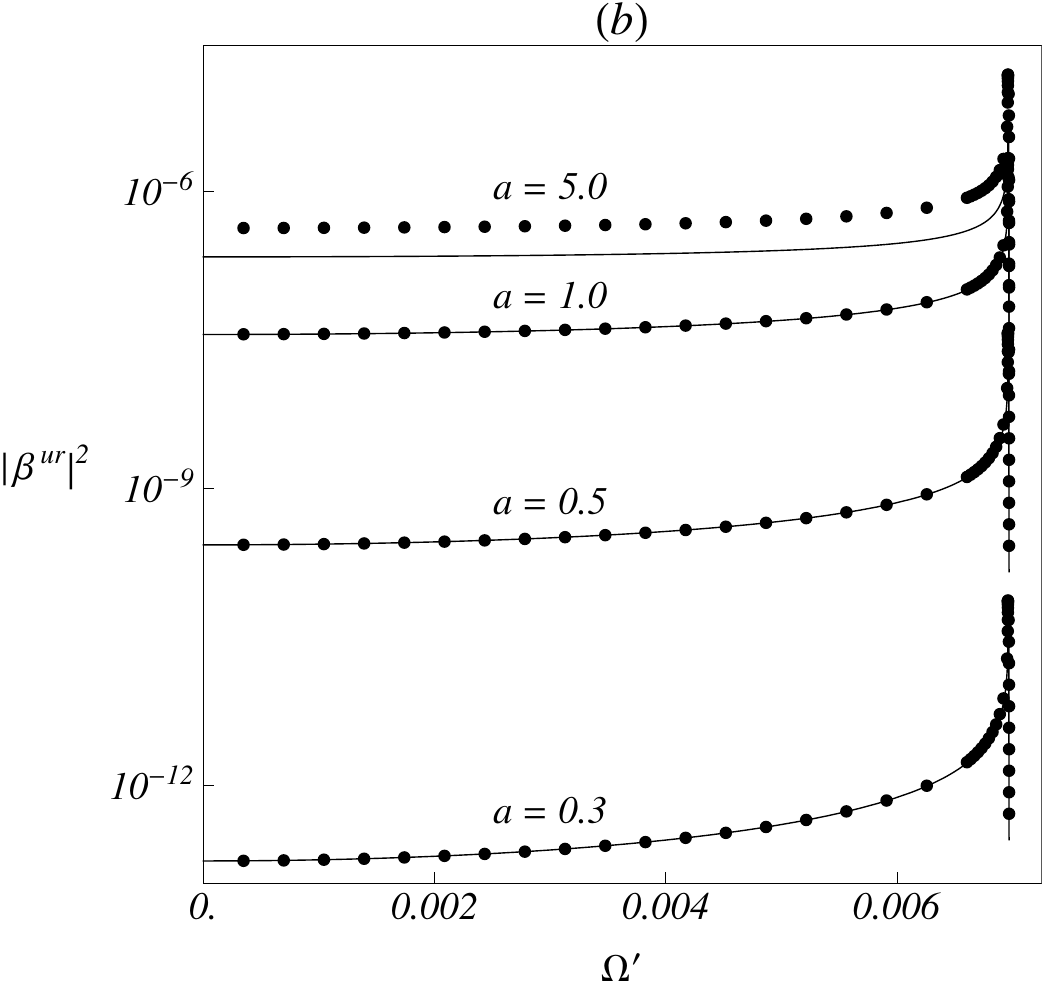}}

\caption[\textsc{Hawking spectra and phase-integral approximation}]{\textsc{Hawking spectra and phase-integral approximation}: The dots
plot the numerically calculated spectra as functions of the normalized
co-moving frequency, Figure $\left(a\right)$ showing the spectra
of the $ul$-$u$ pair whilst Figure $\left(b\right)$ shows the spectra
of the $ur$-$u$ pair. The curves plot the modified phase-integral
predictions of Eqs. (\ref{eq:Hawking_opt_no-horizon_gvh})-(\ref{eq:Hawking_opt_no-horizon_ur}),
and, as before, they are in excellent agreement for $a\lesssim1$.\label{fig:Hawking_freq-dep-chi_WKB}}

\end{figure}

Also worthy of note is that the analytic expressions of Eqs. (\ref{eq:Hawking_opt_no-horizon_gvh})-(\ref{eq:Hawking_opt_no-horizon_ur}),
which were previously found to be in good agreement with numerics,
remain valid here. The plots are compared in Figure \ref{fig:Hawking_freq-dep-chi_WKB}.

\section{Reduction to two solutions\label{sub:Reduction-to-two-solutions}}

Discarding the low-frequency solutions of the dispersion relation
(i.e., the $v$- and $ul$-modes), we are left with only two solutions:
the $u$- and $ur$-modes, lying on either side of the frequency $\omega_{0}$
whose co-moving frequency vanishes. Although we attempted, in the
previous section, to suppress the low-frequency modes, we found that
they tend to remain the dominant component of the radiation. We would
like to find a way to eliminate them entirely, such that the high-frequency
modes are really the only solutions, to check the robustness of the
Hawking spectrum with respect to the presence or absence of the low-frequency
modes.

There is, however, an immediate problem: the full optical wave equation
has exactly two solutions in the absence of dispersion. Introducing
dispersion - and giving the wavenumber $\beta$ a more complicated
dependence on $\omega$ than mere proportionality - serves to increase
the number of solutions. Therefore, we cannot reduce the number of
solutions sufficiently using the full wave equation. Instead, we may
formulate a {}``half'' wave equation that describes purely forward-propagating
waves. This is not easily derived from the optical wave equation (\ref{eq:wave_equation_FIBRES}),
but an analogous equation is derived from the acoustic wave equation
(\ref{eq:acoustic_wave_equation}), which is factorizable into forward-
and backward-propagating components. Our first task, then, is to alter
the optical wave equation slightly so that it can be cast in the same
form as the acoustic wave equation.

\subsection{Analogy between acoustic and optical wave equations}

The acoustic wave equation, fully expanded and in as general a form
as possible, is\begin{equation}
\partial_{t}^{2}\phi+2V\partial_{t}\partial_{x}\phi+\left(\partial_{x}V\right)\partial_{t}\phi+2V\left(\partial_{x}V\right)\partial_{x}\phi+V^{2}\partial_{x}^{2}\phi+F^{2}\left(-i\partial_{x}\right)\phi=0\,.\end{equation}
Similarly for the optical wave equation:\begin{equation}
\partial_{\zeta}^{2}A-2\partial_{\zeta}\partial_{\tau}A-\frac{1}{n_{g}^{2}}\left(\partial_{\tau}\chi\right)\partial_{\tau}A+\left(1-\frac{1}{n_{g}^{2}}\chi\right)\partial_{\tau}^{2}A+\frac{c^{2}}{n_{g}^{2}}\beta^{2}\left(i\partial_{\tau}\right)A=0\,,\end{equation}
where $n_{g}=c/u$ is the group index of the pulse. It is clear that
$\phi$ and $A$, $t$ and $\zeta$, and $x$ and $\tau$ are pairs
of analogous quantities. The terms $F^{2}\left(-i\partial_{x}\right)\phi$
and $\frac{c^{2}}{n_{g}^{2}}\beta^{2}\left(i\partial_{\tau}\right)A$
are also analogous. (The sign of the differential operator is irrelevant
since both $F^{2}\left(k\right)$ and $\beta^{2}\left(\omega\right)$
are even functions of their arguments.) Also, the low-frequency wave
velocity, usually denoted $c$ in the acoustic model, corresponds
to $1/n_{g}$ in the optical model. The term $V^{2}\partial_{x}^{2}\phi$
is clearly analogous to the term $\left(1-\chi/n_{g}^{2}\right)\partial_{\tau}^{2}A$,
so that the flow velocity $V$ - which, we recall, is usually taken
to be negative - corresponds, in the optical model, to $-\left(1-\chi/n_{g}^{2}\right)^{1/2}\approx-1+\chi/\left(2n_{g}^{2}\right)$.
To first order in $\chi$, then, the term $2V\left(\partial_{x}V\right)\partial_{x}\phi$
corresponds to the term $-\left(\partial_{\tau}\chi\right)\partial_{\tau}A/n_{g}^{2}$.
The terms $2V\partial_{t}\partial_{x}\phi+\left(\partial_{x}V\right)\partial_{t}\phi$
would, to first order in $\chi$, correspond to $-2\partial_{\zeta}\partial_{\tau}A+\chi\partial_{\zeta}\partial_{\tau}A/n_{g}^{2}+\left(\partial_{\tau}\chi\right)\partial_{\zeta}A/\left(2n_{g}^{2}\right)$.
Neglecting the latter two terms, the correspondence between the acoustic
and optical wave equations is complete. We conclude that, if we may
neglect the terms $\chi\partial_{\zeta}\partial_{\tau}A/n_{g}^{2}+\left(\partial_{\tau}\chi\right)\partial_{\zeta}A/\left(2n_{g}^{2}\right)$,
then we may write the optical wave equation as\begin{equation}
\left(\partial_{\zeta}+\partial_{\tau}X\right)\left(\partial_{\zeta}+X\partial_{\tau}\right)A+\frac{c^{2}}{n_{g}^{2}}\beta^{2}\left(i\partial_{\tau}\right)A=0\,,\label{eq:optical_wave_eqn_analog}\end{equation}
where we have defined\begin{equation}
X\left(\tau\right)=-1+\frac{1}{2n_{g}^{2}}\chi\left(\tau\right)\,.\label{eq:optical_flow_velocity_analog}\end{equation}
Substituting the differential operators $\partial_{\zeta}$ and $\partial_{\tau}$
for $-i\omega^{\prime}$ and $-i\omega$, respectively, Eq. (\ref{eq:optical_wave_eqn_analog})
becomes\begin{equation}
\left(\omega^{\prime}+\omega X\right)\left(\omega^{\prime}+X\omega\right)A=\frac{c^{2}}{n_{g}^{2}}\beta^{2}\left(\omega\right)A\,.\label{eq:optical_wave_eqn_analog_2}\end{equation}

\subsection{{}``Half'' wave equation}

When $X$ is constant, the wave equation can be Fourier transformed
to give the dispersion relation\begin{equation}
\left(\omega^{\prime}+X\omega\right)^{2}=\frac{c^{2}}{n_{g}^{2}}\beta^{2}\left(\omega\right)\Longrightarrow\omega^{\prime}=-X\omega\pm\frac{c}{n_{g}}\beta\left(\omega\right)\,.\label{eq:optical_dispersion_analog}\end{equation}
In the absence of pulses, or far away from them, $\chi=0$, so from
Eq. (\ref{eq:optical_flow_velocity_analog}) we have $X=-1$ and Eq.
(\ref{eq:optical_dispersion_analog}) becomes $\omega^{\prime}=\omega\pm c\beta\left(\omega\right)/n_{g}$,
where the minus sign corresponds to forward-propagating waves. Singling
out the forward-propagating waves can be achieved by replacing Eq.
(\ref{eq:optical_wave_eqn_analog_2}) with its {}``square root'':\begin{equation}
\left(\omega^{\prime}+X\omega\right)A=-\frac{c}{n_{g}}\beta\left(\omega\right)A\,,\label{eq:square_root_wave_eqn}\end{equation}
or, replacing the differential operators,\begin{equation}
\left(\partial_{\zeta}+X\partial_{\tau}\right)A=i\frac{c}{n_{g}}\beta\left(i\partial_{\tau}\right)A\,.\label{eq:half_wave_eqn}\end{equation}
When $X$ is not constant in $\tau$, Eq. (\ref{eq:half_wave_eqn})
does not give an exact solution of Eq. (\ref{eq:optical_wave_eqn_analog}),
and the norm is no longer conserved. However, the differences from
the full wave equation might be small enough so that Eq. (\ref{eq:half_wave_eqn})
can still give meaningful results.

\subsection{Dispersion relation}

The only remaining task to be done before using the {}``half'' wave
equation is to specify the form of $\beta\left(\omega\right)$ that
gives only two solutions. This can be derived quite simply as follows.
When $X$ is constant, the dispersion relation is $\omega^{\prime}+X\omega=-c\beta\left(\omega\right)/n_{g}$.
In order to have only two solutions for $\omega$, this equation must
be quadratic in $\omega$, and since it already contains $\omega$
to the powers of $0$ and $1$, it follows that $\beta\left(\omega\right)$
must contain either a term in $\omega^{2}$ or a term in $\omega^{-1}$
(possibly alongside terms in $\omega^{0}$ and $\omega^{1}$). Since
$\beta\left(\omega\right)$ must be an odd function, it follows that
its general form is\begin{equation}
\frac{c}{n_{g}}\beta\left(\omega\right)=c_{1}\omega+c_{-1}\omega^{-1}\,.\label{eq:beta_pole_general_form}\end{equation}
In the absence of pulses, or far away from them, the dispersion relation
becomes\begin{equation}
\omega^{\prime}=\left(1-c_{1}\right)\omega-c_{-1}\omega^{-1}\,.\label{eq:dispersion_relation_pole}\end{equation}
Another requirement is that the co-moving frequency $\omega^{\prime}$
should vanish at the frequency $\omega_{0}$:\[
0=\left(1-c_{1}\right)\omega_{0}-c_{-1}\omega_{0}^{-1}\Longrightarrow c_{-1}=\left(1-c_{1}\right)\omega_{0}^{2}\,.\]
Eq. (\ref{eq:beta_pole_general_form}) can now be written\begin{equation}
\frac{c}{n_{g}}\beta\left(\omega\right)=c_{1}\omega+\left(1-c_{1}\right)\frac{\omega_{0}^{2}}{\omega}\,.\label{eq:beta_pole_a}\end{equation}

How is $c_{1}$ to be interpreted? Since the dispersion relation should
be most accurate near $\omega=\omega_{0}$, it is related to the parameter
$\xi\equiv-d\omega^{\prime}/d\omega\mid_{\omega=\omega_{0}}$. Substituting
for $c_{-1}$ in Eq. (\ref{eq:dispersion_relation_pole}), we have\begin{equation}
\omega^{\prime}=\left(1-c_{1}\right)\left(\omega-\frac{\omega_{0}^{2}}{\omega}\right)\,,\label{eq:omega_prime_a}\end{equation}
and, therefore,\begin{equation}
\frac{d\omega^{\prime}}{d\omega}=\left(1-c_{1}\right)\left(1+\frac{\omega_{0}^{2}}{\omega^{2}}\right)\,.\label{eq:group_velocity_a}\end{equation}
When $\omega=\omega_{0}$, this gives $-\xi=d\omega^{\prime}/d\omega\mid_{\omega=\omega_{0}}=2\left(1-c_{1}\right)$.
Substituting in Eqs. (\ref{eq:beta_pole_a}) and (\ref{eq:omega_prime_a}),
we finally have\begin{eqnarray}
\frac{c}{n_{g}}\beta\left(\omega\right) & = & \left(1+\frac{\xi}{2}\right)\omega-\frac{\xi}{2}\frac{\omega_{0}^{2}}{\omega}\,,\label{eq:beta_pole_xi}\\
\omega^{\prime} & = & -\frac{\xi}{2}\left(\omega-\frac{\omega_{0}^{2}}{\omega}\right)\,.\label{eq:omega_prime_xi}\end{eqnarray}

\begin{figure}
\includegraphics[width=0.8\columnwidth]{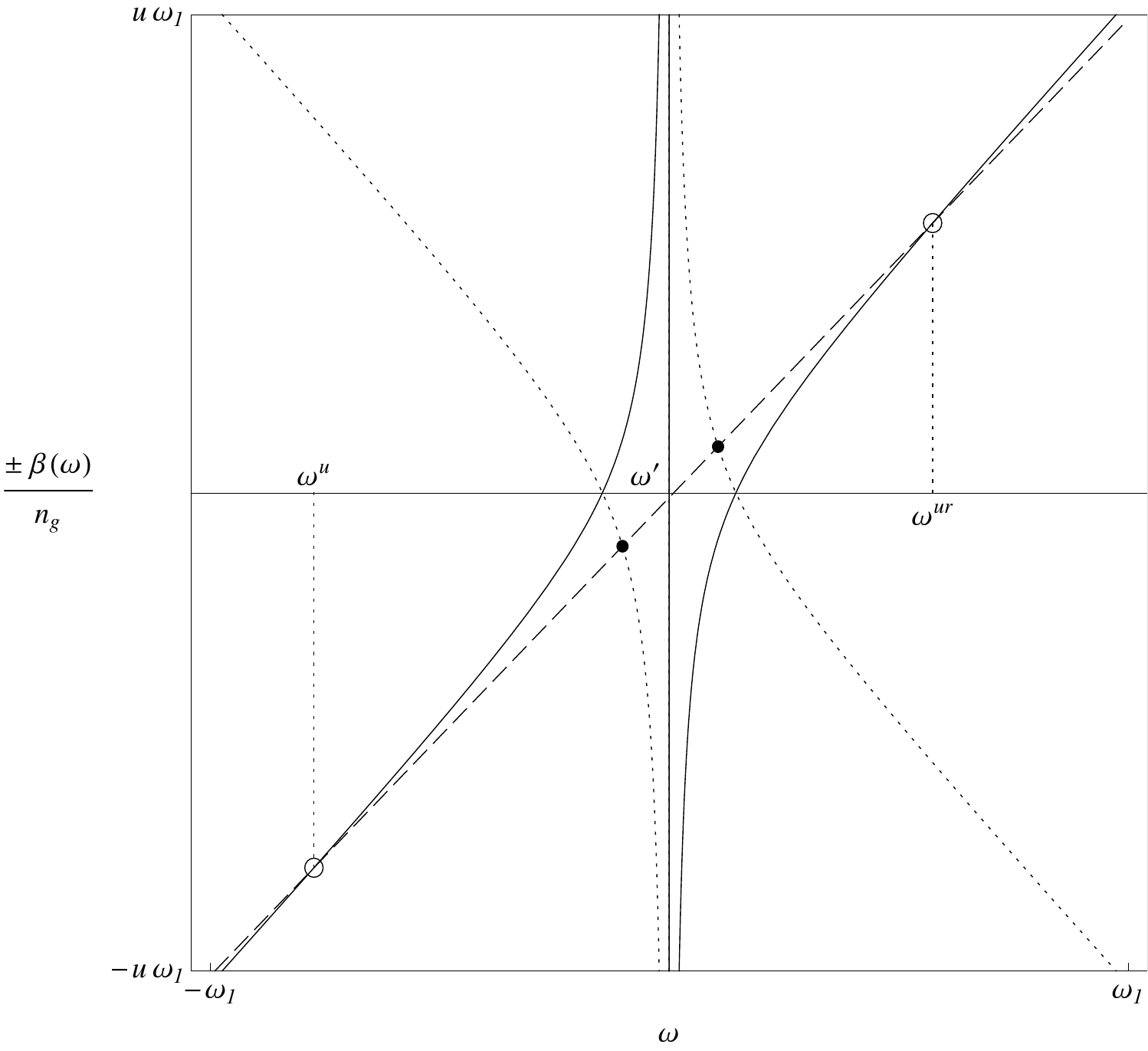}

\caption[\textsc{Two-solution dispersion with positive $\xi$}]{\textsc{Two-solution dispersion with positive $\xi$}: The solutions
of the dispersion relation are the points of intersection between
$u\,\beta\left(\omega\right)$, given by Eq. (\ref{eq:beta_pole_xi}),
and the line $\omega-\omega^{\prime}$. The solid and dotted curves
refer to co- and counter-propagating solutions (not positive- and
negative-norm solutions, as in previous examples). For the {}``half''
wave equation (\ref{eq:half_wave_eqn}), only the solid curve applies.
Note that the full wave equation would always have four real solutions.\label{fig:half_wave_eqn_positive_xi}}

\end{figure}

\begin{figure}
\includegraphics[width=0.8\columnwidth]{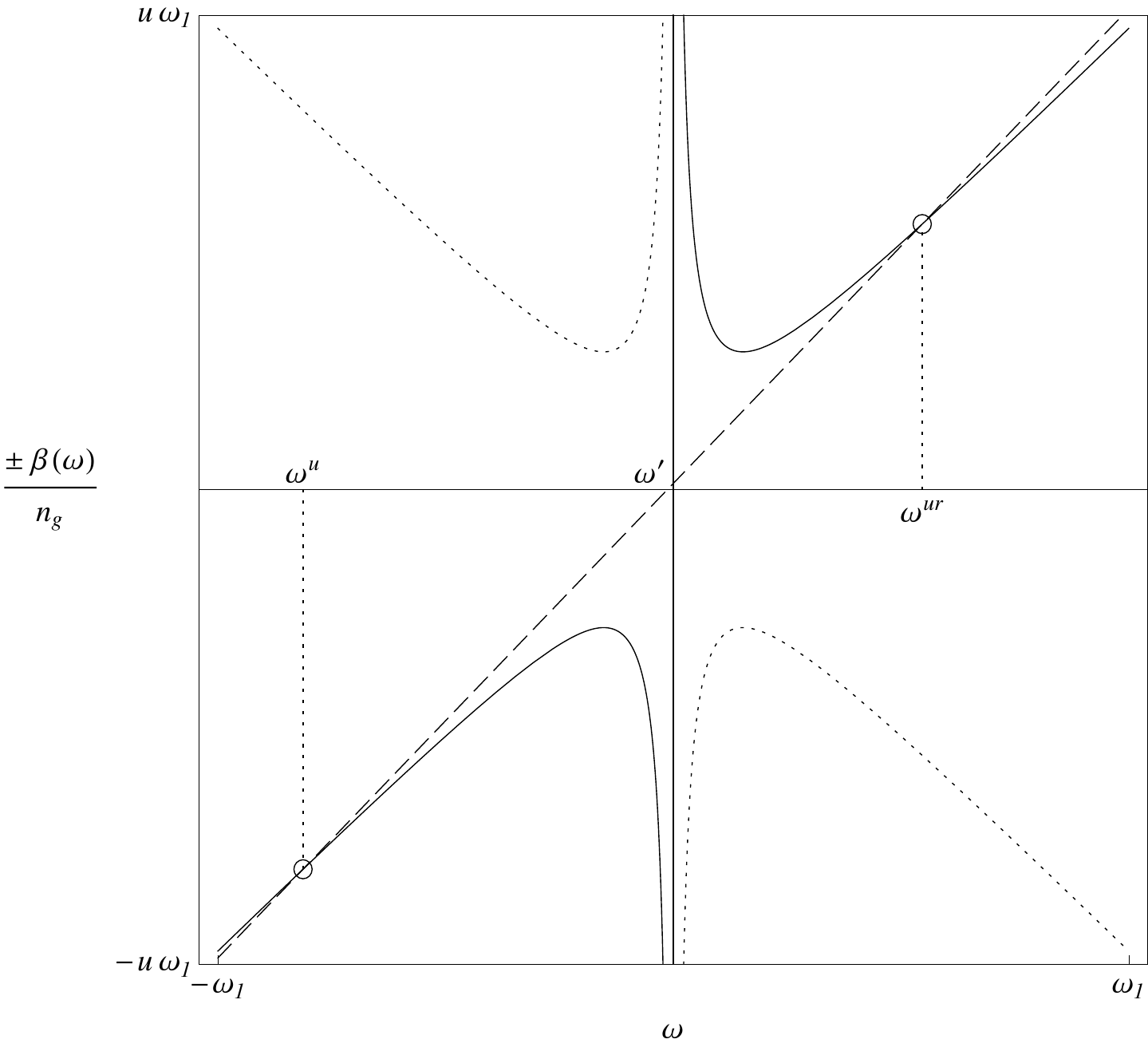}

\caption[\textsc{Two-solution dispersion with negative $\xi$}]{\textsc{Two-solution dispersion with negative $\xi$}: The possibile
solutions of the dispersion relation are the points where $u\,\beta\left(\omega\right)$,
where $\beta\left(\omega\right)$ is given by Eq. (\ref{eq:beta_pole_xi}),
and the line $\omega-\omega^{\prime}$ intersect. The solid and dotted
curves refer to the forward- and backward-propagating solutions (\textit{not},
as in previous examples, the positive- and negative-norm solutions,
which correspond directly to the sign of $\omega-\omega^{\prime}$).
For the {}``half'' wave equation (\ref{eq:half_wave_eqn}), only
the solid curve applies, and there are only the two solutions, corresponding
to the $u$- and $ur$-modes of quartic dispersion (see Fig. \ref{fig:optical_omega_values}).
If $\omega^{\prime}$ is not too large, even the full wave equation
has only two real solutions.\label{fig:half_wave_eqn_negative_xi}}

\end{figure}

The artificiality of this dispersion model yields some peculiar properties.
One of these has to do with the signs of first- and second-order dispersion
(i.e., the first and second derivatives of $\beta\left(\omega\right)$).
First-order dispersion is equal to the reciprocal of the lab-frame
group velocity \cite{Agrawal}. Usually, this is positive: in the
lab frame, the phase and group velocities almost always point in the
same direction. This corresponds, in the above model, to a positive
value of $\xi$, for\[
\frac{c}{n_{g}}\beta^{\prime}\left(\omega\right)=1+\frac{\xi}{2}\left(1+\frac{\omega_{0}^{2}}{\omega^{2}}\right)\,.\]
The second-order dispersion is also usually positive, but there can
exist regimes in which it is negative. (Solitons, for example, can
exist only in regions with negative second-order dispersion, as mentioned
in §\ref{sub:Nonlinear-effects}.) In our case, both first- and second-order
dispersion are positive in the high-frequency regime. However, the
second-order dispersion is given by\[
\frac{c}{n_{g}}\beta^{\prime\prime}\left(\omega\right)=-\xi\frac{\omega_{0}^{2}}{\omega^{3}}\,.\]
Although in reality both are positive, the dispersion model we have
selected forces first- and second-order dispersion to have opposite
signs. Figures \ref{fig:half_wave_eqn_positive_xi} and \ref{fig:half_wave_eqn_negative_xi}
show the dispersion relations corresponding to positive and negative
$\xi$, respectively. We may still use this model, so long as we recognise
that it is only for the sake of verification that it is being used,
and is by no means accurate, especially far from the frequency $\omega_{0}$.

\subsection{Numerical results}

\begin{figure}
\subfloat{\includegraphics[width=0.45\columnwidth]{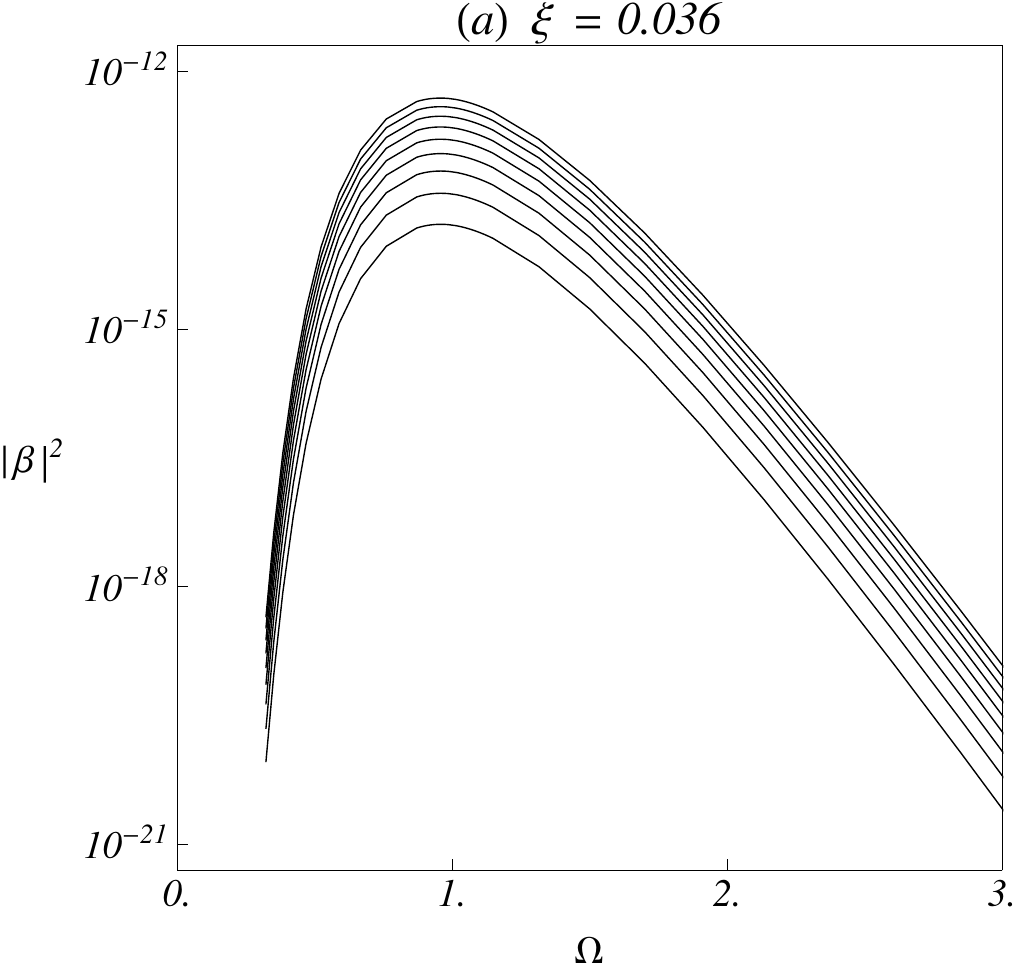}} \subfloat{\includegraphics[width=0.45\columnwidth]{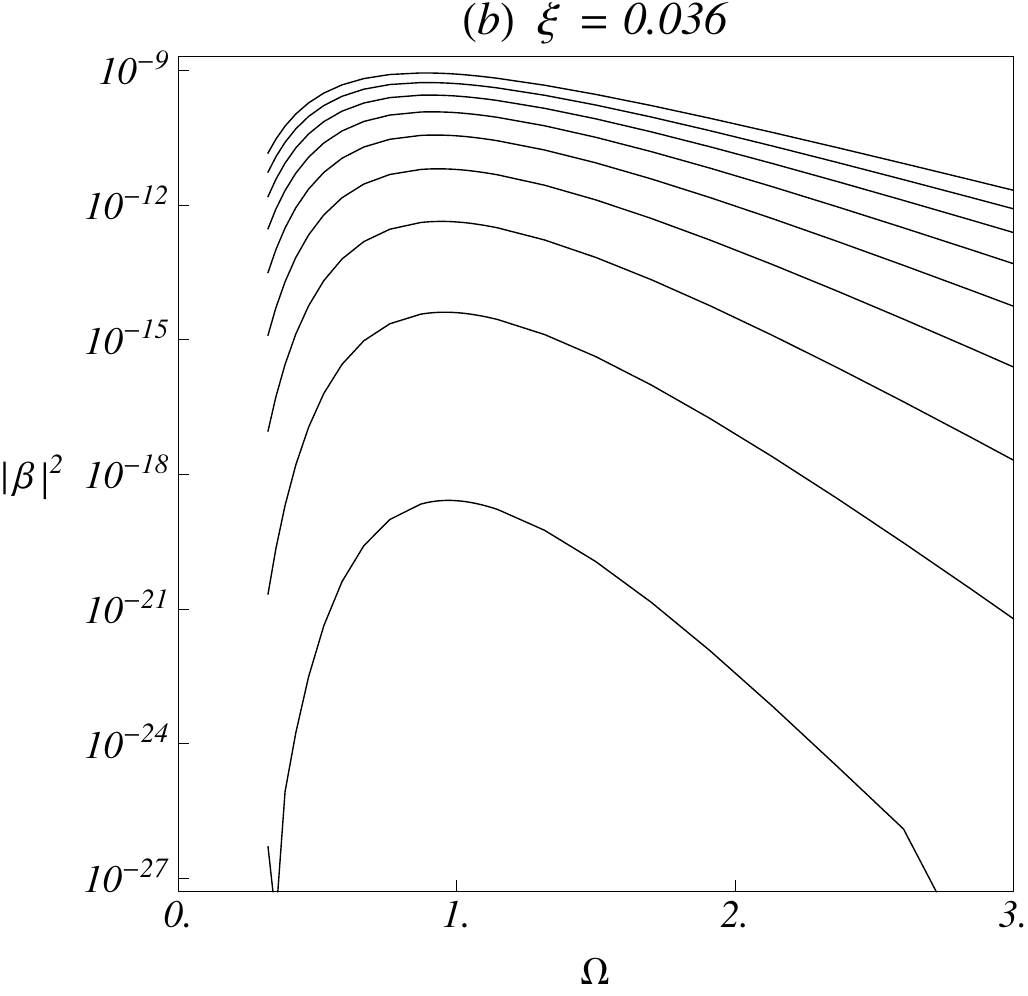}}

\subfloat{\includegraphics[width=0.45\columnwidth]{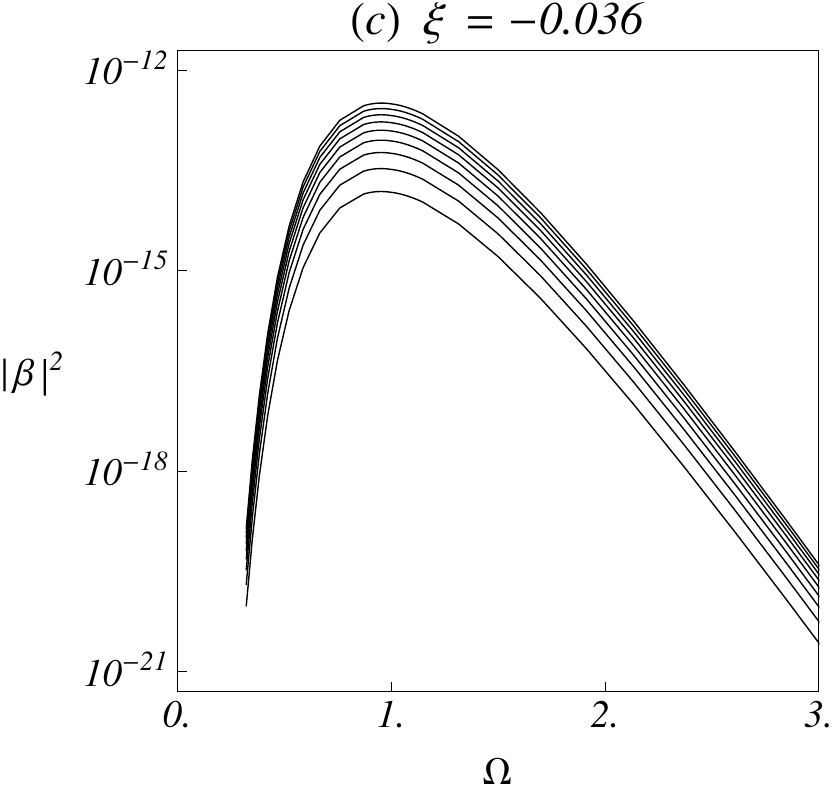}} \subfloat{\includegraphics[width=0.45\columnwidth]{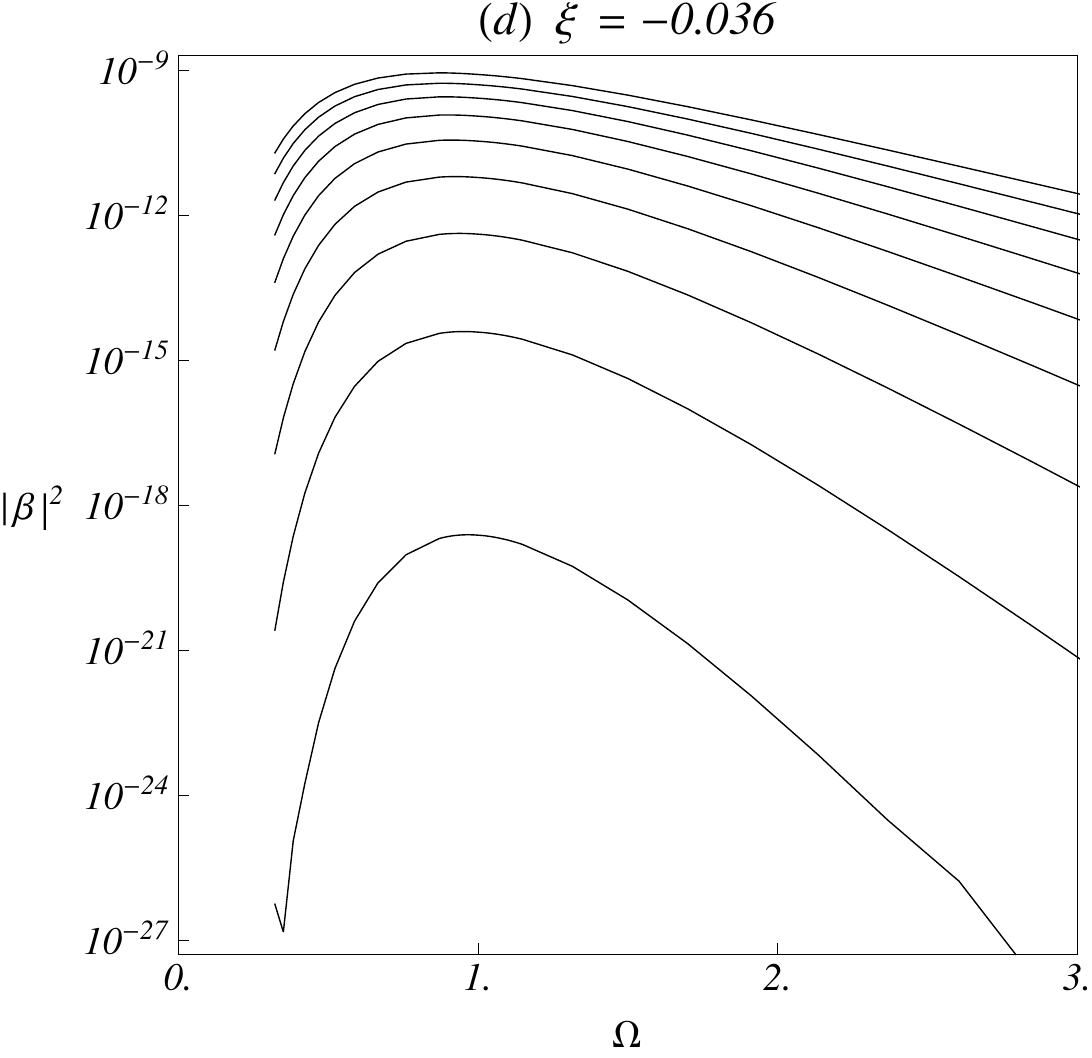}}

\caption[\textsc{Hawking spectra with two-solution dispersion}]{\textsc{Hawking spectra with two-solution dispersion}: These plot
numerically calculated spectra for the dispersion model of Eq. (\ref{eq:beta_pole_xi}).
In Figures $\left(a\right)$ and $\left(b\right)$, $\xi=0.036$,
while in Figures $\left(c\right)$ and $\left(d\right)$, $\xi=-0.036$.
In Figs. $\left(a\right)$ and $\left(c\right)$, $a$ is fixed at
$0.3$ while $h$ varies between $10^{-4}$ and $10^{-3}$; in $\left(b\right)$
and $\left(d\right)$, $h$ is fixed at $10^{-4}$ while $a$ varies
between $0.2$ and $1.0$. Notice that the plots change very little
when the sign of $\xi$ is changed.\label{fig:Hawking_pole-disp_spectra}}

\end{figure}

Some spectra obtained with the dispersion model of Eq. (\ref{eq:beta_pole_xi})
are plotted in Fig. \ref{fig:Hawking_pole-disp_spectra}. Since this
model applies only to frequencies near $\omega_{0}$, we have taken
$\left|\xi\right|=0.036$ in accordance with §\ref{sub:Dispersion-relation}.
Due to the ambiguity in the sign of $\xi$ noted above, we have plotted
the results for both positive and negative $\xi$; note that the change
in sign makes little difference to the results.

One striking feature of the spectra is their width: they are very
broad, reaching frequencies much higher than that obtained by previous
spectra. This is because the dispersion model of Eqs. (\ref{eq:beta_pole_xi})
and (\ref{eq:omega_prime_xi}) does not have a group-velocity horizon
for forward-propagating waves, and therefore has no cut-off frequency.
However, the resulting spectra behave in a similar way to our previous
results as the parameters $h$ and $a$ are varied. The spectra are
almost exactly proportional to $h^{2}$, and since there is no group-velocity
horizon, there is no exception to this rule. Somewhat counter-intuitively,
it is $a$, not $h$, which determines the width of the spectra, and
for lower values of $a$ the height increases exponentially with $a$.
As $a$ increases further, we find that the spectrum tends to a limiting
curve, as expected.

\begin{figure}
\subfloat{\includegraphics[width=0.45\columnwidth]{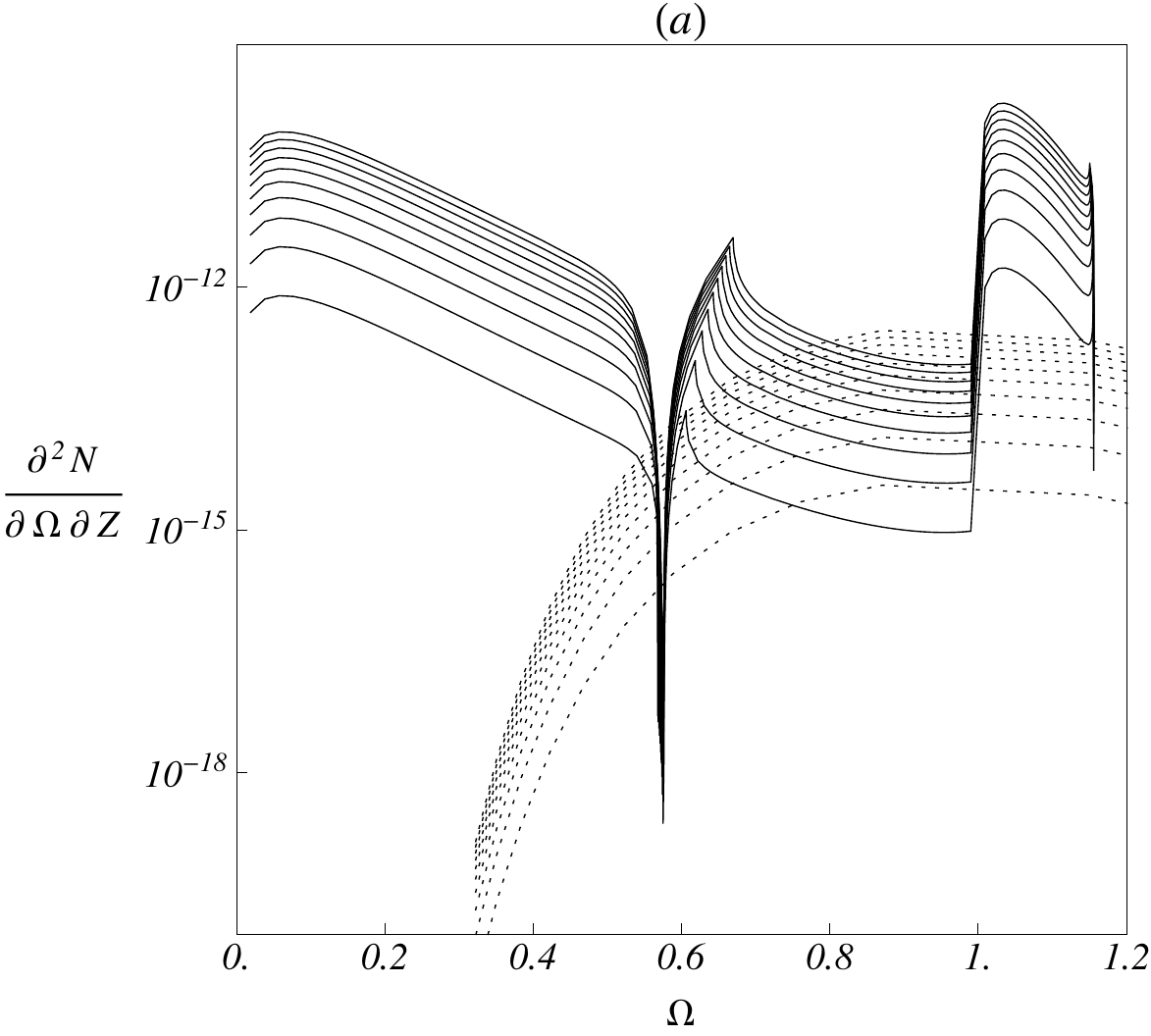}} \subfloat{\includegraphics[width=0.45\columnwidth]{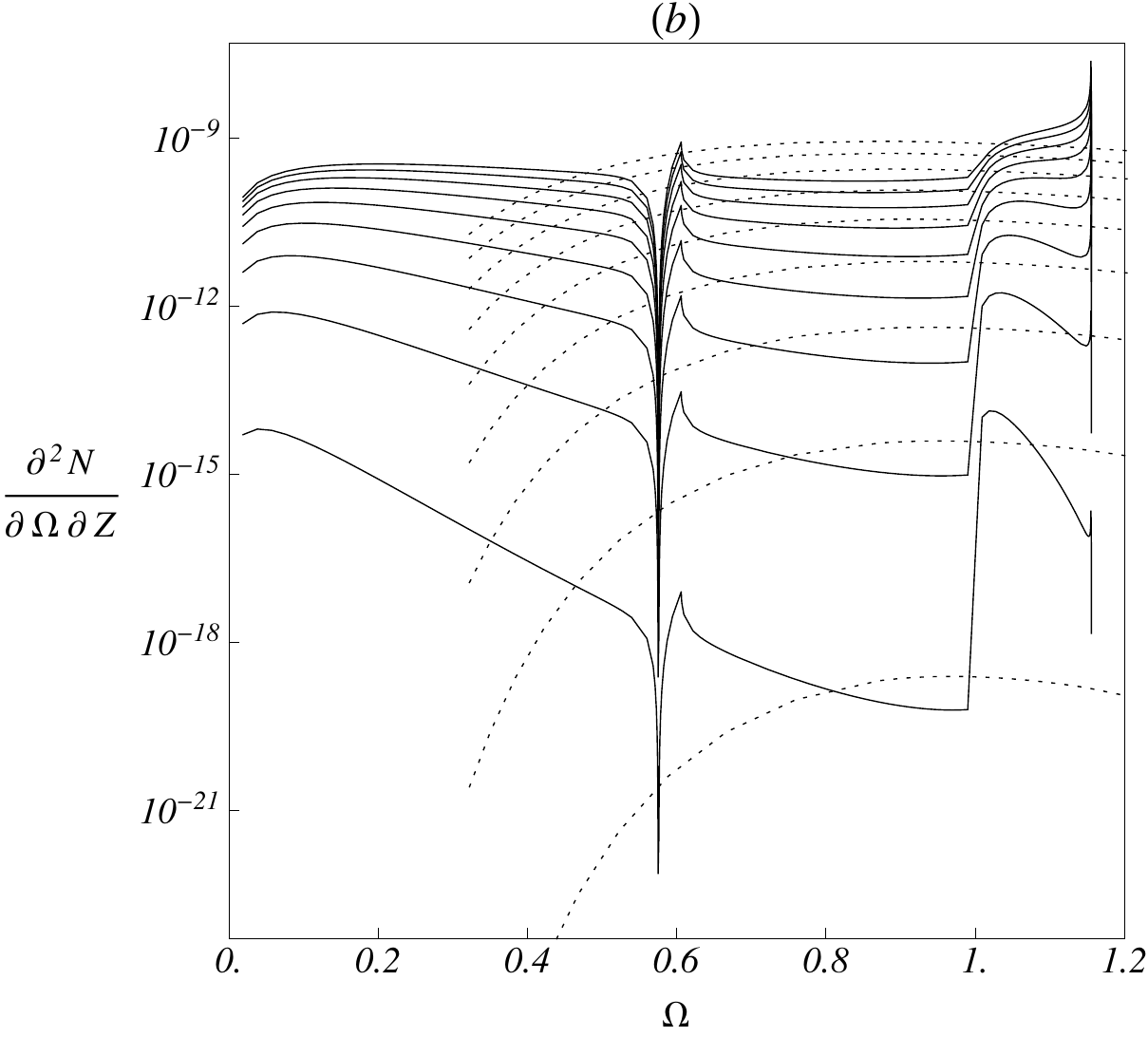}}

\subfloat{\includegraphics[width=0.45\columnwidth]{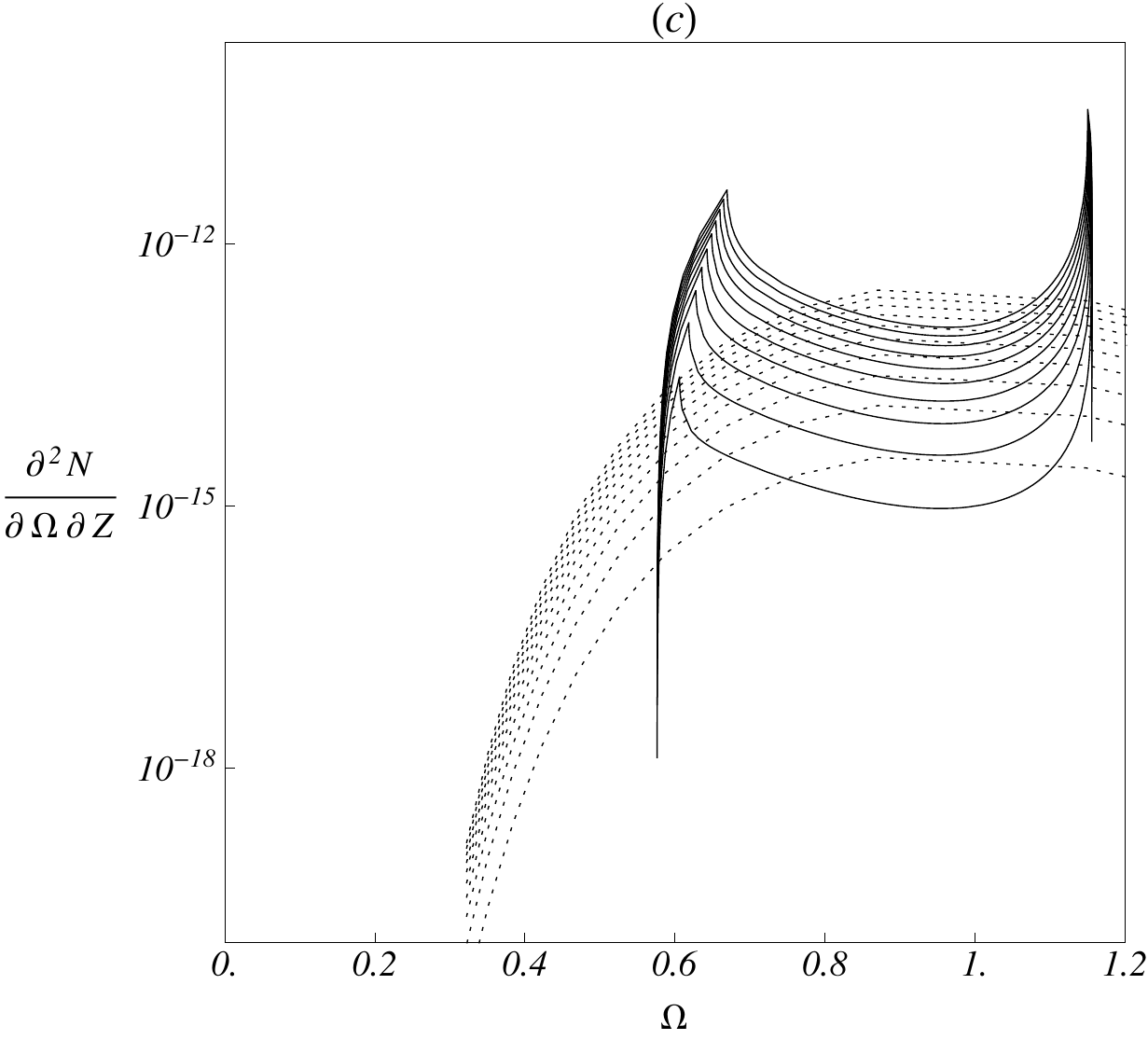}} \subfloat{\includegraphics[width=0.45\columnwidth]{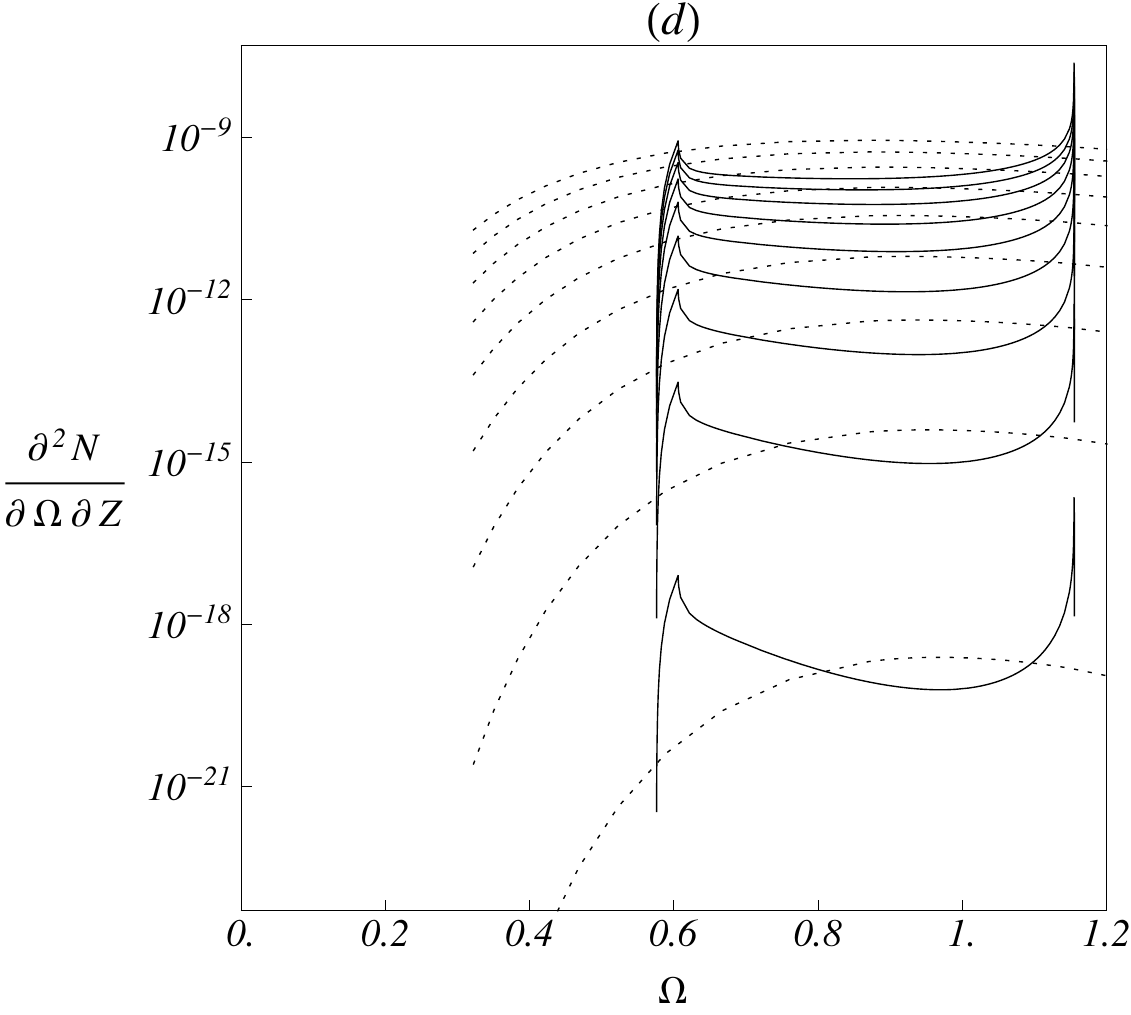}}

\caption[\textsc{Comparing two-solution with quartic dispersion}]{\textsc{Comparing two-solution with quartic dispersion}: The two-solution
spectra of Figure \ref{fig:Hawking_pole-disp_spectra} are shown here
as dotted curves, while the quartic-dispersion spectra of Figures
\ref{fig:Hawking_no-horizon_varying-h}$\left(d\right)$ and \ref{fig:Hawking_no-horizon_varying-a}$\left(d\right)$
are shown as solid curves. Figures $\left(a\right)$ and $\left(b\right)$
show the full spectra, with $h$ varying between $10^{-4}$ and $10^{-3}$
in $\left(a\right)$, and $a$ varying between $0.2$ and $1.0$ in
$\left(b\right)$. Figures $\left(c\right)$ and $\left(d\right)$
show the same curves, but omitting the $ul$-$u$ pair contribution
to the quartic-dispersion spectra.\label{fig:Hawking_pole-disp_comparison}}

\end{figure}

Comparing the results directly with those for quartic dispersion (see
Figure \ref{fig:Hawking_pole-disp_comparison}), we see that, around
the frequency $\Omega=1$ ($\omega=\omega_{0}$), the spectra are
of roughly the same order of magnitude. This concordance is seen more
clearly if we simply remove the $ul$-contribution to the $u$-modes
in the case of quartic dispersion, focusing solely on the coupling
between the $ur$- and the $u$-modes. That this order of magnitude
is reproduced even approximately when the low-frequency modes are
rendered impossible shows that they have little influence on the $ur$-$u$
coupling. It would seem that, even if we must take account of them
to solve the wave equation, we are justified in simply ignoring their
contribution to the creation rate.

\begin{figure}
\includegraphics[width=0.8\columnwidth]{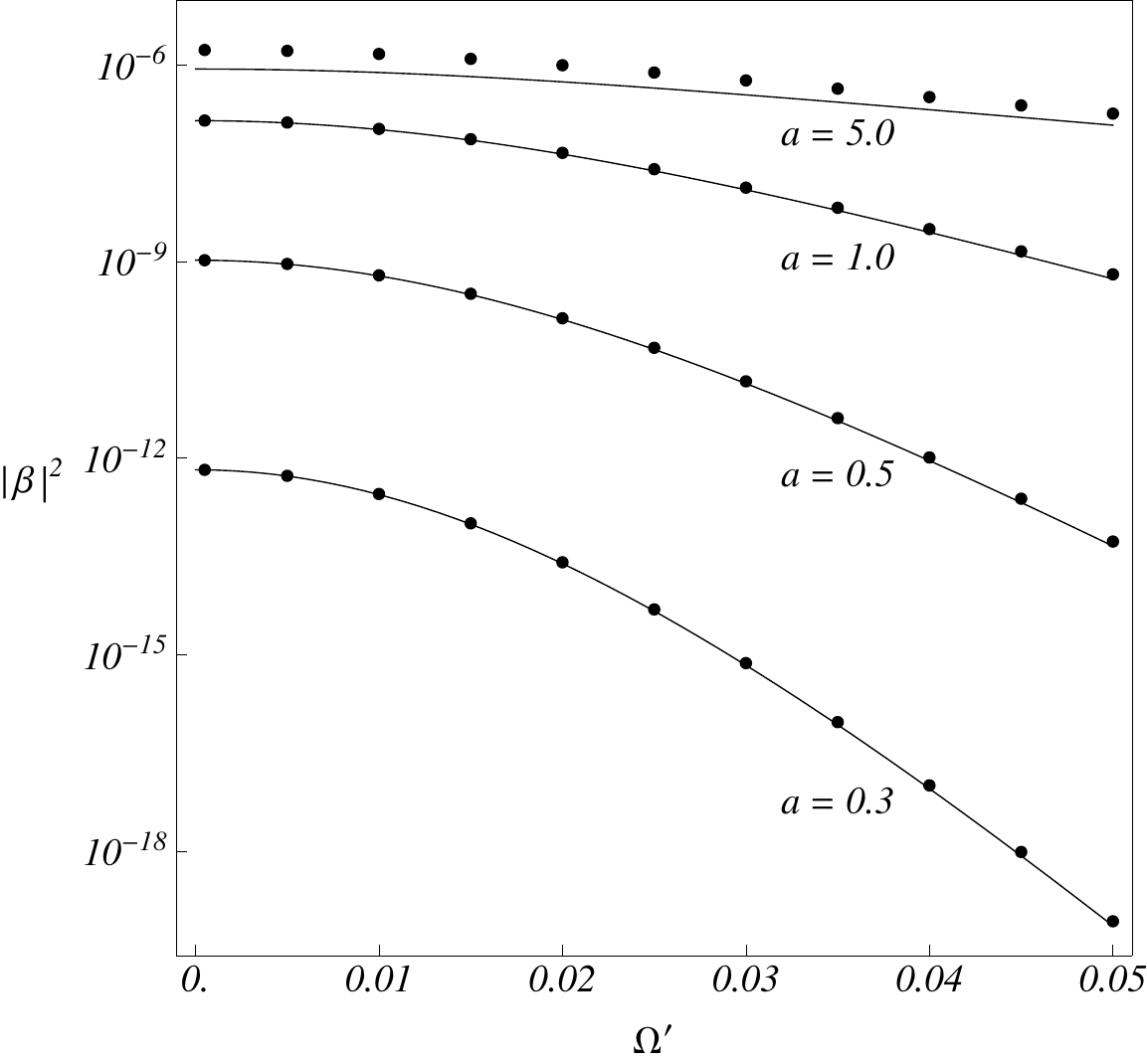}

\caption[\textsc{Two-solution dispersion and phase-integral approximation}]{\textsc{Two-solution dispersion and phase-integral approximation}:
Numerically calculated spectra are shown as dots, while the curves
plot the analytic expression of Eq. (\ref{eq:Hawking_pole-disp_WKB-approx}).
We have taken $h=10^{-4}$ and $\xi=-0.036$, while $a$ varies between
$0.3$ and $5.0$. Note that, as previously, there is very good agreement
for $a\lesssim1$.\label{fig:Hawking_pole-disp_WKB}}

\end{figure}

Again, we may strive for an analytic expression describing the Hawking
spectra, and for this we may turn to the results of previous models
for inspiration. It is found, for low values of $a$, that the spectra
are well-approximated by the formula\begin{multline}
\left|\beta_{\Omega^{\prime}}\right|^{2}\approx\exp\left(-\frac{\pi}{a}\left(\Omega_{R}^{ur}-\Omega_{R}^{u}\right)\right)-\exp\left(-\frac{\pi}{a}\left(\Omega_{L}^{ur}-\Omega_{R}^{u}\right)\right)\\
-\exp\left(-\frac{\pi}{a}\left(\Omega_{R}^{ur}-\Omega_{L}^{u}\right)\right)+\exp\left(-\frac{\pi}{a}\left(\Omega_{L}^{ur}-\Omega_{L}^{u}\right)\right)\,.\label{eq:Hawking_pole-disp_WKB-approx}\end{multline}
Several spectra, together with curves given by Eq. (\ref{eq:Hawking_pole-disp_WKB-approx}),
are plotted in Fig. \ref{fig:Hawking_pole-disp_WKB}.

\section{Concluding remarks\label{sub:Concluding-remarks-LFP-FIBRES}}

The most important result of this chapter is the observation that,
upon suppression of the low-frequency $ul$-modes, the spectra of
the $ur$-modes are to a large extent preserved. Thus, while many
of the spectra of Chapter \ref{sec:Results-for-the-Fibre-Optical-Model}
- particularly those for low values of $a$ - are dominated by $ul$-modes,
we are justified in simply ignoring these, focusing instead only on
the $ur$-$u$ pair. This may be weaker, but it is non-zero, and it
may be very significant for higher values of $a$. The Hawking spectrum, though diminished,
survives.

\pagebreak{}

\part{Appendices\label{par:Appendices}}

\appendix

\chapter{Low-Frequency Temperature in Acoustic Model\label{sec:Appendix_Acoustic}}

In this appendix, we derive the result of Eq. (\ref{eq:limiting_temperature}),
which gives the low-frequency Hawking temperature in the limit of
a discontinuous flow velocity profile with subluminal quartic dispersion;
the normalized dispersion relation is\begin{equation}
f^{2}\left(K\right)=K^{2}-K^{4}\end{equation}
and the flow velocity profile is\begin{equation}
U\left(X\right)=U_{L}\,\theta\left(-X\right)+U_{R}\,\theta\left(X\right)\,,\end{equation}
where $U_{L}<-1$ and $-1<U_{R}<0$. This velocity profile is shown
in Figure \ref{fig:tanh_velocity_profile}, and corresponds to a black
hole horizon.

\begin{figure}
\includegraphics[width=0.8\columnwidth]{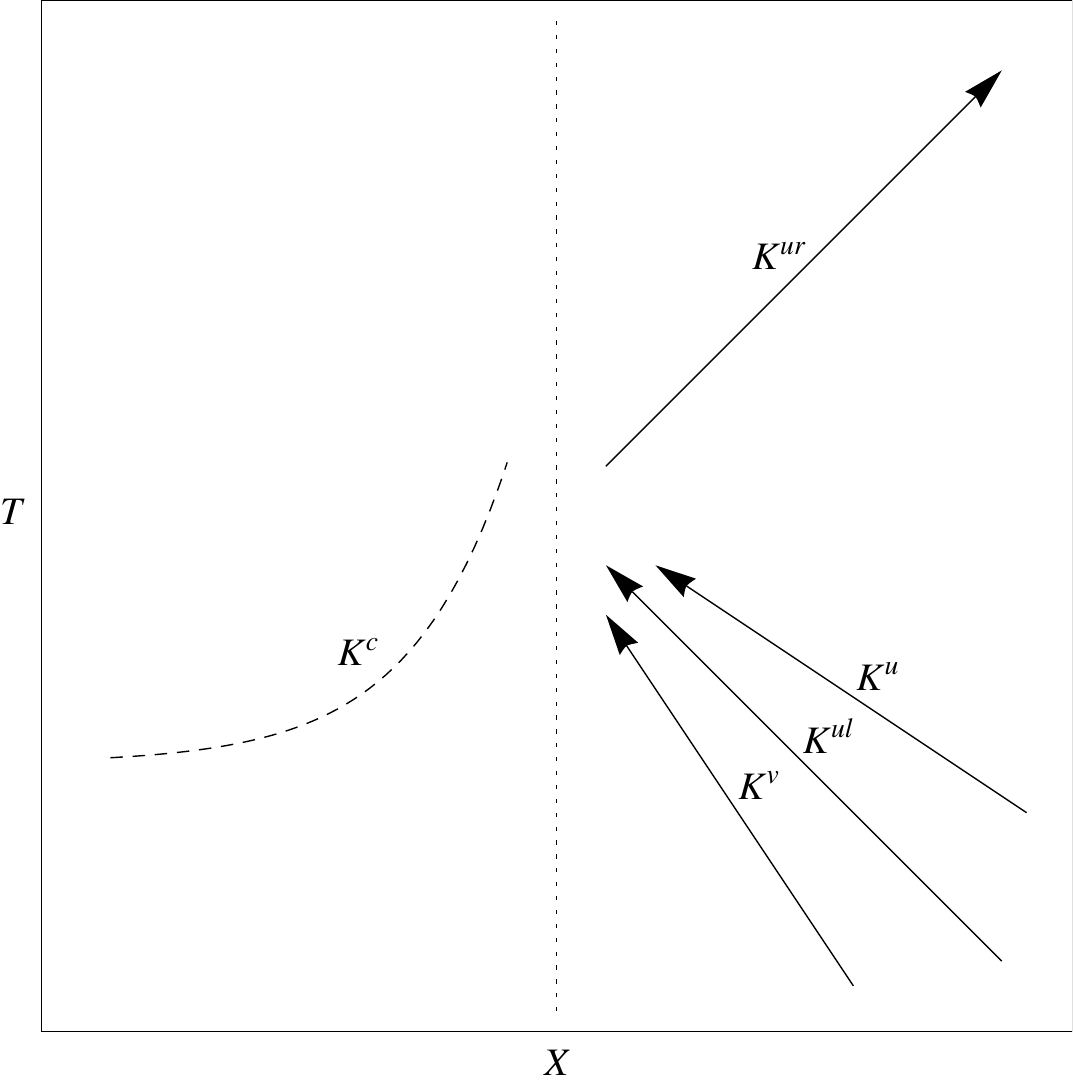}

\caption[\textsc{Space-time diagram of the $K^{ur}$-out mode}]{\textsc{Space-time diagram of the $K^{ur}$-out mode}: This has a
single decaying exponential on the left-hand side, and is the same
mode that is found by numerical integration when the steepness is
not assumed infinite.\label{fig:K_ur_out_mode}}

\end{figure}

The mode to be calculated is the $K^{ur}$-out mode, illustrated in
Figure \ref{fig:K_ur_out_mode}. Since we neglect the coupling to
$v$-modes, the only pair produced is $u-url$, and since we have
a black hole horizon, $K^{ur}$ must be the outgoing wavevector. The
creation rate is determined by the norm of the ingoing $K^{u}$ wavevector.

Given the values of $K$, the coefficients are found by imposing continuity
of $\phi$, $\partial_{X}\phi$ and $\partial_{X}^{2}\phi$, and the
condition of Eq. (\ref{eq:discontinuity_2n-1_derivative}) for $\partial_{X}^{3}\phi$,
at $X=0$. Eq. (\ref{eq:discontinuity_2n-1_derivative}) gives\begin{equation}
\Delta\left(\partial_{X}^{3}\phi\right)=-\left(i\Omega\phi-\left(U_{R}+U_{L}\right)\partial_{X}\phi\right)\left(U_{R}-U_{L}\right)\,.\end{equation}
$\phi$ is expressed as a sum of the plane wave solutions in each
region:\begin{equation}
\phi=\begin{cases}
e^{iK^{c}X}\,, & X<0\\
a^{v}e^{iK^{v}X}+a^{ur}e^{iK^{ur}X}+a^{ul}e^{iK^{ul}X}+a^{u}e^{iK^{u}X}\,, & X>0\end{cases}\,.\end{equation}
The conditions given above can then be expressed in matrix form as
follows:\begin{equation}
\left[\begin{array}{cccc}
1 & 1 & 1 & 1\\
K^{v} & K^{ur} & K^{ul} & K^{u}\\
\left(K^{v}\right)^{2} & \left(K^{ur}\right)^{2} & \left(K^{ul}\right)^{2} & \left(K^{u}\right)^{2}\\
\left(K^{v}\right)^{3} & \left(K^{ur}\right)^{3} & \left(K^{ul}\right)^{3} & \left(K^{u}\right)^{3}\end{array}\right]\left[\begin{array}{c}
a^{v}\\
a^{ur}\\
a^{ul}\\
a^{u}\end{array}\right]=\left[\begin{array}{c}
1\\
K^{c}\\
\left(K^{c}\right)^{2}\\
\left(K^{c}\right)^{3}+\left(U_{R}-U_{L}\right)\left(\Omega-K^{c}\left(U_{R}+U_{L}\right)\right)\end{array}\right]\,.\end{equation}
The solution to this linear equation is\begin{equation}
\left[\begin{array}{c}
a^{v}\\
a^{ur}\\
a^{ul}\\
a^{u}\end{array}\right]=\left[\begin{array}{c}
\frac{\left(K^{ur}-K^{c}\right)\left(K^{ul}-K^{c}\right)\left(K^{u}-K^{c}\right)-\left(U_{R}-U_{L}\right)\left(\Omega-K^{c}\left(U_{R}+U_{L}\right)\right)}{\left(K^{ur}-K^{v}\right)\left(K^{ul}-K^{v}\right)\left(K^{u}-K^{v}\right)}\\
\frac{\left(K^{v}-K^{c}\right)\left(K^{ul}-K^{c}\right)\left(K^{u}-K^{c}\right)-\left(U_{R}-U_{L}\right)\left(\Omega-K^{c}\left(U_{R}+U_{L}\right)\right)}{\left(K^{v}-K^{ur}\right)\left(K^{ul}-K^{ur}\right)\left(K^{u}-K^{ur}\right)}\\
\frac{\left(K^{v}-K^{c}\right)\left(K^{ur}-K^{c}\right)\left(K^{u}-K^{c}\right)-\left(U_{R}-U_{L}\right)\left(\Omega-K^{c}\left(U_{R}+U_{L}\right)\right)}{\left(K^{v}-K^{ul}\right)\left(K^{ur}-K^{ul}\right)\left(K^{u}-K^{ul}\right)}\\
\frac{\left(K^{v}-K^{c}\right)\left(K^{ur}-K^{c}\right)\left(K^{ul}-K^{c}\right)-\left(U_{R}-U_{L}\right)\left(\Omega-K^{c}\left(U_{R}+U_{L}\right)\right)}{\left(K^{v}-K^{u}\right)\left(K^{ur}-K^{u}\right)\left(K^{ul}-K^{u}\right)}\end{array}\right]\,.\label{eq:coefficients_vector_soln}\end{equation}
In accordance with Eq. (\ref{eq:norm_from_coefficient}), the relative
negative-norm component is\begin{equation}
\left|\beta_{\Omega}\right|^{2}=\frac{\left|f\left(K^{u}\right)\right|\left|v_{g}\left(K^{u}\right)\right|\left|a^{u}\right|^{2}}{\left|f\left(K^{ur}\right)\right|\left|v_{g}\left(K^{ur}\right)\right|\left|a^{ur}\right|^{2}}\,.\label{eq:Bogoliubov_coeff_discontinuous_profile}\end{equation}
At low frequencies, we should have $\left|\beta_{\Omega}\right|^{2}\approx T_{\infty}/\Omega$,
where $T_{\infty}$ is the temperature of the radiation. (The subscript
indicates that this is the limiting temperature as the steepness of
the velocity profile becomes infinite.) Our task now, then, is to
express the quantities in Eq. (\ref{eq:Bogoliubov_coeff_discontinuous_profile})
to lowest order in $\Omega$. This requires expressions for the five
possible values of $K$ - $K^{v}$, $K^{ur}$, $K^{ul}$, $K^{u}$
and $K^{c}$ - to lowest order in $\Omega$.

For a constant flow velocity $U$, the solutions for $K$ satisfy
the dispersion relation\[
f^{2}\left(K\right)=K^{2}-K^{4}=\left(\Omega-UK\right)^{2}\,,\]
or, rearranging,\begin{equation}
K^{4}+\left(U^{2}-1\right)K^{2}-2U\Omega K+\Omega^{2}=0\,.\label{eq:quartic_dispersion_relation}\end{equation}
Expressing $K$ as a Taylor series in $\Omega$,\[
K=c_{0}+c_{1}\Omega+c_{2}\Omega^{2}+\ldots\,,\]
and plugging this series into Eq. (\ref{eq:quartic_dispersion_relation}),
we find\begin{multline*}
c_{0}^{2}\left(c_{0}^{2}+U^{2}-1\right)+c_{0}\left(2c_{1}\left(2c_{0}^{2}+U^{2}-1\right)-2U\right)\Omega\\
+\left(2c_{0}^{2}\left(3c_{1}^{2}+2c_{0}c_{2}\right)+\left(c_{1}^{2}+2c_{0}c_{2}\right)\left(U^{2}-1\right)-2Uc_{1}+1\right)\Omega^{2}+\ldots=0\,.\end{multline*}
For the series on the left of this equation to be identically zero
requires that the coefficient of each power of $\Omega$ is equal
to zero:\begin{eqnarray}
c_{0}^{2}\left(c_{0}^{2}+U^{2}-1\right) & = & 0\,,\label{eq:K_Taylor_series_0}\\
c_{0}\left(2c_{1}\left(2c_{0}^{2}+U^{2}-1\right)-2U\right) & = & 0\,,\label{eq:K_Taylor_series_1}\\
\left(2c_{0}^{2}\left(3c_{1}^{2}+2c_{0}c_{2}\right)+\left(c_{1}^{2}+2c_{0}c_{2}\right)\left(U^{2}-1\right)-2Uc_{1}+1\right) & = & 0\,,\label{eq:K_Taylor_series_2}\end{eqnarray}
and so on. From Eq. (\ref{eq:K_Taylor_series_0}), we find that either
$c_{0}=0$ or $c_{0}=\pm\left(1-U^{2}\right)^{1/2}$. (If $\left|U\right|>1$,
then $c_{0}=\pm i\left(U^{2}-1\right)^{1/2}$.) If $c_{0}=0$, Eq.
(\ref{eq:K_Taylor_series_1}) tells us nothing new, and so we must
fall on Eq. (\ref{eq:K_Taylor_series_2}); this reduces to\[
c_{1}^{2}\left(U^{2}-1\right)-2Uc_{1}+1=0\,,\]
a quadratic equation easily solved for $c_{1}$:\[
c_{1}=\frac{1}{2\left(U^{2}-1\right)}\left(2U\pm\sqrt{4U^{2}-4\left(U^{2}-1\right)}\right)=\frac{U\pm1}{U^{2}-1}=\frac{1}{U\mp1}\,.\]
On the other hand, if $c_{0}^{2}=1-U^{2}$ (and $U^{2}\neq1$), then
Eq. (\ref{eq:K_Taylor_series_1}) reduces to\[
c_{1}\left(1-U^{2}\right)-U=0\Longrightarrow c_{1}=\frac{U}{1-U^{2}}\,.\]
To first order in $\Omega$, then, we have\begin{eqnarray}
K^{v} & \approx & -\frac{1}{1-U_{R}}\Omega\,,\label{eq:K_v_first_order}\\
K^{ur} & \approx & \frac{1}{1+U_{R}}\Omega\,,\label{eq:K_ur_first_order}\\
K^{ul} & \approx & \left(1-U_{R}^{2}\right)^{1/2}+\frac{U_{R}}{1-U_{R}^{2}}\Omega\,,\label{eq:K_ul_first_order}\\
K^{u} & \approx & -\left(1-U_{R}^{2}\right)^{1/2}+\frac{U_{R}}{1-U_{R}^{2}}\Omega\,,\label{eq:K_u_first_order}\\
K^{c} & \approx & -i\left(U_{L}^{2}-1\right)^{1/2}-\frac{U_{L}}{U_{L}^{2}-1}\Omega\,.\label{eq:K_c_first_order}\end{eqnarray}
Note that, in Eq. (\ref{eq:K_c_first_order}), we have taken the imaginary
part of $K^{c}$ as negative in order to make it a decreasing exponential
for negative $X$.

From Eq. (\ref{eq:coefficients_vector_soln}), we see that\begin{multline*}
\frac{a^{u}}{a^{ur}}=-\frac{\left(K^{v}-K^{ur}\right)\left(K^{ul}-K^{ur}\right)}{\left(K^{v}-K^{u}\right)\left(K^{ul}-K^{u}\right)}\\
\times\frac{\left(K^{v}-K^{c}\right)\left(K^{ur}-K^{c}\right)\left(K^{ul}-K^{c}\right)-\left(U_{R}-U_{L}\right)\left(\Omega-K^{c}\left(U_{R}+U_{L}\right)\right)}{\left(K^{v}-K^{c}\right)\left(K^{ul}-K^{c}\right)\left(K^{u}-K^{c}\right)-\left(U_{R}-U_{L}\right)\left(\Omega-K^{c}\left(U_{R}+U_{L}\right)\right)}\end{multline*}
Plugging in the expansions of Eqs. (\ref{eq:K_v_first_order})-(\ref{eq:K_c_first_order}),
we find that, to lowest order, $a^{u}/a^{ur}$ is a constant,\[
\frac{a^{u}}{a^{ur}}\approx-\frac{\left(U_{L}^{2}-1\right)^{1/2}\left(\left(U_{L}^{2}-1\right)^{1/2}-i\left(1-U_{R}^{2}\right)^{1/2}\right)}{\left(1-U_{R}\right)\left(1-U_{L}\right)\left(U_{R}-U_{L}\right)}\,,\]
and, therefore,\begin{equation}
\left|\frac{a^{u}}{a^{ur}}\right|^{2}\approx\frac{\left(U_{L}^{2}-1\right)\left(\left(U_{L}^{2}-1\right)+\left(1-U_{R}^{2}\right)\right)}{\left(1-U_{R}\right)^{2}\left(1-U_{L}\right)^{2}\left(U_{R}-U_{L}\right)^{2}}=\frac{\left(1+U_{L}\right)\left(U_{R}+U_{L}\right)}{\left(1-U_{R}\right)^{2}\left(1-U_{L}\right)\left(U_{R}-U_{L}\right)}\,.\label{eq:coefficient_ratio}\end{equation}
We must multiply this by $\left(\left|f\left(K^{u}\right)\right|\left|v_{g}\left(K^{u}\right)\right|\right)/\left(\left|f\left(K^{ur}\right)\right|\left|v_{g}\left(K^{ur}\right)\right|\right)$.
Now, $f\left(K\right)=K\left(1-K^{2}\right)^{1/2}=\Omega-U_{R}K$
and $v_{g}\left(K\right)=U_{R}+f^{\prime}\left(K\right)=U_{R}+\left(1-2K^{2}\right)/\left(1-K^{2}\right)^{1/2}$.
Therefore,\[
f\left(K\right)v_{g}\left(K\right)=U_{R}\left(\Omega-U_{R}K\right)+K\left(1-2K^{2}\right)=U_{R}\Omega+\left(1-U_{R}^{2}\right)K-2K^{3}\,,\]
and so\[
\frac{f\left(K^{u}\right)v_{g}\left(K^{u}\right)}{f\left(K^{ur}\right)v_{g}\left(K^{ur}\right)}=\frac{U_{R}\Omega+\left(1-U_{R}^{2}\right)K^{u}-2\left(K^{u}\right)^{3}}{U_{R}\Omega+\left(1-U_{R}^{2}\right)K^{ur}-2\left(K^{ur}\right)^{3}}\,.\]
Pluggin in the expansions of Eqs. (\ref{eq:K_ur_first_order}) and
(\ref{eq:K_u_first_order}), we find that, to lowest order,\begin{equation}
\frac{f\left(K^{u}\right)v_{g}\left(K^{u}\right)}{f\left(K^{ur}\right)v_{g}\left(K^{ur}\right)}\approx\frac{\left(1-U_{R}^{2}\right)^{3/2}}{\Omega}\,.\label{eq:Bogoliubov_prefactor}\end{equation}
Combining Eqs. (\ref{eq:coefficient_ratio}) and (\ref{eq:Bogoliubov_prefactor}),\begin{equation}
\left|\beta_{\Omega}\right|^{2}\approx\left(1-U_{R}^{2}\right)^{1/2}\frac{\left(1+U_{R}\right)\left(1+U_{L}\right)\left(U_{R}+U_{L}\right)}{\left(1-U_{R}\right)\left(1-U_{L}\right)\left(U_{R}-U_{L}\right)}\frac{1}{\Omega}\,,\end{equation}
whence we may read off the low-frequency temperature, in accordance
with Eq. (\ref{eq:limiting_temperature}):\begin{equation}
T_{\infty}=\left(1-U_{R}^{2}\right)^{1/2}\frac{\left(1+U_{R}\right)\left(1+U_{L}\right)\left(U_{R}+U_{L}\right)}{\left(1-U_{R}\right)\left(1-U_{L}\right)\left(U_{R}-U_{L}\right)}\,.\end{equation}

\pagebreak{}

\chapter{Low-Frequency Temperature in Optical Model\label{sec:Appendix_Optical}}

We now perform an analogous derivation for the optical model. We assume
a discontinuous nonlinearity profile with quartic subluminal dispersion;
the normalized dispersion relation is given by\begin{equation}
B^{2}\left(\Omega\right)=\Omega^{2}+\Omega^{4}\end{equation}
and the nonlinearity profile by\begin{equation}
\chi\left(T\right)=\chi_{L}\,\theta\left(-T\right)+\chi_{R}\,\theta\left(T\right)\,.\end{equation}

\begin{figure}
\includegraphics[width=0.8\columnwidth]{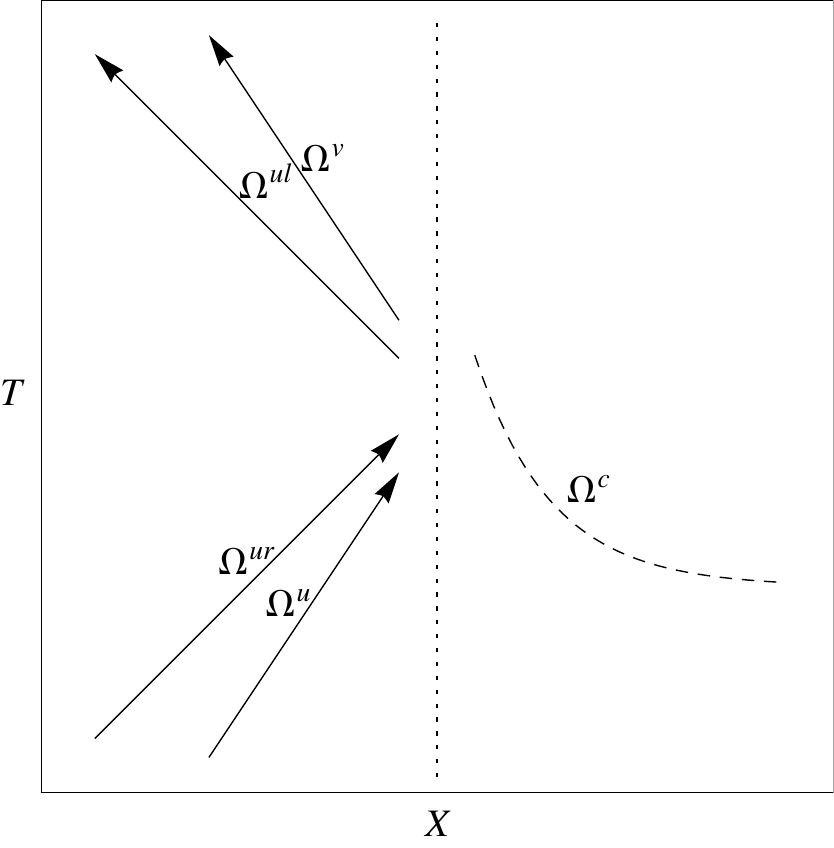}

\caption[\textsc{Space-time diagram of the $\Omega^{ul}$-out mode}]{\textsc{Space-time diagram of the $\Omega^{ul}$-out mode}: This has a single decreasing
exponential on the right-hand side, and is also the mode found by
numerical integration when the steepness of the nonlinearity is not
assumed infinite. (Note that the $v$-mode is actually ingoing, as was pointed out in Figure \ref{fig:co-moving-frame}.)\label{fig:Omega_ul_out_mode}}

\end{figure}

The mode to be solved is the $\Omega^{ul}$-out mode, shown in Figure \ref{fig:Omega_ul_out_mode}, which has a single
decreasing exponential for positive $T$. Ignoring the $v$-mode contribution, the only possible pairing is $u-url$,
and since we have a black hole horizon, $\Omega^{ul}$ is an outgoing
frequency. The creation rate is found via the norm of the ingoing
negative-norm wave with frequency $\Omega^{u}$.

Given the solutions $\Omega$ of the dispersion relation, the coefficients
of the various plane waves are found by imposing the conditions that
$A$, $\partial_{T}A$ and $\partial_{T}^{2}A$ be continuous at $T=0$,
and that $\partial_{T}^{3}A$ has a discontinuity given by Eq. (\ref{eq:discontinuity_in_2n-1_A_derivative}):\begin{equation}
\Delta\left(\partial_{T}^{3}A\right)=\left(\chi_{R}-\chi_{L}\right)\partial_{T}A\,.\end{equation}
$A$ is expressed as a sum of the possible plane waves in each region:\begin{equation}
A\left(T\right)=\begin{cases}
a^{v}e^{-i\Omega^{v}T}+a^{ul}e^{-i\Omega^{ul}T}+a^{ur}e^{-i\Omega^{ur}T}+a^{u}e^{-i\Omega^{u}T}\,, & T<0\\
e^{-i\Omega^{c}T}\,, & T>0\end{cases}\,.\end{equation}
The conditions at $T=0$ can be expressed in matrix form as follows:\begin{equation}
\left[\begin{array}{cccc}
1 & 1 & 1 & 1\\
\Omega^{v} & \Omega^{ul} & \Omega^{ur} & \Omega^{u}\\
\left(\Omega^{v}\right)^{2} & \left(\Omega^{ul}\right)^{2} & \left(\Omega^{ur}\right)^{2} & \left(\Omega^{u}\right)^{2}\\
\left(\Omega^{v}\right)^{3} & \left(\Omega^{ul}\right)^{3} & \left(\Omega^{ur}\right)^{3} & \left(\Omega^{u}\right)^{3}\end{array}\right]\left[\begin{array}{c}
a^{v}\\
a^{ul}\\
a^{ur}\\
a^{u}\end{array}\right]=\left[\begin{array}{c}
1\\
\Omega^{c}\\
\left(\Omega^{c}\right)^{2}\\
\left(\Omega^{c}\right)^{3}+\Omega^{c}\left(\chi_{R}-\chi_{L}\right)\end{array}\right]\,.\end{equation}
This has the solution\begin{equation}
\left[\begin{array}{c}
a^{v}\\
a^{ul}\\
a^{ur}\\
a^{u}\end{array}\right]=\left[\begin{array}{c}
\frac{\left(\Omega^{ul}-\Omega^{c}\right)\left(\Omega^{ur}-\Omega^{c}\right)\left(\Omega^{u}-\Omega^{c}\right)-\Omega^{c}\left(\chi_{R}-\chi_{L}\right)}{\left(\Omega^{ul}-\Omega^{v}\right)\left(\Omega^{ur}-\Omega^{v}\right)\left(\Omega^{u}-\Omega^{v}\right)}\\
\frac{\left(\Omega^{v}-\Omega^{c}\right)\left(\Omega^{ur}-\Omega^{c}\right)\left(\Omega^{u}-\Omega^{c}\right)-\Omega^{c}\left(\chi_{R}-\chi_{L}\right)}{\left(\Omega^{v}-\Omega^{ul}\right)\left(\Omega^{ur}-\Omega^{ul}\right)\left(\Omega^{u}-\Omega^{ul}\right)}\\
\frac{\left(\Omega^{v}-\Omega^{c}\right)\left(\Omega^{ul}-\Omega^{c}\right)\left(\Omega^{u}-\Omega^{c}\right)-\Omega^{c}\left(\chi_{R}-\chi_{L}\right)}{\left(\Omega^{v}-\Omega^{ur}\right)\left(\Omega^{ul}-\Omega^{ur}\right)\left(\Omega^{u}-\Omega^{ur}\right)}\\
\frac{\left(\Omega^{v}-\Omega^{c}\right)\left(\Omega^{ul}-\Omega^{c}\right)\left(\Omega^{ur}-\Omega^{c}\right)-\Omega^{c}\left(\chi_{R}-\chi_{L}\right)}{\left(\Omega^{v}-\Omega^{u}\right)\left(\Omega^{ul}-\Omega^{u}\right)\left(\Omega^{ur}-\Omega^{u}\right)}\end{array}\right]\,.\label{eq:coefficients_vector_soln_optical}\end{equation}
The Bogoliubov coefficient that describes particle creation is given
by\[
\left|\beta_{\Omega^{\prime}}\right|^{2}=\frac{\left|B\left(\Omega^{u}\right)\right|\left|v_{g}\left(\Omega^{u}\right)\right|\left|a^{u}\right|^{2}}{\left|B\left(\Omega^{ul}\right)\right|\left|v_{g}\left(\Omega^{ul}\right)\right|\left|a^{ul}\right|^{2}}\,.\]
At low co-moving frequencies $\Omega^{\prime}$, this should be given
approximately by $T_{\infty}/\Omega^{\prime}$, where $T_{\infty}$
is the low-frequency temperature. (Again, the subscript indicates
that the steepness of the nonlinearity profile has become infinite.)
To find this, we need to express the quantities involved - namely,
the frequencies $\Omega^{v}$, $\Omega^{ul}$, $\Omega^{ur}$, $\Omega^{u}$
and $\Omega^{c}$ - to lowest order in $\Omega^{\prime}$.

For a given co-moving frequency $\Omega^{\prime}$ and constant nonlinearity
$\chi$, the possible lab frequencies satisfy the dispersion relation\[
n_{g}^{2}\left(\Omega^{\prime}-\Omega\right)^{2}=B^{2}\left(\Omega\right)+\chi\Omega^{2}=\left(1+\chi\right)\Omega^{2}+\Omega^{4}\,,\]
or, rearranging,\begin{equation}
\Omega^{4}-\left(n_{g}^{2}-1-\chi\right)\Omega^{2}+2n_{g}^{2}\Omega^{\prime}\Omega-n_{g}^{2}\left(\Omega^{\prime}\right)^{2}=0\,.\label{eq:dispersion_discontinuous_chi}\end{equation}
We express $\Omega$ as a Taylor series in $\Omega^{\prime}$,\[
\Omega=c_{0}+c_{1}\Omega^{\prime}+c_{2}\left(\Omega^{\prime}\right)^{2}+\ldots\,,\]
and plugging into Eq. (\ref{eq:dispersion_discontinuous_chi}) we
find\begin{multline*}
c_{0}^{2}\left(c_{0}^{2}-\left(n_{g}^{2}-1-\chi\right)\right)+2c_{0}\left(c_{1}\left(2c_{0}^{2}-\left(n_{g}^{2}-1-\chi\right)\right)+n_{g}^{2}\right)\Omega\\
+\left(2c_{0}^{2}\left(2c_{0}c_{2}+3c_{1}^{2}\right)-\left(2c_{0}c_{2}+c_{1}^{2}\right)\left(n_{g}^{2}-1-\chi\right)+2n_{g}^{2}c_{1}-n_{g}^{2}\right)\Omega^{2}+\ldots=0\,.\end{multline*}
For this to be identically zero, the coefficient of each power of
$\Omega$ must be equal to zero. Therefore, we have\begin{eqnarray}
c_{0}^{2}\left(c_{0}^{2}-\left(n_{g}^{2}-1-\chi\right)\right) & = & 0\,,\label{eq:Omega_Taylor_series_0}\\
2c_{0}\left(c_{1}\left(2c_{0}^{2}-\left(n_{g}^{2}-1-\chi\right)\right)+n_{g}^{2}\right) & = & 0\,,\label{eq:Omega_Taylor_series_1}\\
\left(2c_{0}^{2}\left(2c_{0}c_{2}+3c_{1}^{2}\right)-\left(2c_{0}c_{2}+c_{1}^{2}\right)\left(n_{g}^{2}-1-\chi\right)+2n_{g}^{2}c_{1}-n_{g}^{2}\right) & = & 0\,,\label{eq:Omega_Taylor_series_2}\end{eqnarray}
and so on. From Eq. (\ref{eq:Omega_Taylor_series_0}), we find $c_{0}=0$
or $c_{0}=\pm\left(n_{g}^{2}-1-\chi\right)^{1/2}$. (If $\chi>n_{g}^{2}-1$,
then $c_{0}=\pm i\left(\chi+1-n_{g}^{2}\right)^{1/2}$.) If $c_{0}=0$,
then Eq. (\ref{eq:Omega_Taylor_series_1}) is trivial, and Eq. (\ref{eq:Omega_Taylor_series_2})
becomes\[
\left(n_{g}^{2}-1-\chi\right)c_{1}^{2}-2n_{g}^{2}c_{1}+n_{g}^{2}=0\,.\]
This is easily solved to give\[
c_{1}=\frac{1}{2\left(n_{g}^{2}-1-\chi\right)}\left(2n_{g}^{2}\pm\sqrt{4n_{g}^{4}-4n_{g}^{2}\left(n_{g}^{2}-1-\chi\right)}\right)=\frac{n_{g}\left(n_{g}\pm\sqrt{1+\chi}\right)}{n_{g}^{2}-1-\chi}=\frac{n_{g}}{n_{g}\mp\sqrt{1+\chi}}\,.\]
On the other hand, if $c_{0}^{2}=n_{g}^{2}-1-\chi$, then Eq. (\ref{eq:Omega_Taylor_series_1})
becomes\[
c_{1}\left(n_{g}^{2}-1-\chi\right)+n_{g}^{2}=0\Longrightarrow c_{1}=-\frac{n_{g}^{2}}{n_{g}^{2}-1-\chi}\,.\]
Therefore, to first order in $\Omega^{\prime}$, we have\begin{eqnarray}
\Omega^{v} & \approx & \frac{n_{g}}{n_{g}+\sqrt{1+\chi_{L}}}\Omega^{\prime}\,,\label{eq:Omega_v_first_order}\\
\Omega^{ul} & \approx & \frac{n_{g}}{n_{g}-\sqrt{1+\chi_{L}}}\Omega^{\prime}\,,\label{eq:Omega_ul_first_order}\\
\Omega^{ur} & \approx & \left(n_{g}^{2}-1-\chi_{L}\right)^{1/2}-\frac{n_{g}^{2}}{n_{g}^{2}-1-\chi_{L}}\Omega^{\prime}\,,\label{eq:Omega_ur_first_order}\\
\Omega^{u} & \approx & -\left(n_{g}^{2}-1-\chi_{L}\right)^{1/2}-\frac{n_{g}^{2}}{n_{g}^{2}-1-\chi_{L}}\Omega^{\prime}\,,\label{eq:Omega_u_first_order}\\
\Omega^{c} & \approx & -i\left(\chi_{R}+1-n_{g}^{2}\right)^{1/2}+\frac{n_{g}^{2}}{\chi_{R}+1-n_{g}^{2}}\Omega^{\prime}\,.\label{eq:Omega_c_first_order}\end{eqnarray}
As before, we have given $\Omega^{c}$ a negative imaginary part so
that it represents a decreasing exponential.

From Eq. (\ref{eq:coefficients_vector_soln_optical}) we find\[
\frac{a^{u}}{a^{ul}}=-\frac{\left(\Omega^{v}-\Omega^{ul}\right)\left(\Omega^{ur}-\Omega^{ul}\right)}{\left(\Omega^{v}-\Omega^{u}\right)\left(\Omega^{ur}-\Omega^{u}\right)}\cdot\frac{\left(\Omega^{v}-\Omega^{c}\right)\left(\Omega^{ul}-\Omega^{c}\right)\left(\Omega^{ur}-\Omega^{c}\right)-\Omega^{c}\left(\chi_{R}-\chi_{L}\right)}{\left(\Omega^{v}-\Omega^{c}\right)\left(\Omega^{ur}-\Omega^{c}\right)\left(\Omega^{u}-\Omega^{c}\right)-\Omega^{c}\left(\chi_{R}-\chi_{L}\right)}\,,\]
and, substituting Eqs. (\ref{eq:Omega_v_first_order})-(\ref{eq:Omega_c_first_order})
in this expression, we find that, to lowest order, $a^{u}/a^{ul}$
has the constant value\[
\frac{a^{u}}{a^{ul}}\approx-\frac{\sqrt{1+\chi_{L}}\left(\chi_{R}+1-n_{g}^{2}\right)^{1/2}}{\left(\chi_{R}-\chi_{L}\right)\left(n_{g}+\sqrt{1+\chi_{L}}\right)}\left\{ \left(\chi_{R}+1-n_{g}^{2}\right)^{1/2}-i\left(n_{g}^{2}-1-\chi_{L}\right)^{1/2}\right\} \,;\]
therefore,\begin{eqnarray}
\left|\frac{a^{u}}{a^{ul}}\right|^{2} & \approx & \frac{\left(1+\chi_{L}\right)\left(\chi_{R}+1-n_{g}^{2}\right)}{\left(\chi_{R}-\chi_{L}\right)^{2}\left(n_{g}+\sqrt{1+\chi_{L}}\right)^{2}}\left\{ \left(\chi_{R}+1-n_{g}^{2}\right)+\left(n_{g}^{2}-1-\chi_{L}\right)\right\} \nonumber \\
 & = & \frac{\left(1+\chi_{L}\right)\left(\chi_{R}+1-n_{g}^{2}\right)}{\left(\chi_{R}-\chi_{L}\right)\left(n_{g}+\sqrt{1+\chi_{L}}\right)^{2}}\,.\label{eq:coefficient_ratio_optical}\end{eqnarray}
We now turn to the factor $\left(\left|B\left(\Omega^{u}\right)\right|\left|v_{g}\left(\Omega^{u}\right)\right|\right)/\left(\left|B\left(\Omega^{ul}\right)\right|\left|v_{g}\left(\Omega^{ul}\right)\right|\right)$.
Using $B\left(\Omega\right)=\Omega\sqrt{1+\chi+\Omega^{2}}=n_{g}\left(\Omega-\Omega^{\prime}\right)$
and \\ $v_{g}\left(\Omega\right)=-1+B^{\prime}\left(\Omega\right)/n_{g}=-1+\left(1+\chi+2\Omega^{2}\right)/\sqrt{1+\chi+\Omega^{2}}/n_{g}$,
we have\[
n_{g}B\left(\Omega\right)v_{g}\left(\Omega\right)=-n_{g}^{2}\left(\Omega-\Omega^{\prime}\right)+\Omega\left(1+\chi+2\Omega^{2}\right)=n_{g}^{2}\Omega^{\prime}-\left(n_{g}^{2}-1-\chi\right)\Omega+2\Omega^{3}\,,\]
and so\[
\frac{B\left(\Omega^{u}\right)v_{g}\left(\Omega^{u}\right)}{B\left(\Omega^{ul}\right)v_{g}\left(\Omega^{ul}\right)}=\frac{n_{g}^{2}\Omega^{\prime}-\left(n_{g}^{2}-1-\chi_{L}\right)\Omega^{u}+2\left(\Omega^{u}\right)^{3}}{n_{g}^{2}\Omega^{\prime}-\left(n_{g}^{2}-1-\chi_{L}\right)\Omega^{ul}+2\left(\Omega^{ul}\right)^{3}}\,.\]
Substituting the first-order Taylor expansions of Eqs. (\ref{eq:Omega_ul_first_order})
and (\ref{eq:Omega_u_first_order}), we find, to lowest order,\begin{equation}
\frac{B\left(\Omega^{u}\right)v_{g}\left(\Omega^{u}\right)}{B\left(\Omega^{ul}\right)v_{g}\left(\Omega^{ul}\right)}\approx\frac{\left(n_{g}^{2}-1-\chi_{L}\right)^{3/2}}{n_{g}\sqrt{1+\chi_{L}}}\frac{1}{\Omega^{\prime}}\,.\label{eq:Bogoliubov_prefactor_optical}\end{equation}
Combining Eqs. (\ref{eq:coefficient_ratio_optical}) and (\ref{eq:Bogoliubov_prefactor_optical}),
we find\begin{equation}
\left|\beta_{\Omega^{\prime}}\right|^{2}\approx\left(n_{g}^{2}-1-\chi_{L}\right)^{1/2}\frac{\sqrt{1+\chi_{L}}\left(\chi_{R}+1-n_{g}^{2}\right)\left(n_{g}-\sqrt{1+\chi_{L}}\right)}{n_{g}\left(\chi_{R}-\chi_{L}\right)\left(n_{g}+\sqrt{1+\chi_{L}}\right)}\frac{1}{\Omega^{\prime}}\,,\end{equation}
from which we read off the temperature\begin{equation}
T_{\infty}=\left(n_{g}^{2}-1-\chi_{L}\right)^{1/2}\frac{\sqrt{1+\chi_{L}}\left(\chi_{R}+1-n_{g}^{2}\right)\left(n_{g}-\sqrt{1+\chi_{L}}\right)}{n_{g}\left(\chi_{R}-\chi_{L}\right)\left(n_{g}+\sqrt{1+\chi_{L}}\right)}\,.\end{equation}
\bibliographystyle{cj}
\bibliography{references}

\end{document}